\documentclass[a4paper,10pt,twoside,bibtotoc,pointlessnumbers,headsepline,openright,cleardoubleplain,smallheadings,abstracton,DIV15,BCOR0.8cm]{scrreprt}

  \usepackage{amsmath}                  
  \usepackage{amssymb}
  \usepackage{physics}
  \usepackage[ansinew]{inputenc}        
  \usepackage{float}                    
  \usepackage{graphicx}                 
  \usepackage[markuppercase]{scrpage2}  
  \usepackage{array}                    
  \usepackage{sublabel}                 
  \usepackage{epsfig}                   
  \usepackage{longtable}                
  \usepackage{afterpage}                
  \usepackage{hhline}                   
  \usepackage{rotating}                 
 \usepackage{./acronym-chris} 
  \usepackage[]{nomencl}                
  \usepackage{url}                      
  \usepackage{paralist}
  \usepackage{stmaryrd}
  \usepackage[format=plain]{caption}
  \usepackage{textcomp}

\usepackage{epstopdf}
\usepackage[english, german]{babel}
\usepackage{longtable}
\usepackage{lscape}
\usepackage{multirow}
\usepackage[noadjust]{cite}
\usepackage{graphicx}
\usepackage{hyperref} 

  \setkomafont{captionlabel}{\sffamily\bfseries}
  \setkomafont{caption}{\normalfont\small}
  \setkomafont{pagehead}{\scshape\itshape\footnotesize}

  \ohead{\pagemark}                   
  \ofoot[\pagemark]{}                 

  \renewcommand{\arraystretch}{1}               

  \renewcommand{\vec}[1]{\mbox{\boldmath$#1$}}  
  \newcommand{\unitvec}[1]{\hat{\vec{#1}}}      

  \def\eref#1{(\ref{#1})}
  \allowdisplaybreaks
  
  \graphicspath{{figures/}}

  \newdimen\intdimi\intdimi=-8.3pt
  \newdimen\intdimii\intdimii=-6.2pt
  \newdimen\intdimiii\intdimiii=-16.3pt
  \def\Oiint{\mathop{\int\kern\intdimi\int\kern\intdimiii\subset\kern\intdimii\supset}}

  \makeglossary

\begin{document}

\selectlanguage{english}
    \ofoot[]{}

\newlength{\titlepagebindingoffset}
\setlength{\titlepagebindingoffset}{0.5cm}
%
%
%
%
%
%
\begin{titlepage}
	
	\sffamily
	
	\begin{center}
		
		\vspace*{2cm}   {\bfseries\LARGE TERAHERTZ ACCELERATION TECHNOLOGY TOWARDS COMPACT LIGHT SOURCES \\[.5ex] }
		
		\vspace{2cm}    {\bfseries\large Schrift \\ }
		{\bfseries\large zur Erlangung der Habilitation \\ }
		{\bfseries\large am Fachbereich Physik \\ }
		{\bfseries\large der Fakult\"{a}t f\"{u}r Mathematik, Informatik und Naturwissenschaften \\ }
		{\bfseries\large der Universit\"{a}t Hamburg \\ }
		
		\vspace{3cm}    {\bfseries\large vorgelegt von \\}
		\vspace{0.5cm}  {\bfseries\Large Arya Fallahi \\}
		
		\vspace{3cm}    {\bfseries\large Hamburg \\ }
		\vspace{0.5cm}  {\bfseries\large April 2018 \\ }
		
	\end{center}
	
\end{titlepage}

\cleardoublepage

    \cleardoublepage                           
    \setheadsepline[\textwidth]{0pt}           
    
    \newpage
    \sffamily
    
    \begin{center}
    	
    	\vspace*{2cm}   {\bfseries\LARGE TERAHERTZ ACCELERATION TECHNOLOGY TOWARDS COMPACT LIGHT SOURCES \\[.5ex] }
    	
    	\vspace{2cm}    {\bfseries\large Schrift \\ }
    	{\bfseries\large zur Erlangung der Habilitation \\ }
    	{\bfseries\large am Fachbereich Physik \\ }
    	{\bfseries\large der Fakult\"{a}t f\"{u}r Mathematik, Informatik und Naturwissenschaften \\ }
    	{\bfseries\large der Universit\"{a}t Hamburg \\ }
    	
    	\vspace{3cm}    {\bfseries\large vorgelegt von \\}
    	\vspace{0.5cm}  {\bfseries\Large Arya Fallahi \\}
    	
    	\vspace{3cm}    {\bfseries\large Hamburg \\ }
    	\vspace{0.5cm}  {\bfseries\large April 2018 \\ }
    	
    \end{center}
    
    \rmfamily
    
    \setheadsepline[\textwidth]{0.5pt}         

  \ihead{ACKNOWLEDGMENTS}
  
  \chapter*{Acknowledgments}
  
  The research studies presented in this habilitation treatment are mainly carried out in the group of ultrafast optics and X-ray sources at the DESY-Center for Free-Electron Laser Science (CFEL) in Hamburg Germany.
  Some parts of the research is carried out in collaboration with colleagues in Research Laboratory of electronics (RLE) at Massachusetts Institute of Technology (MIT) in Boston, USA.
  Evidently, in accomplishment of the presented studies and achievement of the project goals, numerous coauthors from both institutes were involved.
  I would like to avail the opportunity to express my gratitude to every single of these coauthors who contributed to the works presented in this habilitation treatment.
  
  First and foremost, I would like to thank Prof. Franz K\"{a}rtner for his continuous support, brilliant ideas, and guidance throughout this research.
  He provided me and the whole group with invaluable advices and scientific intuition whenever needed.
  From my discussions with him, I earned a priceless insight into details of laser-based experiment and leadership techniques, which helped me a lot to find my way and will certainly be helpful for me in the future.
  I specially appreciate the freedom he gave me to work on my topics of interest, and am highly indebted to his patient attitude during the frustrating moments of the projects.
  
  I am very much grateful to the group leaders, who supported my collaborations in the past years, including Dr. Oliver D. M\"{u}cke, Prof. Dwayne Miller, Prof. Jochen K\"{u}pper, and Prof. Henry Chapmann from CFEL, Prof. Petra Fromme from university of Arizona, Dr. Ralph Assmann from DESY, and Prof. Karl Berggren, Dr. Kyung-Han Hong, and Dr. Luis Fernando Vel\'{a}squez-Garc\'{\i}a from MIT.
  The interesting collaboration on structured photocathodes could not be possible without the dedicated efforts from Dr. Donnie Keathly, Dr. Richard Hobbs, Dr. Michael Swanwick, and Dr. Yujia Yang from MIT.
  The dedicated times from Hong Ye and her industrious attitude was the main element resulting in the realization of velocity-map-imaging instrument.
  I am very much indebted to the collaborations with Dr. Alireza Yahaghi, Dr. Moein Fakhari, and Miguel Arrieta in development of the concepts for ultrafast electron guns.
  Many thanks here to Dr. Ronny Huang for his elaborate efforts to realize the first THz electron gun on the planet.
  I think I will have hard times finding a better collaboration than the one I had with Dr. Dongfang Zhang and Dr. Nicholas Matlis, leading to the realization of the STEAM device.
  Here, I express my deepest and warmest gratitude to both of them.
  The attempts on realizing the THz gun test stand at CFEL benefited very much from the focused studies done by Dr. Grygorii Vaschenko, whom I would like to thank here as well.
  I would like to thank Dr. Liang Jie Wong for the spectacular collaboration on the theory of THz linacs and useful comments on the introduction of this thesis.
  Additionally, I am grateful to Dr. Emilio Nanni for his wonderful first demonstration of THz acceleration.
  Throughout the simulation efforts in this thesis, the software ASTRA was extensively used.
  In this regard, the responsive reaction and quick replies from Dr. Klaus Floettmann to my emails and questions are very deeply appreciated.
  
  Many of the proof-of-principle experiments discussed in this habilitation thesis demanded support from several individuals whom I need to deeply thank here.
  They include the group of scientists working on THz generation schemes in CFEL, including Dr. Koustuban Ravi, Dr. Sergio Carbajo, Dr. Xiaojun Wu, Lu Wang, Dr. Frederike Ahr, and Dr. Spencer Jolly, the group of scientists working on the development of high power lasers for THz generation including Dr. H\"{u}seyin Cankaya, Dr. Michael Hemmer, Dr. Giovanni Cirmi, Dr. Anne-Laure Calendron and Dr. Luis Zapata, the IT support group at CFEL and DESY including Dr. Steve Aplin and Dr. Frank Schleuzen, as well as the team of engineers Thomas Tilp and Matthias Schust.
  Eventually and most importantly, I am greatly in debt to the support, and the endlessly kind helps concerning administrative issues from Mrs. Christine Berber.
  
  I am particularly indebted to my previous two supervisors Prof. Christian Hafner from ETH Zurich and Prof. Mahmoud Shahabadi from University of Tehran, from whom I learned the fundamentals of electromagnetics which was and will be extremely helpful for me.
  I also want to thank my Iranian friends in Hamburg and many others for all the fun times we shared together in the past years.
  My extra special thanks go out to my parents, for their generous help, support, and exciting visits in Hamburg.
  At the end, I like to enclose this chapter of my life with endless gratitude to my wife Dr. Simin Jafarbegloo for her unfailing love and encouraging support during the recent years. 
  
  \cleardoublepage                           

  \renewcommand{\thepage}{}
  \vspace*{6cm}
   {\sffamily\bfseries\large\
   \begin{center}
       The roots of education are bitter, but the fruit is sweet. \\[2ex]\normalsize
       Aristotle
   \end{center}
   }
  \cleardoublepage

    \ohead{\pagemark}                   
    \ofoot[\pagemark]{}                 


  \ihead{ABSTRACT}                      
 \renewcommand{\thepage}{\roman{page}} 
  \setcounter{page}{1}                  
  
  \chapter*{Abstract}
  
  Novel compact accelerating structures are highly favorable compared to conventional devices for studies where high-energy bunches in small but applicable charges are required.
  Increasing the operation frequency and shrinking the accelerating devices is a suitable path for improving the accelerators' performance.
  For this purpose, energy transfer to electrons can be realized in shorter distances, which in turn means introducing higher accelerating gradients.
  Moreover, higher accelerating gradients enable beams with higher quality due to lower emittance growth.
  However, increasing the operation frequency from RF to optical regimes introduces serious challenges in synchronization, stability and acceleration of considerable charge amount.
  Consequently, THz acceleration will likely serve as the optimal operation regime for compact accelerators.
  
  Despite the already-realized high power radiation sources enabling ultrahigh electric fields, increasing the acceleration gradients above the state-of-the-art values is hampered by the damage threshold of materials.
  Recent studies on damage mechanisms in accelerators have revealed the strong dependence of operation threshold on the time duration over which fields are influencing the device.
  Therefore, \emph{fast} accelerating principles based on short excitations need to be developed for further increasing the accelerating gradient.
  The main goal in this habilitation thesis is conceptual developments and proof-of-principle studies for THz acceleration using short pulses.
  The concepts developed here pave the way towards the realisation of cheap and compact particle accelerators with control of particles over ultrashort time scales.
  
  The concepts in this thesis comprise three groups focusing on: (1) fast electron sources, (2) THz injectors, and (3) THz linacs.
  First, the feasibility of ultrafast, high-yield electron emitters based on nanostructured cathodes is demonstrated.
  Benefitting from field enhancement effects, namely tip-enhancement and plasmonic enhancement, laser-induced field emission is realized over large, dense and highly uniform field emitter arrays.
  The theoretical principles of these field emitter arrays are studied and their suitability for pico-Coulomb charge production over femtosecond time-scales is confirmed.
  In the framework of THz injectors, two ground-breaking concepts including ultrafast single-cycle THz guns and segmented THz electron accelerator and manipulator (STEAM) devices are developed and tested.
  The possibility of using transient fields to realize ultrahigh acceleration gradients close to 0.5\,GeV/m is confirmed.
  Specifically, a STEAM device capable of performing multiple high-field operations on the 6D-phase-space of ultrashort electron bunches is demonstrated.
  With this single device, powered by few-micro-Joule, single-cycle, 0.3\,THz pulses, we demonstrated record THz-acceleration of $>30$\,keV, streaking with $<10$\,fs resolution, focusing with $>2$\,kT/m strength, compression to $\sim 100$\,fs as well as real-time switching between these modes of operation.
  Travelling wave THz linacs based on dielectric-loaded metallic waveguides operating under few-cycle excitations are proposed.
  Based on this concept, keV-level energy gain through a linear accelerator using optically-generated THz pulses is demonstrated.
  Moreover, the possibility of electron acceleration to tens of MeV with millijoule level THz pulses is theoretically shown.
  
  The final goal of the above studies is a fully THz-driven compact light source facility, whose start-to-end simulation is fulfilled in this thesis.
  The required THz pulses to excite the light source are categorized under single-cycle and multi-cycle pulses that are generated using laser-driven THz generation concepts.
  The single-cycle THz pulses feed an ultrafast electron gun, whose output is delivered to a THz linac fed by multi-cycle pulses.
  It is shown that 18\,MeV beam energy can be produced using two single-cycle THz beams with 400\,{\textmu}J and one 0.5\,ns 300\,GHz beam with 20\,mJ energy.
  This beam is then transported to an Inverse Compton Scattering (ICS) section, where the 18\,MeV electron beam scatters off a 100\,mJ 1\,{\textmu}m laser beam and generates an X-ray beam with 4\,keV central photon energy and 6$\times$10$^4$ photons per shot.
  
  \cleardoublepage
\selectlanguage{german}
  \ihead{ZUSAMMENFASSUNG}               
  
  \chapter*{Zusammenfassung}

  Neuartige kompakte Beschleunigerstrukturen werden gegen\"{u}ber konventionellen Instrumenten f\"{u}r Untersuchungen, bei denen hochenergetische Teilchen in kleinen, aber anwendbaren Gr\"{o}{\ss}en ben\"{o}tigt werden, stark bevorzugt.
  Es wird oft angenommen, dass eine Erh\"{o}hung der Betriebsfrequenz und das konsequente Schrumpfen der Dimensionen ein geeigneter Weg ist, um die Leistung der Beschleuniger zu verbessern.
  Zu diesem Zweck sollte eine ausreichende Energie\"{u}bertragung auf Elektronen in k\"{u}rzeren Abst\"{a}nden realisiert werden, was wiederum die Einf\"{u}hrung h\"{o}herer Beschleunigungsgradienten bedeutet.
  Zus\"{a}tzlich h\"{o}here Beschleunigungsgradienten erm\"{o}glichen aufgrund des geringeren Emittanzwachstums Elektronenstrahlen mit besserer Qualit\"{a}t.
  Die Erh\"{o}hung der Betriebsfrequenz in den optischen Bereich f\"{u}hrt jedoch zu ernsthaften Problemen in Synchronisation, Stabilit\"{a}t und Beschleunigung der anwendbaren Ladungsmenge.
  Demzufolge steht das optimale Betriebsregime f\"{u}r kompakte Beschleuniger wahrscheinlich im THz-Bereich.
  
  Trotz der Realisierung von Hochleistungsquellen, die hohe elektrische Felder erm\"{o}glichen, ist ein Steigen der Beschleunigungsgradienten oberhalb des Standes der Technik durch die Zerst\"{o}rschwelle von Materialien stark beschr\"{a}nkt.
  Neuere Studien zu Sch\"{a}digungsmechanismen in Beschleunigern haben aufgezeigt, dass die Zerst\"{o}rschwelle von der Zeitdauer, \"{u}ber die Felder das Material beeinflussen, stark abh\"{a}ngen.
  Deswegen m\"{u}ssen \emph{schnelle} Beschleunigungsprinzipien basierend auf kurzen Anregungen entwickelt werden, um die Beschleunigungsgradienten weiter zu erh\"{o}hen.
  Das Hauptziel dieser Habilitationsschrift ist die konzeptionelle Entwicklung und der Proof-of-principle Untersuchungen zur THz-Beschleunigung mit kurzen Impulsen.
  Wir glauben, dass die entwickelten Konzepte den Weg zur Realisierung von billigen und kompakten Teilchenbeschleunigern mit Partikelkontrolle \"{u}ber die ultrakurze Zeitbereiche erleichtern.
  
  Die Konzepte dieser Arbeit bestehen aus drei Gruppen: (1) schnelle Elektronenquellen, (2) THz Injektoren und (3) THz-Linacs.
  Erstens, die Machbarkeit ultraschneller, Hochausbeute-Emittern, die auf Nanostrukturierten Kathoden basieren, wird demonstriert.
  Profitiert von Feldverst\"{a}rkungseffekten, n\"{a}mlich Spitzen- und plasmonische Verst\"{a}rkung, wird laserinduzierte Feldemission \"{u}ber gro{\ss}e, dichte und sehr gleichm\"{a}{\ss}ige Feldemitter-Arrays dargestellt.
  Ihre theoretischen Prinzipien werden untersucht und die Eignung f\"{u}r Piko-Coulomb Ladungserzeugung \"{u}ber Femtosekunden-Zeitskalen wird best\"{a}tigt.
  Im Rahmen von THz-Injektoren, zwei innovative Konzepte wie die ultraschnelle Einzel-Zyklus THz Elektronenkanone und segmented THz electron accelerator and manipulator (STEAM) werden entwickelt und getestet.
  Es wird best\"{a}tigt, dass transiente Felder hohe Beschleungungsgradienten in der N\"{a}he von 0,5\,GeV/m erm\"{o}glichen.
  Insbesondere ist ein STEAM-Ger\"{a}t realisiert, das zu mehreren Hochfeldoperationen auf dem 6D-Phasenraum von ultrakurzen Elektronenpaketen f\"{a}hig ist.
  Mit diesem einzigen Ger\"{a}t, angetrieben von wenigen Mikro-Joule, Einzel-Zyklus, 0,3\,THz Strahlung, demonstrierten wir eine Rekord THz-Beschleunigung von $>30$\,keV, Streaking mit $<10$\,fs Aufl\"{o}sung, Fokussierung mit $>2$\,kT/m St\"{a}rke, Kompression auf $\sim 100$\,fs sowie Echtzeitumschaltung zwischen diesen Betriebsweisen.
  THz-Linacs basierend auf dielektrisch geladenen metallischen Wellenleitern, die mit wenigen Zyklen Anregungen funktionieren, werden vorgeschlagen.
  Basierend auf diesem Konzept, wird keV-Level Energiegewinn durch einen Linearbeschleuniger mit optisch erzeugten THz-Pulsen demonstriert.
  
  Das wertvolle Endziel ist die THz-gesteuerte Kompaktlichtquelle, deren durchg\"{a}ngige Simulation erf\"{u}llt ist.
  Die ben\"{o}tigten THz-Impulse zur Anregung der Lichtquelle werden in Einzel- und Mehrfach-Zyklen Strahlung eingeteilt, die mittels lasergesteuerter THz-Generierungskonzepte erzeugt werden.
  Die Einzel-Zyklus THz-Impulse speisen eine ultraschnelle Elektronenkanone, deren Ausgang an ein von mehrfach-Zyklen-Strahlung angeregten THz-Linac geliefert wird.
  Es wird gezeigt, dass 18\,MeV Elektronenenergie unter Verwendung von zwei 400\,{\textmu}J Einzel-Zyklus und ein 0,5\,ns 20\,mJ mehrfach-Zyklen THz Strahlung in 300\,GHz Zentralfrequenz erzeugt werden kann.
  Dieser Elektronenstrahl wird dann zu einer Inverse Compton Scattering (ICS) Sektion transportiert, in der die 18\,MeV Elektronen einen 100\,mJ 1\,{\textmu}m Laserstrahl treffen und R\"{o}ntgenstrahlung mit 4\,keV zentraler Photonenenergie und $6 \times 10^4$ Photonen pro Laserpuls erzeugen.
  
  \cleardoublepage
  \selectlanguage{english}

  \cleardoublepage
  \setcounter{tocdepth}{3}
  \ihead{CONTENTS}                                  
  \tableofcontents                                  
  \cleardoublepage

  \ihead{LIST OF FIGURES}
  \listoffigures
  \cleardoublepage

  \ihead{LIST OF TABLES}
  \listoftables
  \cleardoublepage

  \ihead{LIST OF ACRONYMS}
  
  \chapter*{List of Acronyms and Abbreviations}
  
  \vspace*{-0.2cm}
  \begin{acronym}[]
  	\acro{ABC}         {Absorbing Boundary Condition}
  	\acro{ACHIP}       {Accelerator on a Chip}
  	\acro{ADE}         {Auxiliary Differential Equations}
  	\acro{AGS}         {Alternating Gradient Synchrotron}
  	\acro{ASU}         {Arizona State University}
  	\acro{ATP}         {Above-Threshold Photoemission}
  	\acro{AXSIS}       {Frontiers in Attosecond X-ray Science: Imaging and Spectroscopy}
  	\acro{CFEL}        {Center for Free Electron Laser Science}
  	\acro{CLIC}        {Cern's Compact Linear Collider}
  	\acro{CLN}         {Congruent Lithium Niobate}
  	\acro{CMOS}        {Complementary Metal Oxide Semiconductor}
  	\acro{COM}         {Center of Mass}
  	\acro{CPS}         {CERN Proton Synchrotron}
  	\acro{CSE}         {Coherent Scattering Emission}
  	\acro{CW}          {Continuous-Wave}
  	\acro{DESY}        {Deutsche Elektronen Synchrotron}
  	\acro{DFG}         {Difference Frequency Generation}
  	\acro{DG}          {Discontinuous Galerkin}
  	\acro{DGTD}        {Discontinuous Galerkin Time Domain}
  	\acro{DLA}         {Dielectric Laser Accelerators}
  	\acro{DOS}         {Density of States}
  	\acro{EM}          {Electromagnetic}
  	\acro{EO}          {Electro-Optic}
  	\acro{ERC}         {European Research Council}
  	\acro{ESRF}        {European Synchrotron Radiation Facility}
  	\acro{FD}          {Finite Difference}
  	\acro{FDTD}        {Finite Difference Time Domain}
  	\acro{FEA}         {Field Emitter Array}
  	\acro{FEL}         {Free-Electron Laser}
  	\acro{FEM}         {Finite Element Method}
  	\acro{FMM}         {Fourier Modal Method}
  	\acro{FN}          {Fowler-Nordheim}
  	\acro{FWHM}        {Full-Wave Half-Maximum}
  	\acro{GE}          {General Electric}
  	\acro{GPT}         {General Particle Tracer}
  	\acro{HIGS}        {High Intensity Gamma-Ray Source}
  	\acro{HSQ}         {Hydrogen SilsesQuioxane}
  	\acro{HZDR}        {Hemlholtz Zentrum Dresden R\"{o}ssendorf}
  	\acro{ICS}         {Inverse Compton Scattering}
  	\acro{IFEL}        {Inverse Free-Electron Laser}
  	\acro{ILC}         {International Linear Collider}
  	\acro{IMFP}        {Inelastic Mean-Free Path}
  	\acro{ITO}         {Indium-doped Tin Oxide}
  	\acro{LASER}       {Light Amplification by Stimulated Emission of Radiation}
  	\acro{LBNL}        {Lawrence Berkeley National Laboratory}
  	\acro{LCLS}        {Linac Coherent Light Source}
  	\acro{LEED}        {Low-Energy Electron Diffraction}
  	\acro{LEP}         {Largest Electron-Positron collider}
  	\acro{LHC}         {Large Hadron Collider}
  	\acro{linac}       {Linear Accelerator}
  	\acro{LPA}         {Laser Plasma Acccleration}
  	\acro{LPWA}        {Laser-Plasma Wakefield Acceleration}
  	\acro{LSPR}        {Localized Surface Plasmon Resonance}
  	\acro{LSS}         {Laser Synchrotron Source}
  	\acro{MCP}         {Micro-Channel Plate}
  	\acro{MIT}         {Massachausset Institut of Technology}
  	\acro{MoM}         {Method of Moments}
  	\acro{NLC}         {Next Linear Collider}
  	\acro{NRF}         {Nuclear Resonance Fluorescence}
  	\acro{NRL}         {Naval Research Laboratory}
  	\acro{NSFD}        {Non-Standard Finite Difference}
  	\acro{OR}          {Optical Rectification}
  	\acro{PC}          {Photoconductive}
  	\acro{PEEM}        {Photo-Emission Electron Microscopy}
  	\acro{PIC}         {Particle In Cell}
  	\acro{PMMA}        {Poly(Methyl-MethAcrylate)}
  	\acro{PPLN}        {Periodically Poled Lithium Niobate}
  	\acro{PPWG}        {Parallel-Plate WaveGuide}
  	\acro{QE}          {Quantum Efficiency}
  	\acro{QPM}         {Quasi Phase-Matching}
  	\acro{QUTIF}       {Quantum Dynamics in Tailored Intense Fields}
  	\acro{RF}          {Radio-Frequency}
  	\acro{RFA}         {Retarding Field Analyzer}
  	\acro{RHIC}        {Relativistic Heavy Ion Collider}
  	\acro{RLE}         {Research Laboratory of Electronics}
  	\acro{RMS}         {Root Mean Square}
  	\acro{SASE}        {Self Amplified Spontaneous Emission}
  	\acro{SEM}         {Scanning Electron Microscope}
  	\acro{SLA}         {Stanford Linear Accelerator}
  	\acro{SLAC}        {Stanford Linear Accelerator Centre}
  	\acro{SLN}         {Stoichiometric Lithium Naiobate}
  	\acro{SMI}         {Spatial Map Imaging}
  	\acro{STEAM}       {Segmented Terahertz Electron Accelerator and Manipulator}
  	\acro{SwissFEL}    {Swiss Free-Electron Laser}
  	\acro{TE}          {Transverse Electric}
  	\acro{TF/SF}       {Total-Field/Scattered-Field}
  	\acro{THz}         {Terahertz}
  	\acro{TM}          {Transverse Magnetic}
  	\acro{TOF}         {Time of Flight}
  	\acro{TPF}         {Tilted-Pulse Front}
  	\acro{TS}          {Thomson Scattering}
  	\acro{TWT}         {Travelling Wave Tube}
  	\acro{UHV}         {Ultra-High Vacuum}
  	\acro{VMI}         {Velocity Map Imaging}
  	\acro{WKB}         {Wentzel-Kramers-Brillouin}
  	\acro{XFEL}        {X-ray Free Electron Laser}
  \end{acronym}
  
  \cleardoublepage


  \renewcommand{\thechapter}{\arabic{chapter}}  

  \automark[section]{chapter}                   
  \ohead[]{\pagemark}                           
  \rehead[]{\leftmark}                          
  \lohead[]{\rightmark}                         

  \cleardoublepage                            
  \setcounter{page}{1}                        
  \pagenumbering{arabic}                      
  
  \chapter{Introduction\label{chap:intro}}
  
  The work presented in this habilitation treatment is a collection of my research efforts and collaborations at the Division of Ultrafast Optics and x-ray Sources within the DESY-Center for Free Electron Laser Science (CFEL).
  The main funding institutes for the presented research were the Helmholtz center Deutsche Elektronen Synchrotron (DESY) and the European Research Council (ERC) under the European Union's Seventh Framework Programme (FP7/2007-2013) through the Synergy Grant ``Frontiers in Attosecond X-ray Science: Imaging and Spectroscopy" (AXSIS) (609920).
  Moreover, this research is also generously supported by the German excellence cluster ``The Hamburg Centre for Ultrafast Imaging - Structure, Dynamics and Control of Matter at the Atomic Scale" of the Deutsche Forschungsgemeinschaft.
  Parts of the research presented here were funded by the Massachausset Institut of Technology (MIT) and performed at the Research Laboratory of Electronics (RLE).
  In addition, helpful supports from the priority program ``Quantum Dynamics in Tailored Intense Fields" (QUTIF) (SPP1840 SOLSTICE) of the Deutsche Forschungsgemeinschaft and the Accelerator on a Chip program (ACHIP) funded by the Gordon and Betty Moore foundation are received.
  Some of the collaborators were funded by the George Foster Fellowship of the Alexander von Humboldt Foundation.
  
  The main focus of this work as apparent from the title is terahertz acceleration, i.e. electron acceleration using terahertz pulses.
  The idea of terahertz acceleration were mainly inspired after the ground-breaking development of high power terahertz sources based on the interaction of high power lasers with nonlinear crystals.
  The available radiations from these sources are short pulses in form of 1-100 radiation cycles, and are fundamentally different from sources used in conventional acceleration technology.
  Using such pulses for particle acceleration requires development of new concepts, simulation tools and proof-of-principle studies, where broadband pulses are used for efficient acceleration.
  Therefore, the presented research is dominated by theoretical and computational studies followed by proof-of-principle experiments confirming the theoretical predictions.
  The developed concepts in this research can be generalized to other frequency regimes than terahertz frequency band.
  Thus, the structures can be considered generally as \emph{fast accelerators} since their operational principles are different from conventional structures.
  The main motivation for studying terahertz (THz) acceleration, as will be discussed later, is establishing a road-map towards compact light sources and free-electron lasers (FEL).
  To this end, studies on radiation physics are unavoidable and are also covered in this collection.
  
  As a result, three technologies with considerable progress in the recent years are playing significant roles in instigating the ideas of the presented research.
  These technologies include \emph{particle accelerators}, \emph{radiation sources}, and \emph{light sources}.
  Throughout this document, radiation sources are referred to as sources from radio-frequency (RF) to optical portions of the electromagnetic spectrum, and light sources are considered as sources with radiation wavelengths in nanometer length scales.
  For example, in a standard facility like synchrotron sources or FEL sources, the output of a radiation source is utilized to feed particle accelerators.
  Accelerators then use this radiation to produce relativistic particles generated required by a light source.
  This introductory chapter elaborates to present the motivation and promises of various technologies chosen in each domain.
  Through a historical review of particle accelerators, and the state-of-the-art research in this domain the benefits of acceleration using short THz pulses are discussed and justified.
  Subsequently, a similar review over the radiation source technology is presented, the motivation for laser-driven THz generation is discussed and the progress as well as challenges towards efficient generation of such pulses are outlined.
  A brief review of the light source technology, its promises and challenges are the focus of the next section in this chapter.
  Eventually, the chapter is enclosed by a detailed presentation of the structure of this habilitation treatment.
  
  \section{Particle Accelerators}
  
  A particle accelerator is by definition an apparatus in which electric or electromagnetic fields are used to propel charged particles to nearly light speed within a well-defined beam.
  The main driving force behind the particle accelerators and indeed the first reason for the start of this technology was high-energy physics \cite{barbalat1994applications}.
  For several decades, the user community of the accelerator facilities was strongly dominated by particle physicists using large accelerators as colliders.
  After the development of synchrotron light sources and later FEL sources, the light source community grew so fast that currently their size among accelerator users is comparable with high-energy physics community.
  In medical physics, there is also an increasing interest in radiation therapy, proton therapy, and radioisotope production for medical diagnostics where accelerators play vital roles.
  Nuclear physicists and cosmologists try to use accelerated particles to investigate the structure, interaction and properties of nuclei and of condensed matter at extremely high temperatures and densities.
  Ion implantation is one example of the many applications of these devices in industry.
  Consequently, accelerators now constitute a separate field of research with professional researchers dedicated to their study, construction and operation.
  The new findings and novel development in particle accelerators as the main tool behind modern physics can often lead to new application eras and revolutions in fields benefitting from accelerator outputs.
  Hence, it is helpful to briefly review the start of this field, where it stands currently and the foreseen future of these devices.
  
  \subsection{Past}
  
  It is very difficult to define a general reference point as ``time null" for the particle accelerator technology.
  The criterion for defining the start point in accelerator research highly affects the time that can be referred to as the birth of this field of activity.
  If the first observation of accelerated particles in a scattering experiment is considered, or the first proposal of an accelerator is taken into account, or even the first operation of an accelerator facility is set as the birth date, then the reference time can change up to 40 years.
  Certainly, this time cannot be after 1932 when Lawrence's cyclotron produced 1.25 MeV protons \cite{lawrence1932production} and Cockcroft \& Walton generator accelerated protons to 400 keV \cite{cockcroft1932experimentsI,cockcroft1932experimentsII}.
  Table\,\ref{AcceleratorHistory} lists the main turning points underpinning accelerator technology during the twentieth century.
  \begin{table}
  	\caption{History line for particle Acelerators} \label{AcceleratorHistory} \centering
  	\renewcommand{\arraystretch}{1.2}
  	\begin{tabular}{|c|p{12cm}|}\hline
  		\textbf{year} & \textbf{achievement} \\ \hline \hline
  		1895 & Lenard performed electron scattering on gasses where less than 100\,keV electrons were detected. \\ \hline
  		1913 & Franck and Hertz excited electron shells by electron bombardment. \\ \hline
  		1906 & Rutherford bombards mica sheet with natural alphas and develops the theory of atomic scattering. \\ \hline
  		1911 & Rutherford published theory of atomic structure. \\ \hline
  		1919 & Rutherford induces a nuclear reaction with natural alphas. \\ \hline
  		1923 & Wider\"{o}e designs the first betatron with the well-known 2-to-1 rule. \\ \hline
  		1924 & Ising proposes resonant acceleration in dirft tubes. \\ \hline
  		1928 & In contrast to Rutherford expectation, Gamov predicts tunneling and perhaps 500\,keV energy suffices for research on nuclei. \\ \cline{2-2}
  		& Wider\"{o}e demonstrates Ising's principle with a 1\,MHz, 25\,kV oscillator to make 50\,keV potassium ions. \\ \cline{2-2}
  		& Encouraged by Rutherford, Cockcroft and Walton start designing an 800\,kV generator. \\ \hline
  		1927 & Wider\"{o}e makes a model betatron but it does not work in practice. \\ \hline
  		1929 & Lawrence, inspired by Wider\"{o}e and Ising, concieves the cyclotron. \\ \hline
  		1931 & Livingston demonstrates the cyclotron by accelerating hydrogen ions to 80\,keV. \\ \hline
  		1932 & Generator reaches 700\,kV and Cockcroft and Walton split lithium atom with only 400\,keV protons. \\ \cline{2-2}
  		& Lawrence's cyclotron produces 1.25\,MeV protons. \\ \hline
  		1940 & Kerst re-invents betatron and builds the first working machine for 2.2\,MeV electrons. \\ \hline
  	\end{tabular}
  \end{table}
  
\begin{figure} \centering
  	$\begin{array}{cc}
  	\includegraphics[draft=false,width=2.0in]{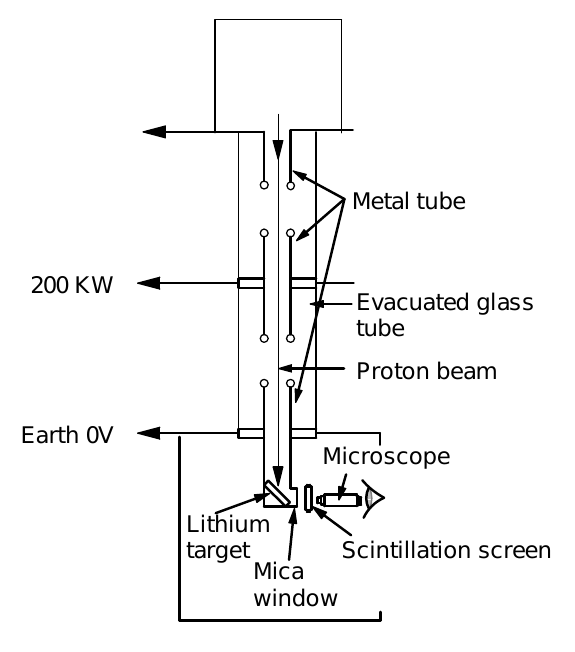} &
  	\includegraphics[draft=false,width=2.0in]{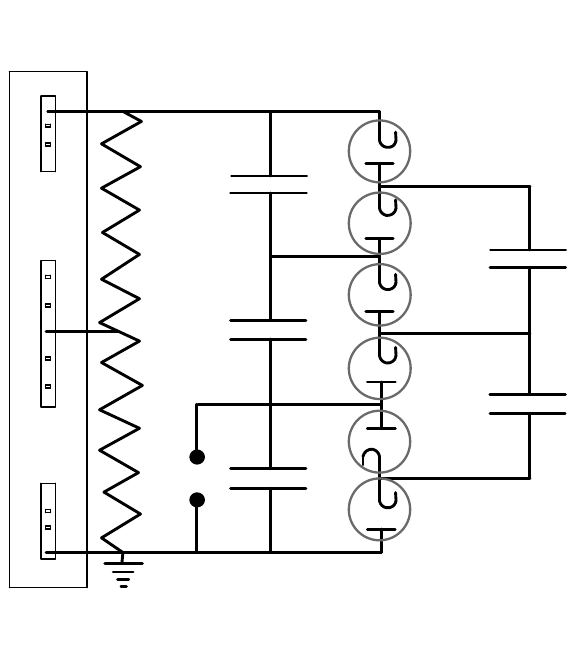} \\
  	(a) & (b)
  	\end{array}$
  	\caption[Cockcraft and Walton's apparatus producing 400\,keV proton beam: (a) the accelerating column, and (b) DC generator]{Cockcraft and Walton's apparatus producing 400\,keV proton beam: (a) the accelerating column, and (b) DC generator (adapted from \cite{bryant1994brief})}
  	\label{CockcroftWaltonGenerator}
  \end{figure}

  The initial developments can be categorized into three separate roots, namely \emph{DC accelerators}, \emph{resonant accelerators}, and \emph{betatrons}.
  The first DC accelerator was realized by Cockcroft and Walton, which was encouraged by Rutherford after Gamov and Gurney predicted tunneling at 500 keV \cite{gamow1928quantentheorie,gurney1928wave}.
  The DC accelerator was later used to split lithium atom and was appraised with the Nobel Prize in 1951 \cite{cockcroft1932experimentsI,cockcroft1932experimentsII}.
  The Cockcroft and Walton generator (Fig.\,\ref{CockcroftWaltonGenerator}) although was the first of its kind, was widely used for many years after as the input stage for larger accelerators.
  The main reason for this wide usage was the high current and low emittance of the output beam.
  Notwithstanding, the device was suffering from a major drawback, which was inability to provide beams of higher energy than the maximum voltage in the generator.
  
\begin{figure} \centering
  	$\begin{array}{cc}
  	\includegraphics[draft=false,width=3.0in]{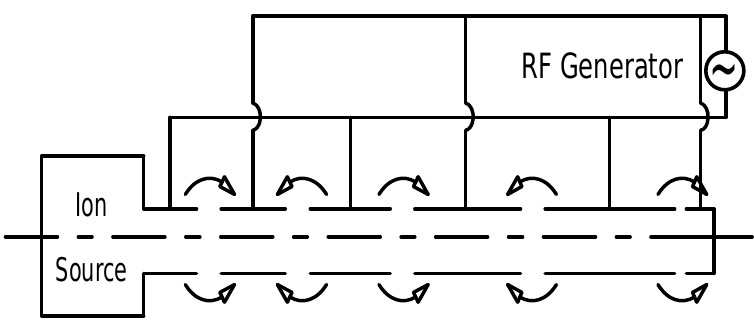} &
  	\includegraphics[draft=false,width=2.5in]{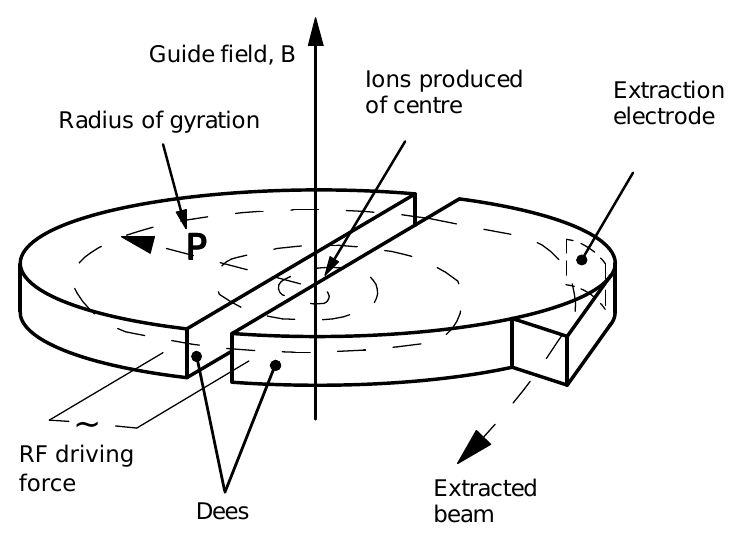} \\
  	(a) & (b)
  	\end{array}$
  	\caption[Resonant accelerators: (a) Ising's RF linac, and (b) Lawrence's cyclotron]{Resonant accelerators: (a) Ising's RF linac, and (b) Lawrence's cyclotron (adapted from \cite{bryant1994brief})}
  	\label{ResonantAccelerator}
  \end{figure}

  The above shortcoming in DC accelerators was the main reason for Ising suggesting the concept of accelerating particles with a linear series of conducting drift tubes in 1924 \cite{ising1924prinzip} and Wider\"{o}e building the proof-of-principle linear accelerator in 1928 \cite{wideroe1928neues}.
  In fact, Ising's original idea can be considered as the beginning of the today linear accelerator (linac) technology.
  Alternate drift tubes are fed by an alternating field source, whose frequency is adjusted so that a particle traversing a gap sees an electric field in the direction of its motion.
  While the particle is inside the drift tube, the field reverses so that it is again directed along the motion at the next gap.
  As the particle gains energy and speed the structure periods must be made longer to maintain synchronism (Fig.\,\ref{ResonantAccelerator}a).
  Building a linac during the 1930's was rather difficult due to technological challenges.
  Ernest Lawrence in 1929, inspired by a written account of Wider\"{o}e's work came up with the idea of the fixed-frequency cyclotron (Fig.\,\ref{ResonantAccelerator}b).
  He built the first model of this device in 1932, which was less than a foot in diameter and could accelerate protons to 1.25 MeV \cite{lawrence1932production}.
  However, the relativistic effects disturb the operation of a cyclotron, limiting the energy of the output beam to sub-relativistic regime.
  
  In parallel to the above developments, betatrons were also proposed by Wider\"{o}e \cite{wideroe1984some} and demonstrated by Kerst \cite{kerst1941acceleration}.
  A betatron uses the circulating electric field around a time-varying magnetic field to transfer energy to particles.
  The widely used term ``betatron oscillation" referring to the particle oscillations inside an alternating magnetic field, originates from this experiment after Kerst and Serber published their paper on particle oscillations in their device \cite{kerst1941electronic}.
  This device is insensitive to relativistic effects and was therefore ideal for accelerating particles to highly relativistic energies.
  The first accelerator delivering particles with energies in the level of hundreds of MeVs was Kerst's betatron which could accelerate electrons to 300 MeV in 1950.
  Although the further progress in betatron development ended in the same year, the device is considered as a reliable and cheap instrument which is continuously built commercially for hospitals and small laboratories.
  
  After the above ground-breaking inventions underpinning the research efforts in accelerator science, constantly new ideas for acceleration mechanisms emerged.
  Some of the suggested ideas revolutionized the technology and brought up new venues for accelerator operations.
  The discovery of the \emph{phase stability principle} in 1944 by Veksler and McMillan led to the invention of \emph{synchrotrons}, which are currently the main facilities serving light source applications \cite{mcmillan1945synchrotron,veksler1945new}.
  The design of synchrotrons were later immensely affected by the proposal of \emph{strong focusing} (also known as \emph{alternating gradient focusing}), which was initially suggested by Christofilos \cite{nicholas1956focussing} and later developed by Courant, Snyder and Livingston \cite{courant1952strong}.
  The use of \emph{superconductivity} in RF accelerator was a prominent step towards accelerating protons to highest energies possible.
  After the progress made in ultra-high frequency technology during World War II, owing to the urgent need for radar technology, the idea of linac structures gained renewed interest.
  Berkeley was the first, with building the \emph{Alvarez accelerator} perfoming linear acceleration of protons up to 32 MeV \cite{alvarez1955berkeley}.
  The first electron linear accelerators were studied at Stanford and MIT in 1946.
  This type of accelerators are currently the largest now in operation facilities delivering a 50 GeV electron beam at the Stanford Linear Accelerator Centre (SLAC).
  
\begin{figure} \centering
  	\includegraphics[draft=false,width=4.0in]{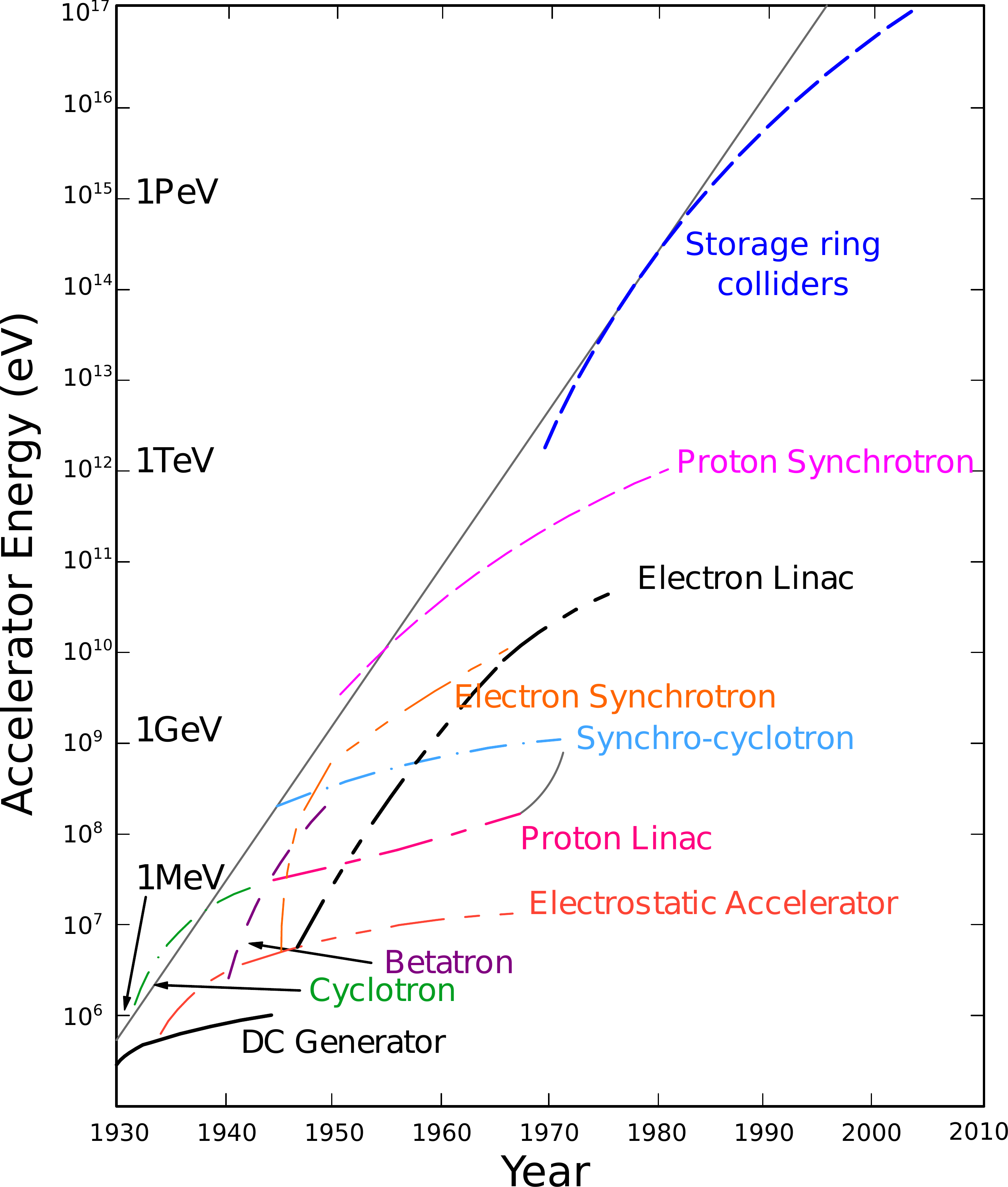}
  	\caption[Livingston Chart]{Livingston Chart: For details on different acceleration concepts refer to \cite{wiedemann2015particle}}
  	\label{LivingstonPlot}
  \end{figure}

  The ``Livingston plot" shown in Fig.\,\ref{LivingstonPlot} reveals the progress that has been made in accelerator energies in the last century.
  This type of plot is called a Livingston chart after M. Stanley Livingston, who first drew up such a chart in the 1960's \cite{livingston1955high}.
  The plot clearly shows that benefitting from novel technologies and ideas, promising increase in the power of accelerators is acquired, higher energy particles are achieved, and unprecedented facilities with new applications are built.
  
  \subsection{Present}
  
  After the developments before 1960's, resonant acceleration was by far the most attractive concept and became the underlying principle for modern accelerators.
  The development of resonant accelerators has progressed along two distinguished paths, which according to the particle trajectories are referred to as \emph{linear accelerators} and \emph{circular accelerators}.
  Particles in linear accelerators travel on a straight line and pass only once through each single accelerator module.
  In contrast, in a circular accelerator, particles traverse a closed orbit periodically for many repetitions, thereby accumulating energy at every passage through the accelerating structure.
  The choice of the accelerator type is mostly determined by the particular application and sometimes the available technology.
  For example, the main advantage of circular accelerators is that a single cavity combined by the confinement action of magnetic fields, is capable of transferring very high energy to particles.
  This is a very efficient scheme where only a relatively small amount of power is required.
  Unfortunately, for light particles the emission of synchrotron radiation can limit the maximum energy achievable, making circular accelerators superior choices for heavy particles and medium energy electrons.
  On the other hand, beam emittance can be maintained smaller in linear accelerators, making them suitable for light source applications.
  Both types continue to be improved and optimized as associated technologies advance.
  
  Synchrotrons as the most frequently implemented circular accelerator structure has been used in many of the today facilities.
  The instruments maintaining the circular orbit of the particles can also be used for storing particles for a long time at fixed energy, which results in the facilities referred to as \emph{storage rings}.
  These configurations are very promising for studying collisions with applications in particle physics.
  This is the reason why all today colliders are in form of synchrotrons or storage rings.
  In 1959, the CERN Proton Synchrotron (CPS) produced 28\,GeV proton beams \cite{regenstreif1959cern1,regenstreif1959cern2} one year before the operation of Brookhaven Alternating Gradient Synchrotron (AGS) delivering 33\,GeV proton beams.
  These two facilities are still in operation and used for applications in particle physics.
  The AGS continued to increase in beam energy through multiple upgrades, and today serves as the injector for the Relativistic Heavy Ion Collider (RHIC) at Brookhaven \cite{hahn1988relativistic}.
  LEP at CERN started operation in 1989 as the Largest Electron-Positron collider with around 104\,GeV beam \cite{bachy1989lep}.
  This facility was later upgraded to the Large Hadron Collider (LHC), the world largest proton collider with 7\,TeV beam installed in the 27\,km long tunnel and started operating in 2008 \cite{evans2007large}.
  
  With time, the technology of RF accelerators became so mature that cascading large number of accelerator modules was routinely feasible.
  This capability led to the development of several facilities based on linear accelerators, where high energy electron beams with low emittances are pursued.
  The Stanford Linear Accelerator (SLA) producing 50\,GeV electron beam, operational since 1966, is the first large scale linear facility built.
  The 3.4\,km long accelerator facility of European x-ray Free Electron Laser (XFEL), which started operating in 2017 and delivers 17.5\,GeV electron beams, is the most recent and advanced facility of this type.
  The future International Linear Collider (ILC) will be the first electron-electron collider that exploits the benefits of linear accelerators in a collision experiment \cite{brau2007international}.
  
  \subsection{Future}
  
  All the previously referenced facilities employ RF accelerators that operate in the RF frequency regime.
  Currently, the delivered power from RF sources is enough to simultaneously fill numerous cavities up to their damage threshold.
  This makes the damage threshold of the accelerator material under intense electric fields, the main obstacle for realizing high accelerating gradients.
  The maximum accelerating gradients in the existing facilities is around 30\,MeV/m at S-band frequency ($\lambda \simeq 10$\,cm).
  With such gradients, kilometer long facilities are inevitable, when GeV-level beams are required.
  It is widely accepted that to further augment the aptitude of the accelerator technology, new approaches should be developed where the achievable accelerating gradients are boosted to higher levels.
  
  Investigating damage mechanisms in accelerators is of key importance to invent new approaches for boosting the achievable accelerating gradient.
  The empirical studies done by Loew and Wang had initially shown that electron field emission, scaling as $f^{0.5}\tau^{-0.25}$ with $f$ the operation frequency, and $\tau$ the pulse duration of the accelerating field, imposes a principal limit on device performance \cite{wang1989rf}.
  However, recent comprehensive studies on breakdown thresholds of various accelerators demonstrated that pulsed heating of the accelerator walls is the dominant factor limiting accelerating gradients \cite{dal2016rf,laurent2011experimental}.
  This conclusion confirmed the observed lower operational gradients in existing facilities when compared with predictions from the previously derived scaling laws.
  The authors concluded that the pulse duration of the accelerating field plays the major role in the breakdown event, since it is directly linked to the pulse energy governing pulsed heating in the device.
  In parallel, the recent detailed studies of breakdown rates in RF accelerators have confirmed this conclusion \cite{wu2017high}.
  The principal outcome of these studies was the value of $E_{max}^6 \tau$ being constant in various facilities operating based on micro-second long pulses in RF regime.
  Therefore, venues for obtaining higher accelerating gradients can be found in short pulse regimes.
  For example, using pico-second long pulses will augment the possible accelerating gradients by at least a factor of ten, making 0.5 GV/m a safe assumption for accelerator design.
  
  Based on the above investigations, one can think of three approaches for pushing the limits of accelerating gradient.
  First, one may try to realize accelerating fields without any surrounding materials or boundaries.
  Such an approach results in the creative field of laser-plasma wakefield acceleration (LPWA) \cite{tajima1979,Walsh1995,Malka1997,Malka2002,geddes2004,Mangles2004,Faure2004,Plettner2005,Leemans2006,Esarey2009,wang2013,leemans2014,litos2014high,steinke2016}.
  In this scheme, a high intensity laser is focused on an atomic gas jet, causing the ionization in the majority of atoms and creating a plasma.
  A resonant motion is thus stimulated in the electrons of the plasma.
  This electron motion breaks the charge balance inside the very dense plasma inducing extremely high gradients in the plasma area surrounding the laser.
  Electrons in the plasma can find the right phase and can be accelerated to high energies.
  Gradients of many tens of GeV/m in few millimeters have been already demonstrated \cite{Leemans2006}, and additionally staging of these interactions have been achieved \cite{steinke2016}.
  High-gradient electron acceleration schemes that directly use laser beams in the complete absence of any materials or boundaries, including plasma, have also been studied and shown to be feasible via ab initio numerical simulations \cite{wong2010direct,wong2017laser,carbajo2016direct}.
  
  Increasing the operation frequency is a proper path from different viewpoints.
  Higher operation frequency shrinks down the device dimensions, which in turn reduces the required energy for particle acceleration.
  This relaxes limitations due to pulsed heating as well.
  Furthermore, higher field emission thresholds as well as easier realization of short pulses in high frequencies assist in obtaining high gradient accelerators.
  Dielectric laser accelerators (DLA) \cite{Peralta2013,England2014,Breuer2013}, as the second solution, and THz-driven linear electron acceleration \cite{Nanni2015,Wong2013,Yoder2005}, as the third one, are promising outcomes of this approach.
  
  Parallel to the aforementioned methods for implementing new forms of particle acceleration, other ideas are also developed and studied in detail towards high-gradient accelerators.
  Inverse Free-Electron Laser (IFEL) \cite{curry2018meter} scheme, inverse Cerenkov accelerator \cite{kimura1995laser}, Laser Plasma Acccleration (LPA) and beam driven acceleration \cite{andonian2012dielectric,antipov2012experimental,gai1988experimental}, where wakefields of a bunch are used to accelerate another bunch are examples of such approaches.
  All of these methods are still in their infancy and each one of them is challenged by specific problems emanated from its operation principles.
  Dedicated research efforts are devoted to solving these problems and mitigating the limitations.
  The future of particle accelerator technology will be shaped by one of these methods, depending on the progress in related source technology and the interest in applications matching each technique.
  
  \section{Radiation Sources}
  
  Particle accelerators need to be fed by high power radiation sources, which are themselves outcomes of advanced technologies with ongoing progress.
  In the review of the particle accelerator history, the development of RF linacs after World War II is a good example for the strong dependence of accelerator development on high power source technology.
  Therefore, it is essential to review the present technology of high power radiation sources, which are either used in accelerators or showing promises for the future accelerator technology.
  
  \subsection{Klystrons}
  
  Kylstron refers to a vacuum tube used to amplify small signals up to high power levels, invented by two brothers Russell and Sigurd Varian of Stanford University \cite{varian1939high}.
  The work of physicist William W. Hansen was instrumental in the development of the klystron according to Varian brothers.
  Their prototype was completed in August 1937, and upon publication in 1939, news of the klystron immediately influenced the work of US and UK researchers working on radar equipment.
  The brothers founded Varian Associates to commercialize the technology and employ it in various applications like linear accelerators in radiation therapy.
  The company is in fact one of the first high-tech companies in Silicon Valley.
  During the second World War, the allied powers relied mostly on klystron technology for the microwave generation in their radar system.
  
\begin{figure} \centering
  	$\begin{array}{cc}
  	\includegraphics[draft=false,width=3.0in]{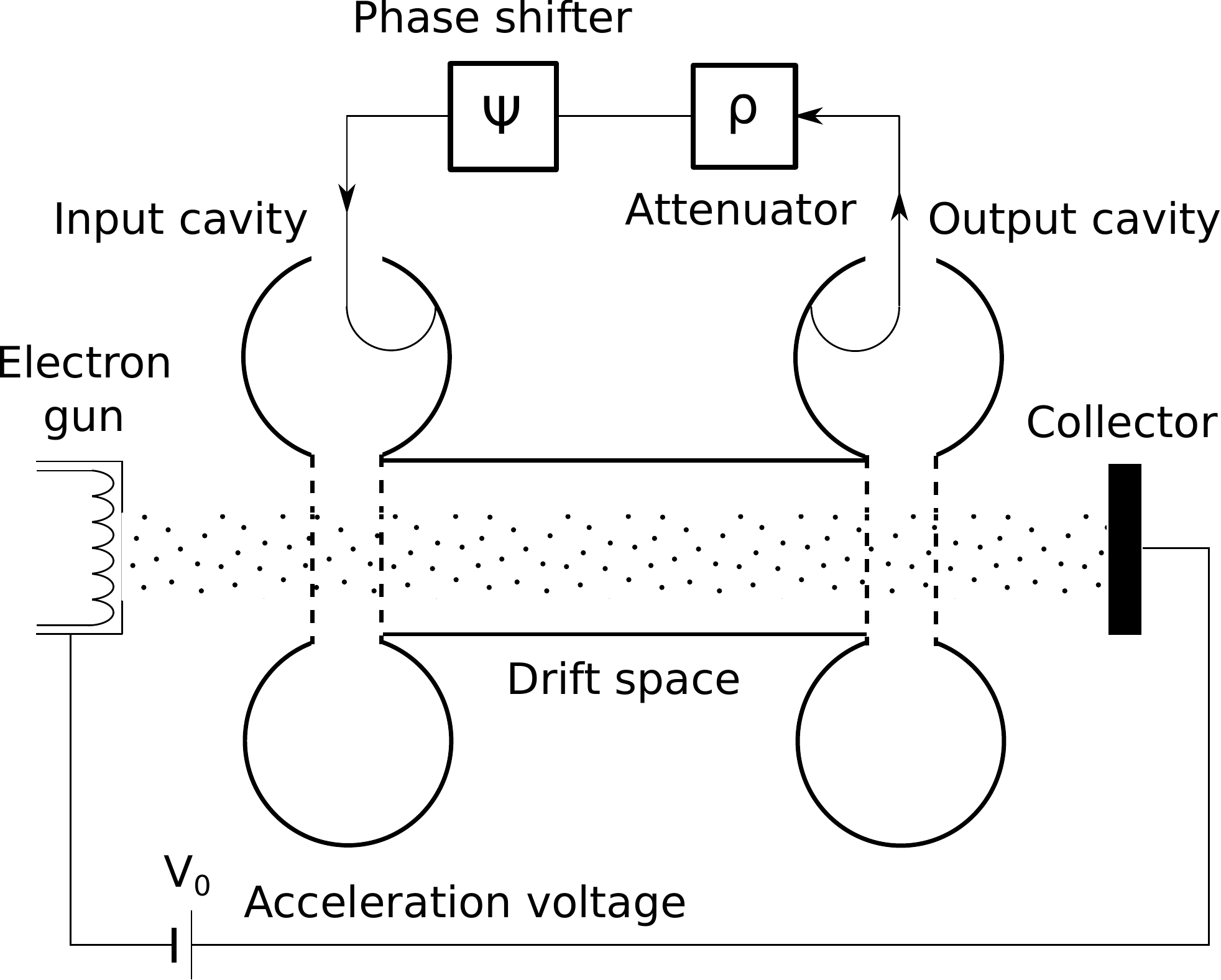} &
  	\includegraphics[draft=false,width=2.8in]{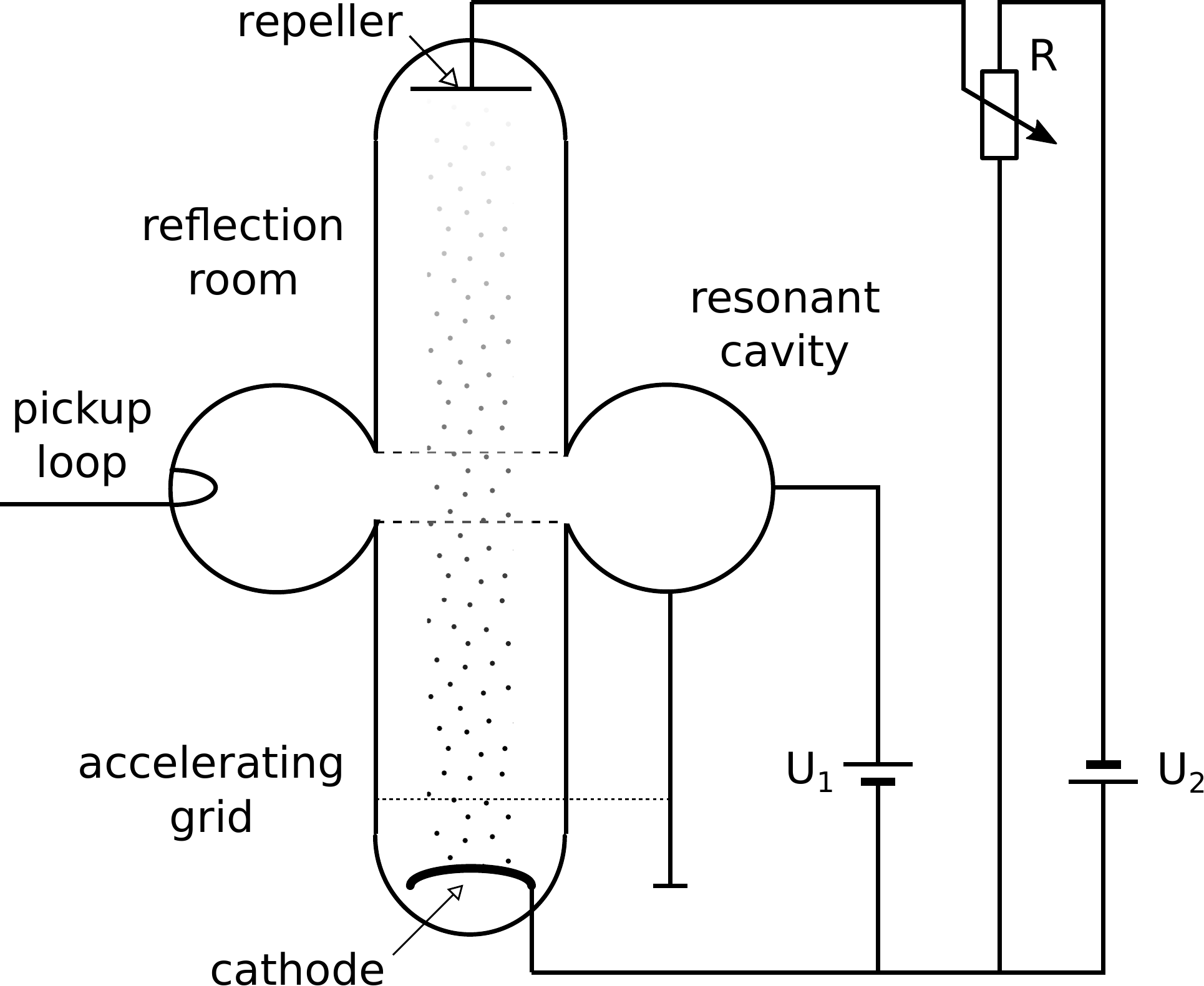} \\
  	(a) & (b)
  	\end{array}$
  	\caption[Schematic illustration of (a) Two-cavity Klystron and (b) repeller Klystron]{Schematic illustration of (a) Two-cavity Klystron (adapted from \cite{shigaev2005chaotic}) and (b) repeller Klystron (adapted from \cite{wolff2012radartutorial})}
  	\label{KlystronSchematic}
  \end{figure}

  Klystron amplifiers have the advantage of coherently amplifying a reference signal, thus their outputs can be precisely controlled in amplitude, frequency and phase.
  This is the reason why they are well-suited for linear particle acceleration technology.
  A schematic illustration of a typical two-cavity klystron amplifier is shown in Fig.\,\ref{KlystronSchematic}a.
  The first cavity is excited by the input signal, which is input-coupled to the cavity by a coaxial-line loop or a waveguide aperture.
  This cavity, acting as a buncher, modulates the velocity distribution of the electron beam produced by a thermionic cathode.
  The second cavity is separated from the buncher by a drift space of length $l$, which should ideally be chosen so that the AC current at the output cavity is a maximum.
  The output cavity is thus excited by the AC signal impressed on the beam in the form of a velocity modulation, which produces an AC current.
  The AC current on the beam is such that the excitation level of the second cavity is much greater than of the buncher cavity.
  This results in an overall amplification of the output signal, extracted from the output cavity.
  A portion of the amplified output is fed back to the buncher cavity in a regenerative manner to obtain self-sustained oscillations.
  
  Another tube based on velocity modulation, and used to generate microwave energy, is the reflex (repeller) klystron, illustrated in Fig.\,\ref{KlystronSchematic}b.
  The reflex klystron contains a reflector plate, named as the repeller, instead of the output cavity used in the other type of klystrons.
  The electron beam is modulated by passing it through an oscillating resonant cavity.
  The feedback required to maintain oscillations within the cavity is obtained by reversing the beam and sending it back through the cavity.
  The electrons in the beam are velocity-modulated before the beam passes through the cavity the second time and will give up the energy required to maintain oscillations.
  The electron beam is turned around by a negatively charged electrode that repels the beam.
  This type of klystron oscillator is called a reflex klystron because of the reflex action of the electron beam.
  
  Klystrons produce microwave power far in excess of that developed by solid state sources.
  They are applied in radar, satellite and wideband high-power communication, radiation oncology, and high-energy physics for particle accelerators and experimental reactors.
  At SLAC, klystrons with outputs in the range of 50\,MW (pulse) and 50\,kW (time-averaged) at frequencies close to 3\,GHz and nanosecond pulse durations are regularly employed.
  Currently, these sources are the standard technology to feed the RF accelerator facilities around the world.
  
  \subsection{Gyrtorons}
  
  Gyrotrons are sources of high-power, coherent radiation capable of generating over one megawatt of continuous-wave (CW) power at wavelengths in the millimeter wave range and powers at the level of many tens of kilowatts well into the terahertz range.
  They can produce high power at millimeter wavelengths because as a fast-wave device its dimensions can be much larger than the wavelength of the radiation.
  This is different from the case of klystrons and generally any other microwave tube based device, where dimensions are tightly determined by the operation wavelength.
  In recent years, major progress is achieved in building and testing pulsed and CW gyrotrons at frequencies above 1.0\,THz.
  
  The electromagnetic radiation in a gyrotron is produced by the interaction of a weakly relativistic gyrating electron beam and a transverse electric (TE) wave close to cut-off in a cavity resonator.
  The gyrotron emission results from the electron cyclotron maser or negative mass instability \cite{lau1984unified}.
  The dependence of an electron's cyclotron frequency on its energy is a relativistic effect.
  From a classical physics point of view, the instability is caused by the dependence of the electron cyclotron frequency on energy, which induces phase bunching
  and coherent emission as the electron beam interacts with an electromagnetic wave.
  The result is a coherent, macroscopic, transverse, cyclotron frequency current that generates transverse electromagnetic (EM) waves.
  
\begin{figure} \centering
  	\includegraphics[draft=false,width=5.0in]{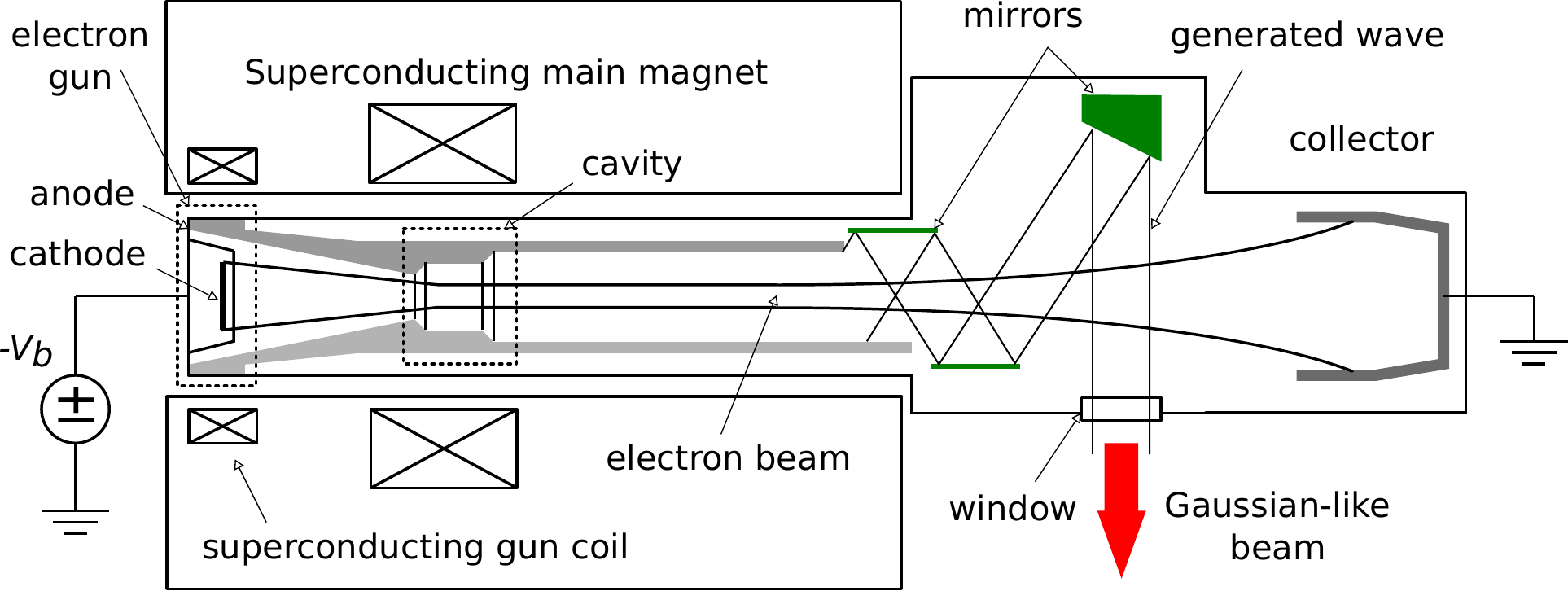}
  	\caption[Schematic Ilustration of a gyrotron]{Schematic Ilustration of a gyrotron (adapted from \cite{booske2011vacuum})}
  	\label{gyrotron}
  \end{figure}

  Fig.\,\ref{gyrotron} illustrates the basic configuration of a modern gyrotron.
  It consists of an electron gun, a cavity resonator, a mode converter, an output window and a collector, and requires a solenoid magnet to give a gyrating motion to the electron beam.
  Electrons emitted from a cathode by the extraction force of the electric field, move in a gradually increasing magnetic field towards the cavity.
  In this movement, the energy of the electron motion along the lines of magnetic field is partially transformed into energy of gyration.
  Electrons that have a cyclotron frequency slightly below the resonant frequency of the microwave excited in the cavity are bunched and decelerated by its transverse electric field.
  As a result, the gyration energy is transferred to the microwave radiation energy by the bunched electrons and induces oscillation.
  The microwave generated within the cavity mode is converted to a wave beam by the mode converter, shaped by some mirrors, and outcoupled through the output window.
  
  Gyrotrons are used for many industrial and high-technology heating applications, such as in nuclear fusion research experiments to heat plasmas and as a rapid heating tool in processing glass, composites, and ceramics.
  They have not still been employed for linear particle acceleration in small waveguide.
  The main obstacle for this application is the still remaining challenges to precisely control the phase of the output radiation.
  However, they are one of the promising choices for realizing small and low-cost accelerators in THz regime \cite{nanni2017prototyping}.
  
  \subsection{Lasers}
  
\begin{figure} \centering
  	\includegraphics[draft=false,width=6.0in]{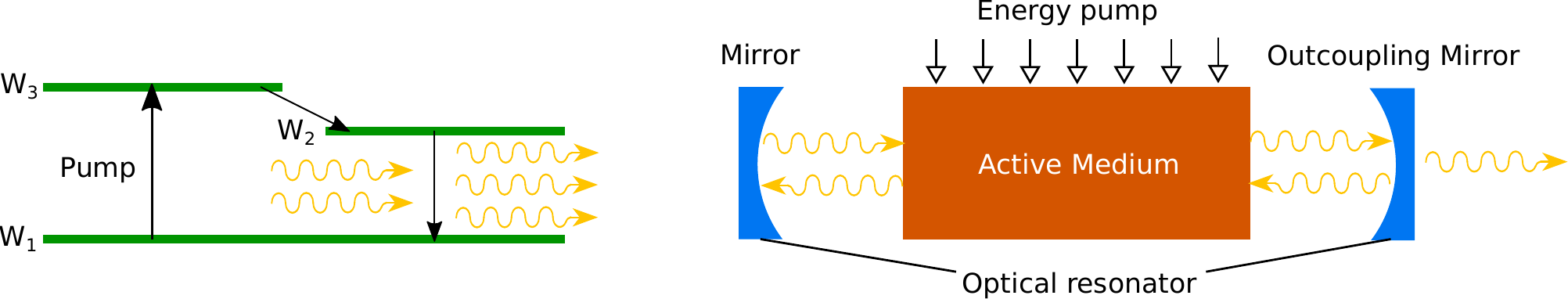}
  	\caption{Laser operation principle.}
  	\label{laserScheme}
  \end{figure}

  Today, one can hardly find a field of science which is not revolutionized by laser technology in the twentieth century.
  The word LASER is an acronym for Light Amplification by Stimulated Emission of Radiation.
  Three basic components constitute the configuration of a conventional laser: the laser medium with at least three energy levels, an energy pump which creates a
  population inversion, and an optical resonator (Fig.\,\ref{laserScheme}).
  The axis of the optical cavity determines the propagation direction of the photons to better than 1\,mrad accuracy.
  In a mono-mode laser exactly one optical eigenmode of the cavity is excited.
  The photons in this mode have all the same frequency $\omega$, the same direction, described by the wave vector $k = (k_x, k_y, k_z)$, the same polarization and the same phase.
  These values are all quantized according to the quantum physics principles, and form a quantum state represented by $\ket{a}$.
  Inside the resonator resides an active medium that contains many atoms in the excited state $W_2$ which can emit radiation of frequency $\omega=(W_2-W1)/\hbar$ by going into the ground state $W_1$.
  In the beginning of the lasing process, with the probability $p_{spon}$ an atom emits its photon by spontaneous emission into any quantum state with arbitrary direction.
  However, only photons in the quantum state $\ket{a}$ travel back and forth between the mirrors and remain in the cavity.
  Any other photon, emitted into other quantum states with a direction different from the resonator axis, will immediately escape from the optical resonator.
  Therefore, the number of photons in state $\ket{a}$ increases with time.
  If already $n$ photons are present in this state, the probability that photon number $(n+1)$ will also go into this state is $(n+1)$ times larger than the probability $p_{spon}$ for emission into any other state:
  \begin{equation}
  p_n = (n + 1)p_{spon} = n p_{spon} + p_{spon} = p_{stim} + p_{spon}
  \label{emissionProbabality}
  \end{equation}
  where, the term $p_{stim}=n p_{spon}$ stands for the stimulated emission probabality, induced by the already existing photons in the quantum state $\ket{a}$, and the term $p_{spon}$ stands for the spontaneous emission which has the same probability $p_{spon}$ for any final state allowed by energy conservation.
  This equation, which can be derived in quantum field theory, is the physical basis of the laser.
  The lasing process starts from a stochastic process, namely spontaneous emission by the excited atoms, and the stimulated emission results in an exponential growth of the light intensity.
  This is the principle of many of the today solid-state and gas lasers with wavelengths ranging from 0.5\,-\,30\,{\textmu}m.
  
  Similar to the majority of science fields, accelerator physics and light sources are strongly affected by the development of high power lasers.
  Fast electron sources based on photoemission from a metallic cathode are one of the long-standing uses of lasers in accelerators \cite{flottmann1997note,dowell2009quantum}.
  This application is currently being further studied to optimize photocathodes \cite{dowell2010cathode}.
  Developing ultrafast electron sources based on field-emission from nano-structures is another research topic with extensive efforts dedicated to it in the past decade \cite{yalunin2011strong}.
  Theoretical and experimental study of this phenomenon is one of the topics focused in this habilitation treatment.
  Recently, several applications such as free electron laser seeding \cite{allaria2012highly}, dielectric laser accelerators \cite{england2014dielectric}, and compact x-ray light sources \cite{kartner2016axsis} have emerged, which highlight the potentials of lasers in the development of future accelerators.
  Furthermore, laser fields are ideal agents to tailor the 6-D phase-space distribution of charged particle beams because they can imprint correlations with extremely high spatio-temporal precision \cite{hemsing2014beam}.
  Laser-plasma wakefield acceleration is a remarkable example of laser utilization to realize record accelerating gradients \cite{Leemans2006}.
  Using lasers as electron wigglers offers the opportunity of producing very high energy photons up to gamma ray regime \cite{phuoc2012all}.
  Use of advanced synchronization schemes in today large research facilities is unavoidable.
  For the stable operation of these facilities, synchronization of various operating modules up to pico-second and femto-second regime is mandatory.
  This demanding precision for synchronization can only be achieved using lasers \cite{xin2017attosecond}.
  
  \subsection{Laser-driven Sources}
  
  As observed from the above review, there exists a frequency range of 0.3 - 10 THz, where neither vacuum electronic sources nor lasers provide solutions for efficient radiation generation.
  This frequency interval cited sometimes as terahertz gap has attracted considerable interest over the past ten years due to numerous important and promising applications.
  Among these applications, molecular spectroscopy, terahertz imaging, terahertz telecommunication, and medical diagnosis using terahertz waves can be referred to.
  The recent developments in electron acceleration and manipulation using terahertz beams, which will be a considerable part of this thesis, have demonstrated the possibility of terahertz acceleration \cite{nanni2015terahertz}.
  Similar to other frequency regimes, high power terahertz sources are a prerequisite for efficient and useful THz acceleration.
  Currently, the demanding power levels useful for electron acceleration is only achieved using optical lasers \cite{kitaeva2008terahertz}, or the so-called laser-driven THz sources.
  
  Different physical approaches are considered now to solve the problem of efficient terahertz generation.
  The electromagnetic field of a terahertz beam can be generated either via the time variation of the local current density, or via the modulation of the local polarization in a crystal.
  The first approach is followed in a variety of sources, like photoconductive (PC) emitters (antennas and switches) \cite{auston1984picosecond} and semiconductor surface emitters.
  Conventionally, a PC emitter is a semiconductor material under a bias voltage and illuminated by visible or near-infrared ultrafast laser pulse with 100-500\,fs pulse duration.
  Absorption of photons results in the transition of electrons from the valence to the conduction band.
  Conduction carriers are accelerated in the external DC bias voltage, and the resultant current surge gives rise to the emission of an electromagnetic pulse.
  The life-time of conduction carriers determines the pulse duration of the emitted radiation.
  If a semiconductor material with picosecond or sub-picosecond carrier recombination time is employed, the spectral peak of the pulse hits the THz range.
  Most semiconductor surfaces emit electromagnetic transient pulses when illuminated by femtosecond visible lasers without formation of a macroscopic current.
  After photoexcitation of the carriers to the conduction band, ultrafast charge transport in the semiconductor generates THz radiation.
  The charge transport can be driven by the intrinsic surface electric field of the semiconductor \cite{zhang1992optoelectronic,dekorsy1993subpicosecond} or by a difference in the mobilities of the electrons and holes \cite{dekorsy1996thz}.
  These mechanisms are the operation principle for semiconductor surface emitters.
  
  The second approach takes advantage from the nonlinear interaction of coherent optical waves with a medium with to second-order optical-to-terahertz susceptibility.
  This category of methods are currently providing the highest generation efficiencies up to percent level optical-to-terahertz energy conversion.
  Employing these techniques in conjunction with the state-of-the-art high power laser technology, enables realization of power and energy levels demanded for electron acceleration.
  The presented monograph relies entirely on the output of such sources for accelerating particles using terahertz fields.
  Hence, a detailed review and discussion on these types of sources are necessary here.
  
  From nonlinear optics, it is known that interaction between two plane waves of frequencies $\omega_1$ and $\omega_2$ in a nonlinear medium with second order susceptibility $\chi^{(2)}$ leads to the generation of the third wave with difference frequency $\Omega = \omega_1 - \omega_2$.
  The amplitude of this difference frequency wave is maximum if the phase-matching condition $\Delta \mathbf{k}=\mathbf{k}_1-\mathbf{k}_2-\mathbf{k}_{\mathrm{THz}} = 0$ is fulfilled.
  This interaction in a one-dimensional framework is usually described by considering the general wave equation, which takes the following form for the amplitude of the THz wave $E(\Omega,x)\exp(-i\Omega t)$:
  \begin{equation}
  \frac{\partial^2 E(\Omega,x)}{\partial x^2}+\epsilon(\Omega)\frac{\omega^2}{c^2}E(\Omega,x)=-\frac{4\pi\Omega^2}{c^2}P^{(nl)}(\Omega,x),
  \label{DFG1DWaveEquation}
  \end{equation}
  where $\epsilon(\Omega)$ stands for the dispersive dielectric constant of the medium and $P^{(nl)}(\Omega,x)$ is the nonlinear polarization, derived from
  \begin{equation}
  P^{(nl)}(\Omega,x) = \chi^{(2)}E_1(\omega_1,x)E_2^*(\omega,x).
  \label{nonlinearPolarization}
  \end{equation}
  To solve the above equation, we consider the following ansatz for the THz wave, which corresponds to the superposition of two forward and backward propagating THz waves:
  \begin{equation}
  E(\Omega,x) = A_f(\Omega,x)e^{(ik_{\mathrm{THz}}x-i\Omega t)} + A_b(\Omega,x)e^{(-ik_{\mathrm{THz}}x-i\Omega t)} .
  \label{FWandBWpropagation}
  \end{equation}
  Solving the wave equation \eref{DFG1DWaveEquation} implicitly neglects the pump depletion effect.
  Under this approximation, the amplitude of the forward and backward propagating waves are calculated as
  \begin{equation}
  A_{f,b} = \frac{i2\phi \Omega^2 L\chi^{(2)}}{k_{\mathrm{THz}}c^2} e^{(-\alpha L/4)}f(\Delta_{f,b}) \left(A_1(\omega_1)A_2^*(\omega_2) \right)
  \label{FWandBWamplitude}
  \end{equation}
  at the initial low-gain stage of the process.
  Here, $L$ is the length of the nonlinear crystal, $\alpha = 2k''_{\mathrm{THz}}$ is equal to the absorption coefficient at the THz frequency $\Omega$, and $f(\Delta_{f,b})$ is derived from
  \begin{equation}
  f(\Delta) = \frac{1}{L}\int\limits_{-L/2}^{L/2}e^{ix\Delta/L}\,dx
  \label{fparameter}
  \end{equation}
  with $\Delta_f = (k_{1x}-k_{2x}-k_{\mathrm{THz}})L$ and $\Delta_b = (-k_{1x}+k_{2x}-k_{\mathrm{THz}})L$.
  From equations \eref{FWandBWamplitude} and \eref{fparameter}, it is easily deduced that the amplitude of the output THz beam is maximized under phase matching condition $\Delta = 0$.
  
  Dispersion of the group velocity $v_g$ is usually negligible for the optical pump.
  In addition, phase matching of the backward propagating wave requires negative group velocity of the beam which is rarely found in materials.
  For same polarization of the two pumping beams, the parameter $\Delta$ for collinear forward-generating processes can be written as
  \begin{equation}
  \Delta_f \approx \left( \frac{\Omega}{v_g} - k_{\mathrm{THz}}(\Omega) \right) L
  \label{deltaParameter}
  \end{equation}
  Considering that the phase velocity of the THz wave is obtained by $v_p(\Omega)=\Omega/k_{\mathrm{THz}}$, the phase-matching condition casts into the following important condition:
  \begin{equation}
  v_g = v_p(\Omega)
  \label{PMcondition}
  \end{equation}
  The above analysis presumes some strong approximations making the predictions invalid for high-gain schemes.
  Nevertheless, the obtained equations suggest the main requirements for efficient terahertz generation, which remain true in any case.
  These requirements include: high second-order susceptibility $\chi^{(2)}$, low THz absorption, and optimal phase matching.
  
  We have introduced these three conditions in the framework of difference frequency generation (DFG) between two optical beams.
  For THz generation, a small difference in the frequencies of each two optical waves corresponding to THz frequency is sufficient.
  Thus, instead of two monochromatic beams with slight difference in frequency, it is enough to have one quasi-monochromatic high-power beam with an appropriate broad spectrum.
  This type of interaction is referred to as optical rectification (OR).
  The spectral bandwidth of a picosecond laser suits for generation of THz waves based on OR phenomenon, whose optimal operation is achieved under the similar three conditions listed above.
  
  The first two conditions are completely dependent on the material used in the source.
  Attractive choices for terahertz generation include Lithium-niobate (LiNbO\textsubscript{3}), Lithium-tantalate (LiTaO\textsubscript{3}), semiconductor crystals like Cadmium-telluride (CdTe), Zinc-telluride (ZnTe), Gallium-arsenide (GaAs), and Gallium-phosphide (GaP), and polymaer materials like DAST (trans-4'-(dimethylamino)-N-methyl-4-stilbazolium tosylate).
  At the same time, employing these materials is beneficial if the phase-matching condition is satisfied.
  The large separation between optical and terahertz frequency typically leads to a difference between optical group velocity and terahertz phase velocity.
  In other words, it disturbs the phase-matching condition.
  Various schemes are proposed and constantly new ideas emerge which elaborate the meeting of phase-matching requirement \cite{kitaeva2008terahertz}.
  Among these schemes, we focus on two types which have shown the most promises for high power terahertz generation, namely \emph{tilted-pulse-front} scheme for generating single-cycle pulses and \emph{quasi-phase matching in periodically poled crystals} for generating multi-cycle pulses.
  
  \textbf{Tilted pulse Front:}  Optical rectification of femtosecond laser pulses with tilted-pulse fronts in lithium niobate has emerged as the most efficient THz generation technique \cite{hebling2002velocity,stepanov2005scaling,stepanov2003efficient,yeh2007generation}.
  This approach produces single-cycle THz fields with optical-to-THz conversion efficiencies close to 1\% at room temperature \cite{fulop2012generation}, and larger than this value at cryogenic temperature \cite{huang2015highly}.
  Consequently, the approach has attracted a lot of interest in the pursuit of mJ-level THz pulse energies.
  
\begin{figure} \centering
  	$\begin{array}{cc}
  	\includegraphics[draft=false,width=3.0in]{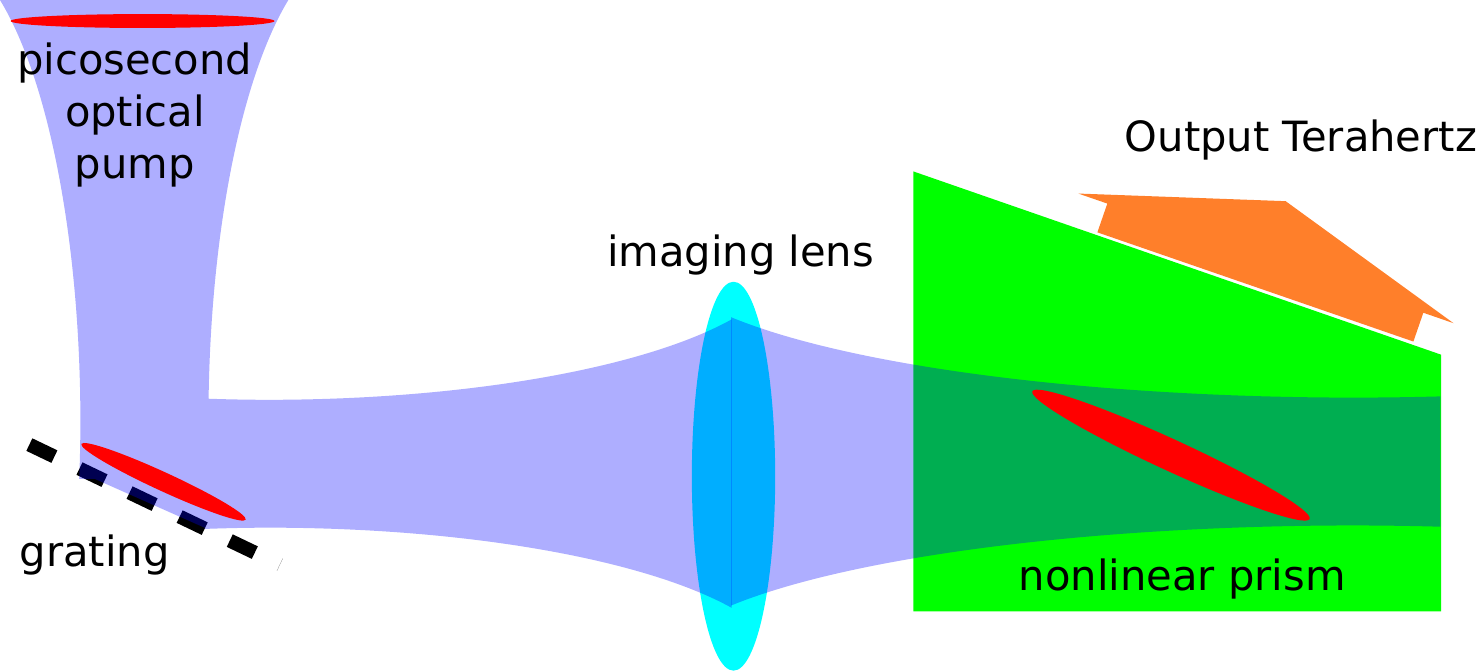} &
  	\includegraphics[draft=false,width=3.0in]{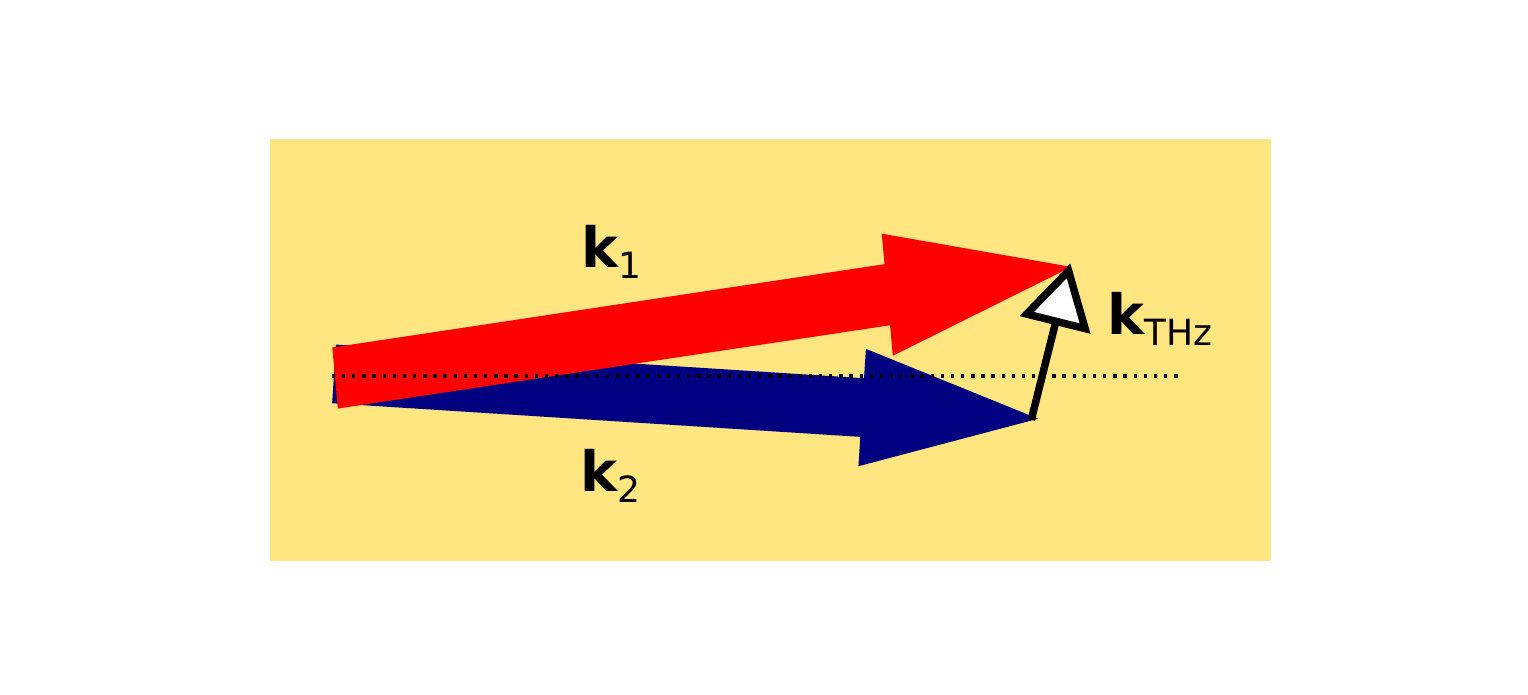} \\
  	(a) & (b)
  	\end{array}$
  	\caption{Tilted pulse front technique: (a) schematic illustration of the setup and (b) phase-matching principle}
  	\label{TPFSetup}
  \end{figure}

  The configuration of the tilted pulse front setup realizes a pump beam with propagation direction at an adjustable angle $\theta$ to the direction of its wave front (Fig.\,\ref{TPFSetup}a).
  This condition is acquired using a grating, through which two dominant diffraction orders are created.
  Using an imaging setup, i.e. a focusing lens, these diffracted beams are made to interfere with each other at the nonlinear crystal.
  The THz beam is emitted in the direction normal to the pump front, in accordance with the Huygens-Fresnel principle.
  The optimal angle of pulse front tilting $\theta$ is determined by the phase-matching condition $\Delta \mathbf{k} = 0$ illustrated in Fig.\,\ref{TPFSetup}b.
  To realize phase-matching, the projection of the optical pulse velocity on the direction of the THz beam, $v_g\cos\theta$ should coincide with the phase velocity $v_p$ of the THz wave.
  Therefore, the phase-matching condition is revised into the following form:
  \begin{equation}
  v_g \cos\theta = v_p(\Omega)
  \label{PMconditionTPF}
  \end{equation}
  Due to the dispersion of the nonlinear material, the pulse front tilting angle is slightly dependent on the frequency spectrum of the pump.
  Additionally, the angle between the wave vectors of the shifted components is necessary for the non-collinear phase matching.
  In \cite{bartal2007toward}, the attainable THz pulse energy was estimated on the base of model calculations for LiNbO\textsubscript{3} of stoichiometric (SLN) and congruent (CLN) composition, GaP, GaSe, and ZnTe.
  For each case, the setup parameters such as pump pulse duration, THz frequency, crystal length and crystal temperature were examined and optimized.
  It was shown that stoichiometric LiNbO\textsubscript{3} under cryogenic conditions is the most promising material for generating up to 2\,THz radiation.
  
\begin{figure} \centering
  	$\begin{array}{cc}
  	\includegraphics[draft=false,width=3.0in]{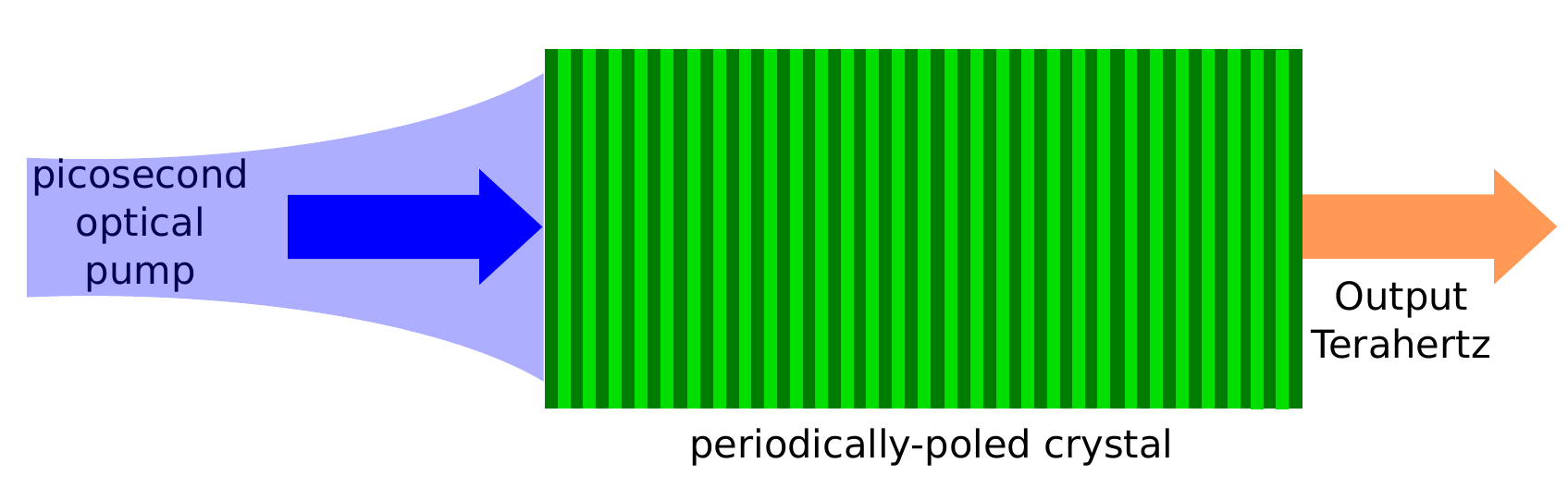} &
  	\includegraphics[draft=false,width=3.0in]{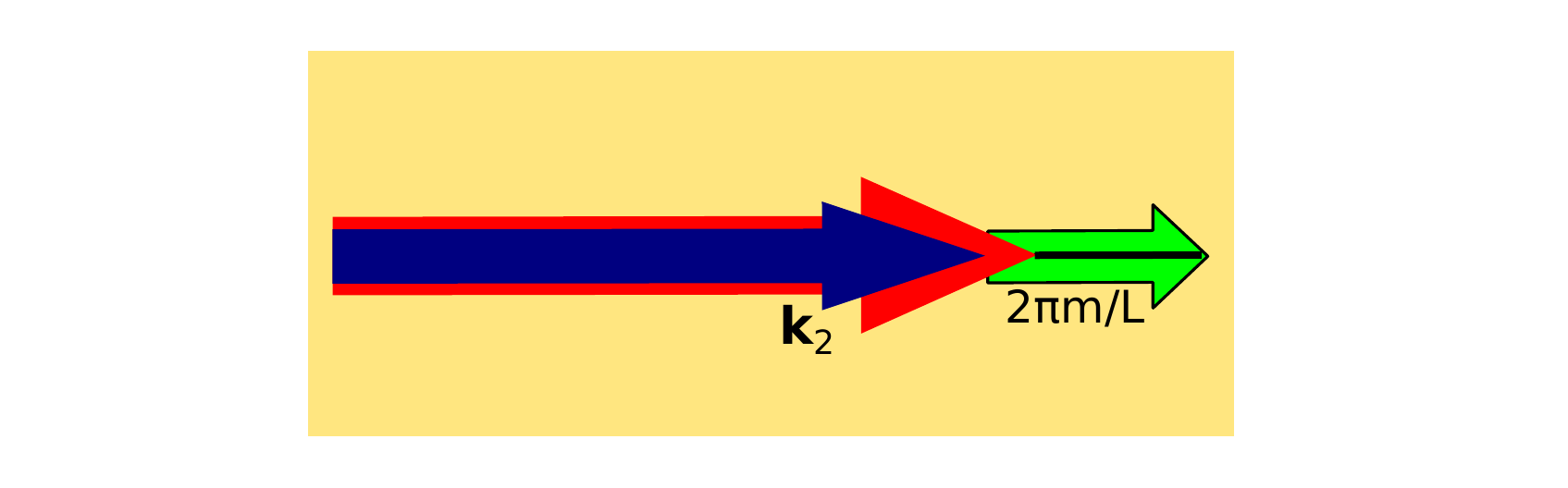} \\
  	(a) & (b)
  	\end{array}$
  	\caption{Quasi phase-matching using periodically poled crystals: (a) schematic illustration of the setup and (b) phase-matching principle }
  	\label{PPLNSetup}
  \end{figure}

  \textbf{Quasi-phase matching in periodically poled crystals:} Another method to fulfill the phase-matching condition over long interaction lengths is using crystal structures with the periodic inversion of crystal axes.
  These structures often referred to as periodically poled crystals maintain a spatial periodic variation of the sign of second-order optical susceptibility, which gives rise to a new condition for the efficient frequency conversion, the so-called quasi phase-matching (QPM) condition.
  A schematic illustration of terahertz generation in a periodically poled crystal is shown in Fig.\,\ref{PPLNSetup}a.
  The amplitude of the generated terahertz beam is found from,
  \begin{equation}
  A_{f,b} = \frac{i2\phi \Omega^2 L}{k_{\mathrm{THz}}c^2} e^{(-\alpha L/4)}\sum\limits_{m=-\infty}^{+\infty} \chi_m f(\Delta_{f,b}+2\pi m) \left(A_1(\omega_1)A_2^*(\omega_2) \right)
  \label{FWandBWamplitudePPLN}
  \end{equation}
  where the term $\chi^{(2)} f(\Delta_{f,b})$ in \eref{FWandBWamplitude} is replaced by a summation over all Fourier harmonics $\chi_m$ of the second-order susceptibility spatial variation $\chi^{(2)}(x)$.
  This equation yields the following new quasi phase-matching condition for efficient terahertz generation:
  \begin{equation}
  \Delta k + \frac{2\pi m}{L} = 0.
  \label{QPMcondition}
  \end{equation}
  
  The idea of using QPM materials for optical rectification and subsequent THz generation was proposed by Lee \cite{lee2000generation}.
  Multi-cycle narrow-band terahertz radiation was produced in periodically poled lithium niobate (PPLN) crystal.
  The authors used femtosecond pulses at 800\,nm and cryogenically cooled (18K) PPLN crystal to reduce THz absorption, and achieved 10\textsuperscript{-5} conversion efficiency.
  Efficient narrow-band terahertz radiation is also achieved in orientation-patterned gallium arsenide \cite{vodopyanov2005terahertz}.
  Equation \eref{FWandBWamplitudePPLN} suggests that THz generation efficiency can be improved after increasing the amplitudes of the optical waves.
  Inspired by this ground, large periodically poled crystals are illuminated by high power laser sources to realize highly efficient narrowband terahertz generation.
  Efficiencies up to 0.1\% and energies of about 1\,{\textmu}J were achieved using optimized femtosecond pump pulses and cryogenic cooling \cite{carbajo2015efficient}.
  But to produce the mJ-level THz pulses needed for acceleration applications, pump pulses up to 1\,J are required, and fs pulses in this case are limited by optical damage and nonlinear effects, especially in longer crystals.
  The idea of chirping (stretching) the pump and overlapping delayed pulse copies to maintain the required difference frequency content was introduced to increase the source efficieny \cite{ahr2017narrowband}.
  Using a pulse sequence is also a promising technique to coherently combine the generated THz waves at different stages of PPLN and boosting the generation efficiency \cite{ravi2016pulse}.
  
\begin{figure} \centering
  	$\begin{array}{cc}
  	\includegraphics[draft=false,width=3.0in]{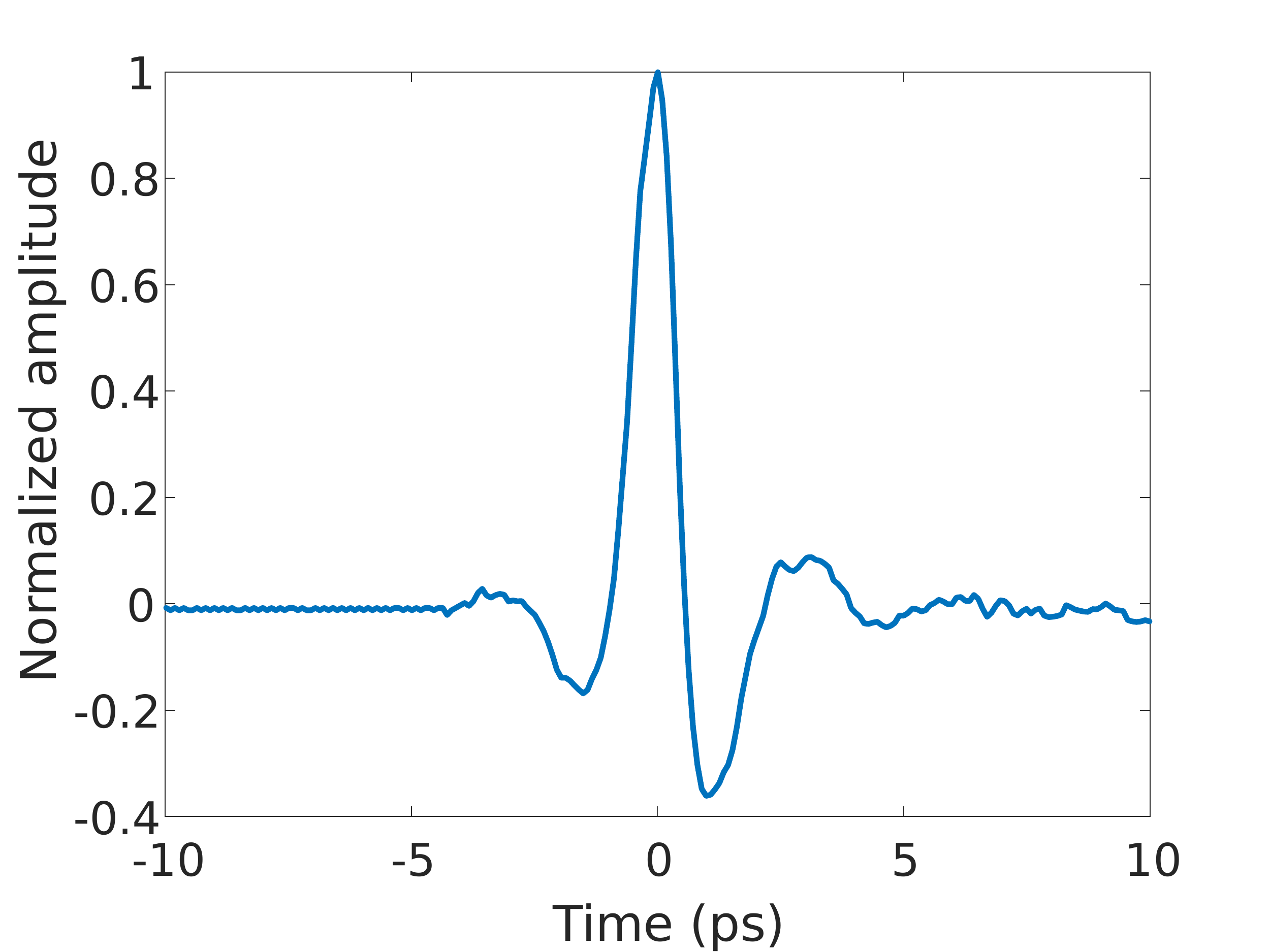} &
  	\includegraphics[draft=false,width=3.0in]{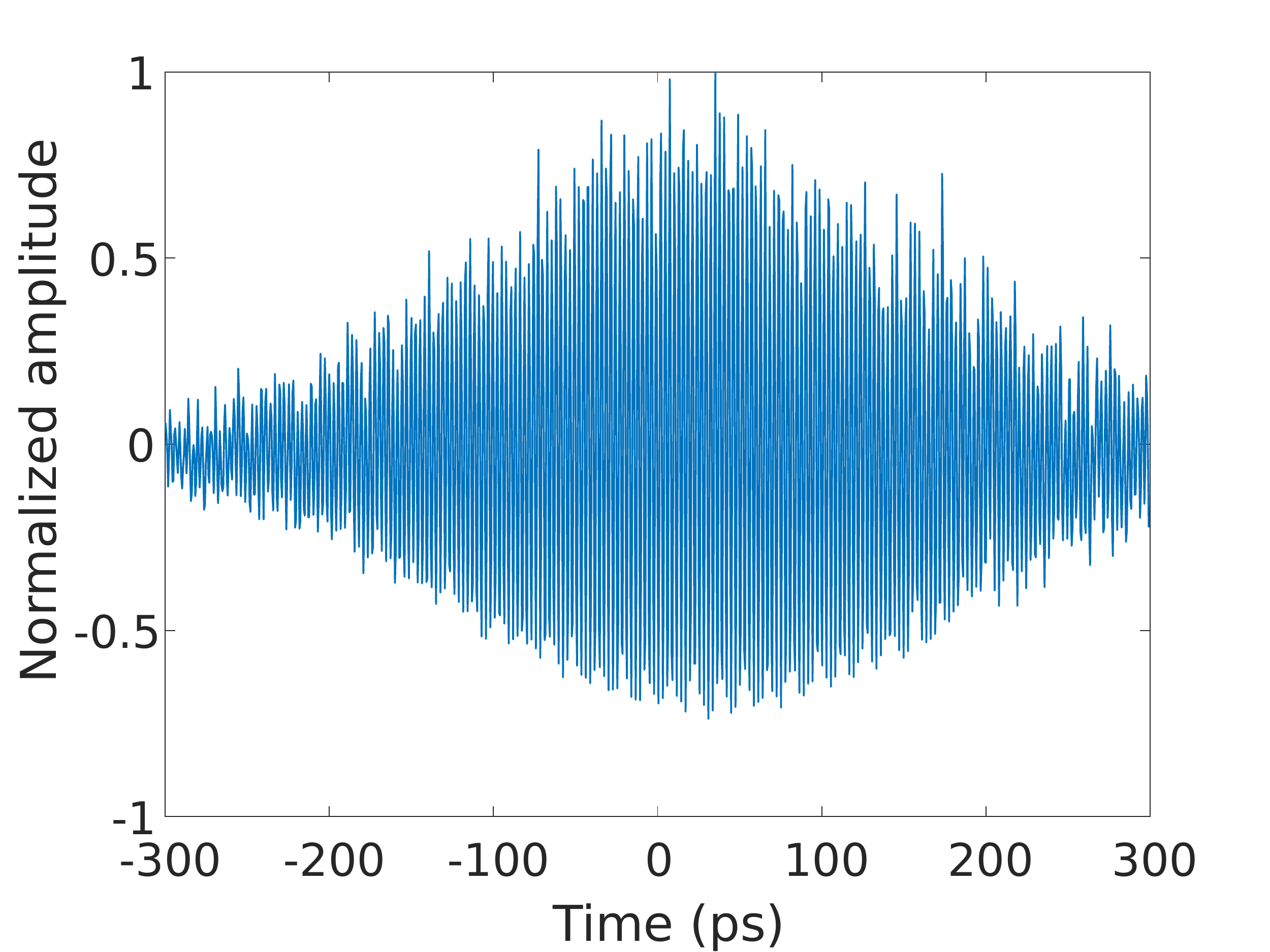} \\
  	(a) & (b)
  	\end{array}$
  	\caption[Temporal signature of the output pulse produced using (a) tilted pulse front and (b) quasi phase-matching techniques]{Temporal signature of the output pulse produced using (a) tilted pulse front and (b) quasi phase-matching techniques \cite{ahr2017narrowband}. }
  	\label{PulseFormats}
  \end{figure}

  The techniques tilted pulse front OR, as well as OR and DFG in PPLN crystals have now reached efficiency levels adequate for generating the radiation energy and power demanded by terahertz accelerators.
  Nonetheless, conventional techniques developed for very narrowband pulses (10\textsuperscript{4}-10\textsuperscript{5}) cycles are not compatible with the output of these sources.
  Tilted pulse front technique produces a single-cycle pulse (Fig.\,\ref{PulseFormats}a) and PPLN based techniques provide pulses which are maximally 100-200 cycles long (Fig.\,\ref{PulseFormats}b).
  This limitation calls for new research and developments aiming at novel accelerators whose inputs are consistent with the output radiation of such sources.
  
  \section{Light Sources}
  
  The search for efficient and compact accelerators is highly supported and instigated by the increasing demand for compact light sources.
  The ultimate goal in the terahertz acceleration research is development of compact light sources, which are low cost and affordable by various research institutes and universities.
  The success of such efforts will solve the existing and frustrating challenges caused by limited number of light sources serving numerous research groups in chemistry, biology, material science and medicine.
  As previously emphasized, the term light source in this monograph refers to source of electromagnetic waves with wavelengths in the range of soft to hard x-ray regime, i.e. 0.01 - 50 nanometers.
  To better comprehend the present state of research and development in this field, the made progress, and future road-map, we present a historical review of the light source technology.
  
  The discovery of x-rays by R\"{o}ntgen in 1895, which was awarded the 1901 Nobel Prize in Physics, opened new venues aiming at scientific use of x-rays to gain new insights into the structure of matter.
  Table\,\ref{XrayAchievements} lists some of the highlights in this domain during the first half of twentieth century.
  \begin{table}
  	\caption{Early history of x-ray research} \label{XrayAchievements} \centering
  	\renewcommand{\arraystretch}{1.2}
  	\begin{tabular}{|c|p{12cm}|}\hline
  		\textbf{year} & \textbf{achivement} \\ \hline \hline
  		1896 & Frost used x-ray to see into human body, Grubbe suggests treating cancer with Patients, and Gage finds the dangerous side-effects of x-ray exposure. \\ \hline
  		1896 & Hall-Edwards performed the first use of x-rays under clinical conditions. \\ \hline
  		1909 & Barkla and Sadler discover characteristic x-ray radiation (1917 Nobel Prize to Barkla). \\ \hline
  		1912 & von Laue, Friedrich, and Knipping observe x-ray diffraction (1914 Nobel Prize to von Laue). \\ \hline
  		1913 & Bragg, father and son, build an x-ray spectrometer (1915 Nobel Prize). \\ \hline
  		1913 & Moseley develops quantitative x-ray spectroscopy and Moseley's Law. \\ \hline
  		1913 & Coolidge designed and manufactured the first high vacuum x-ray tube. \\ \hline
  		1914 & Curie developed radiological cars to do x-ray imaging on injured soldiers in World War I. \\ \hline
  		1916 & Siegbahn and Stenstrom observe emission satellites (1924 Nobel Prize to Siegbahn). \\ \hline
  		1920 & Law enforcement agencies begin using x-ray cameras for inspecting packages and luggage. \\ \hline
  		1921 & Wentzel observes two-electron excitations. \\ \hline
  		1922 & Meitner discovers Auger electrons. \\ \hline
  		1924 & Lindh and Lundquist resolve chemical shifts. \\ \hline
  		1927 & Coster and Druyvesteyn observe valence-core multiplets. \\ \hline
  		1928 & The Internation Committee on Radiological Protection was founded. \\ \hline
  		1931 & Johann develops bent-crystal spectroscopy. \\ \hline
  		1931 & General Electric Company developed the first 1\,MV x-ray generator. \\ \hline
  		1956 & Crick and Watson analyzed DNA structure using x-rays (1962 Nobel prize in Medicine). \\ \hline
  		1962 & Scorpius X-1, the first x-ray source in space, was discovered. \\ \hline
  	\end{tabular}
  \end{table}
  Almost half a century after the x-ray discovery, the broad applications and extensive benefits of this type of radiation urgently called for bright and high quality x-ray sources.
  The attempts towards providing such sources started from a parasitic synchrotron light source as the first generation and advanced to free electron lasers as the fourth generation light source.
  
  \subsection{Synchrotons}
  
  The theoretical basis for predicting synchrotron radiation was carried out in 1897 by Larmor.
  He used classical electrodynamics to derive an expression for the instantaneous total power radiated by an accelerated charged particle.
  The following year, Li\'{e}nard generalized this result to the case of a relativistic particle moving in a circular trajectory.
  Li\'{e}nard's formula showed the radiated power to be proportional to $(E/mc^2)^4/R^2$, where $E$ is particle energy, $m$ is the rest mass, and $R$ is the curvature radius of the trajectory.
  This work was supplemented by Emil Wiechert, so the formalism is generally known as the Li\'{e}nard-Wiechert potentials.
  However, it took time until 1945 until Schwinger developed the classical theory of radiation from relativistic electrons.
  His theory predicted the strongly forward peaked distribution that gives synchrotron radiation its highly collimated property, and the shift of the radiation spectrum to higher photon energies as the electron energy increases.
  
\begin{figure} \centering
  	\includegraphics[draft=false,width=2.5in]{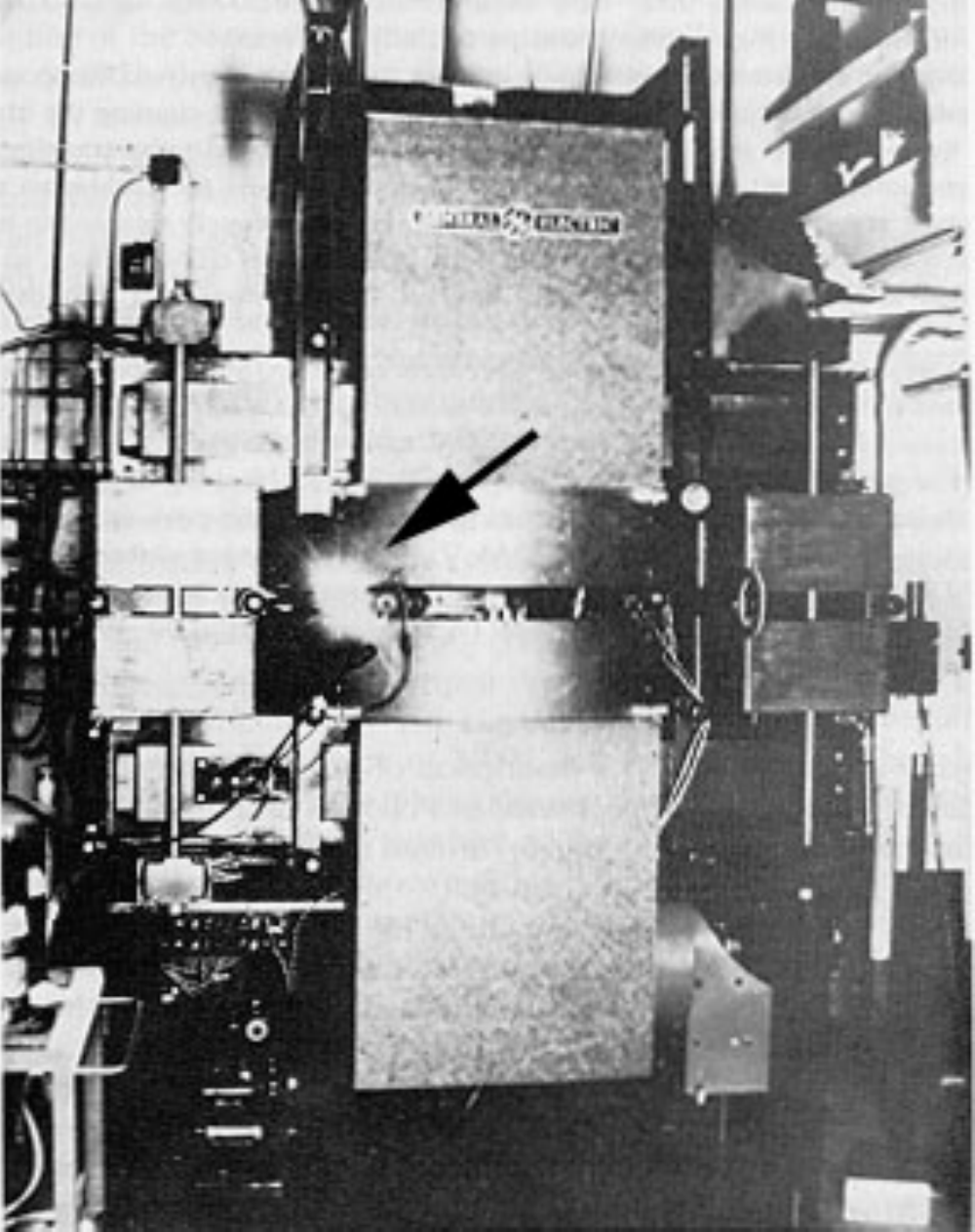}
  	\caption[First Synchrotron radiation observed at General Electric 70\,MeV electron synchrotron.]{First Synchrotron radiation observed at General Electric 70\,MeV electron synchrotron (courtesy of \cite{thompson2001x}).}
  	\label{firstSR}
  \end{figure}

  Encouraged by the discovery of phase stability principle after McMillan's and Veksler's theoretical work, General Electric (GE) launched a new project to build a 70\,MeV electron synchrotron.
  In the design phase of this synchrotron, a transparent coating on the electron tube was devised to enable checking for sparkling in the tube.
  Instead, a gleam of bluish-white light emerging from the electron beam was observed (Fig.\,\ref{firstSR}).
  Langmuir recognized that the light is produced due to synchrotron radiation, which led to the first discovery of this radiation on 24th April 1947 \cite{elder1948radiation}.
  
  In the 1950's and 1960's, particle accelerators originally developed for Nuclear Physics research were modified to allow continuous access to researchers wishing to work with synchrotron radiation.
  A good example of such synchrotron facilities is DESY (Deutsches Elektronen Synchrotron) in Germany.
  From the late 1960's, particle accelerators based on storage rings began to emerge.
  For nuclear and particle physics, a storage ring increases control over where and how the accelerated particles will collide with each other or with a target.
  For scientists interested in using synchrotron radiation, the storage ring makes the production of this radiation continuous, guaranteeing long periods of beam exposure for samples under investigation.
  These were, however, equipment for research in nuclear or particle physics and were not designed or used exclusively for the production of synchrotron light.
  The use of synchrotron radiation under such circumstances was called parasitic operation, and these accelerators are considered as the \emph{first generation light sources}.
  
  The successful usage of synchrotron radiation in experiments by different research groups resulted in the desire for equipments optimized for best radiation production.
  As an outcome, \emph{second generation light sources} emerged.
  In these second generation sources, as in other synchrotron accelerators, light is produced when the electron beam path is curved by magnetic fields of dipole magnets.
  However, the magnetic lattice, required for maintaining the circular path of electrons, is designed to produce the greatest quantity and best quality of synchrotron radiation possible.
  With the quality improvement of synchrotron light, the number of users in various research areas and the number of experimental techniques exploded.
  The novel experiments designed based on synchrotron radiation were continuously wishing for brighter and stronger beam qualities.
  For instance, the studies based on x-ray microscopy, x-ray spectroscopy and crystallography highly benefitted from better spatial and temporal coherence of the beam.
  This demand from the user led to the development of \emph{third generation light sources}.
  
\begin{figure} \centering
  	$\begin{array}{cc}
  	\includegraphics[draft=false,width=3.0in]{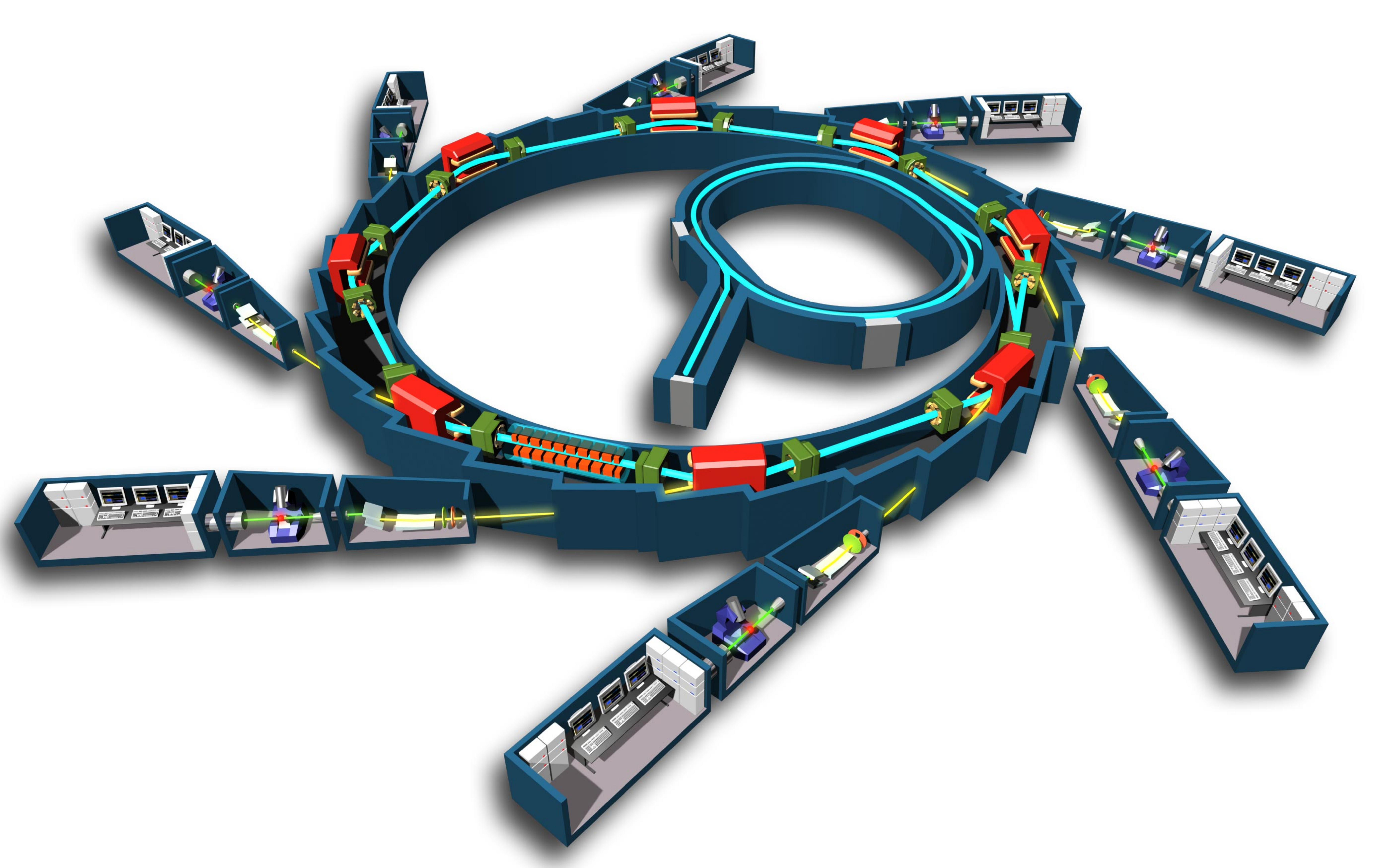} &
  	\includegraphics[draft=false,width=2.6in]{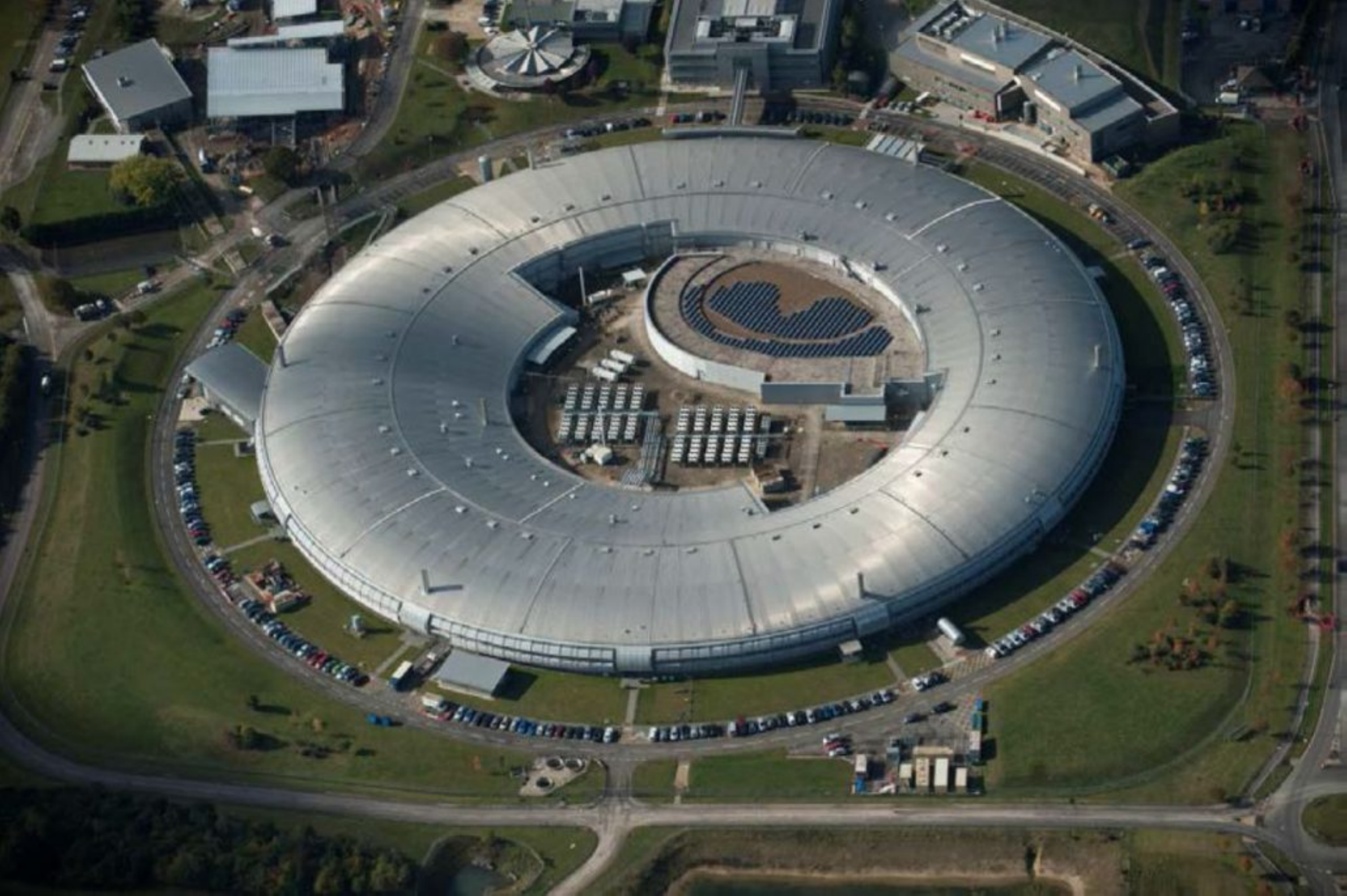} \\
  	(a) & (b)
  	\end{array}$
  	\caption[Third-generation light source: (a) schematic illustration and (b) photo of the Diamond Light Source facility]{Third-generation light source: (a) schematic illustration (produced by J. F. Santarelli, Synchrotron Soleil) and (b) photo of the Diamond Light Source facility (taken from https://www.createmaster.co.uk/project/diamond-light-source-beamline-i-14-emf/).}
  	\label{SynchrtronSchemePhoto}
  \end{figure}

  The third generation light source, which is the present state of advanced synchrotron light sources, takes advantage from a two-ring model and the latest development in accelerator physics to produce electron beams with very low emittance.
  This progress made it possible to use insertion devices such as undulators and wigglers over the electron trajectory.
  The European Synchrotron Radiation Facility (ESRF) in Grenoble was the first of the third-generation hard x-ray sources to operate, coming on line for experiments by users with a 6\,GeV storage ring and a partial complement of commissioned beamlines in 1994.
  The ESRF has been followed by the Advanced Photon Source at Argonne National Laboratory (7\,GeV) in late 1996, and SPring-8 (8\,GeV) in Harima Science Garden City in Japan in late 1997.
  These machines are physically large (850 to 1440 meters in circumference) with a capability for 30 or more insertion devices.
  Fig.\,\ref{SynchrtronSchemePhoto}a schematically shows the basic configuration of the third-generation light source with the accompanying research facility.
  As an example of today synchrotron radiation sources, the photo of Diamond Light Source in England is also shown (Fig.\,\ref{SynchrtronSchemePhoto}b).
  
  \subsection{Free-electron Lasers}
  
  After the world-wide implementation of third generation light sources, the race to develop a new generation of light sources with vastly enhanced performance based on various ideas began.
  The winner of the title \emph{fourth-generation light source} was the hard x-ray free- electron laser based on a very long undulator in a high-energy electron linear accelerator.
  A FEL facility offers peak brightness many orders of magnitude beyond that of the third-generation light sources, as well as pulse lengths of 100\,fs or shorter, with fully coherent characteristics.
  In a free electron laser, relativistic electrons provided from linear accelerators travel through a static undulator and experience a wiggling motion.
  The undulator performance is categorized into two main regimes: (\emph{i}) in a short undulator, each electron radiates as an independent moving charge, which yields an incoherent radiation of electron bunch.
  Therefore, the radiation power and intensity is linearly proportional to the number of electrons.
  (\emph{ii}) For long interaction lengths, the radiated electromagnetic wave interacts with the bunch and the well-known micro-bunching phenomenon takes place.
  Micro-bunching leads to a periodic modulation of charge density inside the bunch with the periodicity equal to the radiation wavelength.
  This effect results in a coherent radiation scaling with the square of the bunch numbers \cite{schmuser2014free,saldin2013physics}.
  Coherent x-ray has shown unprecedented promises in enabling biologists, chemists and material scientists to study various evolutions and interactions with nanometer and sub-nanometer resolutions.
  
  Similar to synchrotron radiation sources, the radiation of a free electron laser is principally based on charge particle emission of radiation when accelerated.
  In synchrotron sources, each electron emits a radiation independently and the radiation immediately leaves the electron trajectory.
  Because of the random distribution of the electrons within a bunch and the large size of the electron bunch compared with the radiated wavelength, the ultimate radiation is incoherent.
  The frequency spectrum in a synchrotron radiation is continuous and covers the interval from zero to the frequencies beyond the cut-off synchrotron frequency $\omega_c$, obtained from
  \begin{equation}
  \omega_c = \frac{3c\gamma^3}{2R}.
  \label{cutoffSRfrequency}
  \end{equation}
  Here, $R$ is the radius of curvature in the bending magnet and $\gamma$ is the Lorentz factor.
  The radiated power in a bending magnet is obtained from
  \begin{equation}
  P_{\mathrm{sync}} = \frac{e^4 \gamma^2 B^2}{6\pi \epsilon_0 c m_e^2}.
  \label{SRpower}
  \end{equation}
  Most of the radiated power is limited to a narrow cone of opening angle $1/\gamma$, which is centered around the instantaneous tangent of the circular orbit.
  As reviewed above, modern synchrotron light sources implement an undulator magnet after the bending magnet to generate a more confined and coherent radiation.
  In an undulator, acceleration of the electrons is realized through a periodic arrangement of magnets with alternating polarization.
  The oscillating magnetic field introduces a wiggling motion to the straight line trajectories of particles, which in turn leads to the generation of an electromagnetic radiation.
  John Madey pioneered the concept of radiation sources from undulators, which is sometimes referred to as low-gain free electron laser \cite{madey1971stimulated}.
  
  It is always helpful to figure out the principle of undulator radiation within the rest frame of relativistic electrons.
  Suppose the undulator period is equal to $\lambda_u$.
  Then, the electron beam, travelling with Lorentz factor $\gamma$, observes this period as $\lambda^*=\lambda_u/\gamma$ and the radiation will have the same wavelength in electron coordinate.
  Once this radiation is transformed to the laboratory frame, the relativistic Doppler effect implies that the observed radiation is equal to ($\lambda^*/2\gamma = \lambda_u/2\gamma^2$).
  For example, when an electron bunch with 500\,MeV energy equivalent to $\gamma\approx1000$ travels through an undulator with period $\lambda_u=1\,cm$, the radiation spectrum will have a central wavelength around $\lambda=5\,nm$.
  
  More accurate treatment of the undulator radiation problem with considering the sinusoidal shape and longitudinal velocity yields the following equation for the radiation wavelength:
  \begin{equation}
  \lambda_l=\frac{\lambda_u}{2\gamma^2}\left(1+\frac{K^2}{2}\right),
  \label{UndulatorWavelength}
  \end{equation}
  where $K=eB_0\lambda_u/(2\pi m_e c)$ is the so-called undulator parameter, with $B_0$ being the peak magnetic field at the center of the undulator.
  Using Larmor equations, the radiated power in one undulator period is found to be
  \begin{equation}
  P_u=\frac{e^4 \gamma^2 B^2}{12\pi \epsilon_0 c m_e^2}.
  \label{UndulatorWavelength}
  \end{equation}
  The initial radiation is incoherent and scales with the number of electrons $N_e$.
  
\begin{figure} \centering
  	$\begin{array}{cc}
  	\includegraphics[draft=false,width=2.2in]{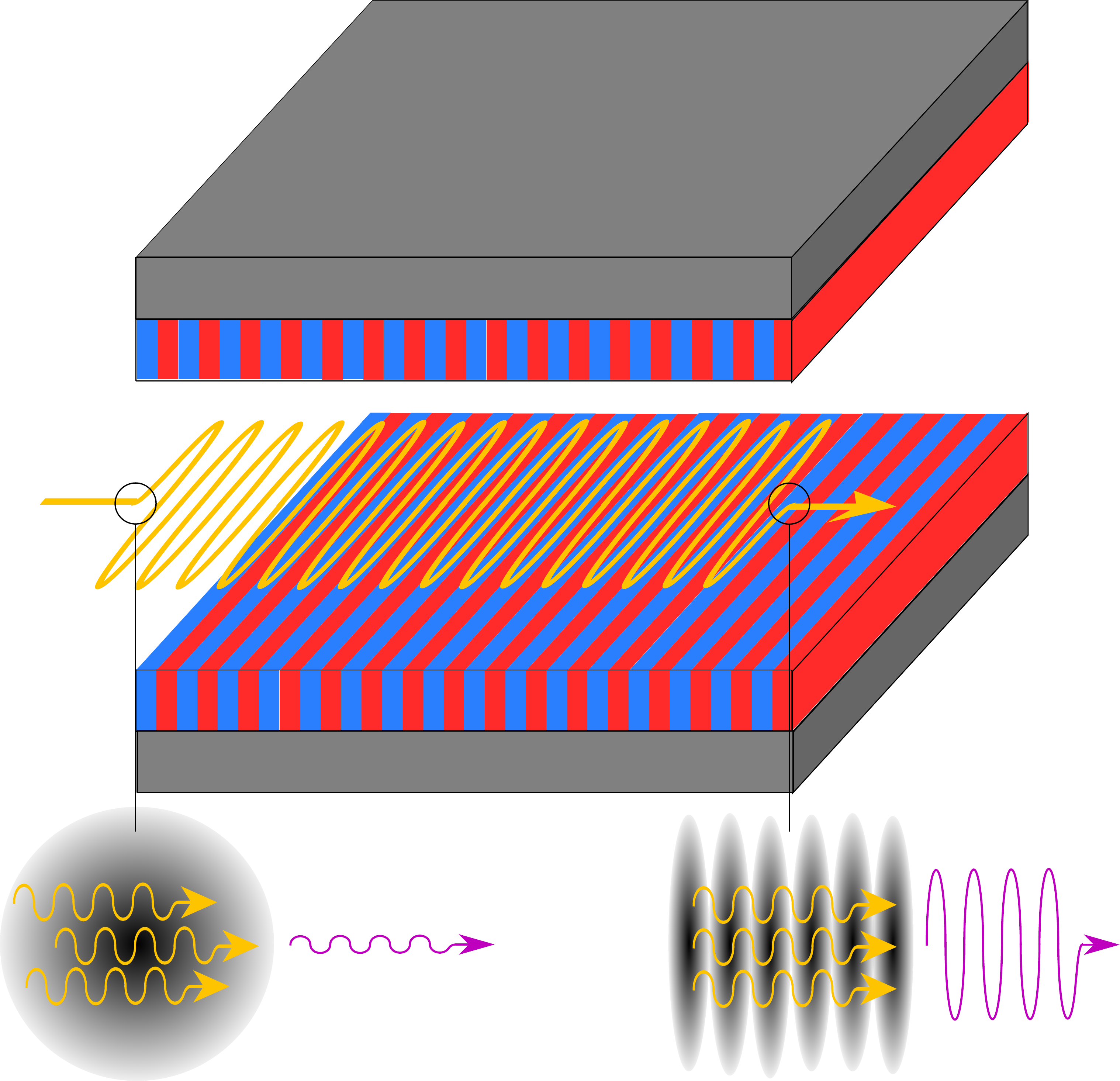} &
  	\includegraphics[draft=false,width=3.0in]{TESLARadiation.pdf} \\
  	(a) & (b)
  	\end{array}$
  	\caption[Free-electron laser operation principle: (a) schematic illustration and (b) radiation gain due to micro-bunching (The data are obtained at the
  	SASE FEL of the TESLA Test Facility]{Free-electron laser operation principle: (a) schematic illustration and (b) radiation gain due to micro-bunching (The data are obtained at the SASE FEL of the TESLA Test Facility \cite{ayvazyan2002generation}).}
  	\label{FELScheme}
  \end{figure}

  The radiation of a set of random emitters can be evaluated by the following equation:
  \begin{equation}
  P_t = P_u N_e \left( 1 + (N_e - 1) f(\omega) \right),
  \label{BunchRadiation}
  \end{equation}
  where $f(\omega)=\int f(t) \exp(i\omega t) dt$ is the Fourier transformation of bunch distribution function.
  In the above equation, the first term is usually referred to as the incoherent radiation and the second term is considered as the coherent term.
  The second term, scaling quadratically with $N_e$, would happen if the Fourier transformation function is large enough.
  In other words, bunches with lengths comparable to the light wavelength can introduce a considerable dominant term.
  Such a condition is not realistic in optical, UV and x-ray regime.
  However, in free electron lasers after a certain propagation length along the undulator, a process of self-organization on the scale of the radiated light wavelength is triggered.
  This process, usually called micro-bunching introduces a strong Fourier coefficient term at the radiation wavelength.
  Consequently, the radiation scales with $N_e^2$, if the undulator is long enough.
  This is the main reason for distinguishing between a short undulator resulting in a low-gain FEL and a long undulator realizing a high-gain FEL (Fig.\,\ref{FELScheme}.
  As mentioned above, the FEL light in the high gain regime is a coherent, monochromatic, polarized, extremely bright and tightly collimated beam.
  These properties are the origin for the extensive interest in high-gain FELs and their promising applications.
  
  The described concept of beam instability and the consequent high gain FEL were first introduced by Pellegrini and Bonifacio \cite{bonifacio1984collective}.
  Twenty years after its proposal, the high-gain FEL theory became the main baseline for designing hard x-ray light sources \cite{pellegrini2016physics,huang2007review}.
  The successful operation of the soft x-ray FLASH facility \cite{ackermann2007operation} in Germany and the spectacular commissioning of the Linac Coherent Light Source (LCLS) at SLAC were the culmination of nearly 35 years of continuous advances in electron beam and FEL physics.
  Other FEL facilities in South Korea, Japan, Italy and Switzerland were also successfully commissioned.
  Recently, the 3.4\,km long facility European-XFEL produced x-ray light with a wavelength of 1.4 nanometre (900 eV).
  With some further development, FEL facilities will play a vital role as the premier source of tunable, intense, coherent photons of either ultra-short time resolution or ultra-fine spectral resolution, from the far infrared to the hard x-ray regime.
  
\begin{figure} \centering
  	\includegraphics[draft=false,width=3.0in]{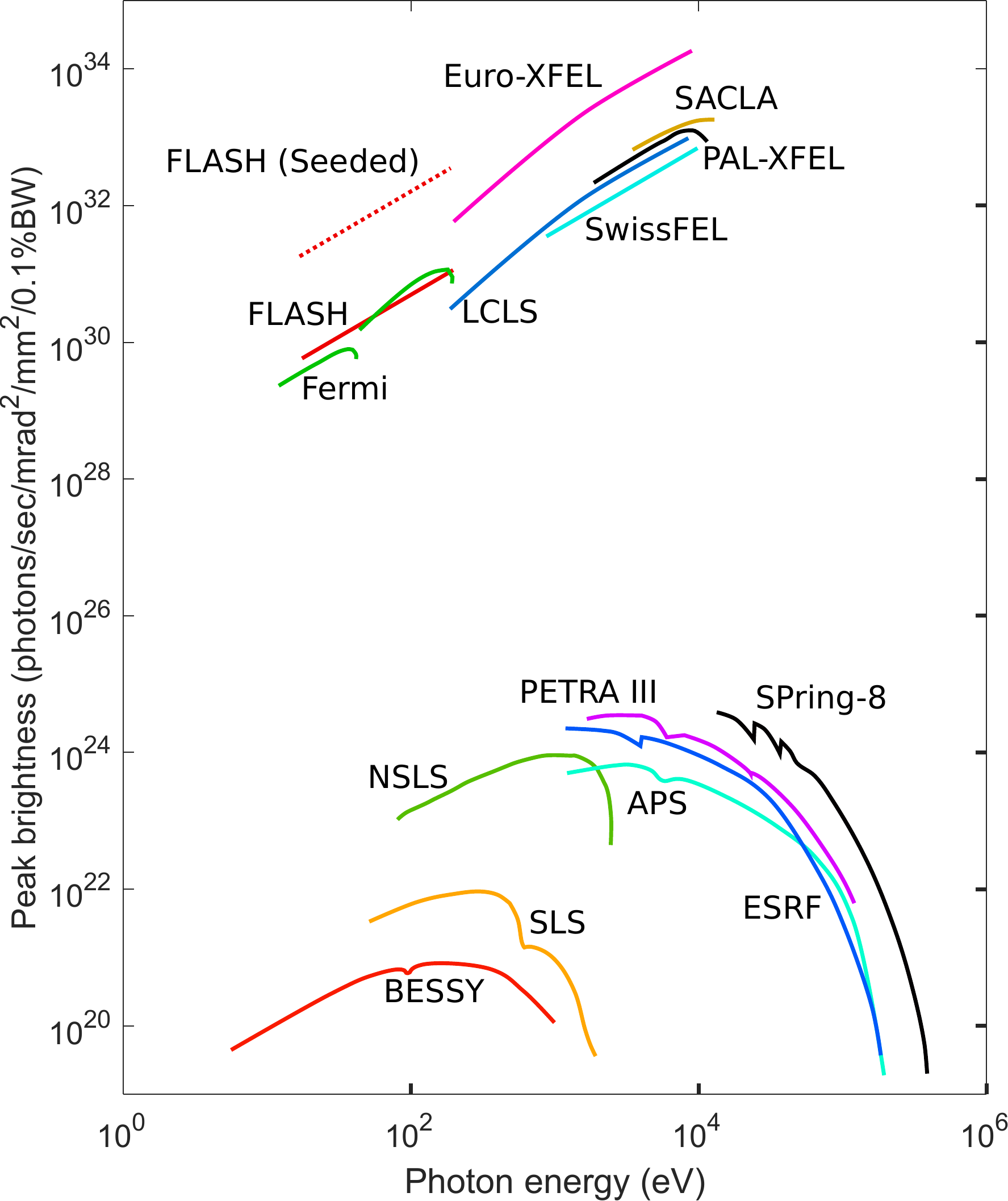}
  	\caption{Peak brightness plotted as a function of photon energy for various light sources around the world.}
  	\label{brightnessComparison}
  \end{figure}

  The spectral brightness of a light source is the ultimate parameter used to assess the quality of the delivered radiation.
  This parameter is defined as the intensity of a radiation source taking into account its spectral purity and opening angle
  \begin{equation}
  B = \frac{\Phi}{4\pi^2 \Sigma_x \Sigma_{\theta x} \Sigma_y \Sigma_{\theta y}}
  \label{Brightness}
  \end{equation}
  where $\Phi$ is the spectral photon flux defined as the number of photons per second and within a given relative spectral bandwidth $\Delta \omega_l/\omega_l$.
  The brightness determines how much monochromatic radiation power can be focused onto a tiny spot on the target.
  Fig.\,\ref{brightnessComparison} compares the brightness of various currently operating light sources around the world.
  The drastically brighter light produced by FEL technology is vividly observed in this plot.
  
  \subsection{Inverse Compton Scattering Sources}
  Besides numerous investigations and studies on the x-ray FELs, research efforts have been devoted to building compact x-ray sources, where novel schemes for generating x-ray radiation in a so-called tabletop setup are examined and verified.
  These efforts can be categorized into two categories: (i) research on compact accelerators, and (ii) compact undulator studies.
  As explained throughout the accelerator part of this chapter, there are currently several pathways towards making miniaturized accelerator modules, like the THz acceleration scheme.
  The second set of approaches aims at using compact undulators like cryogenic undulatos \cite{hara2004cryogenic} and optical undulators, where the oscillations in an electromagnetic wave realize the wiggling motion of electrons \cite{kartner2016axsis,bacci2008compact,chang2013high,gallardo1988theory,bacci2006transverse,graves2012intense}.
  Sources based on optical undulators are typically referred to as \emph{Inverse Compton Scattering} (ICS) or \emph{Thomson Scattering} (TS) sources.
  Several studies have predicted the possibility of achieving FEL like performance using optical undulators \cite{bacci2008compact,gallardo1988theory,bacci2006transverse}.
  
  The first ICS x-ray sources were proposed soon after the discovery of lasers in 1963 \cite{arutyunyan1963compton,milburn1963electron} and experimentally demonstrated one year after the first proposals \cite{kulikov1964compton}.
  Since then, research efforts were devoted to theoretical description of ICS sources \cite{sprangle1992tunable,esarey1993nonlinear,ohgaki1994linearly}.
  The compactness and ability to produce very high energy photons up to the Gamma ray regime \cite{phuoc2012all} is a remarkable peculiarity of ICS sources.
  A comparison between the electron trajectories in a static undulator \cite{schmuser2014free} and in an optical one \cite{esarey1993nonlinear} shows that the equations governing static undulators are the same as optical undulator radiation equations if the following mappings are considered:
  \begin{equation}
  K \leftrightarrow a_0 \qquad \mathrm{and} \qquad \lambda_u \leftrightarrow \lambda_l/2,
  \label{static2OpticMapping}
  \end{equation}
  where $a_0$ and $K$ stand for normalized vector potential and undulator strength parameters, respectively. $\lambda_l$, and $\lambda_u$ denote laser wavelength and undulator period, respectively.
  Therefore, the dominant radiation wavelength in an ICS source is calculated by
  \begin{equation}
  \lambda_X=\frac{\lambda_l}{4\gamma^2}\left(1+\frac{a_0^2}{2}\right).
  \label{UndulatorWavelength}
  \end{equation}
  For example, when an electron bunch with 20\,MeV energy, equivalent to $\gamma\approx40$ interacts with a counter-propagating laser with wavelength $\lambda_l=1$\,{\textmu}m and normalized vector potential $a_0=0.1$, the radiation spectrum will have a central wavelength around $\lambda=0.23\,$nm.
  The possibility of achieving very small radiation wavelengths with MeV-level electron energies is the key feature in ICS sources leading to their compactness.
  They are presently incorporated with conventional particle accelerators to provide high energy photon beams.
  Examples of these sources are Thomson scattering source at Lawrence Berkeley National Laboratory (LBNL) \cite{schoenlein1996femtosecond,leemans1996x}, the High Intensity Gamma-Ray Source (HIGS) at Duke University \cite{carroll1999production,pogorelsky2000femtosecond}, the TREX/MEGA-Ray facility at Lawrence Livermore National Laboratory \cite{albert2010characterization}, Laser Synchrotron Source (LSS) at Naval Research Laboratory (NRL) \cite{ting1995observation}, Thomsan scattering source in Helmholtz Zentrum Dresden R\"{o}ssendorf (HZDR) \cite{jochmann2013high}, and Compact Light Source developed by Lynceantech \cite{eggl2016munich}.
  
  Although the beam brightness and coherence of these compact light sources are not comparable with the photon beams delivered by FELs, they are powerful tools for nuclear resonance fluorescence (NRF), radiography and photo-fission studies for the detection of nuclear materials \cite{bertozzi2008nuclear,albert2011design}.
  Performing Protein crystallography and phase-contrast imaging using ICS sources are also reported \cite{bech2009hard,abendroth2010x}.
  Such applications motivate combination of novel and compact acceleration schemes with the ICS source concept to realize table-top x-ray and Gamma-ray sources.
  In \cite{rykovanov2014quasi}, combination of laser-plasma acceleration with the ICS concept is discussed and examined.
  Using X-band accelerators in burst mode with the ICS scheme is the key concept behind the compact x-ray source at Arizona State University (ASU), producing high energy photon beams at 100\,kHz \cite{graves2014compact}.
  The ultimate goal of the presented research in this habilitation treatment is the conceptual design of a THz-driven ICS source, which uses laser-driven THz sources to perform THz acceleration and eventually produce high energy photon beams through an ICS interaction \cite{kartner2016axsis}.
  This is the simplest version of an x-ray source using a THz based gun and accelerator developed in the AXSIS program at DESY, which will later be further developed towards a coherently emitting source \cite{kartner2016axsis}.
  
  \section{Overview of the Habilitation Thesis}
  
\begin{figure} \centering
  	\includegraphics[draft=false,width=5.0in]{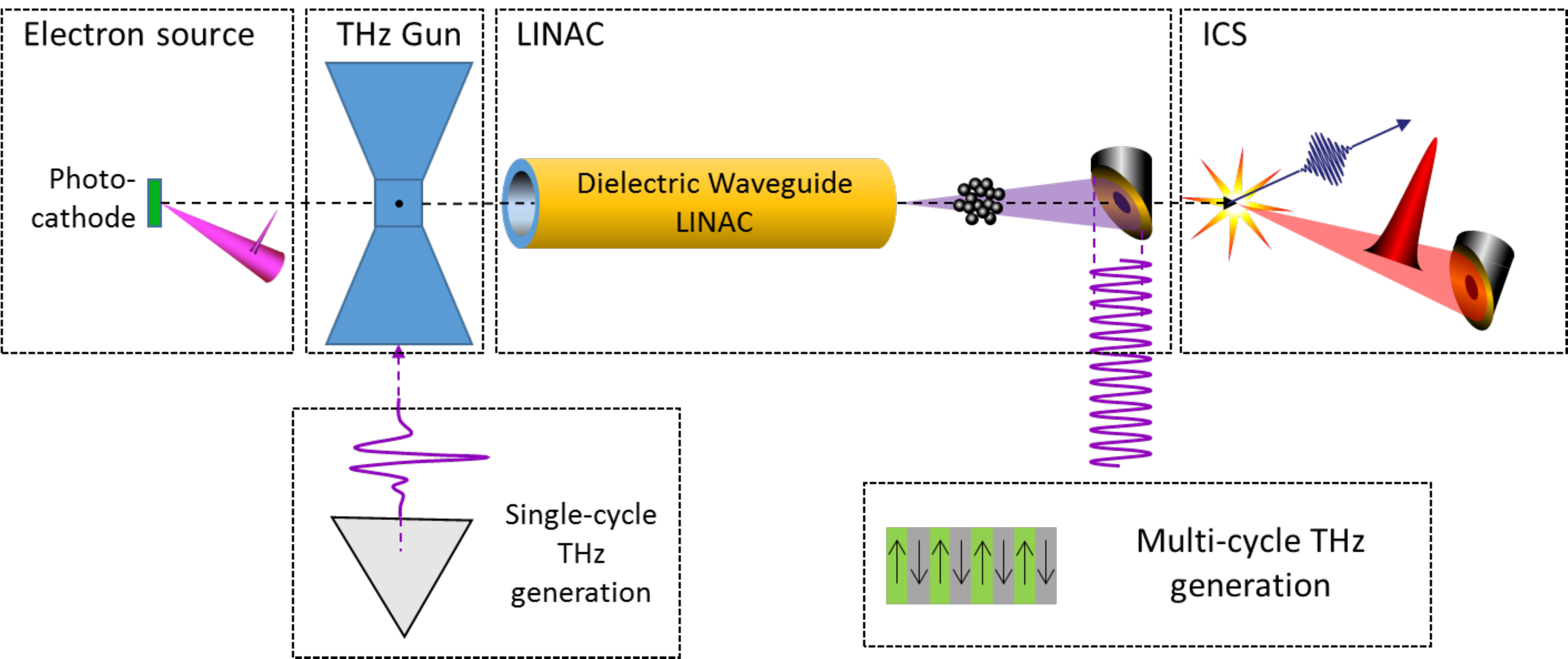}
  	\caption{Schametic layout of a THz-driven x-ray source.}
  	\label{AXSISFacility}
  \end{figure}

  The proceeding chapters of the thesis present the research carried out to develop the required instruments for a THz-driven light source facility.
  A schematic sketch of such a facility is illustrated in Fig.\,\ref{AXSISFacility}.
  The ingredients of the designed facility is similar to any other light source.
  Electrons are created using an electron source and injected to an electron gun (THz gun) which transfers energy to electrons up to relativistic regime.
  These relativistic electrons are received by a THz linac that provides a 15-20\,MeV electron beam.
  Subsequently, the ultra-relativistic particles interact with a counter-propagating laser beam to produce x-ray photons.
  Based on reasons which are thoroughly explained in the next chapters, the electron gun is fed by single-cycle THz pulses and the linac field is provided by a multi-cycle THz generation scheme.
  Each chapter is dedicated to each stage and written in a manner that can be independently studied.
  
  The thesis starts with a simulation chapter, where techniques for precise analysis of components are described.
  First, we introduce a hybrid technique based on the discontinuous Galerkin time domain (DGTD) and the particle in cell (PIC) simulation methods for the analysis of acceleration of charged particles using electromagnetic fields.
  The DGTD algorithm is a three-dimensional, dual-field and fully explicit method for efficiently solving Maxwell equations in the time domain on unstructured grids.
  On the other hand, the PIC algorithm is a versatile technique for the simulation of charged particles in an electromagnetic field.
  We discuss a novel strategy for combining both methods to solve for the electron motion and field distribution when a beam interacts with an electron bunch in a very general geometry.
  The developed software offers a complete and stable numerical solution of the problem for arbitrary charge and field distributions in the time domain on unstructured grids.
  This full-wave technique will be the method for the analysis of particle accelerators, and electron field emitters.
  Second, the analysis method for ICS process is presented.
  The method solves for the particle trajectories using PIC simulations and updates the radiation on an assumed detector in front of the beam according to Li\'{e}nard-Wiechert potential.
  Using this simple technique the complete radiation properties of an ICS interaction is obtained.
  The last part of this chapter revolves around full-wave simulation of free electron lasers.
  The highly sophisticated dynamics involved in a FEL process was the main obstacle hindering the development of general simulation tools for this problem.
  A numerical algorithm based on finite difference time domain/Particle in cell (FDTD/PIC) in a Lorentz boosted coordinate system is presented, which is able to fulfill a full-wave simulation of a FEL process.
  The developed software offers a suitable tool for the analysis of FEL interactions without considering any of the usual approximations.
  A coordinate transformation to bunch rest frame makes the very different length scales of bunch size, optical wavelengths and the undulator period transform to values with the same order.
  Consequently, FDTD/PIC simulations in conjunction with efficient parallelization techniques make the full-wave simulation feasible using the available computational resources.
  Several examples of free electron lasers are analyzed using the developed software, the results are benchmarked based on standard FEL codes and discussed in detail.
  
  In chapter 3, the research activities on electron source analysis, development and characterization are reviewed.
  This chapter consists of three separate parts which focus on conventional flat photocathodes, structured electron sources, and electron source characterization techniques.
  The well-established physics of a flat photocathode and its analysis techniques are firstly described.
  Next, we report the design, modelling, fabrication, and experimental characterization of a novel ultrafast optical field emission cathode comprised of a large, dense and highly uniform array of nano-sharp high-aspect ratio silicon columns.
  Such field emitters offer an attractive alternative to conventional photocathodes, while providing a direct means of structuring the emitted electron beam.
  Another candidate for nano-structured photocathodes is flat gold nano-rods which exploit the plasmonic enhancement of light for ultrafast and efficient emission of structured electron beams.
  The design, fabrication and characterization of Au nanorod optical field emitter arrays are also demonstrated in this section.
  Next, our recently developed methods for electron source characterization are described.
  The techniques consist of velocity-map imaging (VMI) of various electron sources as well as mapping the spatial emission of electrons onto electron-beam resist materials.
  
  Chapter 4 presents a review of the efforts towards miniaturized THz guns.
  This review starts with the concept of producing ultrashort ($\sim$fs) high charge ($\sim$pC) bunches from ultra-compact guns utilizing single-cycle THz pulses.
  It is shown that the readily available THz pulses with energies as low as 20\,{\textmu}J are sufficient to generate multi-10\,keV electron bunches.
  Moreover, It is demonstrated that THz energies of 2\,mJ are sufficient to generate relativistic electron bunches with higher than 2\,MeV energy.
  After the conceptual presentation, an optimized design strategy for these electron guns is outlined.
  We start with designing a gun delivering 400 keV electron beam energy and discuss different techniques to enhance the performance.
  Subsequently, upgrading the design to an 800 keV device is discussed.
  The experimental tests of a single-layer THz gun and a multilayer structure are reviewed in the next section.
  Through these experiments, the feasibility and promise of ultrafast devices for high accelerating gradients are demonstrated.
  we demonstrate an all-optical THz gun yielding peak electron energies approaching 1\,keV, accelerated by $>$300 MV/m THz fields in a novel micron-scale waveguide structure.
  Afterwards, a segmented terahertz electron accelerator and manipulator (STEAM) is introduced, which is capable of performing multiple high-field operations on the 6D-phase space of ultrashort electron bunches.
  With this single device, powered by few-micro-Joule, single-cycle, 0.3 THz pulses, we demonstrate record THz-acceleration of $>$30\,keV, streaking with $<$10\,fs resolution, focusing with $>$2\,kT/m strengths, compression to $\sim$100\,fs as well as real time switching between these modes of operation.
  
  THz linac is the discussed topic in chapter 5.
  The chapter begins with the numerical investigation of the acceleration and bunch compression capabilities of 20\,mJ, 0.6\,THz-centered coherent terahertz pulses in optimized metallic dielectric-loaded cylindrical waveguides.
  In particular, we theoretically demonstrate the acceleration of 1.6\,pC and 16\,pC electron bunches from 1 MeV to 10 MeV over an interaction distance of 20\,mm.
  In addition, the compression of a 1.6\,pC 1\,MeV bunch from 100\,fs to 2\,fs (50 times compression), and a 1.6\,pC 10 MeV bunch from 100 fs to 1.61 fs (62 times) are also theoretically demonstrated.
  As described in the radiation source section, the proposed schemes toward high power THz generation are capable of producing short pulses, which dictates the study of particle acceleration in the pulsed regime rather than continuous-wave regime.
  Consequently, various effects such as group velocity mismatch and group velocity dispersion start to influence the acceleration scenario and impose limits on the maximum energy gain from the pulse.
  Therefore, the chapter elaborates covering design methodologies to optimize the THz linac performance.
  Finally, the chapter is enclosed by experimental demonstration of linear electron acceleration in a THz waveguide.
  
  The last chapter of the thesis tries to utilize the developed concepts to perform a start-to-end simulation of a fully THz-driven table-top x-ray source.
  All of the elements in the source are fed by 1\,{\textmu}m laser technology, offering the unique possibility of inherent synchronization.
  The required terahertz pulses to excite the accelerators are categorized as single-cycle and multi-cycle pulses.
  Two single-cycle 400\,{\textmu}J pulses with central frequencies at 300\,GHz are generated using optical rectification (OR) of picosecond pulses using a titled-pulse-front setup.
  In parallel, four 554\,ps multi-cycle pulses with 10\,mJ energy centered at 300\,GHz are produced using difference frequency generation (DFG) of two 1\,J laser pulses.
  The single-cycle THz pulses feed an ultrafast electron gun, where a 1\,pC electron bunch is generated through photoemission off a flat copper surface excited by a UV laser beam.
  This electron gun delivers a 600\,fC electron bunch with 0.78\,MeV kinetic energy, which is immediately injected into a dielectric-loaded metallic waveguide operating as a linear accelerator (linac).
  At the input of the linac, a coupler is designed which simultaneously couples and combines the four multi-cycle THz beams into a single TM$_{01}$ mode of a dielectric-loaded metallic waveguide.
  To keep the electron bunch confined in the linac, a quadruple lattice is used to control the bunch size over and after the linac.
  The set of linac and quadruples deliver a 360\,fC electron bunch with 19\,MeV beam energy, which is then transported to an inverse Compton scattering (ICS) stage.
  At the ICS interaction point, the 19\,MeV electron beam scatters off a 100\,mJ 1\,{\textmu}m laser beam and generates an x-ray beam with 6.7$\times$10$^4$ photons per shot with photon energies 2\,keV$<E_p<$7\,keV.
  This thesis closes with a summary and outlook to the future possibilities in Chapter 7. 
  
  \chapter{Simulation Techniques in Light Source Technology \label{chap:two}}
  
  \section{Introduction}
  
  The interaction of charged particles with an electromagnetic field occurs in many applications and devices, ranging from radiation sources \cite{saldin2013physics,schmuser2014free} and accelerator physics \cite{wiedemann2015particle,wangler2008rf}, to imaging science and spectroscopy \cite{egerton2011electron}.
  Many of today high power sources such as microwave travelling wave tubes (TWT) and klystrons \cite{pierce1950traveling}, gyrtorons \cite{flyagin1977gyrotron} and magnetrons \cite{collins1948microwave}, synchrotron radiation sources, THz and x-ray FEL are performing based on free electron motion in a properly designed electromagnetic field.
  In accelerator physics, energy transfer to particles is achieved by the action of an electromagnetic wave, either in a cavity or a waveguide.
  Moreover, there exists also a reaction from charged particles to the field in form of radiation.
  The mutual interaction between electrons and a laser beam constitute the fundamentals for developing advanced sources based on inverse Compton scattering \cite{kartner2016axsis,bacci2008compact,chang2013high,gallardo1988theory,bacci2006transverse,graves2012intense,phuoc2012all} and undulator radiation \cite{saldin2013physics,schmuser2014free}.
  In such cases, not only the action of the field on an electron bunch is studied but also the back-action of free charges on the field distribution.
  This type of interaction also plays a major role in the acceleration using wake-fields of particles \cite{andonian2012dielectric}.
  This chapter specializes in simulation techniques used for the analysis of mutual interaction between electromagnetic fields and charged particles.
  
  The numerical algorithms presented in this chapter can be grouped into three categories:
  (\emph{i}) Numerical calculation of particle trajectories inside the numerically simulated electromagnetic field, which is useful for modelling complicated structures where analytical models suffer from inaccurate approximations.
  These techniques are explained in section 2.2 and are later used to simulate structured electron sources and THz guns.
  (\emph{ii}) Numerical calculation of particle trajectories inside analytically calculated electromagnetic field, which is useful for modelling ICS interaction or particle acceleration in waveguides, since analytical formulation of Gaussian beams and waveguide modes are sufficiently accurate.
  This group of algorithms are introduced in section 2.3 and is the base for the simulation of ICS and THz linac in this thesis.
  (\emph{iii}) Numerical calculation of particle trajectories inside analytically calculated electromagnetic field and numerical simulation of particle radiation:
  This type of algorithms, described in section 2.4, is useful for modelling FEL interaction or even ICS interaction including the effect of particle radiation on the bunch distribution.
  
  \section{Simulation of Particle Accelerators}
  
  Due to the broad range of applications, various algorithms are developed to solve charged particle interaction with an electromagnetic wave.
  A widely used technique considers the stream of charges as a current density in the Maxwell's equations and solves for the fields and the electric current simultaneously \cite{talebi2013numerical}.
  This task can be incorporated in many standard algorithms like finite element method (FEM), finite difference time domain (FDTD) as well as method of moments (MoM), and is available in some of the existing commercial software packages \cite{ansoft20093}.
  However, the method treats charge distribution macroscopically and does not support for studying the internal bunch profile evolutions during the interaction.
  Hydrodynamic models for the electron bunch based on distribution functions are developed to mitigate this problem.
  Direct consideration of transient distribution functions in tandem with the electromagnetic fields result in the so-called Maxwell-Vlasov equations, widely studied during 1980's \cite{marsden1982hamiltonian,morrison1980maxwell}.
  A detailed description of solving Maxwell-Vlasov equations for plasma using discontinuous Galerkin (DG) approach is presented in \cite{seal2012discontinuous}.
  These developed models, although very helpful, again consider the cumulative effects of charge distributions and are able to make approximate predictions on the microscopic properties of the bunch.
  In contrast, with PIC codes \cite{dawson1983particle,verboncoeur2005particle} variations in the microscopic bunch parameters can be simulated and are therefore standard computational techniques for simulating beam dynamics.
  
  In PIC simulations, the equations of motion are consecutively updated for a particular electromagnetic field profile.
  The field distribution is obtained by either using analytical formulations or importing some previously solved numerical values.
  The former procedure is often followed in solving for the interaction between optical beams or waveguide modes and a charge distribution, where approximate analytical solutions (e.g. Gaussian and Bessel-Gaussian beams) are available for the fields \cite{salamin2006electron,Wong2013,wong2010direct}.
  Nevertheless, the accuracy of using analytical solutions for such purposes is under debate among scientists \cite{marceau2013validity}.
  The use of numerical field solutions is indeed a standard technique in designing accelerator cavities, where the harmonic solutions of the fields in cavities are used as input to the PIC algorithm.
  However, the method suffers from harsh limitations when short pulses are influencing a bunch, which is of utmost importance in this thesis and has received substantial attention due to the possibilities opened in ultrafast optics \cite{rausch2008controlled} and THz sources \cite{fulop2012generation,huang2015highly}.
  Therefore, the time domain numerical simulation of field propagation when acting on a bunch with considering the microscopic effects is very often encountered in the domains mentioned above.
  
\begin{figure} \centering
  	\includegraphics[draft=false,width=3.0in]{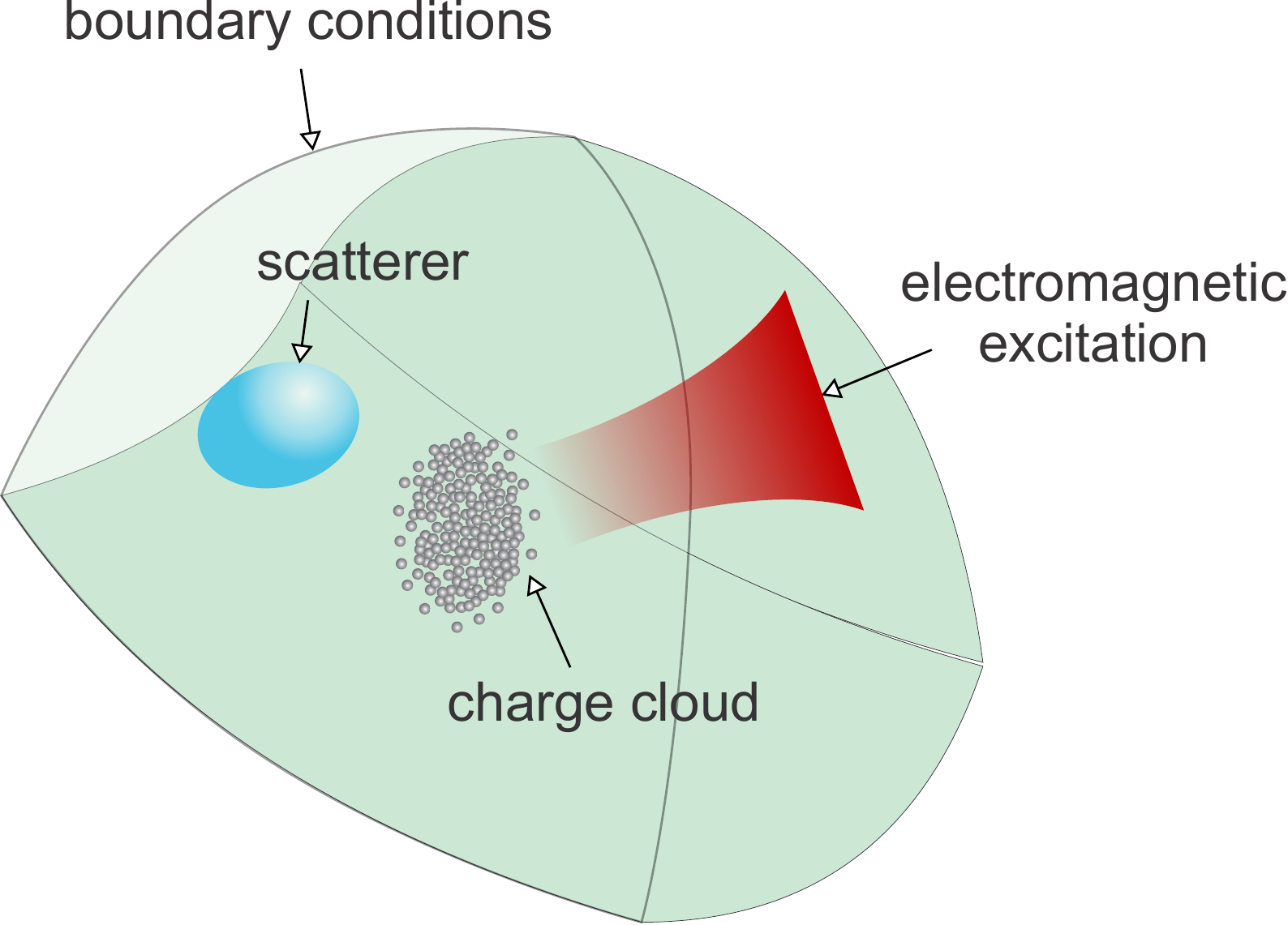}
  	\caption{Schematic illustration of the general problem of light and electron interaction.}
  	\label{lightElectronInteraction}
  \end{figure}

  Fig.\,\ref{lightElectronInteraction} shows a schematic illustration for the general problem of light-electron interaction.
  An arbitrary electromagnetic beam excites fields inside a computational domain with known boundary conditions, including a set of scatterers and an initial charge distribution.
  The algorithm has to solve for the evolution of the electromagnetic fields and charge distribution.
  The goal is to find a solution by providing a time marching mechanism that propagates the electromagnetic fields in a computational domain and simultaneously solves for bunch evolution.
  Firstly, we need to decide on a rigorous time domain method for solving Maxwell's equations.
  FDTD is a superior choice with high efficiency, versatility and flexibility \cite{taflove2005computational}.
  Extensive research efforts have been devoted to develop FDTD/PIC codes leading to software packages like SELFT \cite{kern1996numerische}, MAGIC \cite{goplen1995user}, MAFIA \cite{weiland1996time}, WARP \cite{grote2005warp} and PIConGPU \cite{burau2010picongpu}.
  However, this method suffers from severe limitations being second order accurate in time and space and only amenable to uniform Cartesian grids.
  Techniques such as sub-gridding and split-material voxels have been proposed, without completely solving this problem.
  In this regard, the beginning of twenty-first century  witnessed progress in the discontinuous Galerkin methods for solving time domain electromagnetic equations with high order accuracy and additionally on unstructured grids \cite{hesthaven2002nodal,hesthaven2004high,kabakian2004unstructured,stannigel2009discontinuous,busch2011discontinuous}.
  A complete comparison between the two methods based on finite-difference and discontinuous Galerkin for simulation of nano-photonic systems is presented in \cite{niegemann2009higher}.
  Therefore, we chose DGTD as the Maxwell solver and due to the previously outlined reasons the particle motion is solved by a PIC algorithm leading to the hybrid DGTD/PIC algorithm.
  
  Using DGTD as a kernel for solving Maxwell's equations and coupling it with a PIC algorithm is pioneered by Jacobs and Hesthaven \cite{jacobs2006high}.
  The method introduces equivalent charge $\rho(r)$ and current $J(r)$ densities, and projects them into the computational grid.
  It is also used to model RF accelerators and guns \cite{gjonaj2006accurate}.
  Despite the accuracy in modelling complex geometries, there exist several restrictions in computing particle radiation and wake-fields, which makes the method not suitable for problems involving ultrafast particle acceleration and tiny charge distributions.
  Some examples are (\emph{i}) consideration of smooth functions for charge distribution to avoid Gibbs type phenomena in the field computations, (\emph{ii}) numerical instabilities due to static charge and force build-up, (\emph{iii}) singularities in the field solution of the cell containing the charges, and (\emph{iv}) inability to distinguish between the radiated fields to avoid a charge being affected by its own radiation.
  These effects have been investigated extensively in various studies \cite{jacobs2006high,stindl2011comparison} and several techniques such as hyperbolic divergence cleaning are proposed to mitigate some of the aforementioned problems.
  In \cite{gjonaj2010particle}, advantage was taken from an exception of the above effects, which vanish on uniform Cartesian grids, to simulate plasma wake-field acceleration.
  This solution indeed ignores the outstanding advantage of DG approaches that is the capability of handling unstructured meshes, and is inefficient for problems where many different length scales are involved.
  
  \subsection{Discontinuous Galerkin Time Domain Method}
  For the numerical calculation of field profiles in time, we employ the high order discontinuous Galerkin formulation of Hesthaven and Warburton \cite{hesthaven2002nodal,hesthaven2004high}.
  The method focuses on solving Maxwell's equations for dispersive media
  \begin{equation}
  \nabla \times \vec{H} = \vec{J} + \vec{J}_p + \varepsilon_0 \frac{\partial \vec{E}}{\partial t}, \qquad \mbox{and} \qquad \nabla \times \vec{E} = - \mu_0 \frac{\partial \vec{H}}{\partial t}
  \label{MaxwellEquationDGTD}
  \end{equation}
  where $\vec{J}_p=\partial \vec{P}/\partial t$ stands for the polarization current in the material and $\vec{J}=\vec{J}_0+\sigma \vec{E}$ represents the total flowing current.
  In order to solve this problem, the computational domain is tessellated into M tetrahedral elements $\Omega^m$.
  In each element, the fields and currents are written as an expansion in terms of a set of presumed basis functions
  \begin{equation}
  \vec{Q}^m(\vec{r},t) = \sum_j^N q^m_j(t) \vec{w}_{j}(\vec{r}) \mathrm{,}
  \label{QuantityExpansion}
  \end{equation}
  where $\vec{Q} \in \{\vec{E},\vec{H},\vec{J},\vec{J}_p \}$ represents any of the involved electromagnetic quantities and $N$ is the number of coefficients determined by the order of utilized basis functions.
  Throughout the formulation, the superscripts $m$ are used to refer to the $m$'th element.
  
  The basis functions assumed in this work are the hierarchical polynomial vector basis functions developed by Webb \cite{webb1999hierarchal}.
  The Webb's basis functions are polynomial functions that impose the continuity of expanded quantities over edges and faces of each element, which leads to the so-called edge, face and volume basis functions.
  As shown in \cite{webb1999hierarchal}, for a tetrahedral domain tessellation with polynomial order $n$, $6(n+1)$ edge basis functions, $4(n-1)(n+1)$ face basis functions, and $(n-2)(n-1)(n+1)/2$ volume basis functions are needed and sufficient to achieve a complete expansion basis.
  In other words, the total number of basis functions will be $N = (n+1)(n+2)(n+3)/2$.
  Compared to the nodal basis functions conventionally used in DG approaches \cite{hesthaven2002nodal}, one may refer to our implementation as a modal DG approach, since the coefficients do not correspond to field values at a specific node in the element.
  Due to the accurate inversion of the mass matrices in the DGTD algorithm, the computational cost of nodal and modal basis functions are the same.
  However, the Webb basis functions have benefits in computing the coupling of fields in adjacent elements and the possibility to distinguish between the rotational and irrotational functions.
  More accurately, only specific face and edge functions contribute to the coupling between elements sharing one face, with the contribution from other elements equal to zero.
  This fact results in considerably smaller floating point operations during the time update, thereby motivating the use of Webb basis functions.
  Besides, excluding irrotational basis functions from the expansion offered by Webb basis functions, leads to faster computation without loss in accuracy.
  Such set of functions are well-known as expansion sets with $n+1/2$ orders.
  
  The standard Galerkin method follows the weighted residual approach which satisfies the original equation based on weighted integrals.
  The expanded Maxwell equations (\ref{MaxwellEquationDGTD}) are multiplied with the basis functions and integrated over each element $\Omega^m$, which yields
  \begin{equation}
  \def\arraystretch{2.2}
  \begin{array}{l}
  \displaystyle \varepsilon_0 \sum_j^N S_{ij} \, \frac{de^m_j}{dt} = \sum_j^N  T_{ij} \, h^m_j + \sum_j^N \iint\limits_{\partial \Omega^m} \vec{w}_i \cdot ( \unitvec{n} \times \vec{H} ) \, ds - \sum_j^N S_{ij} \, (j^m_j + j^m_{pj} ) \\
  \displaystyle \mu_0 \sum_j^N S_{ij} \, \frac{dh^m_j}{dt} = - \sum_j^N  T_{ij} \, e^m_j - \sum_j^N \iint\limits_{\partial \Omega^m} \vec{w}_i \cdot ( \unitvec{n} \times \vec{E} ) \, ds
  \end{array}
  \mathrm{,}
  \label{expandedMaxwellEquations}
  \end{equation}
  where
  \begin{equation}
  S_{ij} = \iiint\limits_{\Omega^m} \vec{w}_i \cdot \vec{w}_j \, dv \qquad \mbox{and} \qquad T_{ij} = \iiint\limits_{\Omega^m} ( \nabla \times \vec{w}_i \cdot \vec{w}_j ) \,dv \mathrm{.}
  \label{matrixDefinition}
  \end{equation}
  As observed in the equations, a surface integral appears in the weighted residual formulation.
  This term is the term responsible for the coupling with the adjacent elements.
  The discontinuous nature of the expansion results in different values of tangential field quantities at the surface, if calculated from the expansion in either of the elements.
  This discontinuity leads to the so-called numerical flux through the element surface.
  The clever idea of DG approaches is defining a proper surface field obtained from the values at both elements to acquire a stable and convergent scheme in which the numerical flux tends to zero.
  Based on the upwinding flux theory introduced in \cite{shankar1989time}, the proper definitions of the tangential surface fields are
  \begin{equation}
  \def\arraystretch{2.2}
  \begin{array}{l}
  \displaystyle \unitvec{n} \times \vec{H} = \unitvec{n} \times \frac{Z^m \vec{H}^m + Z^l \vec{H}^l + \unitvec{n} \times (\vec{E}^m - \vec{E}^l) }{Z^m + Z^l} \\
  \displaystyle \unitvec{n} \times \vec{E} = \unitvec{n} \times \frac{Y^m \vec{E}^m + Y^l \vec{E}^l + \unitvec{n} \times (\vec{H}^m - \vec{H}^l) }{Y^m + Y^l}
  \end{array}
  \mathrm{,}
  \label{surfaceFields}
  \end{equation}
  where the superscript $l$ implies that the field value is extracted from the neighboring element.
  Further, we have introduced cell-impedances $Z^m = \sqrt{\mu^m/\varepsilon^m}$ and admittances $Y^m = \sqrt{\varepsilon^m/\mu^m}$.
  Using the above definition, the interelement coupling terms are obtained based on the fields at the local and adjacent elements.
  For the surfaces corresponding to particular boundary conditions such as open, perfect electric or perfect magnetic conductors, proper definitions are available and introduced in \cite{hesthaven2002nodal,hesthaven2004high,kabakian2004unstructured,stannigel2009discontinuous}.
  
  The developed formulation leads to explicit expressions for the time derivative of the coefficient vectors in terms of their values, i.e. the so-called initial value problem $\dot{\vec{q}} = f(\vec{q},t)$, with \vec{q} defined as a vector containing all the coefficients for various quantities.
  The remaining step of integrating the semi-discrete system in time is fulfilled through a fourth-order Runge-Kutta scheme.
  To achieve a stable time marching process the time step for the update must be less than a limit.
  This limit is empirically set according to the criterion introduced in \cite{kabakian2004unstructured}
  \begin{equation}
  \delta t \leq \zeta \delta x / c \mathrm{,}
  \label{timeStepLimit}
  \end{equation}
  where $\delta x$ is the minimum cell size, $c$ the speed of light and $\zeta$ a factor set according to the basis function order which is 1, 1/2, 1/4 and 1/5 for orders from 1 to 4, respectively.
  
  Using the presented DGTD algorithm a general Maxwell solver is developed that solves for the temporal variation of fields for arbitrary excitation and geometries.
  The software is written in C++ and is efficiently parallelized using the Message Passing Interface (MPI) library.
  The problem geometry is drawn and discretized using the Gmsh \cite{geuzaine2009gmsh} software.
  The DUNE \cite{bastian2008genericI,bastian2008genericII} and ALUGRID libraries are utilized for the mesh and grid management
  The space integrals are computed using Gaussian quadratures properly set for the specific order of basis functions.
  Moreover, for considering the dispersive material properties the formulation of auxiliary differential equations (ADE) is employed in conjunction with different material models, including Debye, Drude, and Lorentz.
  The developed DGTD software prepares a platform for propagating electromagnetic fields caused by an external excitation and the particle radiation.
  
\begin{figure} \centering
  	$\begin{array}{cc}
  	\includegraphics[draft=false,width=3.0in]{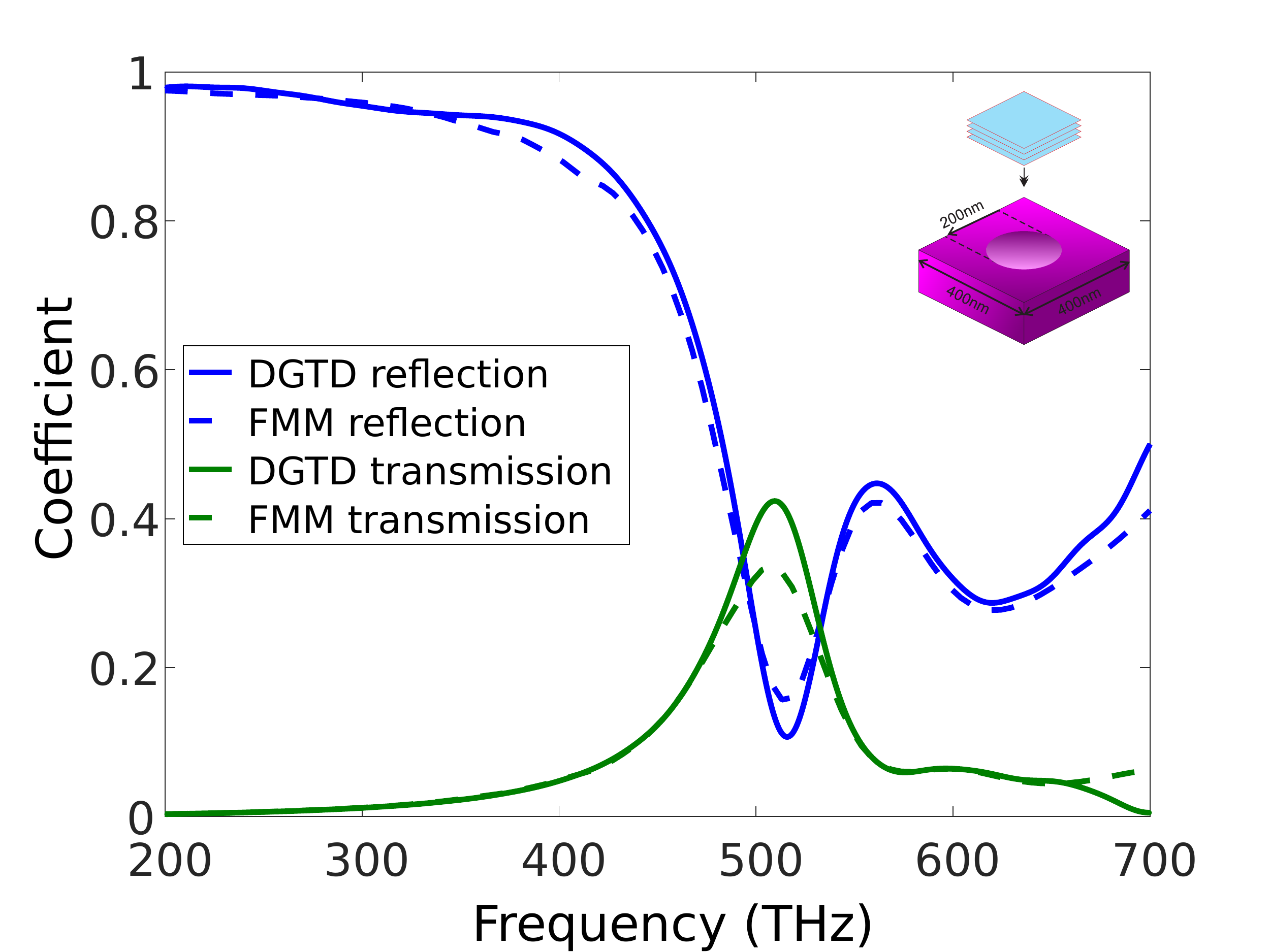} &
  	\includegraphics[draft=false,width=3.0in]{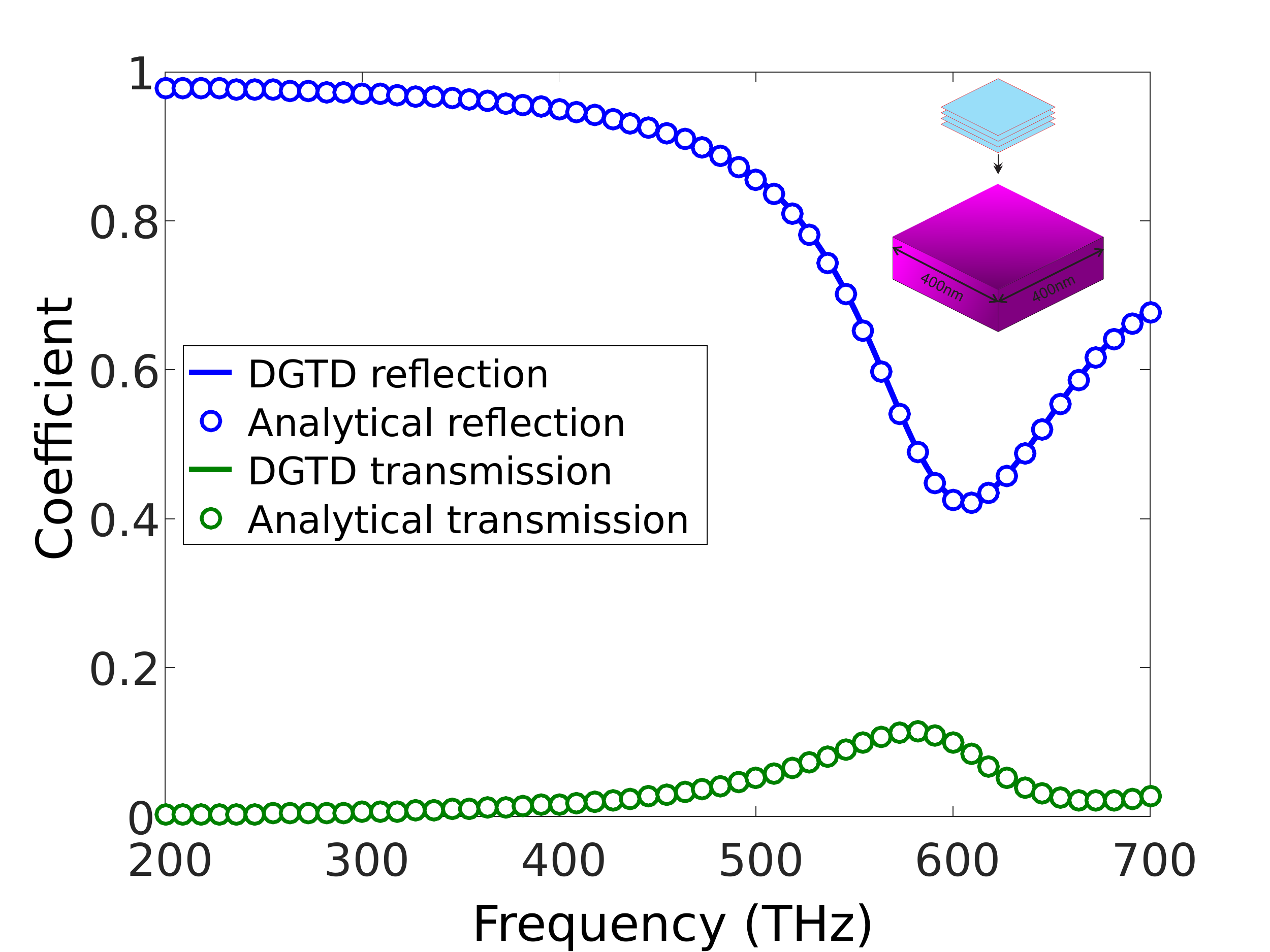} \\
  	(a) & (b)
  	\end{array}$
  	\caption{Normal incidence of a plane wave with Gaussian temporal variations on a gold photonic crystal slab considered for testing the DGTD code: (a) The reflection and transmission coefficient of the photonic crystal slab calculated using DGTD and FMM methods are compared. (b) The same analysis is done for a homogeneous gold slab and compared against analytical results.}
  	\label{DGTDBenchmark}
  \end{figure}

  As an standard process in software development, we examine the implementation through some benchmarks to assess its reliability and accuracy.
  To this end, we consider simple problems whose analytic solutions are available.
  For the DGTD implementation, normal incidence of a plane wave on a photonic crystal slab is analyzed.
  The subset of Fig.\ref{DGTDBenchmark}b shows the considered geometry of the unit cell as well as the plane wave excitation.
  The temporal signature of the plane wave amplitude is assumed to be a Gaussian signal with pulse duration 3\,fs and central wavelength 400\,nm.
  The host material of the photonic crystal slab is assumed to be gold, whose permittivity is obtained from the Drude-Lorentz model proposed in \cite{vial2005improved}.
  We assume periodic boundaries on the sides of the unit cell.
  The two upper and lower boundaries truncate the domain through 1st order absorbing boundary conditions (ABC).
  The total domain is tessellated into 50'083 elements, which leads to 10\,as time steps and 10\,ms computation time for each time update on an 8 core machine with Linux operating system.
  In Fig.\,\ref{DGTDBenchmark}b, we compare the results obtained using the DGTD code and another semi-analytical method, namely Fourier modal method (FMM) developed mainly for the analysis of planar periodic structures.
  In the frequency range of the excitation, good agreement is observed between the results.
  The discrepancy at higher frequencies occurs because of low resolution of the mesh compared to the wavelength.
  In addition, the small difference in reflection and transmission at resonance occurs because of the slow convergence of the FMM results at resonance frequencies.
  To validate this, we repeat the same study for a homogenous gold slab, where analytical solutions are available for both transmission and reflection coefficients, and the effect of boundary truncations are minimal.
  For a homogeneous dielectric slab with complex relative permittivity $\epsilon_r$ and thickness $d$, the reflection an incident plane-wave is derived using transmission-line theory that yield the following equation:
  \begin{equation}
  r = \frac{Z_{in} - Z_0}{Z_{in} + Z_0} \quad \mathrm{with} \quad Z_{in} = Z_s \frac{Z_0 + j Z_s \tan(\beta d)}{Z_s + j Z_0 \tan(\beta d)} \mathrm{,}
  \label{slabanalyticalFormulationReflection}
  \end{equation}
  where $Z_0=\sqrt{\mu_0/\epsilon_0}$ and $Z_s=\sqrt{\mu_0/(\epsilon_r \epsilon_0)}$ are the impedances of vacuum and slab material, respectively.
  $\beta=k_0\sqrt{\epsilon_r}$ is the complex wave propagation number in the slab with $k_0$ being the vacuum wave number.
  Again using the transmission-line theory, the transmission coefficient is obtained from :
  \begin{equation}
  t = ( r + 1 ) \cos (\beta d) + ( r - 1 ) \frac{Z_s}{Z_0} j \sin (\beta d) \mathrm{,}
  \label{slabanalyticalFormulationTransmission}
  \end{equation}
  Fig.\,\ref{DGTDBenchmark}c illustrates the problem and presents numerical and analytical results, which evidence a perfect agreement between the two solutions.
  
  \subsection{Particle In Cell Method}
  
  The PIC method is a general technique used to solve a certain class of partial differential equations encountered in various fields such as fluid dynamics and plasma physics \cite{dawson1983particle}.
  The success of the PIC method for plasma simulation owes to being relatively intuitive and straightforward to implement. %
  The method begins with integrating the equation of motion, which for relativistic electron bunches reads as
  \begin{equation}
  \begin{array}{rcl}
  \displaystyle \frac{\partial}{\partial t} (\gamma m_0 \vec{v} (\vec{r},t)) & = & -e ( \vec{E} (\vec{r},t) + \mu_0 \vec{v} \times \vec{H} (\vec{r},t) ) \\
  \displaystyle \frac{\partial \vec{r} }{\partial t}  & = & \vec{v} (\vec{r},t)
  \end{array}
  \mathrm{.}
  \label{motionEquation}
  \end{equation}
  In traditional PIC implementations, the fields of charges are interpolated to a pre-defined mesh and then a second interpolation returns the field values at the particle locations.
  
  For integrating the equation of motion, we use the same 4th order Runge-Kutta scheme as in DGTD implementation.
  This synergy has the advantage of directly using the calculated field values without the need for time interpolation and maintaining the results of the previous time steps.
  However, reading the DGTD field profile and calculating the values at different charge locations faces a serious difficulty.
  As previously emphasized, the main advantage of DGTD is its capability for handling various geometries, because it is developed for unstructured grids.
  A primitive way to find the tessellation element containing the point of interest in an unstructured grid is to start checking each element based on this criterion and stop the search as soon as the corresponding element is found.
  Imagine the simulation takes 10'000 particles (or macro-particles) into account.
  Additionally, the DGTD simulation contains 100'000 elements.
  Consequently, the described algorithm necessitates one billion element checks in each time step, which drastically increases the computation cost.
  The solution to this problem is the uniform grid mapping algorithm explained as follows.
  
\begin{figure} \centering
  	$\begin{array}{cc}
  	\includegraphics[draft=false,width=2.0in]{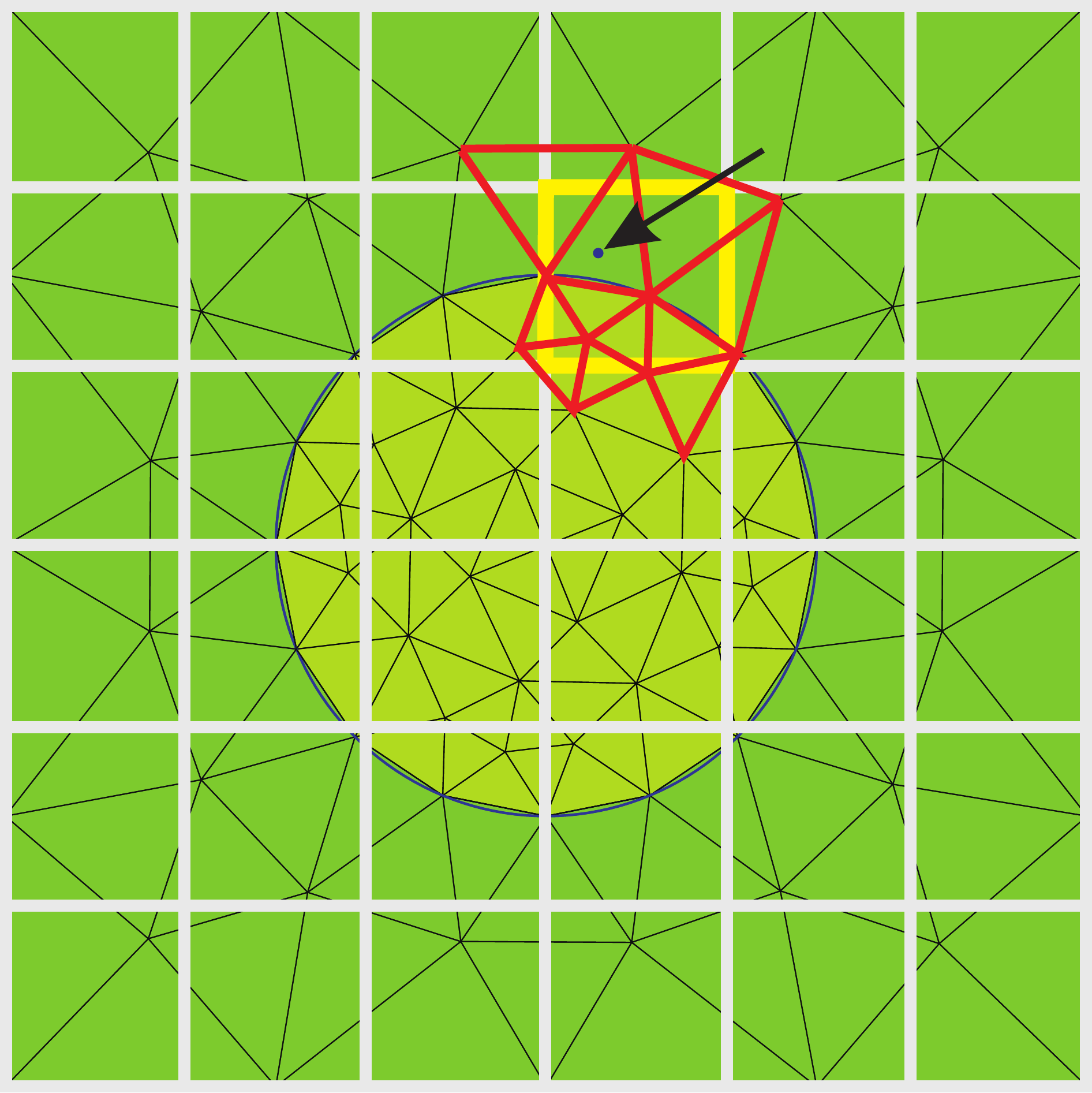} &
  	\includegraphics[draft=false,width=2.0in]{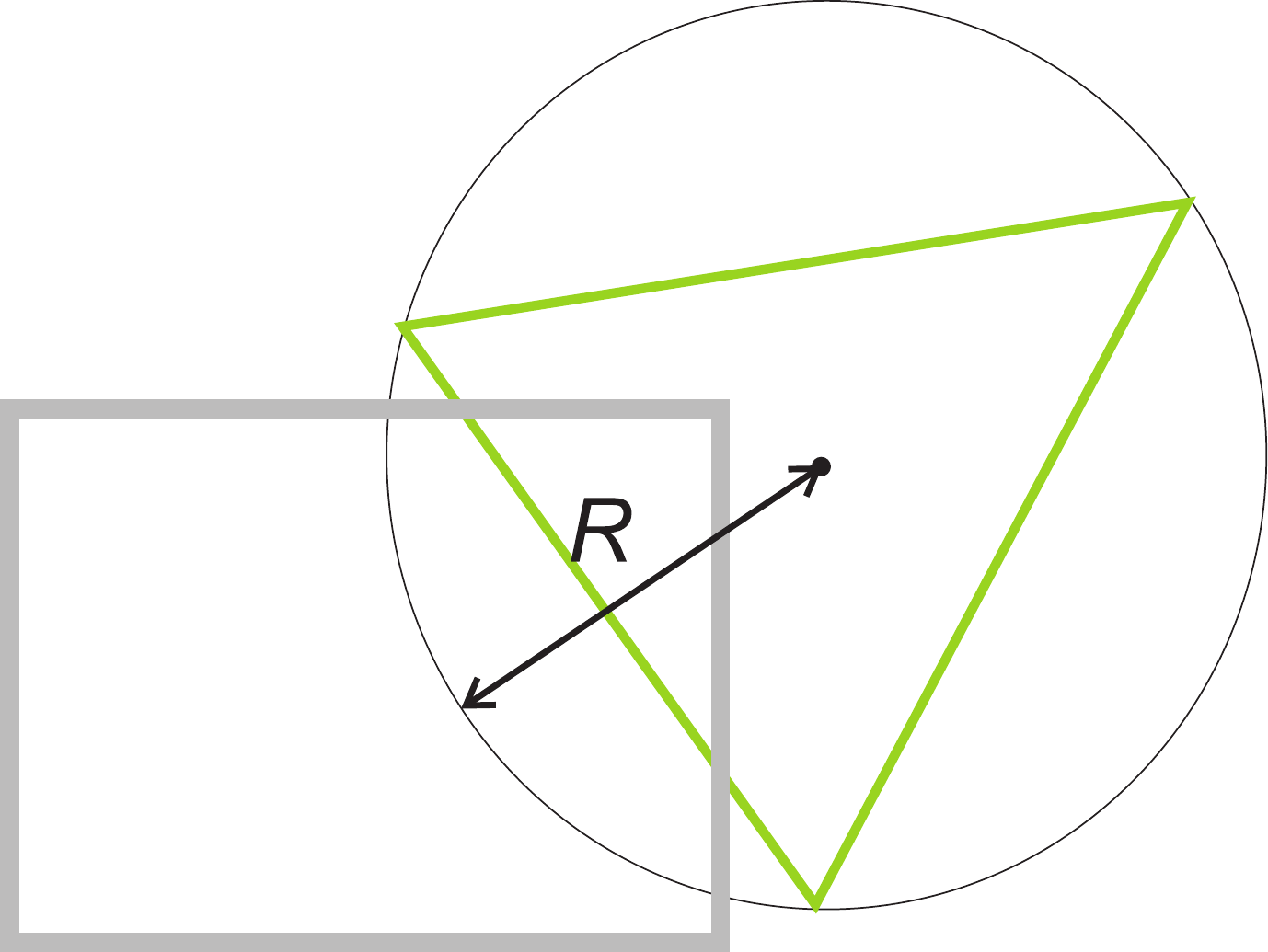} \\
  	(a) & (b)
  	\end{array}$
  	\caption{(a) Illustration of the uniform grid mapping algorithm in 2D: the red triangles are checked to find the triangle containing the charge location. (b) Illustration of the criterion to decide if a pixel in the uniform grid shares a region with a tetrahedral in the unstructured grid.}
  	\label{uniformGridMapping}
  \end{figure}

  The reason for the aforementioned problem in obtaining the field values lies in the unstructured nature of the spatial grid.
  If the space discretization was based on a uniform hexahedral grid, the containing element could be found by using an analytical formulation leading to a much shorter computation time.
  The idea we followed to surmount the problem with an unstructured grid is mapping it on a uniform hexahedral grid.
  Fig.\,\ref{uniformGridMapping}a presents a 2D illustration of this mapping technique.
  First, the whole 2D unstructured mesh is overlapped on a uniform rectangular grid and in a preprocessing step all the triangles which share some common regions with a specific pixel are found.
  Here, the term pixel represents each element of the uniform rectangular grid.
  This analysis provides a map that assigns each pixel in the uniform grid a set of triangles in the unstructured grid.
  By using a simple analytic equation, the pixel containing the location (yellow pixel in Fig.\,\ref{uniformGridMapping}a) is obtained.
  The aforementioned map returns the corresponding triangles (red triangles in Fig.\,\ref{uniformGridMapping}a), which are checked to find the triangle containing the point.
  By following the same procedure in three dimensions, the element of interest is found after checking only few tetrahedrons, which strongly depends on the resolution of the uniform hexahedral compared to the tetrahedral grid.
  
  An important point when constituting the map is the criterion deciding whether a tetrahedron should be maintained in the set for one hexahedron, or in the 2D case, whether a triangle should be maintained in the set for one rectangle.
  A conclusion based on the triangle vertices does not lead to a correct map, because as observed in Fig.\,\ref{uniformGridMapping}b, the two elements can share common regions without the triangle vertices residing in the rectangle.
  A proper conclusion is made using the radius of the circumscribed sphere (or circle in 2D).
  If and only if this sphere has no common region with a hexahedron, the element should be excluded from the corresponding set in the map.
  Using this criterion combined with the introduced search algorithm, a fast procedure for calculating the accelerating field of a charge is achieved which leads to an efficient solution of the system in time domain.
  
\begin{figure} \centering
  	$\begin{array}{cc}
  	\includegraphics[draft=false,width=3.0in]{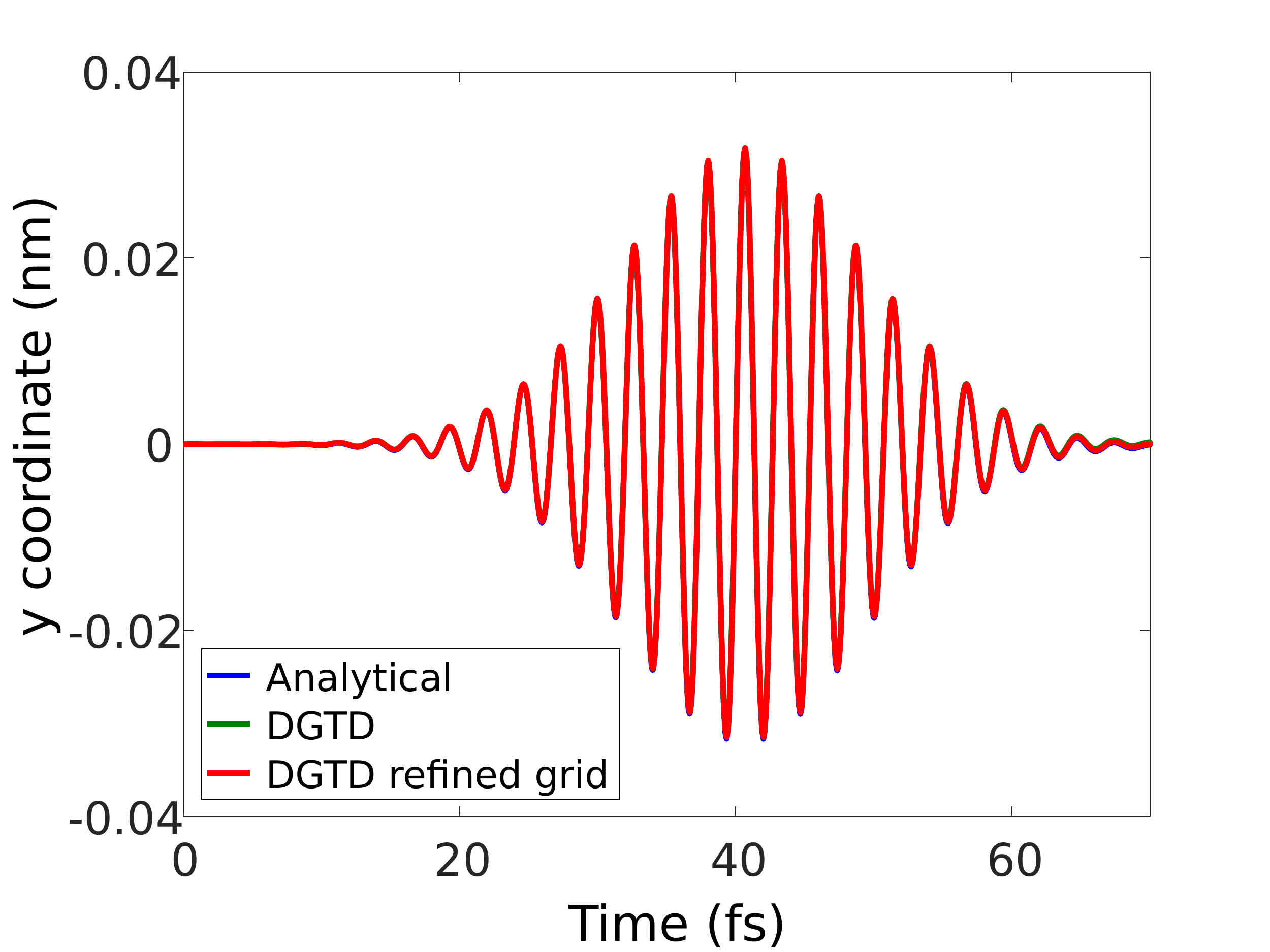} &
  	\includegraphics[draft=false,width=3.0in]{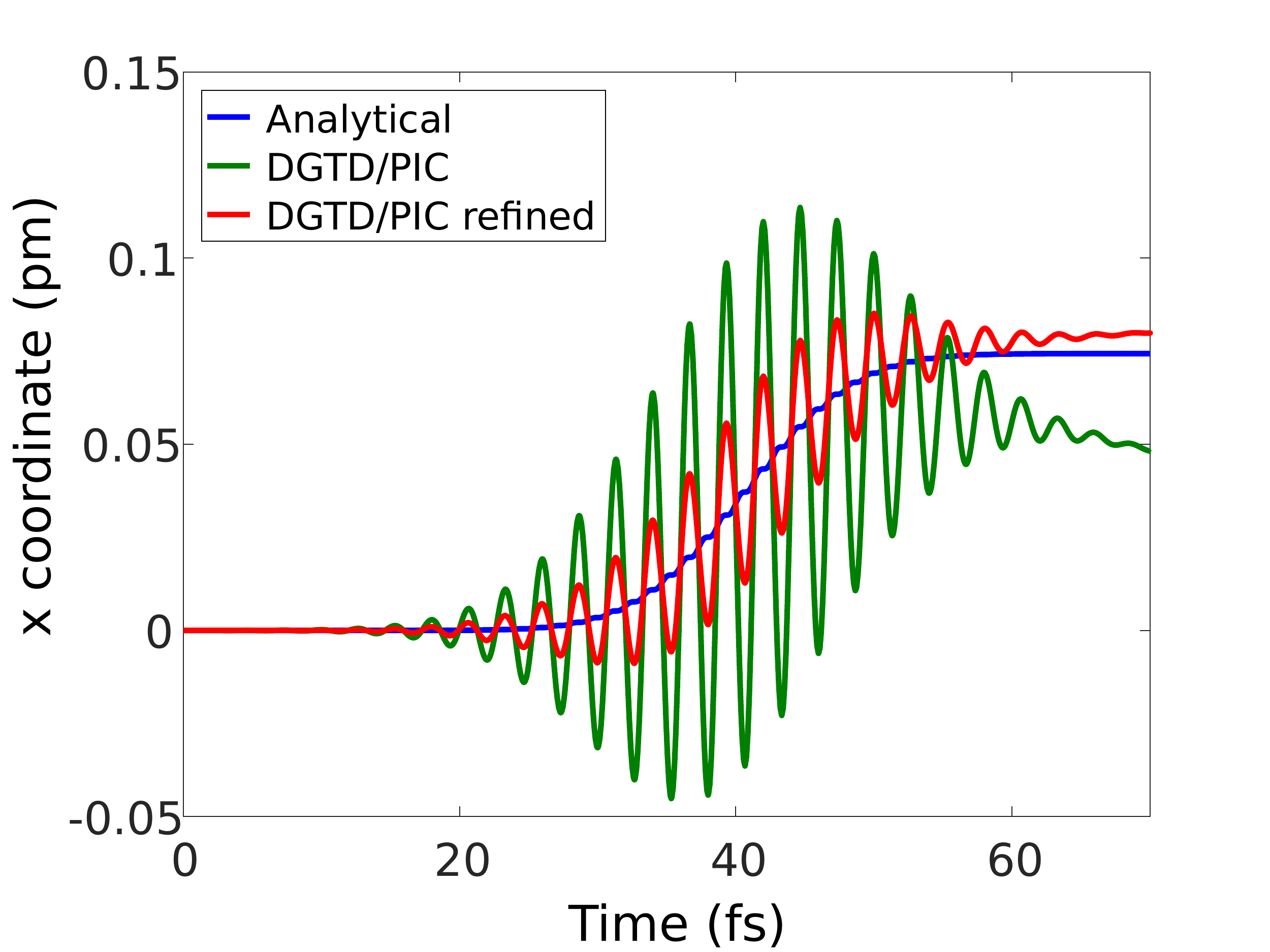} \\
  	(a) & (b)
  	\end{array}$
  	\caption{(a) $x(t)$ and (b) $y(t)$ of an electron residing on the coordinate origin and influenced by an incoming $y$-polarized electromagnetic plane wave. The results, obtained using the DGTD/PIC code are compared with the analytical formulation.}
  	\label{PICBenchmark}
  \end{figure}

  To benchmark the PIC code, we focus on the dynamics of a single charged particle affected by a plane wave.
  Since the plane wave propagation is simulated by the DGTD code, this problem implicitly serves as a benchmark for the DGTD part as well.
  A sphere is considered as the computational domain with 1st order ABC boundaries through which a $y$-polarized plane wave with Gaussian envelope enters and propagates along the $+x$-axis.
  The center wavelength of the incoming pulse is 800\,nm, the pulse duration is assumed to be 15\,fs and the peak field is set to 1\,GV/m.
  An electron resides on the center of the spherical computational domain and moves due to the electromagnetic fields of the plane wave.
  We solve for the position of the electron using the developed DGTD/PIC solver.
  On the other hand, this problem can be solved analytically using the relativistic Hamiltonian of a free particle.
  The vector potential of the considered plane wave reads as
  \begin{equation}
  \vec{A} = - \int\limits_{-\infty}^t E_0 e^{-2 \ln 2 \left( \frac{t-(x-x_0)/c}{\tau} \right)^2} \cos \left( \omega (t-(x-x_0)/c) + \psi_0 \right) \, dt \,\,\unitvec{y}
  \mathrm{,}
  \label{vectorPotential}
  \end{equation}
  where $E_0$ denotes the peak field, $\tau$ is the pulse duration and $\psi_0$ stands for the carrier envelope phase of the signal.
  The time-dependent position of the electron is then obtained as the following:
  \begin{equation}
  \displaystyle x(t) = \frac{e^2}{2\gamma^2 c m^2} \int\limits_{-\infty}^t A_y^2(t) \, dt  \qquad
  \displaystyle y(t) = \frac{e}{\gamma m} \int\limits_{-\infty}^t A_y (t) \, dt \qquad
  \displaystyle z(t) = 0 \mathrm{,}
  \label{electronPosition}
  \end{equation}
  where $\gamma$ represents the Lorentz factor corresponding to the instantaneous energy of the electron.
  The motion along the $y$-axis happens due to the electric field of the plane wave, and the ponderomotive force triggers the motion along the $x$-axis.
  In Fig.\,\ref{PICBenchmark}, we plot and compare the temporal evolution of functions $x(t)$ and $y(t)$ obtained using both the analytical formulation and the DGTD/PIC algorithm.
  A perfect agreement is observed for the variations in $y$ coordinate.
  The motion along x is a second order effect, since it is triggered by the ponderomotive force.
  As seen from Fig.\,\ref{PICBenchmark}, the total amount of particle movement is thousand times less than the wavelength of the plane wave.
  Accurate prediction of this small motion requires very high resolution in the space discretization resulting in large computation cost.
  In Fig.\,\ref{PICBenchmark}, results obtained with one step refinement of the DGTD mesh are also illustrated.
  The comparison of the obtained trajectories evidences the convergence of the results toward the analytical solution.
  As deduced from the curves the DGTD/PIC code is able to predict the oscillatory motion along the $y$-axis with less than 1\% error and the small ponderomotive motion along the $x$-axis with less than 10\% error, which demonstrates the high accuracy of the algorithm and its reliability.
  
  \subsection{Space-charge Calculations}
  
  A big challenge in the simulation using PIC methods is the consideration of particle interactions.
  Various approaches, namely particle-mesh, particle-particle, and their combinations tackle this problem in different ways to achieve either high accuracy or low computation costs \cite{darden1993particle,deserno1998mesh,greengard1987fast}.
  A common point in all these models is neglecting the time delay needed by the charge fields to travel towards other charge positions.
  The reason is the very large memory requirements and consequently hampering of the computations for considering this effect.
  However, this approximation is justified by the fact that the time delay becomes considerable for large charge separations, where the mutual interactions are negligible.
  
  In \cite{fallahi2014field}, we introduced an algorithm based on DGTD/PIC, which enables considering mutual interactions with the time delay effect without making spatial interpolation of the fields.
  All these goals are achieved only with additional computation costs proportional to the number of particles $N$.
  The algorithm is shown to be useful for calculating low-range wake-fields which is important in electron source problems.
  
  The standard DGTD/PIC method \cite{jacobs2006high} considers that the Li\'{e}nard-Wiechert potentials are solutions of the Maxwell equations with the excitations
  \begin{equation}
  \vec{J}(\vec{r},t) = q \vec{v}(\vec{r},t) \delta(\vec{r}-\vec{r}_0) \qquad \mbox{and} \qquad \rho(\vec{r},t) = q \delta(\vec{r}-\vec{r}_0)
  \mathrm{,}
  \label{excitationLienardWiechert}
  \end{equation}
  where $q$ is the charge, $\vec{r}_0$ its location and $\delta(\vec{r})$ denotes the three dimensional delta function.
  On the other hand, the time domain Maxwell equations and particle motion are concurrently solved using DGTD/PIC algorithm.
  Hence, instead of using the ultimate solution in the form of Li\'{e}nard-Wiechert fields, one just adds the excitation current in (\ref{excitationLienardWiechert}) to the Maxwell equations and propagates the space-charge fields together with the incoming light.
  As a result, the fields imported from DGTD that are used to accelerate the charges, account for the mutual interactions as well.
  As explained earlier, there exist several numerical problems when DGTD is combined with PIC through this simple algorithm.
  The method developed in \cite{fallahi2014field} computes Li\'{e}nard-Wiechert fields within one single element and couples them to the propagating fields through surface currents on the boundary in order to alleviate some of the shortcomings of the original DGTD/PIC method.
  For a more detailed presentation of this technique, the reader is referred to \cite{fallahi2014field}.
  Throughout this thesis, we calculate the space-charge effects using particle-particle (or the so-called point-to-point) method and for structured electron source problem, where low-range wake-fields are important the field-based algorithm of \cite{fallahi2014field} is utilized.
  
  \subsection{Field Emission from a Metal Plate}
  
  The developed DGTD/PIC algorithm is a promising method to solve complex, strong-field ultrafast electro-optical problems where no analytic solution exists.
  Hence, it will be widely used for the simulation of structured photocathodes and electron guns.
  One particular problem, that is currently of high interest is the study of ultrafast electron sources based on laser-induced field emission.
  These devices, known as field-emitting cathodes, exploit quantum tunneling in the presence of strong electric fields for the generation of high brightness electron beams.
  Practically, an ultra-short laser pulse illuminates a bulk metal surface and extracts the electrons from the surface.
  The free electrons are then further affected by the existing fields and follow the corresponding trajectory in free-space.
  Since the field emission problem contains propagation and scattering of electromagnetic fields as well as particle motion, the DGTD/PIC algorithm is an appropriate method for simulating this phenomenon.
  In addition, the capability of handling unstructured grids enables one to adaptively increase the mesh resolution at the electron emission points.
  Here, we consider the simple problem of field emission from a flat metal surface.
  More complicated cases will be analyzed in the next chapters and verified with experiments.
  
\begin{figure} \centering
  	$\begin{array}{cc}
  	\includegraphics[draft=false,width=3.0in]{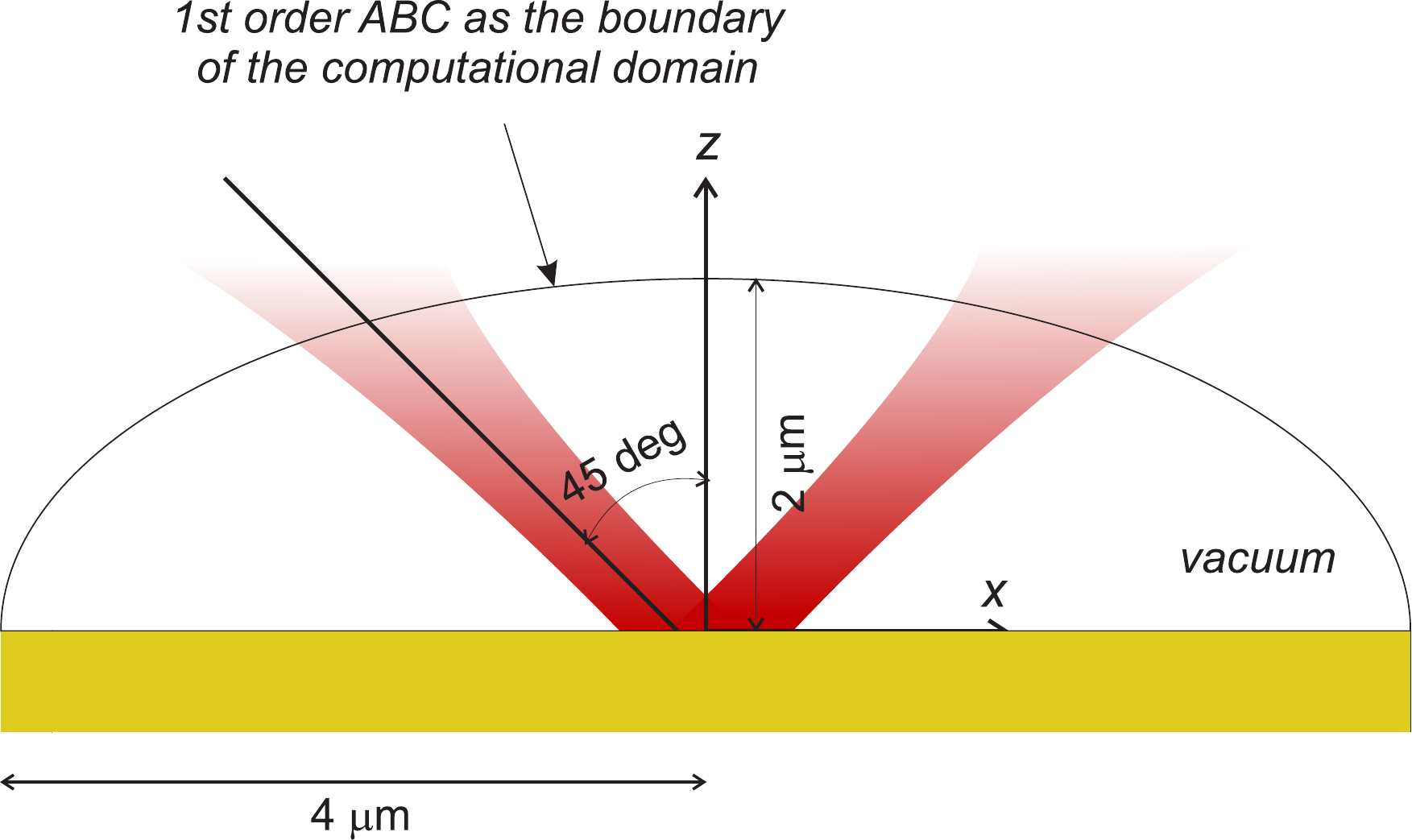} &
  	\includegraphics[draft=false,width=3.0in]{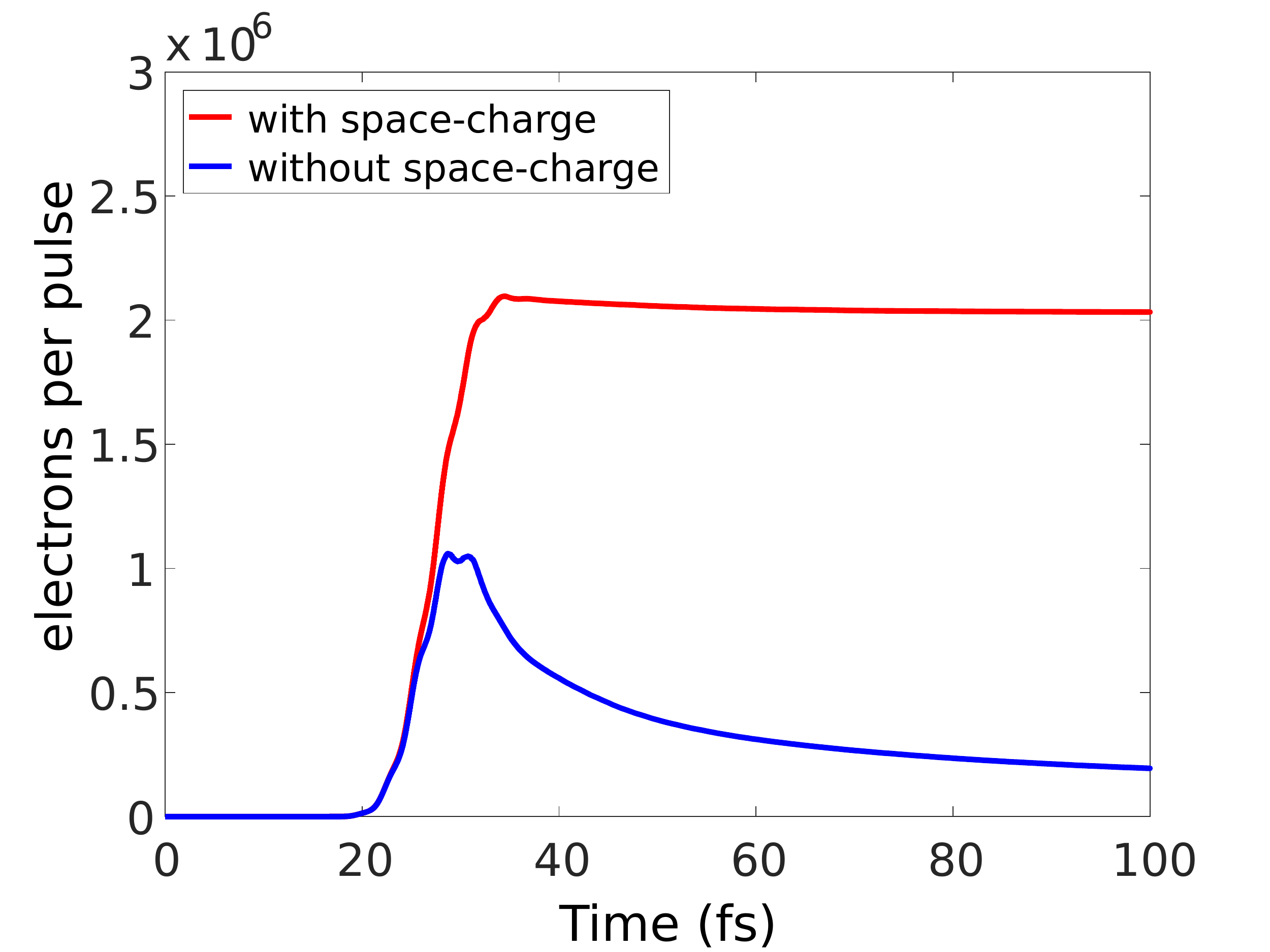} \\
  	(a) & (b)
  	\end{array}$
  	\caption{(a) Geometry of the field-emitter cathode to be studied: A short pulse laser beam impinges on a flat gold plate and simultaneously accelerates the emitted electrons. (b) Charge emitted from the flat surface versus time compared for calculations with and without considering particle radiation.}
  	\label{flatCathode}
  \end{figure}

  Fig.\,\ref{flatCathode}a shows the considered material configuration, computational domain and a excitation visualization.
  We assume a tightly focused Gaussian laser beam with central wavelength $\lambda$=800\,nm, pulse duration $\tau$=10\,fs, Rayleigh radius $w_0$=500\,nm, and peak-field amplitude $E_0$=10\,GV/m that is obliquely ($\theta=45^\circ$) incident on the thin gold surface.
  The gold layer is assumed to be a circular disc with radius 4\,{\textmu}m and thickness 400\,nm.
  The whole computational domain is discretized into 82'137 tetrahedra resulting in 0.6\,s calculation time for field updates at each time step of 2\,as duration.
  Electrons are emitted according to the Fowler-Nordheim emission model for metallic surfaces \cite{nordheim1928electron}.
  The work function is assumed to be $w$=5.1\,eV and the probability for reflection of electrons on the surface, when returning back to the surface is set to $r$=0.3.
  
  The parameters of the interaction is set such that a strong field emission occurs in a short time leading to strong space-charge and wake-field interactions.
  The strong laser pulse is able to extract many electrons from the surface.
  However, due to the mutual interaction between the particles and also the image charge effect, a considerable portion of the emitted charge recombines with the surface.
  This is the well-known space-charge limit in electron guns which can be simulated only correctly, if the mutual interactions are taken into account.
  The image charge effect emerging from the charge distribution interacting with the emitting surface, dictates solving Maxwell's equations including material boundaries.
  
  Numerical modeling of this problem results in the total charge emission versus time shown in Fig.\,\ref{flatCathode}b.
  As seen from the graphs, the maximum emission of charge is happening around the peak of the laser field, which affirms the standard assumptions in the field emission mechanism.
  We compare the field emission results with and without considering particle fields, i.e. space-charge.
  According to the computation without considering particle radiated fields, the short laser pulse leads to emission of about two million electrons from the surface.
  However, when particle radiated fields and the mutual static interactions are considered, the Coulomb blockade effect suppresses tunneling of about half of this charge.
  Afterwards, the attraction from image charge and mutual repulsion of electrons lead to recombination of charge with the gold surface.
  Note that the initial kinetic energy of the electrons immediately after tunneling is assumed to be zero similar to strong field emission from atoms.
  
\begin{figure} \centering
  	\includegraphics[draft=false,width=6.0in]{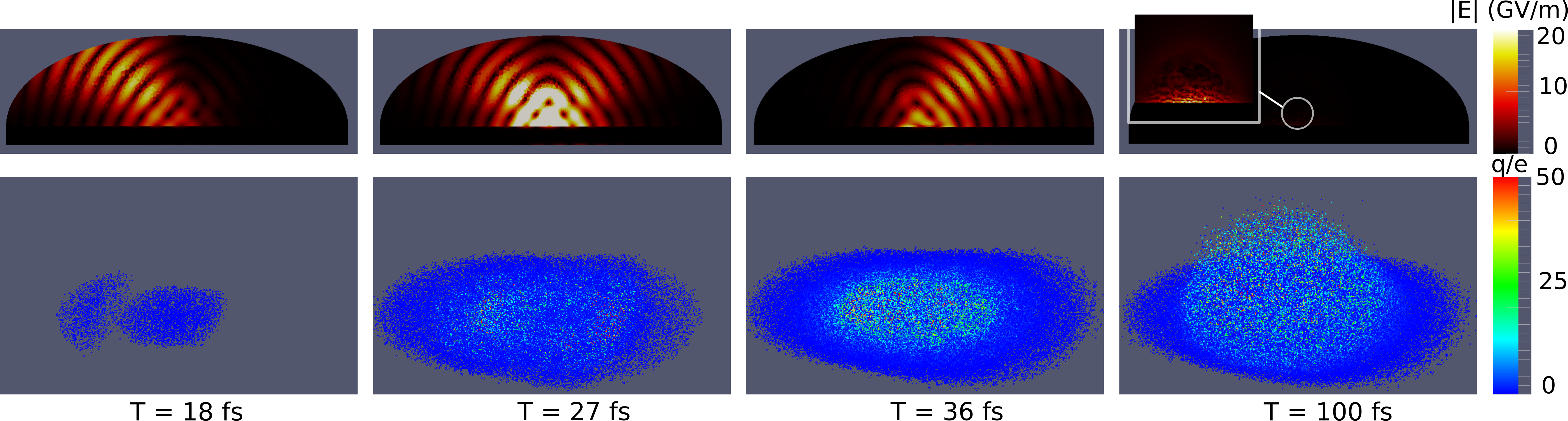}
  	\caption{Snapshots of the incident field and the emitted current from the flat gold emitter: the expansion of the charge cloud with time after the emission process is partially driven by space-charge effects.}
  	\label{flatCathodeProfile}
  \end{figure}

  Fig.\,\ref{flatCathodeProfile} presents snapshots of field and charge distributions above the gold surface.
  The color coded dots in the charge distribution represent a charge cloud that forms in the region.
  Therefore, the charge of each point can be less than the charge of one electron, since it refers to the probability distribution rather than one single electron.
  The figure sequence shows that one side of the bunch is created earlier than the other side due to the oblique incidence of the Gaussian beam.
  Moreover, the zoom on the field distribution in the last snapshot shows the remaining space-charge fields above the surface.
  
  \section{Simulation of ICS Sources}
  
  Several problems in accelerator physics and light source technology require solving particle trajectories within analytically driven field profiles.
  Electron acceleration or beam manipulation in cavities and waveguides are examples of such cases.
  The symmetries in a circular cavity or a cylindrical waveguide lead to simplifications of the field propagation equations, which subsequently result in analytically solvable differential equations.
  Apparently, using numerical solvers, like the DGTD/PIC code described above, for such cases will be very inefficient.
  Therefore, another PIC solver is developed which operates in the same fashion as the PIC solver described above, but imports fields from analytical derivations.
  
  Similar to \eref{motionEquation}, the following equation of motion is taken into account:
  \begin{equation}
  \frac{\partial}{\partial t} \left( \begin{array}{c} \vec{r} \\ \vec{\beta} \end{array} \right) = \left( \begin{array}{c} c\vec{\beta} \\ -\frac{e\sqrt{1-\beta^2}}{m_0 c} (\vec{E}(\vec{r},t)+c\vec{\beta} \times \vec{B}(\vec{r},t) ) \end{array} \right).
  \label{motionEquationAnalytic}
  \end{equation}
  In other words, the time derivative of the pair $(\vec{r},\vec{\beta})$ is written in terms of the instantaneous values of $\vec{r}$ and $\vec{\beta}$ and the electromagnetic fields, i.e. $(\vec{\dot{r}},\vec{\dot{\beta}}) = f(\vec{r},\vec{\beta},t)$.
  This is a straightforward problem to be solved by Runge-Kutta methods.
  Since the electromagnetic fields are derived from the analytic solutions, use of high order Runge-Kutta method is preferred to decrease the computational cost.
  Hence, 5th order Cash-Karp Runge-Kutta method is used to update the equation of motion.
  Since there is no meshing of the computational domain in this scheme, use of mesh-based algorithms to calculate the space-charge effects is not possible.
  Therefore, point-to-point algorithm is used to evaluate the space-charge effect on the bunch properties.
  This method and the corresponding software are used in chapter 5 to design linear accelerator sections, operating based on waveguide modes.
  
  The same hypothesis concerning the analytical evaluation of the field holds for ICS interaction.
  Electromagnetic fields are derived from analytical solution of the paraxial equation for a Gaussian beam, which reads as the following:
  \begin{align}
  & \vec{E}(\vec{r},t) = E_0 \hat{\vec{x}} \frac{w_0}{w(z)}e^{-(r/w(z))^2} e^{-2\ln(2) (t/\tau)^2} \cos\left( \omega t + kz - k\frac{r^2}{2R(z)} + \arctan \left(\frac{z}{z_R}\right) \right) \\
  & \vec{B}(\vec{r},t) = \frac{-E_x(\vec{r},t)}{c} \hat{\vec{y}},
  \label{gaussianBeam}
  \end{align}
  where $w(z)=w_0\sqrt{1+(z/z_R)^2}$, $z_R = \pi w_0^2/\lambda$, and $R(z) = z + z_R^2/z$ are the $1/e$ beam radius at $z$, Rayleigh range and radius of curvature of the beam's wavefronts at $z$, respectively.
  $w_0$ stands for the waist radius of the beam, and $\tau$ is the pulse duration of the ICS laser.
  In addition, $\omega = c/k$, and $k=2\pi/\lambda$ represent the angular frequency and wave number of the ICS light, respectively.
  Consequently, the PIC algorithm with field import from a Gaussian beam function leads to fast calculation of particle trajectories.
  
  Space-charge effects are often neglected in the ICS simulation.
  This assumption originates from the very small interaction time of the electrons with the counter-propagating beam, over which space-charge forces cause negligible changes in the motion of particles.
  Note that in this ICS source, the radiated fields from the bunch are very small compared to the fields of the ICS laser.
  Consequently, the particle trajectories are determined solely by the fields of the laser or optical undulator.
  To simulate radiation in an optical undulator with accounting for the effect of radiated fields on particle trajectories, one needs to use an FEL code discussed in the next section.
  
\begin{figure} \centering
  	$\begin{array}{cc}
  	\includegraphics[draft=false,height=1.2in]{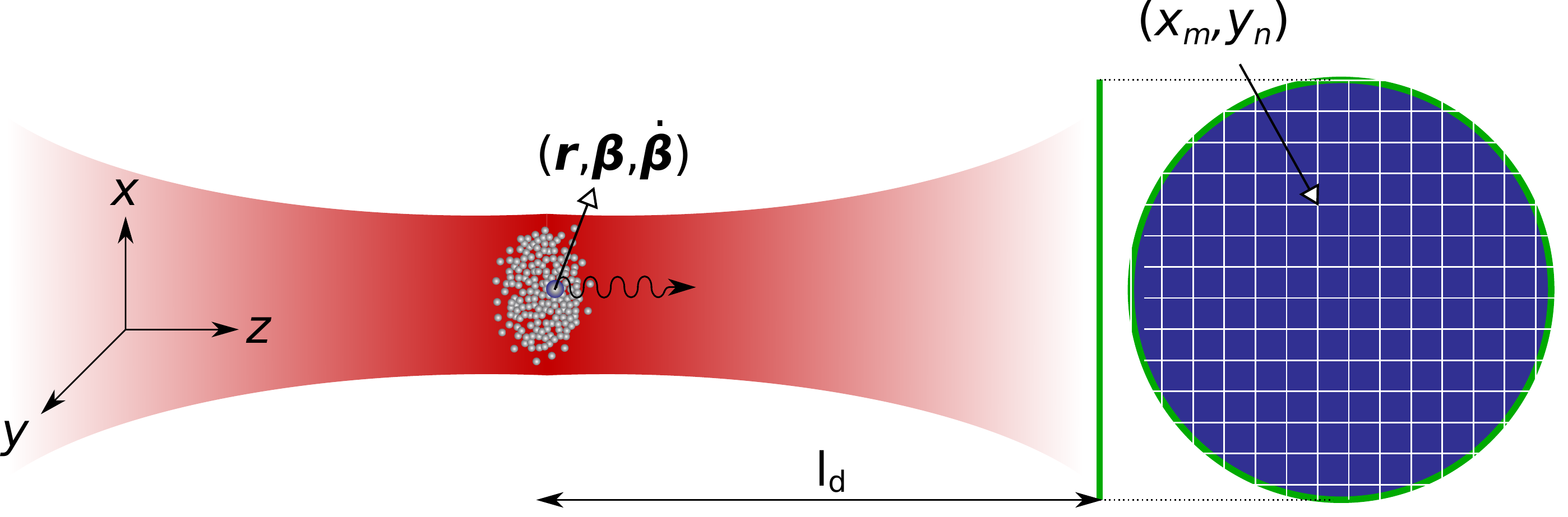} &
  	\includegraphics[draft=false,height=1.2in]{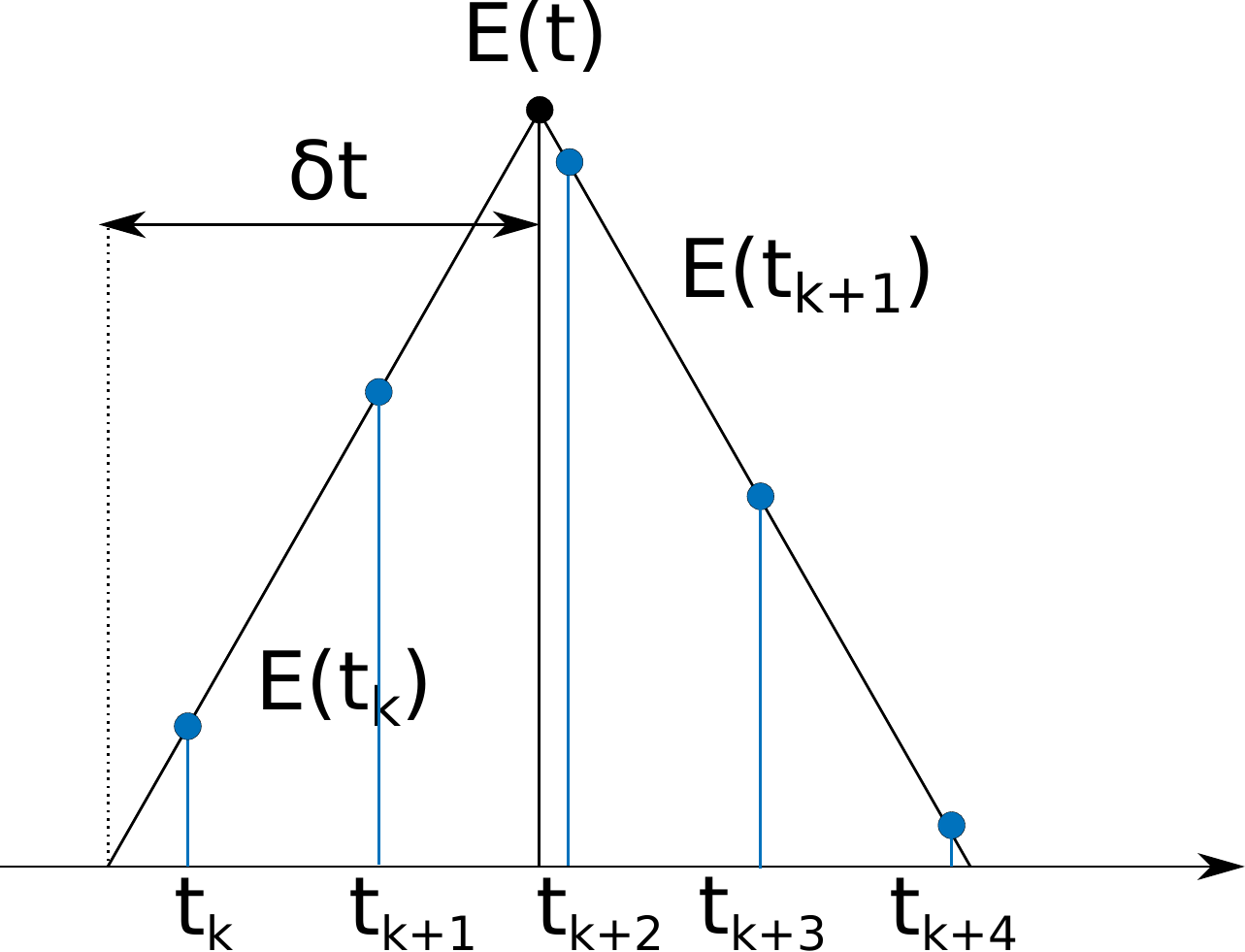} \\
  	(a) & (b)
  	\end{array}$
  	\caption{(a) Simulation setup of the ICS interaction, and (b) Temporal linear interpolation from the delayed time to the sampling time.}
  	\label{detectorICS}
  \end{figure}

  Modelling an ICS interaction is not concluded with determination of the particle trajectories.
  To obtain the radiation properties of the source, the bunch radiation needs to be calculated during the bunch update.
  For this purpose, a detector plate, as shown in Fig.\,\ref{detectorICS}a, is assumed behind the ICS interaction point.
  A 3D matrix $\llbracket F \rrbracket$ is defined, whose first two coordinates correspond to transverse position on the detector, and the third coordinate represents the sampling time of the radiation pattern on the detector.
  Each element of the matrix is a pair of electric and magnetic field at the corresponding instant in time and position.
  More accurately, one can write
  \begin{equation}
  F (m,n,k) = (\vec{E}(\vec{r}_i,t_k),\vec{B}(\vec{r}_i,t_k)),
  \label{detectorMatrix}
  \end{equation}
  with $\vec{r}_i \equiv (x_m,y_n,l_d)$ defined in Fig.\,\ref{detectorICS}a.
  The radiated fields are obtained from the particle parameters according to Li\'{e}nard-Wiechert potentials:
  \begin{align}
  \vec{E}(\vec{r}_i,t+\frac{1}{c}|\vec{r}_i-\vec{r}_p|) & = \frac{1}{4 \pi \varepsilon_0} \left(\frac{q(\unitvec{n} - \vec{\beta})}{\gamma^2 (1 - \unitvec{n} \cdot \vec{\beta})^3 |\vec{r}_i - \vec{r}_p|^2} + \frac{q \unitvec{n} \times \big((\unitvec{n} - \vec{\beta}) \times \dot{\vec{\beta}}\big)}{c(1 - \unitvec{n} \cdot \vec{\beta})^3 |\vec{r}_i - \vec{r}_p|} \right) \\
  \vec{B}(\vec{r}_i,t+\frac{1}{c}|\vec{r}_i-\vec{r}_p|) & = \frac{\unitvec{n}}{c} \times \vec{E}(\vec{r}_i, t+\frac{1}{c}|\vec{r}_i-\vec{r}_p|)
  \label{lienardWichertPotentialICS}
  \end{align}
  where $\unitvec{n}(t) = (\vec{r}_i - \vec{r}_p)/(|\vec{r}_i - \vec{r}_p|)$, with $\vec{r}_p$, $\vec{\beta}$, and $\dot{\vec{\beta}}$ being respectively the position, normalized velocity and normalized acceleration of particle $p$ at time $t$, obtained from \eref{motionEquationAnalytic}.
  It is seen that the field values are obtained at a delayed time,
  \begin{equation}
  t_{\mathrm{delayed}} = t+\frac{|\vec{r}_i-\vec{r}_p|}{c},
  \label{delayedTime}
  \end{equation}
  on the detector, with $\vec{r}_p$ being the particle position.
  The delayed time $t$ does not coincide with the discretized time series $t_k$ in (\eref{detectorMatrix}).
  Therefore, a linear interpolation scheme illustrated in Fig.\,\ref{detectorICS}b is needed to evaluate the contribution of particle $p$ to the radiation element $F(i,j,k)$ representing sampling times $t_k$.
  By summing over all the particles in the bunch and iterating the above algorithm over the entire interaction duration, the radiation of an ICS source is simulated.
  
  Particular care must be exercised in bunch initialization for the simulation of the incoherent radiation in an ICS interaction.
  The assumption of macro-particles results in coherent addition of radiations from electrons considered as one single macro-particle.
  This spurious coherent addition leads to an overestimation of the ultimate photon flux.
  To remedy this problem, a typical approach is to generate an electron bunch with a real number of electrons and fill the 6D phase-space of the bunch according to the cumulative values of the bunch.
  
\begin{figure} \centering
  	$\begin{array}{cc}
  	\includegraphics[draft=false,width=3.0in]{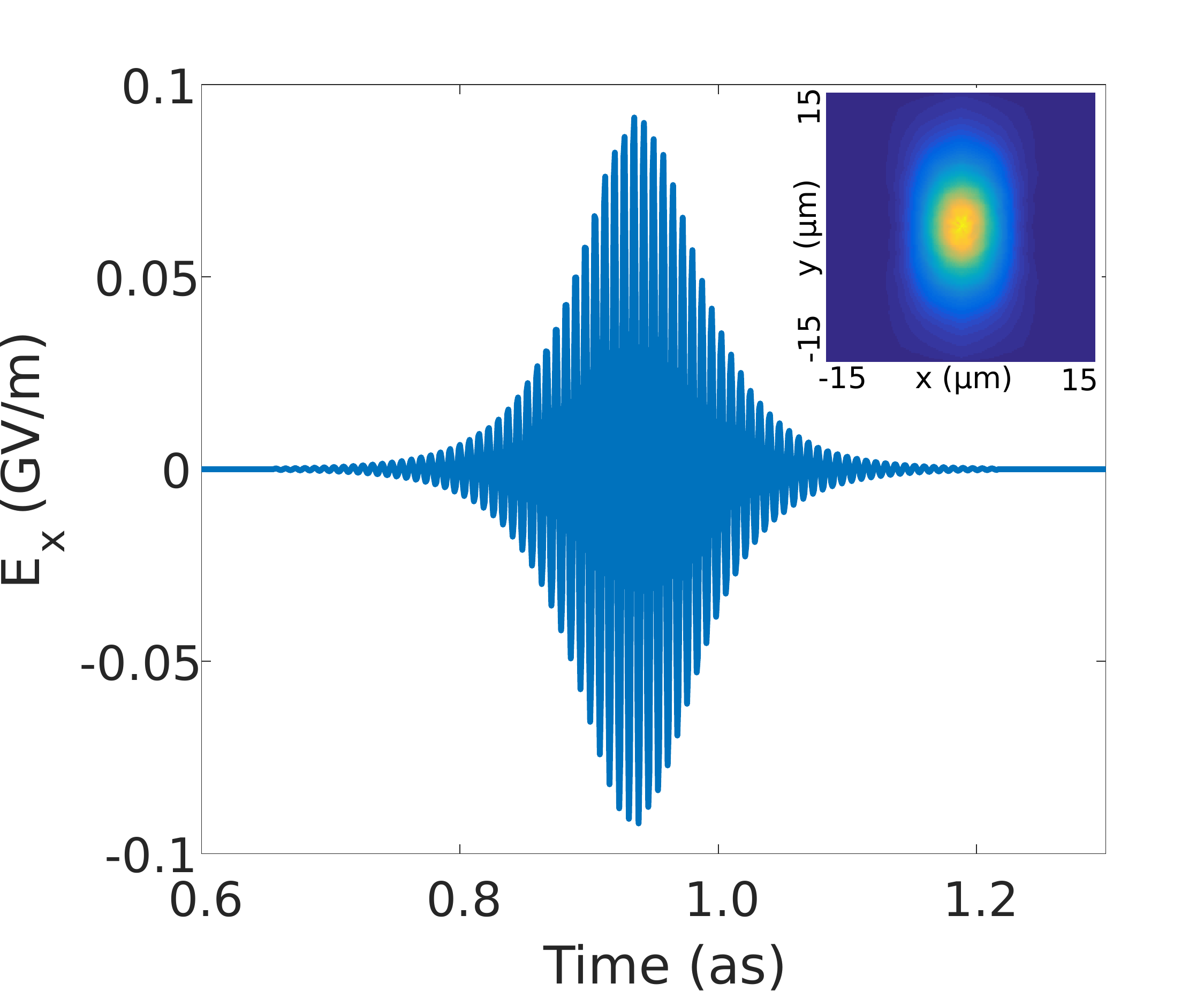} &
  	\includegraphics[draft=false,width=3.0in]{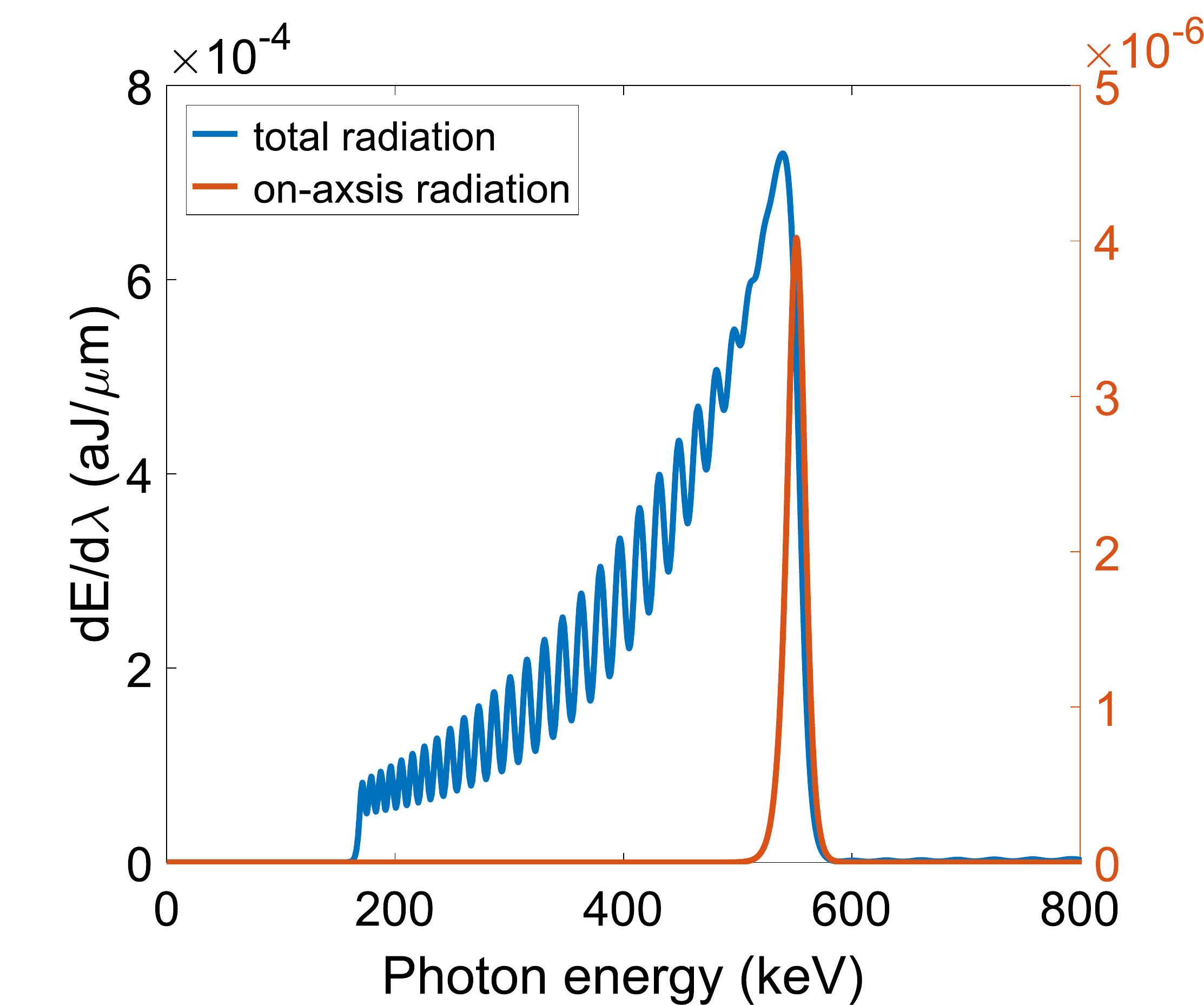} \\
  	(a) & (b)
  	\end{array}$
  	\caption{ICS radiation of a single electron: (a) Electric field of the radiation versus time with the radiation profile 3\,mm away from the interaction point as the subset, and (b) total and on-axis radiation spectrum of the interaction.}
  	\label{ICS-example1}
  \end{figure}

  \textbf{Single electron radiation:} To benchmark the developed code for simulating ICS interaction, we consider an example where a single elctron with energy 152\,MeV hits a counter-propagating laser beam with central wavelength $\lambda = 800$\,nm and pulse duration $\tau = 25$\,fs.
  We assume Rayleigh length of 3\,mm for the beam and pulse energy equal to 10\,mJ.
  Furthermore, it is assumed that the electron motion direction coincides with the laser propagation line, and the collision takes place at laser focal point.
  Fig.\,\ref{ICS-example1}a and Fig.\,\ref{ICS-example1}b illustrate the temporal signature of the field, and the spectrum of the radiation, respectively.
  Two radiation spectra are depicted; the on-axis spectrum accounts for the radiation captured in 50\,{\textmu}rad divergence angle, and the total spectrum corresponds to a 5\,mrad radiation cone.
  It is observed that the on-axis spectrum peaks at 551\,keV which agrees with the expected ICS peak frequency from analytical solution \cite{esarey1993nonlinear}.
  The total efficiency of the interaction defined by $P_{rad}/P_{laser}$ is calculated as $8.8\times 10^{-15}$, which closely agrees with the analytical estimate obtained from $16r_e^2\gamma^2(1+\beta)^2/(3w_0^2 = 9.5 \times 10^{-15}$ \cite{esarey1993nonlinear}.
  The radiation profile on a detector 3\,mm away from the interaction point is also shown as an inset.
  The expected elliptical shape in the radiation profile is evident in the result.
  
\begin{figure} \centering
  	$\begin{array}{cc}
  	\includegraphics[draft=false,width=3.0in]{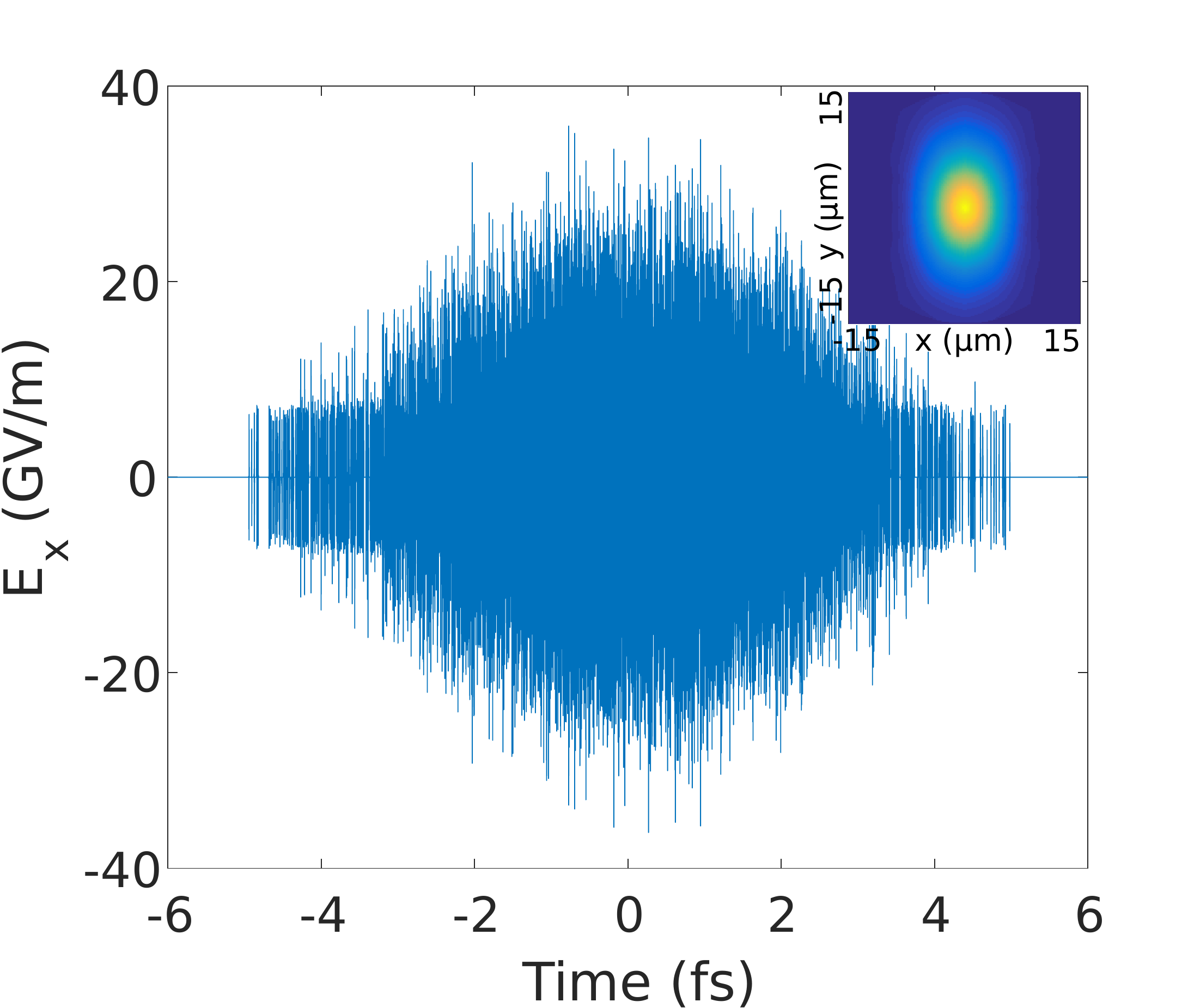} &
  	\includegraphics[draft=false,width=3.0in]{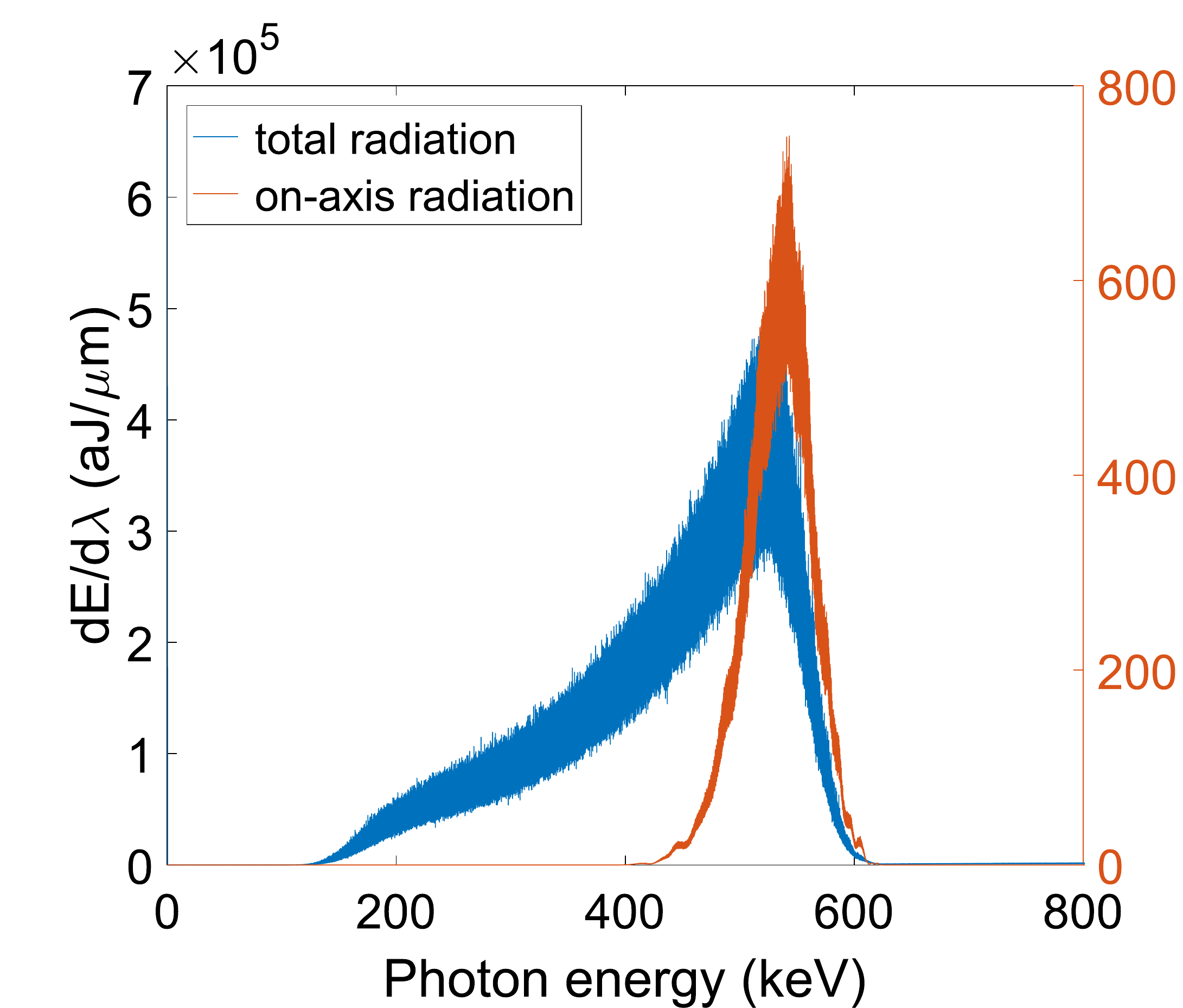} \\
  	(a) & (b)
  	\end{array}$
  	\caption{ICS radiation of an electron bunch: (a) Electric field of the radiation versus time with the radiation profile 3\,mm away from the interaction point as the subset, and (b) total and on-axis radiation spectrum of the interaction.}
  	\label{ICS-example2}
  \end{figure}

  \textbf{Electron bunch radiation:} Now that the ICS code is validated, the radiation produced after interaction of a bunch of electrons with laser beam is simulated.
  The simulation parameters are tabulated in table \ref{ICS-parameters}.
  \begin{table}
  	\caption{Parameters of the ICS interaction}	\label{ICS-parameters} \centering
  	{\footnotesize
  		\begin{tabular}{|c||c|}
  			\hline
  			Bunch mean energy & 152\,MeV \\	\hline
  			Relative energy spread & 2\% \\ \hline
  			Normalized emittance & 0.5\,mm-mrad \\ \hline
  			RMS bunch size & 1.7\,{\textmu}m \\	\hline
  			RMS divergence & 0.1\,mrad \\ \hline
  			Bunch charge & 10\,pC \\ \hline
  			RMS bunch length & 3\,{\textmu}m \\ \hline
  			Laser wavelength & 800\,nm \\ \hline
  			Laser pulse duration & 250\,fs \\ \hline
  			Rayleigh length & 3\,mm \\ \hline
  			Pulse energy & 100\,mJ \\ \hline
  		\end{tabular}
  	}
  \end{table}
  The electron bunch is taken from the typical parameters produced by laser-plasma wakefield acceleration.
  The simulation results are shown in Fig.\,\ref{ICS-example2}, which demonstrates a broader radiation spectrum than the single electron case firstly analyzed.
  
  \section{Simulation of FEL Sources}
  
  Owing to the desire of hard x-ray FEL machines for electrons with ultrarelativistic energies (0.5-1 GeV), these sources are usually giant research facilities with high operation costs and energy consumption.
  Therefore, it is crucial and additionally very useful to develop sophisticated simulation tools, which are able to capture the important features in a FEL radiation process.
  Such tools will be very helpful for designing and optimizing a complete FEL facility and additionally useful for detailed investigation of important effects.
  The last decade had witnessed extensive research efforts aiming to develop such simulation tools.
  As a result, various softwares like Genesis 1.3 \cite{reiche1999genesis}, MEDUSA \cite{biedron19993d}, TDA3D \cite{tran1989tda,faatz1997tda3d}, GINGER \cite{fawley2002user}, PERSEO \cite{giannessi2006overview}, EURA \cite{bacci2008compact}, RON \cite{dejus1999integral}, FAST \cite{saldin1999fast}, CHIMERA (previously PlaRes) \cite{andriyash2015spectral} and PUFFIN \cite{campbell2012puffin} are developed and introduced to the community.
  However, all the currently existing simulation softwares are usually written to tackle special cases and therefore particular assumptions or approximations have been considered in their development \cite{biedron2000multi}.
  Some of the common approximations in FEL simulation are tabulated in table\,\ref{FELapproximations}.
  \begin{table*}[h]
  	\renewcommand{\arraystretch}{1.5}
  	\caption{Common approximations in modelling free electron laser radiation}
  	\label{FELapproximations}
  	\centering
  	{\footnotesize
  		\begin{tabular}{|c|c|c|c|c|c|c|}
  			\hline
  			\multirow{3}{*}{code name} & \multicolumn{6}{|c|}{approximation}  \\
  			\cline{2-7}
  			& steady state & wiggler-average & slow wave & forward & \multirow{2}{*}{no space-charge} & \multirow{2}{*}{slice} \\
  			& approximation & electron motion & approximation & wave & & \\
  			\hline
  			GENESIS 1.3 & optional & \checkmark & \checkmark & \checkmark & \textemdash &  optional \\
  			\hline
  			MEDUSA & optional & \textemdash & \checkmark & \checkmark & \textemdash & \checkmark \\
  			\hline
  			TDA3D & \checkmark & \checkmark & \checkmark & \checkmark & \textemdash & no time-domain \\
  			\hline
  			GINGER & \textemdash & \checkmark & \checkmark & \checkmark & \textemdash & \textemdash \\
  			\hline
  			PERSEO & \textemdash & \textemdash & \textemdash & \checkmark & \checkmark & \textemdash \\
  			\hline
  			CHIMERA & \textemdash & \textemdash & \textemdash & \checkmark & \textemdash & \textemdash \\
  			\hline
  			EURA & \textemdash & \checkmark & \checkmark & \checkmark & \textemdash & \textemdash \\
  			\hline
  			FAST & \textemdash  & \checkmark & \checkmark & \textemdash & \textemdash & \checkmark \\
  			\hline
  			PUFFIN & \textemdash  & \textemdash & \textemdash & \checkmark & \checkmark & \textemdash \\
  			\hline
  		\end{tabular}
  	}
  \end{table*}

  The main goal in the presented research is the analysis of the FEL interaction without considering any of the above approximation.
  The tool could be used for testing the validity of various approximations in different operation regimes and also a reliable approach for preparing the final design of a FEL facility.
  For example, many of the approximations in table\,\ref{FELapproximations}, which sound reasonable for static undulators are not applicable for studying an optical undulator radiation.
  In this regime, due to the various involved length-scales and remarkable impact of the parameter tolerances, having access to a rigorous and robust FEL simulation tool is essential.
  
  One of the difficulties in the x-ray FEL simulation stems from the dramatically large variation of length scales in the involved electromagnetic effects.
  Some of the nominal numbers in a typical FEL simulation are:
  \begin{itemize}
  	\item Size of the bunch: $\sim 100\,$fs or 300\,{\textmu}m
  	\item Undulator period: $\sim 1\,$cm
  	\item Undulator length: $\sim 10-500\,$m
  	\item Radiation wavelength: $\sim 1-100\,$nm
  \end{itemize}
  Comparing the typical undulator lengths with radiation wavelengths immediately communicates the extremely large space for the values.
  This in turn predicts very high computation costs to resolve all the physical phenomena, which is not practical even with the existing supercomputer technology.
  In order to overcome this problem, we exploit Lorentz boosted coordinate system and implement FDTD \cite{taflove2005computational} method combined with PIC simulation in the electron rest frame.
  This coordinate transformation makes the bunch size and optical wavelengths longer and shortens the undulator period.
  Interestingly, these very different length scales transform to values with the same order after the coordinate transformation.
  Consequently, the length of the computation domain is reduced to slightly more than the bunch length making the full-wave simulation numerically feasible.
  We comment that the simulation of particle interaction with an electromagnetic wave in a Lorentz boosted framework is not a new concept.
  The advantage of this technique for the study of relativistic interactions is widely discussed \cite{vay2007,sprangle1979stimulated}.
  The method is currently the standard technique for the simulation of plasma-wakefield acceleration \cite{yu2014modeling,vay2013domain,vay2012novel}.
  Using Lorentz-boosted equations to solve for FEL physics was previously presented in \cite{fawley2009use}, where the code WARP is adapted to simulate a FEL with static undulator.
  In \cite{andriyash2012x}, the dynamics of a FEL based on optical-lattice undulator is described in the electron rest frame.
  Here, we are presenting a software dedicated to the analysis of FEL mechanism by solving principal equations in bunch rest frame.
  
  Along with all the benefits offered by numerical simulation in the Lorentz-boosted framework, there exists a disadvantage emanated from treating quantities different from real three-dimensional fields in the laboratory frame.
  For instance, the field profile along the undulator axis at a certain time does not represent the real radiated field profile, because the fields at various points map to the corresponding values at different time points in the laboratory frame.
  While this feature introduces difficulties in interpreting and investigating the numerical outputs, as discussed in \cite{vay2007} analysis in the moving frame leads to a considerable computational gain motivating the FEL analysis in Lorentz-boosted framework.
  In addition, separate modules and functions can be developed to extract the required plots in stationary frame from the computed values.
  This approach is implemented in our software MITHRA \cite{fallahi2016mithra} to obtain the radiated power.
  The presented study shows how one can numerically simulate a complete FEL interaction using merely Maxwell equations, equation of motion for a charged particle, and the relativity principles, without specific approximations.
  We begin with presenting the whole computational aspects of the numerical method, including the FDTD, PIC, current deposition, Lorentz boosting, quantity initialization, and parallelization.
  Next, different examples of free electron lasers are analyzed and the results are presented in conjunction with some discussions.
  
  \subsection{Numerical Implementation}
  
  In this section, we present the detailed formalism of FDTD/PIC method in the Lorentz boosted coordinate system.
  There are many small yet very important considerations in order to obtain reliable results, which converge to the real values.
  For example, the method for electron bunch generation, particle pusher algorithm and computational mesh truncation need particular attention.
  
  \subsubsection{Finite Difference Time Domain (FDTD)}
  
  FDTD is perhaps the first choice coming to mind for solving partial differential equations governing the dynamics of a system.
  Despite its simple formulation and second order accuracy, there are certain features in this method like explicit time update and zero DC fields, which makes this method a superior choice compared to other algorithms \cite{taflove2005computational}.
  FDTD samples the field in space and time at discrete sampling points and represents the partial derivatives with their discrete counterparts.
  Subsequently, update equations are derived based on the governing differential equation.
  Using these updating equations, a time marching algorithm is acquired which evaluates the unknown functions in the whole computational domain throughout the simulation time.
  In the following, we start with the wave equation which is the governing partial differential equation for our electromagnetic problem.
  
  \textbf{Potential formalism}: The physics of electromagnetic wave and its interaction with charged particles in free space is mathematically formulated through the well-known Maxwell's equations:
  \begin{align}
  \nabla \times \vec{E} &= -\frac{\displaystyle \partial \vec{B}}{\displaystyle \partial t} \\
  \nabla \times \vec{B} &= \mu_0 \vec{J} + \mu_0 \varepsilon_{0} \frac{\displaystyle \partial \vec{E}}{\displaystyle \partial t} \\
  \nabla \cdot \vec{E} &= -\frac{\displaystyle \rho}{\displaystyle \varepsilon_{0}} \\
  \nabla \cdot \vec{B} &= 0
  \end{align}
  These equations in conjunction with the electric current equation $\vec{J}=\rho \vec{v}$ ($\vec{v}$ is the charge velocity) and the Lorentz force equation:
  \begin{equation}
  \label{LorentzForce}
  \vec{F} = q(\vec{E} + \vec{v} \times \vec{B})
  \end{equation}
  are sufficient to describe wave-electron interaction in free space.
  Moving free electrons introduce electric current which enters into the Maxwell's equations as the source.
  Electric and magnetic fields derived from these equations are subsequently employed in the Lorentz force equation to determine the forces on the electrons, which in turn determine their motions.
  As it is evident from the above equations, there are two unknown vectors ($\vec{E}$ and $\vec{B}$) to be evaluated, meaning that six unknown components should be extracted from the equations.
  However, since these two vectors are interrelated and specially because there is no magnetic monopole in the nature ($\nabla \cdot \vec{B}=0 $), one can recast Maxwell's equations in a wave equation for the magnetic vector potential ($\vec{A}$) and a wave equation for the scalar electric potential ($\varphi$):
  \begin{equation}
  \label{HelmholtzA}
  {\vec{\nabla}}^{2}\vec{A} - \frac{1}{c^2} \frac{\displaystyle \partial^2}{\displaystyle \partial t^2}\vec{A} = -\mu_{0} \vec{J}
  \end{equation}
  \begin{equation}
  \label{HelmholtzF}
  {\nabla}^{2}\varphi- \frac{1}{c^2} \frac{\displaystyle \partial^2 \varphi}{\displaystyle \partial t^2} = -\frac{\displaystyle \rho}{\displaystyle \varepsilon_{0}}
  \end{equation}
  where $c=1/\sqrt{\mu_0 \varepsilon_0}$ is the light velocity in vacuum.
  In the derivation of above equations, the Lorentz gauge $\nabla \cdot \vec{A}=-\partial \varphi/c^2\partial t$ is used.
  The original $\vec{E}$ and $\vec{B}$ vectors can be obtained from $\vec{A}$ and $\varphi$ as:
  \begin{equation}
  \label{BvsA}
  \vec{B} = \nabla \times \vec{A}
  \end{equation}
  \begin{equation}
  \label{EvsA}
  \vec{E} = -\frac{\partial \vec{A}}{\partial t}-\nabla \varphi
  \end{equation}
  In addition to the above equations, the charge conservation law written as
  \begin{equation}
  \label{chargeLaw}
  \nabla \cdot \vec{J} + \frac{\partial \rho}{\partial t} = 0,
  \end{equation}
  should not be violated in the employed computational algorithm.
  This is the main motivation for seeking proper current deposition algorithms in the FDTD/PIC methods used for plasma simulations.
  It is immediately observed that the equations (\ref{HelmholtzA}), (\ref{HelmholtzF}), (\ref{chargeLaw}) and the Lorentz gauge introduce an overdetermined system of equations.
  In other words, once a current deposition is implemented that automatically satisfies the charge conservation law, the Lorentz gauge will also hold, provided that the scalar electric potential ($\varphi$) is obtained from (\ref{HelmholtzF}).
  However, due to the space-time discretization and the interpolation of quantities to the grids, a suitable algorithm that holds the charge conservation without violating energy and momentum conservation does not exist.
  The approach that we follow in MITHRA is using the discretized form of (\ref{HelmholtzA}) and (\ref{HelmholtzF}) with the currents and charges of electrons (i.e. macro-particles) as the source and solving for the vector and scalar potential.
  It was shown by Umeda et al. \cite{umeda2003new}, that by using similar weighting functions for both current density ($\vec{J}$) and charge density ($\rho$), and a proper discretization of current density based on positions of the macro-particles according to a Zigzag scheme, a charge conserving deposition scheme is obtained.
  Here, we have implemented the Zigzag scheme to maintain the charge conservation in MITHRA.
  To obtain the fields $\vec{E}$ and $\vec{B}$ at the grid points, we use the momentum conserving interpolation, which will be explained in the upcoming sections.
  
  \textbf{FDTD for Wave Equation:} In cartesian coordinates, a vector wave equation is written in form of three uncoupled scalar wave equations.
  Therefore, it is sufficient to apply our discretization scheme only on a typical scalar wave equation: ${\nabla}^{2}\psi- \partial^2 \psi/c^2\partial t^2 = \zeta$, where $\psi$ stands for $A_l$ ($l \in \{x,y,z\}$); and $\zeta$ represents the term $-\mu_0 J_l$.
  Let us begin with the central-difference discretization scheme for various partial differential terms of the scalar wave equation at the point $(i\Delta x,j\Delta y,k\Delta z,n\Delta t)$.
  In the following equations, $\psi_{i,j,k}^n$ denotes the value of the quantity $\psi$ at the point $(i\Delta x,j\Delta y,k\Delta z)$ and time $n\Delta t$.
  The derivatives are written as follows:
  \begin{align}
  \frac{\partial^{2}}{\partial x^{2}} \psi(x,y,z,t) & \simeq \frac{\psi_{i+1,j,k}^n-2\psi_{i,j,k}^n+\psi_{i-1,j,k}^n}{(\Delta x)^2}
  \\
  \frac{\partial^{2}}{\partial y^{2}} \psi(x,y,z,t) & \simeq \frac{\psi_{i,j+1,k}^n-2\psi_{i,j,k}^n+\psi_{i,j-1,k}^n}{(\Delta y)^2}
  \\
  \frac{\partial^{2}}{\partial z^{2}} \psi(x,y,z,t) & \simeq \frac{\psi_{i,j,k+1}^n-2\psi_{i,j,k}^n+\psi_{i,j,k-1}^n}{(\Delta z)^2}
  \\
  \frac{\partial^{2}}{\partial t^{2}} \psi(x,y,z,t) & \simeq \frac{\psi_{i,j,k}^{n+1}-2\psi_{i,j,k}^n+\psi_{i,j,k}^{n-1}}{(\Delta t)^2}.
  \end{align}
  Combining these four equations, one obtains the value of $\psi$ at instant $(n+1)\Delta t$ in terms of its value at $n\Delta t$ and $(n-1)\Delta t$:
  \begin{align}
  \psi_{i,j,k}^{n+1} = & -\psi_{i,j,k}^{n-1}+ \alpha_1 \psi_{i,j,k}^n + \alpha_2 \psi_{i+1,j,k}^n + \alpha_3 \psi_{i-1,j,k}^n + \alpha_4 \psi_{i,j+1,k}^n + \alpha_5 \psi_{i,j-1,k}^n + \alpha_6 \psi_{i,j,k+1}^n + \alpha_7 \psi_{i,j,k-1}^n \nonumber \\
  & + \alpha_8 \zeta_{i,j,k}^n \nonumber
  \end{align}
  where the coefficients $\alpha_1, \ldots ,\alpha_7$ are obtained from:
  \begin{equation}
  \begin{array}{l}
  \displaystyle
  \alpha_1=2 \left[1-\left(\frac{c \Delta t}{\Delta x}\right)^2-\left(\frac{c \Delta t}{\Delta y}\right)^2-\left(\frac{c \Delta t}{\Delta z}\right)^2\right],
  \qquad
  \alpha_2=\alpha_3=\left(\frac{c \Delta t}{\Delta x}\right)^2,
  \qquad
  \alpha_4=\alpha_5=\left(\frac{c \Delta t}{\Delta y}\right)^2,
  \\ \\ \displaystyle
  \alpha_6=\alpha_7=\left(\frac{c \Delta t}{\Delta z}\right)^2,
  \qquad
  \alpha_8=\left(c \Delta t\right)^2.
  \end{array}
  \end{equation}
  The term $\zeta_{i,j,k}^n$ is the magnitude of the source term at the time $n \Delta t$, which is calculated from the particle motions.
  Usually, one needs a finer temporal discretization for updating the equation of motion compared to electromagnetic field equations.
  If the equation of motion is discretized and updated with $\Delta t_b = \Delta t / N$ time steps, the term $\zeta_{i,j,k}^n$ will be written in terms of the value after each $N$ update:
  \begin{equation}
  \label{currentIntegral}
  \zeta_{i,j,k}^n = -\mu_0 J_l ( n \Delta t ) = -\mu_0 \rho ( n \Delta t ) \frac{\vec{r}^{n+1/2}-\vec{r}^{n-1/2}}{\Delta t}.
  \end{equation}
  As observed in the above equation, the position of particles are sampled at each $n+1/2$ time step, which later should be considered for updating the scalar potential.
  This assumption also results in the calculation of charge density at $n+1/2$ time steps, which should be averaged for obtaining $\rho ( n \Delta t )$.
  
  \textbf{Numerical Dispersion in FDTD:} It is well-known that the FDTD formulation for discretizing the wave equation suffers from the so-called numerical dispersion.
  More accurately, the applied discretization leads to the phase velocity of wave propagation calculated different from (lower than) the vacuum speed of light.
  This may impact the FEL simulation results particularly during the saturation regime, owing to the important role played by the relative phase of electrons with respect to the radiated light.
  Therefore, careful scrutiny of this effect and minimizing its impact is essential for the goal pursued by MITHRA.
  
  To derive the equation governing such a dispersion, we assume a plane wave function for $\psi(x,y,z,t)$ as:
  \begin{equation}
  \psi(x,y,z,t)= e^{-j(k_xx+k_yy+k_zz-\omega t)}
  \end{equation}
  in the discretized wave equation.
  After some mathematical operations, the following equation is obtained for the dispersion properties of central-difference scheme:
  \begin{equation}
  \label{numericalDispersionCD}
  \frac{\sin^2\left(\frac{k_x\Delta x}{2}\right)}{(\Delta x)^2} + \frac{\sin^2\left(\frac{k_y\Delta y}{2}\right)}{(\Delta y)^2} + \frac{\sin^2\left(\frac{k_z\Delta z}{2}\right)}{(\Delta z)^2} = \frac{\sin^2\left(\frac{\omega \Delta t}{2}\right)}{(c\Delta t)^2}.
  \end{equation}
  This equation is evidently different from the vacuum dispersion relation, which reads as
  \begin{equation}
  \label{vacuumDispersionCD}
  k_x^2+k_y^2+k_z^2=\frac{\omega^2}{c^2}.
  \end{equation}
  Comparison of the two equations shows that the dispersion characteristics are similar, if and only if $\Delta x \rightarrow 0$, $\Delta y \rightarrow 0$, $\Delta z \rightarrow 0$, and $\Delta t \rightarrow 0$.
  Another output of the dispersion equation is the stability condition, which is referred to as Courant-Friedrichs-Lewy (CFL) condition \cite{taflove2005computational}.
  The spatial and temporal discretization should be related such that the term $\omega$ obtained from equation (\ref{numericalDispersionCD}) has no imaginary part, i.e. $\sin^2(\omega \Delta t/ 2) < 1$.
  This implies that
  \begin{equation}
  c\Delta t < \left( \sqrt{\frac{\sin^2(\frac{k_x\Delta x}{2})}{(\Delta x)^2} + \frac{\sin^2(\frac{k_y\Delta y}{2})}{(\Delta y)^2} + \frac{\sin^2(\frac{k_z\Delta z}{2})}{(\Delta z)^2}} \right)^{-1} .
  \end{equation}
  The right hand side of the above equation has its minimum when all the sinus functions are equal to one, which leads to the stability condition for the central-difference scheme:
  \begin{equation}
  \label{CFLcondition}
  \Delta t < \left( c\sqrt{\frac{1}{(\Delta x)^2} + \frac{1}{(\Delta y)^2} + \frac{1}{(\Delta z)^2}} \right)^{-1}.
  \end{equation}
  
  As mentioned above, for the FEL simulation, it is very important to maintain the vacuum speed of light along the $z$ direction (throughout this paper $z$ is the electron beam and undulator direction).
  More accurately, if $k_x=k_y=0$, $k_z=\omega/c$ should be the solution of the dispersion equation.
  However, this solution is obtained if and only if $\Delta t = \Delta z /c$, which violates the stability condition.
  To resolve this problem, various techniques are developed in the context of compensation of numerical dispersion.
  Here, we take advantage from the non-standard finite difference (NSFD) scheme to impose the speed of light propagation along $z$ direction \cite{shlager2003comparison,finkelstein2007finite}.
  
  The trick is to consider a weighted average along $z$ for the derivatives with respect to $x$ and $y$, which is formulated as follows:
  \begin{align}
  \frac{\partial^{2}}{\partial x^{2}} \psi(x,y,z,t) & \simeq \frac{\bar{\psi}_{i+1,j,k}^n-2\bar{\psi}_{i,j,k}^n+\bar{\psi}_{i-1,j,k}^n}{(\Delta x)^2}
  \\
  \frac{\partial^{2}}{\partial y^{2}} \psi(x,y,z,t) & \simeq \frac{\bar{\psi}_{i,j+1,k}^n-2\bar{\psi}_{i,j,k}^n+\bar{\psi}_{i,j-1,k}^n}{(\Delta y)^2},
  \end{align}
  with
  \begin{equation}
  \label{NSFDtrick}
  \bar{\psi}_{i,j,k}^n = \mathcal{A} \psi_{i,j,k-1}^n + (1-2\mathcal{A}) \psi_{i,j,k}^n + \mathcal{A} \psi_{i,j,k+1}^n.
  \end{equation}
  Such a finite difference scheme leads to the following dispersion equation:
  \begin{equation}
  \label{NSFDDispersionCD}
  \left( 1 - 4 \mathcal{A} \sin^2(k_z\Delta z/2) \right) \left( \frac{\sin^2(k_x\Delta x/2)}{(\Delta x)^2} + \frac{\sin^2(k_y\Delta y/2)}{(\Delta y)^2} \right) + \frac{\sin^2(k_z\Delta z/2)}{(\Delta z)^2} = \frac{\sin^2(\omega \Delta t/2)}{(c\Delta t)^2}.
  \end{equation}
  It can be shown that if the NSFD coefficient $\mathcal{A}$ is larger than 0.25, and $\sqrt{(\Delta z/\Delta x)^2+ (\Delta z/\Delta y)^2} < 1$, a real $\omega$ satisfies the above dispersion equation for $\Delta t = \Delta z / c$.
  This time step additionally yields $k_z=\omega/c$, for $k_x=k_y=0$.
  
  The value we chose for $\mathcal{A}$ in MITHRA is obtained from
  \begin{equation}
  \label{NSFDCoefficient}
  \mathcal{A} = 0.25 \left( 1 + \frac{0.02}{(\Delta z/\Delta x)^2+ (\Delta z/\Delta y)^2} \right).
  \end{equation}
  The update equation can then be written as
  \begin{align}
  \label{updateEquation}
  \psi_{i,j,k}^{n+1} = & -\psi_{i,j,k}^{n-1} + \alpha'_1 \psi_{i,j,k}^n \nonumber \\
  & + \alpha'_2 ( \mathcal{A} \psi_{i+1,j,k-1}^n + (1-2\mathcal{A}) \psi_{i+1,j,k}^n + \mathcal{A} \psi_{i+1,j,k+1}^n ) \nonumber \\
  & + \alpha'_3 ( \mathcal{A} \psi_{i-1,j,k-1}^n + (1-2\mathcal{A}) \psi_{i-1,j,k}^n + \mathcal{A} \psi_{i-1,j,k+1}^n ) \nonumber \\
  & + \alpha'_4 ( \mathcal{A} \psi_{i,j+1,k-1}^n + (1-2\mathcal{A}) \psi_{i,j+1,k}^n + \mathcal{A} \psi_{i,j+1,k+1}^n ) \nonumber \\
  & + \alpha'_5 ( \mathcal{A} \psi_{i,j-1,k-1}^n + (1-2\mathcal{A}) \psi_{i,j-1,k}^n + \mathcal{A} \psi_{i,j-1,k+1}^n ) \nonumber \\
  & + \alpha'_6 \psi_{i,j,k+1}^n + \alpha'_7 \psi_{i,j,k-1}^n + \alpha'_8 \zeta_{i,j,k}^n.
  \end{align}
  where the coefficients $\alpha'_1, \ldots ,\alpha'_7$ are obtained from:
  \begin{equation}
  \begin{array}{l}
  \displaystyle
  \alpha'_1=2 \left[1-(1-2\mathcal{A})\left( \left(\frac{c \Delta t}{\Delta x}\right)^2+\left(\frac{c \Delta t}{\Delta y}\right)^2 \right)-\left(\frac{c \Delta t}{\Delta z}\right)^2\right],
  \qquad
  \alpha'_2=\alpha'_3=\left(\frac{c \Delta t}{\Delta x}\right)^2,
  \\ \\ \displaystyle
  \alpha'_4=\alpha'_5=\left(\frac{c \Delta t}{\Delta y}\right)^2,
  \qquad
  \alpha'_6=\alpha'_7=\left(\frac{c \Delta t}{\Delta z}\right)^2 - 2\mathcal{A} \left( \left(\frac{c \Delta t}{\Delta x}\right)^2 + \left(\frac{c \Delta t}{\Delta x}\right)^2 \right),
  \qquad
  \alpha'_8=\left(c \Delta t\right)^2.
  \end{array}
  \end{equation}
  To guarantee a dispersion-less propagation along $z$ direction with the speed of light the update time step is automatically calculated from the given longitudinal discretization ($\Delta z$), according to $\Delta t = \Delta z / c$.
  
  \textbf{FDTD for Scalar Potential:} Usually, due to high energy of particles in a FEL process, the FEL simulations neglect the space-charge effects by considering $\varphi \simeq 0$ \cite{andriyash2015spectral}.
  However, this is an approximation which we try to avoid in MITHRA.
  To account for space-charge forces, one needs to solve the wave equation for scalar potential, i.e. (\ref{HelmholtzF}).
  For this purpose, the same formulation as used for the vector potential is utilized to update the scalar potential.
  Nonetheless, since the position of particles are sampled at $t+\Delta t/2$ instants, the obtained value for $\varphi^n$ corresponds to the scalar potential at $(n+1/2)\Delta t$.
  This point should be particularly taken into consideration, when electromagnetic fields $\vec{E}$ and $\vec{B}$ are evaluated.
  
  \textbf{Boundary Truncation:} In order to simulate the FEL problem, we consider a cube as our simulation domain.
  The absorbing boundary condition is also considered for updating the scalar electric potential $\varphi$ at the boundaries.
  Therefore, we introduce the parameter $\xi$, which denotes either $\psi$ or $\varphi$.
  The six boundaries of the cube are supposed to be at: $x=\pm l_{x}/2 $, $y=\pm l_{y}/2 $ and $z=\pm l_{z}/2 $.
  In the following, we only present the formulation for the boundary conditions at $z=\pm l_{z}/2 $.
  The process to extract the equations for the other four boundaries will be exactly similar.
  
  \emph{First Order ABCs:} The partial differential equations implying first order ABCs at $ z=\pm l_{z}/2 $ are:
  \begin{equation}
  \mp \frac{\partial^2 \xi}{\partial z \partial t} - \frac{1}{c}\frac{\partial^2 \xi}{\partial t^2}=0
  \end{equation}

  \emph{Second Order ABCs:} The partial differential equations implying second order ABCs at $ z=\pm l_{z}/2 $ are:
  \begin{equation}
  \mp \frac{\partial^2 \xi}{\partial z \partial t} - \frac{1}{c}\frac{\partial^2 \xi}{\partial t^2} - \frac{c}{2}\frac{\partial^2 \xi}{\partial x^2} - \frac{c}{2}\frac{\partial^2 \xi}{\partial y^2}=0
  \end{equation}
  
  Particular attention should be devoted to the implementation of Mur second order absorbing boundary condition at edges and corners.
  Separate usage of the above equations for second order case encounters problems in the formulation.
  On one hand, unknown values at grid points outside the computational domain appear in the equations, and on the other a system of overdetermined equations will be obtained.
  The solution to this problem is to discretize all the involved boundary conditions at the center of the cubes (for corners) or squares (for edges).
  A simple addition of the obtained equations cancels out the values outside the computational domain and returns the desired value meeting the considered absorbing boundary condition.
  The first and second order Mur boundary condition gradually lose their accuracy in absorbing the incident field when large angles of incidence are involved.
  For this reason, in an FEL simulation using MITHRA, boundaries need to be considered far enough from the radiating particles to decrease the effect of boundary truncation on the simulation accuracy.
  
  \subsubsection{Particle In Cell (PIC)}
  
  Particle in cell (PIC) method is the standard algorithm to solve for the bunch dynamics within an electromagnetic field distribution.
  The method discretizes the bunch 6D distribution function as an ensemble of macro-particles, takes the time domain data of the electric and magnetic fields, and updates the macro-particle position and momentum using a proper particle-pusher technique.
  We comment that the electromagnetic fields in the motion equation are the total fields in the computational domain, which in a FEL problem is equivalent to the superposition of undulator field, radiated field and the seeded field in case of a seeded FEL problem.
  Often considering all the individual particles involved in the problem ($\sim 10^6-10^9$ particles) leads to high computation costs and long simulation times.
  The clever solution to this problem is the macro-particle assumption, through which an ensemble of particles ($\sim 10^2-10^4$ particles) are treated as one single entity with charge to mass ratio equal to the particles of interest, which are here electrons.
  The relativistic equation of motion for electron macro-particles then reads as
  \begin{equation}
  \label{equationOfMotion}
  \frac{\partial}{\partial t} (\gamma m \vec{v}) = -e(\vec{E}+ \vec{v} \times \vec{B}), \qquad \mathrm{and} \qquad \frac{\partial \vec{r}}{\partial t} = \vec{v},
  \end{equation}
  where $\vec{r}$ and $\vec{v}$ are the position and velocity vectors of the electron, $e$ is the electron charge and $m$ is its rest mass.
  $\gamma$ stands for the Lortenz factor of the moving particle.
  
  \textbf{Update Algorithm:} There are numerous update algorithms proposed for the time domain solution of \eref{equationOfMotion}, including various Runge-Kutta and finite difference algorithms.
  Among these methods, Boris scheme has garnered specific attention owing to its interesting peculiarity which is being simplectic.
  Simplectic update algorithms are update procedures which maintain the conservation of any parameter in the equation which obey a physical conservation law.
  In contrast to the acceleration problem, in a FEL problem effect of the magnetic field on a particle motion plays the most important role.
  Therefore, using a simplectic algorithm is essential to obtain reliable results.
  This was the main motivation to choose the Boris scheme for updating the particle motion in MITHRA.
  
  We sample the particle position at times $m\Delta t_b$, which is represented by $\vec{r}^m$ and the particle normalized momentum at times $(m-\frac{1}{2})\Delta t_b$ which is written as $\gamma \vec{\beta}^{m-1/2}$.
  Then, by having $\vec{r}^m$ and $\gamma \vec{\beta}^{m-1/2}$ as the known parameters and $\vec{E}^m_t$ and $\vec{B}^m_t$ as the \emph{total} field values imposed on the particle at instant $m\Delta t$, the values $\vec{r}^{m+1}$ and $\gamma \vec{\beta}^{m+1/2}$ are obtained as follows:
  \begin{align}
  \arraycolsep=1.0pt\def\arraystretch{2.0}
  \vec{t}_1 & = \gamma \vec{\beta}^{m-1/2} - \frac{e\Delta t_b \vec{E}^m_t}{2mc} \nonumber \\
  \vec{t}_2 & = \vec{t}_1 + \alpha \vec{t}_1 \times \vec{B}^m_t \nonumber \\
  \vec{t}_3 & = \vec{t}_1 + \vec{t}_2 \times \frac{2 \alpha \vec{B}^m_t}{1+\alpha^2 \vec{B}^m_t \cdot \vec{B}^m_t} \label{BorisScheme} \\
  \gamma \vec{\beta}^{m+1/2} & = \vec{t}_3 - \frac{e\Delta t_b \vec{E}^m_t}{2mc} \nonumber \\
  \displaystyle \vec{r}^{m+1} & = \vec{r}^l + \frac{c \Delta t_b \gamma \vec{\beta}^{m+1/2} }{\sqrt{1+\gamma \vec{\beta}^{m+1/2} \cdot \gamma \vec{\beta}^{m+1/2} } } \nonumber
  \end{align}
  where $\alpha = -e\Delta t_b / (2 m \sqrt{1+\vec{t}_1 \cdot \vec{t}_1})$.
  $\vec{E}^m_t = \vec{E}^m_{ext} + \vec{E}^m$ and $\vec{B}^m_t = \vec{B}^m_{ext} + \vec{B}^m$ are total fields imposed on the particle, which are equal to the superposition of the radiated field with the external fields, i.e. the undulator or the seed fields.
  In order to figure out the derivation of the equations \eref{BorisScheme}, the reader is referred to \cite{boris1970acceleration,boris1970relativistic}.
  As seen from the above equations, the electric and magnetic fields at time $m \Delta t_b$ and the position $\vec{r}$ of the particle are needed to update the motion.
  In the next section, the equations to extract these values from the computed values of the magnetic and scalar potential are presented.
  Note that to achieve a certain precision level, the required time step in updating the bunch properties ($\Delta t_b$) is usually much smaller than the time step for field update ($\Delta t$).
  In MITHRA, there exists the possibility for setting different time steps for PIC and FDTD algorithms.
  
  \textbf{Field Evaluation:} As described in the FDTD section, the propagating fields in the computational domain are evaluated by solving the wave equation for the magnetic vector potential, i.e. \eref{HelmholtzA}.
  To update the particle position and momentum, one needs to obtain the field values $\vec{E}^m$ and $\vec{B}^m$ from the potentials $\vec{A}$ and $\varphi$.
  For this purpose, the equations \eref{BvsA} and \eref{EvsA} need to be discretized in a consistent manner to provide the accelerating field with lowest amount of dispersion and instability error.
  First, the values of magnetic and scalar potentials at $t+\Delta t/2$ are used to evaluate the electromagnetic fields at the cell vertices.
  Subsequently, the field values are interpolated to the particle location for updating the equation of motion.
  An important consideration at this stage is compatible interpolation of fields from the cell vertices with the interpolations used for current and charge densities.
  Similar interpolation algorithms should be followed to cancel the effect of self-forces on particle motion.
  
  Using the equation \eref{BvsA}, the magnetic field $\vec{B}^n_{i,j,k}$ at cell vertex $(i,j,k)$ is calculated as follows:
  \begin{equation}
  {B_x}^n_{i,j,k} = \frac{1}{2} \sum_{l=n,n+1} \left( \frac{{A_z}_{i,j+1,k}^l-{A_z}_{i,j-1,k}^l}{2\Delta y} - \frac{{A_y}_{i,j,k+1}^l-{A_y}_{i,j,k-1}^l}{2\Delta z} \right),
  \end{equation}
  \begin{equation}
  {B_y}^n_{i,j,k} = \frac{1}{2} \sum_{l=n,n+1} \left( \frac{{A_x}_{i,j,k+1}^l-{A_x}_{i,j,k-1}^l}{2\Delta z} - \frac{{A_z}_{i+1,j,k}^l-{A_z}_{i-1,j,k}^l}{2\Delta x} \right),
  \end{equation}
  \begin{equation}
  {B_z}^n_{i,j,k} = \frac{1}{2} \sum_{l=n,n+1} \left( \frac{{A_y}_{i+1,j,k}^l-{A_y}_{i-1,j,k}^l}{2\Delta x} - \frac{{A_x}_{i,j+1,k}^l-{A_x}_{i,j-1,k}^l}{2\Delta y} \right).
  \end{equation}
  Similarly, equation \eref{EvsA} is employed to evaluate the electric field at the cell vertices.
  The electric field $\vec{E}^n_{i,j,k}$ is obtained from the following equations:
  \begin{align}
  {E_x}^n_{i,j,k} = & \left( -\frac{{A_x}_{i,j,k}^{n+1}-{A_x}_{i,j,k}^n}{\Delta t} - \frac{\varphi_{i+1,j,k}^n-\varphi_{i-1,j,k}^n}{2\Delta x} \right), \\
  {E_y}^n_{i,j,k} = & \left( -\frac{{A_y}_{i,j,k}^{n+1}-{A_y}_{i,j,k}^n}{\Delta t} - \frac{\varphi_{i,j+1,k}^n-\varphi_{i,j-1,k}^n}{2\Delta y} \right), \\
  {E_z}^n_{i,j,k} = & \left( -\frac{{A_z}_{i,j,k}^{n+1}-{A_z}_{i,j,k}^n}{\Delta t} - \frac{\varphi_{i,j,k+1}^n-\varphi_{i,j,k-1}^n}{2\Delta z} \right).
  \end{align}
  
\begin{figure}
  	\centering
  	\includegraphics[height=2.0in]{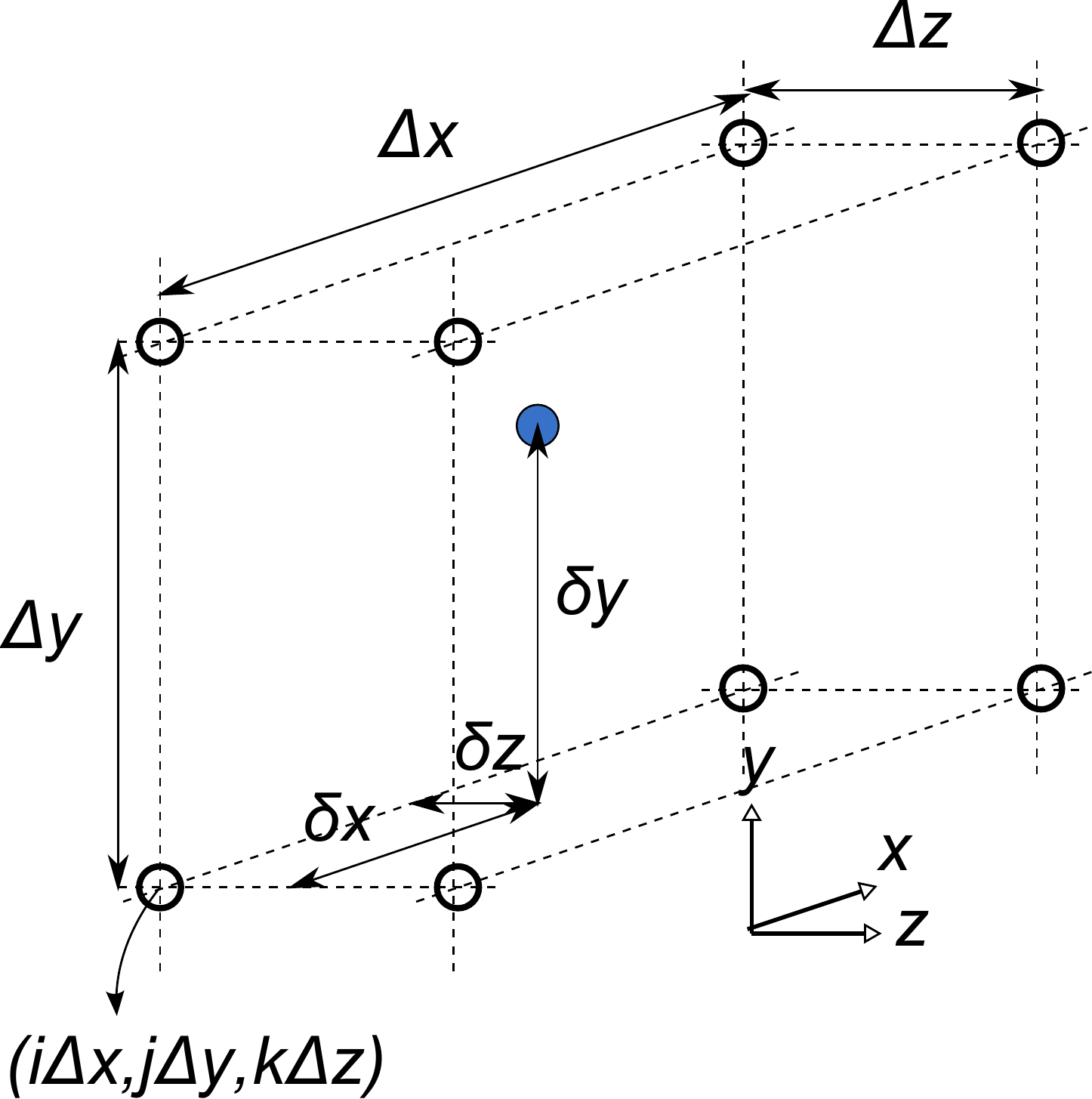}
  	\caption{Schematic illustration of the parameters used to locate a particle within the computational domain.}
  	\label{FDTDPICScheme}
  \end{figure}

  Suppose that a particle resides at the cell $ijk$ with the grid point indices shown in Fig.\,\ref{FDTDPICScheme}.
  As illustrated in Fig.\,\ref{FDTDPICScheme}, the distance to the corner $(i\Delta x, j\Delta y, k \Delta z)$ is assumed to be $(\delta x, \delta y, \delta z)$.
  We use a linear interpolation of the fields from the vertices to the particle position to calculate the imposed field.
  If $\varsigma$ denotes for a component of the electric or magnetic field, i.e. $\varsigma \in \{ E_x, E_y, E_z, B_x, B_y, B_z \}$, one can write
  \begin{equation}
  \label{fieldInterpolation}
  \varsigma^p = \sum\limits_{I,J,K} \left(\frac{1}{2} + (-1)^I \left| \frac{1}{2} - \frac{\delta x}{\Delta x} \right| \right)
  \left(\frac{1}{2} + (-1)^J \left| \frac{1}{2} - \frac{\delta y}{\Delta y} \right| \right)
  \left(\frac{1}{2} + (-1)^K \left| \frac{1}{2} - \frac{\delta z}{\Delta z} \right| \right) \varsigma_{i+I,j+J,k+K},
  \end{equation}
  where $I$, $J$, and $K$ are equal to either 0 or 1, producing the eight indices corresponding to the eight corners of the mesh cell.
  
  \textbf{Current Deposition:} Once the position and momentum of all the particles over the time interval $\Delta t $ is known, one needs to couple the pertinent currents into the wave equation \eref{HelmholtzA}.
  As described before, this coupling over time is implemented through the equation \eref{currentIntegral}.
  The remaining question is how to evaluate the related currents on the grid points, i.e. the method for performing a spatial interpolation.
  To maintain consistency, we should use a similar interpolation scheme as used for the field evaluation.
  This assumption leads to the following equation for spatial interpolation.
  \begin{equation}
  \label{chargeIntegral}
  {\rho^p}_{i+I,j+J,k+K} = \rho \left(\frac{1}{2} + (-1)^I \left| \frac{1}{2} - \frac{\delta x}{\Delta x} \right| \right) \left(\frac{1}{2} + (-1)^J \left| \frac{1}{2} - \frac{\delta y}{\Delta y} \right| \right) \left( \frac{1}{2} + (-1)^K \left| \frac{1}{2} - \frac{\delta z}{\Delta z} \right| \right)
  \end{equation}
  where $\rho$ is the charge density attributed to each macro-particle, namely $q/(\Delta x \Delta y \Delta z)$.
  ${\rho^p}_{i,j,k}$ is the charge density at the grid point $(i,j,k)$ due to the moving particle $p$ in the computational mesh cell (Fig.\,\ref{FDTDPICScheme}).
  $I$, $J$, and $K$ are equal to either 0 or 1, which produce the eight indices corresponding to the eight corners of the mesh cell.
  The total charge density $\rho_{i,j,k}$ will be a superposition of all the charge densities due to the moving particles of the bunch.
  We have removed the superscripts corresponding to the time instant, to avoid the confusion due to different time marching steps $\Delta t$ and $\Delta t_b$.
  The above interpolation is carried out at each update step of the field values.
  One can consider the above interpolation equations as a rooftop charge distribution centered at the particle position and expanding in the regions $(-\Delta x < x < \Delta x, -\Delta y < y < \Delta y, -\Delta z < z < \Delta z)$.
  Eventually, equation \eref{currentIntegral} is used to calculate the corresponding current densities.
  
  The combination of equation \eref{currentIntegral} and \eref{chargeIntegral} should maintain the charge conservation law (equation \eref{chargeLaw}) in a discretized space.
  For this purpose, the projection from position vectors $\vec{r}$ to the Cartesian components in \eref{currentIntegral} should be done using the so-called ZigZag scheme proposed in \cite{umeda2003new}.
  According to this scheme when a particle moves from the point $(x_1,y_1,z_1)$ to $(x_2,y_2,z_2)$, the motion is divided into two separate movements, namely (i) from $(x_1,y_1,z_1)$ to $(x_r,y_r,z_r)$, and (ii) from $(x_r,y_r,z_r)$ to $(x_2,y_2,z_2)$.
  The coordinates of the relay point $(x_r,y_r,z_r)$ are obtained from the following equation:
  \begin{align}
  x_r = & \min\left[ \min(i_1 \Delta x,i_2 \Delta x) + \Delta x, \max\left( \max(i_1 \Delta x,i_2 \Delta x), \frac{x_1+x_2}{2} \right) \right] \nonumber \\
  y_r = & \min\left[ \min(j_1 \Delta y,j_2 \Delta y) + \Delta y, \max\left( \max(j_1 \Delta y,j_2 \Delta y), \frac{y_1+y_2}{2} \right) \right] \\
  z_r = & \min\left[ \min(k_1 \Delta z,k_2 \Delta z) + \Delta z, \max\left( \max(k_1 \Delta z,k_2 \Delta z), \frac{z_1+z_2}{2} \right) \right] \nonumber
  \end{align}
  where $(i,j,k)$ with indices 1 and 2 represent the cell numbers containing the initial and final points, respectively.
  Since potential $\vec{A}$ and $\varphi$ are obtained from current and charge in exactly similar ways (update equations), if charge and current obey the charge conservation, the gauge condition will be automatically satisfied.
  In other words, if the initial potentials satisfy the gauge condition, solving equations \eref{HelmholtzA}, \eref{HelmholtzF}, and \eref{chargeLaw} results in potential distributions at time $t$ which also satisfy the gauge condition.
  The only requirement is that both potentials are discretized and updated in the same way.
  
  \subsubsection{Quantity Initialization}
  
  The previous two sections on FDTD and PIC algorithms present a suitable and efficient framework for the computation of interaction between charged particles and propagating waves.
  However, the initial conditions are always required for a complete determination of the problem of interest.
  For a FEL simulation, the initial conditions corresponding to the FEL input are given to the FDTD/PIC solver.
  For example, in case of a Self Amplified Spontaneous Emission (SASE) FEL, the initial fields are zero and there is no excitation entering the computational domain, whereas for a seeded FEL, an outside excitation should be considered entering the computational domain.
  The explanation of how such initializations are implemented in MITHRA is the goal in this section.
  
  One novel feature of the method, followed here, is the solution of Maxwell's equations in the bunch rest frame.
  It can be shown that a proper coordinate transformation yields the matching of all the major parameters in a FEL simulation, namely bunch length, undulator period, undulator length, and radiation wavelength.
  \begin{figure}
  	\centering
  	\includegraphics[width=6.0in]{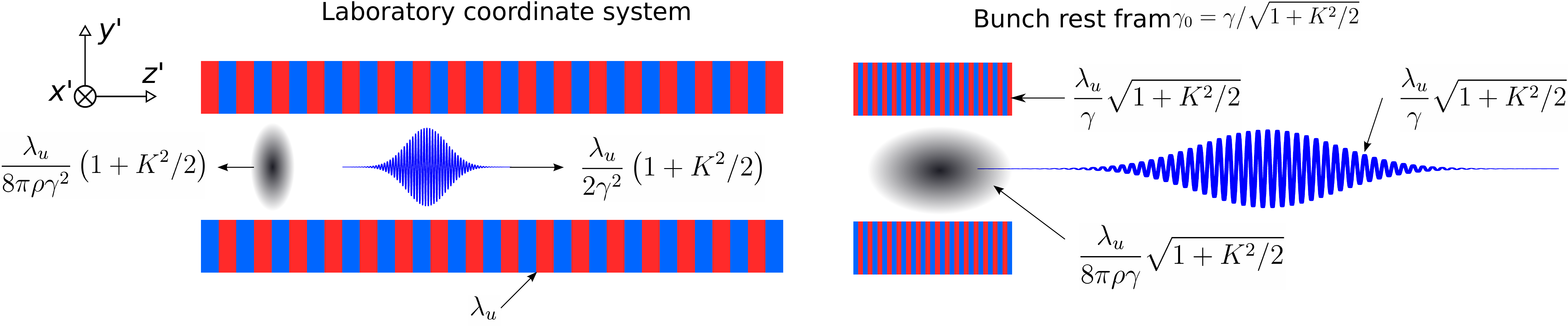}
  	\caption{Schematic illustration of the Lorentz boosting to transform the problem from the laboratory frame to the bunch rest frame.}
  	\label{LorentzBoostingScheme}
  \end{figure}
  Fig.\,\ref{LorentzBoostingScheme} schematically describes the advantage of moving into the bunch rest frame.
  In a typical FEL problem, the FEL parameter $\rho_{FEL}$ is about $10^{-3}$.
  Therefore, simulation of FEL interaction with a bunch equal to the cooperation length of the FEL ($L_c=\lambda_l/(4 \pi \rho_{FEL})$, with $\lambda_l$ being the radiation wavelength) requires a simulation domain only 100 times larger than the wavelength.
  This becomes completely possible with the today computer technology and constitutes the main goal of MITHRA.
  In this section, the main basis for Lorentz boosting the simulation coordinate is described first.
  Afterwards, the relations for evaluating the undulator fields in the Lorentz boosted framework are presented.
  Finally, the electron bunch initialization in the Lorentz-boosted framework is discussed.
  
  \textbf{Lorentz Transformation:} It is known from the FEL theory that a bunch with central Lorentz factor equal to $\gamma$ moves in an undulator with an average Lorentz factor equal to $\gamma_0=\gamma/\sqrt{(1+K^2/2)}$, where $K=eB\lambda_u/(2\pi m c)$ is the undulator parameter determining the amplitude of the wiggling motion.
  Consequently, a frame moving with normalized velocity $\beta_0 = \sqrt{1-1/\gamma_0^2}$ is indeed the bunch rest frame, where the volume of the computational domain stays minimal.
  Transforming into this coordinate system necessitates tailoring the bunch and undulator properties.
  For this purpose, the Lorentz length contraction, time dilation and relativistic velocity addition need to be employed.
  
  In MITHRA, the input parameters are all taken in the laboratory frame and the required Lorentz transformations are carried out based on the bunch energy.
  The required transformations for the computational mesh are as the following:
  \begin{equation}
  \Delta z = \Delta z' \gamma_0, \qquad \Delta t = \Delta t' / \gamma_0,  \qquad \Delta t_b = \Delta t'_b / \gamma_0,
  \label{LorentzTransformB}
  \end{equation}
  where the prime sign stands for the quantities in the laboratory frame.
  The quantities without prime are values in the bunch rest frame, which are used in the FDTD/PIC simulation.
  With the consideration of the above transformations, the length of the total computational domain along the undulator period and the total simulation time is also transformed similarly.
  
  In addition to the data for the computational mesh, the properties of the electron bunch also change after the Lorentz boosting.
  This certainly affects the bunch initialization process which is thoroughly explained in the next section.
  An electron bunch in MITHRA is initialized and characterized by the following parameters:
  \begin{enumerate}[(i)]
  	\item Mean electron position: $(\bar{x}_b, \bar{y}_b, \bar{z}_b)$,
  	\item Mean electron normalized momentum: $(\overline{\gamma \beta}_x, \overline{\gamma \beta}_y, \overline{\gamma \beta}_z)$,
  	\item RMS value of the electron position distribution: $(\sigma_x, \sigma_y, \sigma_z)$,
  	\item RMS value of the electron normalized momentum distribution: $(\sigma_{\gamma \beta_x}, \sigma_{\gamma \beta_y}, \sigma_{\gamma \beta_z})$.
  \end{enumerate}
  As mentioned previously, the above parameters are entered by the user in the laboratory frame.
  To transform the given values to the bunch rest frame the position related parameters are changed as
  \begin{equation}
  \begin{array}{ccc}
  \bar{x}_b = \bar{x}'_b, & \bar{y}_b = \bar{y}'_b, & \bar{z}_b = \displaystyle \frac{\bar{z}'_b}{\gamma_0 (1 - \bar{\beta'}_z \beta_0)} , \\
  \sigma_x = \sigma'_x, & \sigma_y = \sigma'_y, & \sigma_z = \displaystyle \frac{\sigma'_z}{{\gamma_0 (1 - \bar{\beta'}_z \beta_0)}}.
  \end{array}
  \end{equation}
  To transfer the momentum related quantities, we assume that the main contribution to the Lorentz factor is the momentum along $z$ direction or the undulator period.
  In other words, $(\overline{\gamma \beta}_x, \overline{\gamma \beta}_y, \overline{\gamma \beta}_z) = \gamma ( \bar{\beta}_x, \bar{\beta}_y, \bar{\beta}_z )$, with $\gamma = 1/\sqrt{1-\bar{\beta}_z^2}$.
  Similarly, the RMS values can also be written as $(\sigma_{\gamma \beta_x}, \sigma_{\gamma \beta_y}, \sigma_{\gamma \beta_z}) = \gamma (\sigma_{\beta_x}, \sigma_{\beta_y}, \sigma_{\beta_z})$.
  Using the relativistic velocity transformation \cite{JacksonClassical}, the transformation equations for the above values are found as follows:
  \begin{align}
  \gamma & = \gamma' \gamma_0 (1 - \bar{\beta'}_z \beta_0), \\
  ( \bar{\beta}_x, \bar{\beta}_y, \bar{\beta}_z ) & = ( \bar{\beta'}_x, \bar{\beta'}_y, \sqrt{1 - 1/\gamma^2} ), \\
  (\sigma_{\gamma \beta_x}, \sigma_{\gamma \beta_y}, \sigma_{\gamma \beta_z}) & = (\sigma'_{\gamma \beta_x}, \sigma'_{\gamma \beta_y}, \sigma'_{\gamma \beta_z} ) \gamma_0 (1 - \bar{\beta'}_z \beta_0). \label{LorentzTransformE}
  \end{align}
  Equations \eref{LorentzTransformB}-\eref{LorentzTransformE} provide a sufficient set of equations to perform the Lorentz boost to the bunch rest frame.
  
  \textbf{Field Initialization:} The utilized FDTD/PIC algorithm solves the Maxwell's equation coupled with the motion equation of an ensemble of particles.
  Therefore, in addition to the field values, particle initial conditions should also be initialized.
  For a SASE-FEL problem, the initial field profile is zero everywhere, whereas for a seeded FEL the initial seed should enter the computational domain through the boundaries.
  In both cases, the external field which is the undulator field should separately be initialized.
  
  \emph{Undulator Field:} By solving the Laplace equation for the magnetic field, the undulator field in the laboratory frame is found to be as the following (Fig.\,\ref{LorentzBoostingScheme}) \cite{schmuser2014free}:
  \begin{equation}
  \label{undulatorField}
  B'_x = 0, \qquad B'_y = B_0 \cosh(k_uy')\sin(k_uz'), \qquad B'_z = B_0 \sinh(k_uy')\cos(k_uz'),
  \end{equation}
  where $B_0$ is the maximum transverse field of the undulator.
  To calculate the undulator field in the bunch rest frame, first the position is transformed to laboratory frame $(x',y',z')$ through the Lorentz boost equations.
  Afterwards, the field is evaluated using the equation \eref{undulatorField}.
  Eventually, these fields are transformed back into the bunch rest frame.
  The above approach, although adds few mathematical operations for the calculation of undulator fields, it enables straightforward implementation of various realistic effects, like fringing fields of the entrance section and non-gaussian field profiles.
  
  An important consideration in the initialization of undulator field is the entrance region of the undulator.
  A direct usage of the equation \eref{undulatorField} with zero field for $z'<0$ causes an abrupt variation in the particles motion, which results in a spurious coherent radiation.
  In fact, in a real undulator, there exists fringing fields at the undulator entrance, which remove any abrupt transition in the undulator field and consequently the particle radiations \cite{sagan2003magnetic}.
  To the best of our knowledge, the fringing fields are always modeled numerically and there exists no analytical solution for the problem.
  Here, we approximate the fringing fields by a gradually decreasing magnetic field in form of a Neumann function.
  The coefficients in the function are set such that the particles do not gain any net transverse momentum and stay in the computational domain as presumed.
  The undulator field for $z'<0$ in the laboratory frame is obtained as the following:
  \begin{equation}
  B'_x = 0, \qquad B'_y = B_0 \cosh(k_uy')k_uz'e^{-(k_uz')^2/2}, \qquad B'_z = B_0 \sinh(k_uy')e^{-(k_uz')^2/2},
  \label{fringingField}
  \end{equation}
  The same transformations as in \eref{undulatorField} can be used to approximate the fringing field values in the bunch rest frame.
  
  \emph{Seed Field:} External excitation of free electron laser process using a seed mechanism has proved to be advantageous in terms of output spectrum, photon flux and the required undulator length \cite{pellegrini2016physics,schmuser2014free}.
  Such benefits have propelled the proposal of seeded FEL schemes.
  To simulate such a mechanism, MITHRA uses the TF/SF (total-field/scattered-field) technique to introduce an external excitation into the computational domain.
  When seeding is enabled by having a non-zero seed amplitude, the second and third points (after the boundary points) constitute the scattered and total field boundaries, respectively.
  Therefore, during the time marching process, after each update according to equation \eref{updateEquation} the excitation terms are added to the fields at TF/SF boundaries.
  For example for the TF/SF boundaries close to $z=z_{min}$ plane, the field values to be used in the next time steps are obtained as the following:
  \begin{align}
  & \text{ SF boundary: }  \psi_{i,j,k}^{'n+1} = \psi_{i,j,k}^{n+1} + \mathcal{A} ( \alpha'_2 f_{i+1,j,k+1}^n  + \alpha'_3 f_{i-1,j,k+1}^n + \alpha'_4 f_{i,j+1,k+1}^n + \alpha'_5 f_{i,j-1,k+1}^n ) + \alpha'_6 f_{i,j,k+1}^n, \nonumber \\
  & \text{ TF boundary: }  \psi_{i,j,k}^{'n+1} = \psi_{i,j,k}^{n+1} - \mathcal{A} ( \alpha'_2 f_{i+1,j,k-1}^n  + \alpha'_3 f_{i-1,j,k-1}^n + \alpha'_4 f_{i,j+1,k-1}^n + \alpha'_5 f_{i,j-1,k-1}^n ) - \alpha'_7 f_{i,j,k-1}^n,
  \end{align}
  where $f_{i,j,k}^n$ is the excitation value at time $n\Delta t$ and position $(i\Delta x,j\Delta y,k\Delta z)$.
  The excitation value is calculated based on the imposed seed fields, which are usually either a plane wave or a Gaussian beam radiation.
  
  \textbf{Electron Bunch Generation:} As described previously, the evolution of the electron bunch is always simulated by following the macro-particle approach, where an ensemble of particles are represented by one sample particle.
  This typically reduces the amount of computation cost for updating the bunch properties by three or four orders of magnitude.
  Due to the high sensitivity of a FEL problem to the initial conditions, correct and proper initialization of these macro-particles play a critical role in obtaining reliable results.
  In computational accelerator physics, different approaches are introduced and developed for bunch generation.
  Some examples are random generation of particles, mirroring macro-particles at different phases to prevent initial average bunching factors, and independent random filling of different coordinates to prevent unrealistic correlations \cite{reiche2000numerical}.
  Among all the different methods, using the sophisticated methods to load the bunch in a "quasi-random" manner seem to be the most appropriate solutions.
  The Halton or Hammersley sequences, as generalizations of the bit-reverse techniques, are implemented in MITHRA for particle generation.
  These sequences compared to random based filling of the phase space avoid the appearance of local clusters in the bunch distribution.
  In addition, the uniform filling of the phase space prevents initial bunching factor of the generated electron bunch, making it well-suited for FEL simulations.
  
  For details on the nature of Halton sequences, the reader is referred to the specialized documents.
  By having the above uniform distributions, the 6D phase space of the initial bunch can be filled according to the desired bunch properties.
  
  In MITHRA, different schemes for the user is implemented to generate the initial electron bunch.
  The main requirements for initializing the bunches is to generate 1D and 2D set of numbers with either uniform or Gaussian distributions.
  Suppose $x_1$ and $x_2$ are two uncorrelated number sequences produced by the Halton algorithm.
  A 1D uniform distribution $y_1$ with average $y_{m1}$ and total width $y_{s1}$ is found by the following transformation:
  \begin{equation}
  \label{uniform1D}
  y_1 = y_{s1} (x_1 - \frac{1}{2}) + y_{m1}.
  \end{equation}
  Such a distribution is used when a bunch with uniform current profile ($z$ distribution of particles) is to be initialized.
  On the other hand, a 1D Gaussian distribution is needed when radiation of a bunch with Gaussian current profile is modeled.
  To generate bunches with Gaussian distribution, we employ Box-Muller's theory to extract a sequence of numbers with Gaussian distribution from two uncorrelated uniform distributions.
  Based on this theory, a 1D Gaussian distribution $y_2$ with average $y_{m2}$ and deviation width $y_{s2}$ is found by the following transformation:
  \begin{equation}
  \label{gaussian1D}
  y_2 = y_{s2} \sqrt{-2 \ln x_1} \cos(2\pi x_2) + y_{m2}.
  \end{equation}
  
  Similar to the undulator fields, an abrupt variation in the bunch profile results in an unrealistic coherent scattering emission (CSE), which happens if the uniform bunch distribution is directly initialized from equation (\ref{uniform1D}).
  CSE is avoided by imposing smooth variations in the particle distribution.
  For this purpose, a small Gaussian bunch with the same density as the real bunch and a width equal to an undulator wavelength is produced.
  The lower half of the bunch (particles with smaller $z$) is transferred to the tail and the other half is placed at the head of the uniform bunch.
  Hence, a uniform current profile with smooth variations at its head and tail is created.
  
  The transverse coordinates of the bunches are initialized using 2D distributions.
  In MITHRA, a 2D Gaussian distribution is assumed for transverse coordinates.
  To generate such a distribution, two independent sets of numbers $x_1$ and $x_2$ are generated based on Halton sequence.
  The desired 2D Gaussian distribution with average position $(y_{m3},y_{m4})$ and total deviation $(y_{s3},y_{s4})$ is produced as the following: 
  \begin{align}
  \label{gaussian2D}
  \displaystyle y_3 & = y_{s3} \sqrt{-2 \ln x_1} \cos(2\pi x_2) + y_{m3}, \nonumber \\
  \displaystyle y_4 & = y_{s4} \sqrt{-2 \ln x_1} \sin(2\pi x_2) + y_{m4}.
  \end{align}
  Such algorithms are similarly used to generate the distribution in particle momenta.
  The only difference is that for initializing a distribution in momentum merely Gaussian profiles are considered in transverse and longitudinal coordinates.
  
  Free electron laser radiation should start from a nonzero initial radiation.
  This radiation can be in form of an initial seed field, initial modulation in the bunch, or the radiation from bunch shot noise.
  The implementation of seeding through an external excitation using TF/SF boundaries was described in field initialization section.
  Here, we explain how an initial bunching factor, $<e^{jk_uz}>$, is introduced to the electron bunch profile.
  
  For this purpose, we follow the procedure proposed in \cite{penman1992simulation} and \cite{reiche2000numerical}.
  A small variation $\delta z$ is applied to a particle distribution, which is generated using the above formulations.
  $\delta z$ for each particle is obtained from
  \begin{equation}
  \label{bunchingFactor}
  \delta z = \xi \gamma_0 k_u b_f \sin (2 \xi \gamma_0 k_u z),
  \end{equation}
  where $b_f$ is the given bunching factor of the distribution, and $\xi=1+\bar{\beta}_z/\beta_0$ accounts for the change in the bunch longitudinal velocity after entering the undulator.
  The introduced variation to the bunch coordinates, i.e. $z \rightarrow z+\delta z$, yields a bunch with all the given particle and momentum distributions and the desired bunching factor, $b_f$.
  
  \subsubsection{Parallelization}
  
  The large and demanding computation cost needed for the simulation of the FEL process even in the Lorentz boosted coordinate frame necessitates solving the problem on multiple processors to achieve reasonable computation times.
  Therefore, efficient parallelization techniques should be implemented in the FDTD/PIC algorithm to develop an efficient software.
  Traditionally, there are two widely used techniques to run a computation in parallel on several processors: (1) \textit{shared} memory, and (2) \textit{distributed} memory parallelization.
  In the shared memory parallelization or the so-called multi-threading technique, several processors run a code using the variables saved in one shared memory.
  This technique is very suitable for PIC algorithms because it avoids the additional costs of communicating the particle position and momenta between the processors.
  On the other hand, distributed memory technique distributes the involved variables among several processors, solves the problem in each processor independently and communicates the required variables whenever they are called.
  The distributed memory technique is very suitable for FDTD algorithm due to the ease of problem decomposition beyond various machines.
  The advantage is fast reading and writing of the data and the possibility to share the computational load between different machines.
  
  Choosing a suitable parallelization scheme for the hybrid FDTD/PIC algorithm depends on both problem size and machine implementations.
  One can also implement hybrid techniques to take advantage from different features of both shared and distributed algorithms.
  After checking both algorithms on a variety of machines and different FEL problems, we concluded that using distributed memory often leads to more efficient and faster computation.
  
\begin{figure}
  	\centering
  	\includegraphics[width=4.0in]{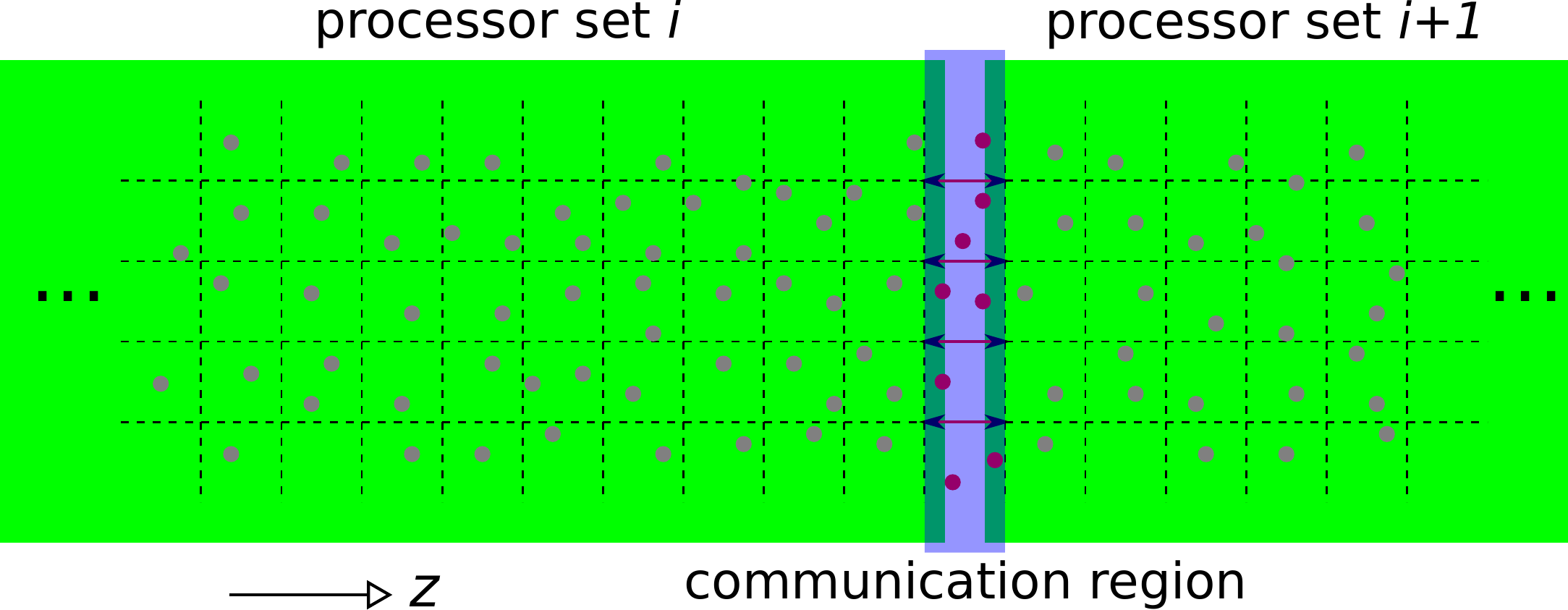}
  	\caption{Schematic illustration of the domain decomposition used for distributed memory parallelization in MITHRA}
  	\label{parallelization}
  \end{figure}

  To parallelize the computation among $N$ sets of processors, the whole computational domain is divided into $N$ domains along $z$ (undulator period) axis.
  In each time update of the field, the field values at the boundaries of each domain are communicated with the corresponding processor set.
  To parallelize the PIC solver, we define a communication domain which as shown in Fig.\,\ref{parallelization}, is the region between the boundaries of each processor.
  After each update of the particles position, it is checked if the particle has entered a communication domain.
  In case of residing in the communication region, the master processor, which is the processor containing the particle in the previous time step, communicates the new coordinates to the slave processor, which is the processor sharing the communication region with the master one.
  Through this simple algorithm, both parallelization schemes function simultaneously to achieve the fastest computation feasible and compatible with an available computing machine.
  
  \subsection{Results}
  
  The goal in this section is more accurate evaluation of the pros and cons of the developed FDTD/PIC algorithm through the presented examples.
  For example, the computation time, numerical stability and numerical convergence and more importantly the reliability of the results are studied based on some standard examples.
  
  \subsubsection{Example 1: Infrared FEL}
  
  \textbf{Problem Definition:} As the first example, we consider an infrared FEL with the parameters tabulated in table \ref{example1}, which is inspired by the numerical analysis presented in \cite{tran1989tda}.
  \begin{table}
  	\caption{Parameters of the Infrared FEL configuration considered as the first example.}
  	\label{example1}
  	\centering
  	{\footnotesize
  		\begin{tabular}{|c||c|}
  			\hline
  			FEL parameter & Value \\ \hline \hline
  			Current profile & Uniform \\ \hline
  			Bunch size & $(260\times260\times100)$\,{\textmu}m \\ \hline
  			Bunch charge & 29.5\,pC \\ \hline
  			Bunch energy & 51.4\,MeV \\	\hline
  			Bunch current & 88.5\,A \\ \hline
  			Longitudinal momentum spread & 0.01\% \\ \hline
  			Normalized emittance & 0.0 \\	\hline
  			Undulator period & 3.0\,cm \\ \hline
  			Magnetic field & 0.5\,T \\ \hline
  			Undulator parameter & 1.4 \\ \hline
  			Undulator length & 5\,m \\ \hline
  			Radiation wavelength & 3\,{\textmu}m \\ \hline
  			Electron density & $2.72\times10^{13} 1/\text{cm}^3$ \\ \hline
  			Gain length (1D) & 22.4\,cm \\ \hline
  			FEL parameter & 0.006 \\ \hline
  			Cooperation length & 39.7\,{\textmu}m \\ \hline
  			Initial bunching factor & $0.01$ \\ \hline
  		\end{tabular}
  	}
  \end{table}
  The bunch distribution is assumed to be uniform in order to compare the results with one-dimensional FEL theory.
  For the same purpose, the transverse energy spread is considered to be zero and a minimal longitudinal energy spread is assumed.
  
  In the mesh definition, the transverse size of the computational domain is considered almost 10 times larger than the bunch transverse size.
  In the contrary, the longitudinal size of the mesh is only three times larger than the bunch length.
  This needs to be considered due to the failure of absorbing boundary conditions for the oblique incidence of the field.
  Furthermore, the bunch and undulator both have tapering sections (equation \eref{fringingField}) to avoid abrupt transitions producing coherent scattering emission (CSE).
  
\begin{figure}[t]
  	\centering
  	$\begin{array}{cc}
  	\includegraphics[width=3.0in]{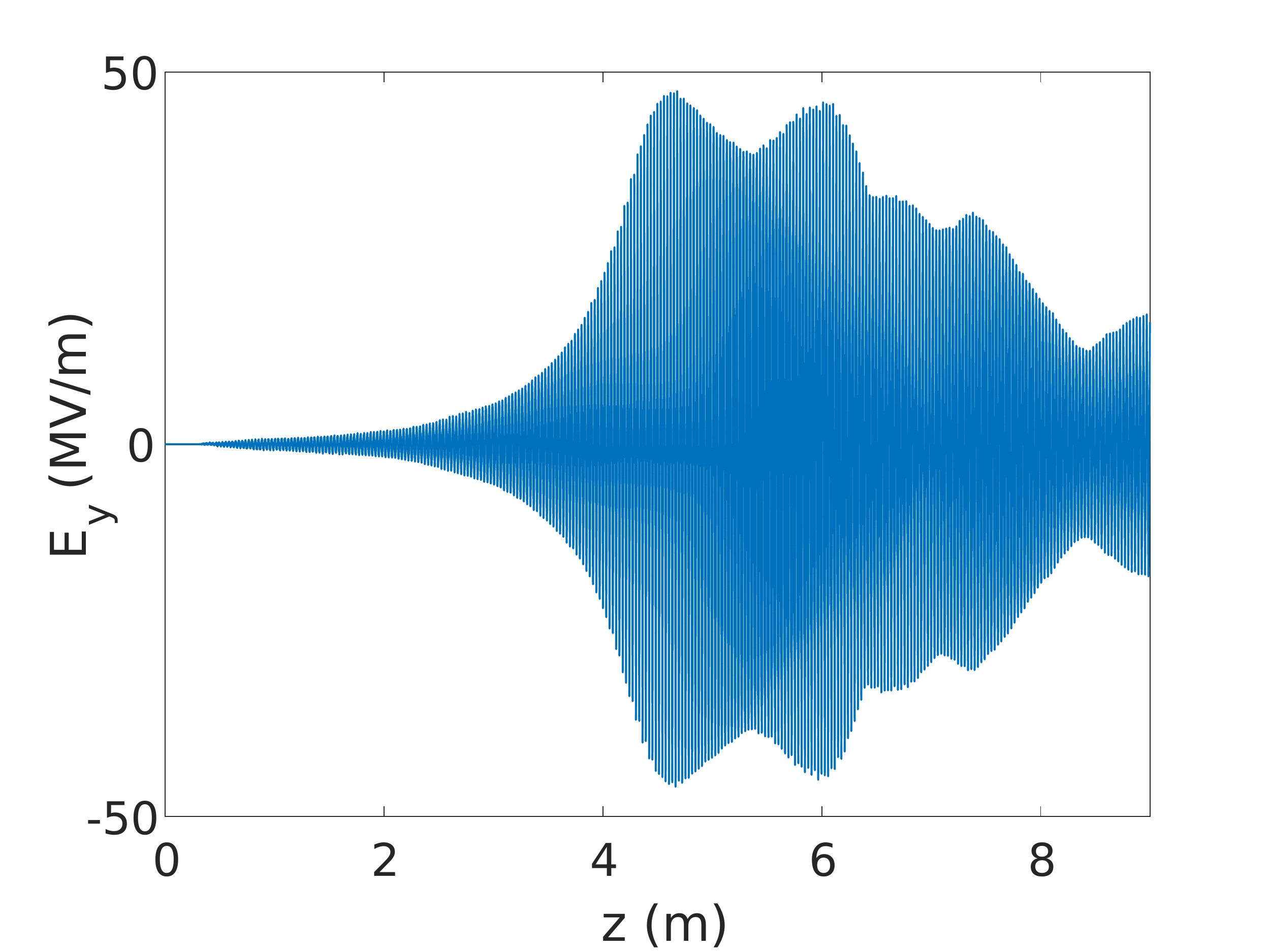} &
  	\includegraphics[width=3.0in]{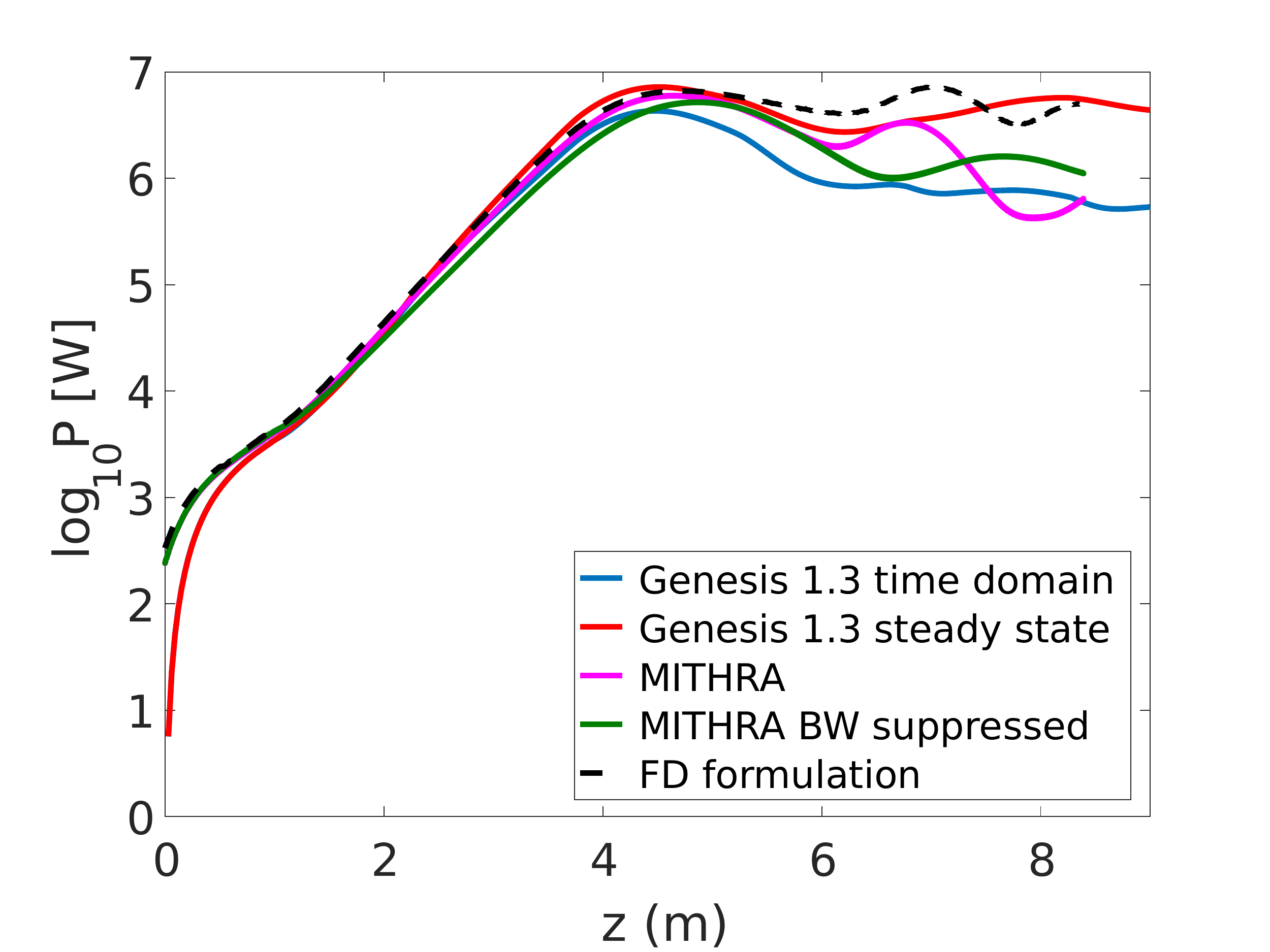} \\
  	(a) & (b)
  	\end{array}$
  	\caption{(a) The transverse field $E_y$ at 110\,{\textmu}m distance from the bunch center and (b) the total radiated power calculated at 110\,{\textmu}m distance from the bunch center in terms of the traveled undulator length.}
  	\label{power-example1}
  \end{figure}

  \textbf{Simulation Results:} In the beginning, we neglect the space-charge effect only to achieve a good assessment of MITHRA simulation results.
  The investigation of space-charge effect will be performed in the second step.
  Fig.\,\ref{power-example1}a shows the transverse electric field sampled at 110\,{\textmu}m in front of the bunch center.
  The logarithmic plot of the radiated power for different propagation lengths ($z$) is also depicted in Fig.\,\ref{power-example1}b.
  We comment that the full-wave analysis offered by MITHRA obtains the total radiated field as a superposition of forward, backward and near-field radiation components.
  In an FEL simulation, one is often interested in the forward radiation component, which can only be extracted at a distance in front of the radiation source, namely the electron bunch.
  This is the main reason for illustrating the radiated power and field at 110\,{\textmu}m in front of the bunch center.
  
  According to the 1D FEL theory the gain length of the considered SASE FEL configuration is $L_G=22.4\,$cm.
  The gain length calculated from the slope of the power curve is $L_G=22\,$cm.
  There exists also a good agreement in the computed saturation power.
  The beam energy according to the data in table \ref{example1} is 1.52\,mJ which for the bunch length of 100\,{\textmu}m corresponds to $P_{beam}=4.55$\,GW beam power.
  The estimated saturation power according to the 1D theory is equal to $P_{sat} = \rho P_{beam} = 2.7$\,GW.
  The saturation power computed by MITHRA is $2.6$\,GW.
  
  We have also performed a comparison study between the obtained results from MITHRA and the code Genesis 1.3, which is presented in Fig.\,\ref{power-example1}b.
  As observed, both codes produce similar results in the initial state and the gain regime.
  Nonetheless, there exists a considerable discrepancy between the calculated radiated power in the saturation regime.
  The illustrated results in Fig.\,\ref{power-example1}b show that the steady state and time domain analyses using Genesis do not produce similar results.
  This shows that the bunch is not long enough to justify the steady state approximation, and dictates a time domain analysis for accurate simulation.
  However, the results obtained by MITHRA at saturation do not match with the Genesis results even in the time domain.
  
  The origin of such a discrepancy is described as follows:
  As explained previously, Genesis 1.3 and all the existing softwares for FEL simulation neglect the backward radiation of the electrons.
  Such an approximation is motivated by the inherent interest in forward radiation in the FEL process.
  However, the backward radiation although is seldom used due to its long wavelength, it influences the motion of electrons, the charge distribution and in turn the FEL output.
  The influence of low-frequency backward radiation on the performance of free electron lasers has been already studied in a 1D regime \cite{maroli2000effects}.
  The effect becomes stronger in the saturation regime, when the electrons are microbunched and the FEL forward radiation is a strong function of the particles distribution.
  To demonstrate this effect, we changed the parallelization algorithm of the field-solver so that the propagated fields are only coupled along the FEL propagation direction.
  This trick suppresses the propagation of backward radiation.
  The results of such an analysis is also shown in Fig.\,\ref{power-example1}b, which shows a relatively better agreement with time domain simulation results returned by Genesis 1.3.
  The still existing discrepancy is attributed to the different formulations of FDTD and TDA algorithms as well as the introduced tapers in bunch current and undulator fields.
  
  There exists a discrepancy between MITHRA and Genesis 1.3 results at the beginning of the undulator.
  The reason for this discrepancy in the initial radiation is that MITHRA initializes the bunch outside the undulator.
  After passing through the fringing fields of the undulator, CSE happens which causes MITHRA to show the beginning of radiation from a value different from zero, whereas in Genesis 1.3 and in many of the typical FEL codes radiation starts from zero.
  We preferred such an operation basis in MITHRA to consider for the CSE effect in real FEL simulations.
  Furthermore, in Fig.\,\ref{power-example1}b, we compare the results obtained using the NSFD implemented in MITHRA and standard FD scheme.
  As observed, formulation based on FD predicts higher radiation power compared to NSFD.
  This effect happens due to the smaller phase velocity of light when wave propagation follows dispersion equation \eref{numericalDispersionCD}.
  The result is slower phase slippage of electron bunch over the radiation and consequently later saturation of the radiation.
  
\begin{figure*}
  	\centering
  	\includegraphics[width=6.0in]{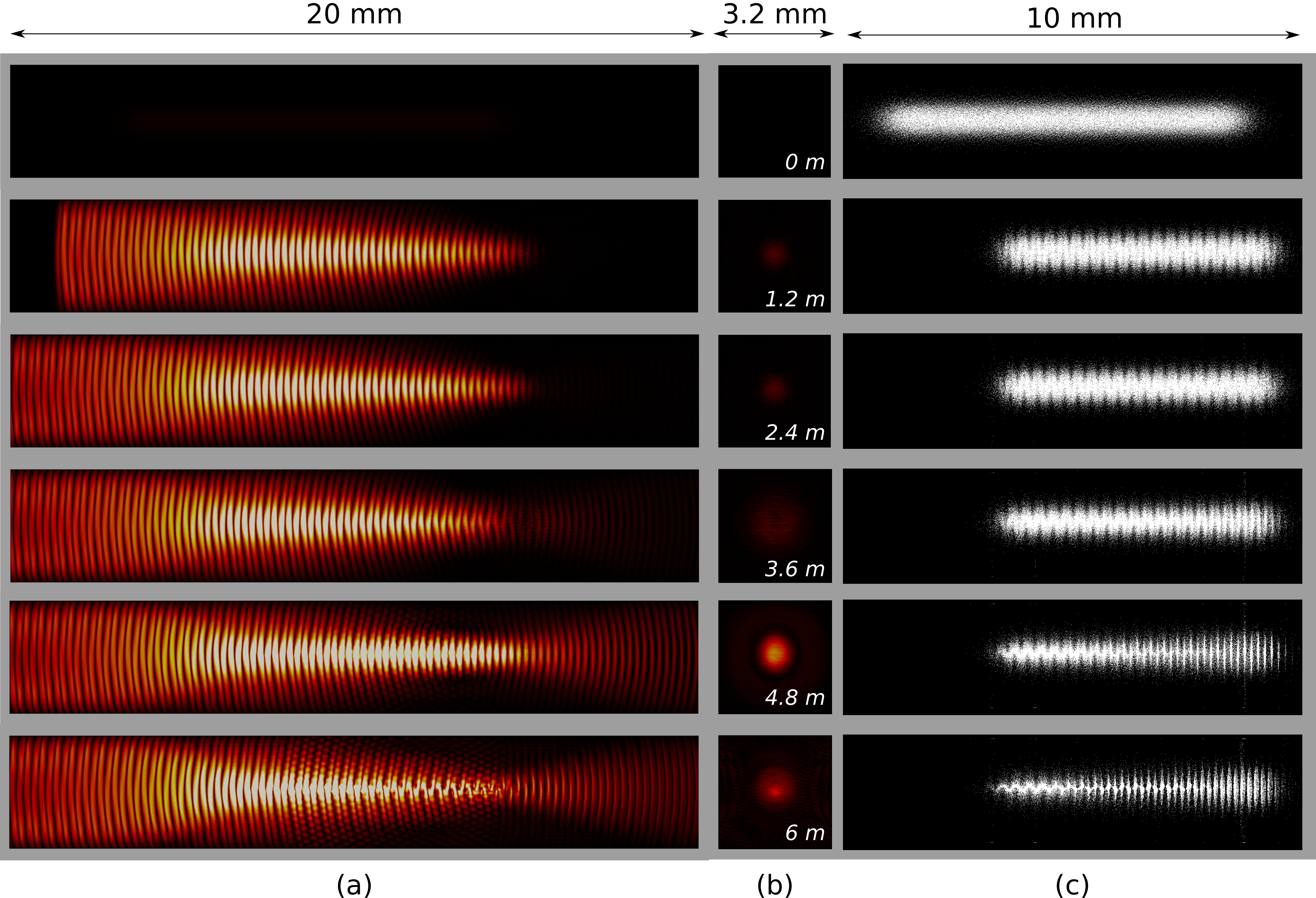}
  	\caption{Snapshots of the radiated field profile taken at (a) $x=0$ and (b) z=60\,{\textmu}m plane and (c) the bunch profile viewed from the $x$ axis.}
  	\label{profile-example1}
  \end{figure*}

  As a 3D electromagnetic simulation, it is always beneficial to investigate the electromagnetic field profile in the computational domain.
  Using the field visualization capability in MITHRA, snapshots of the field profile at different instants and from various view points are provided.
  In Fig.\,\ref{profile-example1}, snapshots of the radiated field profile at different time instants are illustrated.
  The emergence of lasing radiation at the end of the undulator motion is clearly observed in the field profile.
  We believe that the transverse modulations observed in the field profile at the end of FEL interaction are due to the transverse discretization of the spatial domain.
  
  Snapshots of the bunch profile are also presented beside the field profile.
  The main FEL principle which is the lasing due to micro-bunching of the electron bunch is observed from the field and bunch profiles.
  The first two snapshots evidence a considerable change in the bunch length, which occurs due to the entrance in the undulator.
  The bunch outside of the undulator with Lorentz factor $\gamma$ travels faster than the bunch inside the undulator with Lorentz factor $\gamma/\sqrt{1+K^2/2}$.
  Therefore, after the entrance to the undulator the bunch length becomes shorter.
  This effect may not be easily observed in real laboratory frame, but is significant in electron rest frame.
  In addition, it is observed that some of the macro-particles escape the bunch after propagation throughout the undulator.
  This effect is observed after space-charge effects are included in the simulation, which introduces intense transverse forces, particularly in the regions where bunch distribution is dense due to micro-bunching effect.
  
  \textbf{Convergence Analysis:} The convergence rate of the results is one important characteristic used to assess a numerical algorithm.
  In our FEL analysis, there are several parameters introduced by the numerical method which may affect the final result.
  These parameters include (1) number of macro-particles ($n$), (2) time step for updating equation of motion ($\Delta t_b$), (3) longitudinal mesh size ($l_z$), (4) transverse mesh size ($l_x=l_y$), (5) longitudinal discretization ($\Delta z$) and (6) transverse discretization ($\Delta x = \Delta y$).
  Studying the convergence of the results is crucial to acquire an estimate for the uncertainty in the reported values.
  Here, this task is accomplished by sweeping over the above parameters and plotting the error function defined as the following:
  \begin{equation}
  \label{errorDefinition}
  \mathrm{error} = \frac{\int_{z_i}^{z_f} | P(z)-P_0(z) | \mathrm{dz}}{\int_{z_i}^{z_f} P_0(z) \mathrm{dz}},
  \end{equation}
  where $z_i$ and $z_f$ are the beginning and end of the undulator, respectively and $P_0$ is the reference simulation result which is chosen as the results with the highest resolution.
  
\begin{figure}[t]
  	\centering
  	$\begin{array}{ccc}
  	\includegraphics[draft=false,width=2.0in]{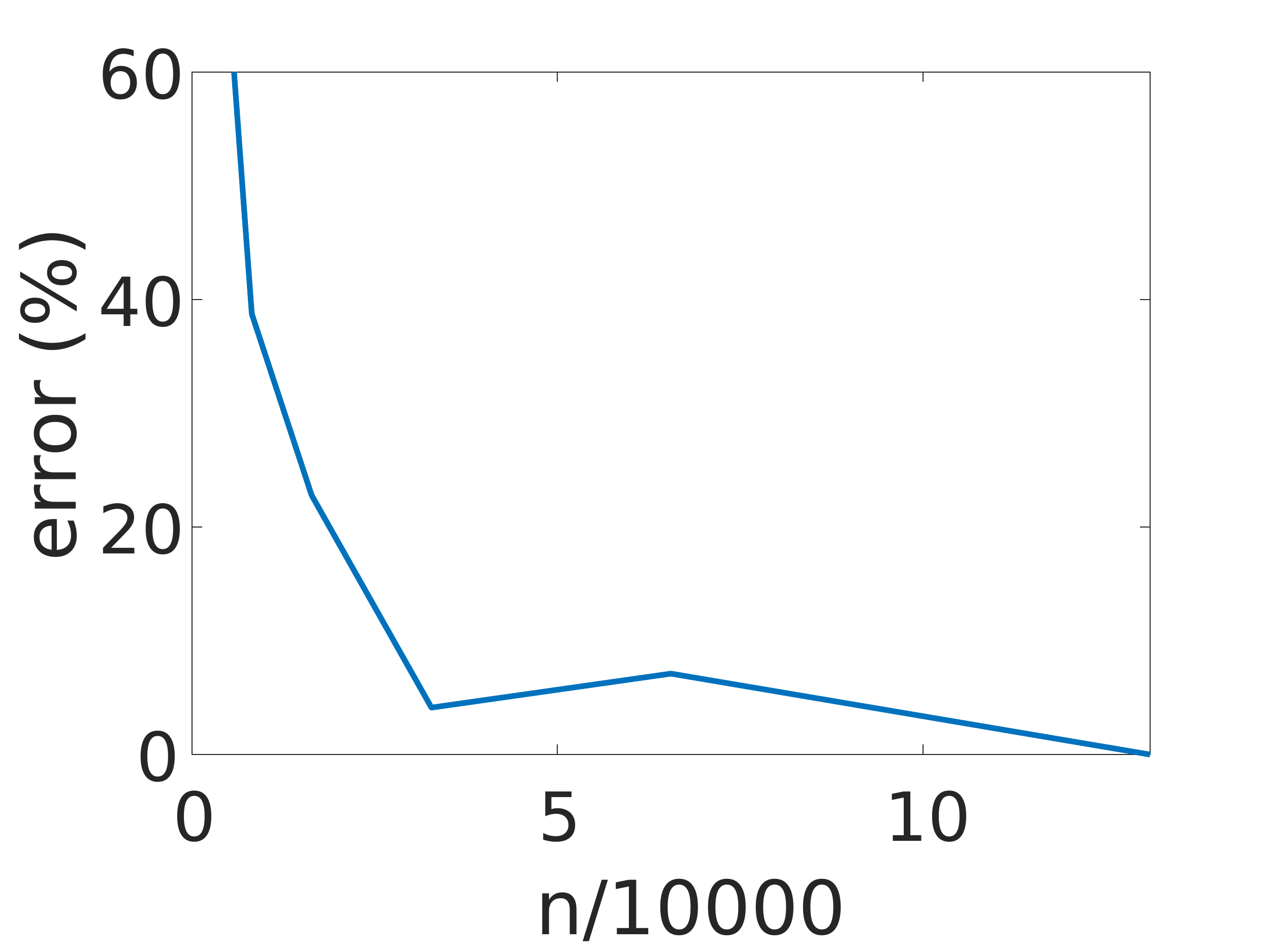} &
  	\includegraphics[draft=false,width=2.0in]{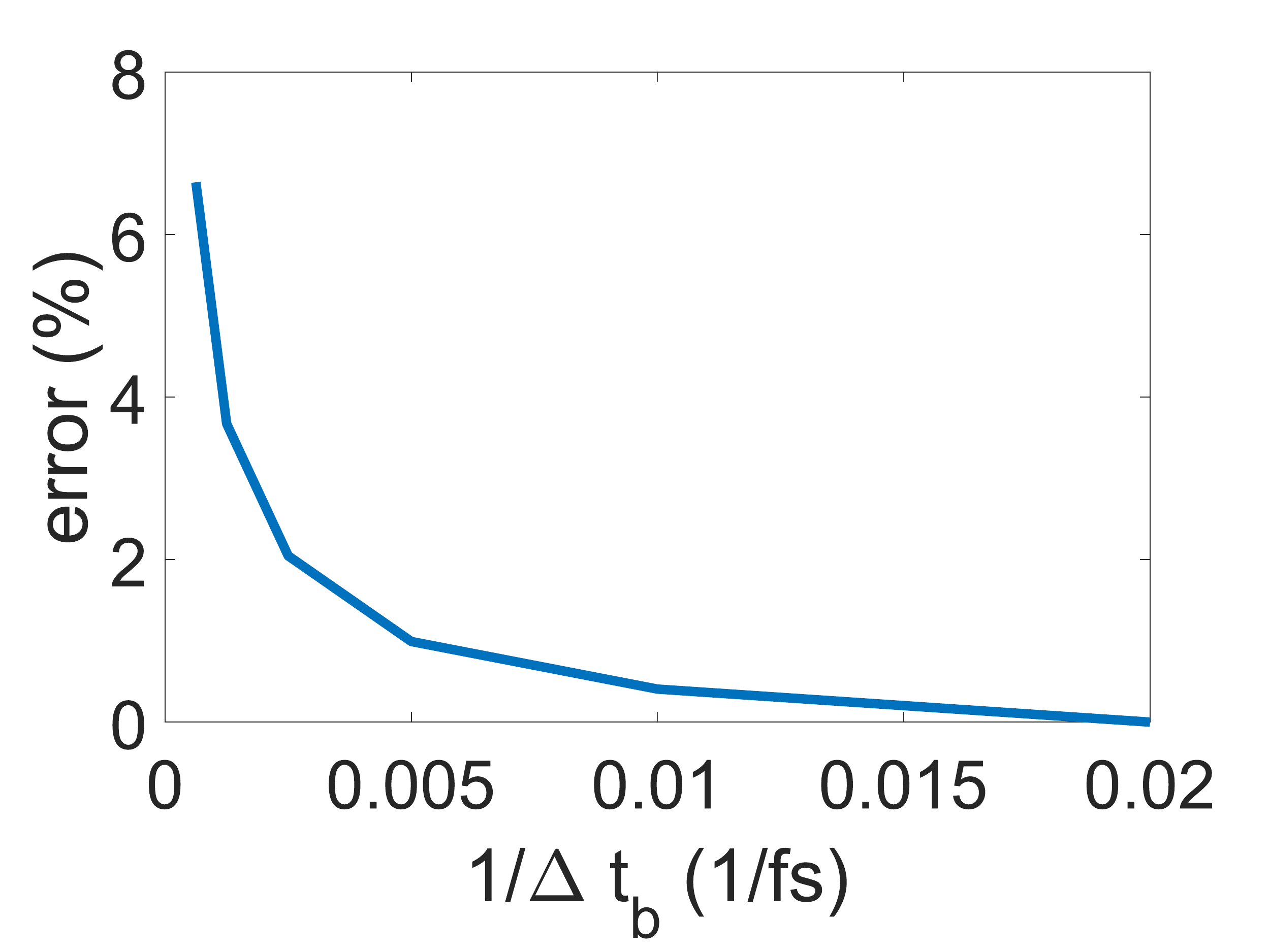} &
  	\includegraphics[draft=false,width=2.0in]{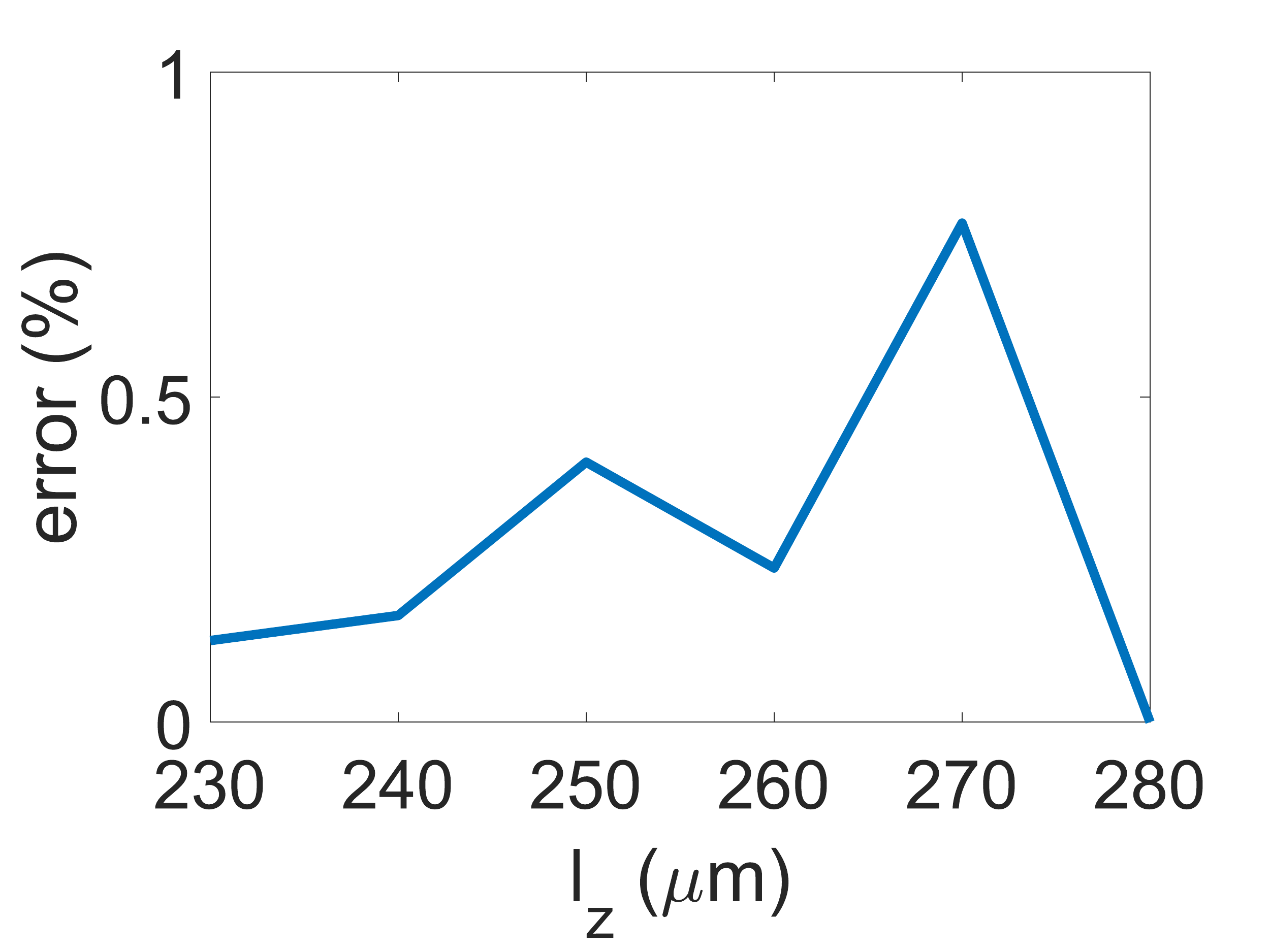} \\
  	(a) & (b) & (c) \\
  	\includegraphics[draft=false,width=2.0in]{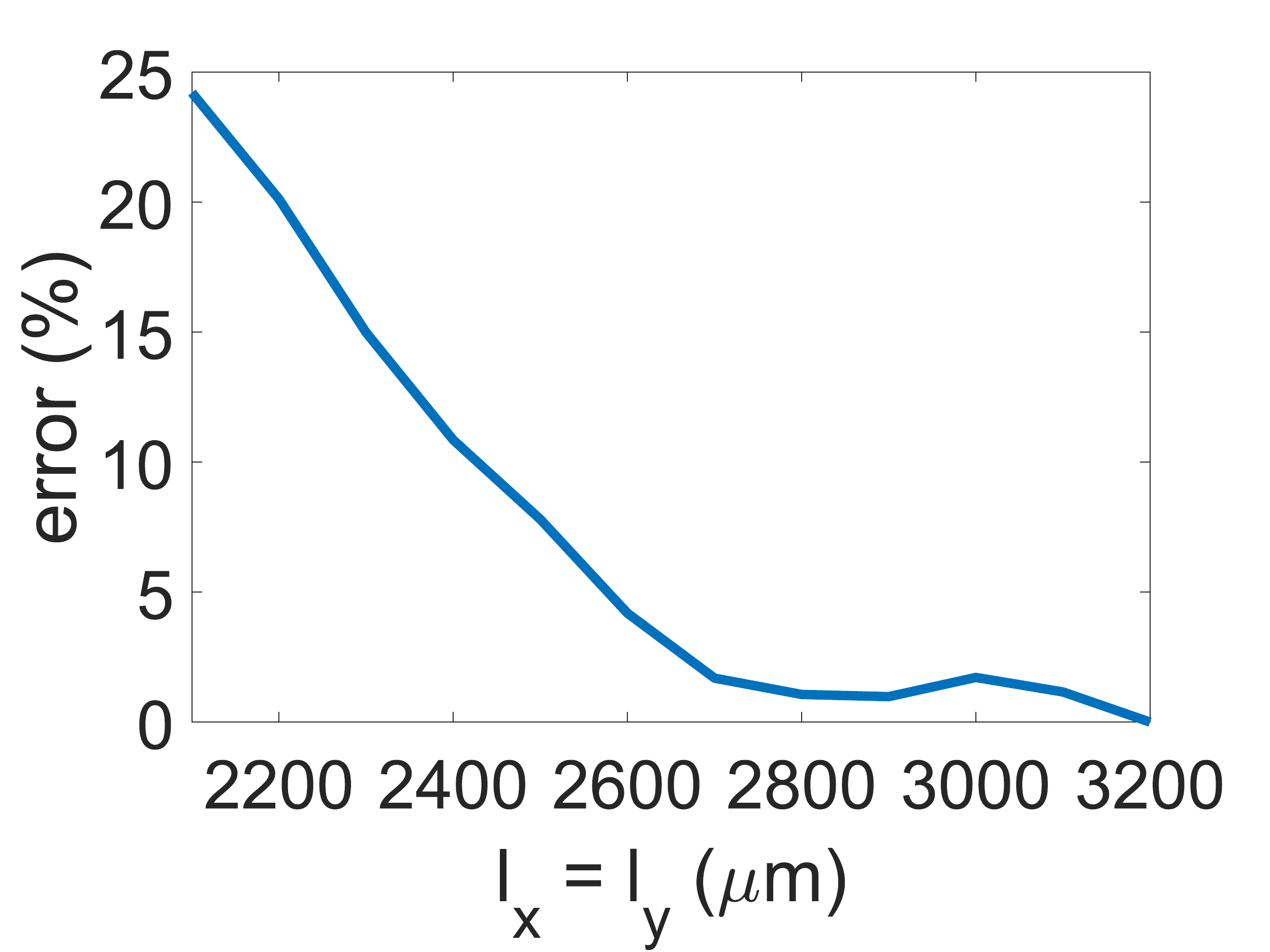} &
  	\includegraphics[draft=false,width=2.0in]{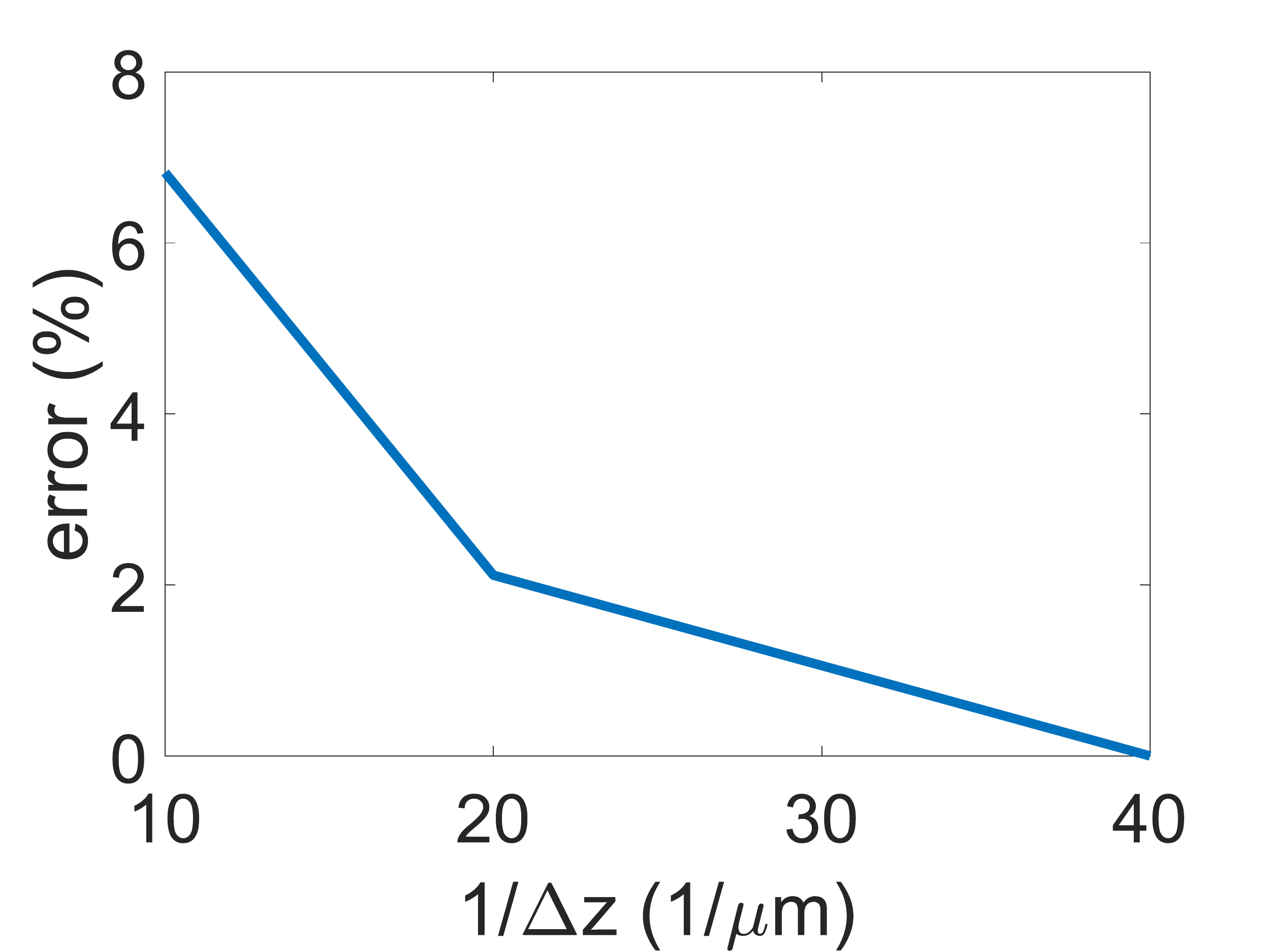} &
  	\includegraphics[draft=false,width=2.0in]{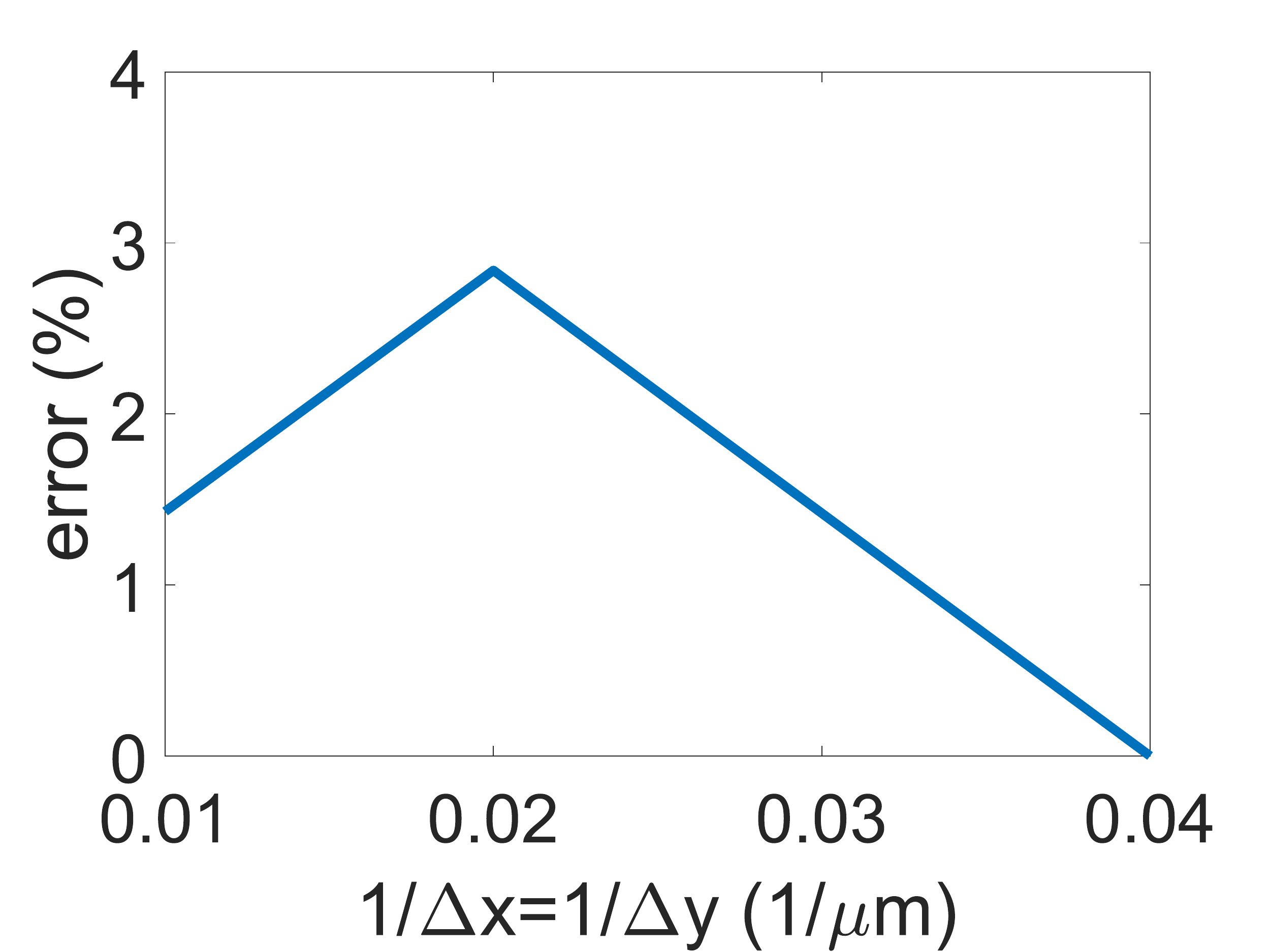} \\
  	(d) & (e) & (f)
  	\end{array}$
  	\caption{Convergence study for the different involved parameters in the considered FEL simulation: (a) $n$, (b) $\Delta t_b$, (c) $l_z = c \Delta t$, (d) $l_x=l_y$, (e) $\Delta z$ and (f) $\Delta x = \Delta y$}
  	\label{convergenceStudy}
  \end{figure}

  In Fig.\,\ref{convergenceStudy} the convergence study is shown for the aforementioned parameters. %
  Generally, accuracy of less than 3\% is achieved by using the initially suggested values.
  
\begin{figure}[t]
  	\centering
  	$\begin{array}{cc}
  	\includegraphics[width=3.0in]{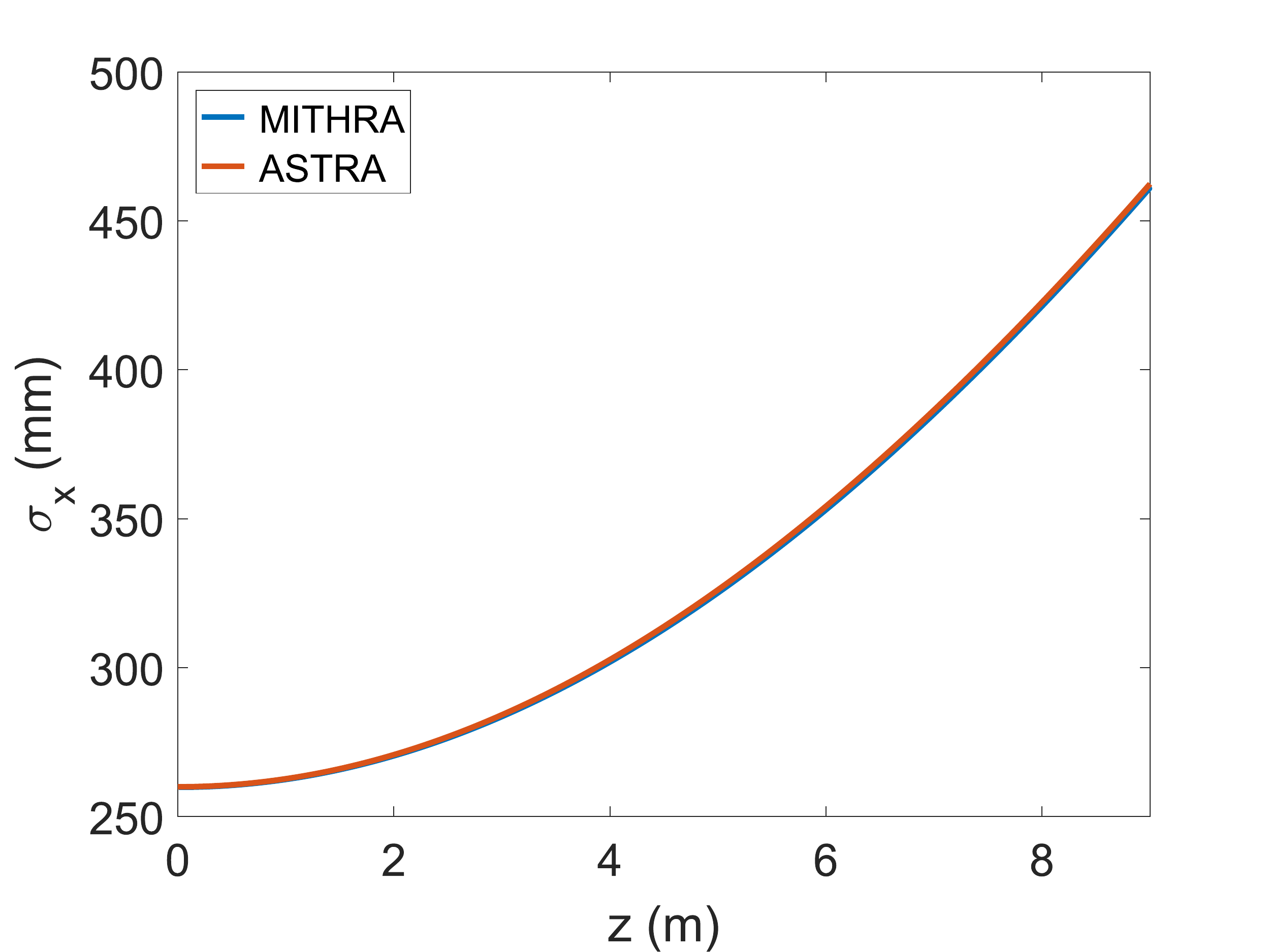} &
  	\includegraphics[width=3.0in]{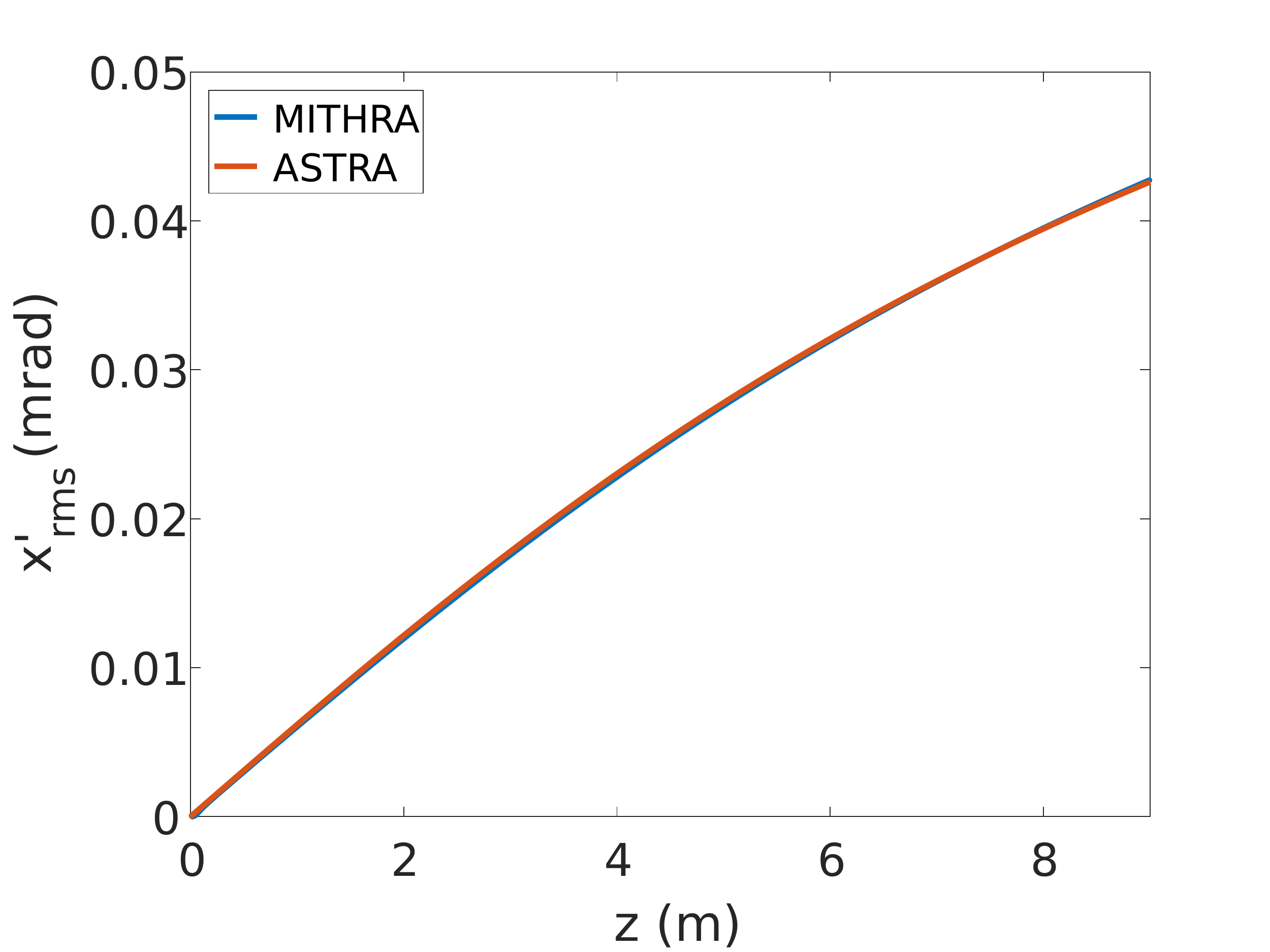} \\
  	(a) & (b)
  	\end{array}$
  	\caption{The evolution of the bunch assumed in the infra-red FEL example after a free-space propagation: (a) bunch size and (b) bunch divergence are obtained using MITHRA and ASTRA and compared against each other.}
  	\label{ASTRAvsMITHRA}
  \end{figure}

  \textbf{Space-charge effect:} A promising benefit offered by MITHRA is the assessment of various approximations used in the previously developed FEL codes.
  As an example, the algorithm used in the TDA method to evaluate the space-charge effect can be examined and verified using this code.
  The TDA method implemented in Genesis 1.3 software considers a periodic variation of space-charge force throughout the electron bunch \cite{tranFEL,reiche2000numerical}.
  This assumption is implicitly made when electric potential equation is solved in a discrete Fourier space over one slice.
  The truncation of this Fourier series is equivalent to the periodic repetition of the simulation domain.
  However, a simple investigation of bunch profiles shown in Fig.\,\ref{profile-example1}c shows that a periodic assumption for the electron distribution may be a crude approximation.
  In addition, this assumption is favored by the FEL gain process and potentially decreases any detrimental influence of the space-charge fields on the FEL radiation.
  On the other hand, the algorithm in TDA method considers longitudinal space-charge forces and neglects transverse forces, which is merely valid in high energy electron regimes.
  To make sure that such effects are modeled correctly in MITHRA, we have performed comparisons with particle transport code ASTRA \cite{flottmann2011astra} for free-space propagation problem.
  The results of this verification are shown in Fig.\,\ref{ASTRAvsMITHRA}.
  Perfect agreement was observed between the two softwares.
  
\begin{figure}[t]
  	\centering
  	$\begin{array}{cc}
  	\includegraphics[width=3.0in]{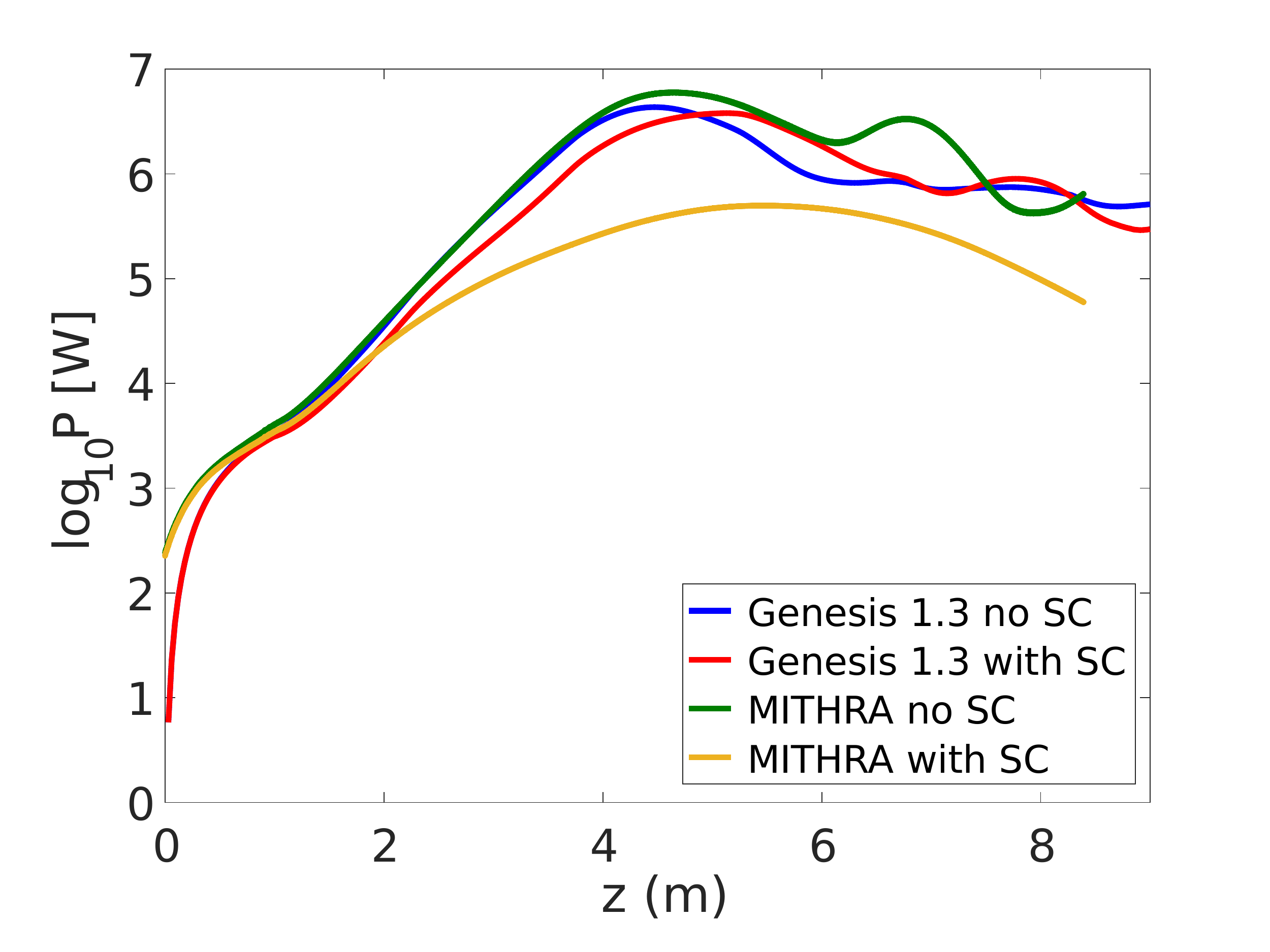} &
  	\includegraphics[width=3.0in]{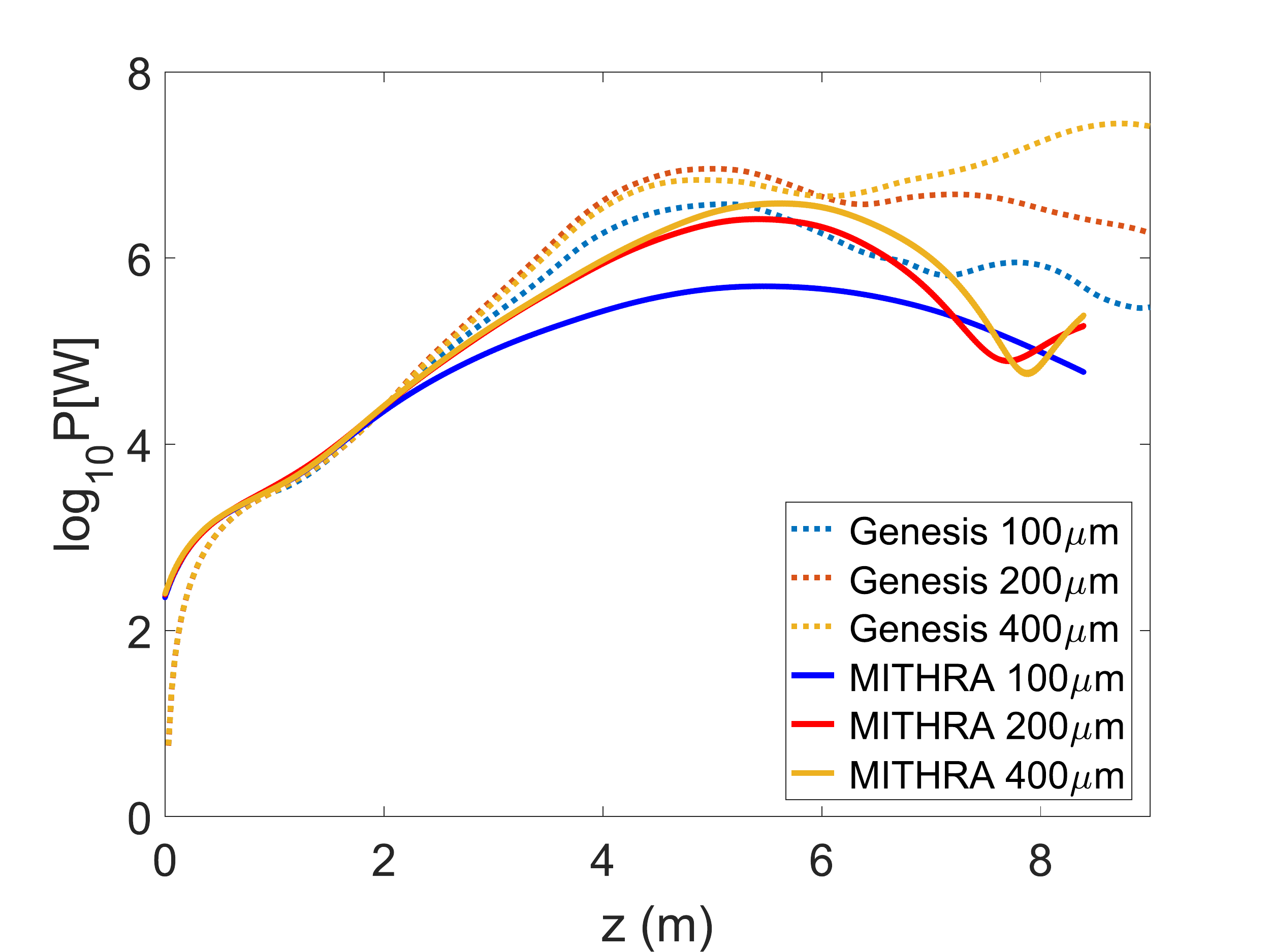} \\
  	(a) & (b)
  	\end{array}$
  	\caption{The total radiated power calculated at 110\,{\textmu}m distance from the bunch center in terms of the traveled undulator length (a) with and without space-charge consideration and (b) various lengths of the bunch with space-charge assumption.}
  	\label{spaceChargeEffect}
  \end{figure}

  In Fig.\,\ref{spaceChargeEffect}a, we are comparing the solution of the FEL problem using MITHRA and Genesis 1.3 with and without considering the space-charge effect.
  As observed in the results, the effect of space-charge on the radiation gain predicted by MITHRA is much stronger than the same effect predicted by Genesis.
  This is attributed to the assumption of periodic variations in the space-charge force made in TDA algorithm.
  If such a hypothesis is correct, the observed discrepancy should reduce once the radiation from a longer bunch is simulated, because the accuracy of periodicity assumption increases for longer bunches.
  Indeed, this is observed after repeating the simulation for longer electron bunches with similar charge and current densities.
  The results of such a study are illustrated in Fig.\,\ref{spaceChargeEffect}b.
  
  \subsubsection{Example 2: Seeded UV FEL}
  
  \textbf{Problem Definition:} As the second example, we consider a seeded FEL in the UV regime to verify the implemented features for simulating a seeded FEL.
  \begin{table}
  	\caption{Parameters of the UV seeded FEL configuration considered as the second example.}
  	\label{example2}
  	\centering
  	{\footnotesize
  		\begin{tabular}{|c||c|}
  			\hline
  			FEL parameter & Value \\ \hline \hline
  			Current profile & Uniform \\ \hline
  			Bunch size & $(95.3\times95.3\times150)$\,{\textmu}m \\ \hline
  			Bunch charge & 54.9\,pC \\ \hline
  			Bunch energy & 200\,MeV \\	\hline
  			Bunch current & 110\,A \\ \hline
  			Longitudinal momentum spread & 0.01\% \\ \hline
  			Normalized emittance & 0.97\,{\textmu}m-rad \\	\hline
  			Undulator period & 2.8\,cm \\ \hline
  			Magnetic field & 0.7\,T \\ \hline
  			Undulator parameter & 1.95 \\ \hline
  			Undulator length & 15\,m \\ \hline
  			Radiation wavelength & 0.265\,{\textmu}m \\ \hline
  			Electron density & $2.52\times10^{14} 1/\text{cm}^3$ \\ \hline
  			Gain length (1D) & 8.9\,cm \\ \hline
  			FEL parameter & 0.015 \\ \hline
  			Cooperation length & 3.34\,{\textmu}m \\ \hline
  			Initial bunching factor & $0.0$ \\ \hline
  			Seed type & Gaussian beam \\ \hline
  			Seed focal point & 70\,cm \\ \hline
  			Seed beam radius & 183.74\,{\textmu}m \\ \hline
  			Seed pulse length & infinite \\ \hline
  			Seed power & 10\,kW \\ \hline
  		\end{tabular}
  	}	
  \end{table}
  The parameters of the considered case are taken from \cite{giannessi2006overview}, which are tabulated in table \ref{example2}.
  The bunch distribution is again assumed to be uniform with a long current profile ($\sim$1000 times the radiation wavelength) in order to compare the results with the steady state simulations.
  For the same reason, the seed pulse length is considered to be infinitely long, i.e. a continuous wave pulse.
  The transverse energy spread is calculated for a bunch with normalized transverse emittance equal to 0.97\,mm\,mrad.
  Because of the very long bunch compared to the previous example, the number of required macro-particles to obtain convergent results is 8 times larger.
  Furthermore, the stronger undulator parameter dictates a smaller time step for the simulation of bunch dynamics.
  Note that MITHRA, takes the bunch step value as an initial guess, it automatically adjusts the value based on the calculated time step for mesh update.
  
\begin{figure}[t]
  	\centering
  	$\begin{array}{cc}
  	\includegraphics[width=3.0in]{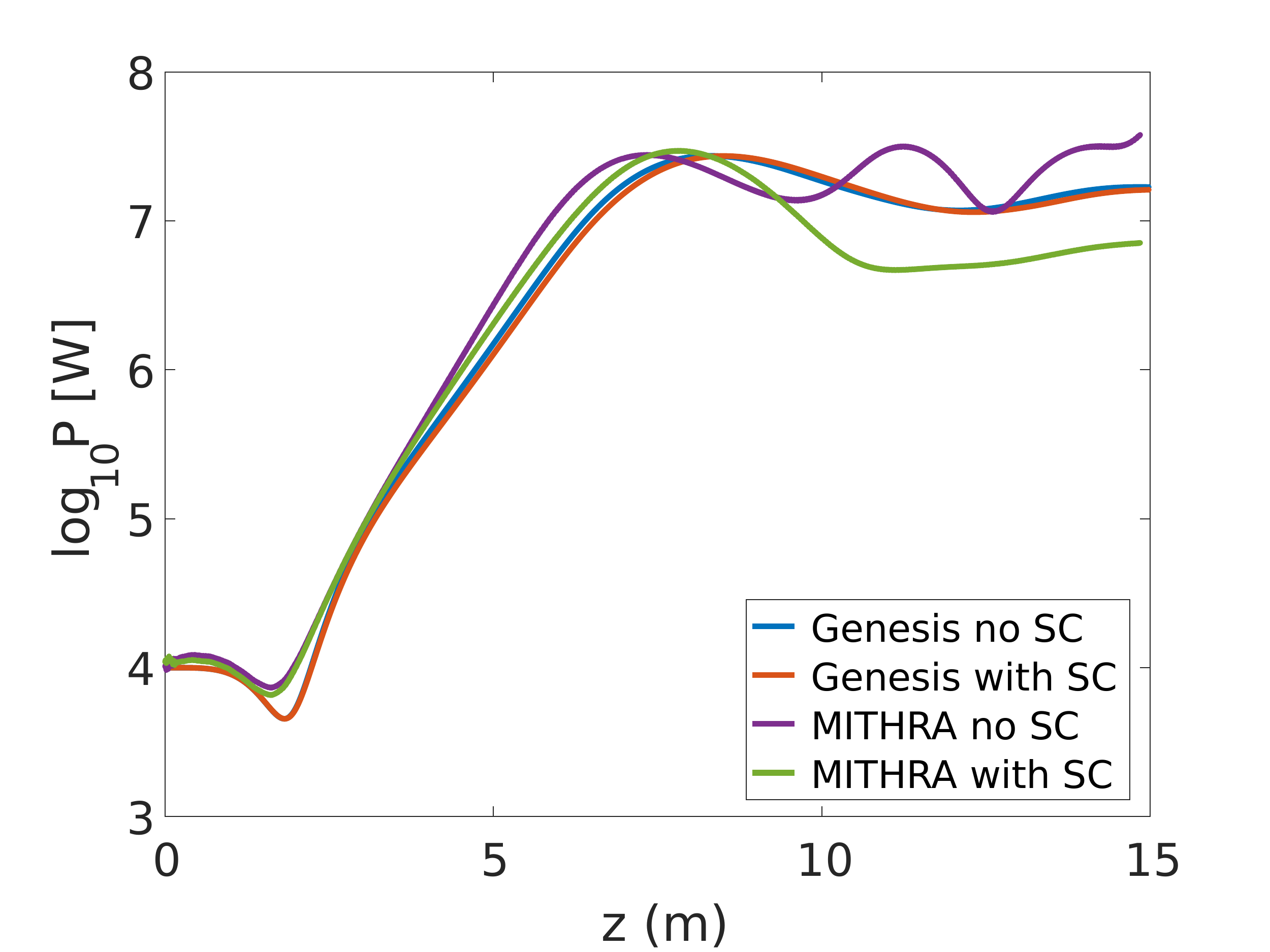} &
  	\includegraphics[width=3.0in]{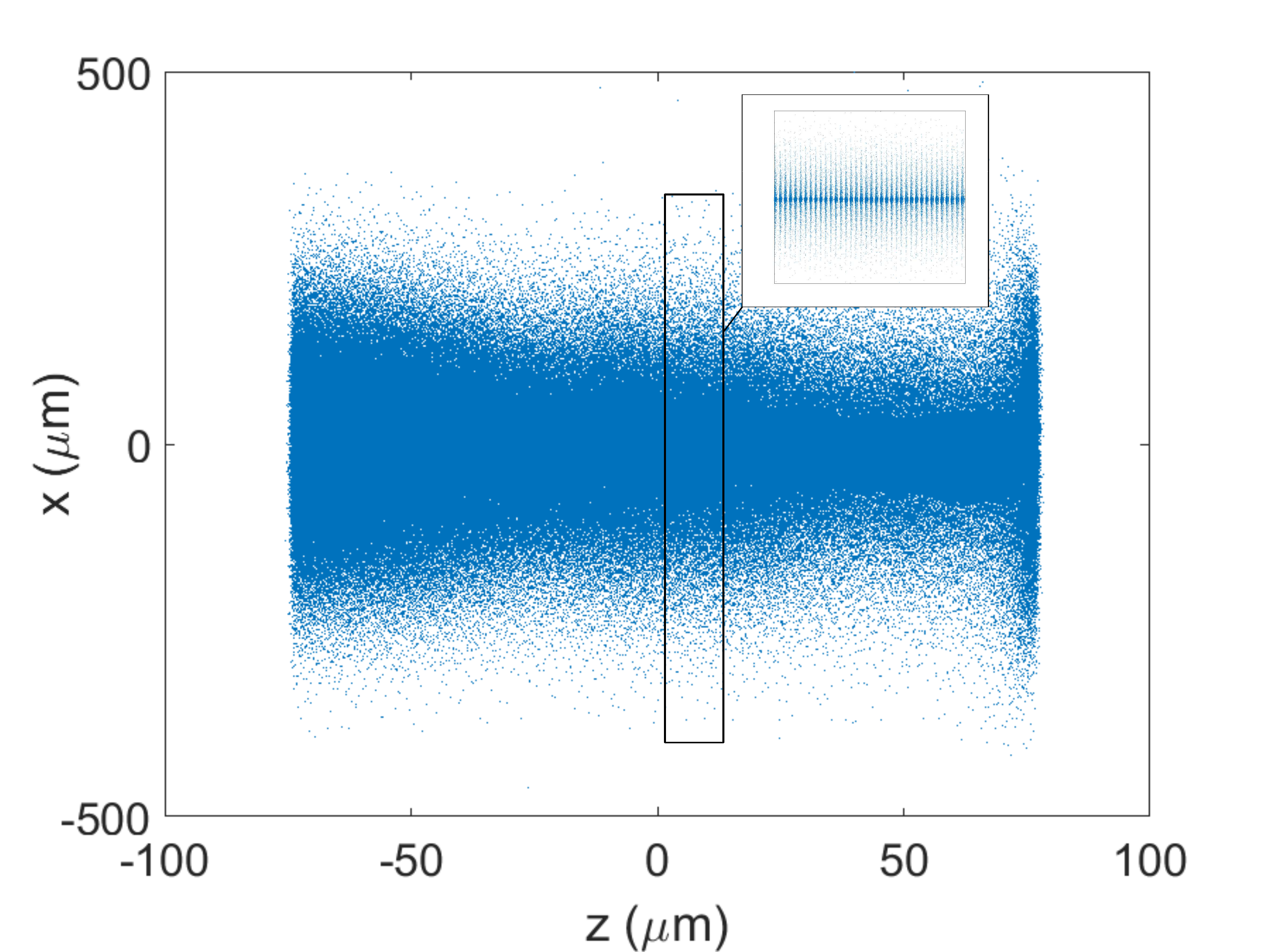} \\
  	(a) & (b)
  	\end{array}$
  	\caption{(a) The total radiated power measured at 80\,{\textmu}m distance from the bunch center in terms of the traveled undulator length and (b) the bunch profile at 12\,m from the undulator begin.}
  	\label{power-example2}
  \end{figure}

  \textbf{Simulation Results:} Fig.\,\ref{power-example2}a shows the radiated power in terms of traveled undulator distance computed using MITHRA and Genesis.
  As observed again in this example, the results agree very well in the seeded and gain regime, with notable discrepancies in the saturation regime.
  In Fig.\,\ref{power-example2}b, the bunch profile after 12\,m propagation in the undulator is also depicted.
  The micro-bunching of the large bunch is only visible once a zoom into a part of the bunch is considered.
  The investigation of the results with and without considering space-charge effect shows that in the seeded and gain intervals, space-charge plays a negligible role.
  However, in the saturation regime the effect of space-charge predicted by MITHRA is stronger than the effect predicted by Genesis.
  
  \subsubsection{Example 3: Hard X-ray FEL}
  
  \textbf{Problem Definition:} In the third example, simulation of a problem with parameter sets corresponding to the short pulse regime of the hard x-ray FEL source in the LCLS facility is pursued.
  \begin{table}[h]
  	\caption{Parameters of the hard x-ray FEL configuration considered as the third example.}
  	\label{example3}
  	\centering
  	{\footnotesize
  		\begin{tabular}{|c||c|}
  			\hline
  			FEL parameter & Value \\ \hline \hline
  			Current profile & Uniform \\ \hline
  			Bunch size & $(30.0\times30.0\times0.8)$\,{\textmu}m \\ \hline
  			Bunch charge & 20.0\,pC \\ \hline
  			Bunch energy & 6.7\,GeV \\	\hline
  			Bunch current & 7.5\,kA \\ \hline
  			Longitudinal momentum spread & 0.1\% \\ \hline
  			Normalized emittance & 0.2\,{\textmu}m-rad \\	\hline
  			Undulator period & 3.0\,cm \\ \hline
  			Undulator parameter & 3.5 \\ \hline
  			Undulator length & 40\,m \\ \hline
  			Radiation wavelength & 0.62\,nm \\ \hline
  			Gain length (1D) & 0.92\,m \\ \hline
  			FEL parameter & 0.0015 \\ \hline
  			Cooperation length & 19.3 nm \\ \hline
  			Initial bunching factor & $0.0033$ \\ \hline
  		\end{tabular}
  	}	
  \end{table}
  The parameters considered in this example are tabulated in table \ref{example3}.
  
\begin{figure}
  	\centering
  	\includegraphics[width=3.0in]{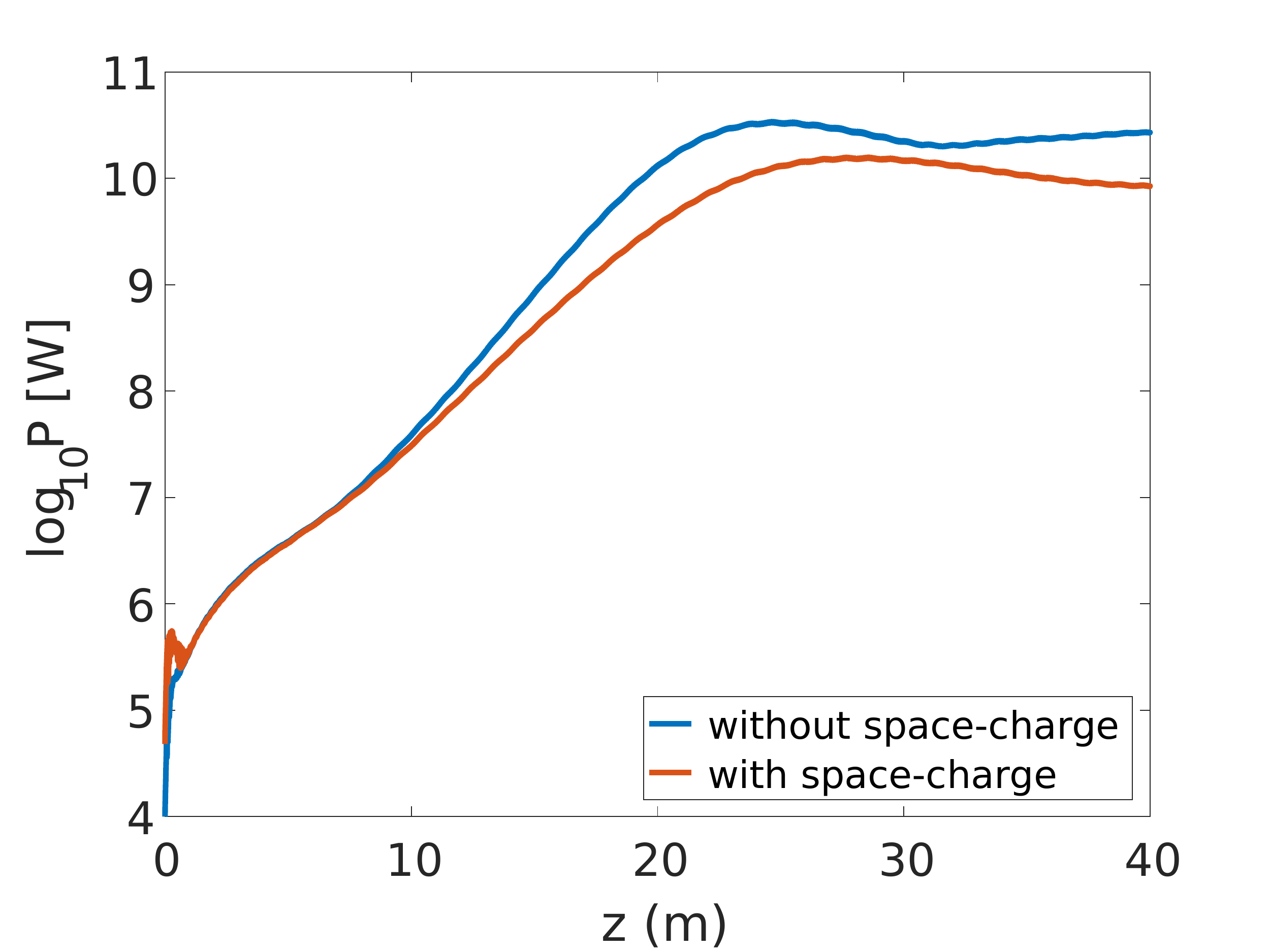} \\
  	\caption{Total radiated power measured at 30\,nm distance from the bunch center in terms of the traveled undulator length for the hard x-ray FEL source as the third example.}
  	\label{power-example3}
  \end{figure}

  \textbf{Simulation Results:} Fig.\,\ref{power-example3}a shows the computed radiated power in terms of traveled undulator distance with and without consideration of space-charge effects.
  According to the 1D FEL theory, the FEL gain length for this example is around 0.92 m, which predicts saturation after around 18 m of undulator length.
  However, due to 3D effects this saturation length is slightly longer than the predictions of 1D FEL theory.
  Here, saturation length of about 22 m is observed for a space-charge free simulation.
  In addition, the space-charge effect seems to be considerable after 10 m of undulator propagation, which contradicts with the typical assumptions that such effects are negligible for multi-GeV beams.
  This large space-charge effect, not observed in the previous examples, is occurring due to the very short bunch length, which intensifies the Coulomb repulsion forces at the head and tail of the bunch.
  A rough estimate of the Coulomb field leads to 1 V/m electric field, which in 10 meters of free propagation adds a displacement about 8 nm to the relativistic electrons.
  This value being ten times larger than the radiation wavelength confirms the strong effect of space-charge forces.
  In the beginning of the radiation, some artifacts are observed for the calculation with space-charge effect.
  These artifacts are emerging because of the unrealistic modeling of fringing fields at the entrance of the undulator, which leads to nonzero magnetic field divergence.
  To correct for such effects, numerical distribution of the fringing fields should be taken into account.
  We comment that these results are merely simulating the interaction tabulated in table \ref{example3}.
  In real FEL implementation, several focusing stations between undulator sections are realized to prevent the growth of the bunch size, thereby minimizing the space-charge effects in the FEL operation.
  
  \subsection{Example 4: Optical Undulator}
  
  Using the software MITHRA, one can perform a full-wave simulation of inverse Compton scattering (ICS) or the so-called optical undulator.
  The possibility of lasing or the so-called micro-bunching in an electron beam due to an interaction with a counter-propagating laser beam has been under debate for several years.
  A full-wave analysis of such an interaction definitely gives valuable physical insight to this process.
  Note that the classical treatment of this interaction within MITHRA does not allow for consideration of quantum mechanical effects.
  It is known that the radiation of photons results in a backward force on electrons which leads to a change in their momenta.
  In the spontaneous radiation regime, the ratio $\rho_1 = \hbar\omega/\gamma mc^2$, representing the amount of quantum recoil due to each photon emission, quantifies this effect.
  In the FEL gain regime, $\rho_2 = (\hbar\omega/2 \rho_{FEL} \gamma mc^2)^2$, with $\rho_{FEL}$ being the FEL parameter, estimates the level of quantum recoil influence on the gain process \cite{bonifacio2006quantum,bonifacio2005quantum}.
  The use of classical formulation for optical undulators is only valid if $\rho_1 \ll 1$ and $\rho_2 \ll 1$.
  
\begin{figure}[t]
  	\centering
  	\includegraphics[width=3.0in]{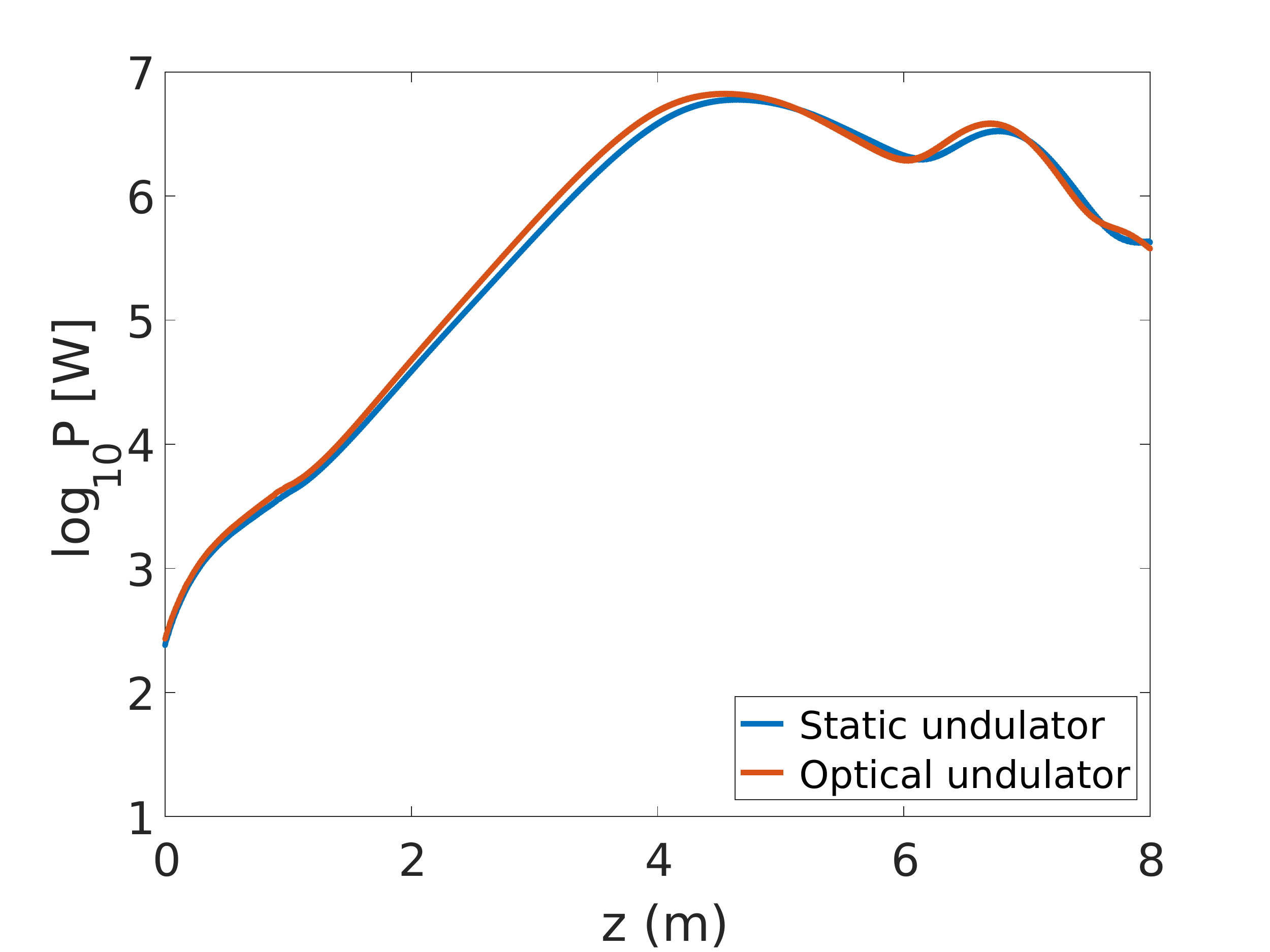}
  	\caption{The total radiated power calculated at 110\,{\textmu}m distance from the bunch center in terms of the traveled undulator length compared for two cases of an optical and static undulator.}
  	\label{ICS-benchmark}
  \end{figure}

  Before embarking on the analysis and interpretation of results for a typical ICS experiment, a benchmark to validate the analysis of optical undulators using FDTD/PIC is presented.
  It is known that electron trajectories in a static undulator with undulator parameter $K$ and periodicity $\lambda_u$ are similar to the trajectories in an electromagnetic undulator setup with normalized vector potential $a_0=K$ and wavelength $\lambda_l=2\lambda_u$ \cite{esarey1993nonlinear}.
  We take the first SASE FEL example in table \ref{example1} into account and analyze the same configuration but with an equivalent optical undulator, namely a plane wave with normalized vector potential $a_0=1.417$, and wavelength $\lambda_u = 6$\,cm.
  Fig.\,\ref{ICS-benchmark} illustrates a comparison between the radiated infrared light for the static and optical undulator cases.
  The very close agreement between the two results validates the implementation of optical undulators in MITHRA.
  
  The parameters of the optical undulator setup, considered as the fourth example, are tabulated in table \ref{example4}.
  Since we observe deviations from the predictions of one-dimensional FEL theory in our simulations, we have not listed the parameters calculated using the 1D FEL theory.
  We believe the discrepancies are originated from the small number of electrons in each 3D wave bucket, i.e. only 2 electrons.
  This intensifies the 3D effects in the bunch interaction with counter-propagating laser pulse.
  We comment that for the listed parameters $\rho_1=2\times10^{-4}$ and $\rho_2=0.003$, which are much smaller than errors caused by space-time discretization.
  \begin{table}[h]
  	\caption{Parameters of the FEL configuration with optical undulator considered as the third example.}
  	\label{example4}
  	\centering
  	{\footnotesize
  		\begin{tabular}{|c||c|}
  			\hline
  			FEL parameter & Value \\ \hline \hline
  			Current profile & Uniform \\ \hline
  			Bunch size & $(60\times60\times144)$\,nm \\ \hline
  			Bunch charge & 0.45\,fC \\ \hline
  			Bunch energy & 15\,MeV \\	\hline
  			Bunch current & 0.93\,A \\ \hline
  			Longitudinal momentum spread & 0.1\% \\ \hline
  			Normalized emittance & 1.75\,nm-rad \\	\hline
  			Laser wavelength & 1\,{\textmu}m \\ \hline
  			Laser strength parameter & 1.0 \\ \hline
  			Pulse duration & 4\,ps \\ \hline
  			Laser pulse type & flat-top \\ \hline
  			Radiation wavelength & 0.41\,nm \\ \hline
  			Electron density & $5.4\times10^{18} 1/\text{cm}^3$ \\ \hline
  			Initial bunching factor & $0.0$ \\ \hline
  		\end{tabular}
  	}	
  \end{table}
  
\begin{figure}
  	\centering
  	$\begin{array}{cc}
  	\includegraphics[width=3.0in]{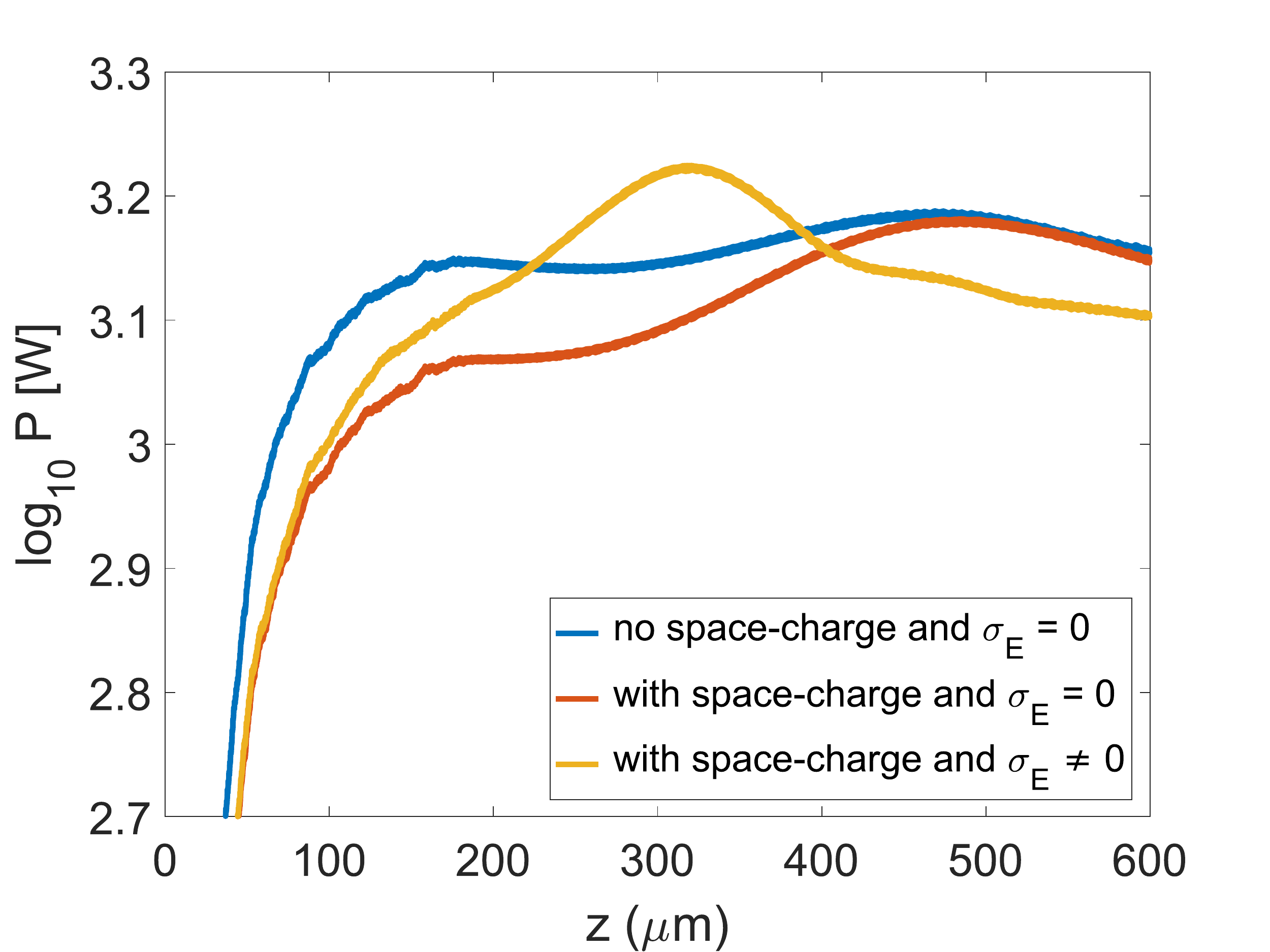} &
  	\includegraphics[width=3.0in]{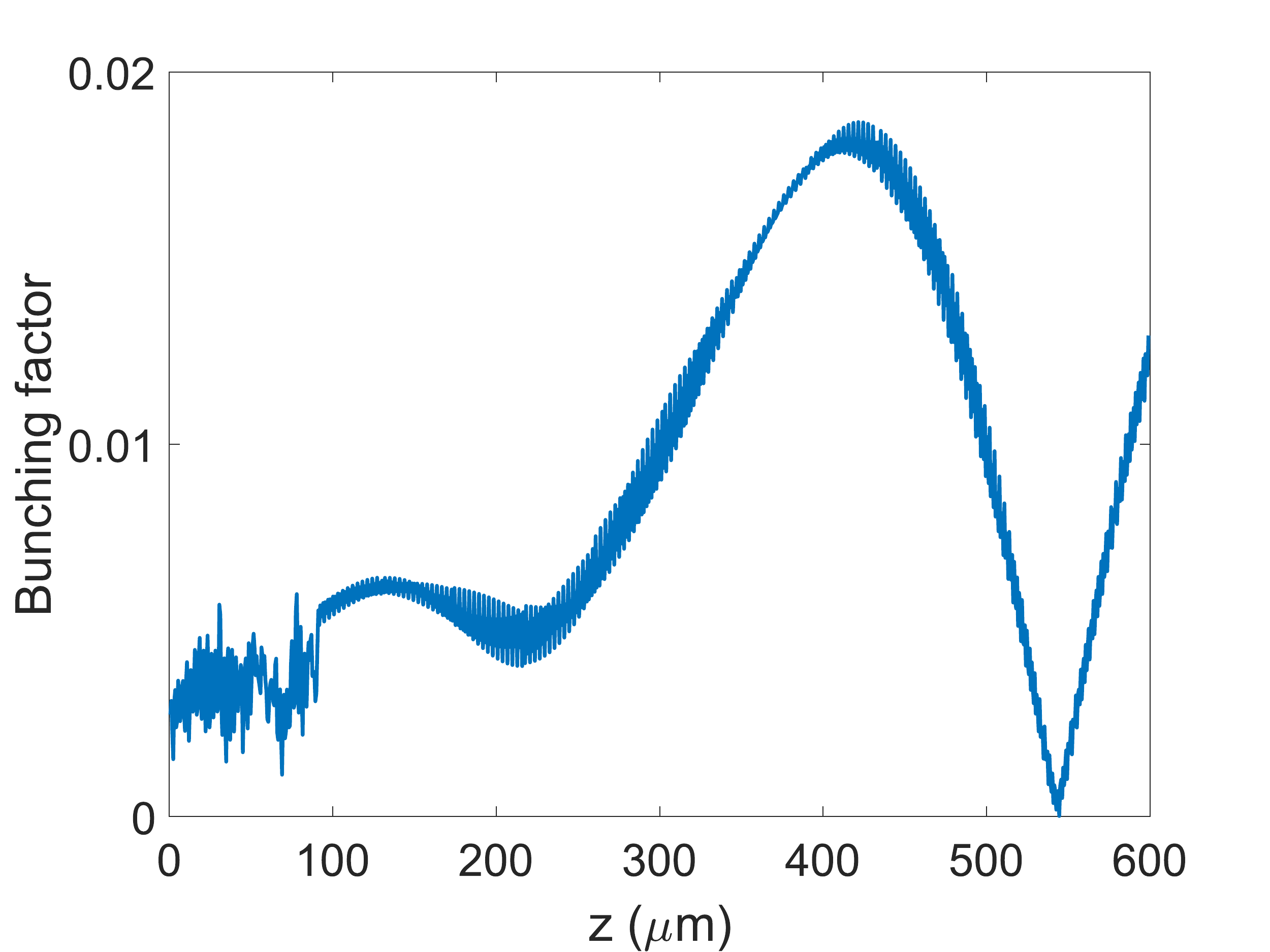} \\
  	(a) & (b)
  	\end{array}$
  	\caption{(a) The total radiated power measured at 82\,nm distance from the bunch center in terms of the traveled distance, and (b) mean bunching factor over the whole bunch during the ICS interaction.}
  	\label{power-example4}
  \end{figure}

  Fig.\,\ref{power-example4}a shows the radiated power in terms of traveled undulator distance detected 82\,nm away from the bunch center.
  The effects of space-charge and energy spread ($\sigma_E$) of the bunch are illustrated through different comparisons.
  It is observed that the gain obtained in this regime is very small compared with a usual static undulators.
  The reason for this effect is the very large shot noise in the bunch because of the low number of particles in each micro-bunch.
  Note that in this simulation, each electron is modeled as one single particle without any initial modulation in the bunch distribution.
  The strong shot noise due to 3D effects causes a strong initial radiation, which reaches the expected saturation power after a low gain.
  The case with zero energy spread corresponds to initialization of the bunch particles with exactly the same momentum.
  
  To show that the micro-bunching effect takes place in this regime as well, the bunching factor of the electron beam is depicted in Fig.\,\ref{power-example4}b.
  The bunching of the electrons due to the ICS interaction is clearly observed in the plot of bunching factor.
  This increase in the bunching factor disappears when particle radiation is suppressed in the code.
  Therefore, one can conclude that despite the low-gain, micro-bunching takes place.
  This micro-bunching process is expected to increase the longitudinal coherence of the output radiation.
  However, due to very strong initial radiation and the noise caused by 3D effects no coherent amplification is observed.
  According to the depicted power and pulse shape, total number of emitted photons is approximately equal to $4.2\times10^3$.
  
\begin{figure}
  	\centering
  	\includegraphics[width=3.0in]{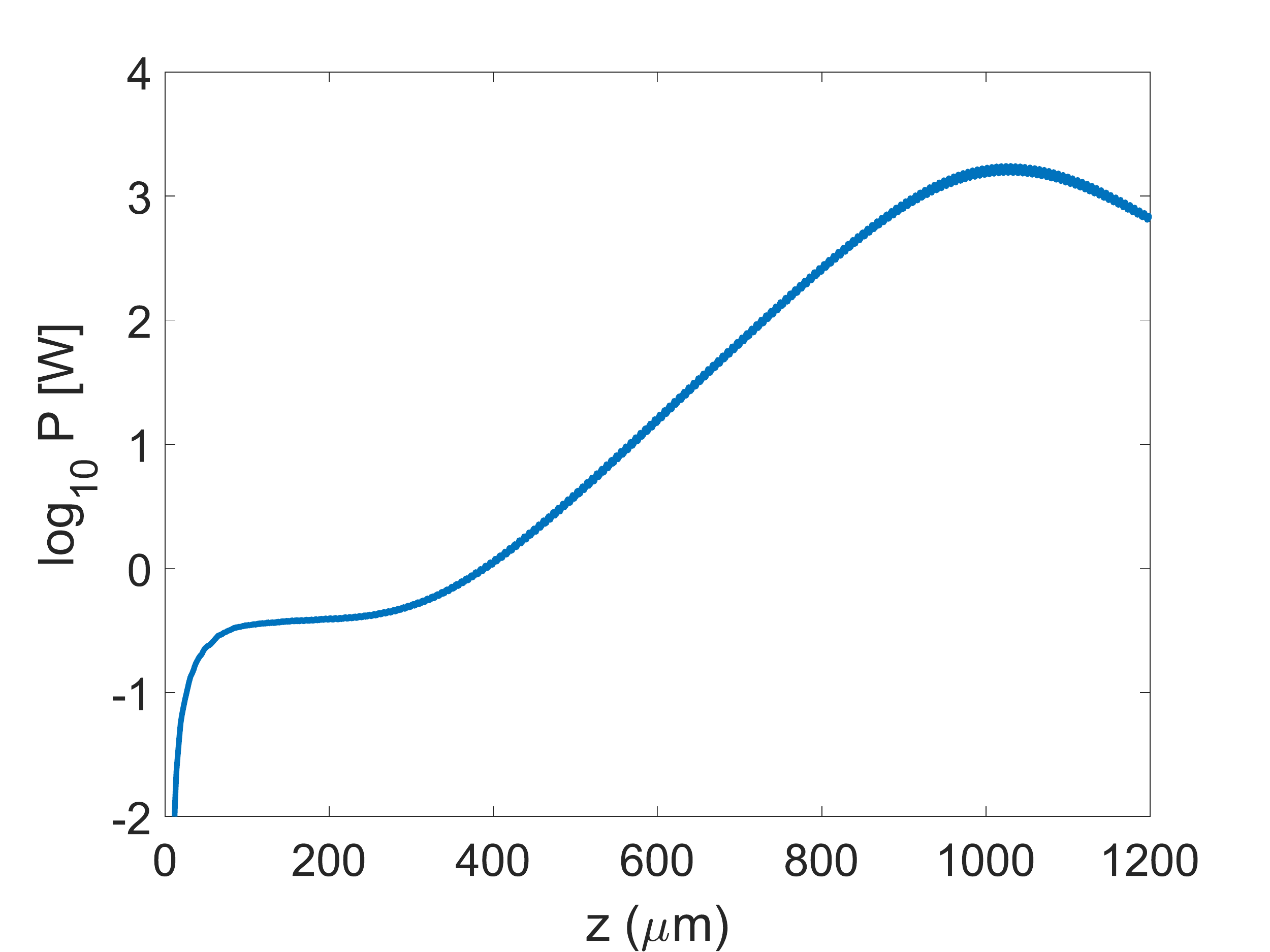} \\
  	\caption{The total radiated power measured at 82\,nm distance from the bunch center in terms of the traveled distance for an imaginary bunch where each electron is represented by a cloud of 1000 particles.}
  	\label{power-example4imaginary}
  \end{figure}

  To demonstrate the presented hypothesis related to the micro-bunching of bunches with low number of electrons per wavelength bucket, we perform an \emph{unreal} simulation, where each electron is presented by 1000 particles.
  The thousand particles are distributed evenly throughout each wavelength bucket in order to drastically reduce the shot noise level.
  In this case, each particle represents a charge 1000 times smaller than the charge of one electron.
  In addition, we assume an initial bunching factor equal to 0.001 for the input bunch to trigger the stimulated gain.
  In Fig.\,\ref{power-example4imaginary}, the radiation of such a charge configuration is depicted.
  The results reveal the radiation start from much lower powers, possibility of achieving the radiation gain and saturating in the same power level as above.
  
  This behavior and the contrast with simulation based on real number of electrons are described as follows: According to the FEL principle, the radiation gain is obtained due to periodic arrangement of radiators.
  Usually this principle is described as the initial incoherent radiation being proportional to $N$, whereas the coherent radiation of the modulated beam being proportional to $N^2$.
  Now, the question is how this $N$ value is defined.
  The world is three-dimensional and $N$ should be the number of electrons in a 3D coherence volume.
  When this number is two, the FEL principle predicts that the gain should be only twofold.
  However, 1D FEL theory does not consider particles in the 3D coherence volume and takes particles in 1D coherence length into account.
  Therefore, modeling such a problem strongly affected by 3D effects suffers from inaccurate assumptions in 1D FEL theory.
  The visualization of the fields also shows strong oscillations which disagrees with slow-varying approximation considered in typical FEL codes.
  The reason why this gain is becoming large when the number of macro-particles is increased (Fig.\,\ref{power-example4imaginary}) originates from the above effect.
  Indeed the number of radiators per coherence volume $N$ is so large that the cumulative 3D effects are vanished.
  This results in a low initial noise and clear observation of the radiation gain.
  Consequently, the presented simulation by MITHRA agrees with the basic FEL principle, according to which low number of electrons per coherence volume prevents achieving the radiation gain, even if the electron bunch is micro-bunched.
  
  \section{Conclusion}
  
  This chapter began with the introduction of a hybrid method for numerical simulation of the interaction between an electromagnetic field and a bunch of charged particles.
  The proposed procedure combines the DGTD and PIC methods for the analysis of combined field propagation and particle motion.
  A uniform grid mapping algorithm is developed to efficiently import the computed fields from the DGTD to the PIC code.
  The developed software based on the DGTD/PIC technique is verified through various benchmarks and finally applied to problem of electron field-emission from a metal cathode.
  This software offers an efficient tool for simulating laser particle interaction and can serve as a powerful simulation tool for designing particle accelerators.
  We will use this software for designing THz guns and simulating bunch emission from electron sources.
  
  The next section focused on the development of a PIC code based on analytical field formulation, which will be used for simulation of light sources based on ICS interaction.
  The simulation method ignores the effect of particle radiation on the charge distribution and computes the radiation of the bunch from the particle trajectories.
  This tool will be used for evaluating the bunch dynamics in THz linacs and simulating the ICS interaction at the final stage of the THz driven light source in chapter 6.
  
  In the last section, using the FDTD algorithm to solve time domain Maxwell's equations combined with the PIC method to solve for the particle trajectories, software MITHRA is developed for the full-wave numerical modeling of a FEL process.
  The algorithm takes advantage of the drastically reduced computational costs when Lorentz transformations are employed to solve the problem in the bunch rest frame.
  The developed software provides a proper tool for accurate analysis of free electron lasers.
  In addition, novel schemes to develop new radiation sources can be tested using this tool.
  Several benchmarks were presented to show the reliability of the results, including an infra-red FEL and a UV SASE FEL.
  The agreement between the results obtained by MITHRA and the standard software Genesis 1.3 shows the reliability of the results produced by both softwares.
  Furthermore, the software is utilized to simulate an inverse Compton scattering interaction including the effect of particle radiation on the bunch distribution.
  
  \chapter{Electron Source Technology \label{chap:three}}
  
  \section{Introduction}
  
  The focus of this chapter is the first step in any light source, which is the electron source, i.e. an electron emitter.
  An electron emitter refers to a source providing electron bunches with a precise kinetic energy, being used in various instruments ranging from television sets and computer displays to electron microscopes and particle accelerators.
  The most common electron sources, referred to as \emph{cathodes} function based on extraction of electrons from a bulk material by overcoming the surface barrier that binds them to the cathode.
  In the case of metals this barrier is called the work function.
  In terms of emission mechanism, electron emitters are categorized into: thermionic, photocathode, cold emission and plasma source.
  \emph{Thermal emitters} use temperature to overcome the surface barrier whereas \emph{photo-emitters} use photon absorption.
  Emitters that exploit quantum tunneling in the presence of high electric fields to go through the barrier are known as \emph{field emitters}.
  The term \emph{thermal field emitter} refers to electron emitters that use an electric field to decrease the barrier and make thermal emission possible at lower temperatures.
  
  Among the different mentioned types of emitters, the high brightness and improved energy resolution of a field emitter enable the creation of a source that is ultrafast and monochromatic, making this type the natural candidate for advanced ultrafast applications \cite{vecchione2009gallium}.
  Furthermore, the total intensity within a focused spot from a field emission source is known to exceed that from other conventional sources in the limit that quantum effects become important \cite{goldstein2003scanning}.
  By taking advantage from the high beam brightness with the narrow electron energy spread in field emission cathodes, high resolution electron microscopes are realized \cite{crewe1966scanning,spence2009high}.
  Other applications of field emission cathodes include x-ray tubes, RF vacuum electronic oscillators and amplifiers, and field emission displays \cite{schwoebel2006field,whaley2002experimental}.
  
  Field-emission from solid surfaces in the presence of static electric fields has been of technological interest for many years.
  Recently, there has been a growing interest in field-emission in the presence of strong, oscillating electric fields in the optical regime \cite{keathley2012strong,kruger2011attosecond,hommelhoff2006field,ropers2007localized,hommelhoff2006ultrafast}.
  This interest has been inspired by the numerous applications of bright, low-emittance and short electron bunches produced by \emph{laser-induced field-emission} and the growing availability of high-power, short duration optical sources.
  Moreover, the plasmonic effects in the light-metal interaction play a very beneficial role for focusing the optical beam in sub-wavelength dimensions and thereby achieving strongly enhanced field values.
  Electrical gating of a field emitter array (FEA) can switch the field emission beam down to few picoseconds range \cite{tsujino2011sub}.
  Even shorter electron pulses with femtosecond duration can be generated via laser-induced field emission by irradiating sharp metallic emitter tips with near infrared femtosecond laser pulses and applying a DC bias potential \cite{hommelhoff2006field}.
  Because of the small emission area, applying the single needle cathode to high current applications such as RF power amplifiers or short wavelength free-electron lasers is not possible.
  The remedy to this problem is using field emitter arrays which are often developed by advanced nanofabrication techniques.
  
  We embark on discussing the electron source technology by briefly reviewing the physics and modelling techniques of conventional flat photocathodes.
  Next, our research efforts on realizing ultrafast cathodes based on laser-induced field-emission is presented.
  In parallel to the electron source development, attempts on accurate and reliable characterization is vital towards successful implementation of new electron emitters.
  The discussion on two characterization techniques under development in our group is the focus of the last section in this chapter.
  
  \section{Flat Photocathodes}
  
\begin{figure} \centering
  	\includegraphics[draft=false,width=3.0in]{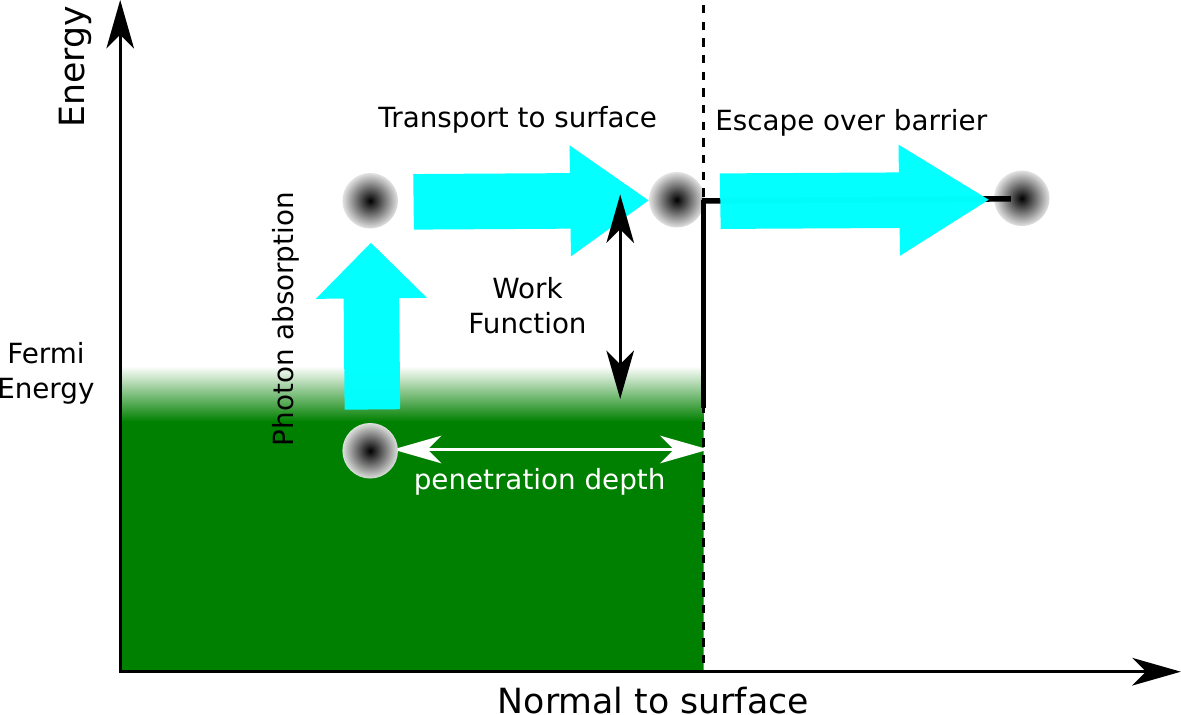}
  	\caption{Schematic illustration of the three step model in photo-electric emission.}
  	\label{photoelectricEffect}
  \end{figure}

  The conventional source of electrons in almost every particle accelerator facility consists of a flat metallic surface illuminated by an ultraviolet photon beam.
  The electrons are then emitted due to the photo-electric effect, which can be described by the three steps of the Spicer model \cite{berglund1964photoemissionT}:
  \begin{itemize}
  	\item Photon absorption by the electron
  	\item Electron transport to the surface
  	\item Escape through the barrier
  \end{itemize}
  Fig.\,\ref{photoelectricEffect} schematically illustrates the emission mechanism.
  Electrons needs to pass through a barrier equal to the work function minus Schottky energy, which accounts for a lower barrier when a large external field is applied to the material surface.
  Usually, this Schottky effect significantly increases the quantum efficiency by lowering the emission barrier.
  To obtain a good estimate for the quantum efficiency, one can assume that all the electrons absorbing photons escape the barrier.
  Then, the process of transport to the surface takes place.
  Throughout this transport, electron-electron scattering as well as electron-phonon scattering change the energy distribution of emitted electrons.
  Ultimately, the escape through the barrier occurs if the electron momentum normal to the surface is sufficiently large.
  
  The detailed and through analysis of each step yields the quantum efficiency of the emission process as well as the 6D phase-space distribution of the emitted charge.
  For details of such analyses, the reader is referred to \cite{dowell2009quantum}.
  The 6D phase-space distribution of electrons is the important and relevant outcome for our light source development goal.
  The widely accepted model for photoemission injects particles on the photoemission area over the pulse duration of the laser beam.
  The spatial profile of the emitted electrons resembles that of the laser beam.
  The momentum distribution of the particles forms a hemisphere with radius
  \begin{equation}
  p_m = \sqrt{2m_e(n\hbar\omega - \phi_{eff})/3}
  \label{hemisphereRadius}
  \end{equation}
  where $n$ is the total number of photons required for absorbing enough energy to pass over the barrier, $\hbar\omega$ is the incident photon energy, $\phi_{eff}$ is the effective work function including the possible reduction due to Schottky effect.
  In addition to the absolute mean energy value obtained from \eref{hemisphereRadius}, the energy distribution of the emitted electrons has a spread around the mean value.
  This energy spread is mainly determined by the cathode temperature and the bandwidth of the incident laser.
  Both theoretical formulation and experimental inspection of the photoemission phenomenon has shown negligible energy spreads caused by the above effects.
  The main contribution to the emittance of the emitted bunch emanates from the excess energy of electrons over the potential barrier, i.e. $(n\hbar\omega - \phi_{eff})$.
  The normalized transverse emittance of the beam can be calculated via \cite{dowell2009quantum}
  \begin{equation}
  \epsilon_{x,y}=\sigma_{x,y} \sqrt{\frac{n\hbar\omega - \phi_{eff}}{3 m_e c^2}}
  \label{photoCathodeEmittance}
  \end{equation}
  The detailed photoemission model is implemented in the ASTRA bunch generator and will be used in our source development studies \cite{flottmann2011astra}.
  
  \section{Field Emitter Arrays}
  
  \subsection{Silicon nanotips}
  
\begin{figure} \centering
  	$\begin{array}{cc}
  	\includegraphics[draft=false,height=2.0in]{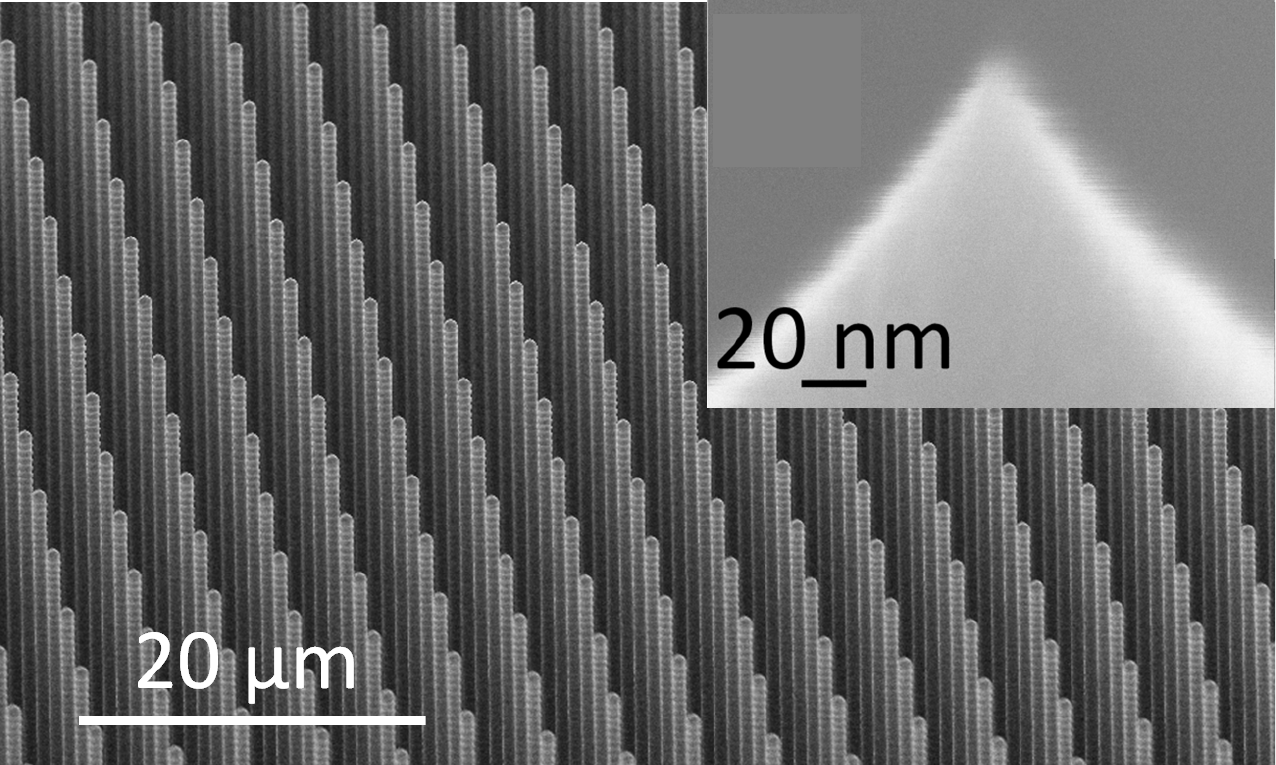} &
  	\includegraphics[draft=false,height=2.0in]{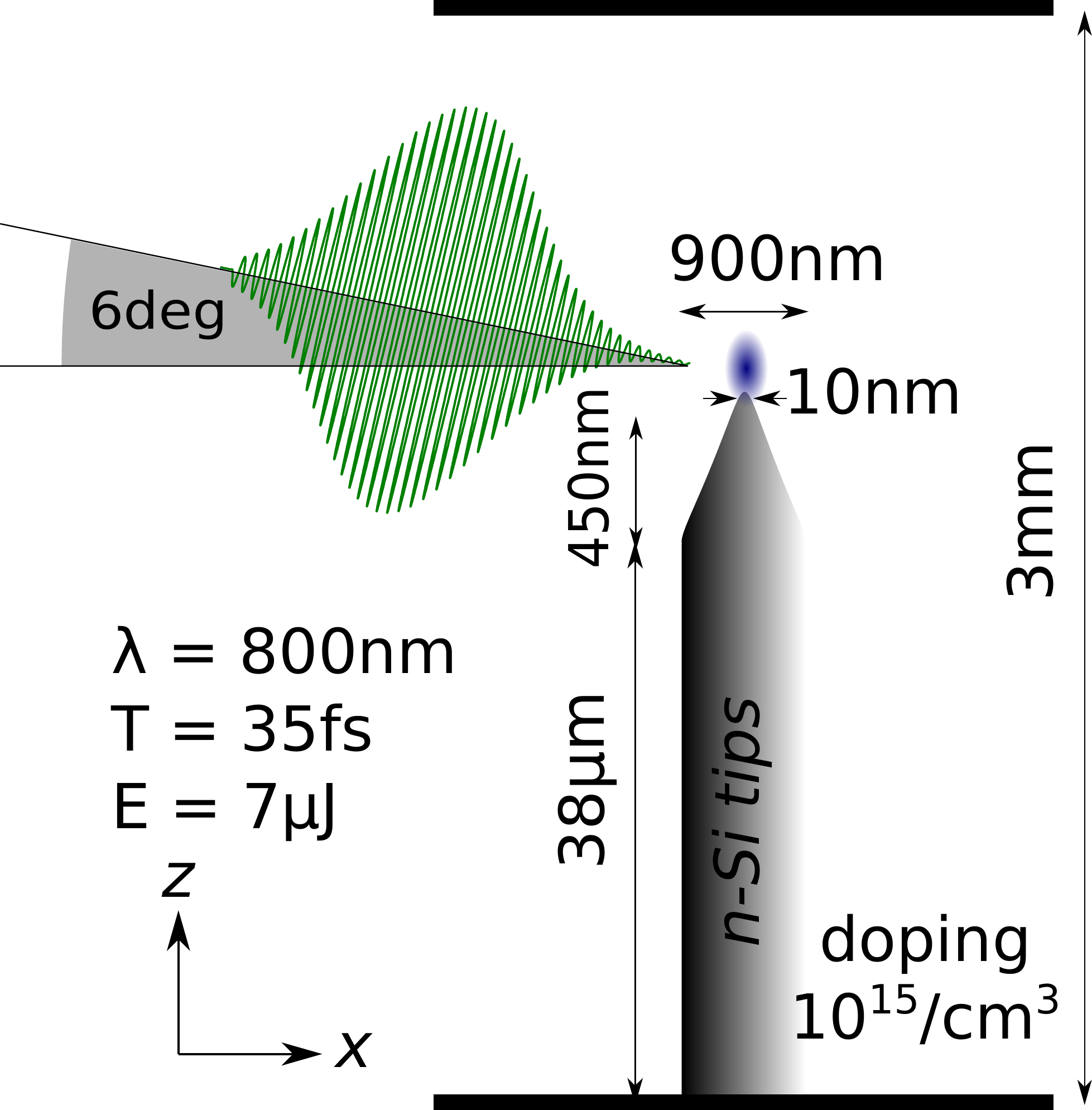} \\
  	(a) & (b)
  	\end{array}$
  	\caption{Images and schematic of emitter structure: (a) Scanning electron microscope (SEM) image of uniform array of high-aspect-ratio Si columns with 5\,{\textmu}m pitch, with SEM close-up of a single tip as a subfigure, and (c) Schematic of a single 800\,nm pulse interacting with a single silicon tip. $\lambda$ is the laser wavelength, $T$ is the pulse duration, and $E$ is laser energy.}
  	\label{siliconNanoTip}
  \end{figure}

  State-of-the-art ultrafast cathodes are flat surfaces that use highly reactive materials to lower the work function and increase the quantum efficiency of single-photon absorption for ultraviolet (UV) pulses; these devices have short lifetimes and need to be fabricated and operated in ultrahigh vacuum \cite{guo2013band}.
  Multiphoton and strong-field emission cathodes are an attractive alternative to circumvent these issues.
  Strong-field electron tunneling from solids without damage \cite{hommelhoff2006field,bormann2010tip,schertz2012field,dombi2013ultrafast,piglosiewicz2014carrier,yanagisawa2011energy,keathley2012strong} occurs when the electric field of high intensity
  optical pulses interacts with field enhancing structures to bend down the potential barrier at the surface such that the electron's tunneling time is shorter than one optical cycle \cite{keldysh1965ionization}, with the potential for attosecond electron pulse generation \cite{kruger2011attosecond}.
  Much of the previous work on nanostructured multiphoton and strong-field emission cathodes has focused on single metal tips that are serially manufactured \cite{herink2012field}.
  Here, wafer-level semiconductor batch fabrication techniques are used to create massively multiplexed arrays of nanosharp high-aspect-ratio single-crystal silicon pillars with high uniformity ($>$100'000 tips, 4.6 million tips/cm$^2$, 4.4\,nm average radius of curvature with a standard deviation of 0.6\,nm), resulting in greatly enhanced array electron emission (Fig.\,\ref{siliconNanoTip}) \cite{swanwick2014nanostructured}.
  This also enables the generation of attosecond electron pulses at the tip surface with considerable charge, i.e., hundreds of fC from a single cycle near-IR drive source when used along with a THz source for charge extraction \cite{fallahi2016short}.
  A high-aspect-ratio silicon column topped by a nanosharp tip achieves electron emission at low power by greatly enhancing the incident electric field.
  The massive multiplexing of the pillars with low tip radii spread drastically increases the total current emission and also structures the emission as a series of planar arrays of electron bunches.
  In a field emitter array, a broad tip radii distribution causes severe array sub-utilization because the emission current has an exponential dependence on the local surface electric field at the tips, and hence field factor of the emitters, which is inversely proportional to the tip radii.
  For ultrafast electron source applications achieving a homogeneous charge distribution in the bunch is strongly desired, thereby making small variations in the tip dimensions essential.
  Here, a fabrication process is developed that attains small tip variation across the array as a result of the diffusion-limited oxidation step that sharpens the tips.
  We identified the current-voltage-optical excitation parameter range where strong-field emission occurs and observed that charge effects are negligible \cite{swanwick2014nanostructured}.
  
  Beyond increasing the spectral efficiency of a planar silicon cathode, the multiphoton process of emission and ultimately strong-field emission using near-IR pulses lead to a natural localization of electron emission.
  For instance, assuming a three-photon absorption process is necessary to liberate an electron, with a factor of 10 field enhancement occurring only near the end of the tip, electron emission there increases by a factor of 1 million with respect to regions with no enhancement.
  This leads directly to a nanoscale confinement of electron emission \cite{ropers2007localized}, circumventing the need of extra complications in fabrication, such as the use of a mask layer to create structured electron beams.
  Furthermore, by pushing the local electric field intensities at the tip surface high enough using near-IR pulses, the tunneling, or strong-field regime of emission is achieved (Fig.\,\ref{siliconNanoTip}).
  This opens up applications to attosecond science, as the physics describing this emission implies that the electrons are being emitted over a narrow subcycle region of the driving pulse's electric field.
  The tips could thus be used as a near-field (i.e., near the tip) attosecond probe with increased signal yield due to multiplexing \cite{baum2007attosecond,barwick2009photon}.
  
\begin{figure} \centering
  	$\begin{array}{cc}
  	\includegraphics[draft=false,width=3.0in]{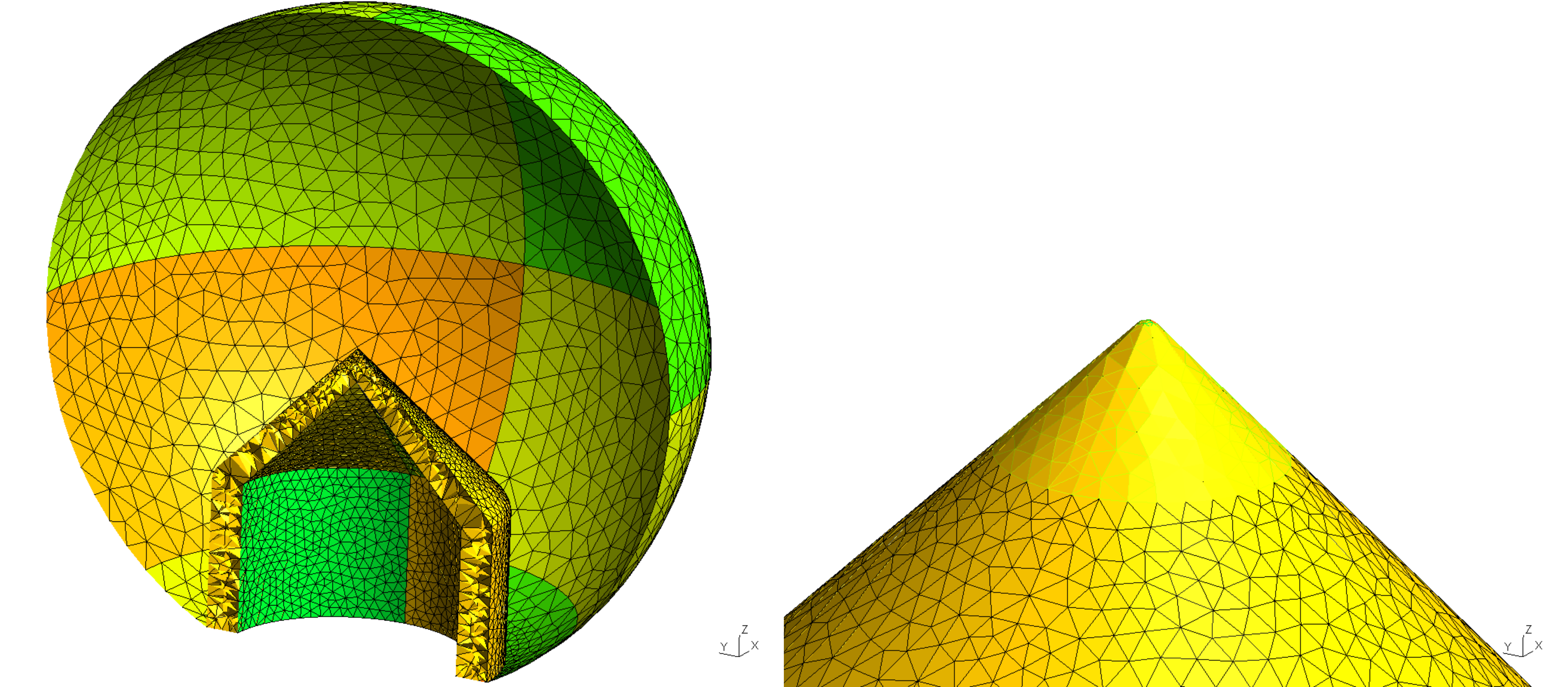} &
  	\includegraphics[draft=false,width=3.0in]{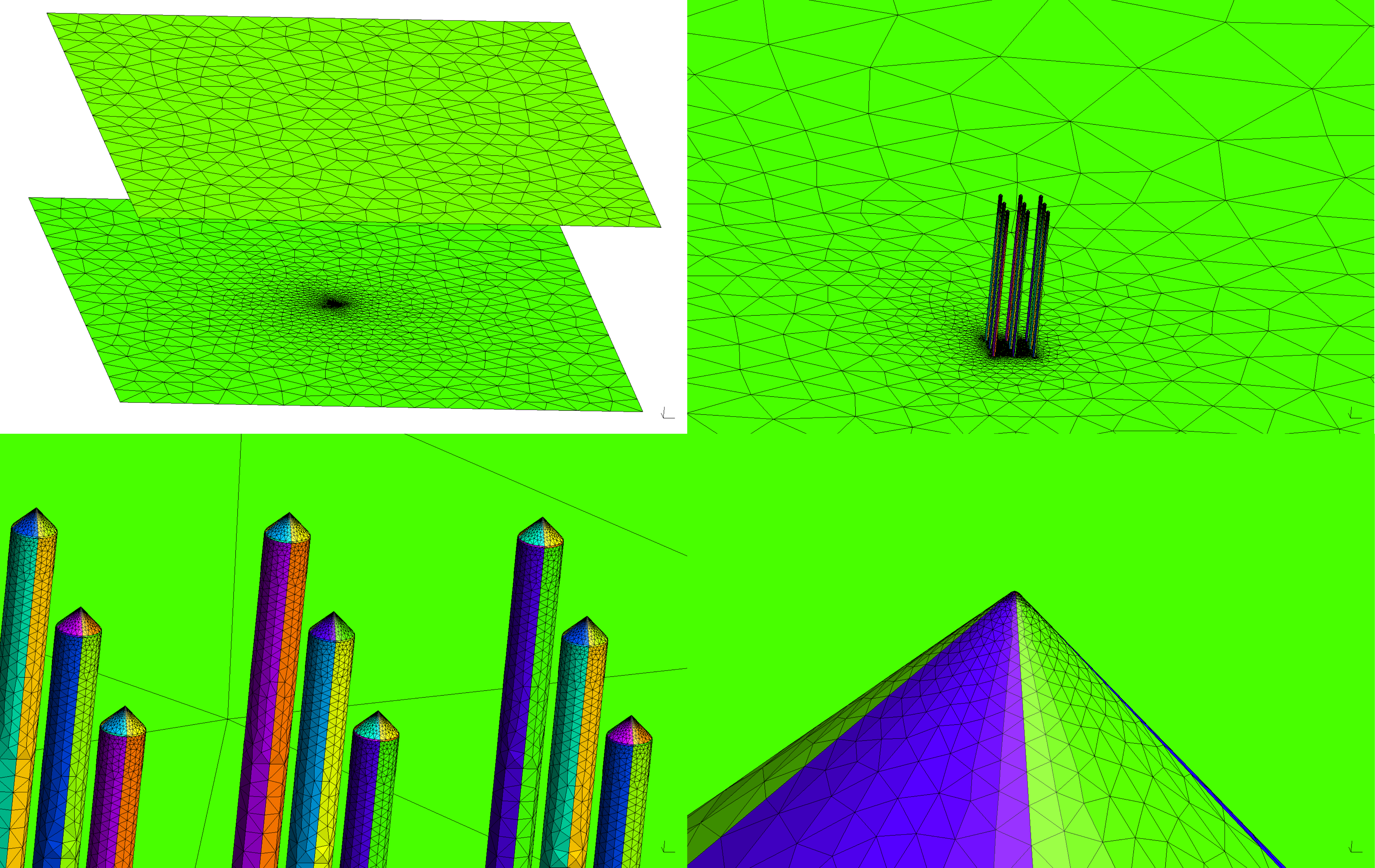} \\
  	(a) & (b)
  	\end{array}$
  	\caption{Cross-sections of the tetrahedral mesh used for solving the (a) Maxwell and (b) Poisson equation on the nanorods.}
  	\label{nanotipSimulationMesh}
  \end{figure}

\begin{figure} \centering
  	$\begin{array}{ccc}
  	\includegraphics[draft=false,height=1.5in]{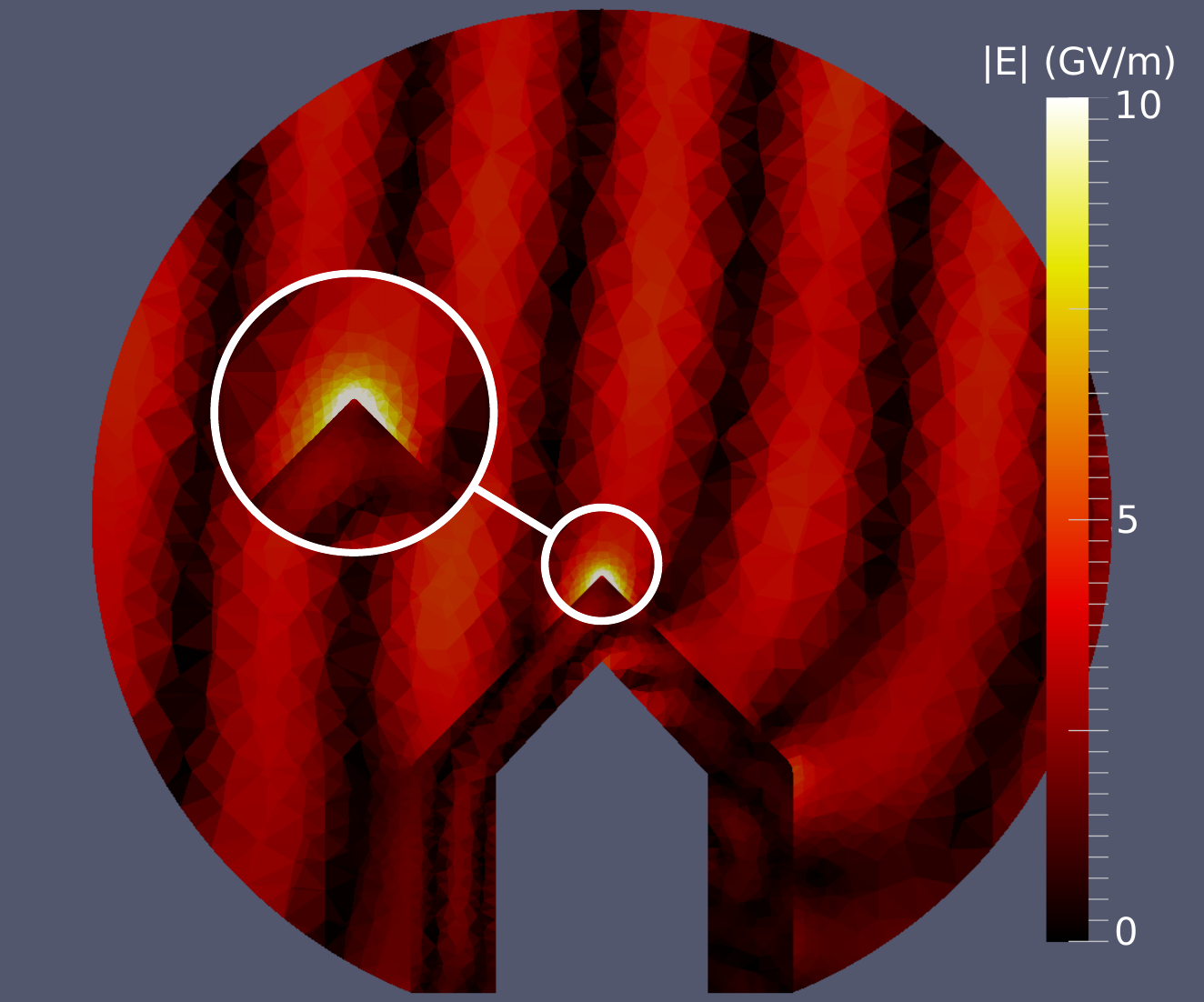} &
  	\includegraphics[draft=false,height=1.5in]{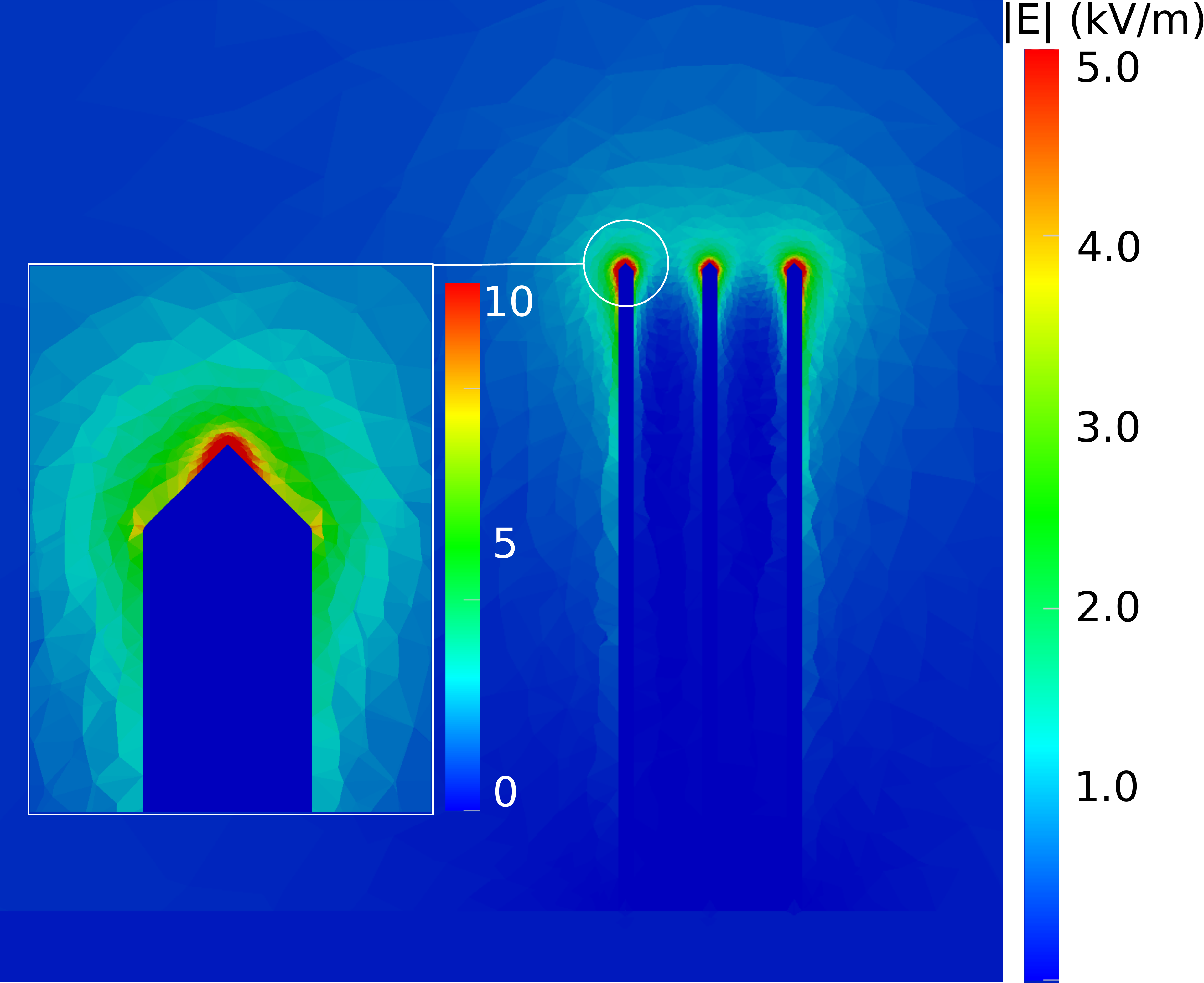} &
  	\includegraphics[draft=false,height=1.5in]{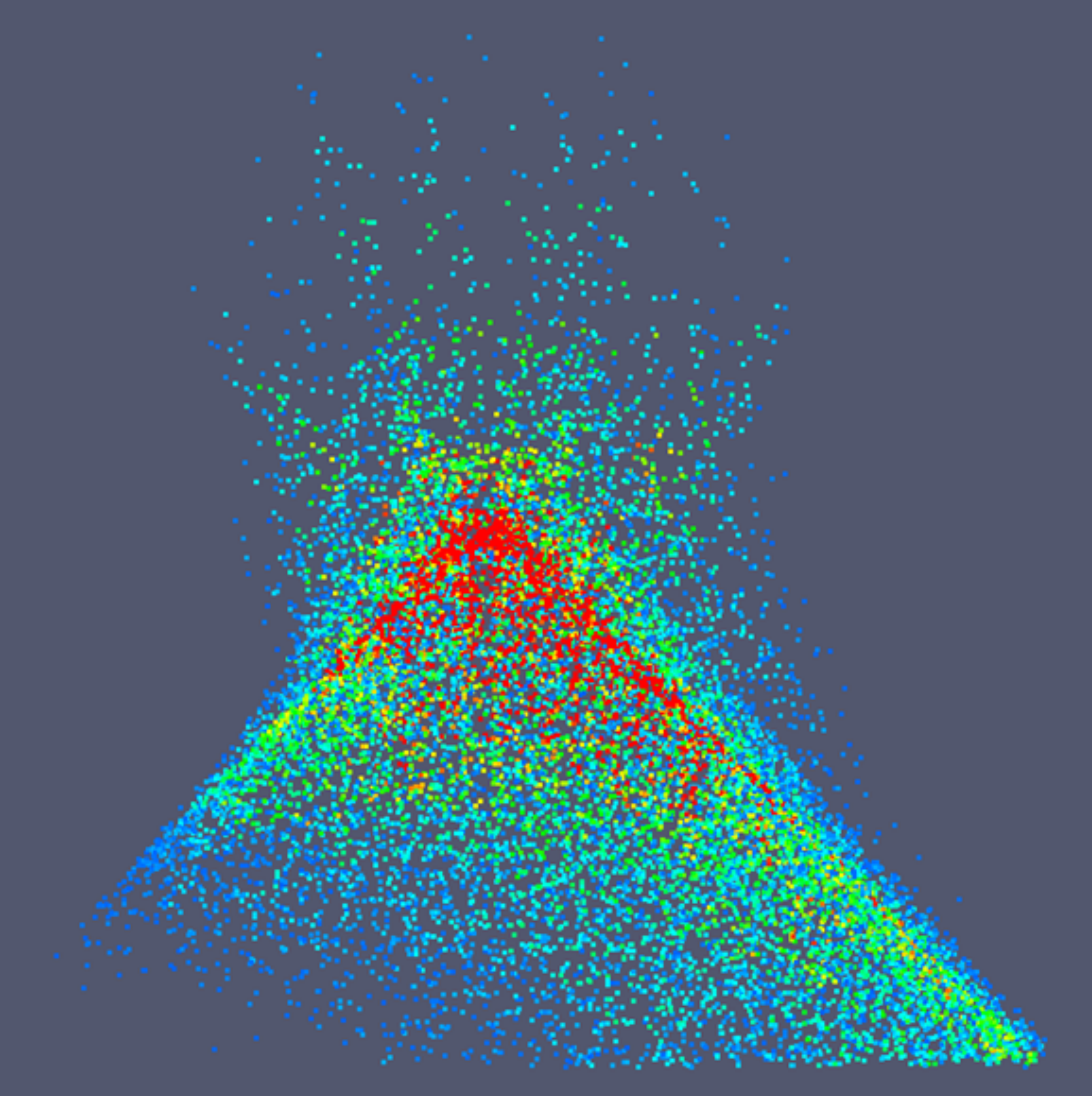} \\
  	(a) & (b) & (c)
  	\end{array}$
  	\caption{(a) The computed field profile using DGTD algorithm. (b) The computed static field profile using FEM Poisson solver. (c) snap-shot of the emitted charge cloud from the tip.}
  	\label{nanotipSimulationProfile}
  \end{figure}

\begin{figure} \centering
  	$\begin{array}{cc}
  	\includegraphics[draft=false,width=3.0in]{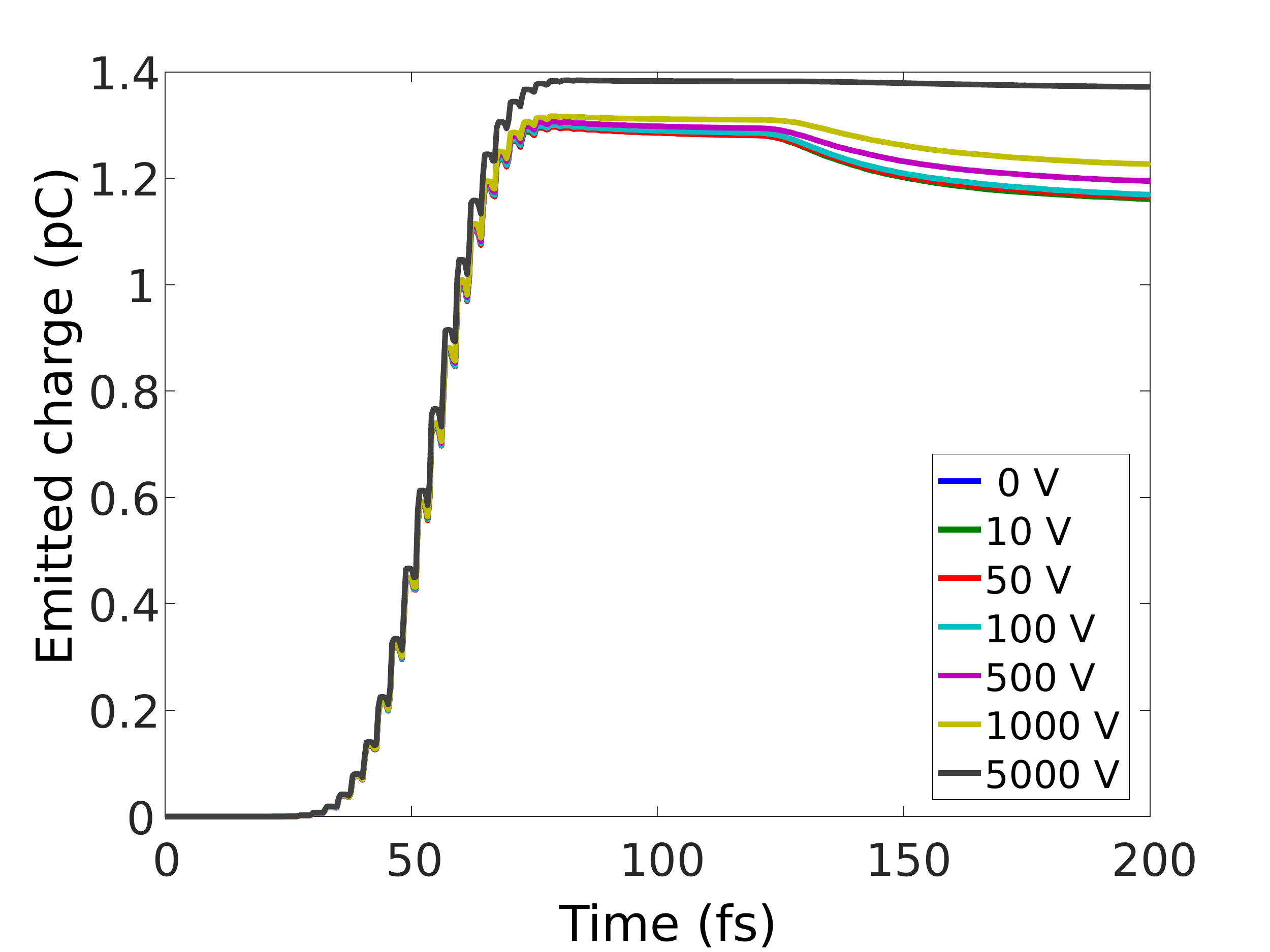} &
  	\includegraphics[draft=false,width=3.0in]{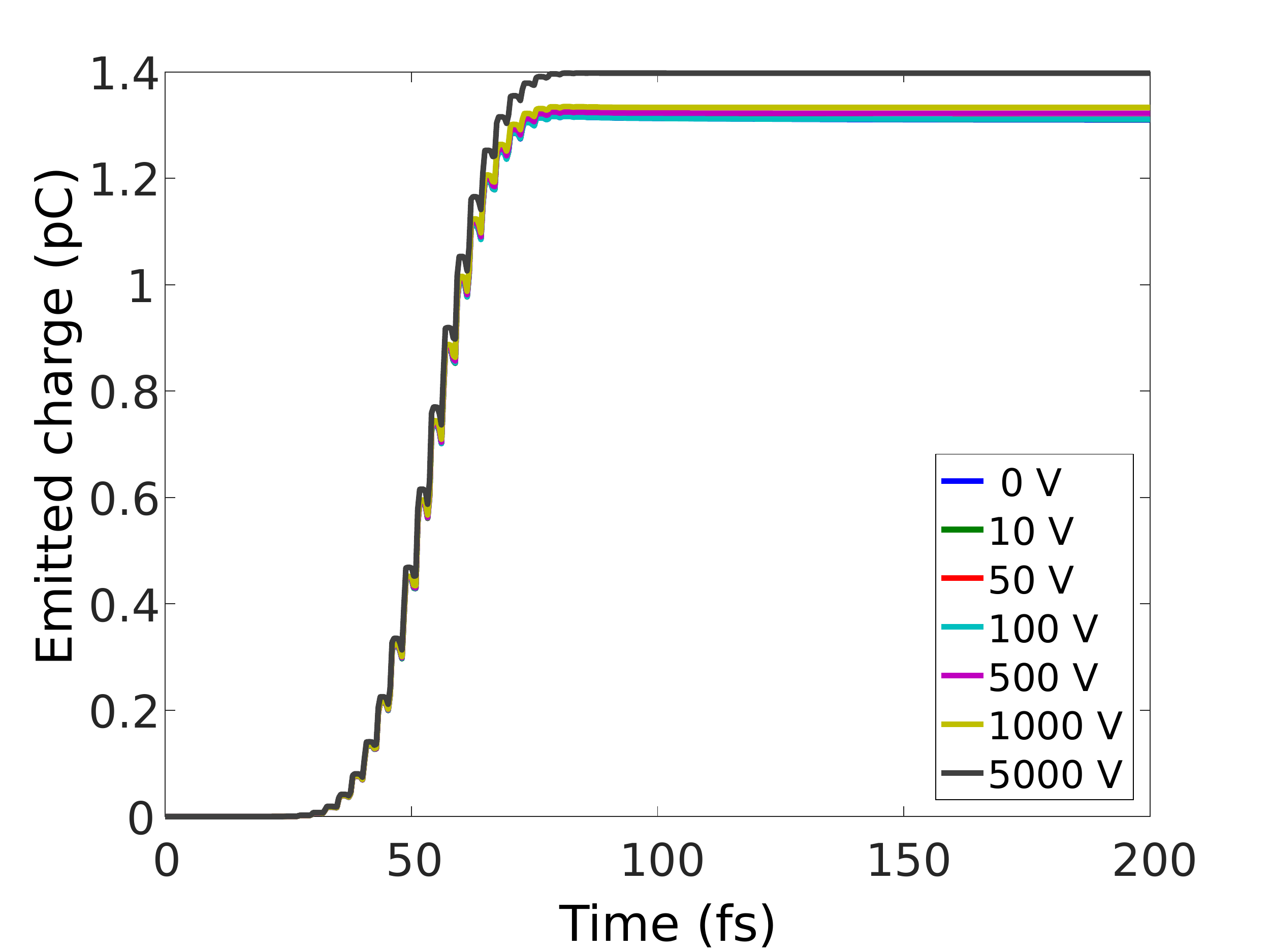} \\
  	(a) & (b)
  	\end{array}$
  	\caption{Emitted current from a single tip multiplied by 2200 versus time for different anode voltages: (a) without and (b) with considering space-charge and image charge effects. If the electron returns to the surface, it is assumed to have a recombination rate of 0.7. The applied DC field reduces the collision rate of the emitted electrons to the surface, thereby increasing the total amount of emitted charge. This is more prevalent when image charge effects are considered}
  	\label{nanotipSimulationResult}
  \end{figure}

  Expanding on the experimental results of single tip emission \cite{bormann2010tip,piglosiewicz2014carrier,yanagisawa2011energy,kruger2011attosecond,herink2012field} we begin by modeling the emission from a single tip in the time domain using 35\,fs 800\,nm pulses at 6 degrees grazing incidence (Fig.\,\ref{siliconNanoTip}b, Fig.\,\ref{nanotipSimulationMesh}, Fig.\,\ref{nanotipSimulationProfile} and Fig.\,\ref{nanotipSimulationResult}).
  The model computes the field enhancement of the laser and static anode bias voltage, while also performing particle tracing.
  The simulations are based on the Fowler-Nordheim (FN) \cite{nordheim1928electron} model of electron emission, applicable in the tunneling regime \cite{yalunin2011strong}, accounting for space-charge and Coulomb-blockade effects by adding the fields of a moving charge to the time domain Maxwell solver \cite{paarmann2012coherent,hoffrogge2014tip}.
  The simulation challenges are the range of length and time scales involved, the electron dynamics in the presence of static as well as ultrahigh frequency fields, the electron-electron interaction (i.e., space-charge effects), and the Coulomb blockade of the electron bunch induced on the surface.
  Fig.\,\ref{nanotipSimulationResult}a and b show the modeled current from a single tip (multiplied by 2200 to match the number of tips illuminated in the experiment) with and without space-charge, respectively.
  The effect of space charge is small, reducing the total emitted electrons by 14\% for a 10\,V bias compared to a bias of 5\,kV across a 3\,mm anode to cathode gap.
  Particle tracing shows that this is due to a rapid spread of electrons leaving the tip, thus reducing space-charge effects that would otherwise drive the electrons back to the cathode.
  As seen from the obtained emitted charge simulations, the space-charge consideration leads to a small recombination rate for the electrons (i.e. electrons reabsorbed by the cathode), which vanishes with an increase in anode voltage.
  This confirms the intuition used by Bormann et al. \cite{bormann2010tip} to describe the high current yields observed from a single Au tip and is within the range of fields experimentally tested, far below the damage threshold.
  
\begin{figure}[h] \centering
  	$\begin{array}{cc}
  	\includegraphics[draft=false,width=2.7in]{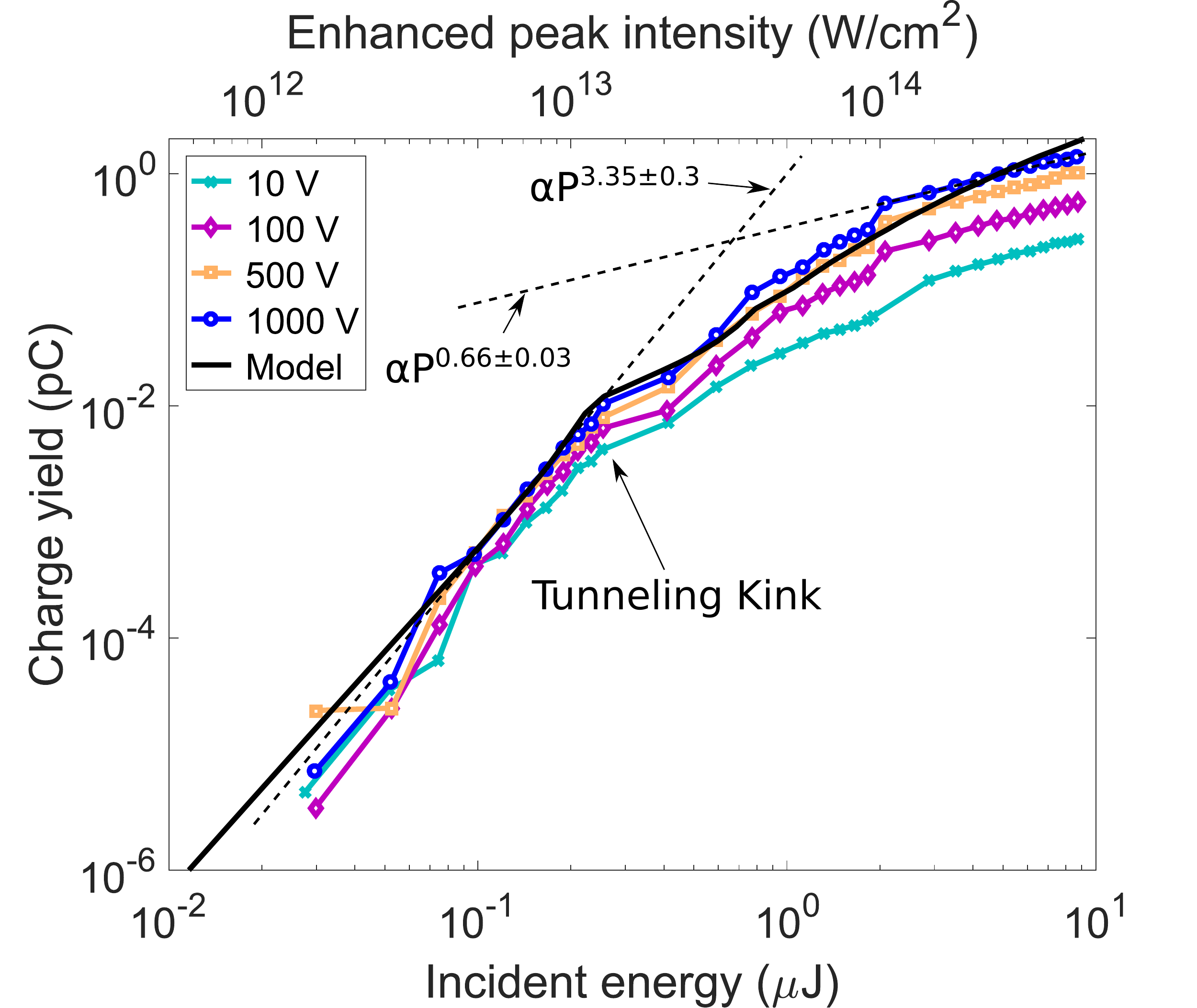} &
  	\includegraphics[draft=false,width=2.7in]{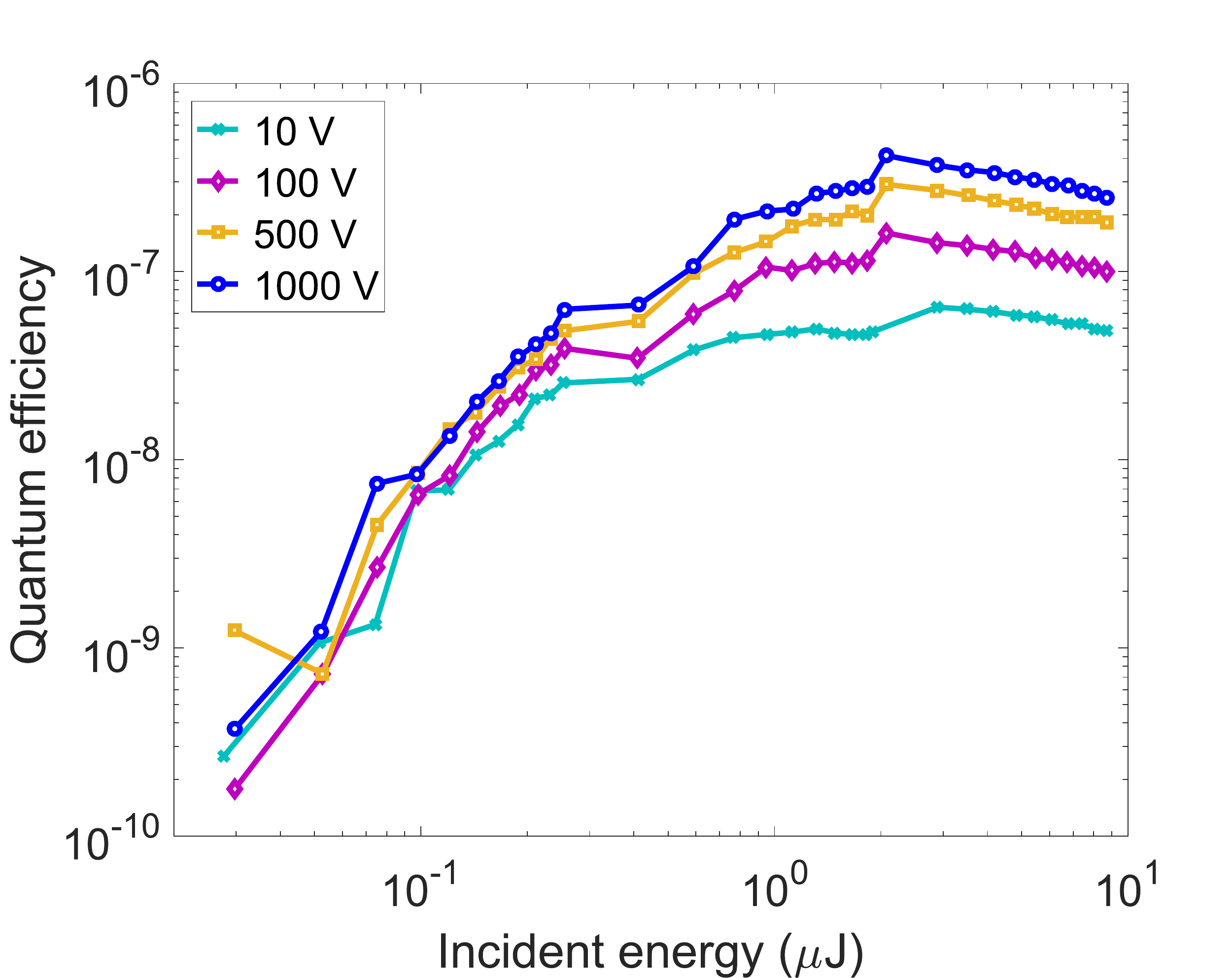} \\
  	(a) & (b) \\
  	\includegraphics[draft=false,width=2.7in]{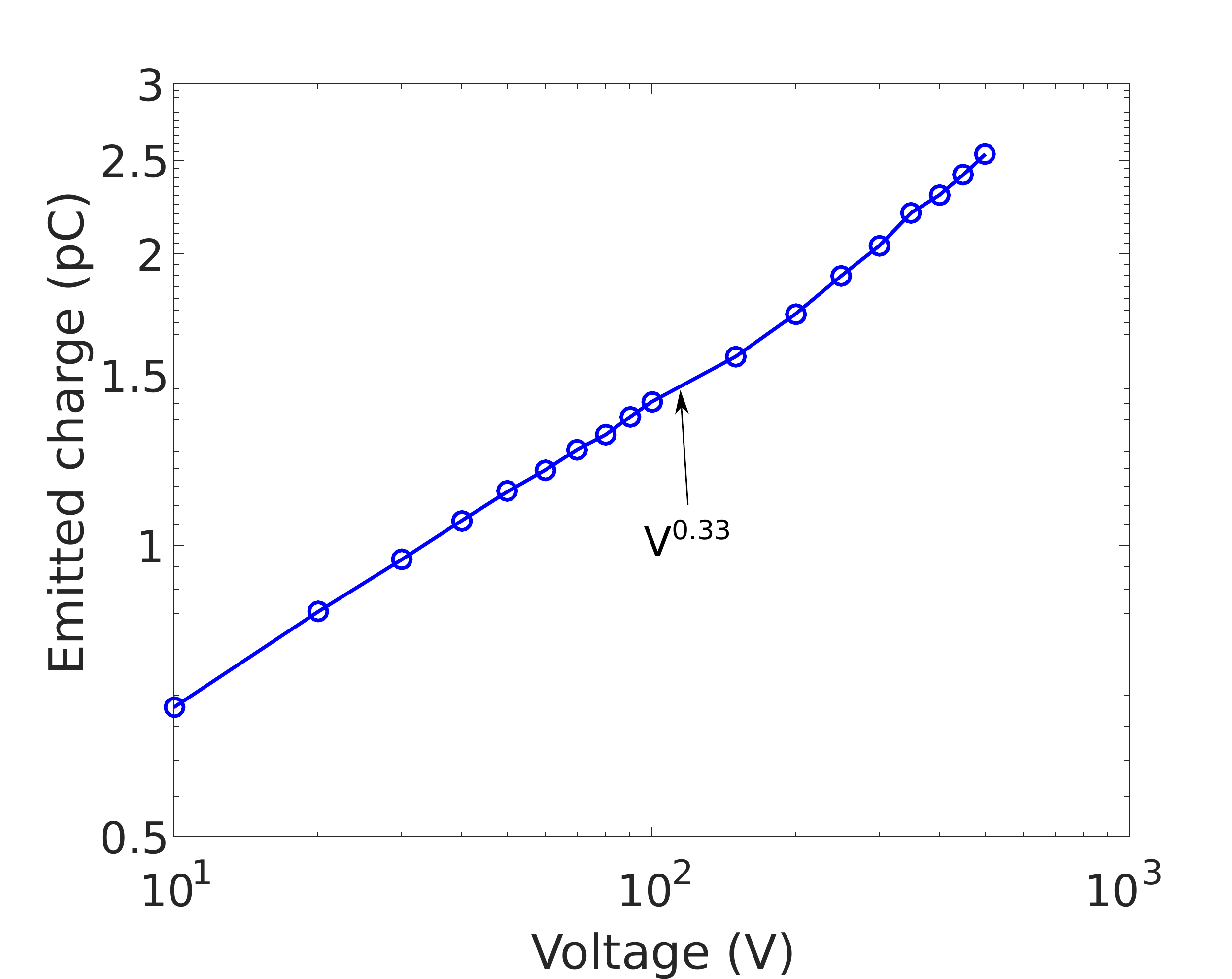} &
  	\includegraphics[draft=false,width=2.7in]{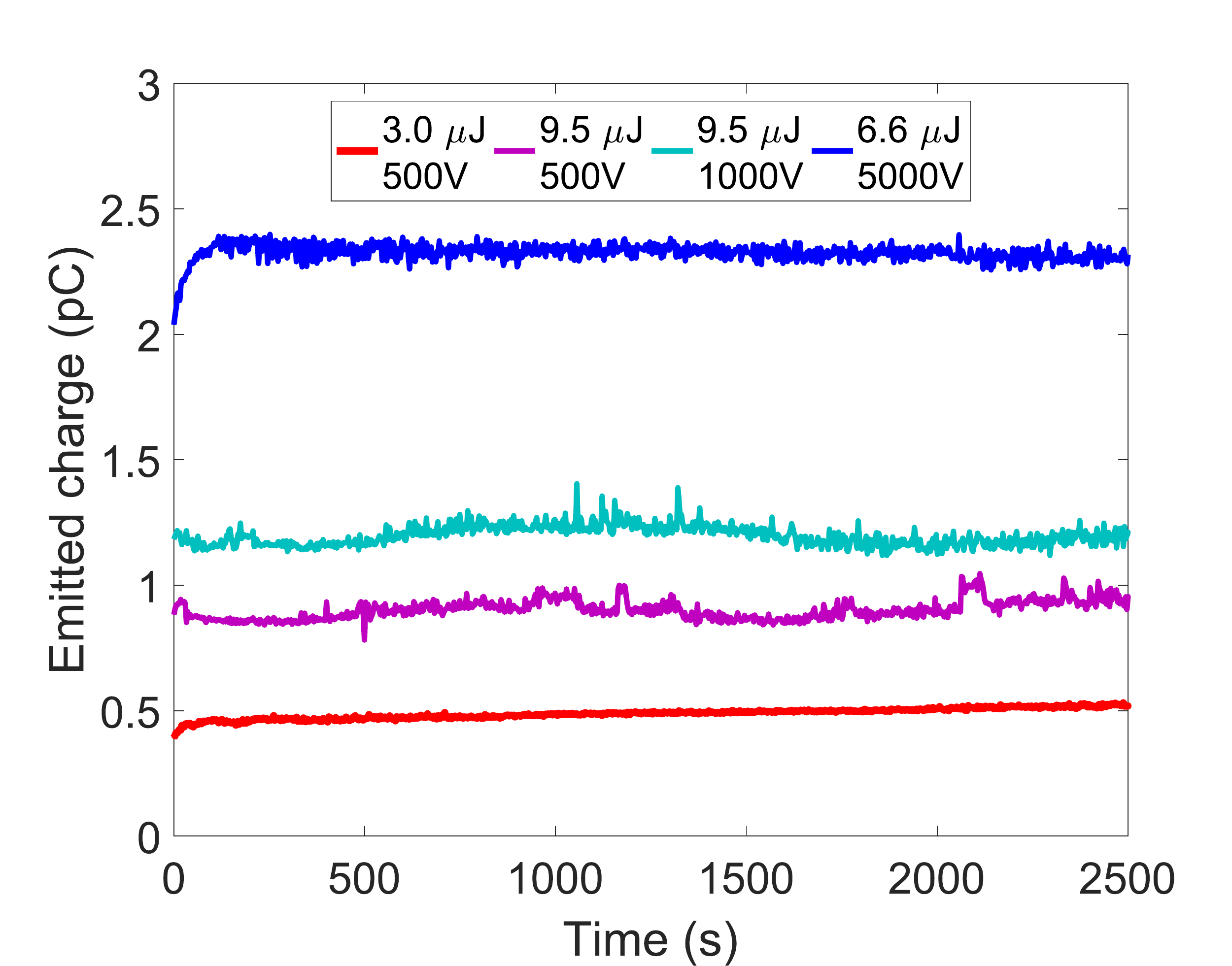} \\
  	(c) & (d) \\
  	\includegraphics[draft=false,width=2.7in]{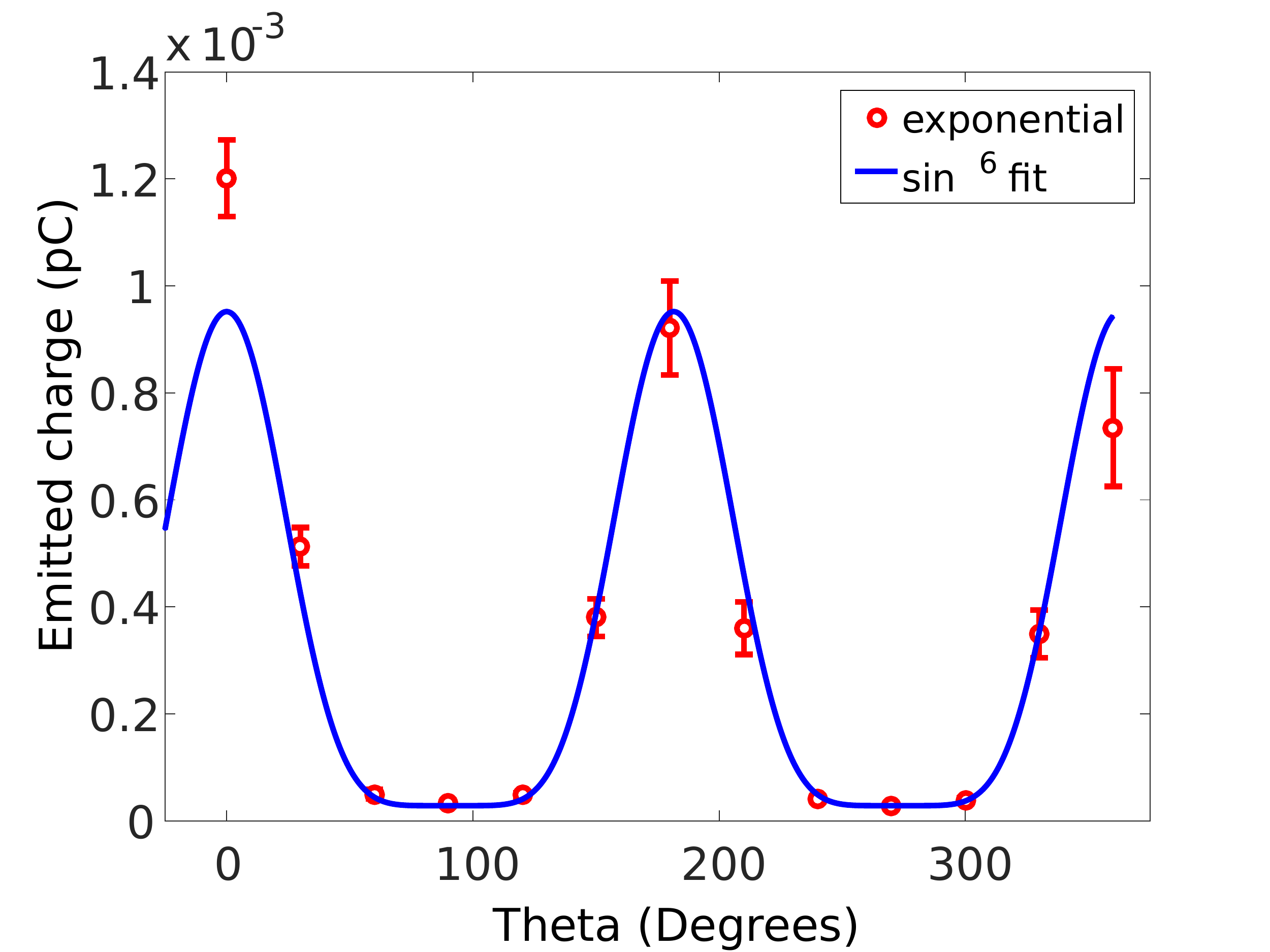} &
  	\includegraphics[draft=false,width=2.7in]{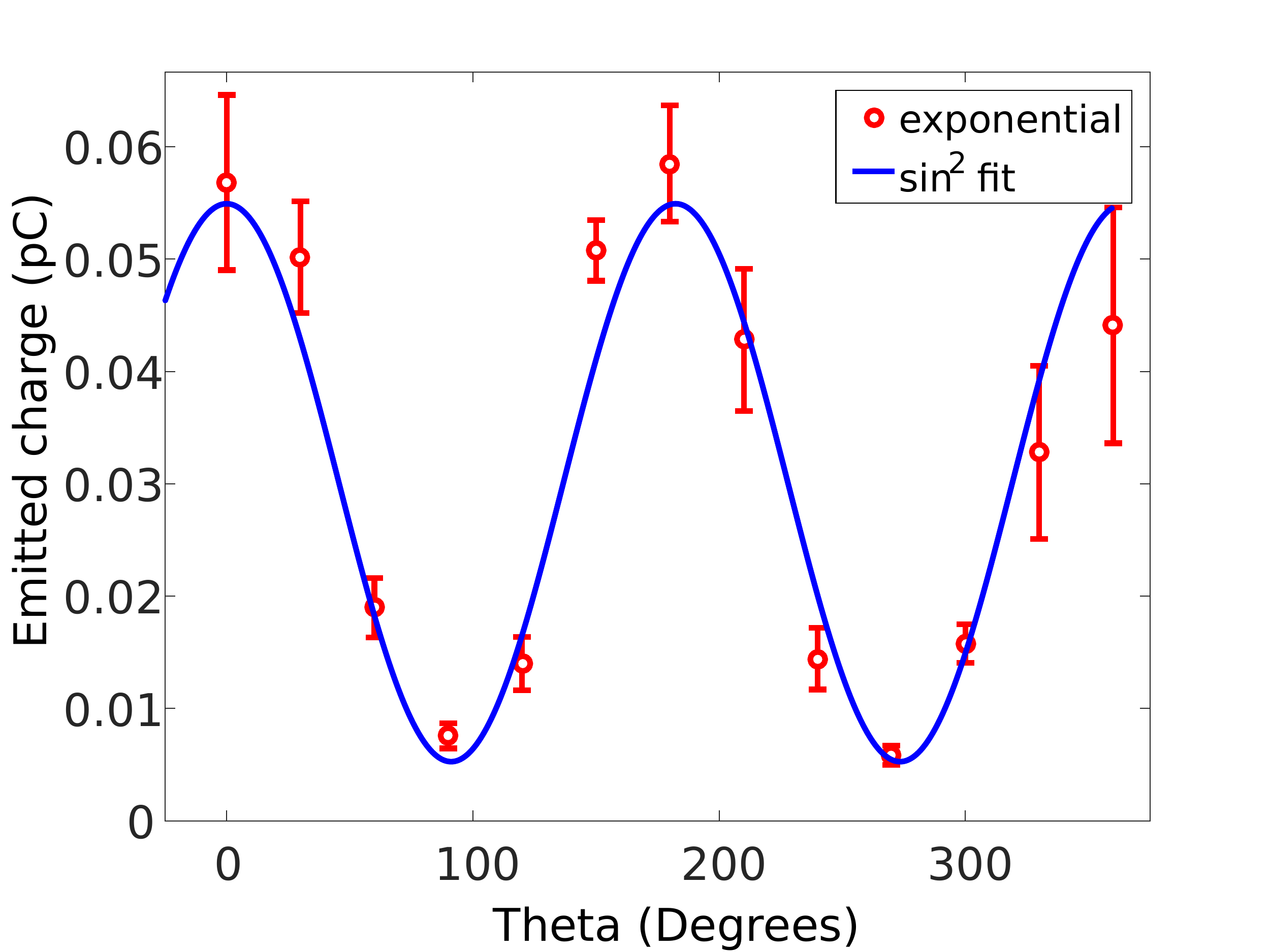} \\
  	(e) & (f)
  	\end{array}$
  	\caption{Experimental results: (a) Emitted charge and (b) overall quantum efficiency (QE) as a function of laser pulse energy for various anode bias voltages. A 3-photon emission growth is observed at low intensities, followed by a tunneling kink at an enhanced peak intensity near $1\times10^{13}$\,W/cm$^2$. (c) Log-log plot of emitted charge versus bias-voltage at fixed incident pulse energy of 10.8\,{\textmu}J. The $<1$ voltage dependence slope indicates that the emission is not fully space-charge limited. (d) Stability of emitted current from the photocathode over time showing stable output. In (e) and (f) the polarization was changed continuously from $0^{\circ}$ (along the tip) to $360^{\circ}$ to show the effect of polarization on charge yield. The bias voltage was maintained at 10\,V for these measurements. (e) At an enhanced peak intensity of $5.7\times10^{12}$\,W/cm$^2$, the current increases with the third power of pulse energy. (f) At an enhanced peak intensity of $27.5\times10^{12}$\,W/cm$^2$, $\gamma$ is less than 2 and the emission increases linearly with intensity.}
  	\label{nanoTipExperiment}
  \end{figure}

  Our experimental results indicate that the high electric field of the ultrashort laser pulses combined with the field enhancement of the nanosharp high-aspect-ratio silicon tip array resulted in large current emission at small laser energies, pico-Coulomb emitted charge at microjoule incident energy (Fig.\,\ref{nanoTipExperiment}a).
  While the overall quantum efficiency (Fig.\,\ref{nanoTipExperiment}b), calculated simply as the number of electrons emitted per incident photon, does not exceed $10^{-6}$, this is high considering that only a very small fraction of the emitter surface is utilized, as the tip diameter is sub-10\,nm and the tip spacing is 5\,{\textmu}m (i.e., about one 3 millionth of the total array area emits electrons).
  The emission is also greatly enhanced as compared to our measurements on planar Si, yielding just 1\,fC of charge for 5\,{\textmu}J incident energy corresponding to a QE of $10^{-10}$.
  Assuming the same field enhancement at the tip, it is estimated that the overall QE could be further increased by more than 1 order of magnitude by reducing the tip spacing.
  At low energy ($<0.2$\,{\textmu}J), the charge yield has a slope of $\sim 3.4 \pm 0.3$ on a log-log scale (i.e., $\propto P^{3.4}\,\text{pC/{\textmu}\text{J}}^{3.4}$ as shown in Fig.\,\ref{nanoTipExperiment}a, where $P$ is the pulse energy).
  This matches closely to the expected slope of 3 for a three-photon absorption process, given that the electron affinity of Si is 4.05\,eV, and the photon energy at 800\,nm is 1.55\,eV (Later in this section, the slope is meant to be the power law dependence).
  Around 0.2\,{\textmu}J, there is a kink in the log-log plot that is observed at all bias levels.
  For the case of single tips, it has been observed that such a bend-over in current yield occurs near a Keldysh parameter of $\gamma = \sqrt{\Phi/2U_p} \approx 2$ \cite{bormann2010tip,piglosiewicz2014carrier,yalunin2011strong}, where $\Phi$ is the material work function, and $U_p$ is the ponderomotive potential of the local laser field:
  \begin{equation}
  U_p = \frac{q^2 F_0^2}{4 m \omega^2},
  \label{ponderomotivePotential}
  \end{equation}
  where $q$ is the electron charge, $F_0$ the peak electric field strength, $m$ the electron mass, and $\omega$ the angular frequency of the laser.
  
  To simulate total electron yield as a function of incident intensity, a model based on strong-field perturbation theory \cite{yalunin2011strong,hoffrogge2014tip} was implemented.
  Averaging effects due to both pulse duration and beam shape on the emitter surface were accounted for.
  This result is compared to the experimental data in Fig.\,\ref{nanoTipExperiment}a.
  A good fit is obtained across all incident laser pulse energies for the highest bias voltages, where a deviation is only observed at the highest incident pulse energies.
  An electric field enhancement factor of $\sim 10.5$ was used to account for an enhanced peak intensity at the tip relative to the incident peak intensity of the pulse, which is in good agreement with the 9.4 enhancement factor found with the electromagnetic model (Fig.\,\ref{nanotipSimulationProfile}a).
  For peak intensities beyond the kink (i.e., $\gamma < 2$), electrons begin to tunnel into vacuum faster than they can oscillate back into the tips \cite{bormann2010tip,keldysh1965ionization}, and the emission follows a time-averaged Fowler-Nordheim model \cite{yalunin2011strong}.
  This regime is commonly referred to as the strong-field or tunnelling regime of electron emission.
  As shown later, while the initial slope change is due to the transition to the tunneling emission regime, the final charge yield at the highest incident energies is reduced from the expected value by as much as 80\% for a bias of 10\,V due to the onset of a space-charge induced virtual cathode (Fig.\,\ref{nanoTipExperiment}a).
  
  To ensure that the emission is due to electric field enhancement at the tip and not just an increase in surface area (i.e., extra emission along the shank of the tips), the charge yield is measured while rotating the polarization angle ($\theta$) at a fixed bias of 10\,V (Fig.\,\ref{nanoTipExperiment}e-f).
  For both low and high pulse energies, peak emission occurs when the polarization is parallel to the axis of the tip, and minimum emission for the orthogonal polarization.
  For the case of high pulse energies, the polarization followed a $\sin^2(\theta)$ dependence, while for low energies that of a $\sin^6(\theta)$ dependence, corresponding to the tunneling and multi-photon emission regions, respectively.
  
\begin{figure} \centering
  	$\begin{array}{cc}
  	\includegraphics[draft=false,height=2.0in]{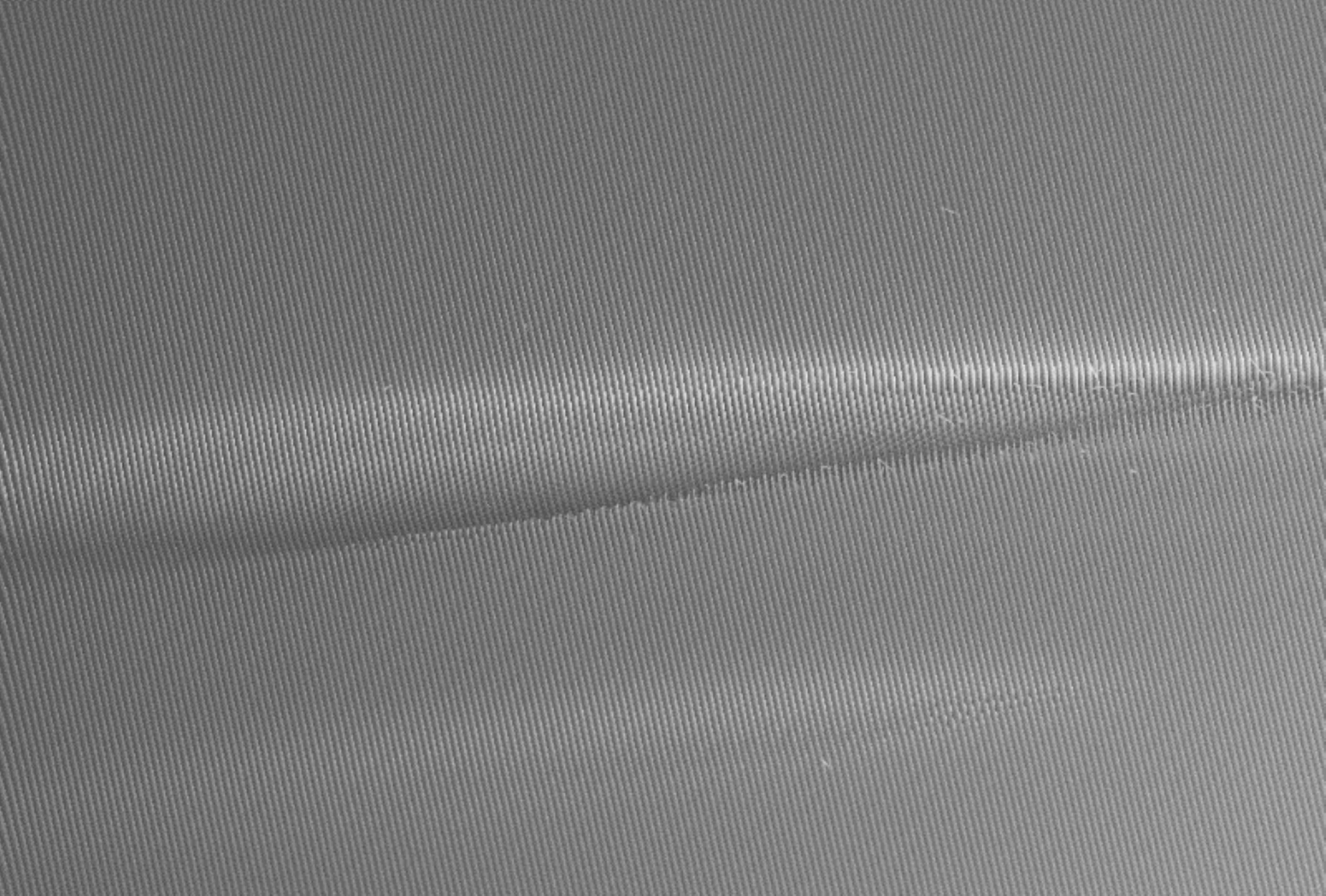} &
  	\includegraphics[draft=false,height=2.0in]{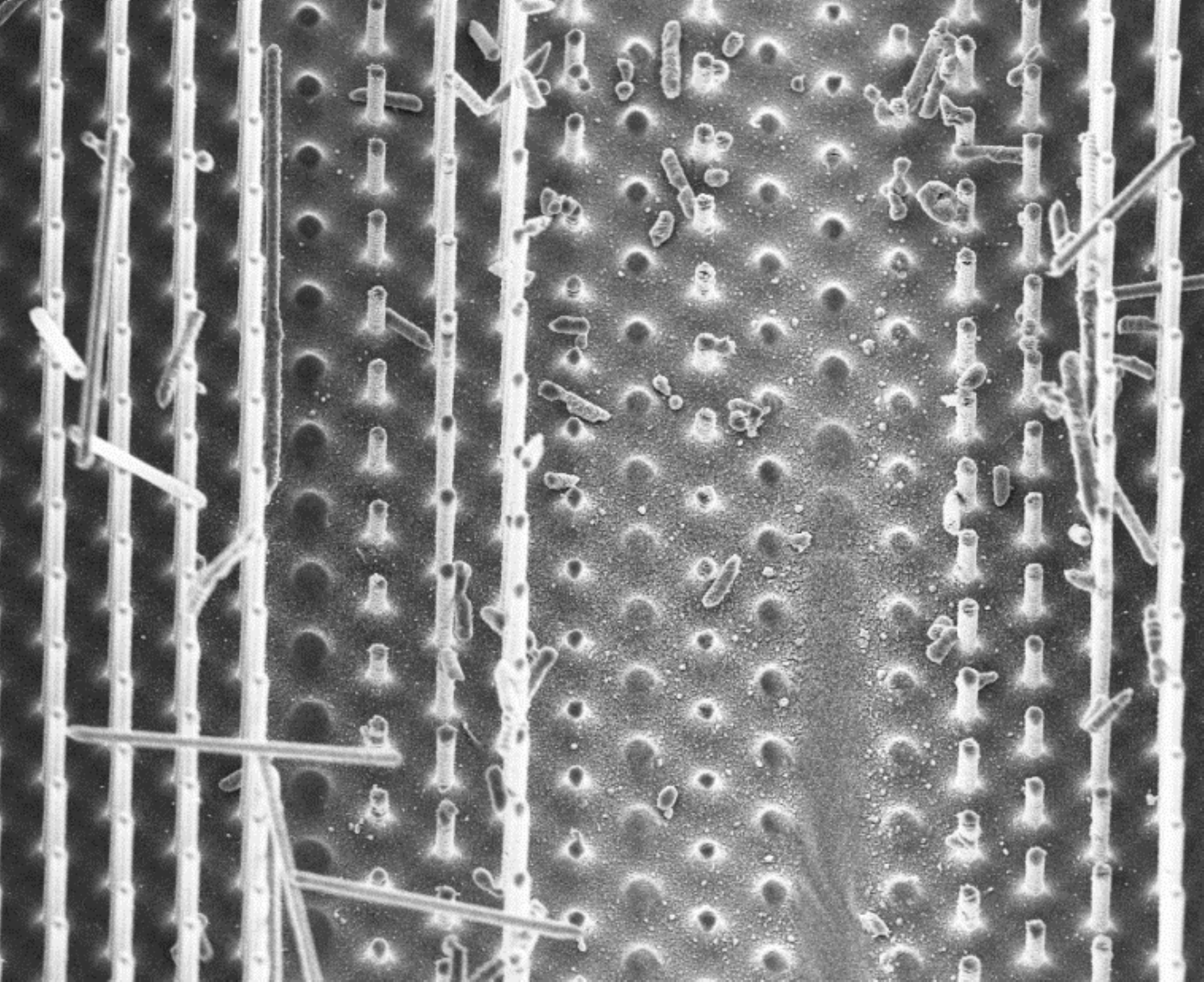} \\
  	(a) & (b)
  	\end{array}$
  	\caption{(a) SEM of high aspect-ratio single-crystal Si pillars with uniform nanosharp tips damaged after exposed to 20.5\,{\textmu}J laser pulses. (b) SEM close up of the damaged area.}
  	\label{semPhotos}
  \end{figure}

  Fig.\,\ref{nanoTipExperiment}d shows four different cathode currents on four different sample locations with the beam being unblocked at time 0\,s at an unexposed area.
  All four curves show stable current emission after 8 million pulses, which is important for electron source applications where surviving millions of cycles at high charge output is required.
  The lowest curve in Fig.\,\ref{nanoTipExperiment}d has lower noise than the others and is slightly rising.
  This is because field emission current is highly nonlinear with respect to the photon energy, and the activation of the cathode takes longer at lower fields \cite{guo2013band}.
  The highest curve in Fig.\,\ref{nanoTipExperiment}d shows an average 2.3\,pC electron emission per pulse.
  SEM images taken after exposing the tips to 8 million pulses showed no measurable difference between the non-exposed tips and the exposed tips for laser energy pulses below 10\,{\textmu}J.
  For laser pulses with 10.8\,{\textmu}J energy, some dulling of the tips is observed, and the tip radii spread is also increased.
  When the energy of the laser pulses is over 20\,{\textmu}J, the tips are ablated, leaving a 85\,{\textmu}m $\times$ 1800\,{\textmu}m mark in the samples that matches the laser spot size (Fig.\,\ref{semPhotos}).
  
\begin{figure} \centering
  	$\begin{array}{cc}
  	\includegraphics[draft=false,width=3.0in]{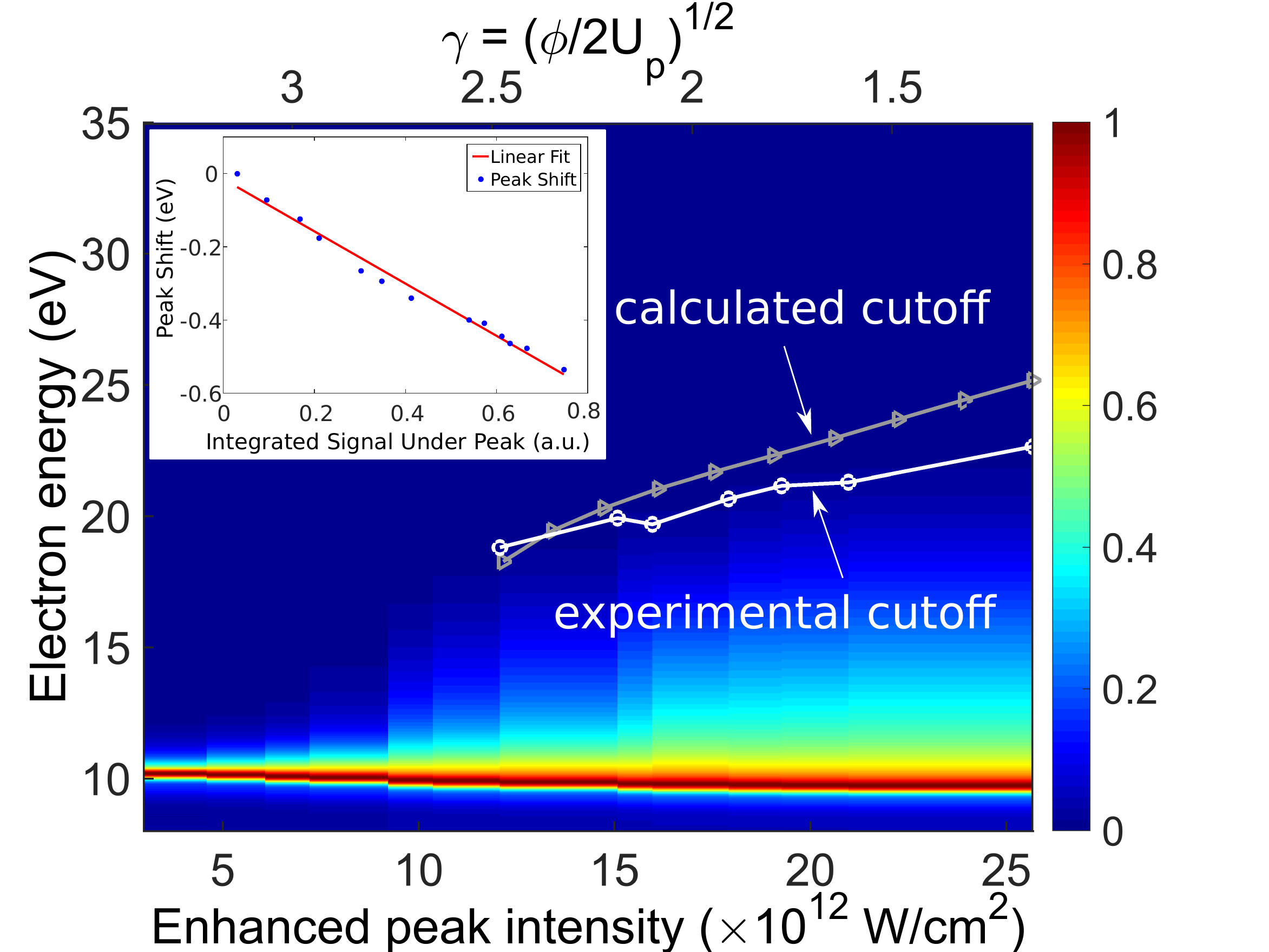} &
  	\includegraphics[draft=false,width=3.0in]{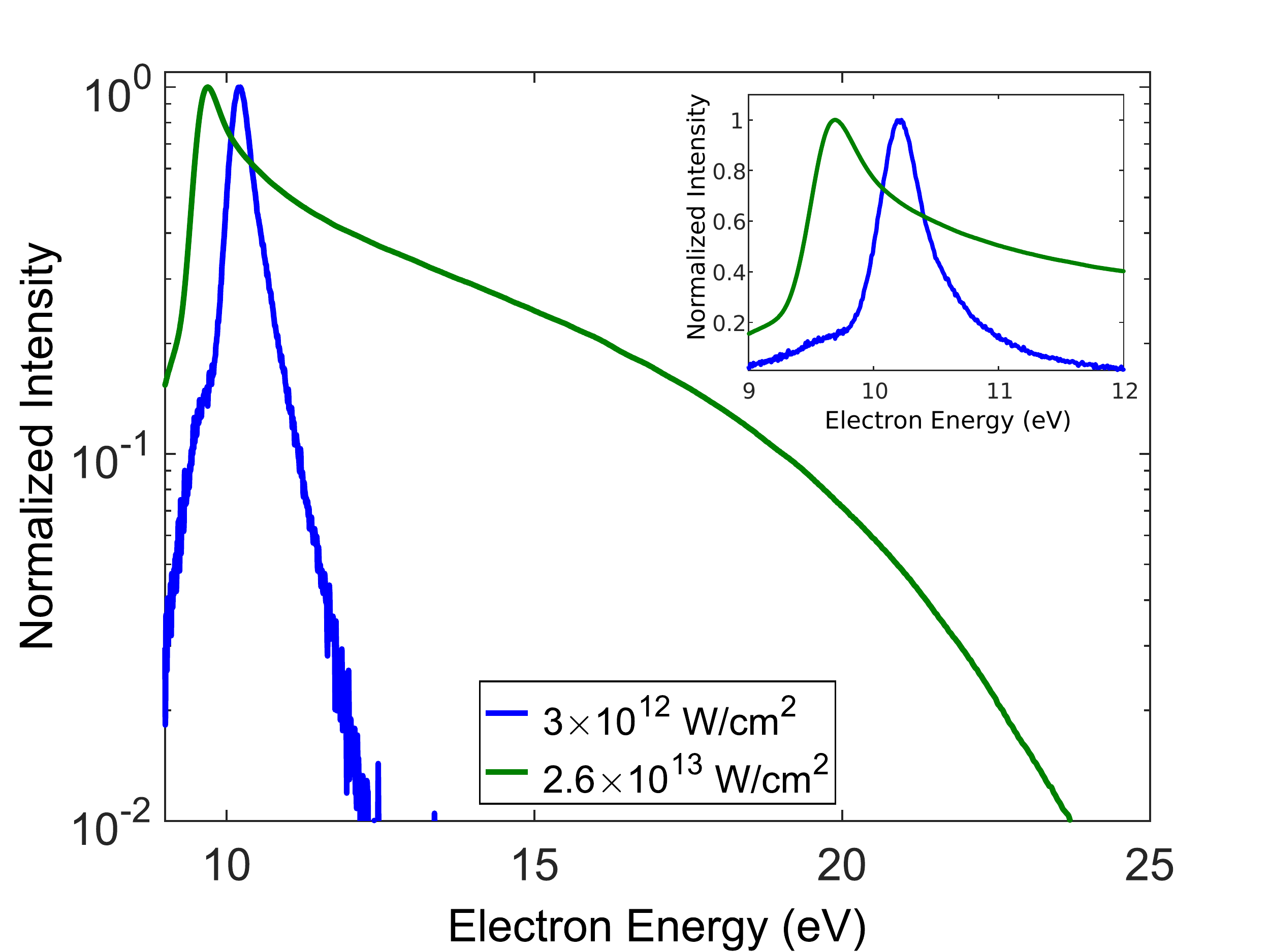} \\
  	(a) & (b)
  	\end{array}$
  	\caption{Electron energy spectra. In (a) normalized spectra are plotted as a function of increasing pulse energy to show the extension of the high-energy plateau as laser energy is increased. The 10\,eV offset in the spectra is due to the applied bias between the sample and entrance aperture of the spectrometer. For intensities beyond the observed current kink, the experimental and calculated cutoffs are plotted. The calculated cutoff uses a Simple-man model incorporating the simulated electric field profile along the axis of the tip. The inset shows the energy shift of the peak near 10\,eV as a function of current underneath the peak region of the spectra. In (b) spectra are selected from the lowest and highest pulse energy points tested. The inset contains a zoom on the low-energy peak of the spectra on a linear scale.}
  	\label{energySpectra}
  \end{figure}

  A second hallmark of strong-field, or tunneling emission is found in the measured electron energy spectra, shown as a function of increasing laser intensity in Fig.\,\ref{energySpectra}.
  Due to a narrow emission window near the peak of the electric field and subsequent acceleration and rescattering with the tip, the electron spectrum develops an asymmetric structure having a sharp, low-energy direct electron peak followed by a broad plateau extending out to high energies \cite{piglosiewicz2014carrier,yanagisawa2011energy,kruger2011attosecond,herink2012field}.
  Classically speaking, an electron born within a laser field alone can at most be accelerated to an energy of $2U_p$.
  However, with the inclusion of a rescattering boundary, the plateau cutoff energy can exceed $10U_p$ \cite{paulus1995classical}.
  
  An electron time-of-flight (TOF) spectrometer was used to study the electron energy spectra as a function of incident drive intensity at 800\,nm.
  The results of this scan show a sharp ($<1.5$\,eV FWHM peak width) low-energy peak with a high-energy plateau extending to around 12\,eV beyond the low-energy peak at the highest pulse energy tested.
  To ensure the high-energy plateau is indeed due to laser acceleration after emission, space-charge broadening must be ruled out.
  The single-tip modeling results described previously (Fig.\,\ref{nanotipSimulationResult}) show that pulse spreading indeed occurs with the inclusion of space-charge; however the high-energy plateau was still dominated by laser accelerated electrons for charge yields exceeding 1\,pC, while the yield in the spectral measurements shown in Fig.\,\ref{energySpectra} did not exceed 50\,fC.
  While this rules out such effects in the single-tip limit, the substrate and neighboring charges may also influence the spectra.
  Femtosecond electron pulse spreading from a planar cathode due to electron-electron interactions has been shown both theoretically and experimentally to scale as the square root of the number of particles in the electron bunch and be inversely proportional to the electron bunch radius \cite{passlack2006space,siwick2002ultrafast}.
  Passlack et al. \cite{passlack2006space} experimentally demonstrated that for an electron pulse with a group velocity corresponding to 0.18\,eV, the pulse broadening did not exceed 300\,meV for more than 75'000 electrons per pulse and an initial bunch radius of 350\,{\textmu}m.
  Accounting for the differences in initial bunch radius based on the beam profile used, even a conservative estimate does not indicate broadening the electron pulse by more than 2\,eV at the highest yield measured.
  
  To determine the plateau cutoff extension described by laser acceleration, a semiclassical model to analyze the cutoff scaling as a function of laser intensity was used.
  The enhanced peak intensity was calibrated by matching the current scaling measured at the spectrometer to the measurements in Fig.\,\ref{nanoTipExperiment}a.
  Using this calibrated peak intensity, a cutoff scaling analysis was then performed by using the well-established Simple-man model \cite{dombi2013ultrafast,piglosiewicz2014carrier,herink2012field} for enhanced peak laser intensities at and exceeding the observed kink in current yield (i.e. the tunneling regime).
  The details of the calculation are outlined in \cite{keathley2012strong}, where we have replaced the dipolar decay function describing the electric field profile with the simulated profile from the electromagnetic simulations described earlier.
  Also, the oxide layer is assumed to be negligible.
  To account for the DC bias in the simulation, the solution is shifted by 10\,eV.
  Space-charge is neglected in this calculation, following the previous discussion.
  The cutoff value was defined as being the energy where the condition $I(E) = 0.1I(E/2)$ is satisfied, where $I$ is the spectrum intensity and $E$ the electron energy.
  The results are overlaid with the energy spectra in Fig.\,\ref{energySpectra}a and compared to the measured cutoff values using the same condition.
  The calculated cutoff values are offset to slightly higher energies, with a slightly increased slope relative to the measured values.
  Overall, the agreement between the measured and the predicted cutoff values is reassuring given that the peak intensity was calibrated using the kink in current yield, an entirely separate measurement, rather than as a free parameter to achieve the best fit.
  
  The absolute value of the cutoff using this method is sensitive to the exact spectral shape.
  Since the semiclassical model results in spectra having much steeper cutoff than those observed experimentally, it is difficult to find an absolute match between calculated and measured values.
  However, the difference in slope is more interesting as this points to a deviation between the modeled and actual field decay away from the tip apex.
  The calculations here already show a reduced slope for the modeled field decay as compared to the case of a homogeneous electric field due to the fact that the electron excursion starts to be on the order of the field decay length, resulting in a minimum adiabaticity parameter $\delta \approx 15$ \cite{herink2012field}.
  While not deep into the subcycle regime where the adiabaticity parameter is much less than 1 and the cutoff scales linearly with the field \cite{herink2012field}, the overall reduction in cutoff energy can be quite severe much before this regime is reached.
  
\begin{figure} \centering
  	$\begin{array}{cc}
  	\includegraphics[draft=false,width=3.0in]{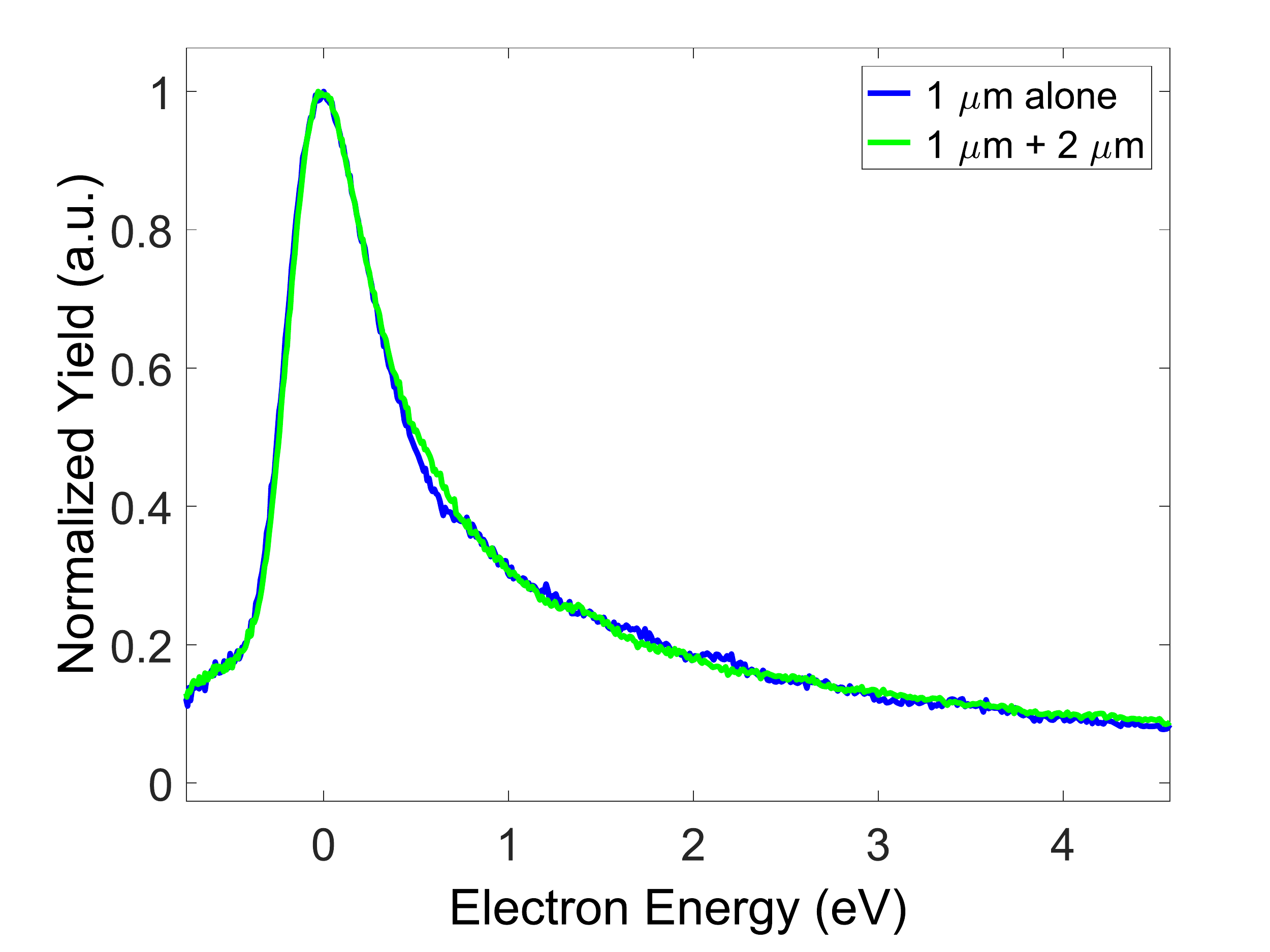} &
  	\includegraphics[draft=false,width=3.0in]{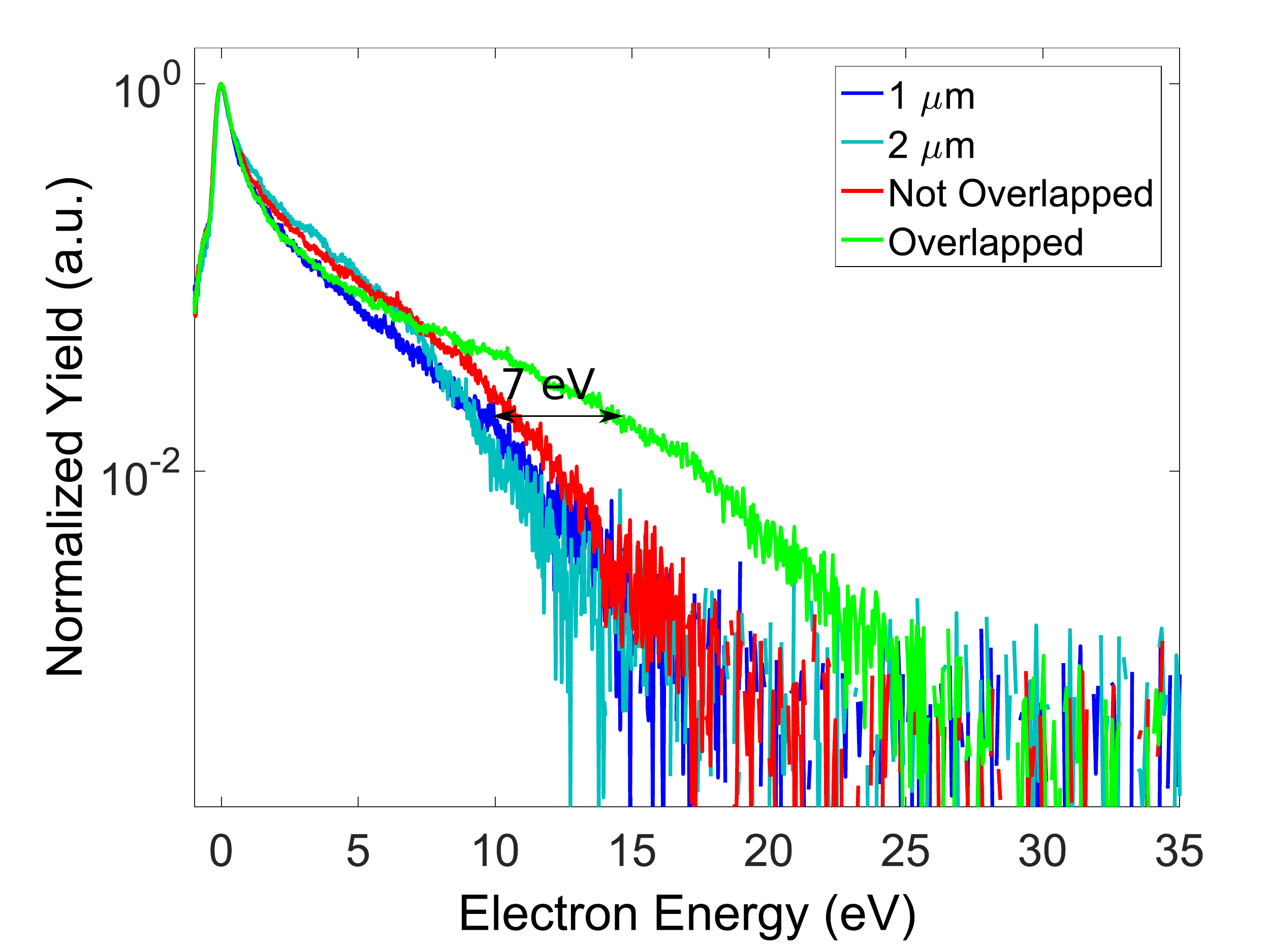} \\
  	(a) & (b)
  	\end{array}$
  	\caption{Comparison of the two-color results. (a) Spectra taken with the 1\,{\textmu}m pulse alone (blue), and overlapped in time with the 2\,{\textmu}m pulse (green) on a linear scale centered near the direct peak. (b) The spectra from 1\,{\textmu}m alone (blue), 2\,{\textmu}m alone (teal), 1\,{\textmu}m with 2\,{\textmu}m not overlapped in time (red) and overlapped in time (green) respectively. In (b), all curves are plotted on a log scale to emphasize the shift in cutoff energy of the lower yield plateau. In both (a) and (b), the spectra are all shifted such that their peaks are centered at 0\,eV and are normalized to 1.}
  	\label{cutoffScalingStudy}
  \end{figure}

  Laser-induced cutoff scaling also indicates that the emission process is prompt with respect to the driving electric field, as the laser can only accelerate electrons that are present within the duration of the laser pulse itself.
  Preliminary cross correlation electron emission measurements using two-color pulses further demonstrate the prompt nature of the electron emission and laser-induced spectral shaping (Fig.\,\ref{cutoffScalingStudy}a).
  Measuring electron emission as a function of delay between a 1\,{\textmu}m pulse and 2\,{\textmu}m pulse incident on the tips resulted in a sharp current spike tens of femtoseconds in duration that shows no evidence of a long-lifetime pedestal on either side.
  Furthermore, when the pulses were overlapped, the cutoff was extended by around 7\,eV (Fig.\,\ref{cutoffScalingStudy}b), while the bandwidth of the low energy spectral peak was effectively unchanged.
  This is shown to be due to an increase in ponderomotive acceleration in the presence of a two-color field \cite{paulus1995classical}.
  Such results indicate the possibility of tailoring laser waveforms to engineer emitted electron spectra.
  
  Another feature that stands out from the electron spectra is the slight loss in energy of the main spectral peak as the intensity is increased.
  If the effect is solely due to changing ponderomotive potential, then the shift should vary linearly with peak intensity, which was not found to be the case.
  The single tip model at the beginning of the paper indicates that image charge effects from the tip alone can contribute significantly to electron deceleration and recombination with the tip surface at pico-Coulomb level yields across the entire array.
  In recent years, observations of peak shifts in photoemission due to image-charge effects have been studied in detail across a variety of emission levels \cite{arafune2004energy,zhou2005space}.
  Zhou et al. \cite{zhou2005space} show experimentally and theoretically that the image charge-related shift from a planar conductive sample is due mostly to the amount of charge in the bunch, and shifts the mean energy linearly with respect to the total number of electrons in the bunch.
  A simple analysis shows that the main spectral peak indeed shifts linearly to lower energies with respect to the number of charges in the peak, but not the total charge (Fig.\,\ref{energySpectra}a inset).
  This interpretation is also consistent with the idea that the fast moving electrons quickly escape the low-energy bunch after laser acceleration, and contribute minimally to the peak shift.
  
  At higher incident energies, the anode bias voltage had a significant effect on the emitted charge (Fig.\,\ref{nanoTipExperiment}a).
  At 9.3\,{\textmu}J incident energy, the emitted charge is 0.27\,pC with 10\,V anode bias and 1.4\,pC with a 1000\,V anode bias.
  From finite element modeling, the local electric field from the anode bias ($\sim 3$\,MV/m at 500\,V anode bias) is 3-4 orders of magnitude lower than the peak field from the incident laser ($\sim 6$\,GV/m before enhancement at $\sim 11$\,{\textmu}J pulse), meaning there should be no noticeable increase in electric field and emitted current due to
  the increase in anode bias.
  Also, the slight reduction in emission due to space-charge effects around a single emitter, as outlined in Fig.\,\ref{nanotipSimulationResult}, does not account for the extent of anode bias dependence observed experimentally.
  With this in mind, it is necessary to turn to a macroscopic space-charge model that accounts for the charge from neighboring emitters.
  
  To model the emission across the entire pulse energy range tested, a method employing strong-field perturbation theory was used, which has been successful in modeling such emission from atomic systems \cite{milovsevic2006above} and more recently nanotips \cite{bormann2010tip,yalunin2011strong}.
  The equation describing the ionization process is given, to first order, as
  \begin{equation}
  I \propto \frac{1}{\hbar} \int\limits_0^\infty | M_p^{(1)}|^2 dp,
  \label{ionizationProbability}
  \end{equation}
  where
  \begin{equation}
  M_p^{(1)} = \frac{-i}{\hbar} \int\limits_{-\infty}^{\infty} dt e^{iS_p(t)/\hbar} \bra{p+qA(t)} qzF(t) \ket{\Psi_0},
  \label{transitionAmplitude}
  \end{equation}
  Here, $S_p$ is the action due to the laser field and is given by
  \begin{equation}
  S_p(t)=\frac{1}{2m} \int\limits^t dt \{ p+qA(t) \}^2
  \label{laserFieldAction}
  \end{equation}
  In \eref{transitionAmplitude} and \eref{laserFieldAction}, $A(t)$ is the magnetic vector potential, $\Psi_0$ the ground state before excitation, $p$ the final momentum, $F(t)$ the electric field of the laser pulse, $M_p^{(1)}$ the transition amplitude of the electron to final momentum $p$, and $m$ the electron mass.
  In \eref{transitionAmplitude}, $F(t)$ is taken to be a Gaussian pulse having a 35\,fs FWHM in intensity.
  There are a few assumptions in this expression.
  First, supply is not accounted for, which means that the expression is only accurate when the electron emission is not limited by the electron supply from the conductor.
  Second, the emission is assumed to be dominated by electrons located just above the conduction band of Si, having an electron affinity of 4.05\,eV.
  Lastly, the electric field inside the conductor is neglected, meaning the spatial part of the matrix element is only calculated over the vacuum half-space.
  
\begin{figure} \centering
  	$\begin{array}{c}
  	\begin{array}{cc}
  	\includegraphics[draft=false,width=3.0in]{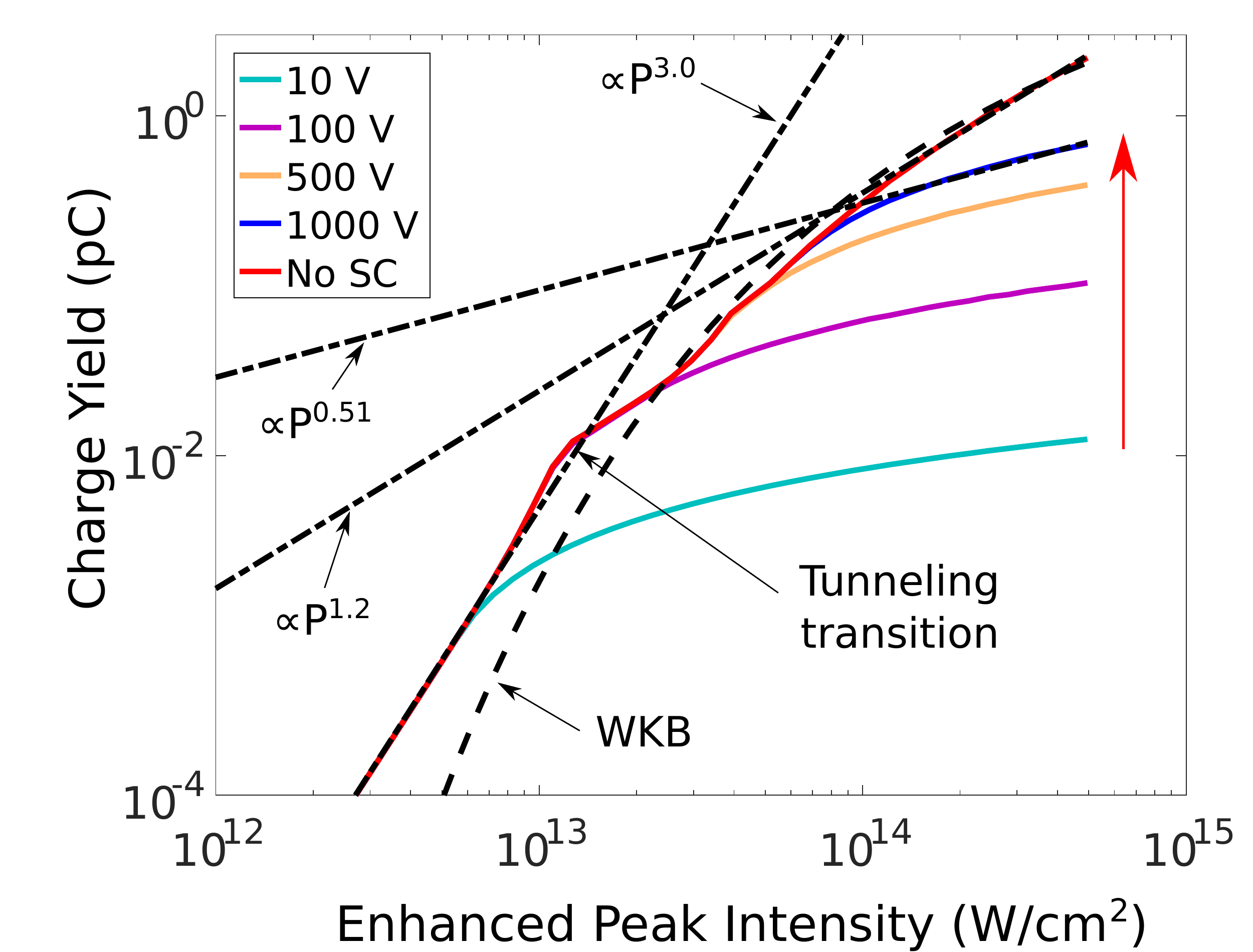} &
  	\includegraphics[draft=false,width=3.0in]{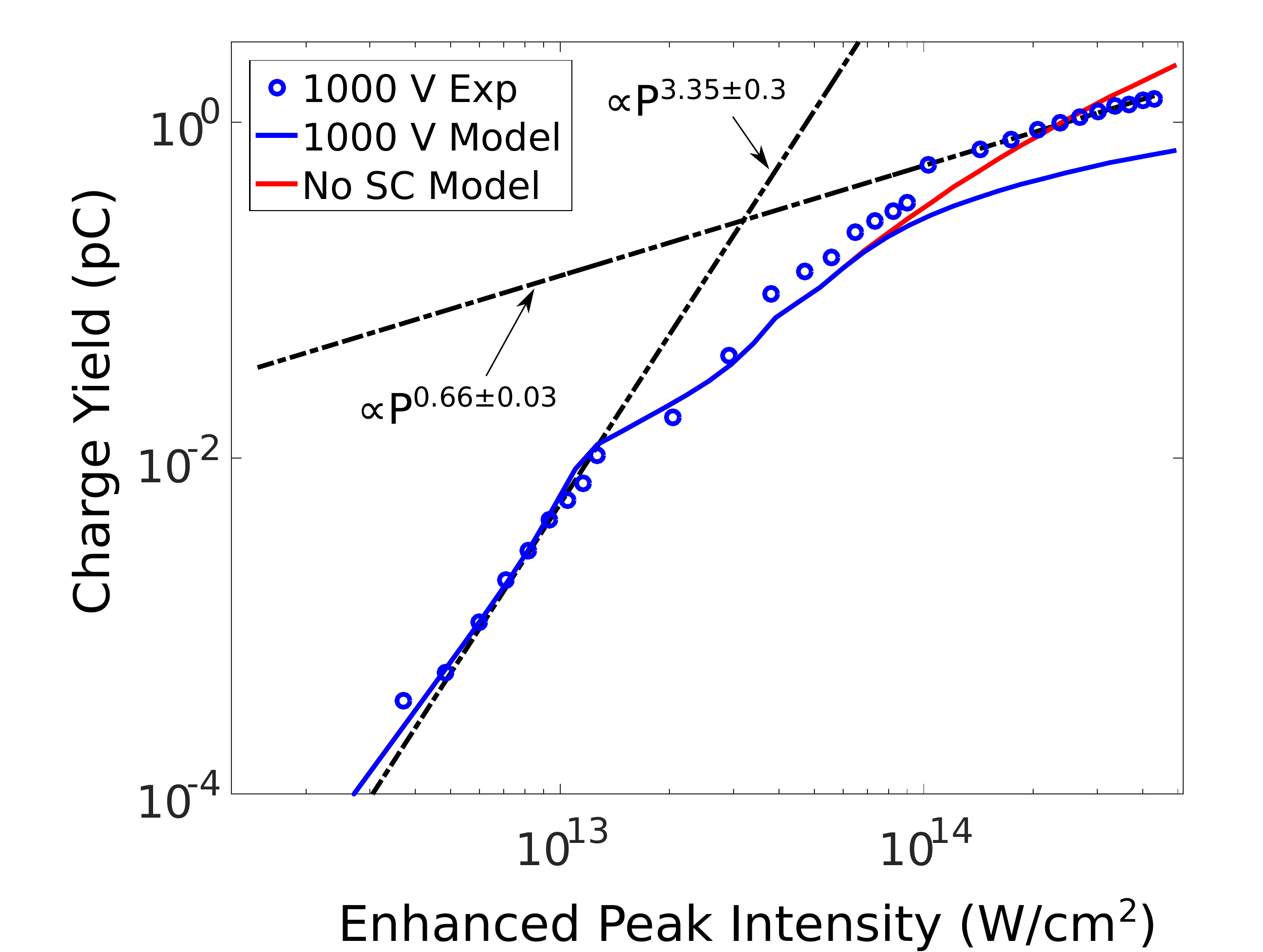} \\
  	(a) & (b)
  	\end{array} \\
  	\includegraphics[draft=false,width=5.5in]{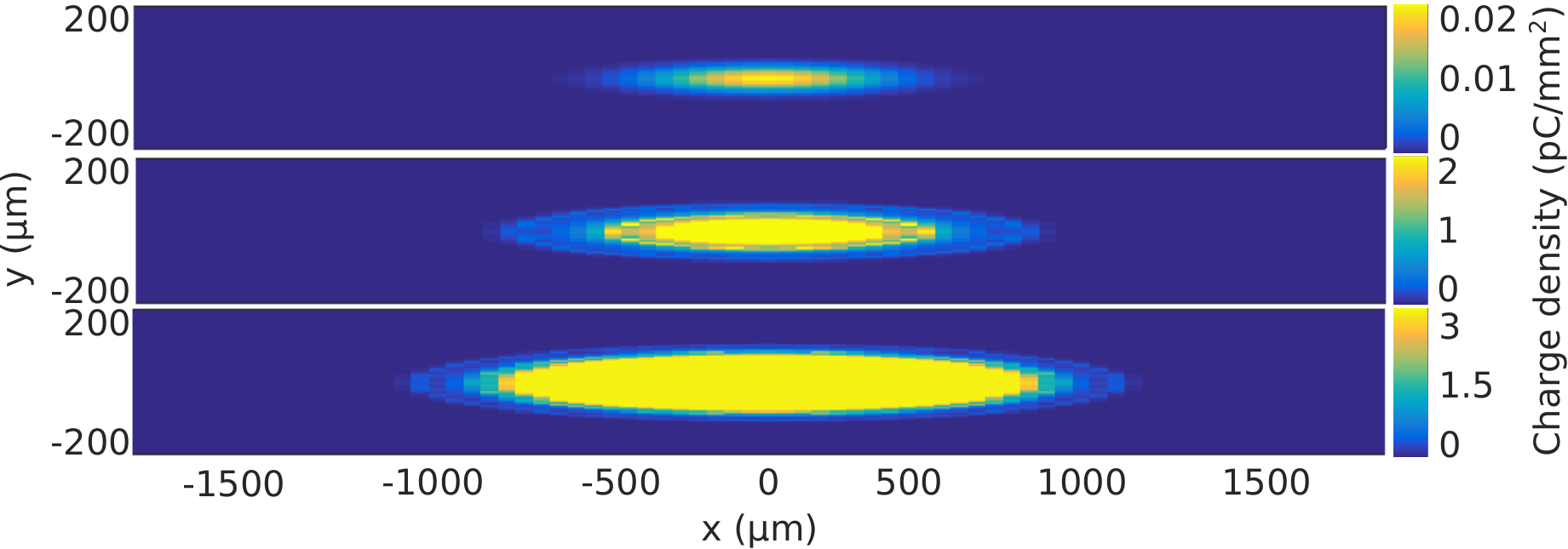} \\
  	(c)
  	\end{array}
  	$
  	\caption{Modeled and calculated results: (a) Model results of the emitted current as a function of laser pulse energy and bias voltage. No SC refers to the case where no space-charge effects were considered ($Q_c \rightarrow \infty$). The red arrow indicates increasing $Q_c$, i.e. increasing virtual cathode limit, as a function of bias. As expected, the data fits well to a spatially and temporally averaged WKB model after the tunneling kink. (b) The experimental data at 1\,kV is compared to the model with and without consideration of the virtual cathode limit. (c, d, e) Calculated electron beam profiles at the anode plate for 0.1, 1.6, and 10\,{\textmu}J pulse energies with a bias voltage of 1000\,V. Due to saturation at $Q_c$, the model shows that the effective emitter area grows, and the electron bunch develops a top-hat shape}
  	\label{nanoTipEmissionModel}
  \end{figure}

  A key feature of this emission model is accounting for the possibility of multi-photon emission as a function of laser energy at low intensities, unlike a pure FN model that underestimates electron emission in this region by several orders of magnitude.
  Furthermore, the model includes a description of how the emission transitions from multi-photon absorption to quasi-static tunneling.
  As a consequence of the electron being emitted into a strong laser field, the ponderomotive potential, $U_p$, adds to the effective work function of the boundary \cite{schenk2010strong}.
  This means that as the laser strength increases, higher order photon absorption becomes necessary in order to liberate an electron.
  When this happens, the slope of the yield bends over, matching more closely to a time-averaged Wentzel-Kramers-Brillouin (WKB) tunneling emission rate (Fig.\,\ref{nanoTipEmissionModel}), justifying the use of the FN model at high intensities.
  
  To model the charge density across the surface of the cathode, a Gaussian driver beam of 90\,{\textmu}m FWHM was projected across the emitter surface at an angle of $84^{\circ}$ matching the experimental beam size and angle.
  The beam spot was then divided into grids having a constant intensity, and the strong field emission model described earlier was used to determine the differential current from each grid point.
  For the case of a 1\,kV bias, aside from the very highest intensities, the model describes the total current yield profile with surprising accuracy (Fig.\,\ref{nanoTipEmissionModel}b) as only the field enhancement and constant pre-factor were used for fitting.
  
  From the single tip model, we find that at high laser fields the transverse size of the charge cloud rapidly diverges after the total charge emission (supplementary material in \cite{swanwick2014nanostructured}).
  The results show that less than 1\,ps after the charge emission (compared to 1\,ns of flight time to reach the anode at a 3\,mm spacing), the cloud transverse size is as large as the lattice constant, i.e. 5\,{\textmu}m
  This could easily lead to a global space-charge effect where the extraction bias is screened by the emitted current bunch.
  The simplest approach to account for global space-charge is to assume that the charge from each emitter converges into a sheet of charge just above the cathode surface after emission.
  To account for space charge, the differential charge from each spatial grid point was not allowed to exceed the critical charge, given by
  \begin{equation}
  Q_c = \frac{V A_g \epsilon_0}{d}
  \label{criticalCharge}
  \end{equation}
  where $V$ is the bias voltage, $A_g$ the grid area, $\epsilon_0$ the permittivity of free space, and $d$ the anode-to-cathode spacing.
  The space-charge limit provided in \eref{criticalCharge} is more applicable than the Child-Langmuir current limit for ultrafast cathodes, where the current is bunched into a thin sheet rather than spread across the entire anode-cathode gap \cite{valfells2002effects}.
  While this analysis ignores the near-field tip enhancement, this should decay back to the solution for a planar sheet of charge within hundreds of nanometers of the tip surface, making expression in \eref{criticalCharge} relevant for electron transport in the vacuum.
  The induced virtual cathode voltage used to determine $Q_c$ also naturally accounts for an induced image charge potential (assuming a planar image charge surface).
  The results of this model effectively describe an electron pulse that first saturates in the center, while the wings continue to increase, thus leading to a larger effective spot size and a top hat profile.
  Model results for no space-charge limit (i.e., $Q_c \rightarrow \infty$), 10\,V, 100\,V, 500 and 1000\,V DC bias are given in Fig.\,\ref{nanoTipEmissionModel}.
  
  The model results match many of the features observed in the experimental data.
  At low intensities, the emission follows a 3-photon absorption process, where there is no dependence on bias voltage.
  At higher intensities, when the current yield goes beyond the space-charge limit in the center of the laser spot, the current begins to saturate depending on the bias voltage applied (Fig.\,\ref{nanoTipEmissionModel}c).
  Beyond a 500\,V bias voltage, the bend over due to a transition to the tunneling regime is prevalent, followed by a very gradual transition to space-charge saturation at a higher current yield.
  However, overall the model seems to underestimate the current limit for each voltage.
  For example, without the space-charge limit, the ultimate slope at the highest current yield is $\sim -1.2$.
  This value is calculated as $\sim -0.51$ for a 1\,kV bias in the model as compared to $\sim -0.66$ for the experimental data.
  Also, in the experimental results, the tunneling kink was clearly observed for every bias level.
  As noted in Valfells et al. \cite{valfells2002effects}, the major limitation of this calculation is that it does not account for the initial velocity of the charge leaving the cathode.
  This leads to an underestimation of the total current limit as the initial energy of the electrons lead to larger beam radii and the requirement for higher potentials to prevent their escape.
  From the electron spectroscopy results (Fig.\,\ref{energySpectra}a-b), a large spread in longitudinal energy was observed, and the single tip particle tracing indicates significant transverse momentum, both of which explain charge yields exceeding the limit imposed by \eref{criticalCharge}.
  
  \subsection{Gold nanorods}
  
  \subsubsection{Introduction}
  
  The tip geometry, as the ground for the above inspected electron source, takes advantage of the field enhancement at the apex of a tip to realize strong field electron emission.
  An alternative method for locally enhancing the field of an optical beam is plasmonic enhancement in nanoparticles.
  Nanoparticles exhibiting localized surface plasmon resonances (LSPR) are useful for nanooptics applications that require optical-field enhancement.
  The local enhancement of optical fields at the nanoscale by the collective oscillation of electrons (plasmons) in such nanostructures when illuminated at resonant wavelengths enables the use of these nanostructures for various applications.
  Examples are surface-enhanced Raman spectroscopy \cite{nie1997probing}, high-resolution imaging \cite{imura2005near}, nanochemistry \cite{christopher2011visible}, metamaterials \cite{zhao2013tailoring,zhao2012twisted,fafarman2012chemically}, sensor \cite{mayer2011localized,anker2008biosensing,lal2007nano}, optoelectronics \cite{koller2008organic}, nanolithography \cite{volpe2012near}, and photocathode \cite{dombi2013ultrafast} applications.
  Plasmonic nanoparticle arrays are of particular interest for use as ultrafast, high-brightness photoelectron emitters in next-generation x-ray free electron lasers (XFELs) enabling ultrafast x-ray imaging, and diffraction, as well as time-resolved electron microscopy and spectroscopy experiments.
  
  XFELs, as well as other electron-emission applications, rely on critically on photocathode performance.
  Metallic photocathodes for example are desirable due to their relative insensitivity to contamination, which allows their operation under poorer vacuum conditions than high-efficiency alkali halides.
  As such, there has been a drive to improve the efficiency of metallic photocathodes such as Au and Cu \cite{li2013surface,polyakov2013plasmon,aeschlimann1995observation}.
  Moreover, the performance of XFELs relies on the ability to first generate nanometer scale density modulations in the electron beam, which can then be used to coherently emit x-rays \cite{dowell2010cathode,barletta2010free,freund2012three}.
  A compact coherent x-ray source based on a modulated electron beam produced by a nanostructured photocathode has been recently proposed \cite{graves2012intense}.
  Consequently, the development of nanostructured photocathodes is key to improving next-generation ultrafast, coherent x-ray sources.
  
  Electron emission has previously been demonstrated from arrays of plasmonic nanoparticles and nanostructured plasmonic surfaces \cite{dombi2013ultrafast,li2013surface,polyakov2013plasmon,nagel2013surface,douillard2008short}.
  Dombi et al. and Nagel et al. have both demonstrated electron emission and acceleration within the surface-plasmon enhanced near-field of plasmonic particles lying in-, and out-of-plane of the substrate, respectively \cite{dombi2013ultrafast,nagel2013surface}.
  Douillard et al. have also previously investigated electron emission from multipolar plasmonic particles by photoemission electron microscopy (PEEM) \cite{douillard2008short}.
  Prior work on ultrafast photoemission from plasmonic nanoparticle arrays focused on the energy spectra of electrons produced from such particles, rather than the quantitative charge-yield.
  Additionally, the nanoparticles studied were of dimensions significantly larger than those studied in the present work.
  Furthermore, while quantitative studies of charge-yield from plasmonic Au photocathodes have been performed recently \cite{polyakov2013plasmon}, they were restricted to a range of laser intensities where the emission mechanism lay firmly within the multiphoton absorption regime.
  Thus, a quantitative investigation of photocathode performance in the strong-field regime is presently lacking.
  
  Here, scaling the critical dimensions of the emitters, fabricated by high resolution electron-beam lithography, into the sub-20\,nm regime is discussed \cite{hobbs2014high}.
  Furthermore, the effects of substrate, and traditional adhesion-promoting layers such as Ti, on charge yield from overlying Au nanorods are investigated.
  The effects of various elements, playing important roles in photocathode operation, such as laser intensity, applied DC field, angle of linear polarization, and nanorod array density, on charge yield are also studied.
  
  Fabrication of Au nanorods with sub-20\,nm critical dimensions will allow greater localization of the electron emission site, which is of interest for creating nanostructured electron beams as discussed above.
  Investigation of the effect of the Au/substrate interface on electron emission from arrays of plasmonic Au nanorods prepared by electron-beam lithography will also be key to optimizing the efficiency of such electron sources.
  The existence of a substrate not only shifts the spectral position of the Au nanorod LSPR, but also modifies the optical near-field distribution due to mode hybridization \cite{halas2011plasmons}.
  Higher index substrates lead to a more pronounced red-shift of the LSPR, and stronger field localization at the interface between the plasmonic nanostructure and substrate \cite{knight2009substrates,wu2009finite,schweikhard2011polarization,zhang2011substrate,hutter2012interaction}.
  A strong optical field enhancement at the nanorod/vacuum interface rather than the nanorod/substrate interface is preferred for photocathode applications to reduce electron scattering from the substrate; thus, a low index, electrically conductive substrate, is preferred.
  Moreover, the effects of conventional adhesion-promoting metals such as Ti, used in the preparation of Au nanorods by electron-beam lithography, on photoelectron yield are worthy of investigation.
  Previously, such metallic layers have been shown to reduce the Q-factor of the LSPR within overlying Au nanorods due to increased damping of the resonance \cite{habteyes2012metallic}.
  Lastly, the scaling of emission current as a function of nanorod array density, laser-intensity and applied anode bias will be key to understanding the factors affecting charge-yield from plasmonic photocathodes, such as space-charge, electron emission mechanism and optical field enhancement.
  A better understanding of the factors affecting charge-yield from these photocathodes may then allow us to generate more efficient electron sources for next-generation, ultrafast metrology.
  
  \subsubsection{Results and discussion}
  
\begin{figure} \centering
  	$\begin{array}{cc}
  	\includegraphics[draft=false,width=3.0in]{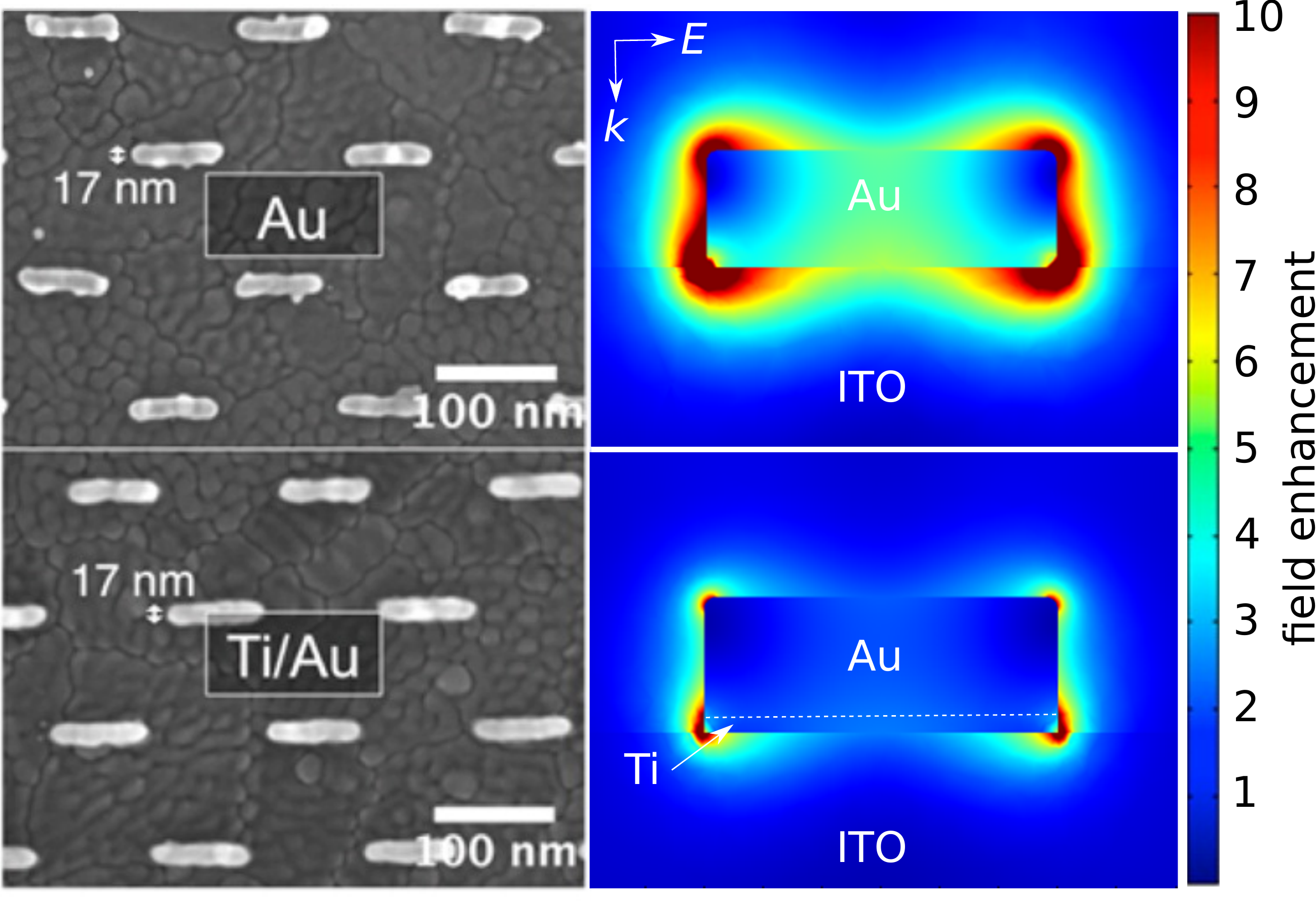} &
  	\includegraphics[draft=false,width=3.0in]{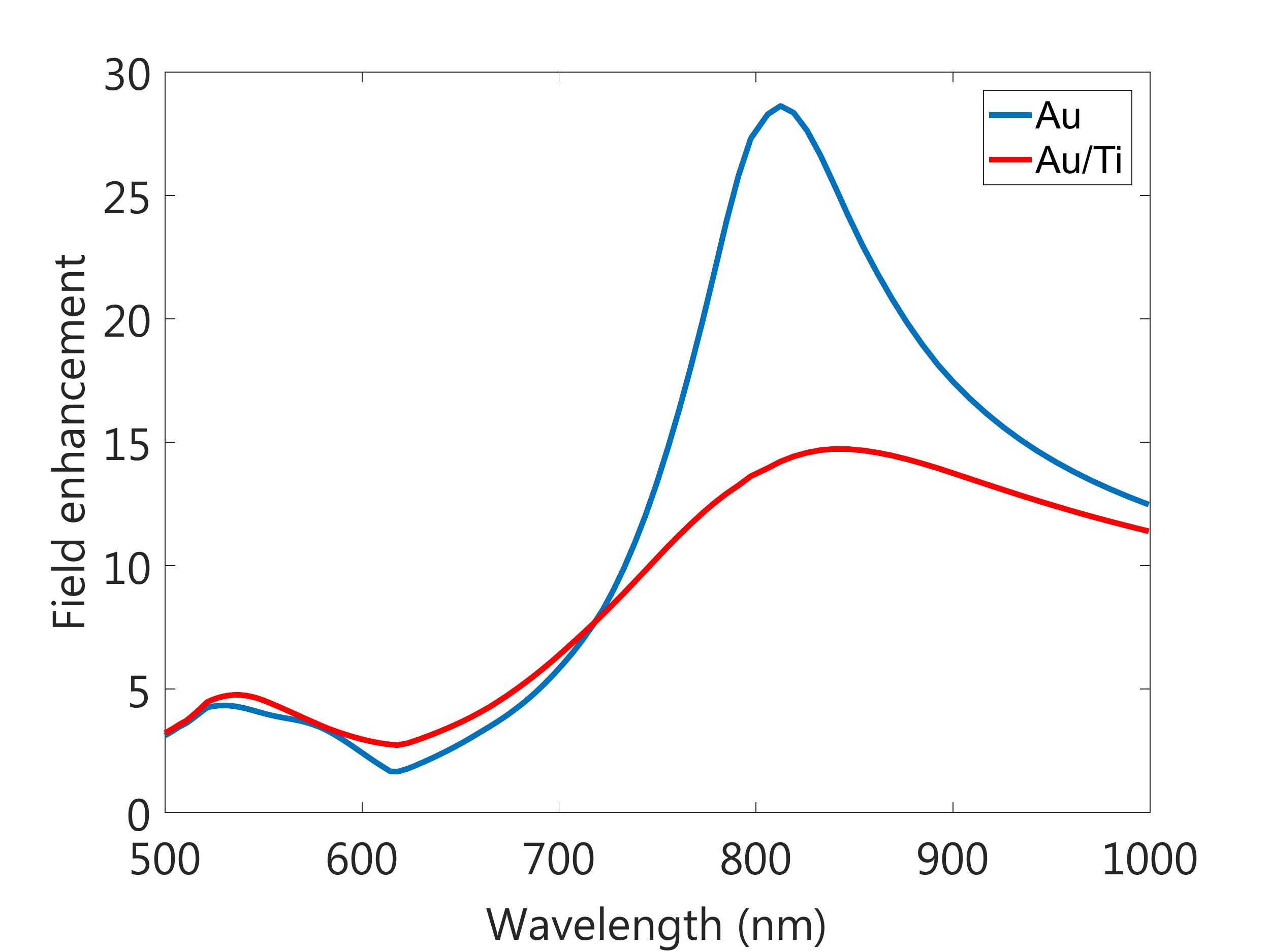} \\
  	(a) & (b) \\
  	\includegraphics[draft=false,width=3.0in]{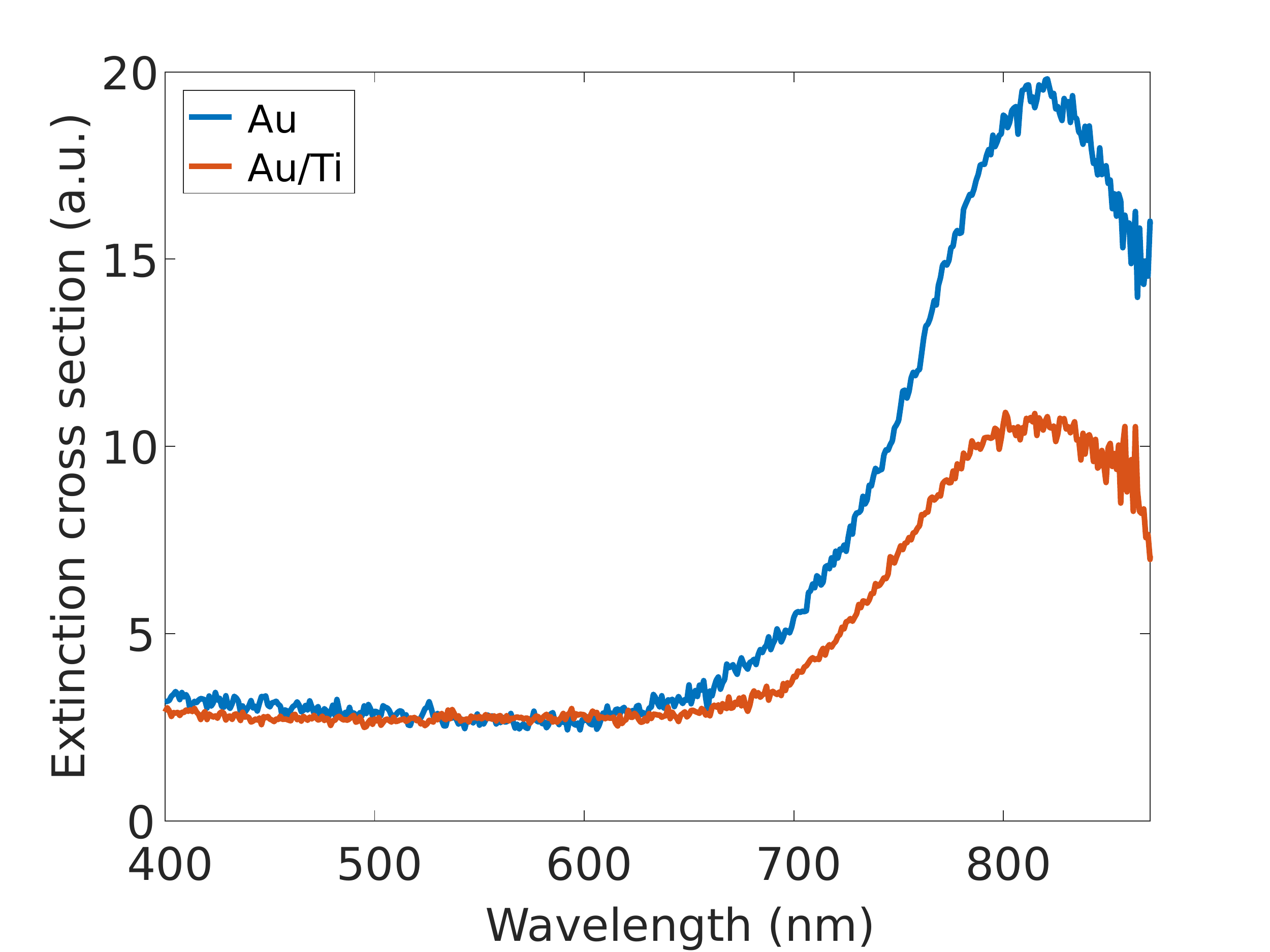} &
  	\includegraphics[draft=false,width=3.0in]{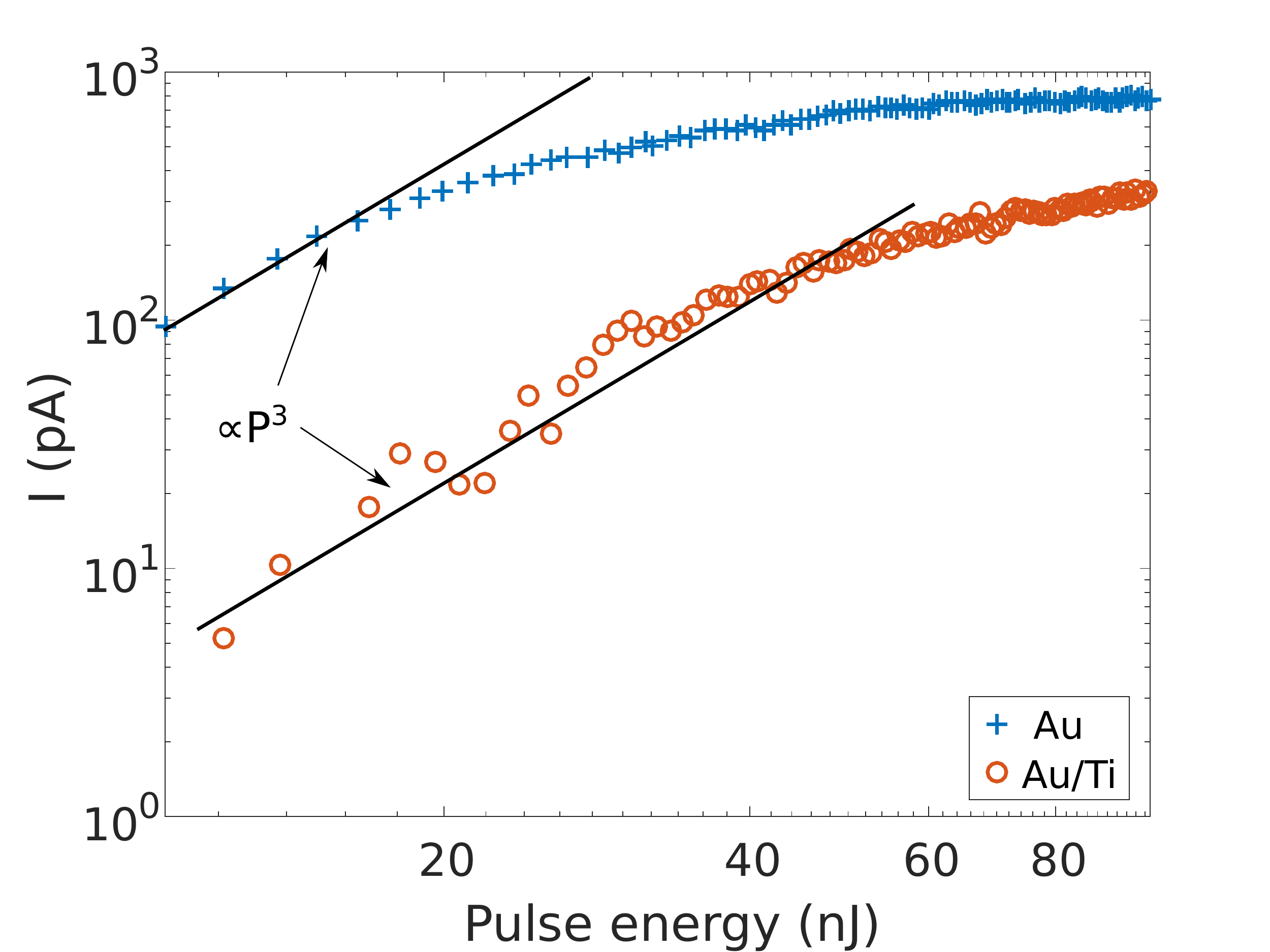} \\
  	\\
  	(c) & (d)
  	\end{array}$
  	\caption{(a) SEM images of Au nanorods prepared without (top) and with (bottom) a 3\,nm Ti adhesion layer on 80\,nm ITO on a Si substrate, and the corresponding cross section of simulated spatial distribution of near-field enhancement (color scale is saturated). (b) Simulated field-enhancement spectra for Au nanorods prepared with (red line) and without (black line) a Ti layer. (c) Optical extinction spectra acquired for a 200\,nm pitch array of Au nanorods prepared with (red line) and without (black line) a 3\,nm Ti adhesion promotion layer. The spectra show a doubling of the extinction for Au nanorods prepared without Ti. (d) Log-log plot of emission current vs pulse energy with an applied anode bias of 1\,kV, for a 400\,nm pitch square array of Au nanorods, prepared without (open black squares) and with (open red circles) a Ti adhesion layer. Both arrays display emission current scaling with the 3rd power of laser pulse-energy (intensity) for low values of pulse-energy, as indicated by the color coordinated lines overlaid on each data set. At 12.1\,nJ the emission current from the Au nanorod array is 26 times that of the Ti/Au nanorod array as indicated on the plot.}
  	\label{nanorodExperimentModel}
  \end{figure}

  Here, effects of a Ti adhesion layer on optical near-field enhancement, and hence on photocathode performance, are inspected.
  Fig.\,\ref{nanorodExperimentModel} shows SEM images, results of near-field simulations, optical extinction spectra and photoemission measurements, for plasmonic Au nanorod arrays with sub-20\,nm critical dimensions, fabricated both with and without a Ti adhesion-promoting layer.
  
  Fig.\,\ref{nanorodExperimentModel}a shows SEM images of high-density Au nanorod arrays, prepared with and without a Ti adhesion layer on an indium-doped tin oxide (ITO)-coated sapphire substrate.
  For details of nanorod array fabrication, the reader is referred to \cite{hobbs2014high}.
  The image highlights the ability to fabricate nanorods with dimensions in the sub-20\,nm regime in the absence of an adhesion-promoting layer such as Ti.
  The simulation results of near-field enhancement in the vicinity of an Au nanorod on an ITO substrate both with, and without, a Ti adhesion-layer are also shown.
  The results clearly show a stronger near-field enhancement for the case of Ti-free Au nanorods.
  An ITO substrate was selected for this work due to its low index and relatively high electrical conductivity.
  The simulated field-enhancement spectra (Fig.\,\ref{nanorodExperimentModel}b) show that the peak field-enhancement for a Ti-free Au nanorod is approximately twice that of an Au nanorod with a Ti layer.
  The power absorption in Au is significantly higher in the absence of a Ti layer.
  Thus, removal of the Ti layer and fabrication of Au nanorods directly on an ITO substrate should lead to improved charge yield and quantum efficiency due to reduced damping of the LSPR.
  Notably, the measured (Fig.\,\ref{nanorodExperimentModel}c) and simulated absorption spectra for the Au nanorod arrays studied in this work display broad bandwidths ($\sim 100$\,nm), which may thus support shorter optical pulses than the 35\,fs (40\,nm bandwidth) pulses used here, and consequently may be of interest for production of sub-10\,fs electron pulses.
  
  Fig.\,\ref{nanorodExperimentModel}c shows optical extinction spectra acquired for 200\,nm pitch arrays of Au nanorods prepared with (red line), and without (black line), a Ti adhesion promotion layer.
  The presence of a 3\,nm Ti layer led to a halving of the optical extinction.
  Assuming that the extinction cross-section is proportional to the optical intensity, or equivalently the square of the optical field, then a 4-fold reduction in the extinction cross-section would have been expected based on the peak-field simulation results shown in Fig.\,\ref{nanorodExperimentModel}b.
  Possible causes for the discrepancy between simulation and experiment include the fact that the Ti deposited in the experiment is likely to contain a significant amount of oxygen, thus reducing damping of the surface plasmon resonance with respect to pure Ti metal, which was used in the simulation.
  A reduction in plasmon damping would result in a greater optical field enhancement and thus an increased optical extinction cross-section.
  Moreover, the simulation results in Fig.\,\ref{nanorodExperimentModel}b are representative of the peak optical field at the nanorod apexes, however, the integrated field over the entire rod would be more representative of the contribution to optical extinction.
  Fig.\,\ref{nanorodExperimentModel}d shows a log-log plot of emission current versus laser pulse-energy for nanorod arrays prepared with, and without Ti.
  We have consistently observed enhanced emission from Au nanorod arrays prepared without an additional metallic adhesion promoter such as Ti.
  The results shown in Fig.\,\ref{nanorodExperimentModel}d demonstrate a 26-fold increase in emission current, at an incident pulse energy of 12.1\,nJ, for a 400\,nm pitch square array of Ti-free Au nanorods, compared to an identical array prepared with a 5\,nm Ti layer.
  The log-log plot of emission current versus pulse-energy shows that both arrays display a slope commensurate with a 3-photon process at 12.1\,nJ.
  The observed 26-fold increase in emission current thus suggests that the optical field is enhanced 1.7 times more by Ti-free Au rods than equivalent TiAu nanorods, which is in good agreement with the predicted doubling of field-enhancement from the simulation results shown in Fig.\,\ref{nanorodExperimentModel}b.
  Fig.\,\ref{nanorodExperimentModel}d also shows that the emission current deviates from 3-photon scaling with increasing pulse-energy.
  The observed deviation from 3-photon scaling with increasing pulse-energy may be attributed to the onset of space-charge-limited current and formation of a virtual cathode, or to a fundamental change in the electron emission mechanism.
  
\begin{figure} \centering
  	$\begin{array}{cc}
  	\includegraphics[draft=false,width=3.0in]{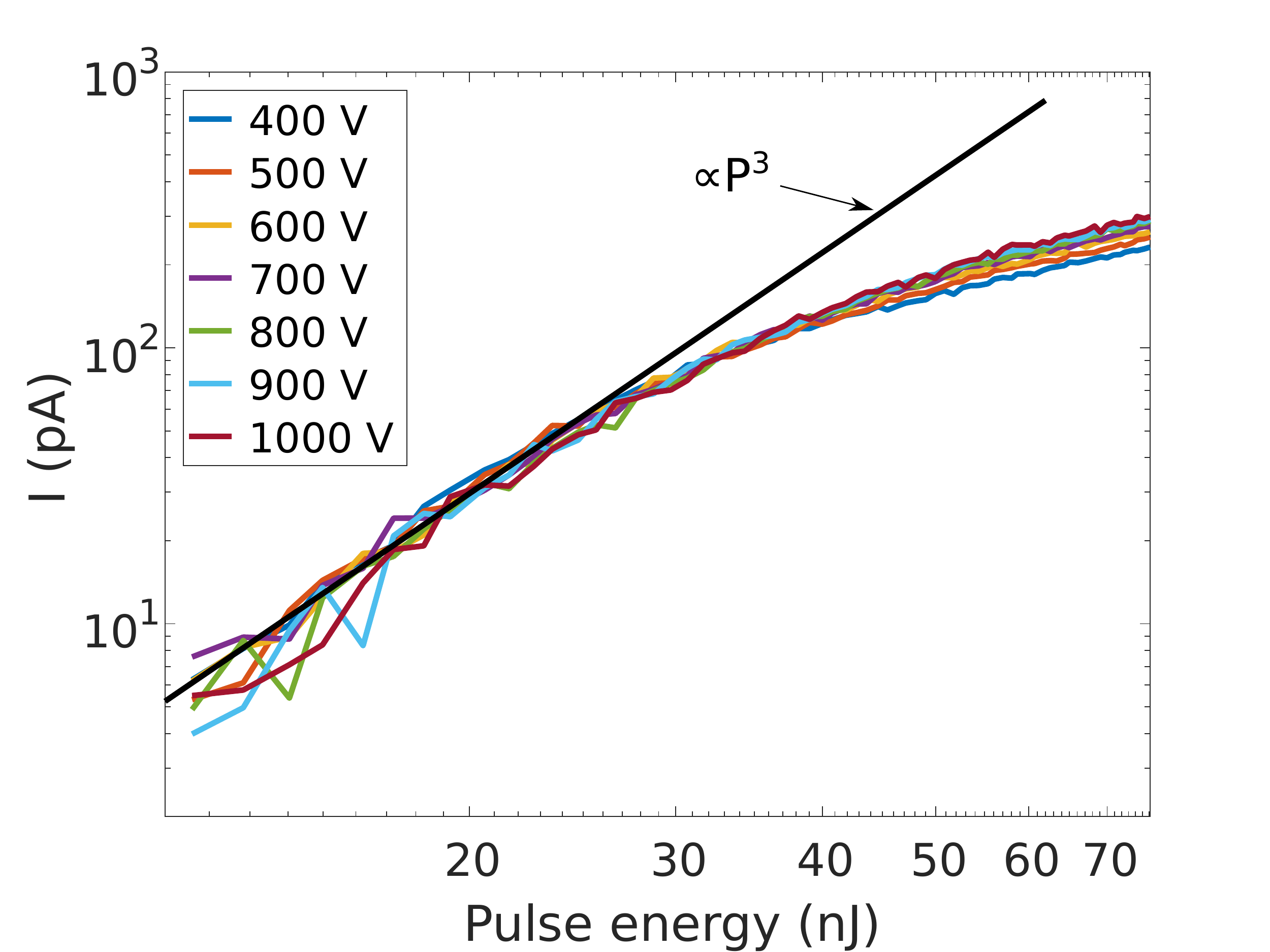} &
  	\includegraphics[draft=false,width=3.0in]{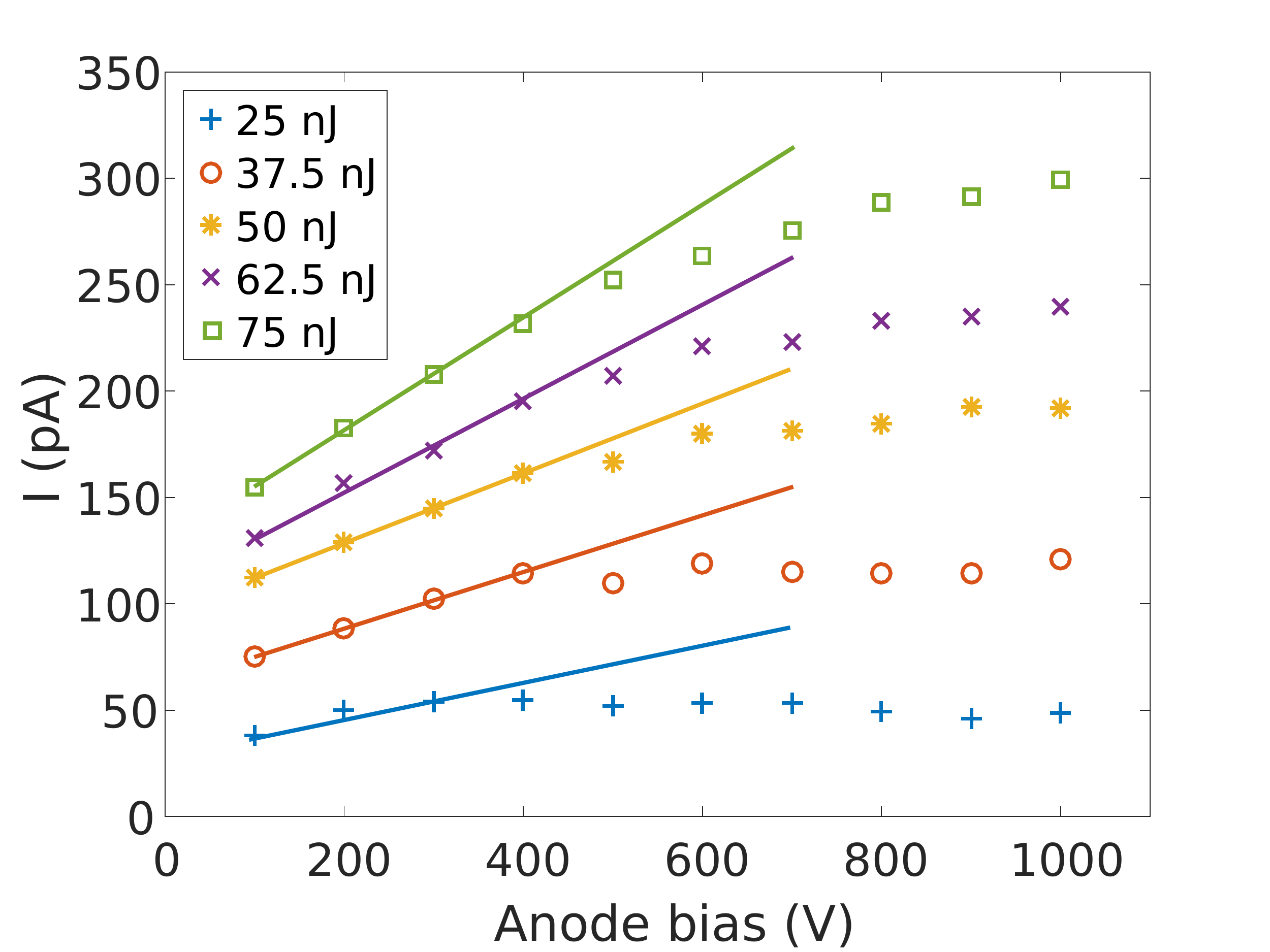} \\
  	(a) & (b)
  	\end{array}$
  	\caption{(a) Log-log plot of emission current vs pulse energy ($P$) for a 1\,{\textmu}m pitch square array of Au nanorods for various applied anode bias values. Emission current scales as $P^3$ up to a pulse energy value of 27\,nJ (dashed line). (b) Plot of emission current (measured at the cathode) vs applied anode bias for the same Au nanorod array used in (a), at various pulse-energy values.}
  	\label{nanorodCurrent}
  \end{figure}

  Au nanorod arrays with various pitches have been studied in this work.
  Space-charge effects, as discussed later in the text, are particularly pronounced for higher density arrays with pitches of 200\,nm or less due to the associated increase in charge density produced.
  Consequently, to first understand fundamental emission characteristics in the absence of global space-charge effects, we have investigated low-density arrays of Au nanorods.
  Fig.\,\ref{nanorodCurrent} displays results of the dependence of emission current on both laser intensity (pulse energy) and on applied anode bias (static DC field).
  
  Fig.\,\ref{nanorodCurrent}a shows a log-log plot of emission current versus incident laser-pulse energy for a 1\,{\textmu}m pitch square array of Au nanorods.
  Emission current is seen to scale with the third power of pulse-energy at low intensity consistent with an electron emission mechanism based on the absorption of 3 photons (total energy 4.53-4.77\,eV).
  The work-function ($\phi$) of Au has been reported as being in the range of 4.7-5.3\,eV \cite{hansen1994work,lide2012crc}.
  Consequently, the 3-photon scaling observed here is indicative of a work function for the Au nanorod arrays of less than 4.8\,eV.
  A 3-photon scaling for a work function larger than 4.8\,eV may also be achieved by photo-field emission, whereby an electron from Au is excited to an intermediate state below the vacuum barrier, from which it then tunnels to vacuum \cite{hommelhoff2006field}.
  The emission current is observed to deviate from the 3-photon scaling behavior at a pulse-energy of 27\,nJ (12.1\,GW/cm$^2$ before plasmonic enhancement) irrespective of applied anode bias in the 400-1000\,V anode bias range.
  Were this deviation due to space-charge effects, a shift in the deviation point would be expected, as its position should depend on anode bias.
  We can therefore conclude that this deviation instead represents a fundamental change in the emission process, which has been previously attributed to a transition from multiphoton emission to direct strong-field emission in studies of single-tip emitters illuminated with ultrafast infrared pulses \cite{bormann2010tip,keathley2012strong,swanwick2014nanostructured}.
  Previously, the Keldysh parameter ($\gamma$) has been used to estimate the magnitude of the optical field required to support strong-field emission, where $\gamma<2$ may describe quasi-static tunneling emission in the strong field regime, and the transition to tunneling behavior usually occurs in the range $1 < \gamma < 2.33$ \cite{keathley2012strong,kruger2011attosecond,swanwick2014nanostructured}.
  In our system, a 27\,nJ pulse-energy is equivalent to an optical field of 0.3\,GV/m.
  A plasmonic field enhancement factor of 40 has been numerically simulated at the Au nanorod surface for a 1\,{\textmu}m pitch square array (Fig.\,\ref{nanorodExperimentModel}a).
  Consequently, an optical field of 12.1\,GV/m is expected at the Au nanorod surface or equivalently a Keldysh parameter of $\gamma = 1.5$.
  Thus, the simulated field-enhancement factor, and the experimentally observed intensity at which deviation occurs from multiphoton emission scaling, supports a transition in the emission mechanism from multiphoton emission to strong-field tunneling at a pulse-energy of 27\,nJ.
  
  Fig.\,\ref{nanorodCurrent}b shows a plot of emission current in terms of anode bias for five different pulse-energy values.
  This Figure shows that the emission current depends on the anode bias for low bias values, while emission current seems to be independent of anode bias for higher values.
  In the low-bias regime emission current scales linearly with anode bias, which is consistent with space-charge limited current ($I_\mathrm{SCL}$) as defined by the single
  sheet model \cite{valfells2002effects}
  \begin{equation}
  I_\mathrm{SCL} = \frac{\epsilon_0 A V f_\mathrm{laser}}{d}.
  \label{ISCL}
  \end{equation}
  Here $\epsilon_0$ is the vacuum permittivity, $A$ is the area of the sheet of emitted charge, $V$ is the bias voltage, $f_\mathrm{laser}$ is the repetition rate of the laser (3\,kHz), and $d$ is the effective anode-cathode spacing ($\sim 1$\,mm).
  Noticeably, the slope of the linear, current versus anode bias plot is observed to increase with pulse-energy.
  The slope of the space-charge limited data in Fig.\,\ref{nanorodCurrent}b should be related directly to the area of the emitted sheet of charge by \eref{ISCL}.
  The area of the sheet of charge should in turn depend on the spatial distribution of laser intensity, which is related directly to the laser pulse-energy for a radially symmetric Gaussian beam.
  We have found that the observed increase in slope corresponds to the expected increase in the effective area of the laser beam with increasing pulse energy.
  For example, for a Gaussian beam ($w_0 = 76.3$\,{\textmu}m, FWHM = 90\,{\textmu}m), the area of the beam with a threshold optical field of 9\,GW/cm$^2$ increases by a factor of 2 as the pulse energy is doubled from 37.5 to 75\,nJ.
  Similarly, the slope of the linear region of the plot in Fig.\,\ref{nanorodCurrent}b increases by a factor of 2 from 0.13 to 0.26\,pA/V, or equivalently from 5.77$\times 10^{-9}$ to 1.15$\times 10^{-8}$\,m$^2$, when the pulse energy is increased from 37.5 to 75\,nJ.
  An optical field of 0.39\,GV/m is equivalent to the peak optical field for a 45\,nJ pulse-energy in our system, suggesting that the onset of space-charge limited current occurs at this incident pulse energy for the $\sim 1$\,MV/m static field employed in this work.
  In the high-bias regime, emission current is no longer space-charge limited and is seen to flatten out.
  For example, emission current appears to behave independently of the applied anode bias for bias values greater than 600\,V at pulse-energy values of 50\,nJ (22.5\,GW/cm$^2$ before plasmonic enhancement) or less.
  However, at 75\,nJ (33.7 GW/cm$^2$ before plasmonic enhancement), the emission current has not yet saturated at an anode bias of 1\,kV, suggesting that the emission current remains influenced by space-charge at this pulse energy.
  
\begin{figure} \centering
  	$\begin{array}{cc}
  	\includegraphics[draft=false,width=3.0in]{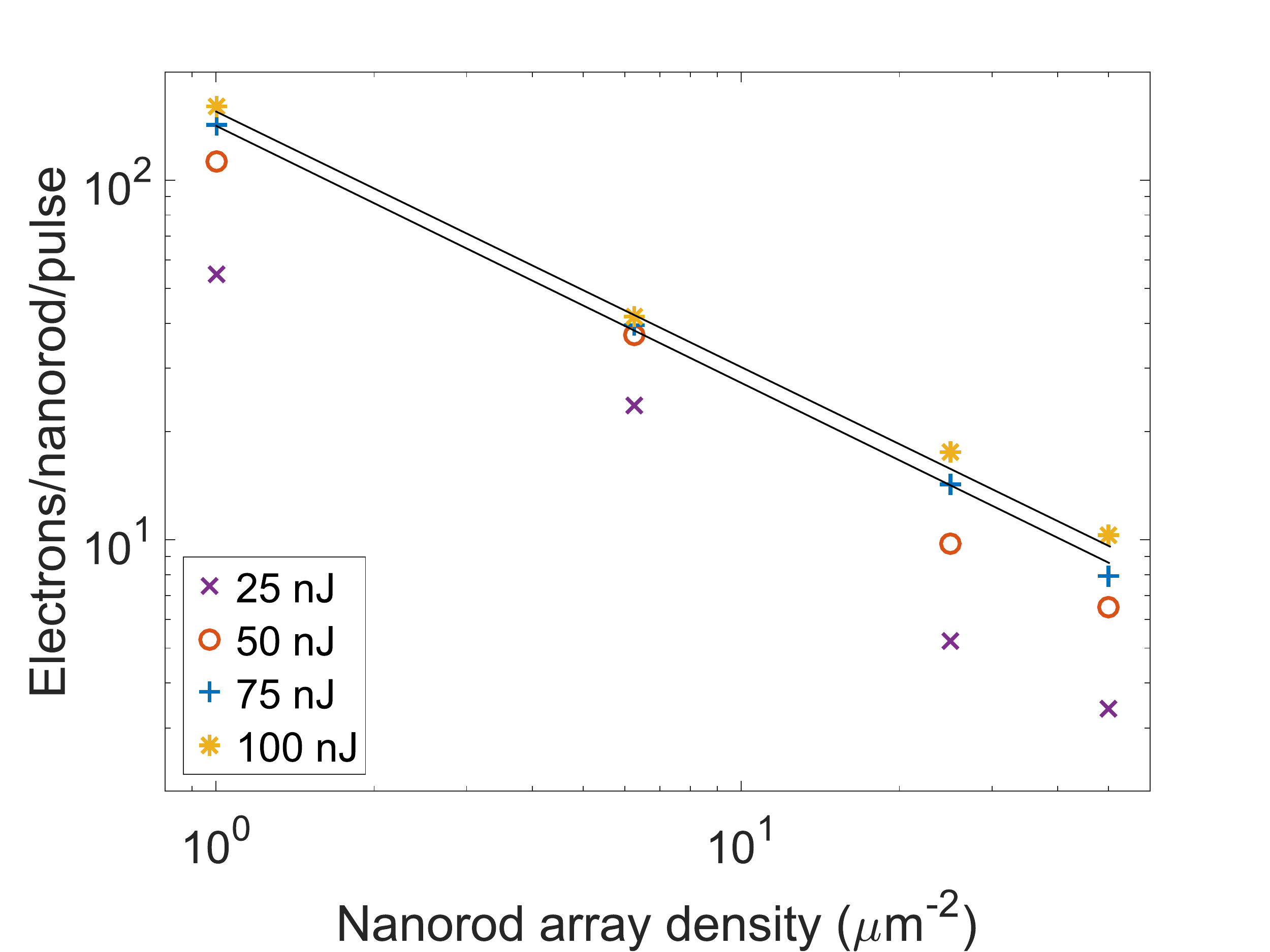} &
  	\includegraphics[draft=false,width=3.0in]{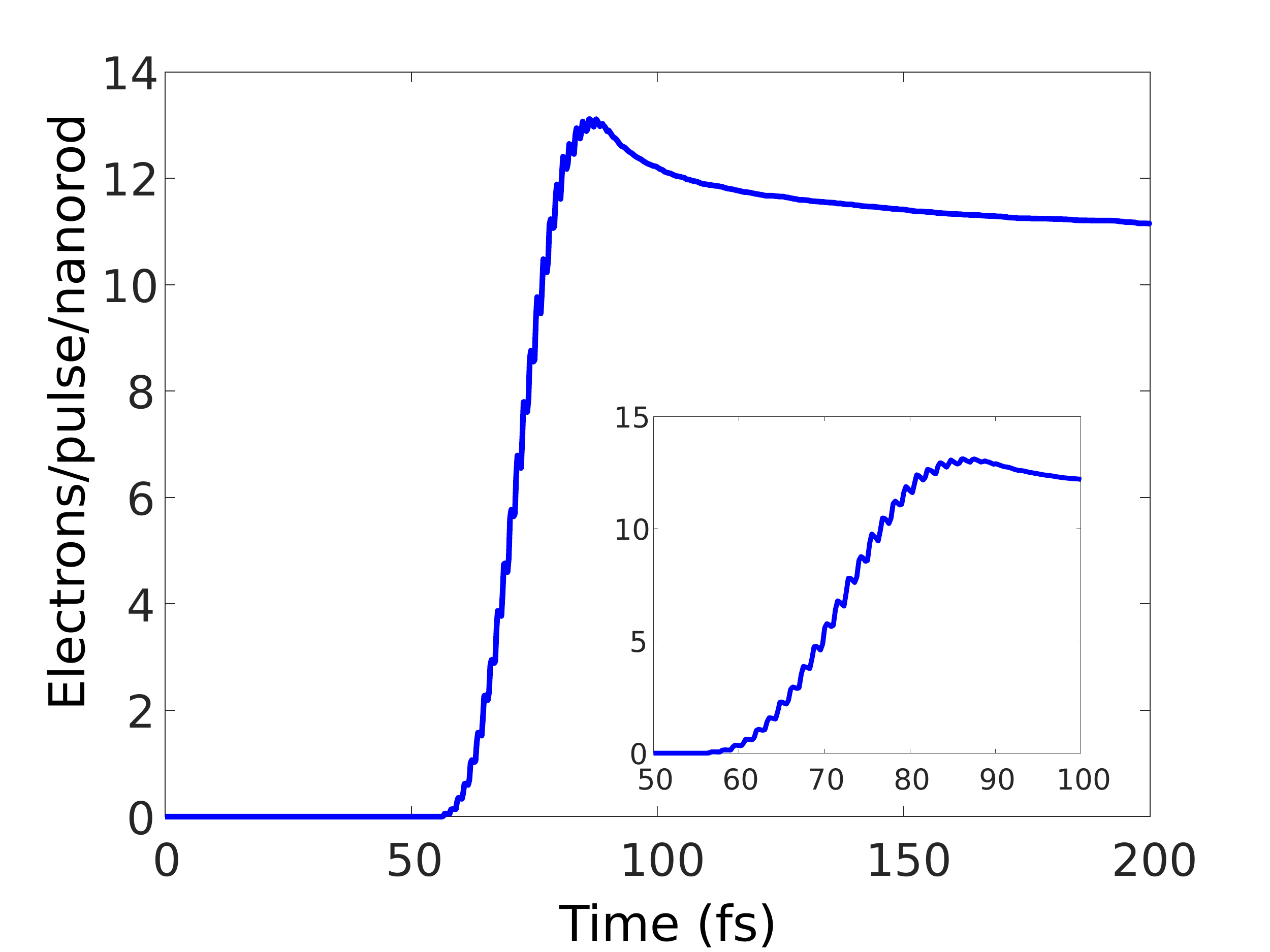} \\
  	(a) & (b)
  	\end{array}$
  	\caption{(a) Log-log plot of emitted charge yield per nanorod per pulse vs nanorod array density for four different pulse energies and a fixed anode bias of 1000\,V. The color-coordinated lines represent power law fits to the data for 75 and 100\,nJ pulse-energy. (b) Simulated temporal evolution of accumulated electron yield per nanorod for
  		a 200\,nm pitch square array (25 nanorods/{\textmu}m$^2$) of Au nanorods illuminated with a 35\,fs, 75\,nJ pulse centered at 70\,fs. The simulation results assume a strong-field tunneling mechanism of emission, which is expected for a 75\,nJ pulse-energy.}
  	\label{nanorodchargeYield}
  \end{figure}

  The effect of nanorod array density on the average charge yield per nanorod, per optical pulse, has also been investigated.
  The charge yield per nanorod is expected to decrease with increasing array density due to (1) an increased effect of space-charge as the electron sources are pushed closer together, and (2) increased charge screening due to near-field coupling within the nanorod array, resulting in a reduction in nanorod field-enhancement.
  We have observed emission of more than 200 electrons per nanorod per 35\,fs optical pulse from a 1\,{\textmu}m pitch, square array, with an incident pulse energy of 120\,nJ, and applied anode bias of 1\,kV.
  Moreover, we have observed a power-law relationship between charge-yield and nanorod array density at high pulse-energy values, as shown in Fig.\,\ref{nanorodchargeYield}a.
  The data does not follow a power-law relation at low incident pulse-energy ($<50$\,nJ) as emission from high-density arrays is space-charge limited even at low laser-intensity, while emission from lower density arrays is not.
  Emission is space-charge limited for all array densities studied when higher incident pulse energy ($>75$\,nJ) is applied.
  At high incident pulse energy, the charge yield per nanorod per optical pulse ($Q$) is related to the array density ($p$) by the relation $Q = p^{-0.7}$.
  An array of emitters producing uniform circular disks of charge would be expected to exhibit a relation $Q = p^{-1}$ due to Coulombic effects in the space-charge limited regime.
  The observed $Q = p^{-0.7}$ relation may be due to an asymmetric charge distribution produced by the nanorods thus leading to asymmetric space-charge effects in the nanorod arrays.
  
  Fig.\,\ref{nanorodchargeYield}b presents particle-in-cell simulation results for electron emission from Au nanorods.
  The analysis is fulfilled for 200\,nm pitch square arrays of Au nanorods illuminated with a 35\,fs, 75\,nJ pulse centered at a time of 70\,fs.
  Results predict a charge-yield of 14 electrons per nanorod for a single pulse, which agrees well with the experimentally obtained charge-yield of 14 electrons.
  The temporal evolution of electron yield predicts that the electrons are emitted mainly within the central 20\,fs of the pulse.
  In the rising edge of the plot, fast oscillations are observed with a period of 1.33\,fs, which corresponds to a half-cycle of 800\,nm light.
  Consequently, these oscillations are due to the periodic emission from each pole of the dipole emitter as the optical field changes in sign with every half-cycle.
  The charge-yield from each nanorod is observed to peak at a time of 85\,fs before declining slightly to a steady yield of $\sim 11$ electrons per nanorod.
  The observed decline in charge-yield is due to the space-charge field causing electrons close to the cathode surface to be pushed back to the substrate.
  This causes a slow recombination of the electrons, which becomes weaker at stronger anode bias voltages.
  
  To investigate the stability of emission current from Au nanorod arrays, we have measured the emission current from an array of Ti-free Au nanorods, identical to that shown in Fig.\,\ref{nanorodExperimentModel}a, for over 5 million pulses.
  Emission current was measured using an incident pulse-energy of 120\,nJ and applied anode bias of 1\,kV.
  The mean emission current was 2.7\,nA, with a standard deviation of 30\,pA.
  SEM analysis of the Au nanorod array following extended emission at 120\,nJ pulse energy, showed that a small region of nanorods exhibited damage in a circular area with $\sim 1$\,{\textmu}m radius.
  The observed damage can be attributed to the Gaussian intensity distribution in the laser beam, which may induce field evaporation and electromigration of Au at the center of the Gaussian spot where the optical field is strongest.
  
\begin{figure} \centering
  	$\begin{array}{cc}
  	\includegraphics[draft=false,width=3.0in]{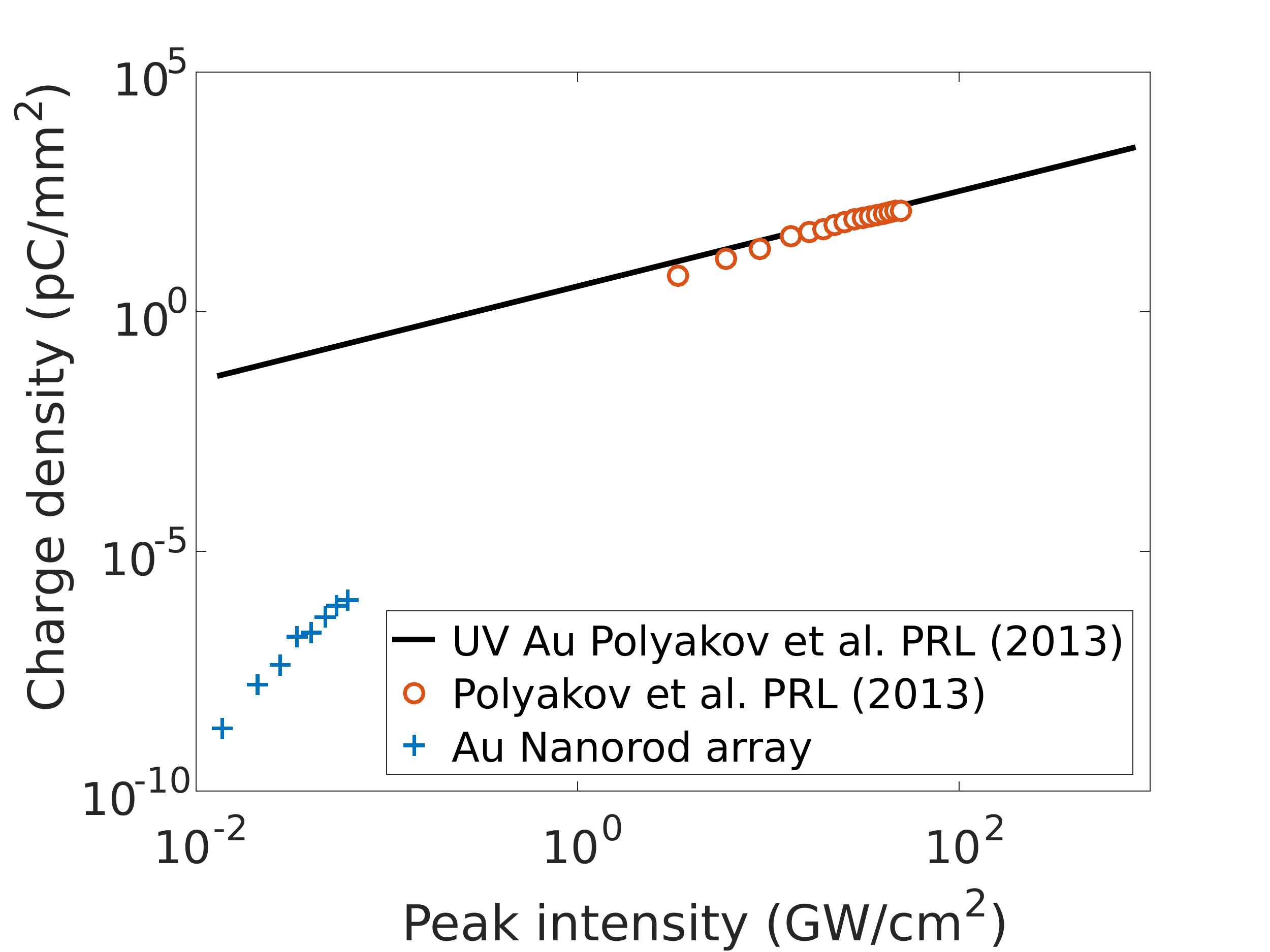} &
  	\includegraphics[draft=false,width=3.0in]{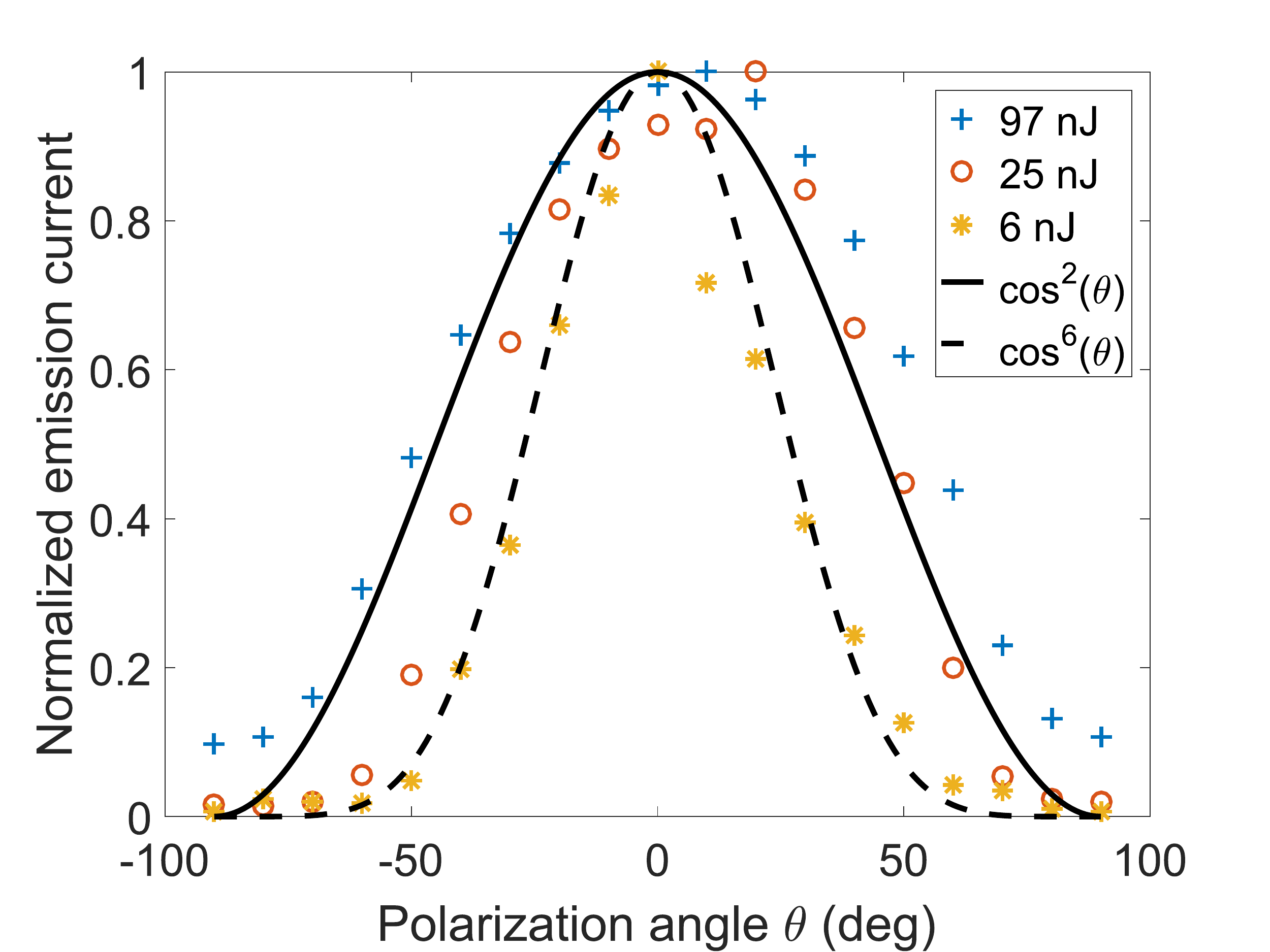} \\
  	(a) & (b)
  	\end{array}$
  	\caption[(a) Plot of emitted charge density vs peak laser intensity, for a 200\,nm pitch Au nanorod array (red circles) using a 1\,kV anode bias and 35\,fs optical pulse at a 3\,kHz repetition rate. The plot also shows emitted charge density for a plasmonic photocathode developed by Polyakov et. al. (black squares), which scales with the 4th power of incident laser intensity, suggesting a 4-photon absorption process. (b) Plot of normalized emission current vs linear polarization angle ($\theta$) at three different values of incident pulse energy. Emission current is highest when the linear polarization is aligned to the long-axis of the nanorod as shown schematically. Emission current follows a $\cos^6(\theta)$ dependence at low pulse-energy, which is equivalent to a 3-photon process. The polarization dependence broadens to a $\cos^2(\theta)$ relation at higher pulse-energy, which is consistent with the transition to strong optical field-emission.]{(a) Plot of emitted charge density vs peak laser intensity, for a 200\,nm pitch Au nanorod array (red circles) using a 1\,kV anode bias and 35\,fs optical pulse at a 3\,kHz repetition rate. The plot also shows emitted charge density for a plasmonic photocathode developed by Polyakov et. al. (black squares) \cite{polyakov2013plasmon}, which scales with the 4th power of incident laser intensity, suggesting a 4-photon absorption process. (b) Plot of normalized emission current vs linear polarization angle ($\theta$) at three different values of incident pulse energy. Emission current is highest when the linear polarization is aligned to the long-axis of the nanorod as shown schematically. Emission current follows a $\cos^6(\theta)$ dependence at low pulse-energy, which is equivalent to a 3-photon process. The polarization dependence broadens to a $\cos^2(\theta)$ relation at higher pulse-energy, which is consistent with the transition to strong optical field-emission.}
  	\label{nanorodPolarization}
  \end{figure}

  Electron emission from Au nanorod arrays was found to depend strongly on the angle of linear polarization of the incident optical pulse, as depicted in Fig.\,\ref{nanorodPolarization}b.
  Emission current was observed to follow a $\cos^6(\theta)$ dependence on polarization angle at low intensity, which is consistent with the 3-photon scaling shown at low intensity in Fig.\,\ref{nanorodCurrent}a.
  Additionally, the polarization dependence transitions to a $\cos^2(\theta)$ dependence at higher intensity, which is in good agreement with the observed transition in the electron emission mechanism from that based on multi-photon absorption to quasi-static tunneling emission.
  
  Recently, Polyakov et al. have observed photoelectron emission from a plasmonic Au photocathode triggered by 60\,fs, linearly polarized pulses, from an 805\,nm Ti:sapphire laser \cite{polyakov2013plasmon}.
  They observed a charge-yield, which scaled as the fourth power of incident laser intensity as represented by the open black squares in Fig.\,\ref{nanorodPolarization}a.
  Polyakov et al. hypothesized that this scaling may continue to laser intensities as high as 50\,GW/cm$^2$, at which point their photocathode, which is triggered by an 800\,nm laser, may outperform a planar Au photocathode operating under UV illumination (black line Fig.\,\ref{nanorodPolarization}a).
  In this work, we have seen that a transition from multi-photon emission scaling to strong-field tunneling can occur at a laser intensity of 12.1\,GW/cm$^2$ (27\,nJ pulse energy), while 3-photon scaling has been measured for intensities as low as $\sim 1$\,GW/cm$^2$.
  Consequently, we suggest that plasmonic photocathodes can generate enhanced optical fields sufficient to support strong-field tunneling emission at laser intensities $\sim 10$\,GW/cm$^2$, and thus that such photocathodes do not display electron emission characteristic of multi-photon absorption at laser intensities for which it was previously predicted.
  
  The maximum charge density emitted from a Au nanorod array photocathode in this work was observed for 200\,nm pitch arrays of Ti-free, Au nanorods.
  Fig.\,\ref{nanorodPolarization}a shows a plot of charge density emitted per pulse versus peak laser intensity for such an array of Au nanorods (open red circles).
  Charge-yield from high density, 200\,nm pitch, Au nanorod arrays is still limited by space-charge effects, even at the highest applied anode bias values used in this work.
  The observed space-charge suppression of emission current may be alleviated at increased static field.
  For example, integration of the photocathode within an RF gun capable of producing fields of $>10\,$MV/m will allow demonstration of an Au nanorod array photocathode, excited by 800\,nm light, with a QE that may surpass that of the equivalent UV photoemission process.
  
  In this work, a QE of $1.2\times10^{-5}$ has been measured for 200\,nm pitch Au nanorods illuminated with 800\,nm light at an intensity of 10\,GW/cm$^2$ from the data in Fig.\,\ref{nanorodPolarization}.
  The QE for Au illuminated with UV light (266\,nm) has been reported as $4.5\times10^{-5}$ \cite{srinivasan1991photoemission}.
  The transmitted laser intensity was measured as $\sim 90$\% for a 200\,nm pitch array of Au nanorods at the laser focus for a laser intensity of 34\,GW/cm$^2$.
  Thus, an internal QE can be calculated as $1.2\times10^{-4}$ considering $\sim 10$\% of the incident photons as scattering from the nanorod array to produce photoelectrons.
  When the 10\% power conversion efficiency of 800\,nm wavelength light to 266\,nm wavelength light by third harmonic generation, and the factor of 3 difference in energy between the IR and UV light are taken into consideration, plasmonic Au nanorod arrays triggered by 800\,nm wavelength light can be considered as $\sim 100$ times more efficient than UV-triggered bulk Au photocathodes.
  Furthermore, as has been discussed, application of an increased static bias to lift the space-charge limit would further improve the QE for this system.
  
  \section{Electron Source Characterization}
  
  In addition to current yield or the corresponding quantum efficiency discussed in the previous sections, the spatio-temporal properties of the electron bunch emitted from a source play significant roles in the application of interest, e.g. light source operation.
  It is widely accepted that precise control of the x-ray pulse characteristics, including spectral coverage and temporal and spatial beam profiles are of utmost importance for various applications.
  These parameters are directly influenced by the properties of the electron bunch generating the x-ray pulses.
  Therefore, the accurate characterization of the electron beam quality is essential for to improve of the underlying electron source technology.
  Due to its importance for the performance of accelerators a multitude of techniques have been developed during the last years to measure the transverse \cite{ye2015velocity,bainbridge2014velocity} as well as the longitudinal energy spread.
  A detailed summary and the state of the art can be found elsewhere \cite[and references therein]{lee2015review}.
  Moreover, high quality electron bunches are instrumental in experiments where materials are studied \emph{via} electron diffraction \cite{ihee2001direct,siwick2003atomic,gulde2014ultrafast}.
  In electron diffractive imaging, accurate measurement of the bunch profile is mandatory to obtain reliable imaging results.
  In what follows, two newly developed techniques for characterization of the emitted electron bunch using velocity map imaging \cite{ye2017velocity}, and for exploring emission mechanisms using a novel spatial mapping method \cite{hobbs2017mapping} are presented.
  The two methods provide a complete framework for assessment of different electron source technologies.
  
  \subsection{Velocity Map Imaging Spectrometer}
  
  \subsubsection{Introduction}
  
  The key measure in electron-beam quality is electron-beam emittance, i.e. the transverse phase-space distribution of the generated electron bunches.
  To quantify electron beam emittance as a function of photocathode composition and emission mechanisms, we demonstrate a velocity-map-imaging (VMI) spectrometer that allows us to directly access the transverse velocity distribution (the term \emph{velocity} refers to the vector quantity) of photoemitted electrons, enabling the measurement of root-mean-square (RMS) normalized emittance from various cathodes.
  Usually, emission mechanisms are classified as thermionic emission, photoemission, or tunneling emission under extraordinarily high electric fields.
  More recently, nanostructured and plasmonic photocathodes used with multiphoton or strong-field optical emission have been used as improved electron
  sources \cite{kruger2011attosecond,herink2012field,mustonen2011five,putnam2017optical,tsujino2016measurement,kartner2016axsis,polyakov2013plasmon,li2013surface}.
  Both, the experimental characterization and the theoretical description of the electron emittance from such cathodes is highly important, motivating the direct VMI measurements developed here.
  
  As a first proof-of-principle example, we report on quantitative measurements of multiphoton emission from a 400\,nm thick Au thin film at room temperature, which was excited with 45\,fs laser pulses centered at 800\,nm.
  Furthermore, these measurements allowed us to benchmark the performance of this new experimental setup.
  Quantum-yield-dependent measurements were performed by recording the events of electrons impinging on the detector when varying the average laser power and the polarization angle, respectively.
  These experimental results confirm that four-photon emission occurs from the planar Au surface.
  In our experiments, the 2D transverse velocity/momentum distribution of photoemitted electrons was directly imaged onto the detector.
  An experimental 3D energy distribution was reconstructed from the measured 2D VMI data using a mathematical algorithm and compared to the theoretically derived 3D-space energy distribution from the Berglund-Spicer photoemission model \cite{berglund1964photoemissionT,berglund1964photoemissionE,krolikowski1969photoemission,krolikowski1970photoemission}.
  The very good agreement of our experimental results with the theoretical model demonstrates the applicability of VMI for the characterization of the RMS normalized emittance of photoelectron emitters.
  
  \subsubsection{Experimental Setup}
  
  The velocity-mapping technique maps the velocity coordinates of particles onto a 2D detector without, to first order, the influence of the particles' spatial coordinates.
  To achieve this, a configuration of electrostatic lenses, in the simplest case using three parallel electrodes, is employed to spatially tailor the electric fields \cite{eppink1997velocity,chichinin2009imaging}.
  The setup can also be used to image and magnify the spatial coordinates while suppressing the effect of velocity coordinates, which is then referred to as spatial-map imaging (SMI) \cite{eppink1997velocity,stei2013high}.
  The spectrometer demonstrated here aims to characterize the electron-RMS-normalized emittance \emph{via} characterizing the average spread of electron coordinates in position-and-momentum phase space.
  
\begin{figure}
  	\centering
  	\includegraphics[draft=false,width=4.5in]{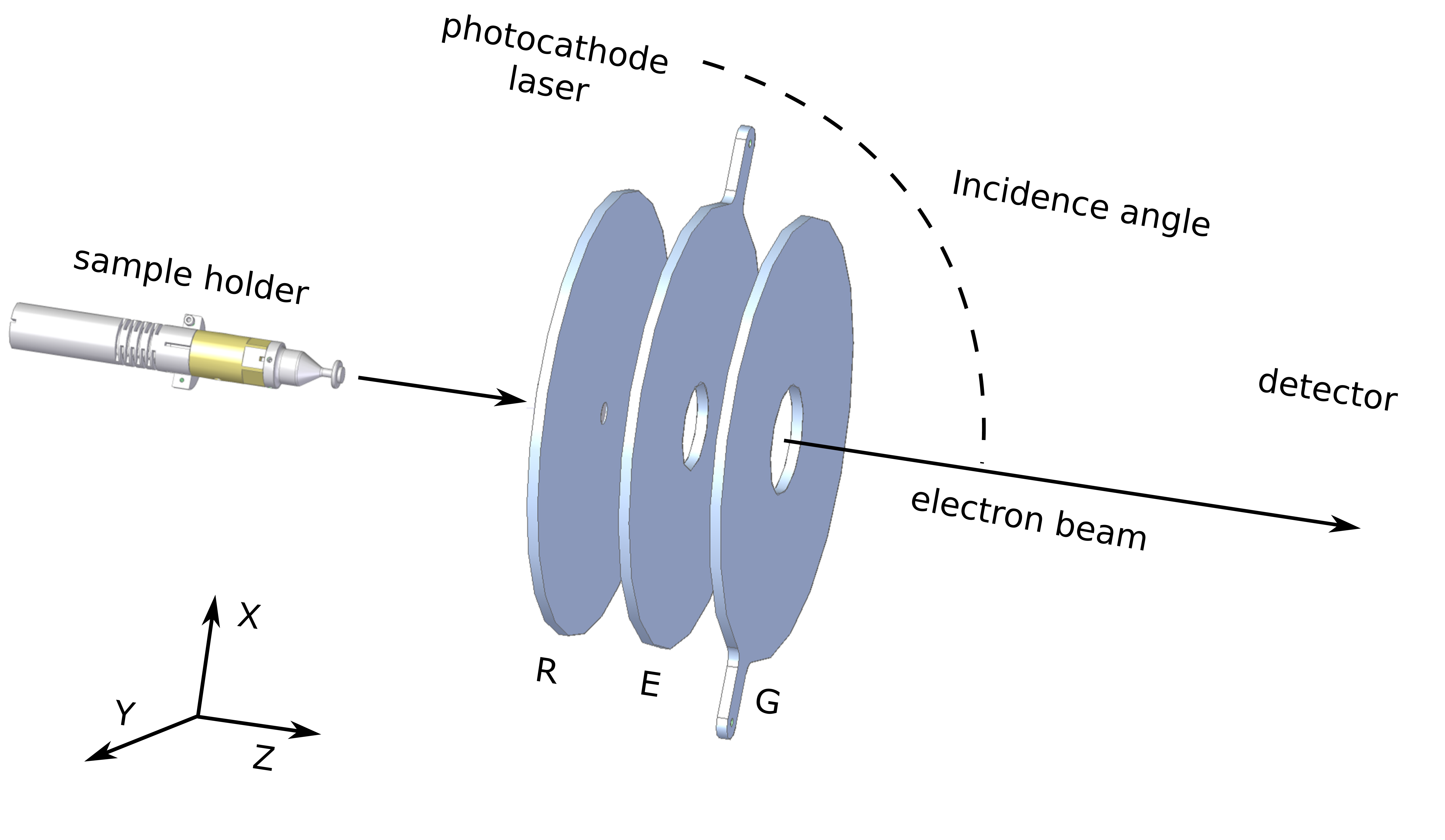}
  	\caption{Schematic of the velocity-map-imaging (VMI) spectrometer consisting of three parallel electrodes, R: repeller, E: extractor, G: ground. The sample is mounted on the top of the holder, which can be retracted from this main chamber into a load-lock chamber.}
  	\label{VMISetup}
  \end{figure}

  The schematic of the spectrometer is shown in Fig.\,\ref{VMISetup}.
  The sample is mounted on the top of the sample holder, which can be retracted into a load-lock chamber.
  The load lock is designed for exchanging the sample without disturbing the ultra-high vacuum (UHV) condition in the imaging system.
  When performing the electrostatic imaging experiments, the sample holder is transferred into the main chamber and brought in contact with the repeller plate to make sure they are at the same electric potential.
  The main chamber, maintained at $10^{-9}$\,mbar, contains a stack of three cylindrically symmetric plates, labeled repeller (R), extractor (E) and ground (G) electrodes in Fig.\,\ref{VMISetup}.
  They are arranged in parallel, separated by 15\,mm, and, with applied potentials, served as the electrostatic lens.
  This is followed by a $\sim 0.5$\,m drift tube, which ends with a detector assembly consisting of a double microchannel plate (MCP, Chevron configuration), a phosphor screen (P46) with a diameter of 40\,mm, and a CMOS camera (Optronis CL600$\times$2) for recording images of the electron distributions.
  The full configuration is shielded against stray fields by a {\textmu}-metal tube.
  A 800-nm 45-fs Ti:Sapphire laser amplifier with a 3\,kHz repetition rate was used to illuminate the sample at a glancing incidence angle of $\sim 84^{\circ}$, with a laser focal intensity spot size of $\sim 17\times160$\,{\textmu}m$^2$ RMS on the sample.
  In our experiments, electron-distribution images are read out at a repetition rate of 1\,kHz, limited by the camera-acquisition frame rate.
  The average number of electrons emitted per pulse is on the order of one or less, which excludes space charge effects that were reported before \cite{petite1992origin}.
  
  To calibrate and optimize the spectrometer field configuration for both SMI and VMI, a fixed potential of -6\,kV was applied to both the repeller plate and the sample holder while the ground plate was grounded.
  While scanning the extractor voltage from -5.8\,kV to -4.3\,kV, we observed the focusing of the electron bunch depending on the extractor voltage \cite{muller2015electron}.
  This behavior is explored based on the RMS of the electron bunch size in the $x$ and $y$ directions on the detector.
  The SIMION \cite{manura2011simion} software is used to simulate the electric field configuration and to calculate the electron trajectories from a 2D Gaussian source with $\sigma_x=140$\,{\textmu}m and $\sigma_y=15$\,{\textmu}m, yielding an RMS behavior curve that fits the experimental results.
  SMI is obtained at the minimum RMS size, i.e. at an extractor voltage of -5560\,V, corresponding to a magnification factor of 7.5.
  From the measured SMI data, the RMS size is analyzed to be $\sigma_x=158$\,{\textmu}m and $\sigma_y=20$\,{\textmu}m, which is in good agreement with the simulated electron-bunch size and the laser-focal-spot size.
  Importantly, in this experiment, hot spots due to sample surface roughness can conveniently be observed and located in SMI mode.
  Therefore, we are able to find suitably flat areas without hot spots within the laser-spot size, which can then be used for velocity mapping.
  The extractor voltage for VMI conditions is found at -4790\,V according to the SIMION simulations and the calibration factor of velocity-per-pixel is 8014~m/(s$\cdot$pixel) on the detector.
  In order to minimize field distortions, the sample front surface should be placed in the same plane as the repeller front surface.
  However, samples of different thickness lead to a position offset with reference to the repeller front, which strongly influences the field configuration.
  Therefore, the extractor voltage for operating in SMI and VMI mode are optimized by voltage adjustments of $[50,-50]$\,V and $[400,-200]$\,V, respectively, to correct for a position offset of $[-0.5,0.5]$\,mm.
  In this case, the necessary re-adjustment of the potential right after exchanging a sample is quickly accomplished.
  
\begin{figure}
  	\centering
  	$\begin{array}{cc}
  	\includegraphics[draft=false,width=3.0in]{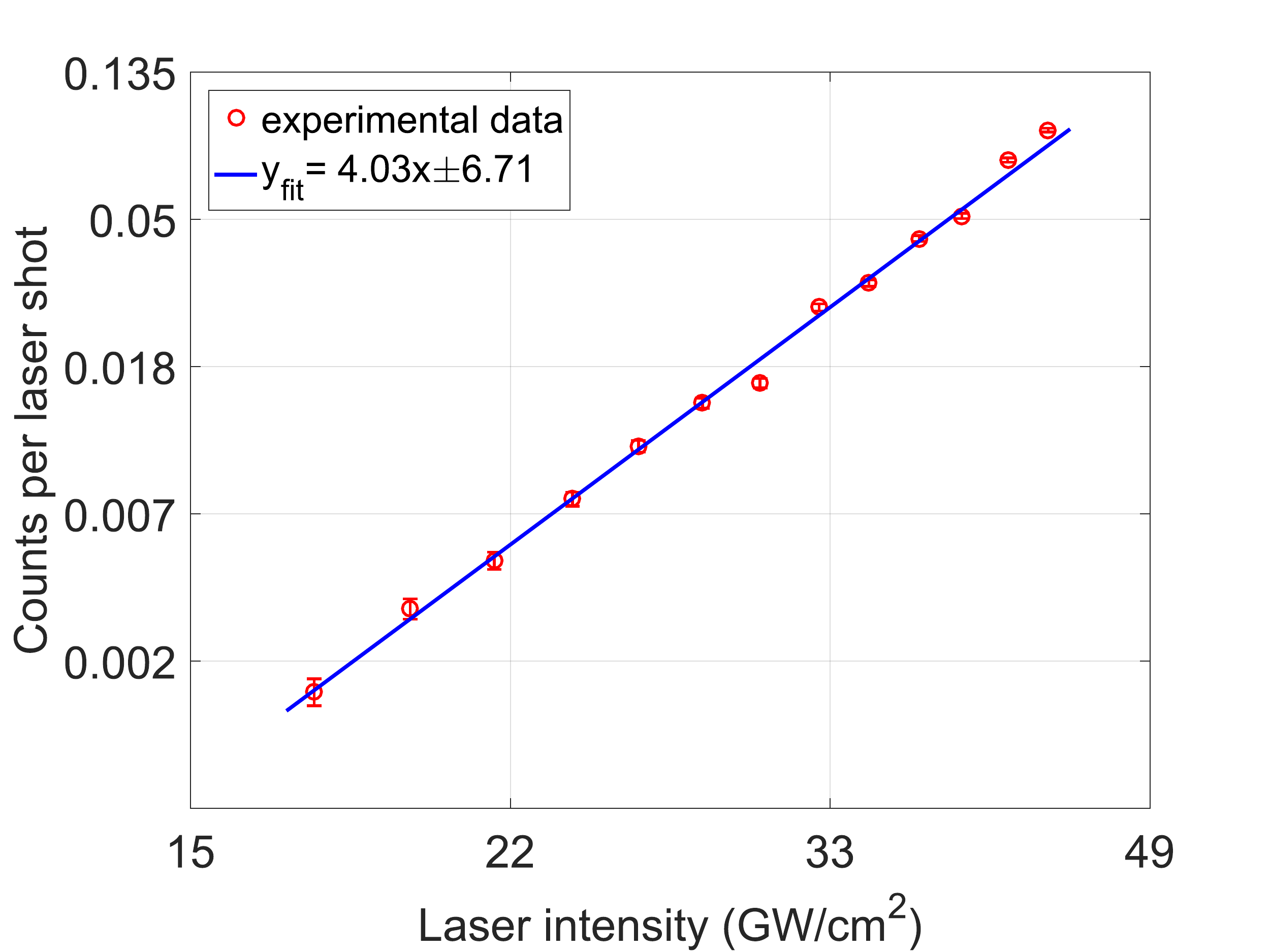} &
  	\includegraphics[draft=false,width=3.0in]{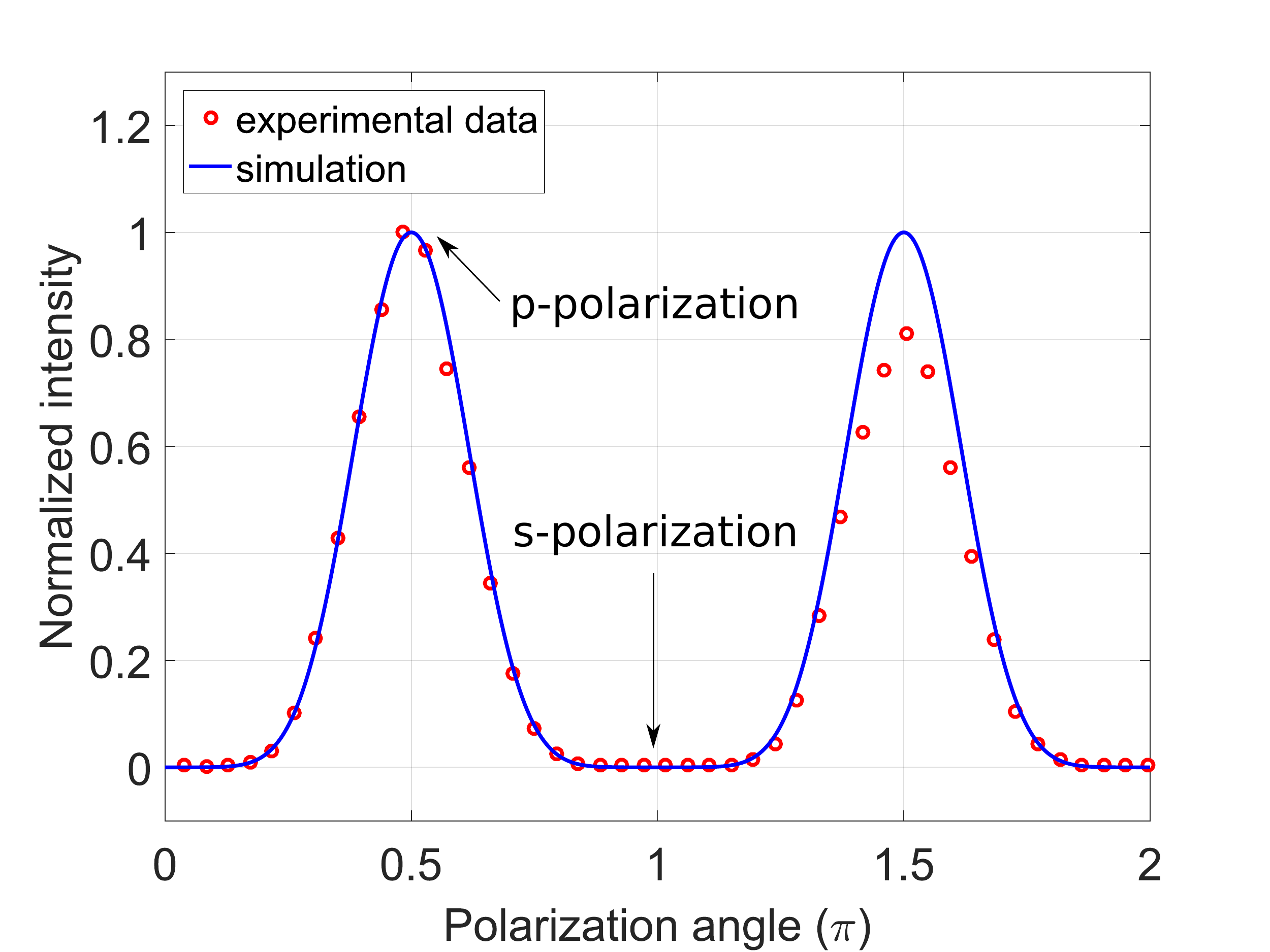} \\
  	(a) & (b)
  	\end{array}$
  	\caption{Counts of the photoemitted electrons as function of (a) average laser power and (b) laser polarization angle. The experimental data for polarization angles $>\pi$ are of reduced quality due to experimental instabilities, e.g. in drifting laser pointing.}
  	\label{laserDependence}
  \end{figure}

  \textbf{Experimental Results:} Fig.\,\ref{laserDependence}a shows the photoemitted electron yield per laser shot as a function of incident laser peak intensity on a logarithmic scale.
  The error bars show the statistical errors of the photoemitted electron counts.
  The blue line reflects the results of a linear regression analysis that yielded a slope of $c_x\approx4.03$, with a coefficient of determination $R^2\approx0.997$.
  
  The Fowler-Dubridge model for the $n$-th order photoelectric current can be written in a generalized form as \cite{bechtel1975four}
  \begin{equation}
  J \propto A(1-R)^n \, I^n \, F\left(\frac{nh\nu-e\phi}{kT}\right)
  \label{FDmodel}
  \end{equation}
  where $n$ is the number of photons, $h$ is the Planck constant, $A$ is the Richardson coefficient, $R$ the reflection coefficient from the metal surface, $I$ the incident light intensity, $\phi$ the metal work function, and $F(x)=\int_0^\infty\ln(1+e^{-(y+x)})dy$ the Fowler function.
  
  The experimental data in Fig.\,\ref{laserDependence}a follows a power law with a slope of $\sim 4$, in agreement with a 4-photon emission process according to the nonlinear photoelectric effect, which indicates that simultaneous absorption of 4 photons (photon energy 1.55\,eV at 800\,nm) has to take place to overcome the metal work function $W$ \cite{damascelli1996multiphoton}, which is reported as 5.31-5.47\,eV for Au \cite{lide2012crc}.
  As shown in Fig.\,\ref{laserDependence}b, varying the laser polarization angle, the photoemitted electron intensity reaches a maximum when the laser is p-polarized (electric field normal to the sample surface), and appears minimum when it is s-polarized.
  For multiphoton emission at a certain incident light intensity, the electron yield mostly depends on the bulk absorption coefficient, expressed as term $(1-R)^n$ in the Fowler-Dubridge model \cite{damascelli1996multiphoton}.
  $R$ is calculated by Fresnel equations with $n_1=1$ and $n_2=0.189+i4.71$ \cite{polyanskiy2014refractive} at an incidence angle of 84$^{\circ}$.
  The plotted $(1-R)^4$ curve fits very well with the data, which proves again the 4-th order multiphoton process.
  
\begin{figure}
  	\centering
  	$\begin{array}{cc}
  	\includegraphics[draft=false,width=3.0in]{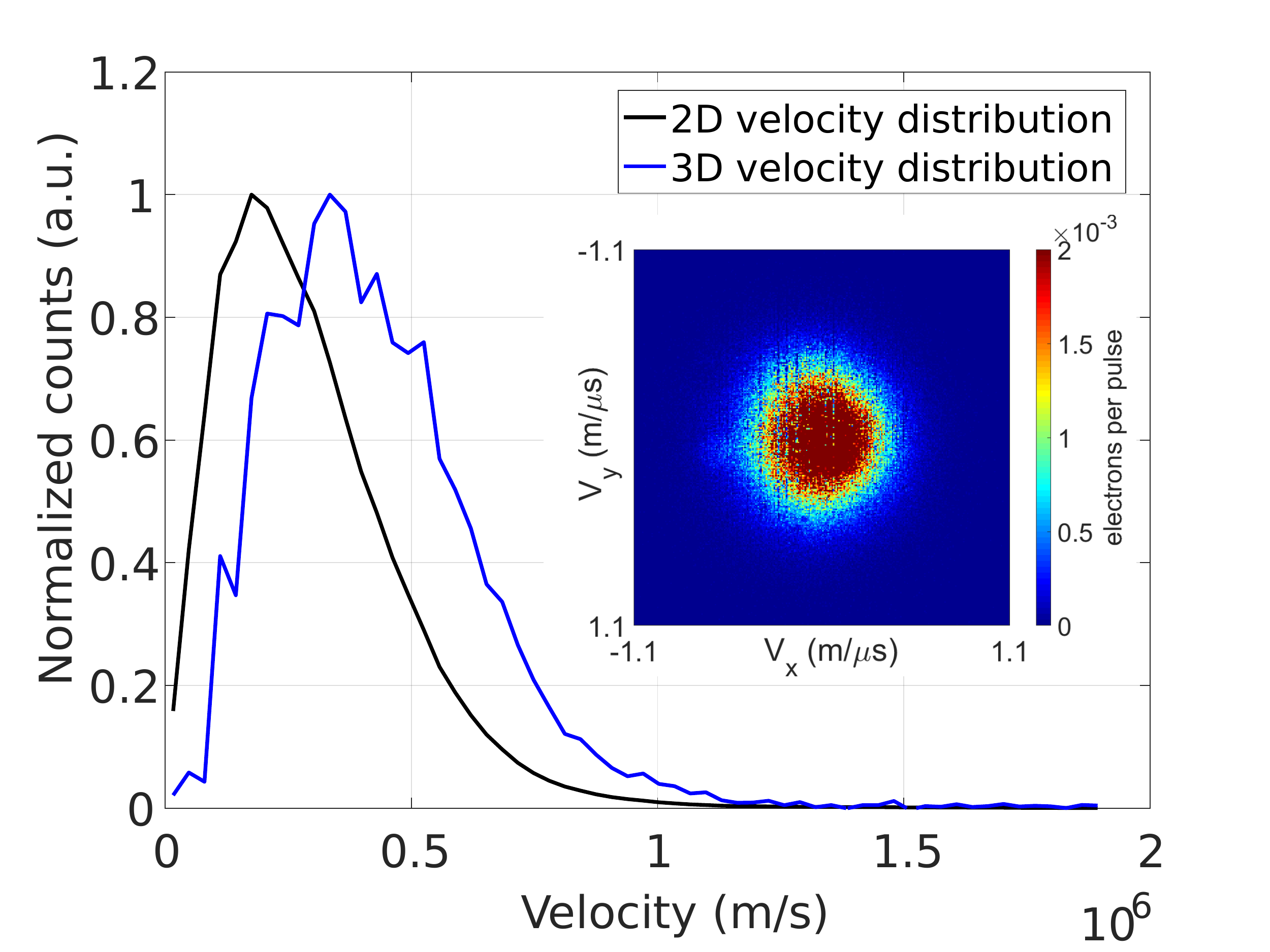} &
  	\includegraphics[draft=false,width=3.0in]{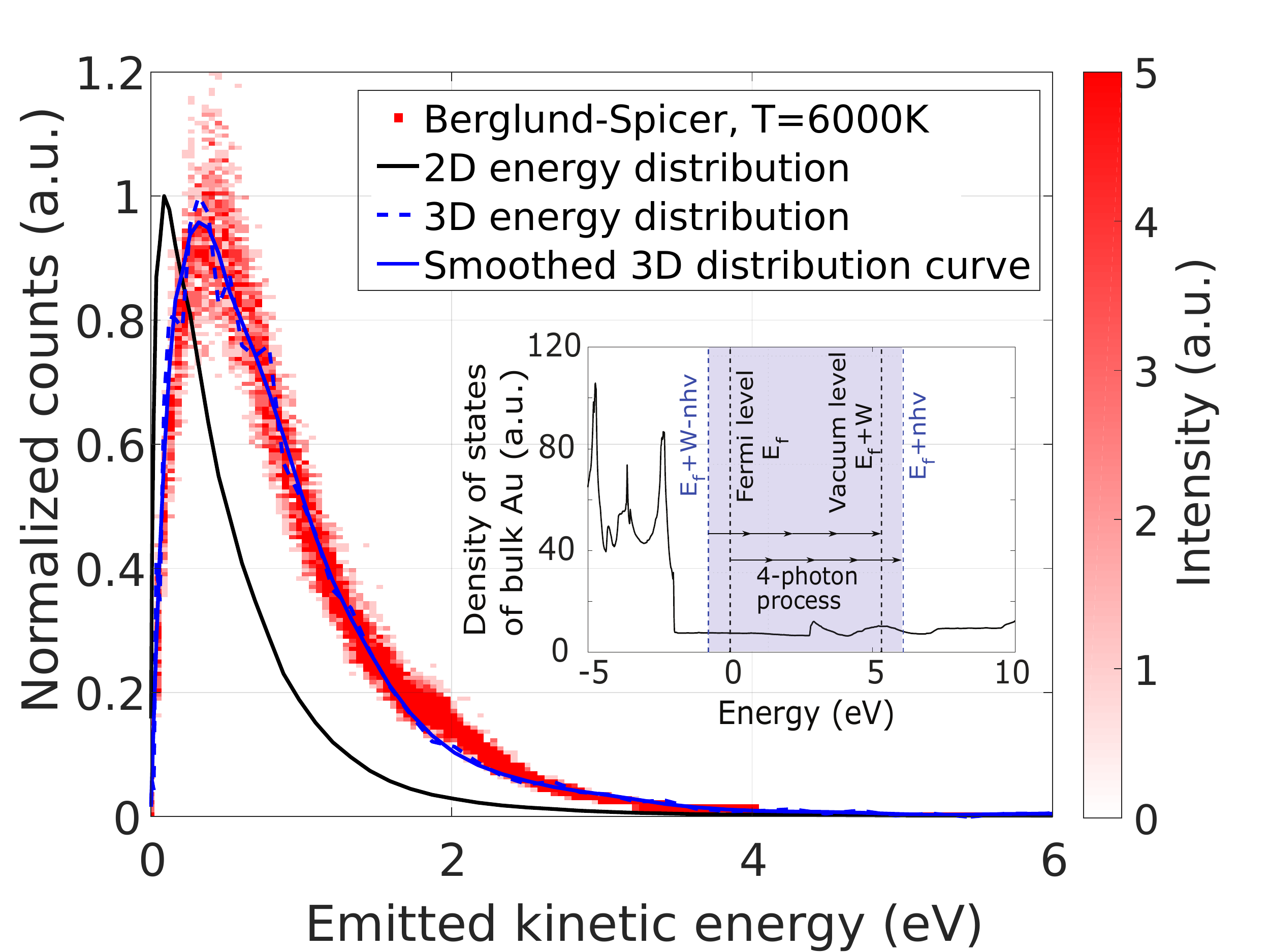} \\
  	(a) & (b)
  	\end{array}$
  	\caption{(a) Projected 2D (black curve) and reconstructed 3D (blue curve) radial velocity distribution of the measured velocity-map image that is shown in the inset. (b) Reconstructed kinetic-energy distribution and its simulation using the Berglund-Spicer model assuming an electron temperature of 6000~K. The color bar of the 2D histogram represents the probabilities of photoelectron kinetic energies due to the photon-energy spectrum of the laser. The inset shows the density of states calculated for bulk Au, which is used in the Berglund-Spicer model simulation. The blue area depicts the four-photon-ionization range.}
  	\label{electronDistribution}
  \end{figure}

  A velocity-map image from a planar Au surface is shown in the inset of Fig.\,\ref{electronDistribution}a.
  The image was integrated over $6\times10^4$ laser shots with an energy of $\sim 50$\,nJ, corresponding to a peak intensity of $4\times10^{10}\,\text{W/cm}^2$ on the cathode.
  Generally, in laser-induced multiphoton emission the emitted electron velocity vectors exhibit cylindrical symmetry along the direction normal to the sample surface.
  Therefore, the center of mass (COM) of the image is set as coordinate origin.
  The corresponding angle-integrated radial velocity distribution of the projected electrons is plotted in Fig.\,\ref{electronDistribution}a as black line.
  To allow for comparison with the theoretical model, the 3D velocity/energy distribution is required.
  Introducing a novel mathematical method similar to the Onion Peeling algorithm \cite{dasch1992one}, we are able to reconstruct the momentum/energy distribution when the angular distribution of emitted electrons is known.
  Fortunately, for multiphoton emission, the intensity of photoemitted electrons at various angles $\theta$ can be derived from the Berglund-Spicer model \cite{berglund1964photoemissionE} as
  \begin{equation}
  I(\theta) \propto \aleph^2 \cos\theta\cdot\frac{1}{1+\alpha \, l(E)} \cdot \frac{1}{\sqrt{1-\aleph^2\cdot\sin^2\theta}}
  \label{BSmodel}
  \end{equation}
  where $\alpha$ is the optical absorption coefficient, $l(E)$ is the electron-electron scattering length for an electron of kinetic energy $E$, and $\aleph$ expresses the electron analogy of refraction at the vacuum-metal boundary \cite{dowell2009quantum}.
  For a small $\aleph$ (our case, $\aleph=0.275$), i.e. an incident photon energy $nh\nu$ comparable to the work function $W$, the equation can be simplified to $I(\theta)\propto\cos\theta$ \cite{poole1972photoelectron,pei2002angular}.
  Therefore, the 3D velocity distribution can be reconstructed based on the presumed distribution function.
  
  The reconstructed velocity distribution is plotted as blue line in Fig.\,\ref{electronDistribution}a, and the smoothed energy distribution shown in Fig.\,\ref{electronDistribution}b.
  The energy distribution of the emitted electrons shows an energy spread of $\sim 1$\,eV, which corresponds to the energy difference between a four-1.55\,eV-photon excitation and the Au work function of 5.31\,eV.
  
  \subsubsection{Discussion}
  
  The Berglund-Spicer three-step model is employed as the analytic expression for the kinetic energy distribution of the photoemitted electrons.
  As the model is derived for single-photon emission, it is implied in our analysis that the electrons at an initial energy state $E_0$ absorb a sufficient number of photons instantaneously, rather than sequentially, to be pumped to a higher energy state $E=E_0+nh\nu$.
  The kinetic energy distribution for single-photon emission \cite{berglund1964photoemissionT} is adapted to multiphoton emission as
  \begin{equation}
  N(E) dE \propto \frac{K \, C(E) \, \alpha}{\alpha+1/l(E)} dE \times\left[1+4\left(\frac{E-E_f}{nh\nu}-1+\ln\frac{nh\nu}{E-E_f}\right)\right]
  \label{N(E)dE}
  \end{equation}
  where $E_f$ is the Fermi energy of Au, $C(E)=0.5(1-\sqrt{W/E})$ for $E \geq W$ is a semiclassical threshold function, and $l(E)$ is the electron-electron scattering length, which is proportional to $E^{-3/2}$.
  The absorption coefficient $\alpha$ is calculated from the extinction coefficient $k=4.71$ as $\alpha=4\pi k/\lambda$ and taken as a constant $\alpha=7.7\times10^{5}\,\text{cm}^{-1}$ independent of electron energy.
  $K$ is a correction factor related to both $C(E)$ and $\alpha\,l(E)$, which is between 0.5 and 1.
  To evaluate \eref{N(E)dE}, the probability of a photon carrying energy $h\nu$ is calculated from the measured laser spectrum in the range from 760 to 850\,nm.
  To overcome the barrier of 5.31\,eV, an electron is assumed to always absorb four photons ($n\equiv4$).
  Absorption of different photon energies leads to slight differences of the quantum yields at a certain emitted kinetic energy as one can see from Fig.\,\ref{electronDistribution}b.
  The main consequence of absorbing photons with various energies is the spectral/intensity broadening, which is illustrated by the 2D-histogram in Fig.\,\ref{electronDistribution}b.
  The temperature of the Fermi-Dirac distribution has been adjusted such that the mean of the histogram matches our experimental three dimensional energy distribution.
  We mention that \eref{N(E)dE} only includes the emitted electrons that experience none or one electron-electron scattering process during transport to the metal-vacuum surface.
  Electron-electron scattering is dominant over electron-phonon scattering and reshapes the energy distribution on a short timescale, i.e. during an ultrashort laser pulse.
  
  The density of states (DOS), i.e. the number of states available for electrons at a certain energy level, is shown in the inset of Fig.\,\ref{electronDistribution}b.
  During the photoemission process, an energy state $E_0$ is first occupied by an electron, which is then excited to a higher energy state $E$, which was empty.
  As fermions, electrons obey the Pauli exclusion principle.
  In thermal equilibrium, the possibility of electrons to occupy an available energy state is given by the Fermi-Dirac distribution $f_\text{FD}$.
  However, excitation of a metal with ultrashort strong laser pulses initially creates a non-equilibrium distribution that then thermalizes via electron-electron scattering towards a Fermi-Dirac distribution.
  In gold, this thermalization occurs on a timescale of hundreds of femtoseconds \cite{farm1993direct,fann1992electron}.
  Subsequently, the electrons cool down by dissipating energy into the lattice via electron-phonon scattering occurring on a longer picosecond timescale.
  In the following discussion, where we employ the Berglund-Spicer model in our analysis, we assume that the electronic system can be described by a Fermi-Dirac distribution with quasi-equilibrium electron temperature $T_e$.
  Hence, the appropriate densities of states and Fermi-Dirac distributions are multiplied with the energy distribution as $N(E)dE\,f_\text{FD}(E_0)\,\text{DOS}(E_0)\,(1-f_\text{FD}(E))\,\text{DOS}(E)$, resulting in the spectrum shown in Fig.\,\ref{electronDistribution}b.
  
  The best fit with our reconstructed experimental energy distribution is obtained for an electron temperature of 6000\,K.
  This is comparable to previously observed electron temperatures of 7000\,K in surface-enhanced multiphoton emission from copper \cite{aeschlimann1995observation}.
  The high energy tail of the spectrum indicates that very ``hot'' electrons are photoemitted by the femtosecond laser pulse, consistent with the high excess energy deposited into the electronic system.
  For the tail up to 4\,eV above-threshold photoemission (ATP), i.e. the absorption of one (or more) extra photon occurring together with the four-photon process, might need to be taken into account \cite{banfi2005experimental}.
  Moreover, for our experimental conditions, we can neglect tunnel ionization, which could result in high energy emitted electrons.
  Taking into account Fresnel losses, we estimate the absorbed peak intensity for the recorded image, Fig.\,\ref{electronDistribution}a, to be $\sim 4 \times 10^{9} $\,W/cm$^2$.
  This implies a Keldysh parameter $\gamma=\sqrt{W/2U_p} \approx 17 \gg 1$, which is well in the multiphoton emission regime; here, $U_p \propto\lambda^2 I$ is the ponderomotive energy with laser wavelength $\lambda$ and intensity $I$.
  
  Since both, the measured quantum yield and the momentum distribution, are in quantitative agreement with the Fowler-Dubridge and Berglund-Spicer models, as one would expect from multiphoton emission from a planar Au cathode, the VMI spectrometer has successfully been implemented as a tool to characterize the photoemitted electrons from cathodes, especially to directly measure the transverse momentum distribution.
  Assuming there is no correlation between the location of emission and the transverse momentum, the RMS-normalized emittance $\varepsilon_n$ is defined as
  \cite{dowell2009quantum}
  \begin{equation}
  \varepsilon_{n_{\zeta}}=\frac{\sqrt{\langle\zeta^2\rangle\langle{p_{\zeta}}^2\rangle}}{m_0c} \text{,~with~} \zeta\in\{x,y\}
  \end{equation}
  where $\langle\zeta^2\rangle$ is the spatial spread and $\langle{p_{\zeta}}^2\rangle$ is the momentum spread of the electron bunch.
  From the velocity map image shown in the inset of Fig.\,\ref{electronDistribution}a, the RMS-normalized emittance of the planar Au photocathode irradiated by 45\,fs 800\,nm laser pulses with a focal spot size of $\sigma_x=161$\,{\textmu}m and $\sigma_y=17$\,{\textmu}m is characterized to be $\varepsilon_{n_x}=128$\,nm$\cdot$rad and $\varepsilon_{n_y}=14$\,nm$\cdot$rad in the $x$ and $y$-directions, respectively.
  To decrease the intrinsic normalized emittance, in principle, one needs to decrease either the emission area or the momentum spread.
  The former can be intuitively decreased by an extremely tight focal spot size or sharp tip surface, which geometrically limits the emission area.
  To reduce the momentum/energy spread, choosing a proper material with appropriate work function and irradiating it by a laser beam with matched photon energy, for example the photoemission of Cu under 266-nm laser irradiation, is expected to help.
  Further reduction is expected when entering the strong-field emission regime, where the electrons are considered to adiabatically tunnel through the surface barrier with zero initial momentum and are then driven by the instantaneous optical field \cite{corkum1993plasma,kruger2011attosecond}.
  Under these conditions electrons are expected to be emitted with a relatively small divergence angle and significantly lower transverse momentum spread.
  
  In order to characterize future low emittance sources, high resolution emittance measurements are mandatory.
  The presented spectrometer has the potential to measure the initial spatial- and momentum-distribution of the electrons and, therefore, the emittance, in high resolution.
  The transverse energy resolution of the velocity mapping $dE=mv_\mathrm{2D}dv_\mathrm{2D}$ is linearly increasing with the transverse velocity $v_\mathrm{2D}$ in a VMI spectrometer.
  In our case, the spatial resolution of the detector, the Chevron MCP, is 100\,{\textmu}m.
  This matches the resolution provided by a single camera pixel ($dv_\mathrm{2D}=8014\,\text{m}/(\text{s}\cdot\text{pixel})$).
  Therefore, the transverse energy resolution of the spectrometer is given by \text{$0.2\,\text{meV}\leq dE\leq 90\,\text{meV}$}.
  The lower boundary corresponds to the resolution in the detector center, whereas the upper boundary is the resolution at the edge.
  Therefore, compared to other techniques, our spectrometer has an unprecedented transverse energy resolution in the center.
  For our current settings the maximum detectable transverse energy is on the order of 10\,eV.
  This results in a relative resolution of $<$1\% at the edge of the detector, again given by the resolution of the detector (or camera).
  It should be noted that the current transverse energy resolution could in principle simply be increased by using a larger detector, longer drift region, and a higher resolution camera.
  A three times better resolution of \text{$0.07\,\text{meV}\leq dE \leq 30\,\text{meV}$} is obtainable, i.e. with a 12\,cm diameter detector, a 1.5\,m drift tube, and a high resolution camera.
  The current spatial resolution in SMI mode was given by 100\,{\textmu}m/7.5=13\,{\textmu}m on the cathode which was sufficient to measure the initial distribution of the electrons.
  With the same changes on the setup, as discussed above, a $3\times$ better spatial resolution on the cathode can be reached.
  This results in an overall resolution in the emittance given by 0.5\,nm$\cdot$rad with the possibility to improve it to $\sim 0.06$\,nm$\cdot$rad for future experiments.
  
  A comparison of existing methods to characterize ultra-low-emittance photocathode is presented in \cite{lee2015review}.
  The different methods all agree that the apparatus and corresponding transfer functions have to be modeled.
  The transverse energy resolution is typically worse than the one obtained here.
  The most outstanding advantage of the VMI spectrometer is that an entire single Newton sphere is captured at once and various Newton spheres are simply superimposed.
  This implies that the mapping is non-destructive in the sense that no filter functions like retarding voltages in combination with pinholes need to be applied as for the 2D energy analyzer \cite{karkare20152}.
  Therefore, it avoids slow electrons with their extremely stray-field-sensitive trajectories.
  Free expansion, reported as the simplest method by far \cite{lee2015review}, is the closest technique to the VMI spectroscopy demonstrated here.
  This technique is conceptually the analog to the early ion imaging experiments before the invention of the velocity-map-imaging spectrometer \cite{eppink1997velocity}.
  However, the commonly present electrode grids lead to transmission reduction, severe trajectory deflection, and blurring due to the non-point-source geometry.
  In addition, the incident laser in the demonstrated free expansion setup \cite{feng2015novel}, was focused onto the sample through the grid, which seriously deforms the starting intensity distribution.
  The high energy resolution of the VMI in comparison with the free expansion technique is attributed to the inhomogeneous electric field in the spectrometer.
  This allows, in first order, to get rid of the spatial contribution in the velocity coordinates.
  Therefore, a single measurement is sufficient to obtain the velocity map without contributions from the initial source distribution.
  Furthermore, non-cylindrically-symmetric-velocity distributions, obtained from for example nanotips, can be measured as well.
  As a final touch, operating the VMI spectrometer under SMI conditions allows the mapping of the initial source distribution, which circumvents the modeling of the active laser-matter interaction area.
  Overall, the simplicity of the VMI spectrometer and the super-short measurement times, typically only a few minutes, enables the easy integration into more sophisticated electron sources.
  
  \subsection{Mapping Electron Emission on Materials}
  
  In addition to the properties of the emitted electron bunch, the emission mechanism in the electron source is also a very important feature.
  Understanding the emission mechanism through experimental characterization is essential towards designing and optimizing advanced electron sources.
  The principles of operation in flat (i.e. unstructured photocathodes) has been deeply studied and currently well understood \cite{dowell2009quantum}.
  The developed models are widely tested in experiments, and their predictions are confirmed to high precisions \cite{berglund1964photoemissionT,berglund1964photoemissionE}.
  These models are now the ground for electron source design based on flat photocathodes.
  Similarly, to characterize the emission mechanism in structured cathodes, theoretical models are required, which should be developed based on experimental observations.
  
  Mapping field-driven electron emission on plasmonic nanoparticles to measurable observables will be key to engineering plasmonic structures for ultrafast electron sources.
  Moreover, identifying and differentiating between electronic processes occurring within individual plasmonic nanoparticles will allow for greater control of reaction pathways on their surfaces.
  In \cite{hobbs2017mapping}, we demonstrated that two common electron-beam lithography resists, poly(methyl-methacrylate) (PMMA) and hydrogen silsesquioxane (HSQ), can be used to image both electron emission and energy transfer on the surface of plasmonic nanoparticles.
  Here, we present the characterization results of the gold rod and triangular emitters using PMMA based technique.
  
  PMMA has been used in earlier works as an imaging layer for electrons.
  For example, a PMMA layer was integrated on the anode of a vacuum diode device to image electrons emitted by DC field emission from ZnO nanowires \cite{sun2012polymethyl}.
  Moreover, recently Volpe et al. \cite{volpe2012near} and Dregely et al. \cite{dregely2013vibrational} have used PMMA and HSQ respectively to map the optical near-field of plasmonic Au nanoantennas.
  Volpe et al. observed that PMMA decomposed within the plasmonic hotspots of Au nanorods illuminated at a resonant wavelength of 800\,nm.
  They also found that the volume of decomposed PMMA scaled with the fourth power of the laser intensity, which suggests an exposure mechanism based on four-photon absorption.
  Jiang and Gordon \cite{jiang2013nonlinear} later showed that the mechanism proposed by Volpe et al. may not produce sufficient PMMA decomposition to support lithographic activity.
  Additionally, neither the work of Volpe et al., nor that of Jiang and Gordon, considered an exposure mechanism based on electron emission from the plasmonic nanoparticles.
  However, as we know from previous works \cite{hobbs2014high,putnam2017optical,dombi2013ultrafast,hobbs2014highdensity,schweikhard2011polarization,nagel2013surface} femtosecond lasers can drive significant electron emission currents from plasmonic nanoparticles and produce extremely high peak current densities in such particles.
  
  In what follows, it is shown that the electron emission from Au nanoantennas excited by femtosecond laser pulses is sufficient to expose PMMA.
  Moreover, it is observed that the spatial distribution of the exposed PMMA is in agreement with the expected distribution of emitted electrons based on simulations of both the optical near-field and the trajectories of electrons emitted by strong-field tunneling emission.
  In contrast to Volpe et al., a laser source with a central wavelength of 1200 nm is used to excite the LSPRs in Au nanorods.
  As such, cumulative absorption of six photons would be required to expose PMMA by multiphoton absorption (given that PMMA absorbs weakly at wavelengths longer than 200\,nm); the reduced probability of six-photon absorption is exploited to enable nanometer mapping of electron emission in PMMA.
  This work shows that lithographic materials may provide a number of opportunities to study the roles of plasmons in energy transfer at the nanometer length scale and femtosecond time scale.
  
\begin{figure}
  	\centering
  	$\begin{array}{cc}
  	\includegraphics[draft=false,width=2.0in]{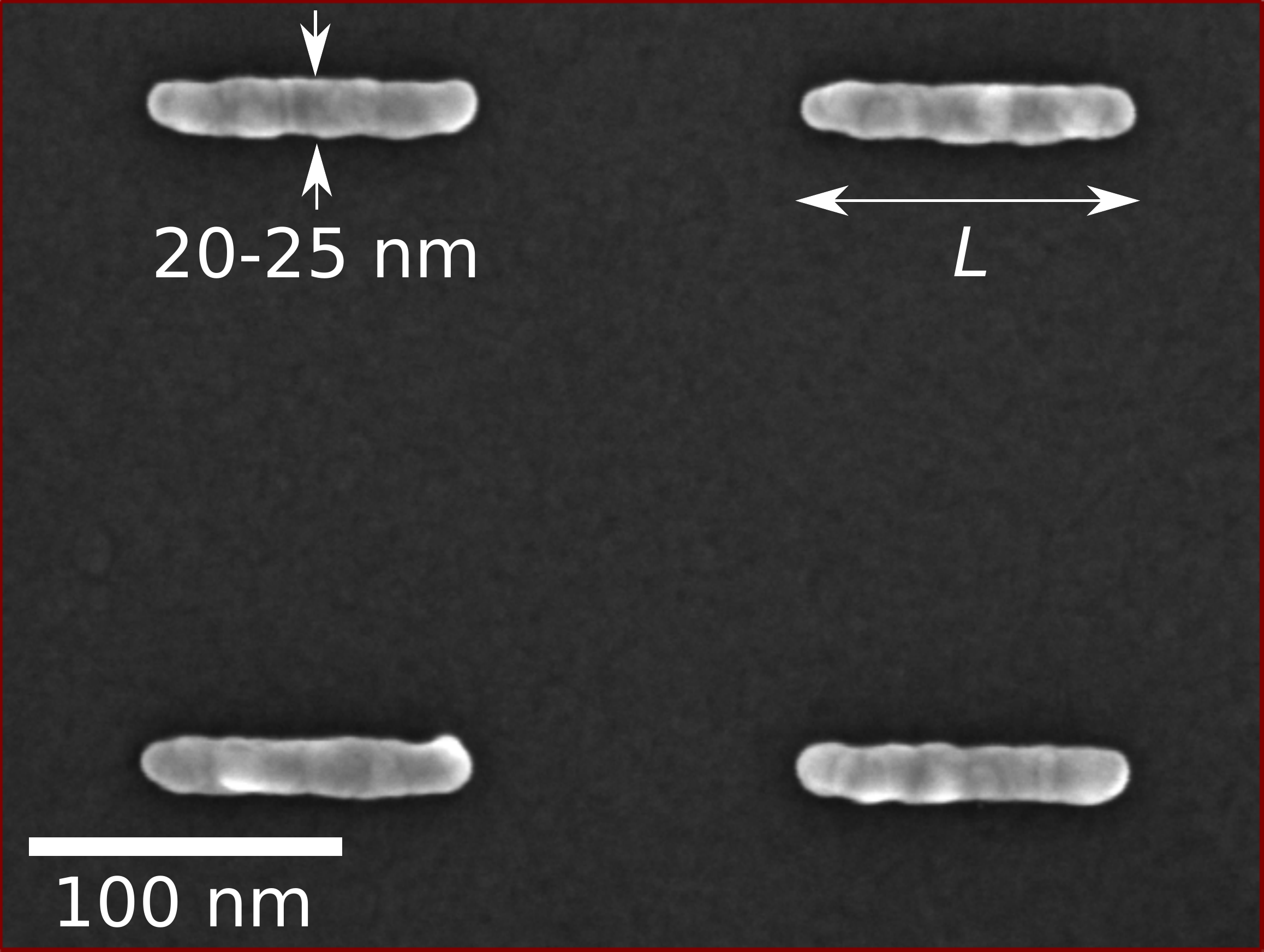} &
  	\includegraphics[draft=false,width=2.0in]{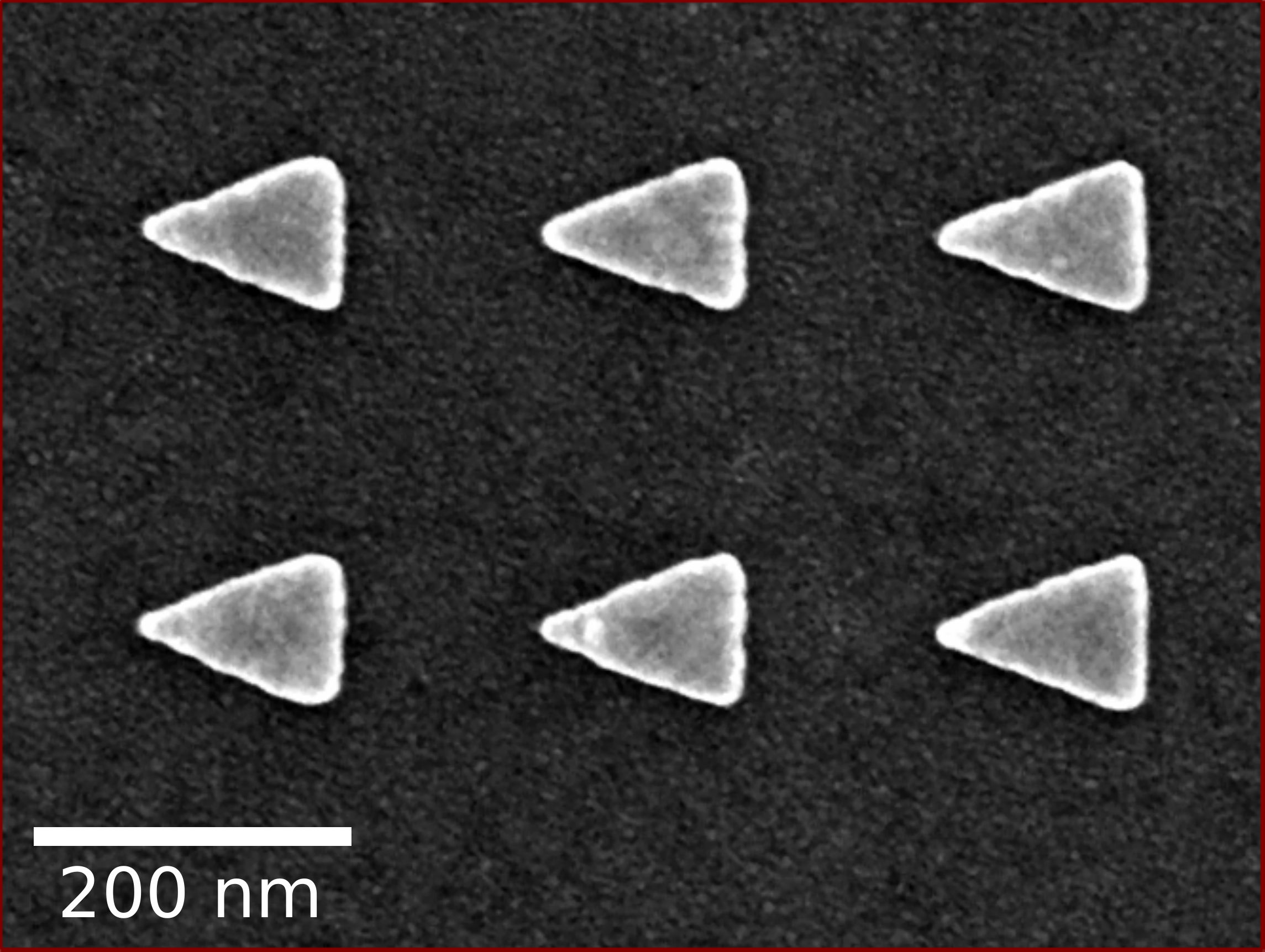} \\
  	(a) & (b) \\
  	\includegraphics[draft=false,width=2.0in]{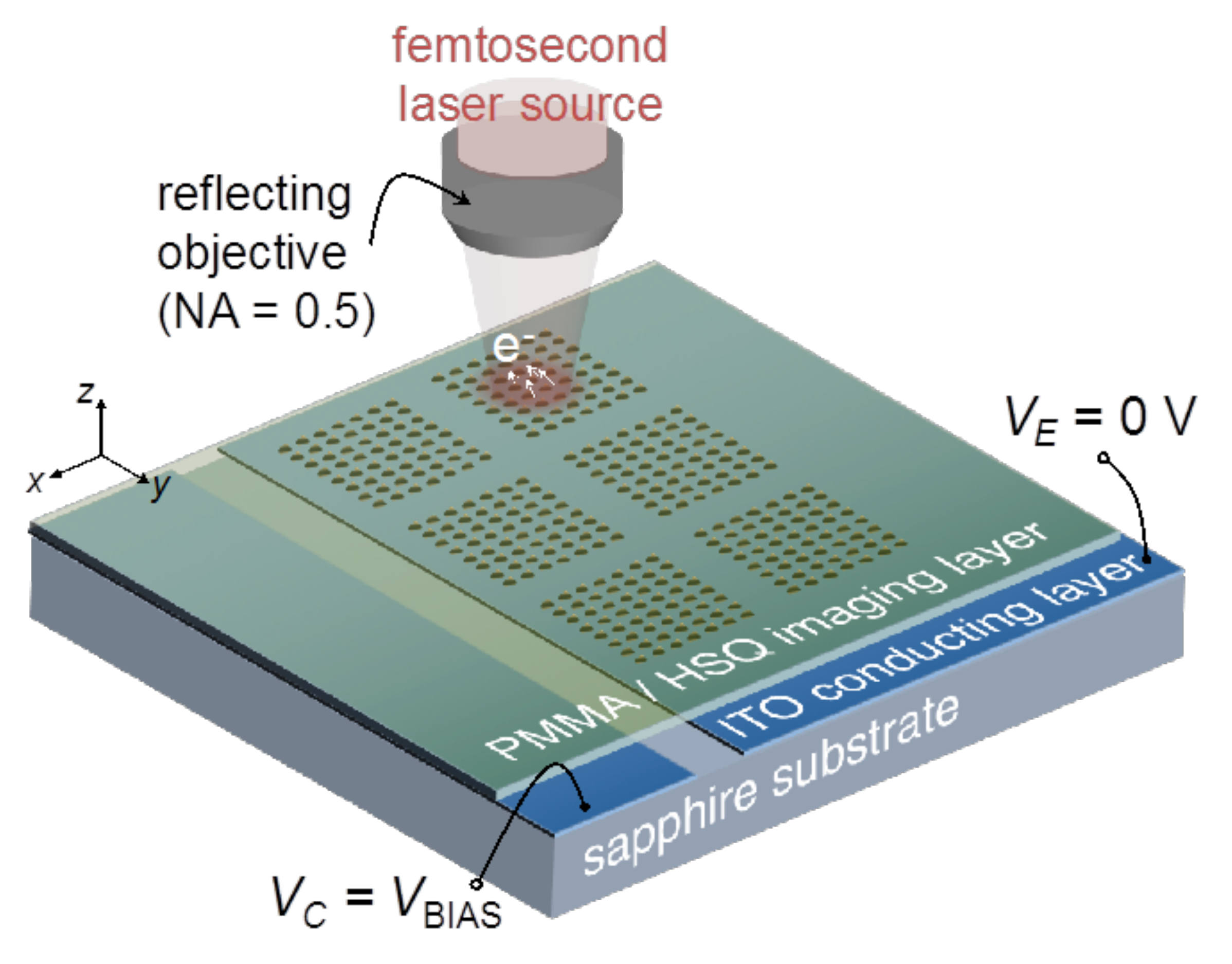} &
  	\includegraphics[draft=false,width=2.0in]{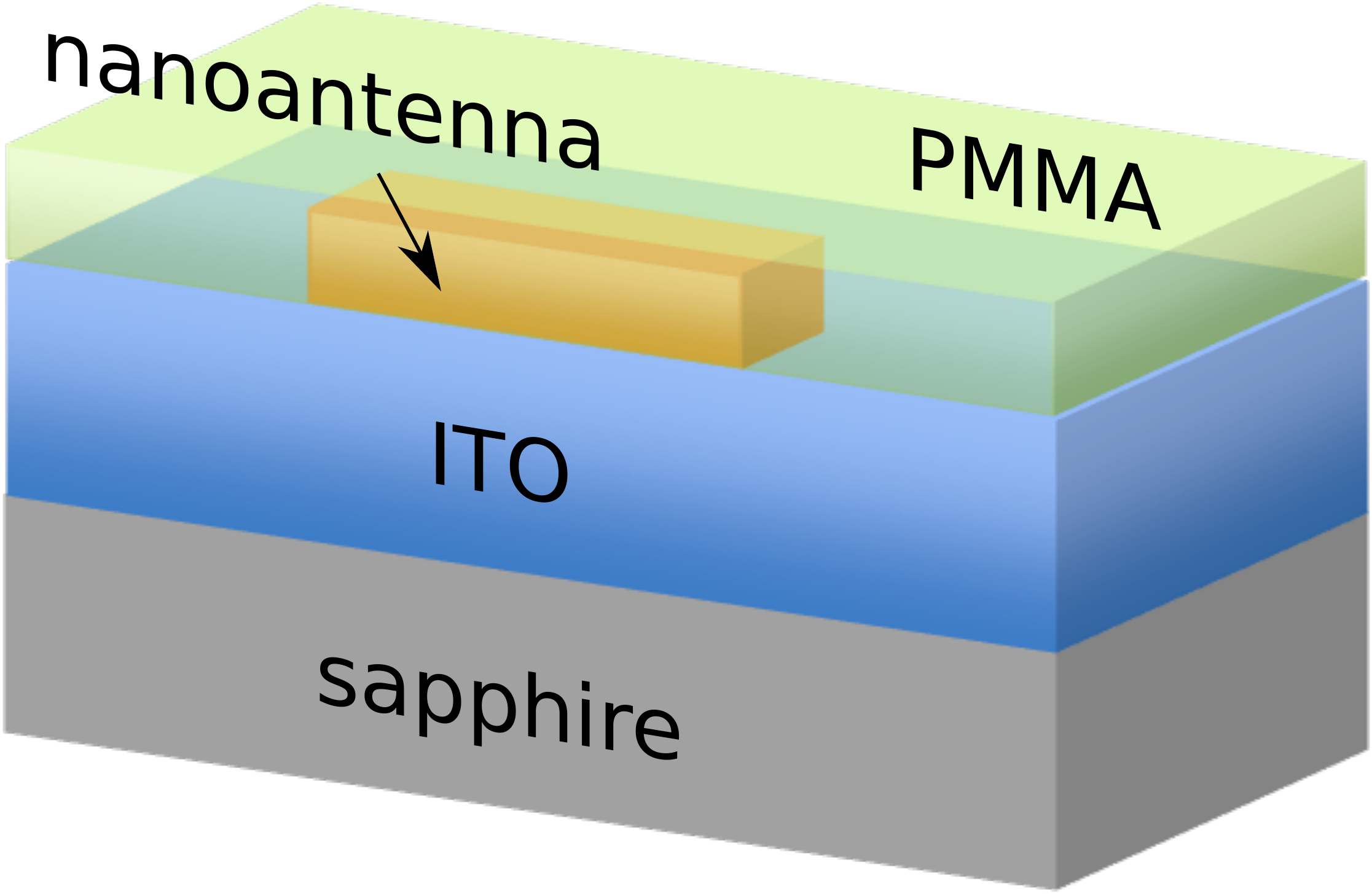} \\
  	(c) & (d) \\
  	\end{array}$
  	\caption{Nanoantennas and experimental approach. (a,b) Top-down SEM micrographs of an array of Au nanorod (a) and nanotriangle (b) antennas fabricated by electron-beam lithography. (c) Schematic representation of experimental setup. The 10\,fs pulses of linearly polarized light with a central wavelength of 1.2\,{\textmu}m and a bandwidth of 400\,nm were focused to a $1/e^2$ diameter of 5.2\,{\textmu}m on the nanoantenna arrays, which were fabricated on an ITO-coated (blue regions) sapphire substrate (gray regions). A 5\,{\textmu}m gap was etched in the ITO layer as shown to allow a bias to be applied between the nanoantenna emitter array (the emitter electrode: blue region on the right; $V_\mathrm{E} = 0$) and the collector electrode (blue region on the left; $V_\mathrm{C} = V_\mathrm{BIAS}$). (d) Schematic showing a nanoantenna coated with a 20-nm-thick layer of PMMA (semitransparent green region), which acts as an imaging layer for emitted electrons.}
  	\label{setupPMMA}
  \end{figure}

  Fig.\,\ref{setupPMMA} outlines the experimental approach used in this work.
  Arrays of Au nanoantennas such as those shown in Fig.\,\ref{setupPMMA}a and b were fabricated on ITO-coated sapphire substrates by electron-beam lithography.
  The nanoantenna arrays were prepared without a conventional adhesion promoting layer such as Ti or Cr as these layers have been shown to detrimentally affect optical field enhancement \cite{hobbs2014high,madsen2017observing}.
  Photoemission from the nanoantenna arrays was measured using the setup shown schematically in Fig.\,\ref{setupPMMA}c (the setup is further described in our recent paper \cite{putnam2017optical}).
  The nanoantennas fabricated in this work typically produced $\sim 1$ electron per nanoantenna per laser pulse or equivalently, 1-10\,C/cm$^2$ ($10^4-10^5$\,electrons/nm$^2$) at the poles of the nanoantennas.
  We coated the nanoantenna arrays with a thin film of electron-beam resist PMMA.
  The resist was exposed by illuminating the nanoantennas with the femtosecond laser source (12.5\,mW average power, $\sim 10$\,fs pulse, 78.4\,MHz repetition rate, 10\,s exposure time, $\sim 0.6$\,MJ/cm$^2$ average fluence per exposure).
  Following exposure, the exposed PMMA resist is removed using an appropriate developer (see \cite{hobbs2017mapping}).
  
\begin{figure}
  	\centering
  	$\begin{array}{c}
  	\begin{array}{cc}
  	\includegraphics[draft=false,height=2.0in]{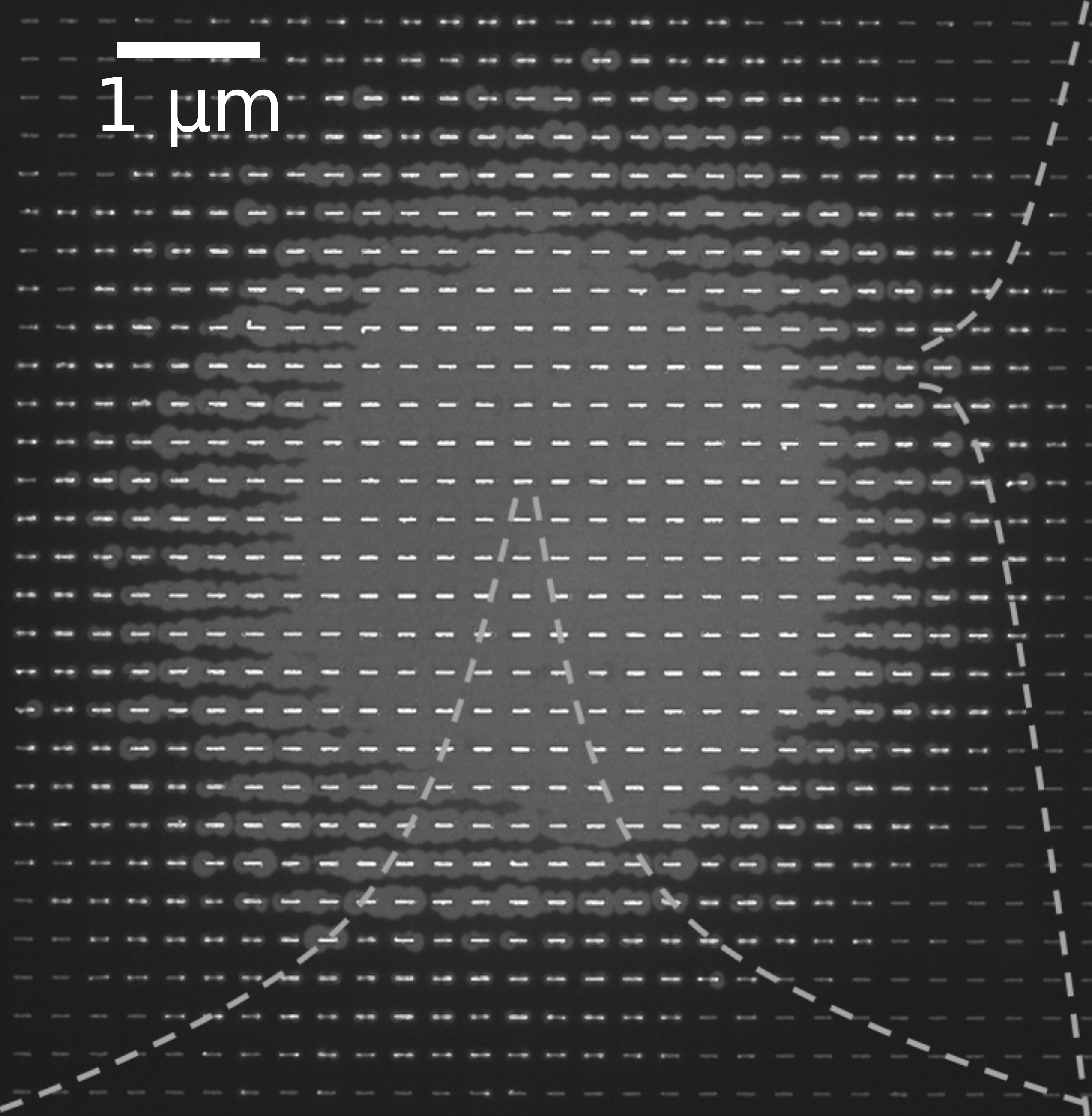} &
  	\includegraphics[draft=false,height=2.0in]{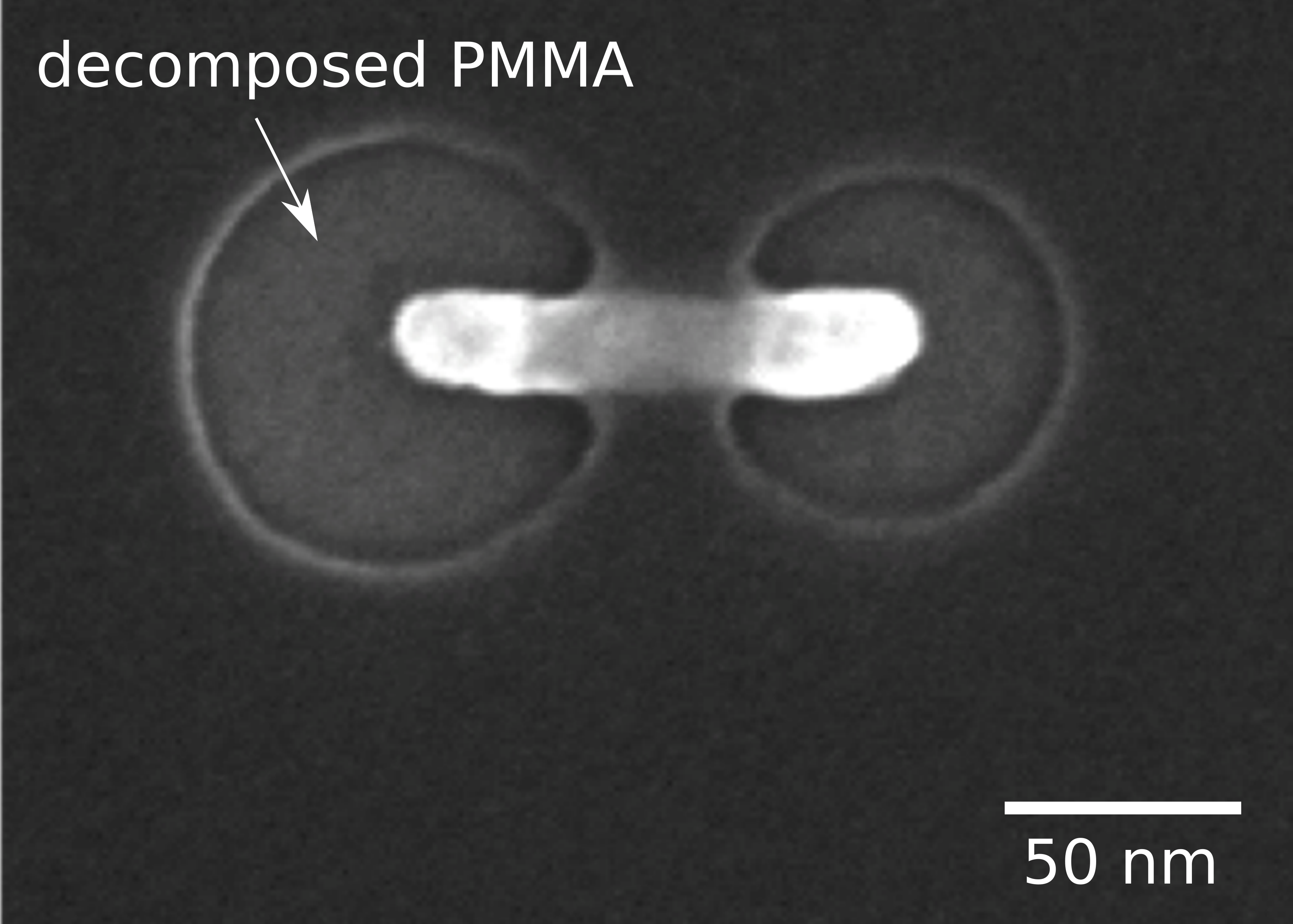} \\
  	(a) & (b)
  	\end{array} \\
  	\begin{array}{ccc}
  	\includegraphics[draft=false,height=2.0in]{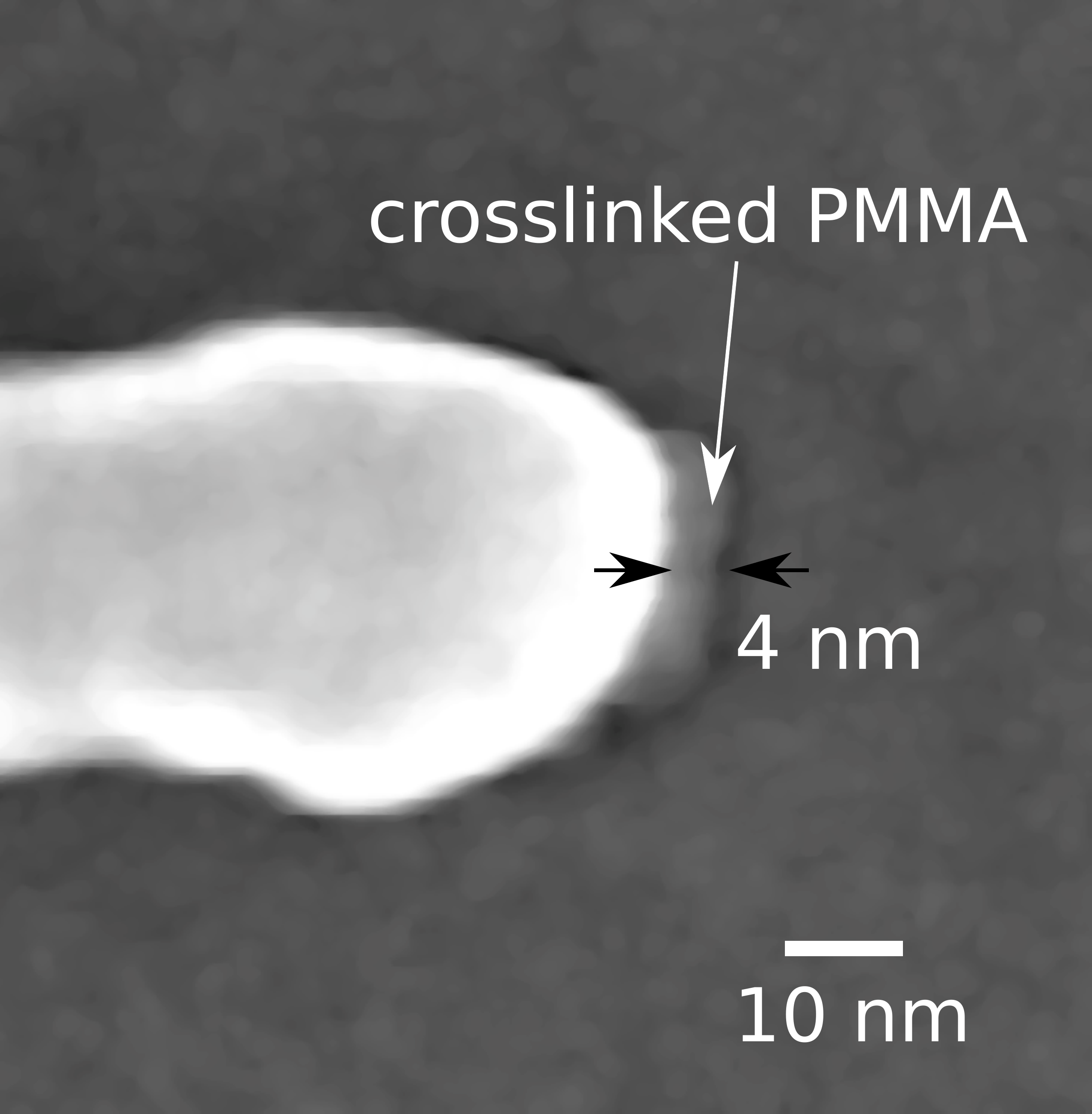} &
  	\includegraphics[draft=false,height=2.0in]{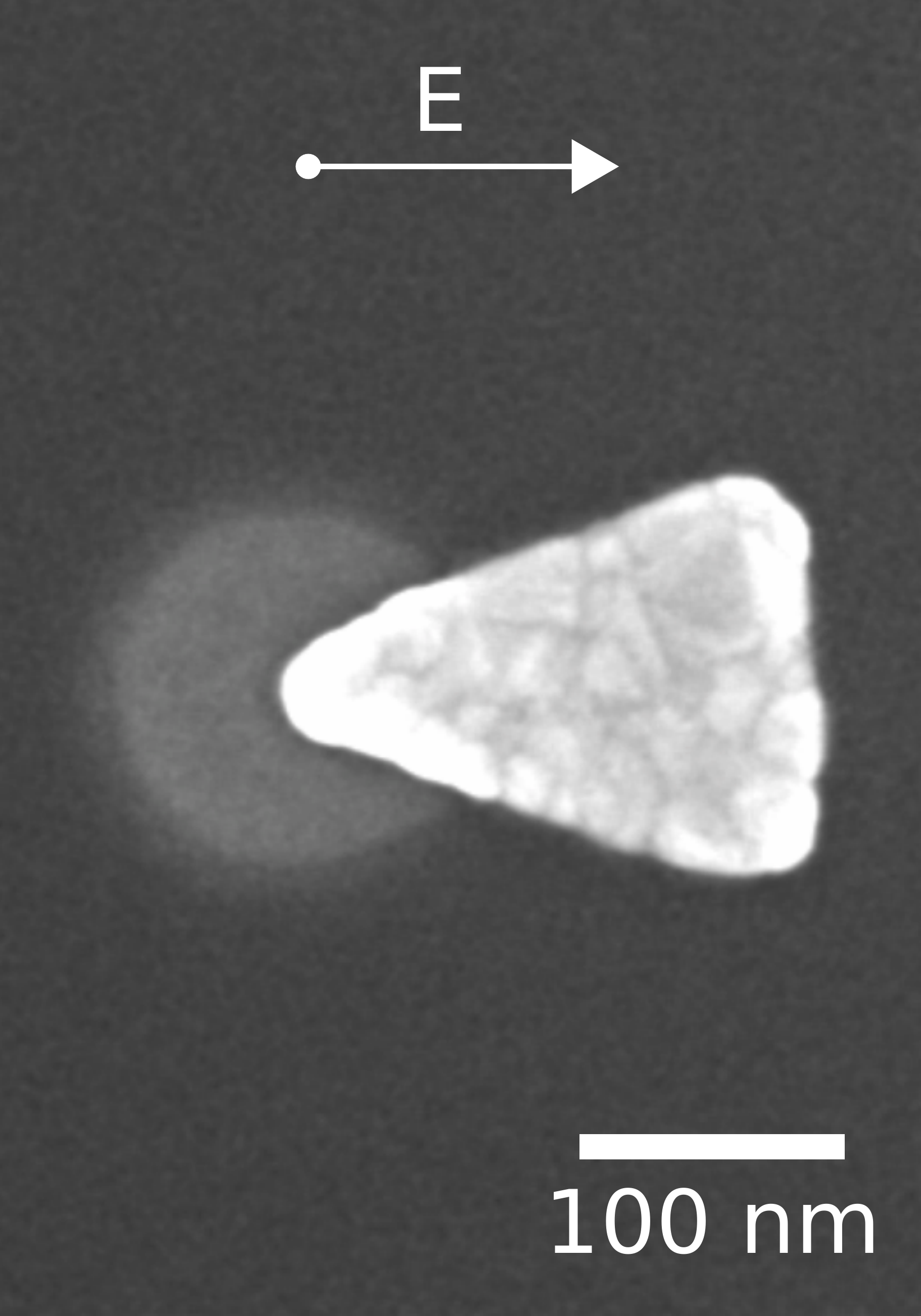} &
  	\includegraphics[draft=false,height=2.0in]{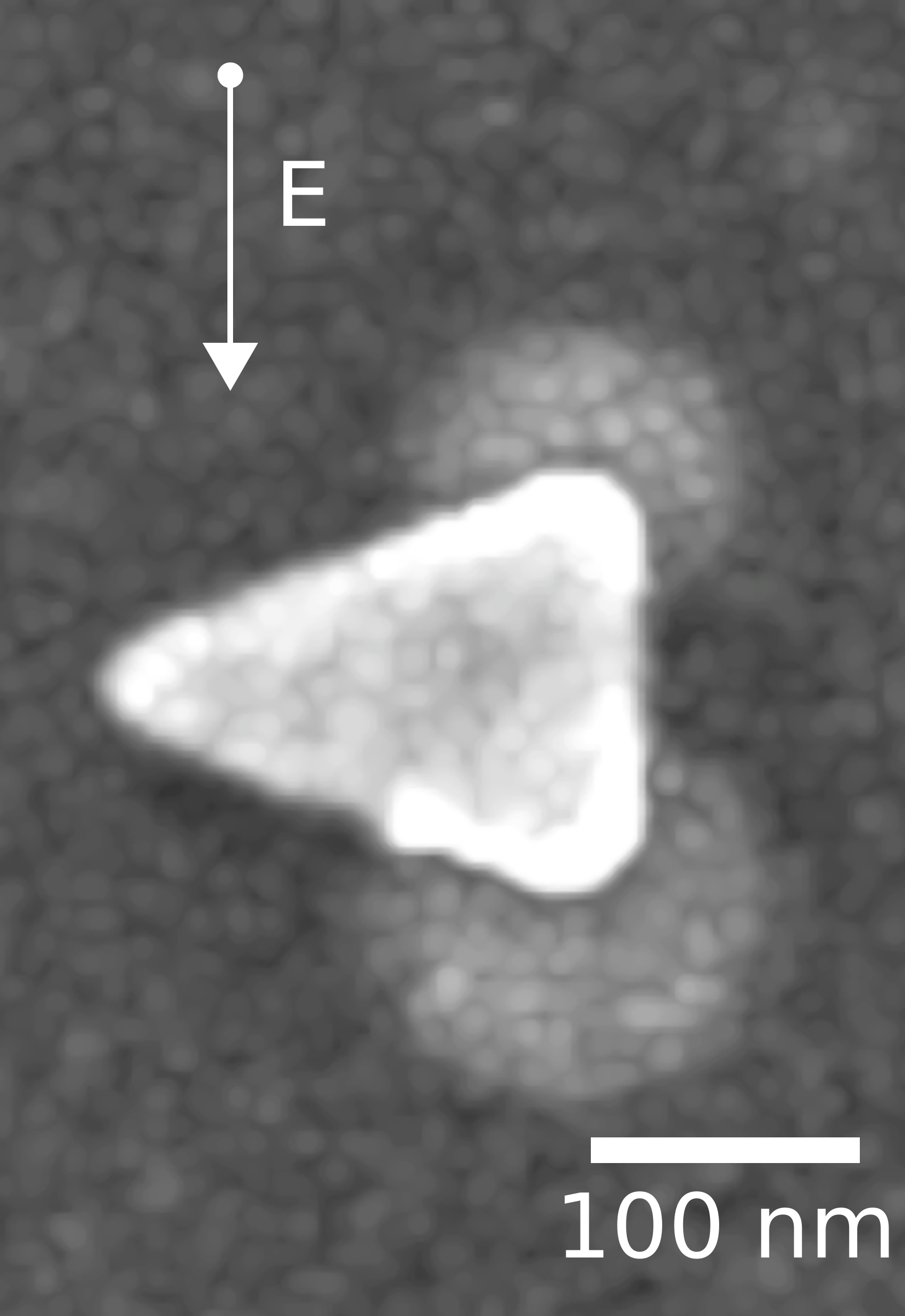} \\
  	(c) & (d) & (e)
  	\end{array}
  	\end{array}$
  	\caption{Imaging electron emission from plasmonic nanoantennas with PMMA. (a) Top-down SEM micrograph of a 260-nm-pitch Au nanorod array coated with a 20-nm-thick layer of PMMA. The coated array was previously exposed to $\sim 10^9$ femtosecond laser pulses and the PMMA was subsequently developed. (b) SEM micrograph of a nanorod antenna near the periphery of the exposed region shown in panel (a). The image shows regions of exposed and developed PMMA at the poles of the nanoantenna. (c) SEM micrograph of a pole of a nanorod near the center of the exposed region in panel a. A narrow strip of material believed to be cross-linked PMMA is present at the apex of the nanorod. (d, e) SEM
  		micrographs of exposed PMMA around nanotriangles illuminated with $\sim 10^9$ femtosecond laser pulses. The polarization of these pulses was aligned (d) parallel to and (e) orthogonal to the long-axis of the nanotriangle antennas.}
  	\label{SEMimages}
  \end{figure}

  Fig.\,\ref{SEMimages} shows SEM images of both nanorod and nanotriangle antennas that were coated with 20-25\,nm of PMMA, exposed with the femtosecond laser source, and
  developed to remove the exposed PMMA.
  We used a low temperature development process to improve the contrast of the features produced in PMMA.
  Fig.\,\ref{SEMimages}a shows a low-magnification SEM image of the exposed area of a 260-nm-pitch square array of 130-nm-long Au nanorod antennas after developing the exposed PMMA.
  The nanorod arrays were exposed to $\sim 8 \times 10^8$ laser pulses each having an energy of 0.16\,nJ.
  Fig.\,\ref{SEMimages}b shows an SEM image of a nanorod near the periphery of the exposed area shown in Fig.\,\ref{SEMimages}a.
  The nanorod shows well-defined regions of exposed PMMA at its poles where the optical fields peaked and thus where electron emission is expected to be most efficient.
  
  Fig.\,\ref{SEMimages}a and b show that PMMA is exposed at distances of more than 50\,nm from the poles of the Au nanoantennas.
  This is a rather surprising result considering that the emitted electrons are expected to have kinetic energies of $<$10\,eV and intuitively might be expected to propagate shorter distances ($<$50\,nm) in PMMA based on universal inelastic mean-free path (IMFP) curves \cite{seah1979quantitative}.
  However, low-energy electrons have been observed to expose resists at distances of $\sim 100$\,nm from their point of origin; for example, Duan et al. \cite{duan2010sub} measured the point-spread function of a 30\,keV electron beam in HSQ on a 50-nm-thick freestanding membrane and observed resist exposure $\sim 100$\,nm from the point of exposure.
  Their result suggests that low-energy secondary electrons are capable of exposing resist at significant distances from the point of exposure when the dose is sufficiently high.
  Moreover, while the IMFP of electrons typically decreases with decreasing electron energy, reports suggest that the IMFP rises steeply when the electron energy drops below 10\,eV \cite{seah1979quantitative,saldin1994mean,akkerman1999characteristics}.
  
  An important property of PMMA is that it is a dual-tone electron-beam resist; PMMA behaves as a positive-tone resist at low doses and as a negative-tone resist when exposed to high doses of electrons \cite{duan2010sub}.
  Briefly, when exposed to low doses of electrons, for example, $\sim 10$\,electron/nm$^2$ for 30\,keV electrons, the long polymer chains in PMMA decompose into smaller, lower molecular weight chains that may be readily removed by an appropriate solvent (the developer).
  However, when PMMA is exposed to higher electron doses, for example, $\sim 10^4$\,electron/nm$^2$ for 30\,keV electrons, the low-molecular-weight fragments of PMMA can cross-link to form higher molecular weight species (typically referred to as negative-tone PMMA in the field of electron-beam lithography) \cite{duan2010sub}.
  A local incident laser intensity of $\sim 0.1$\,GW/cm$^2$ was required to observe positive tone exposure of PMMA in the hot spots of nanoantennas in this work.
  
  Fig.\,\ref{SEMimages}c shows an SEM image of a nanorod near the center of the exposed region displayed in Fig.\,\ref{SEMimages}a.
  A narrow strip of material resembling cross-linked negative-tone PMMA is present at the apex of the nanorod shown.
  This negative-tone PMMA indicates that this location was where the PMMA received the highest electron-exposure dose.
  A local incident laser intensity of $\sim 50$\,GW/cm$^2$ was required to observe negative-tone exposure of PMMA for nanoantennas excited along their long axis in this work.
  Additionally, in Fig.\,\ref{SEMimages}d and e, we show the polarization dependence of the PMMA exposure for Au nanotriangle antennas.
  Fig.\,\ref{SEMimages}d shows an SEM image of an Au nanotriangle exposed with the linear polarization of our source aligned with the long axis of the triangle; the resulting PMMA exposure appears at the apex of the long axis.
  Fig.\,\ref{SEMimages}e shows an SEM image of a nanotriangle exposed with orthogonal polarization; here, the exposed PMMA appears at the other apexes of the triangle.
  Negative tone exposure of PMMA was not observed for nanotriangles excited by light having linear polarization aligned to the short axis of triangles.
  
\begin{figure}
  	\centering
  	$\begin{array}{cc}
  	\begin{array}{c}
  	\includegraphics[draft=false,width=3.0in]{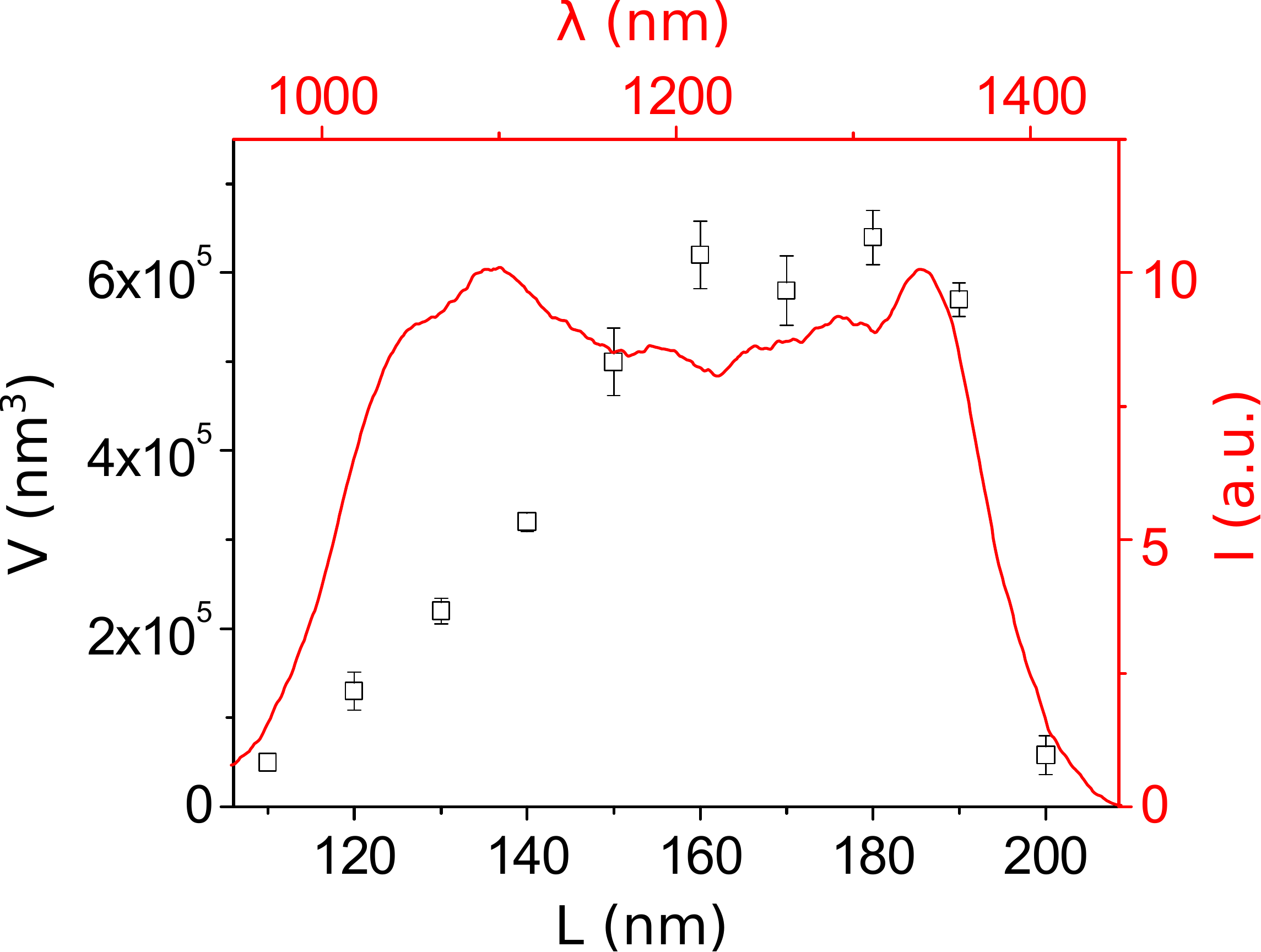} \\
  	(a)
  	\end{array} &
  	\begin{array}{c}
  	\includegraphics[draft=false,width=2.0in]{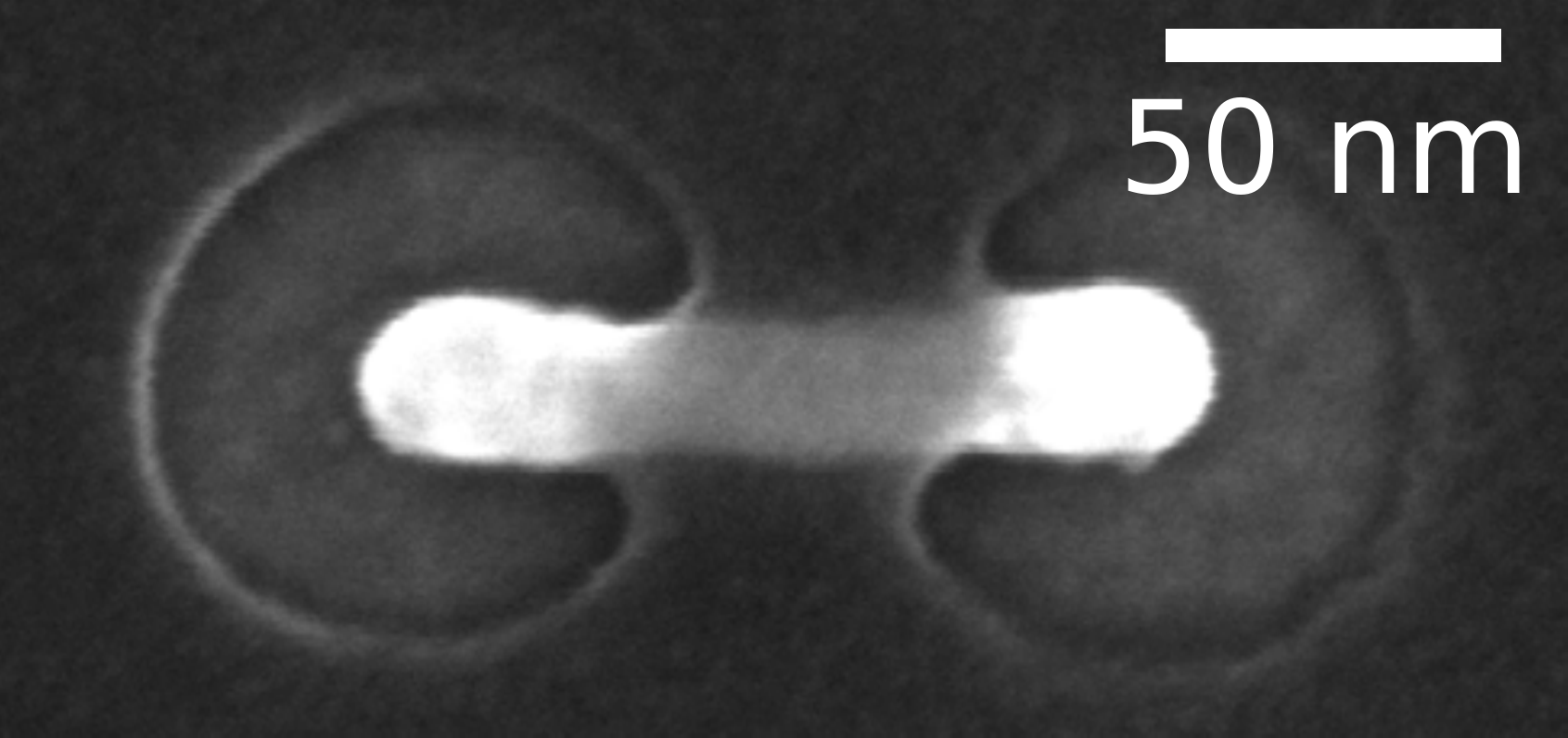} \\
  	(b) \\
  	\includegraphics[draft=false,width=2.0in]{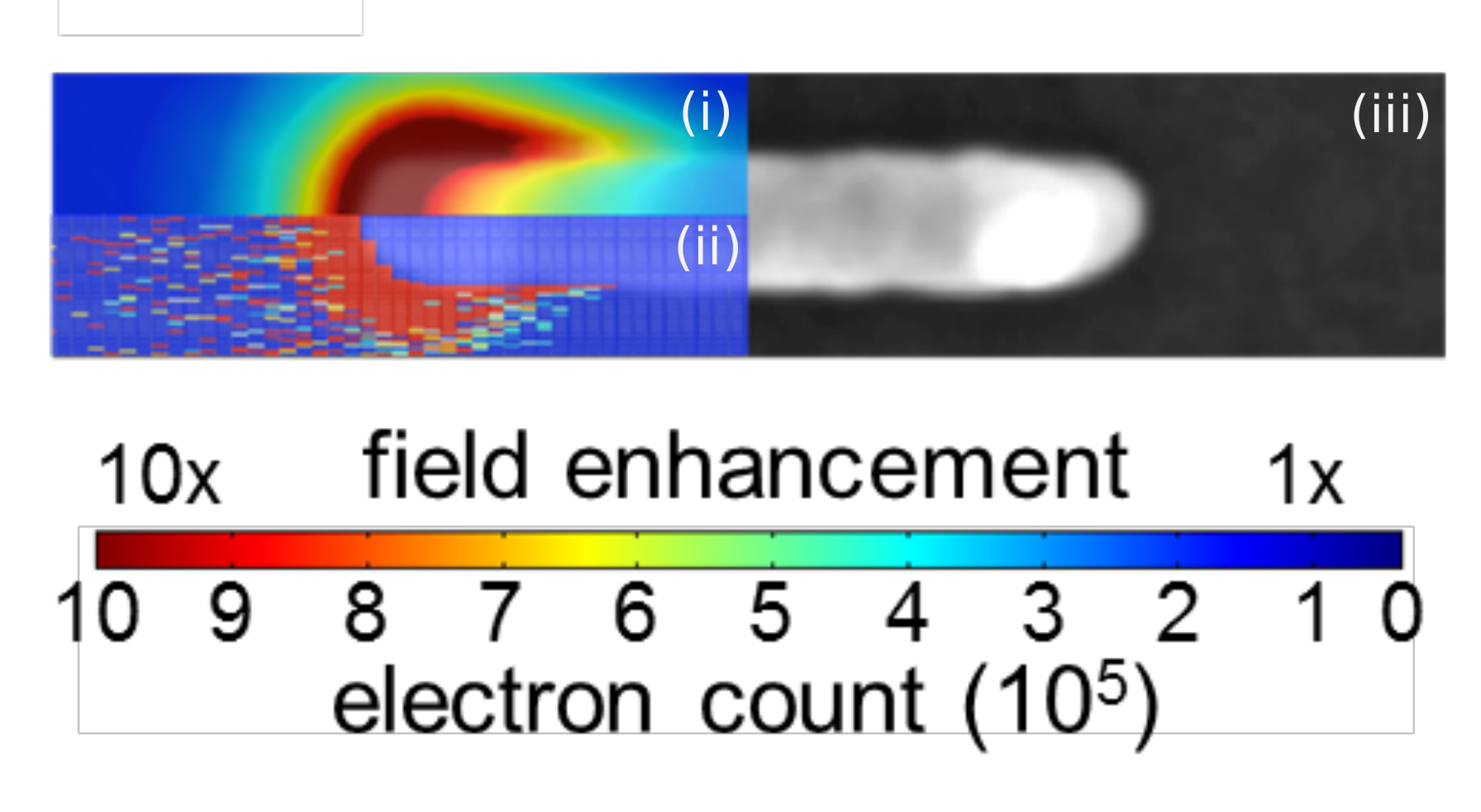} \\
  	(c)
  	\end{array}
  	\end{array}$
  	\caption{PMMA exposure versus the nanorod resonance, field enhancement at the nanorods, and spatial distribution of emitted photoelectrons. (a) Black data represent a plot of the mean volume ($V$) of developed PMMA at the pole of a nanorod antenna of length ($L$); the volume was measured for nanorods a distance of $\sim 2.5-3.0$\,{\textmu}m from the center of the laser spot. Error bars represent the maximum and minimum values of $V$ for each nanorod length. The red curve represents the normalized intensity ($I$) spectrum of the laser used to excite electron emission. (Note that the pitch of each nanorod array was $4L$.) (b) Example of an SEM micrograph used to measure the volume of developed PMMA. (c) (i) A simulated map of the magnitude of optical field enhancement at the pole of 130-nm-long nanorod antenna (scale is saturated at a field-enhancement factor of 10, while the field-enhancement peaks at 25 near the nanorod apex). (ii) A map of the simulated number of electrons passing through a plane at the air-ITO interface. (iii) SEM micrograph of a 130-nm long Au nanorod (scale bar 50\,nm).}
  	\label{simulationPMMA}
  \end{figure}

  To better understand the observed PMMA exposure and correlate the distribution of exposed PMMA to simulations of the local optical-field enhancement and photoelectron distribution, we investigated the dependence of PMMA exposure on the LSPR of Au nanorods.
  The results of these investigations are summarized in Fig.\,\ref{simulationPMMA}.
  Fig.\,\ref{simulationPMMA}a shows (black data points) the volume ($V$) of exposed and developed PMMA at the poles of Au nanorods 2.5-3.0\,{\textmu}m from the center of the exposed spot (estimated by inspection of SEM images).
  We measured $V$ for 5-10 poles for each nanorod length ($L$) and plotted these values against $L$.
  The error bars on the values of $V$ represent the maximum and minimum values measured in the region 2.5-3.0\,{\textmu}m from the center of the laser spot.
  The plot also shows the spectrum of the femtosecond laser source (red trace).
  We aligned the upper and lower $x$-axes of the plots in Fig.\,\ref{simulationPMMA}a by simulating the extinction spectra of nanorods having the lengths shown and mapping the value of nanorod length to the corresponding wavelength of the simulated LSPR peak.
  The plot in Fig.\,\ref{simulationPMMA}a shows that PMMA exposure was only observed for nanorods having a LSPR overlapping the laser spectrum.
  
  Fig.\,\ref{simulationPMMA}b shows an example SEM image of an Au nanorod inspected for use in the preparation of the plot shown in Fig.\,\ref{simulationPMMA}a.
  Regions of exposed and developed PMMA are clearly seen at the poles of the nanorod antenna.
  Fig.\,\ref{simulationPMMA}c shows an SEM image of a 130-nm-long, 20-nm-wide Au nanorod antenna, and a color map of the simulated optical field enhancement is overlaid in the upper left quadrant of the image.
  The regions of exposed PMMA produced by our experiment and shown in the SEM image in Fig.\,\ref{simulationPMMA}b overlap well with the regions displaying strong field enhancement (shown in Fig.\,\ref{simulationPMMA}c).
  The results of the simulated optical near field were used to estimate optical-field-driven photoemission currents from these plasmonic nanoantennas using a Fowler-Nordheim model for electron emission \cite{hobbs2014high,swanwick2014nanostructured,fallahi2014field}.
  The trajectories of emitted electrons were additionally simulated using a particle-in-cell model to produce a map of the distribution of electrons colliding with the ITO substrate after emission from the plasmonic antenna \cite{fallahi2014field}.
  Simulations of electron trajectories were performed in vacuum without the overlying PMMA layer for simplicity; however, simulations of the antenna near-field did include the PMMA layer to best represent the local field profile.
  A color map of the number of emitted electrons recombining with the substrate is overlaid in the lower left quadrant of the SEM image in Fig.\,\ref{simulationPMMA}c.
  The electron-emission simulations suggest that $10^4-10^5$\,electrons/nm$^2$ were incident on the substrate near the apex of the nanorod.
  This estimate is in good agreement with the measured emission currents of $10^4-10^5$\,electrons/nm$^2$.
  Both the simulated and measured electron doses calculated at the poles of the plasmonic nanoantennas are commensurate with the doses required to expose PMMA with electrons in the 1-50\,eV energy range.
  For example, McCord and Pease patterned PMMA using a scanning probe lithography technique with 20\,eV electrons and observed positive-tone behavior in PMMA at doses of $10^3-10^5$\,electrons/nm$^2$ \cite{mccord1988lift}.
  
  \section{Conclusion}
  
  This chapter reviewed the accomplished research on the electron source development and characterization.
  The discussion started with a study on high aspect ratio nanotips.
  The results showed an ultrashort electron pulse emission from large structured field emission cathodes including micro and macroscopic effects, enabling many new avenues for physics and engineering.
  We clearly demonstrated the transition from the multiphoton to the strong-field tunneling regime across a massive array of tips, while accounting for space-charge effects.
  Modeling of the current yield at high field strengths demonstrated rapidly diverging electron trajectories coming from a highly localized volume near the tip apex, with little space-charge reduction of current yield.
  However, as the electrons form a current sheet above the emitter, extraction was found to be limited by the formation of a virtual cathode.
  Such space-charge limitations can be easily mitigated when used in a RF photoinjector due to the high RF extraction field.
  The cathodes are also fabricated in standard CMOS processes and are stored in air at standard conditions before testing at high vacuum (10$^{-8}$ Torr), which is a major advantage over reactive low work function cathodes that are fabricated and stored in ultrahigh vacuum conditions.
  Tip arrays are highly uniform, standard deviation of less than 1\,nm tip radius of curvature, from die to die on a wafer but also from wafer to wafer.
  Emitter multiplexing has major advantages over reactive low work function cathodes or cathodes made of just a single tip because of the confined structured electron beam that the emitter arrays produce.
  
  Next, based on a detailed study on gold nanorods fabricated on ITO substrates, we proposed that the structured photocathodes may be sufficiently robust for use in XFEL
  systems when operated using a laser-intensity below the damage threshold ($\sim 45$\,GW/cm$^2$) and under a sufficiently strong static-field ($>$10\,MV/m).
  Under such conditions, Au nanorod arrays triggered by ultrafast pulses of 800\,nm light, may outperform equivalent UV-triggered Au photocathodes, while also offering nanostructuring of the electron pulse produced from such a cathode, which is of interest for future XFEL development where nanostructured electron pulses may facilitate more efficient and brighter XFEL radiation.
  Moreover, Au nanorods triggered by 800\,nm light at intensities above 12\,GW/cm$^2$ may emit electrons by a strong-field tunneling mechanism, and thus may support production of attosecond electron bursts, which are key to the development of attosecond science.
  Further investigations are required to maintain the initial levels of confinement of electrons in both space and time possible at the emitter surface, into a propagating nanostructured pulse-train.
  As a result, the Au nanorod photocathodes developed in this work represent an additional step toward the development of analytical tools with attosecond temporal resolution.
  
  The studies on electron sources proceeded with efforts towards the reliable characterization of such particle sources.
  We demonstrated an electron spectrometer with VMI and SMI capabilities, which intuitively allows for high-resolution measurements of the RMS-normalized emittance of photocathodes through the direct observation of the transverse position and momentum distributions.
  We verified and benchmarked the capabilities of the instrument in a proof-of-concept experiment, in which we characterized the photoemitted electrons from a 400\,nm
  thin Au film.
  For ultrashort femtosecond laser pulses with a peak intensity lower than $10^12$\,W/cm$^2$ and a central wavelength of 800\,nm, which corresponds to $\gamma = 1$, multiphoton emission is shown to be the dominant contribution to the entire electron current.
  
  We intend to utilize this technique for the emittance characterization of electron bunches strong-field emitted from nanotips under optical field irradiation.
  Such devices should show superior emittance \cite{herink2012field,tsujino2016measurement}.
  Moreover, the small radii of the sharp tips realize a field enhancement, which dramatically lowers the laser power required for entering the strong-field regime and thus avoids damaging of the cathodes.
  Our ongoing work aims at the characterization of electron emission from nanostructured array emitters, which are predicted to provide high-current low emittance coherent electron bunches in the strong-field emission regime.
  The demonstrated imaging spectrometer will thereby foster the further development of the XFELs and ultrafast electron microscopy and diffraction \cite{tsujino2016measurement,sciaini2011femtosecond} and also open up new opportunities in the study of correlated electron emission from surfaces \cite{hattass2008dynamics} and of vacuum nanoelectronic devices \cite{evtukh2015vacuum}.
  
  Complete study of the electron source definitely requires studies on emission mechanism of the electrons.
  While this study has reached a mature state in conventional flat photocathodes, ongoing research is taking place in pursuit of characterizing the emission properties in nanostructured photocathodes.
  We have shown that electron beam resists can be used to map electron emission from plasmonic nanoantennas with nanometer-scale resolution.
  The doses required to expose PMMA via electrons emission from plasmonic antennas are consistent with those previously measured in low-electron-energy scanning-probe lithography.
  Our simulations of the spatial distribution of the optical near-field and emitted electrons are in good agreement with the observed features of the exposed PMMA.
  These results suggest the possibility of controlling hot electron distributions via nanostructure geometry, and such control presents an opportunity to engineer plasmonic nanoantennas tailored for specific photochemical applications by controlling the location and energy of hot electrons transferred from metallic plasmonic nanoantennas to molecular species at their surface.
  Moreover, as has been highlighted previously, the ability to controllably pattern the surfaces of nanoparticles is desirable for applications in sensing and catalysis.
  As a result, the method outlined in this work may also provide a route to the production of nanoparticles with surface patterns that can be controlled by the methods described here and used for the development of new photocatalyst and optoelectronic systems.
  
  \chapter{Terahertz Gun}
  
  \section{Introduction}
  
  Over the last two decades, ultrabright electron injectors have given rise to new devices for high resolution study of structural dynamics, where the direct observation of atomic motion governing structural transitions is the ultimate dream \cite{siwick2003atomic,dwyer2006femtosecond,zewail20064d}.
  A scrutiny of the electron beam equations and their comparison with wave equations show that emittance in an electron beam plays the same fundamental role as wavelength in an electromagnetic wave.
  For example, the divergence of an electron beam is directly obtained from its emittance value.
  The resolution limit in electron diffractive imaging is affected by this parameter.
  In a FEL, emittance ($\varepsilon$) of the electron beam should be better than the radiated x-ray wavelength ($\varepsilon \leq \lambda/4\pi$) to achieve the optimal FEL performance \cite{schmuser2014free}.
  The effect of this property on the electron beam is the main reason behind the substantial research efforts on photocathodes and photo-injectors to improve the electron beam quality by reducing its emittance \cite{dowell2010cathode}.
  Such achievements often introduce breakthroughs in the pertinent science fields and enable unconventional technologies as well as investigations in fundamental science.
  As a result, numerous approaches are proposed to decrease the emittance of electron beams provided by electron guns.
  The conservation of emittance in relativistic regime preserves this parameters and transfers the value up to the interaction point.
  Many of the proposed techniques to lower the emittance try to reach this goal by increasing the accelerating gradient, thereby reducing the time allowed at sub-relativistic regime for emittance growth.
  
  The achievable accelerating gradient in an injector is known to be the main limiting factor governing the emittance and consequently the length of the output bunch.
  In a conventional particle accelerator, the electrical breakdown of metals introduces a strong limitation on the accelerating fields which are typically 10-100\,MV/m \cite{wang1985measurements,wang1989rf,loew1988rf}.
  This fact turns out to be the major limit determining the maximum accelerating gradients in many large scale facilities like SLAC \cite{wang1985measurements}, CERN's compact linear collider (CLIC) \cite{linssen2012physics} and the design of the next linear collider (NLC) \cite{phinney2002next}.
  Moreover, the small accelerating gradient dictates long accelerator lengths, making it the main impediment in developing compact and therefore lower cost devices employing beams of particles with relativistic energies.
  The desire to realize compact accelerators has spurred much research into the use of alternative acceleration schemes, such as dielectric laser accelerators (DLA) \cite{Peralta2013,England2014,Breuer2013}, laser plasma acceleration (LPA) \cite{tajima1979,Malka1997,Malka2002,geddes2004,Mangles2004,Faure2004,Leemans2006,wang2013,leemans2014,steinke2016}, and THz acceleration \cite{Nanni2015,Wong2013,Yoder2005}.
  
  The empirical studies done by Loew and Wang \cite{loew1988rf,wang1989rf} had initially shown that electron field emission, scaling as $f^{1/2}/\tau^{1/4}$ with $f$ the operation frequency, and $\tau$ the pulse duration of the accelerating field, is the main reason for electrical breakdown \cite{Kilpatrick1957}.
  The above approximate scaling behavior justified research towards higher operating frequencies and ultrafast schemes to achieve compact accelerators \cite{Braun2003}.
  However, the recent comprehensive study on breakdown thresholds of various accelerators \cite{laurent2011experimental,dal2016rf,Dolgashev2010} demonstrated that pulsed heating of the accelerator walls is the dominant factor limiting accelerating gradients.
  This conclusion confirmed the observed lower operational gradients in existing facilities when compared with predictions from the previously derived scaling laws.
  The authors concluded that the pulse duration of the accelerating field plays the major role in the breakdown event, since it is directly linked to the pulse energy governing the pulsed heating in the device.
  The same outcome was also achieved in the framework of CLIC project, where breakdown ratios of different accelerator modules where tested over long operation times \cite{wu2017high}.
  Therefore, focusing efforts on efficient acceleration using short pulses opens new potentials to realize high gradients, which in turn leads to low-emittance bunches and compact devices.
  
  Generally, there is a \emph{conceptual gap} between standard accelerator technology and ultrafast science.
  Microwave and millimeter-wave technology, used in conventional accelerators, are very well developed for producing CW radiation.
  Therefore, accelerators are mostly designed with narrowband excitations.
  Examples are the widely used cascaded cavities which operate based on a resonant behavior and traveling wave accelerators, in which fields of a guided mode are employed for acceleration \cite{wangler2008rf,wiedemann2015particle}.
  Hence, direct usage of a standard accelerator geometry excited by a short pulse laser incurs wasting a large portion of input energy.
  In \cite{fallahi2016short}, we introduced novel structures that aim to accelerate particles from rest using short pulse excitation, which we like to call \emph{single-cycle ultrafast electron guns}.
  The review of the concept, design and implementation of such devices constitute the materials in this chapter.
  
  \section{Ultrafast Single-Cycle Guns}
  
  The last decade has witnessed extensive efforts on acceleration of electrons using optical pulses \cite{Peralta2013,England2014,Breuer2013,tajima1979,Leemans2006,wang2013,leemans2014,steinke2016}.
  However, the acceleration schemes based on optical pulses suffer from the difficulties caused by the short optical wavelengths.
  Some examples are emittance growth of the electron beam, increased energy spread in the bunch, and challenging timing synchronization for optical acceleration.
  On the other side, research in THz pulse generation using optical rectification has led to single-cycle pulses \cite{Hoffmann2011,Fulop2010,Fulop2011,Huang2013}.
  The achieved performance in this process has reached percent level optical to THz conversion efficiency \cite{Huang2013,Schneider2014}.
  Considering that picosecond lasers are necessary for single-cycle THz generation, which have been developed to much higher average power and pulse energies than fs type of lasers, THz acceleration using single-cycle pulses has become a viable option.
  The first sub-keV devices are already realized and the predictions are evidenced based on experimental results \cite{huang2016terahertz}.
  Nonetheless, this scheme similarly demands broadband devices which function based on short pulse excitations.
  
  This section introduces concepts for accelerating particles using single-cycle THz pulses.
  The considered temporal profile of the excitation is a single-cycle pulse described by
  \begin{equation}
  f(t)=A_0 \exp (-2 \ln2 (t-t_0 )^2 / \tau^2) \cos(\omega(t-t_0)+\phi_0 ),
  \label{singleCyclePulse}
  \end{equation}
  where $A_0=A_0(x)$ and $t_0=\pm x/c$ stand for the position dependent amplitude and pulse-center, respectively and $\phi_0$ is the carrier envelope phase of the signal.
  $\omega=2 \pi f_0$ denotes the angular frequency of the signal and $\tau=1/f_0$ is the pulse duration of the single-cycle pulse.
  The temporal signature of such a single-cycle pulse is shown in Fig.\,\ref{singleCyclePulse}.
  \begin{figure}
  	\centering
  	\includegraphics[draft=false,width=3.0in]{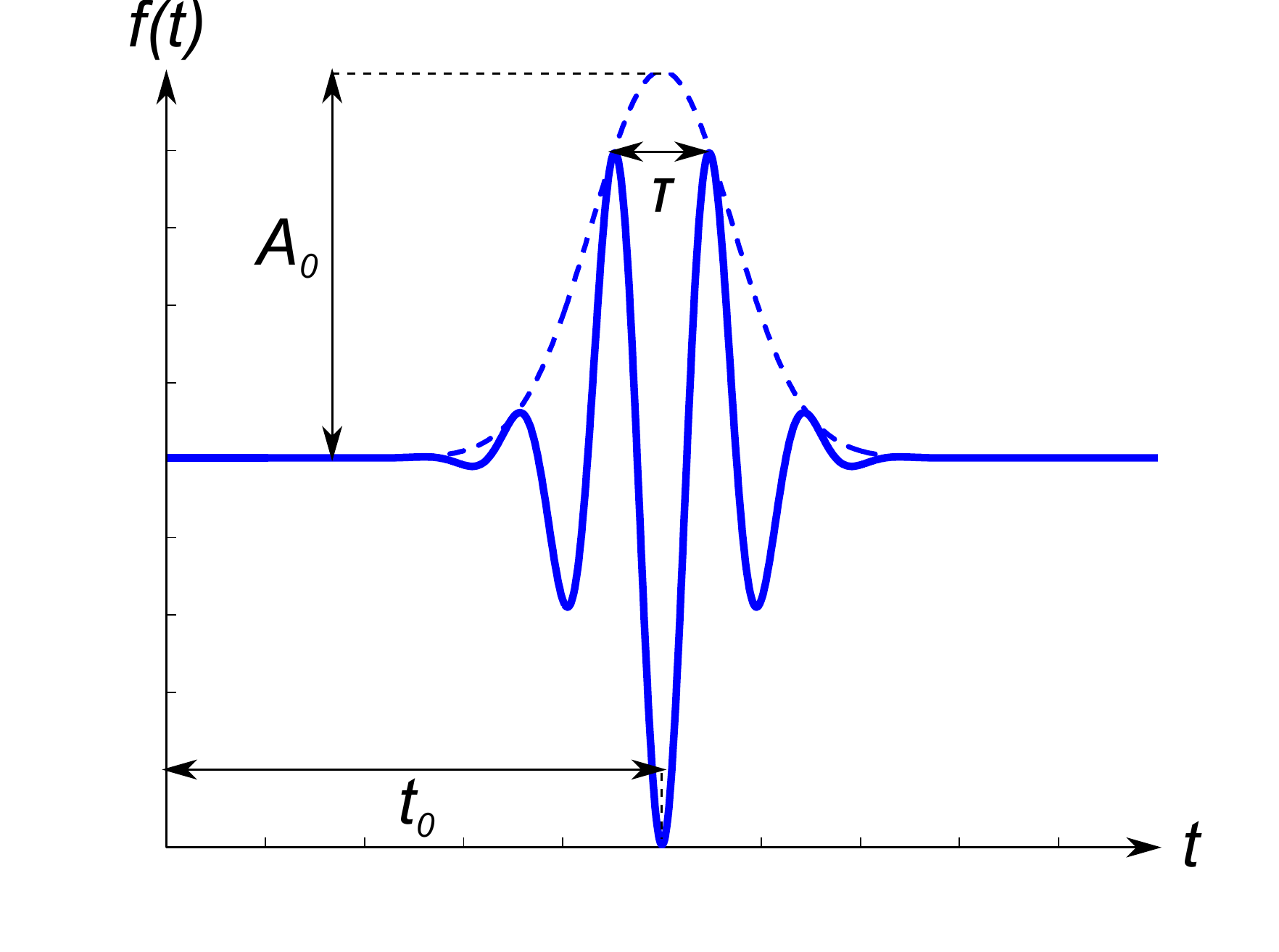}
  	\caption{Temporal signature of a single cycle pulse considered as the excitation in this paper.}
  	\label{singleCyclePulse}
  \end{figure}
  Note that the above solution is an approximate solution for a single-cycle pulse and suffers from the inaccuracy of containing a non-zero DC component.
  However, the error is 0.001 of the peak-field, which is negligible in our study.
  Although we have considered the illustrated single-cycle pulse, the principle also works for few-cycle pulses, at the expense of additional energy.
  We present structures which are useful in two different regimes of (\emph{i}) low energy and (\emph{ii}) high energy THz beams.
  
  Detailed numerical simulations of the introduced structures play a central role in the presented research.
  For this purpose, we use the DGTD/PIC method introduced in chapter 2.
  This software enables capturing all the involved field diffraction effects through the 3D full-vector time-domain solution and reliable computation of the electron trajectories.
  All the bunch evolution calculations in this study are carried out with the consideration of space-charge effects which is simulated using a point-to-point algorithm \cite{fallahi2014field}.
  For initialization of macro-particles in the guns, we have used the ASTRA photoemission model \cite{flottmann2011astra,dowell2009quantum}.
  Note that ASTRA does not simulate the particle acceleration within transient fields and is merely used here for bunch generation.
  
  \subsection{Low-energy Single-cycle Ultrafast Electron Gun}
  
  \subsubsection{Parallel Plate Gun}
  
  Based on the recent demonstration of 1\% level optical to THz conversion efficiencies \cite{Huang2011}, 2-mJ level slightly sub-ps pulses can safely generate 20-{\textmu}J level single-cycle THz pulses typically at 300\,GHz central frequency.
  If this beam is focused down to the diffraction limit, the total electric field at the focus with $2\lambda$ spot size ($\lambda$ is the central wavelength) is about 50\,MV/m.
  In this field, initially at rest electrons are able to move maximally $\delta x= eE\tau/m\omega \simeq 7.5$\,{\textmu}m, being 2-3 orders of magnitude smaller than the wavelength.
  Consequently, the electrons are affected by both accelerating and decelerating cycles, leading to an inefficient acceleration process.
  To acquire an efficient acceleration scheme, two goals must be achieved: 1) The accelerating field should be enhanced in order to lengthen the amplitude of electron vibration, and 2) the electron should leave the pulse before the decelerating cycle begins.
  
\begin{figure}
  	\centering
  	$\begin{array}{c}
  	\includegraphics[draft=false,width=4.5in]{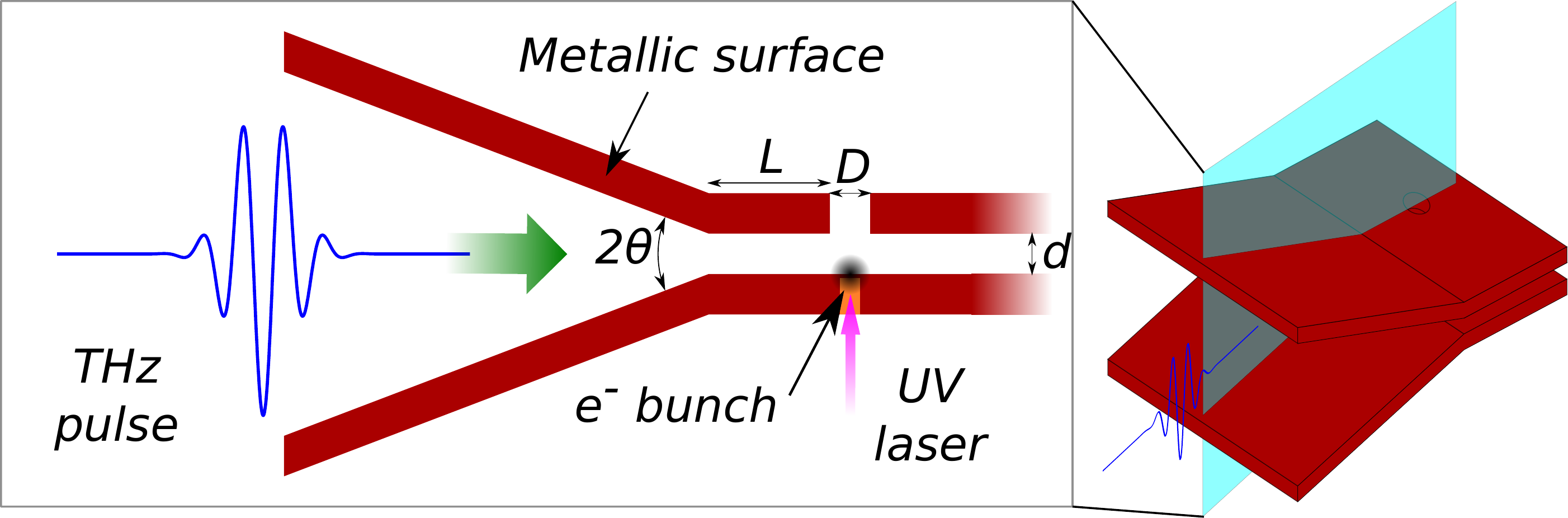} \\
  	(a) \\
  	\begin{array}{cc}
  	\includegraphics[draft=false,width=3.0in]{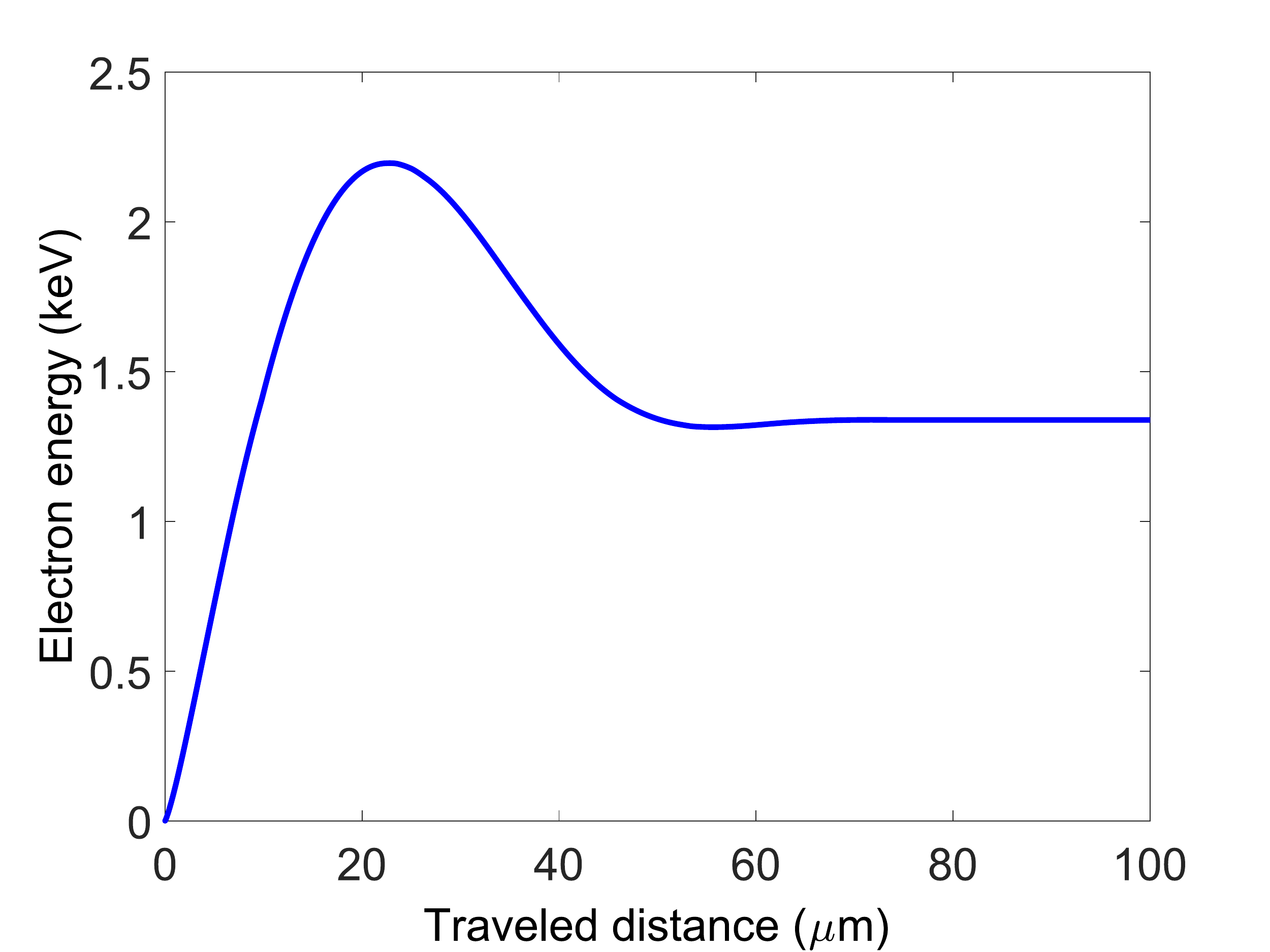} &
  	\includegraphics[draft=false,width=3.0in]{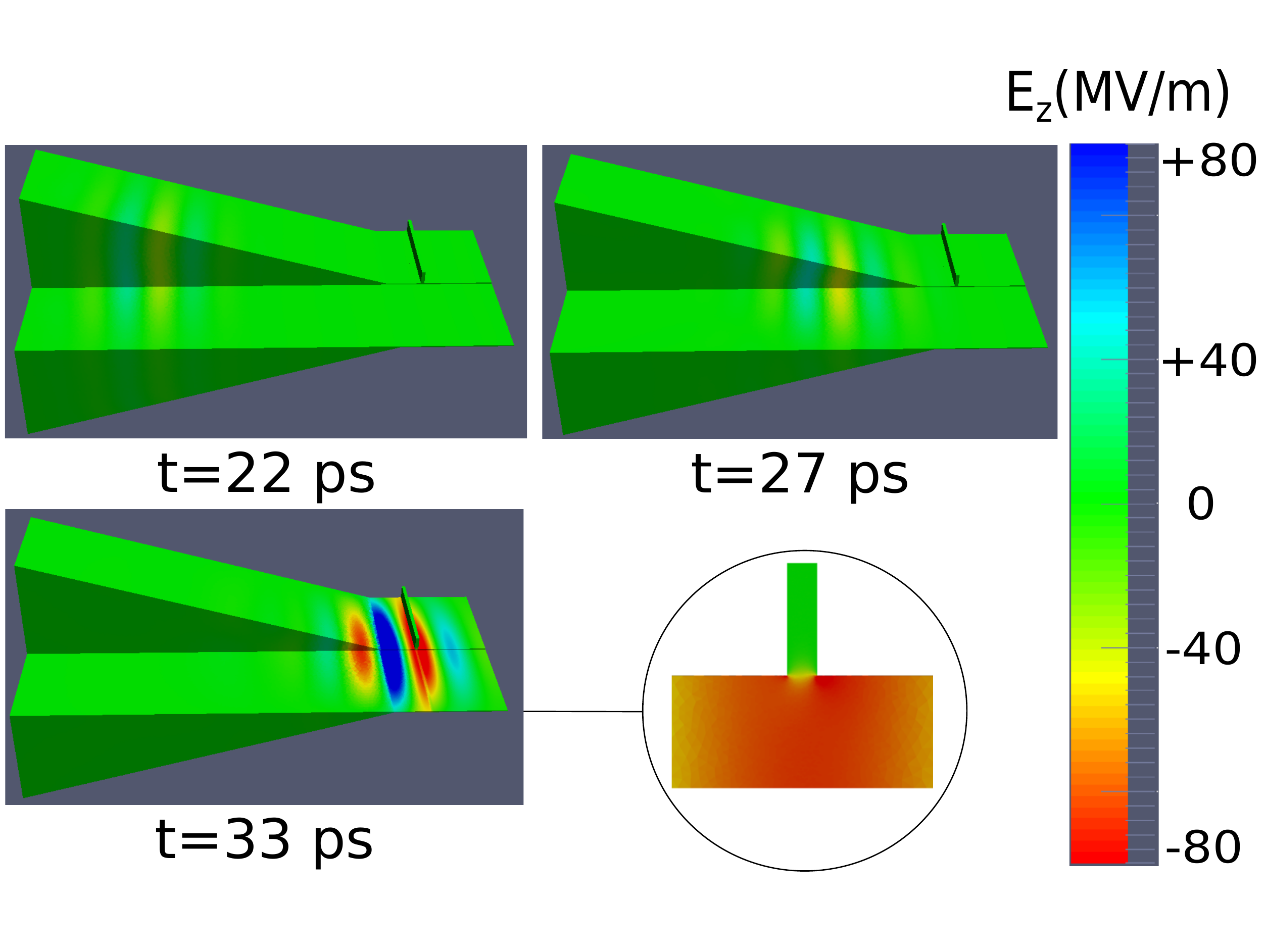} \\
  	(b) & (c) \end{array}
  	\end{array}$
  	\caption{(a) Schematic illustration of the planar single cycle THz gun, (b) Energy of an electron released at the instant with $E_z=50\,\text{MV/m}$ versus the traveled distance, (c) Snapshots of the accelerating field ($E_z$) profile over two half-space cuts of the gun at $t_1=22\,\text{ps}$, $t_2=27\,\text{ps}$, and $t_3=33\,\text{ps}$.}
  	\label{lowEGunConceptAcceleration}
  \end{figure}

  Fig.\,\ref{lowEGunConceptAcceleration}a schematically shows the idea for a two-dimensional (planar) device that pursues the above goals.
  Two metallic plates form a structure like a 2D horn receiver antenna to focus the incoming linearly polarized THz beam below the diffraction limit.
  Although the insertion loss of the incident energy is unavoidable, the confined THz beam travels through the region between the two plates and reaches the injection point of electrons.
  The photocathode laser (usually a UV laser synchronized with the THz pulse) releases an electron bunch from the cathode surface, when the accelerating field of the THz pulse arrives at the injection point.
  The electrons are then accelerated by the incoming THz beam and leave the acceleration region after a distance $d$, i.e. the separation between the two plates.
  The value of $d$ is designed so that the electrons experience merely the accelerating half-cycle of the THz pulse.
  The input THz beam is a 20-{\textmu}J single-cycle Gaussian pulse with central frequency $f_0=300\,\text{GHz}$ and focused to a spot size diameter equal to 2\,mm ($2\lambda$).
  We assume $L=500$\,{\textmu}m and $D=50$\,{\textmu}m and optimize the other dimensions for best acceleration, which yields $\theta=16^{\circ}$ and $d=30$\,{\textmu}m.
  The designed gun realizes a three-fold enhancement in the peak accelerating field ($E_z^{max}=150\,\text{MV/m}$).
  Using the PIC simulations, it is demonstrated that an electron released at the instant with accelerating gradient $E_0=50\,\text{MV/m}$ ($\phi = 20^{\circ}$) gains $1.33\,\text{keV}$ energy when leaving the acceleration region.
  A gun with similar parameters was recently demonstrated with energies up to 0.8\,keV, which is discussed later in this chapter \cite{huang2016terahertz}.
  Fig.\,\ref{lowEGunConceptAcceleration}b and c show the energy of the electron along its acceleration path as well as the profile of the accelerating field ($E_z$) at three different time points.
  
  \begin{figure}
  	\centering
  	$\begin{array}{c}
  	\includegraphics[draft=false,width=4.0in]{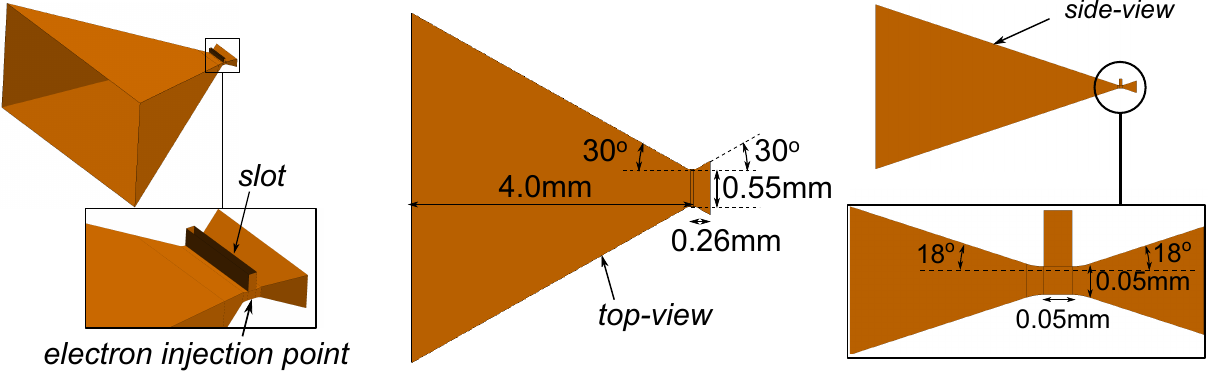} \\
  	(a) \\
  	\begin{array}{cc}
  	\includegraphics[draft=false,width=2.9in]{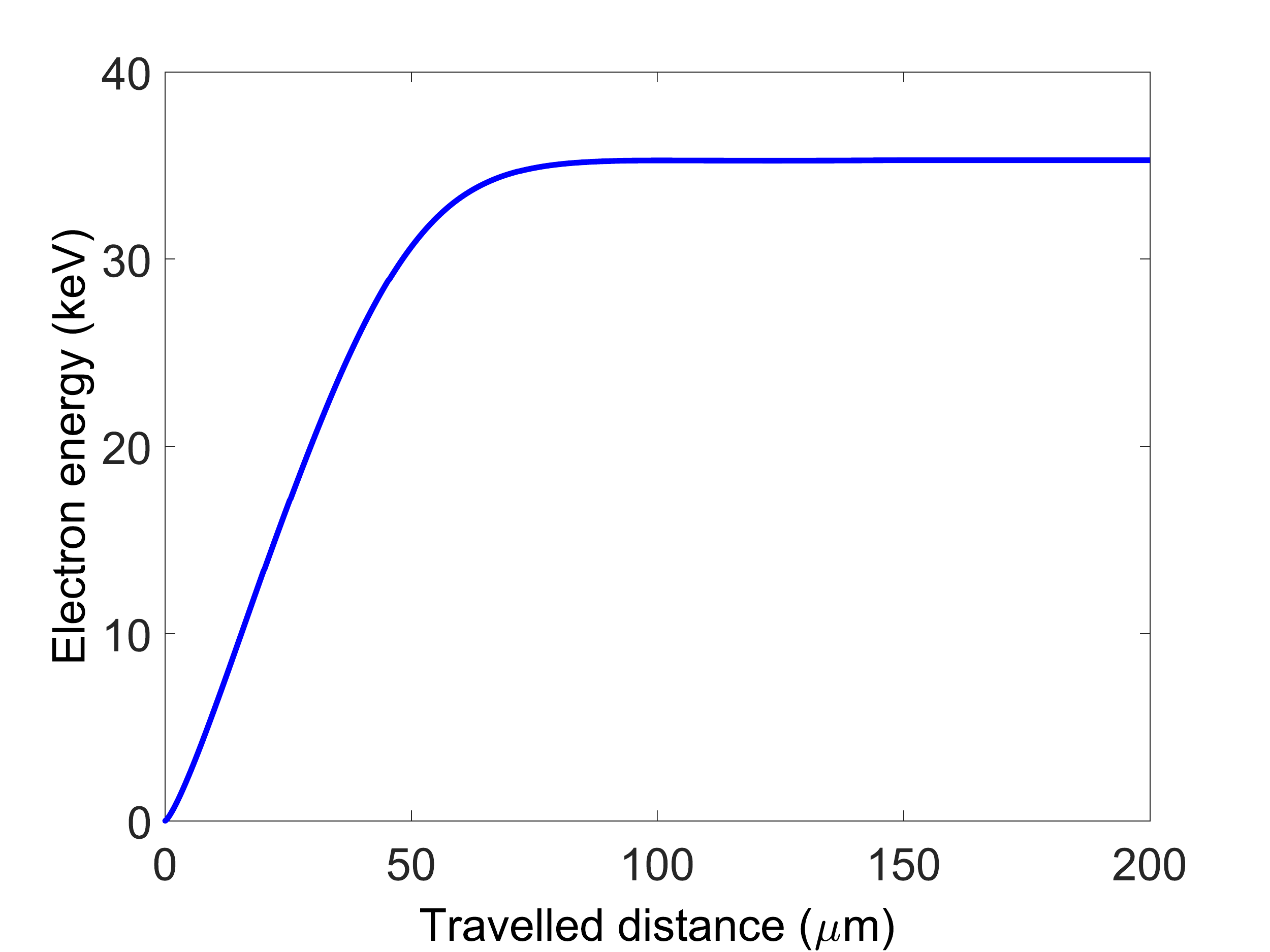} &
  	\includegraphics[draft=false,width=3.0in]{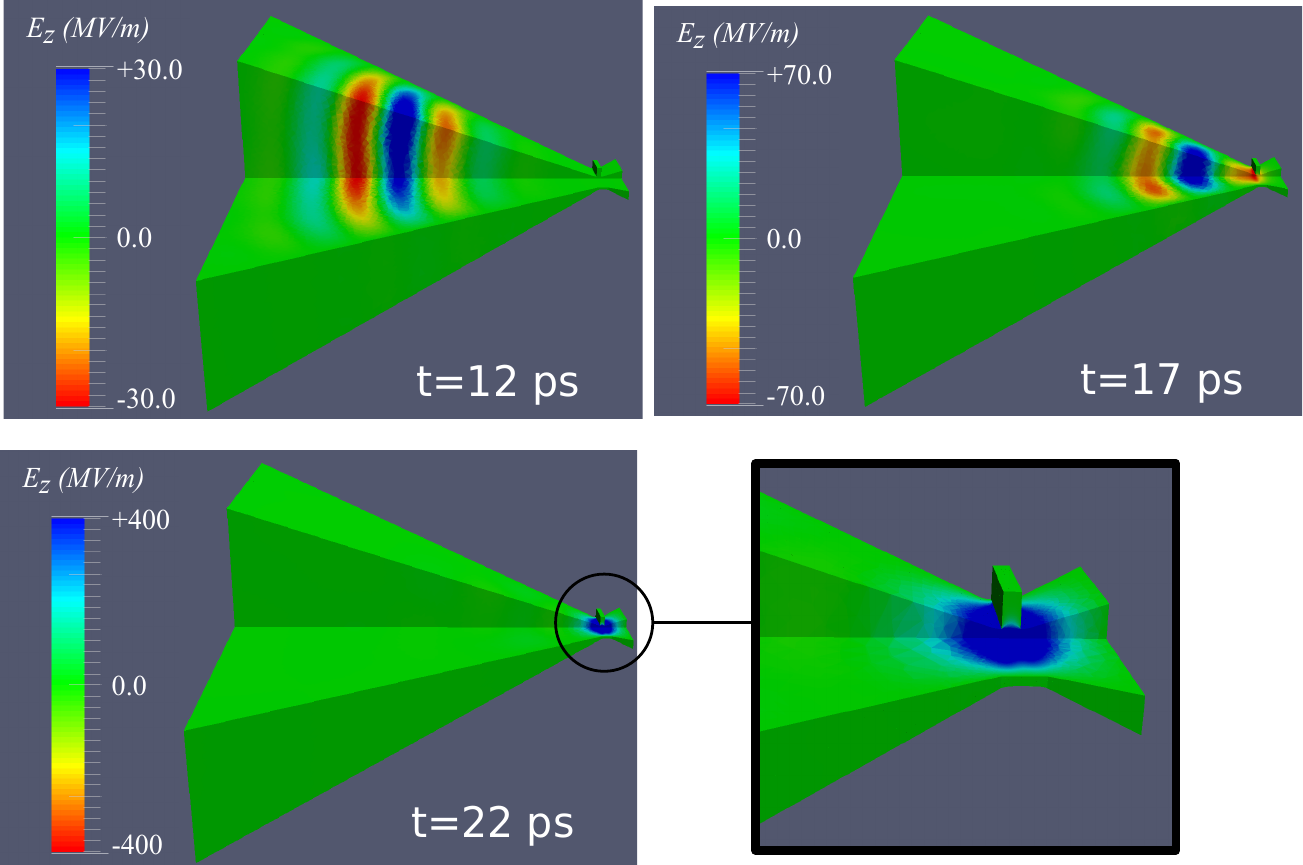} \\
  	(b) & (c) \end{array}
  	\end{array}$
  	\caption{(a) Schematic illustration of the low energy ultrafast electron gun, (b) Energy of the accelerated electron versus the traveled distance, and (c) Snapshots of ($E_z$) profile over two half-space cuts of the gun. Note the change in the color map scaling for different snapshots.}
  	\label{lowEGunFinal}
  \end{figure}
  
  \subsubsection{Horn Gun}
  
  There exist several techniques to enhance the efficiency of the above concept in planar devices:
  (1) In a planar device, the focusing of the THz beam is carried out only in the vertical plane, i.e. the E-plane. The same focusing can also be introduced in the H-plane to further enhance the accelerating field.
  (2) The focusing in the H-plane introduces cut-off frequencies to the wave propagation. Therefore, the length of the injection region should be reduced to a fraction of the wavelength to enable the \emph{tunneling} of the accelerating field into the acceleration point.
  (3) Adding a reflector at the receiving side of the structure with $\lambda/4$  distance from the electron injection point, causes the preceding decelerating half cycle to be inverted and added to the accelerating cycle upon reflection. Therefore, the total accelerating gradient at the injection point is enhanced.
  (4) The same cut-off frequency effect holds also for the receiving side of the structure. Consequently, structuring the right side of the gun similar to the left side enhances the tunneling and thereby increases the accelerating gradient.
  By taking the above considerations into account, an ultrafast electron gun driven by a low-energy single-cycle THz pulse is designed as shown in Fig.\,\ref{lowEGunFinal}a.
  The structure consists of two oppositely standing horn couplers (with different lengths), which realize a high accelerating gradient within a single cycle.
  Thus, the design is named as a \emph{horn gun}.
  The energy of an electron at rest, injected at the instant with vertical field $E_z=50\,\text{MV/m}$, in terms of travel distance as well as snapshots of the accelerating field profile in the device are shown in Fig.\,\ref{lowEGunFinal}b and c.
  The simulations evidence an enhancement of the accelerating gradient by a factor of 15, leading to a peak acceleration field of $782\,\text{MV/m}$.
  The final energy of the electron leaving the gun is $35.3\,\text{keV}$, being ideal for electron diffraction imaging.
  
  \begin{figure}
  	\centering
  	$\begin{array}{ccc}
  	\includegraphics[draft=false,width=2.0in]{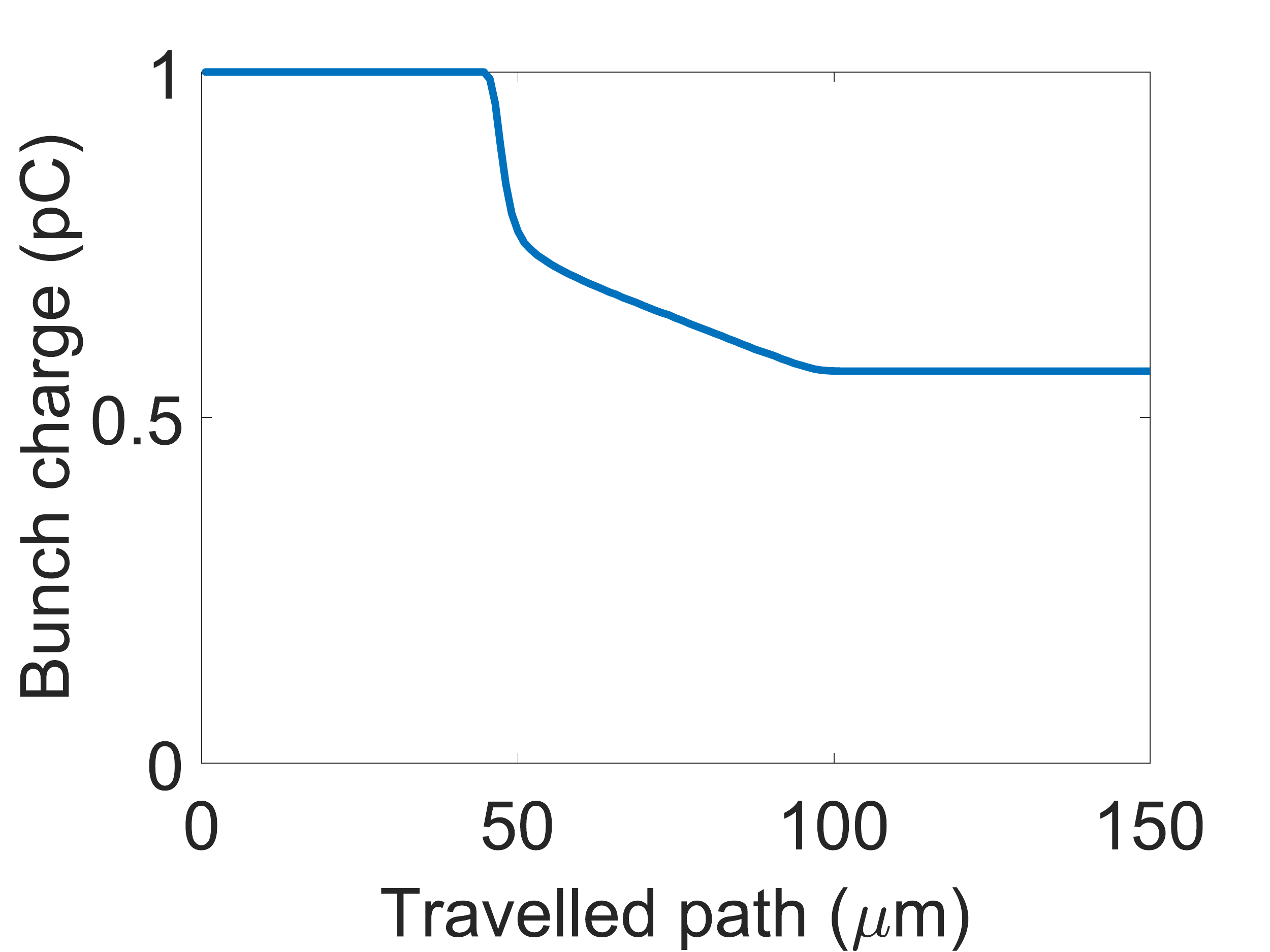} &
  	\includegraphics[draft=false,width=2.0in]{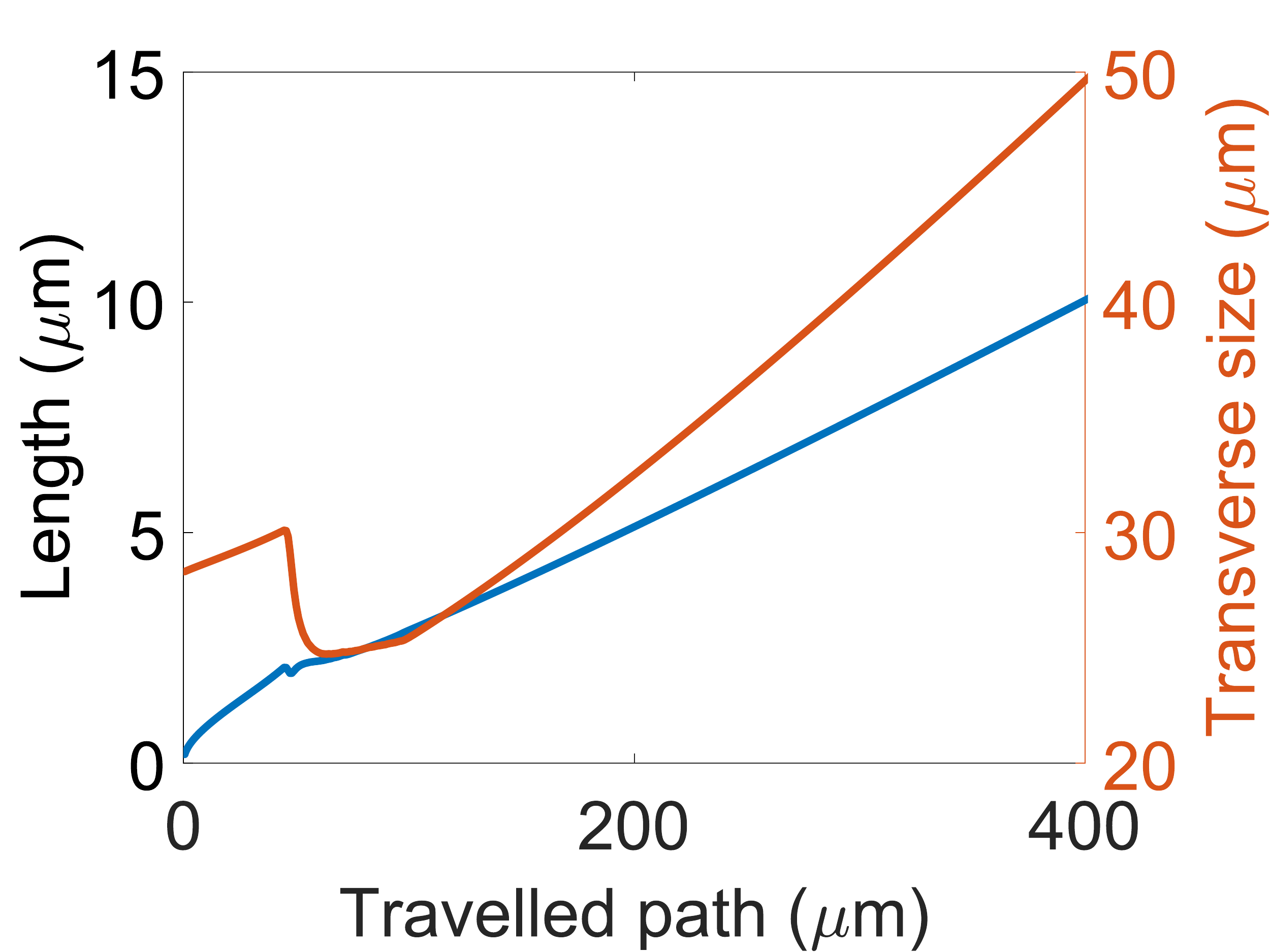} &
  	\includegraphics[draft=false,width=2.0in]{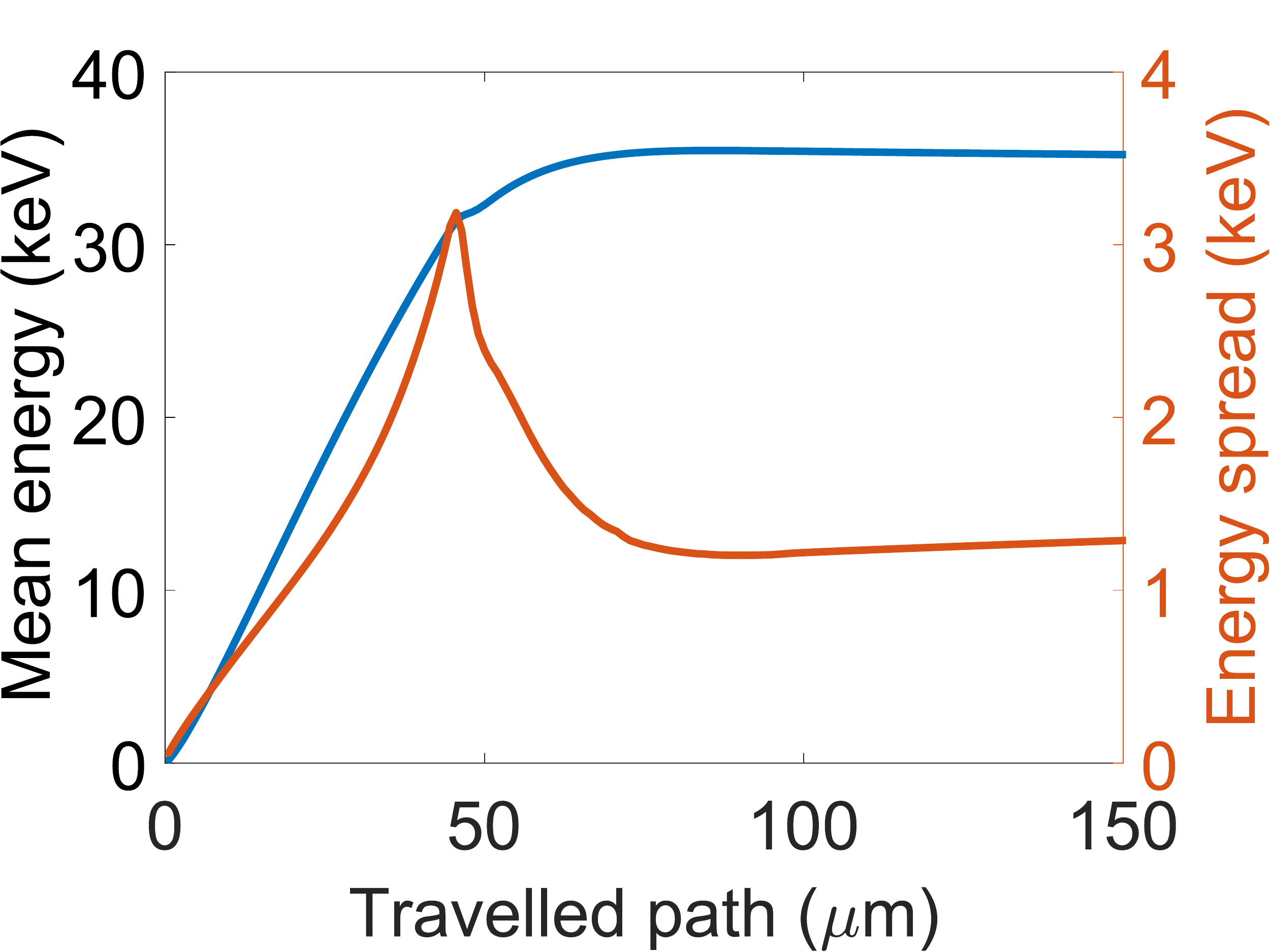} \\
  	(a) & (b) & (c) \\
  	\includegraphics[draft=false,width=2.0in]{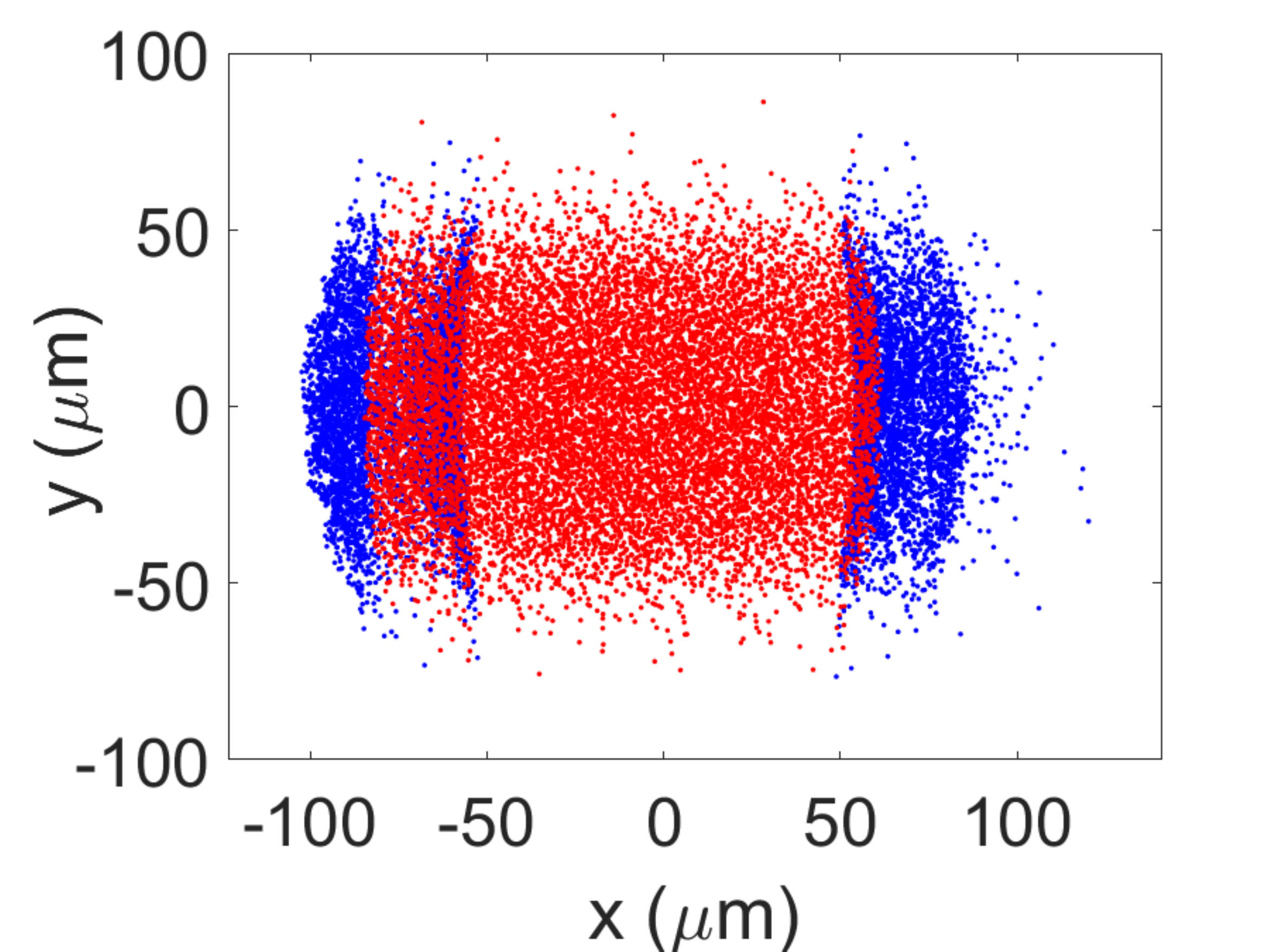} &
  	\includegraphics[draft=false,width=2.0in]{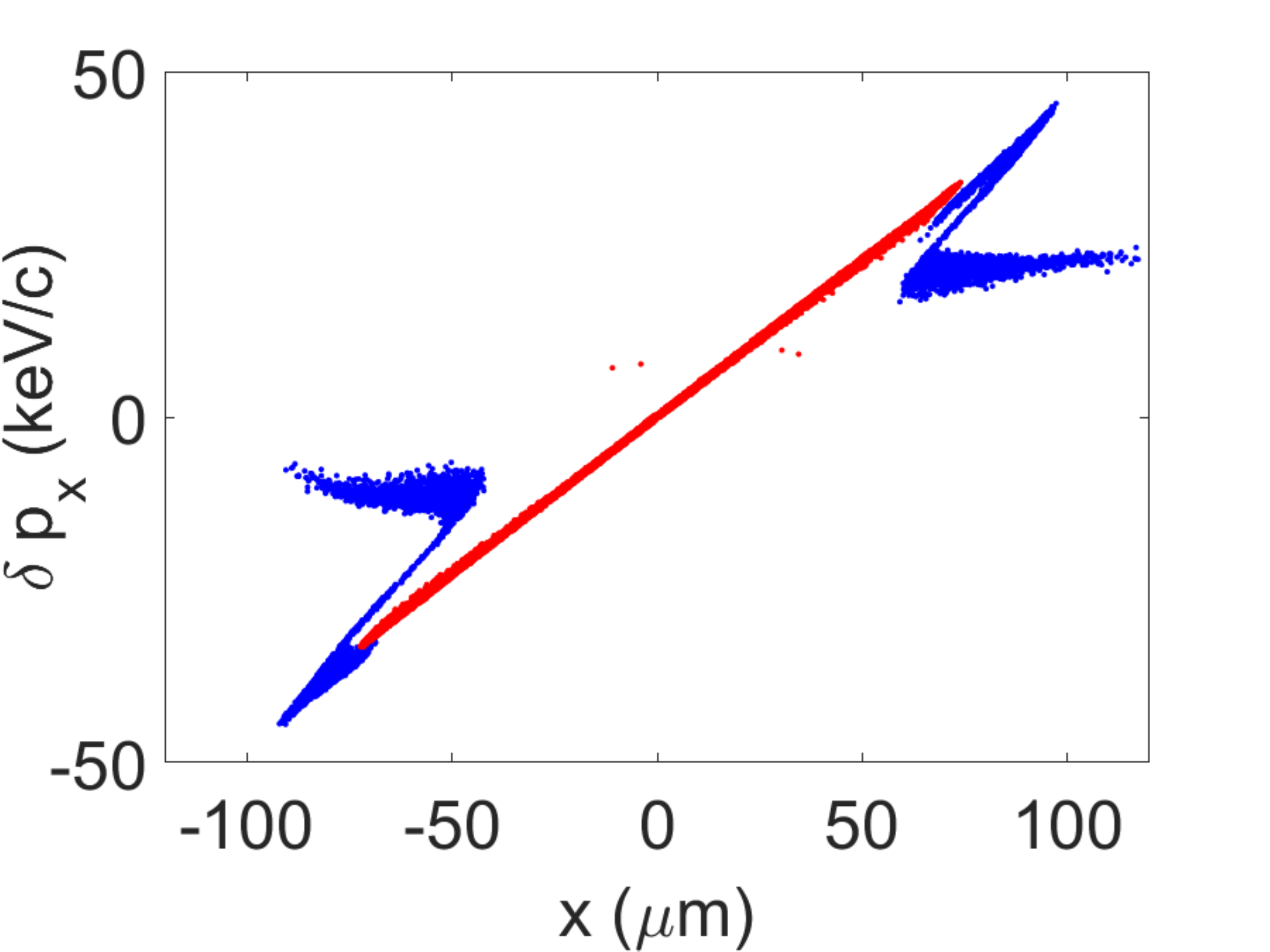} &
  	\includegraphics[draft=false,width=2.0in]{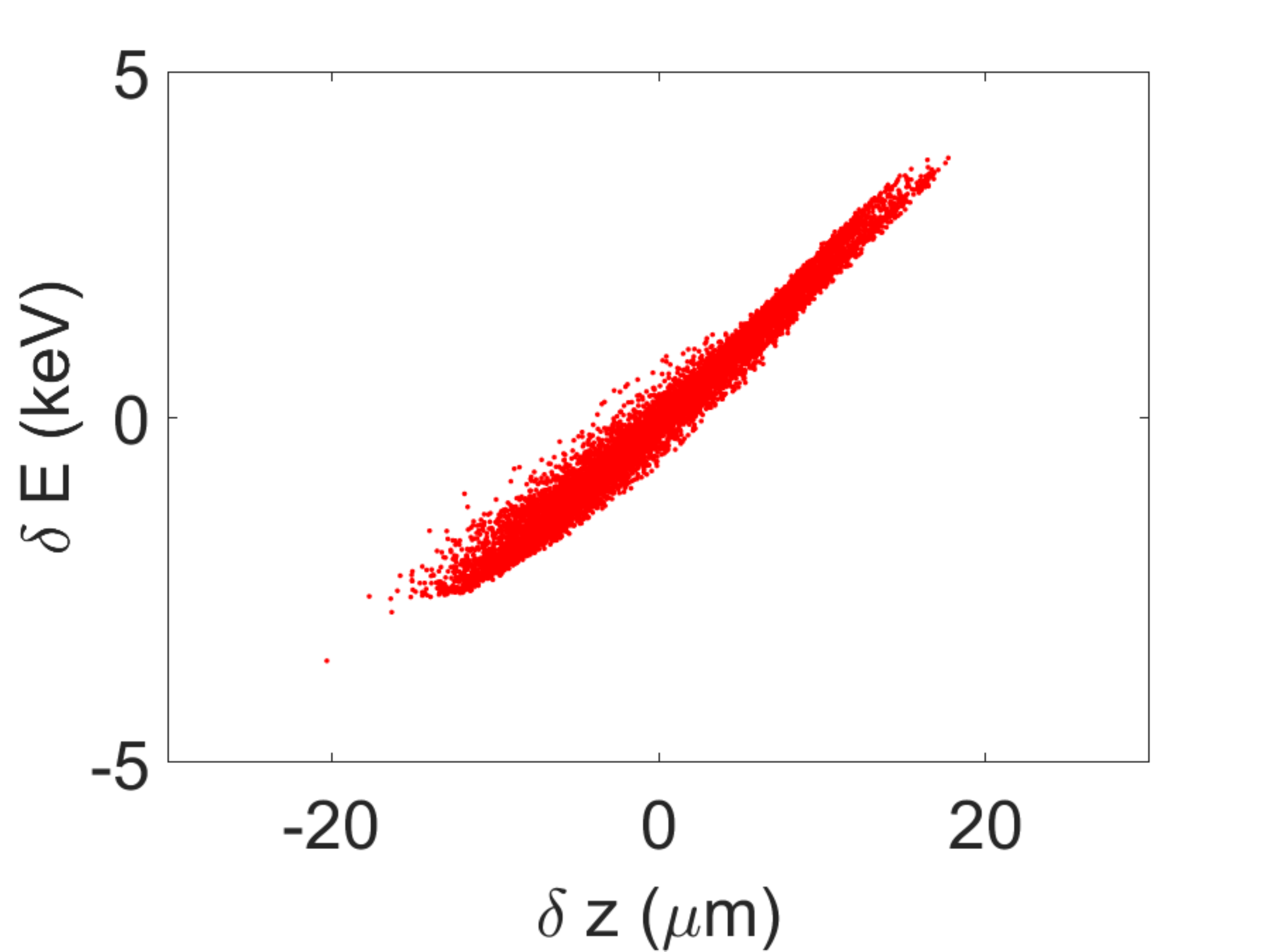} \\
  	(d) & (e) & (f)
  	\end{array}$
  	\caption{Acceleration of 1\,pC photoemitted bunch in the high energy THz gun: (a) bunch charge, (b) size and (c) energy in terms of travelled path and (d) shape, (e) transverse phase space, and (f) longitudinal phase-space 100\,{\textmu}m from gun exit are depicted (red dots: output particles; blue dots: trapped particles).}
  	\label{lowEGunFinalAcceleration}
  \end{figure}
  For the study of bunch evolution, we assume a copper cathode excited by a UV laser pulse with pulse duration equal to 40\,fs and spot size diameter 40\,{\textmu}m.
  The UV laser energy is chosen such that 1\,pC of charge is released which is modeled by 20'000 macro-particles.
  The evolution of the bunch properties as well as snapshots of the bunch profile are shown in Fig.\,\ref{lowEGunFinalAcceleration}.
  The mean energy of the bunch increases to 35\,keV with an energy spread of about 3\%, which happens due to the large spot size of the injected bunch compared to the THz wavelength (1\,mm).
  It is observed that because of the collisions of the electrons with the metallic boundaries due to the transverse momentum of electrons (Fig.\,\ref{lowEGunFinalAcceleration}e), only 57\% of the photo-emitted electrons are extracted out of the gun.
  This effect shows the limitation on the bunch size and correspondingly the amount of charge which can be accelerated with \emph{good} quality using the proposed THz gun.
  Our simulations show that placing the introduced gun within a longitudinal magnetic field, i.e. producing the so-called magnetized beam, enhances the aptitude of the gun in terms of accelerated bunch charge.
  For this purpose, structures producing 1\,T-level magnetic fields are required \cite{fallahi2016short}.
  The final normalized emittances of the bunch are $(\varepsilon_{n,x},\varepsilon_{n,y},\varepsilon_{n,z})$=(0.02,0.06,0.013)\,mm$\cdot$mrad and the output bunch length is about 95\,fs.
  
\begin{figure}
  	\centering
  	$\begin{array}{cc}
  	\includegraphics[draft=false,width=2.5in]{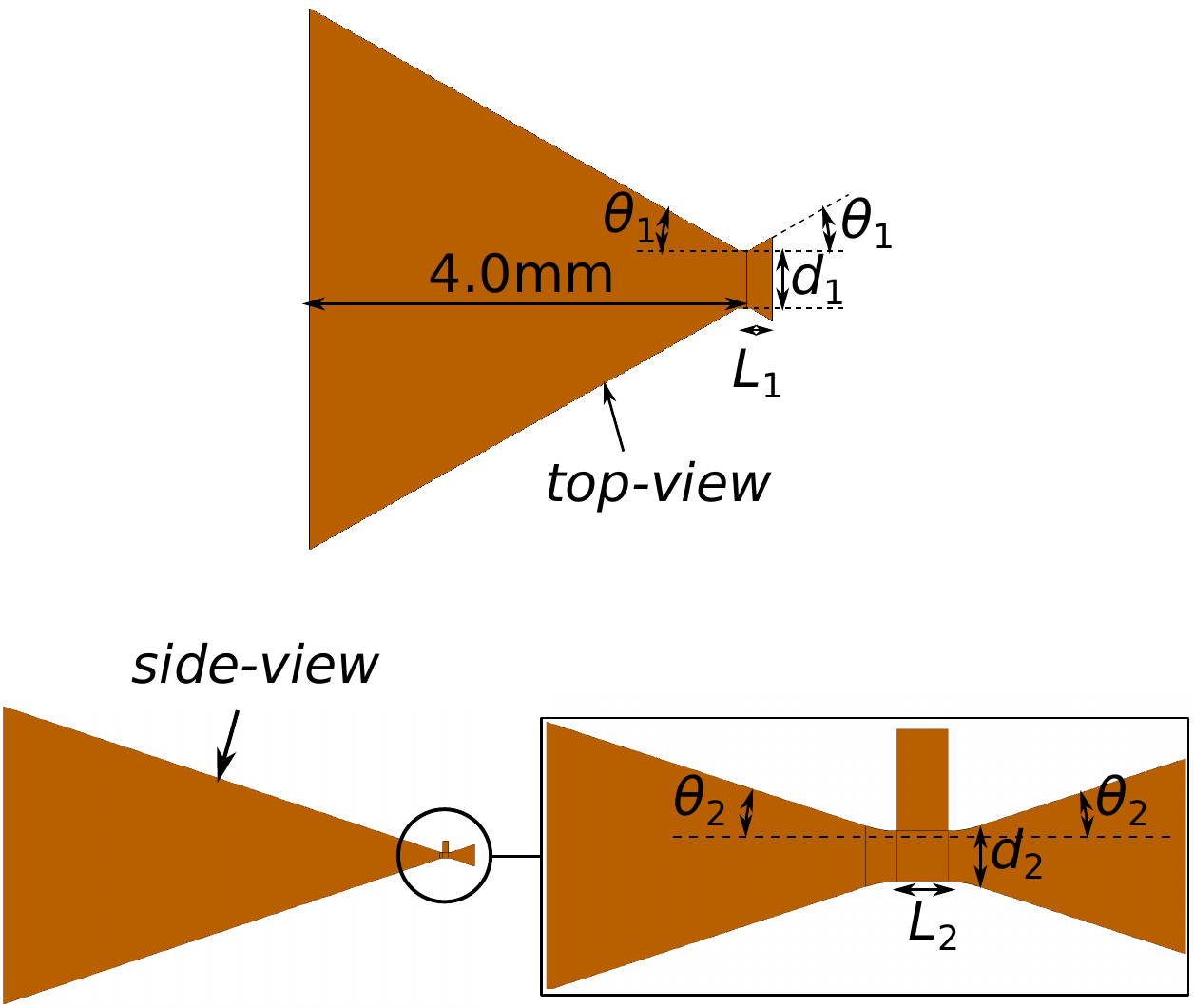} &
  	\includegraphics[draft=false,width=3.0in]{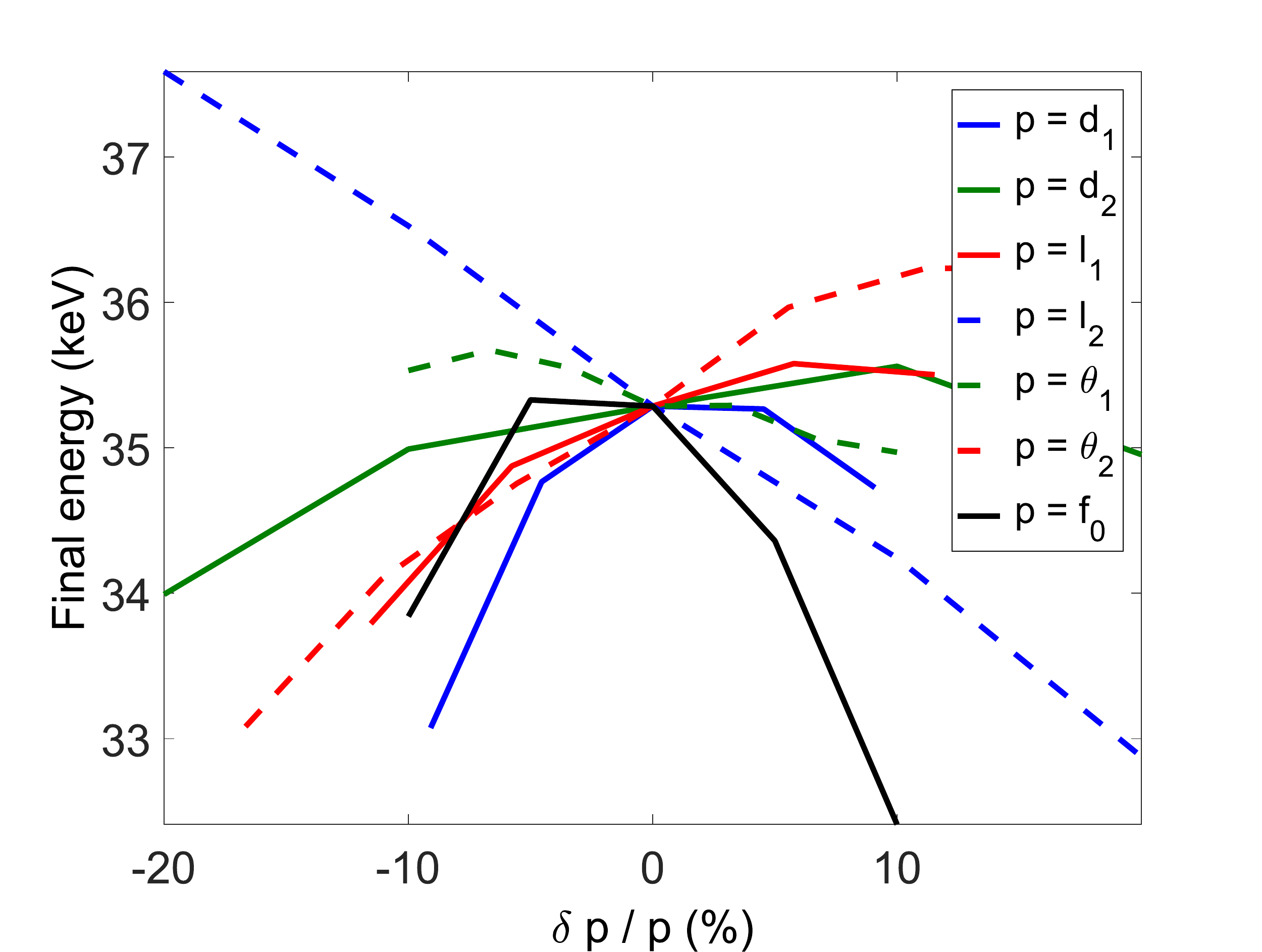} \\
  	(a) & (b) \\
  	\end{array}$
  	\caption{Results of the sensitivity study for the horn gun: (a) schematic illustration of the dimensions studies, and (b) final energy of the electron bunch in terms of the relative variation in the studied dimensions. The legend on the left of the plot introduces the studied parameters. }
  	\label{lowEGunSensitivity}
  \end{figure}

  In addition to the bunch acceleration study, a sensitivity analysis of the introduced gun is also of utmost importance.
  For this purpose, we change each of the values independently by $\pm$10\% from the optimum design, study the THz propagation, inject electrons at the instant when the accelerating field is 50\,MV/m, and evaluate the acceleration performance.
  The results of this analysis evidence maximum 3\% change in the final energy as the sensitivity of the device to implementation errors (Fig.\,\ref{lowEGunSensitivity}).
  This is a very promising sensitivity compared to conventional electron guns, which has its origin in the broadband operation of the device.
  
  The concept of the THz-driven horn gun is also tested as a source of the ultra-short electron bunches.
  To characterize the low-energy, low-charge beam produced by such a gun tailored diagnostic devices were developed and commissioned at a test-stand chamber in CFEL (DESY).
  Results of the first experiments on the production and characterization of the electron beam presented in \cite{vashchenko2017characterization} demonstrate the feasibility of this concept.
  
  \subsection{High-energy Single-Cycle Ultrafast Electron Guns}
  
  \subsubsection{Two-dimensional Concept}
  
  In the above designs, it was observed that optimum focusing of the THz beam with only 20\,{\textmu}J energy leads to peak fields as large as 0.8\,GV/m on the electron emitter, which is close to the damage threshold of copper and other metals \cite{dal2016rf}.
  This means that increasing the energy of the input THz beam to achieve higher acceleration rates is not realistic.
  However, today's THz generation technology has realized higher THz energy levels \cite{Schneider2014}.
  Consequently, an issue is illuminated; how can one achieve efficient acceleration using high energy short pulses without surpassing the damage threshold?
  Here, we try to answer this question by introducing structures which operate based on single-cycle THz beams with around 2\,mJ energy at 300\,GHz central frequency.
  
  For this purpose, two important points must be taken into account:
  (\emph{i}) The electron may gain relativistic energy, which intensifies the effect of the transverse magnetic field of the THz pulse. This effect causes a push from the THz pulse along its propagation direction and leads to a curved trajectory for the electron motion.
  (\emph{ii}) A high-energy THz beam should not be focused to small spot-sizes to avoid dark current excitation.
  As a consequence, to achieve an efficient acceleration, matching the phase front of the THz beam with the electron trajectory is essential.
  
\begin{figure}
  	\centering
  	\includegraphics[draft=false,width=6.0in]{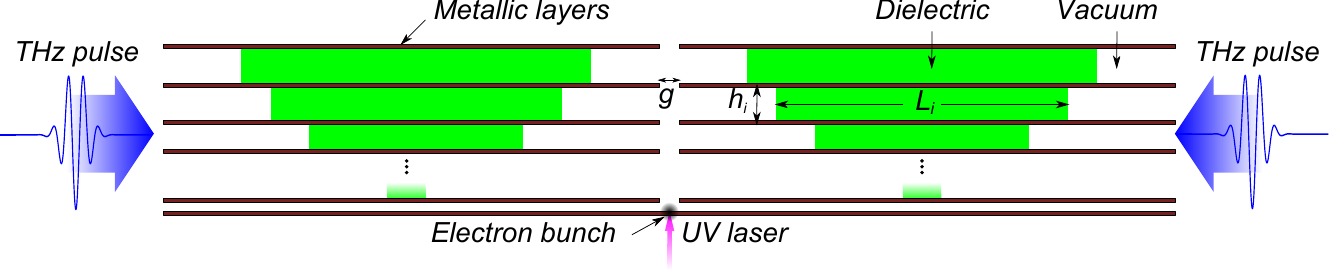}
  	\caption{Schematic illustration of the high-energy ultrafast gun concept.}
  	\label{2DConcept}
  \end{figure}

  The configuration illustrated in Fig.\,\ref{2DConcept} is a 2D presentation of the concept devised to solve the above two problems.
  \begin{figure}
  	\centering
  	$\begin{array}{cc}
  	\includegraphics[draft=false,width=2.7in]{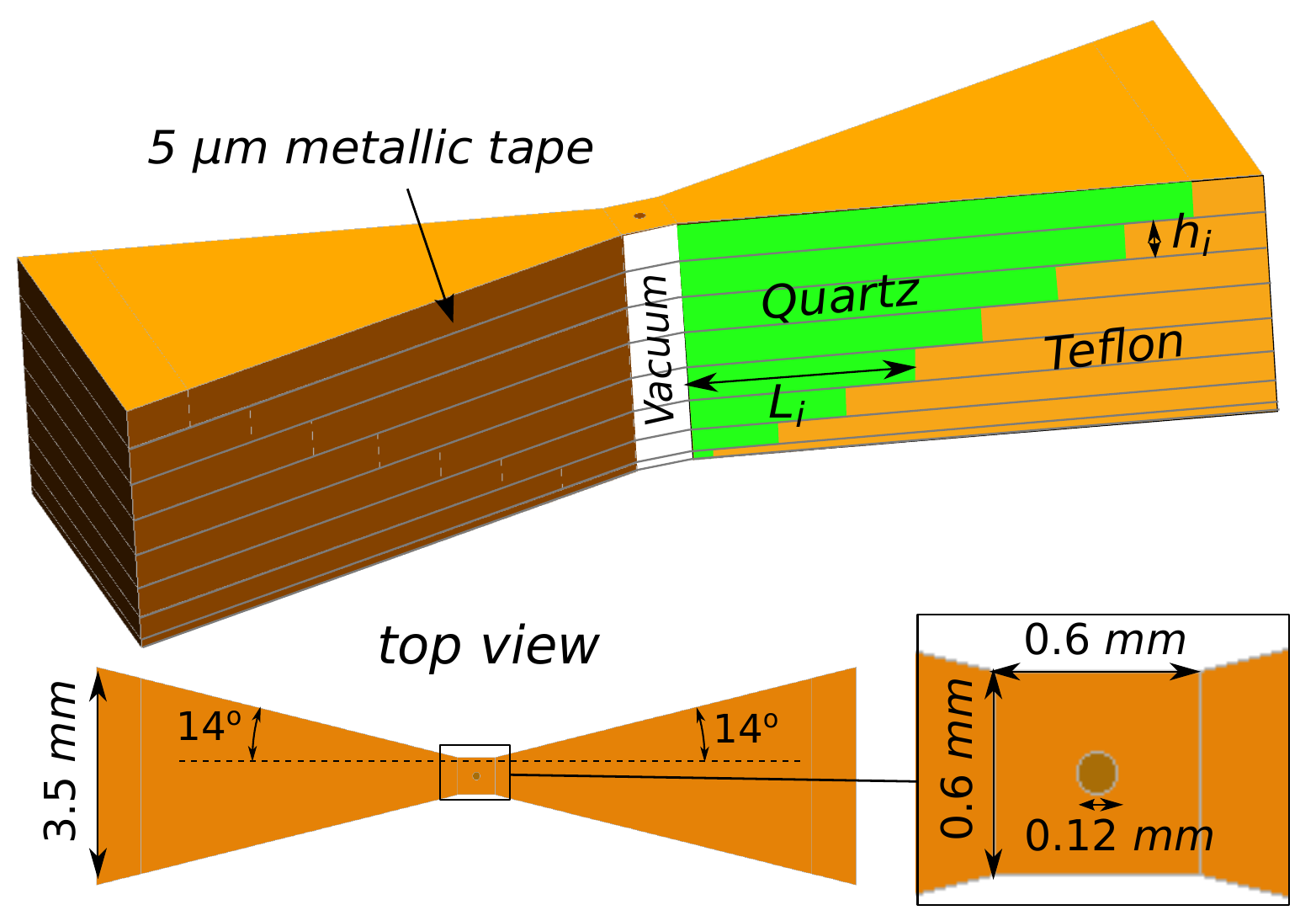} &
  	\includegraphics[draft=false,width=2.7in]{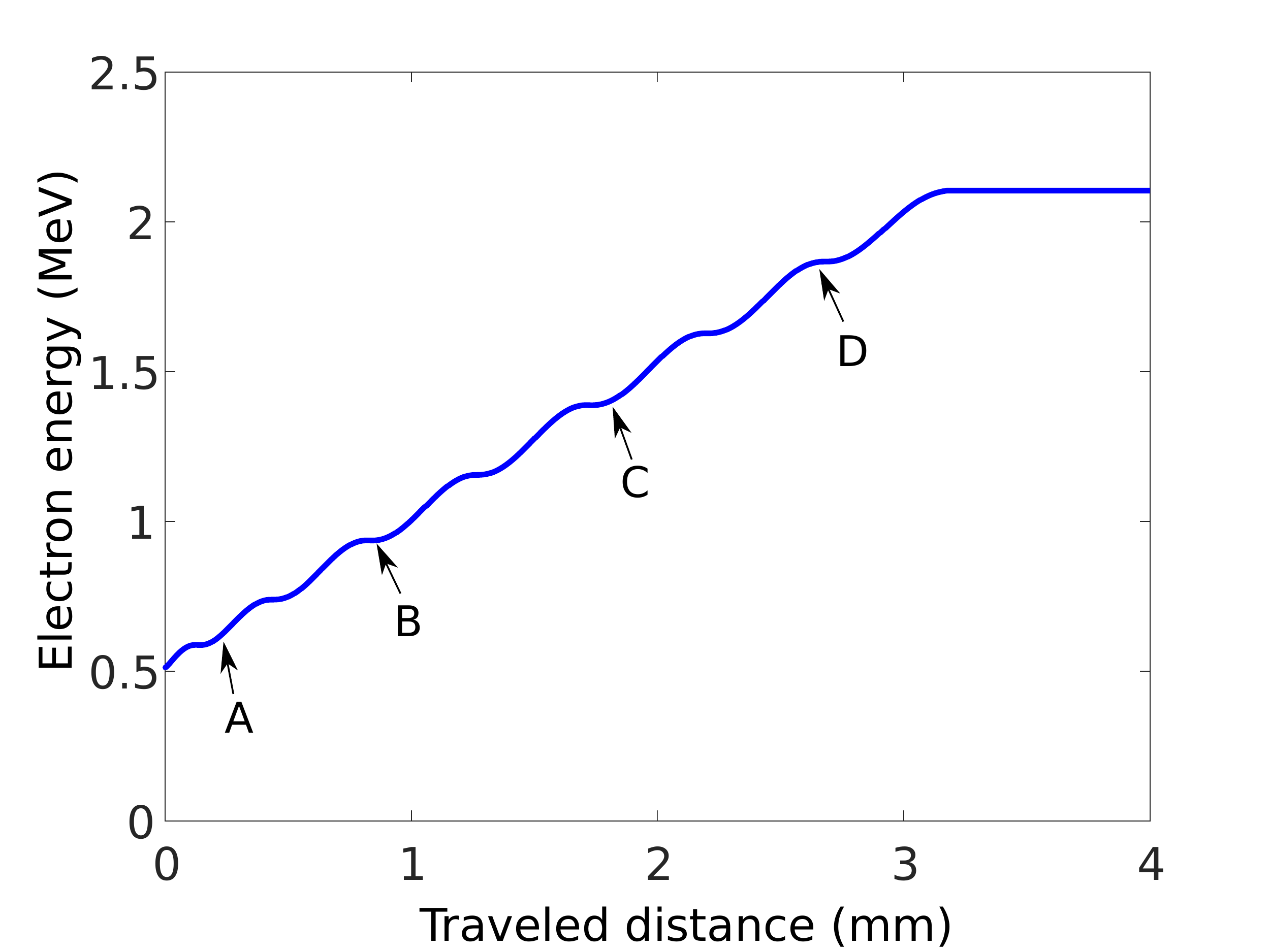} \\
  	(a) & (b) \\
  	\includegraphics[draft=false,width=2.7in]{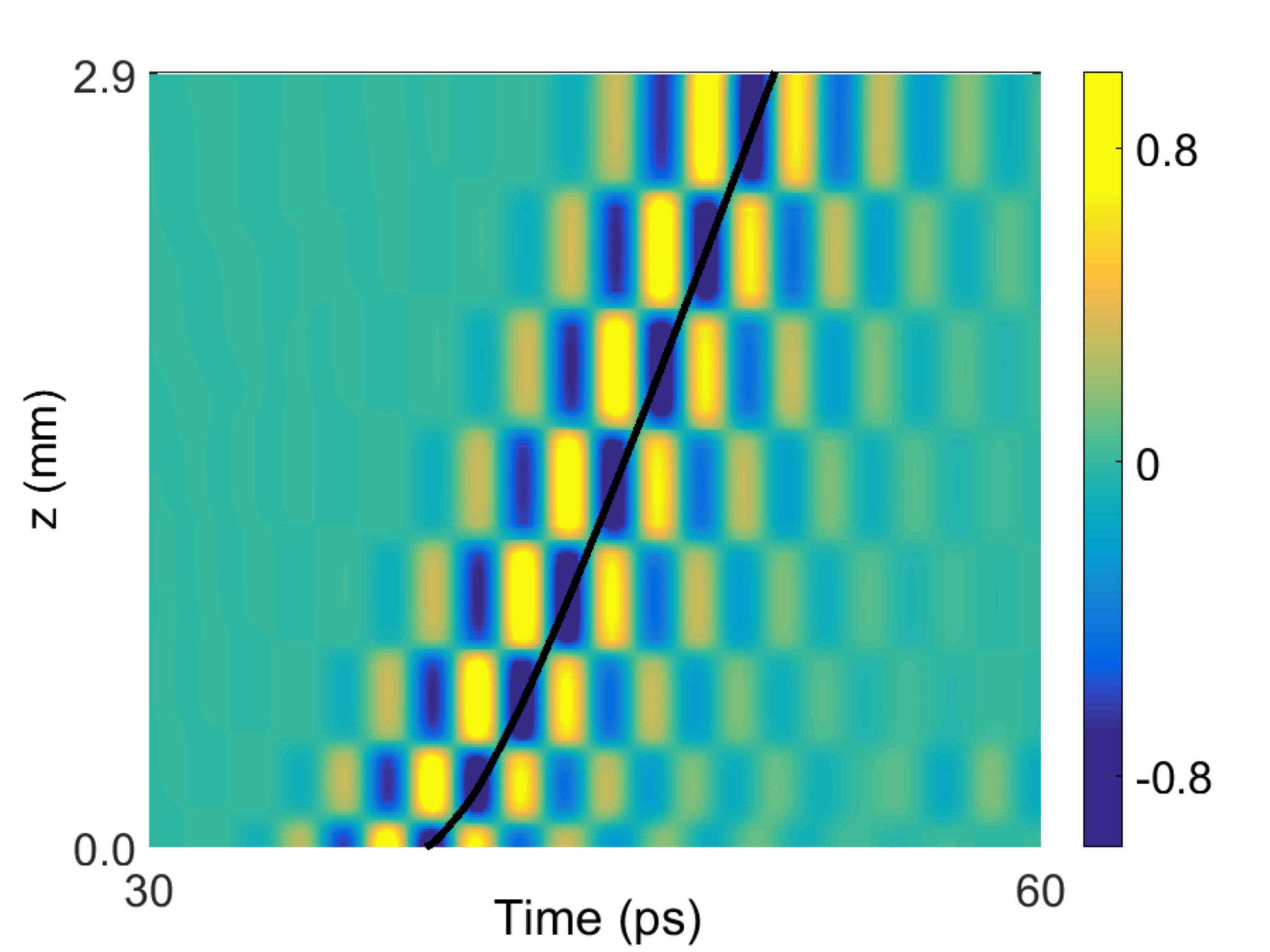} &
  	\includegraphics[draft=false,width=2.7in]{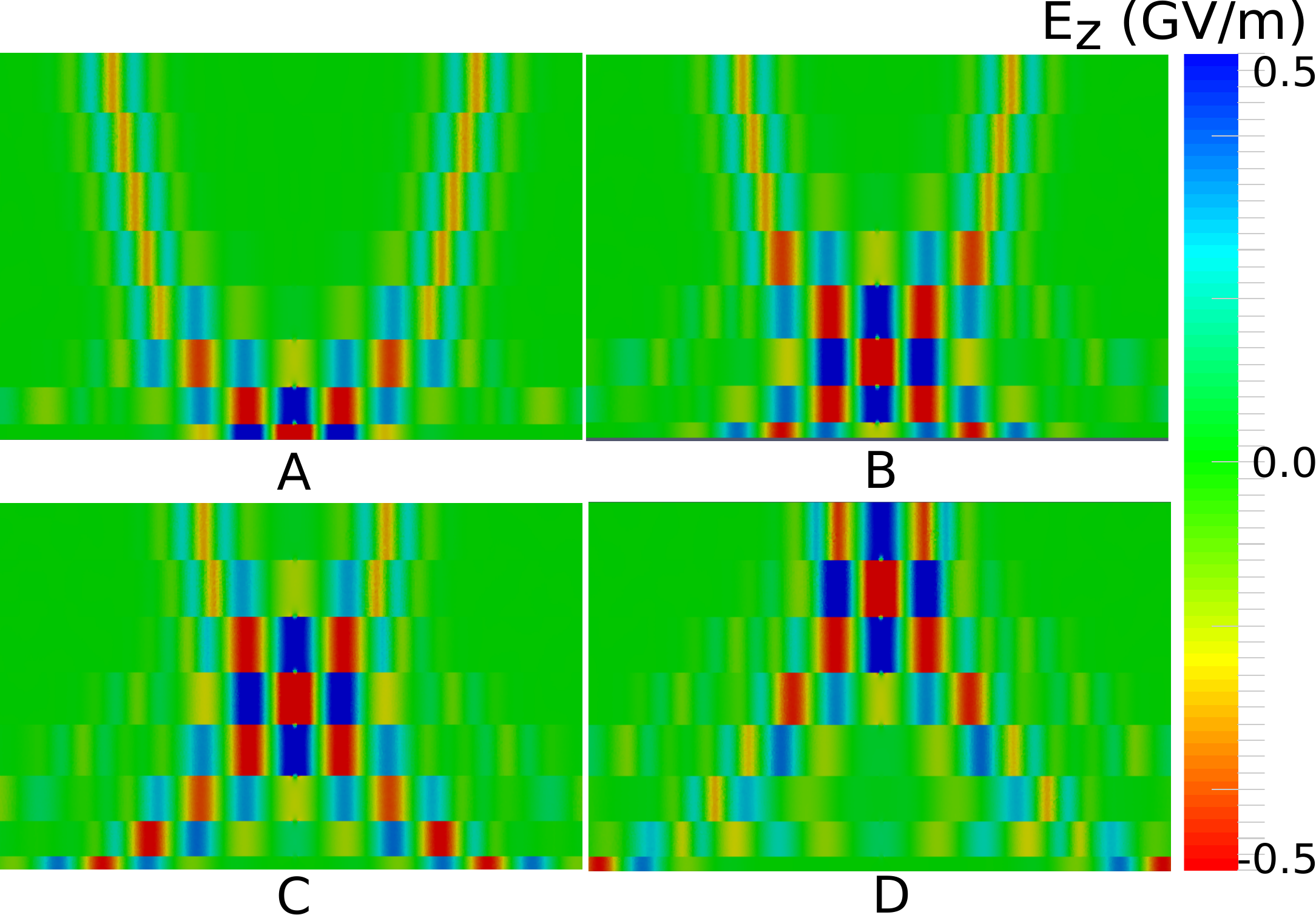} \\
  	(c) & (d) \\
  	\end{array}$
  	\caption{(a) schematic illustration of the optimized ultrafast electron gun, (b) electron energy $E_e(z)$ versus the traveled distance, (c) electron energy versus position of the electron superposed on the accelerating field map $E_z(z,t)$, and (d) snapshots of the field profile at four different instants labeled in (b).}
  	\label{highEGunConceptAcceleration}
  \end{figure}
  First, two linearly polarized THz beams are symmetrically coupled into the multilayer structure in order to cancel out the magnetic field where they overlap.
  Second, the phase front of the THz beam is divided into several parts, which are isolated from each other using thin metallic surfaces.
  In each layer, dielectric inclusions are added to delay the arrival time of the pulse to the acceleration region.
  By properly designing the filling factor of dielectrics and the thickness of each layer, continuous acceleration of electrons from rest throughout the whole phase front can be achieved.
  
  \subsubsection{Multilayer Gun}
  
  As learned from the study on low energy guns, focusing the incoming excitation in the transverse plane enhances the acceleration efficiency.
  Furthermore, to avoid suspended thin metallic films in vacuum (Fig.\,\ref{2DConcept}), we consider quartz ($\epsilon_r=4.41$) and teflon ($\epsilon_r=2.13$) for delaying the arrival time.
  The structure shown in Fig.\,\ref{highEGunConceptAcceleration}a is the ultrafast THz gun designed for operation based on two single-cycle THz Gaussian beams with each 1.1\,mJ energy and central frequency 300\,GHz.
  Without loss of generality, the beam profile is flat top along the acceleration axis and Gaussian in the transverse direction with 2\,mm spot size.
  For a completely Gaussian beam, individual couplers should be designed to couple the beam energy into the gun input (see supplementary material in \cite{fallahi2016short}).
  We assume that two linearly polarized plane waves with the aforementioned temporal signature and peak field $0.5\,$GV/m illuminate the multilayer gun from both sides. %
  An eight layer configuration is designed for the considered excitation with the thickness of each layer $h_i=\{0.1,0.27,0.35,0.40,0.42,0.43,0.44,0.45\}$\,mm, and the length of the quartz inclusions $L_i=\{0.2,0.865,1.55,2.25,2.95,3.7,4.4,5.1\}$\,mm.
  The size of the acceleration channel is considered to be $g=120$\,{\textmu}m.
  The simulation results shown in Fig.\,\ref{highEGunConceptAcceleration}b demonstrate acceleration of an electron from rest to 2.1\,MeV.
  Similar to the previous cases, the electron is released at the point with $E_z=50$\,MV/m.
  The realization of phase front matching with the electron motion is shown by snapshots of the field profile in Fig.\,\ref{highEGunConceptAcceleration}d and Fig.\,\ref{highEGunConceptAcceleration}c, showing the accelerating field $E_z(t,z)$ superposed on the particle position $z_e(t)$.
  The small average momentum of the particles in the bottom layers causes small travelling distances within one half-cycle.
  Therefore, the thicknesses of the layers need to be smaller than the top layers to achieve the required synchronism.
  This effect leads to the observed gradual increase in the energy gain in different layers (Fig.\,\ref{highEGunConceptAcceleration}b).
  
\begin{figure}
  	\centering
  	$\begin{array}{ccc}
  	\includegraphics[draft=false,width=2.0in]{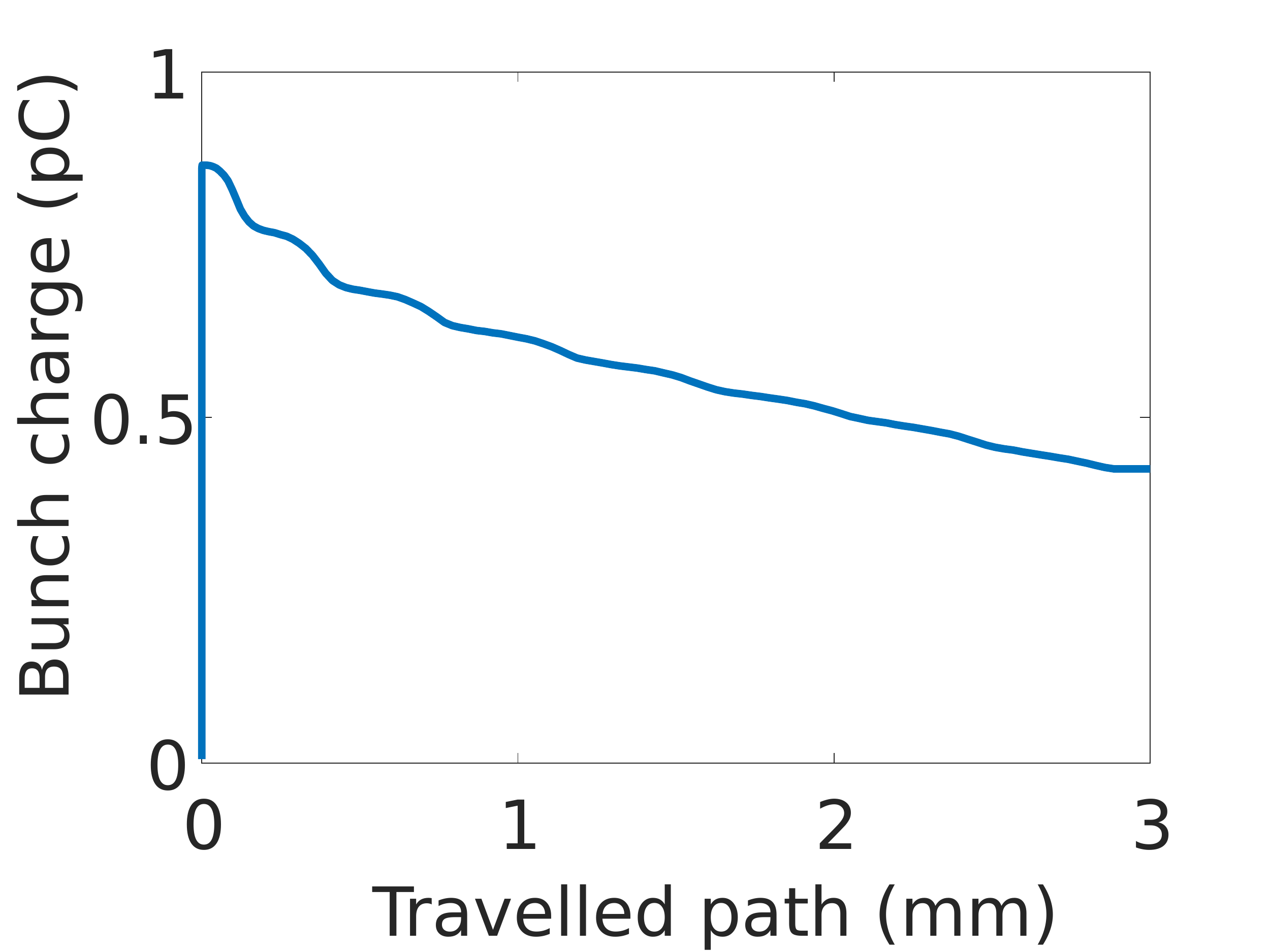} &
  	\includegraphics[draft=false,width=2.0in]{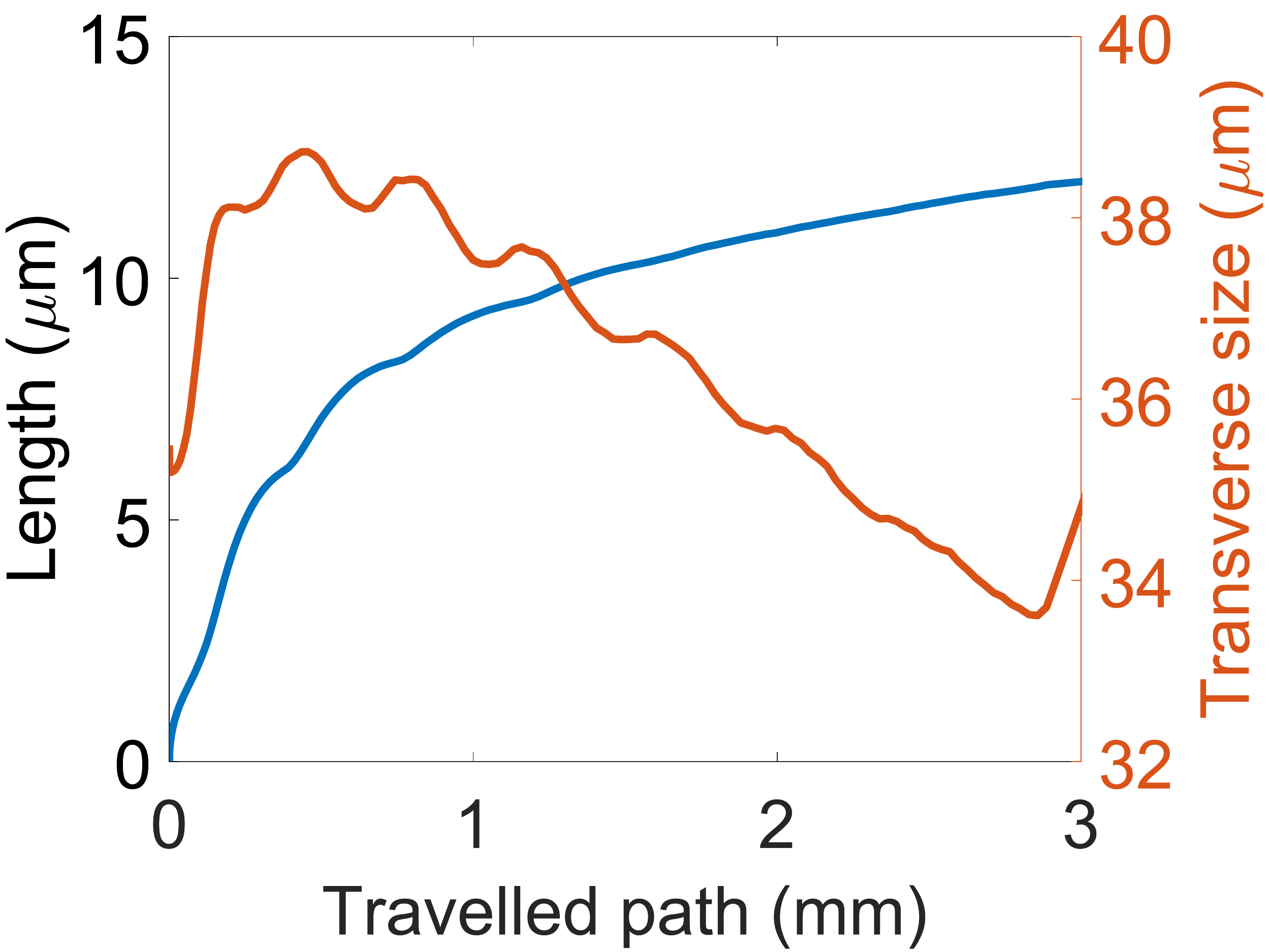} &
  	\includegraphics[draft=false,width=2.0in]{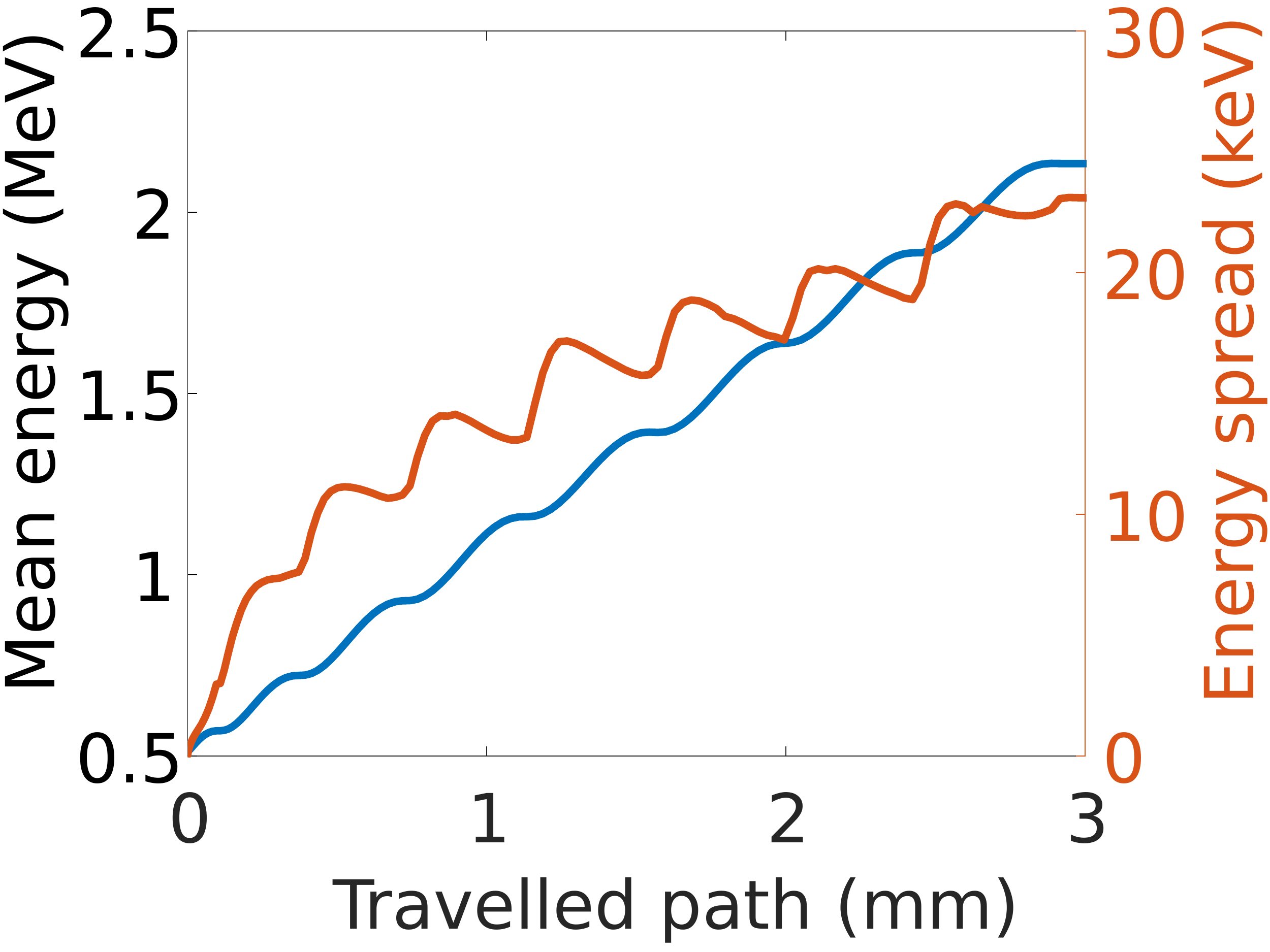} \\
  	(a) & (b) & (c) \\
  	\includegraphics[draft=false,width=2.0in]{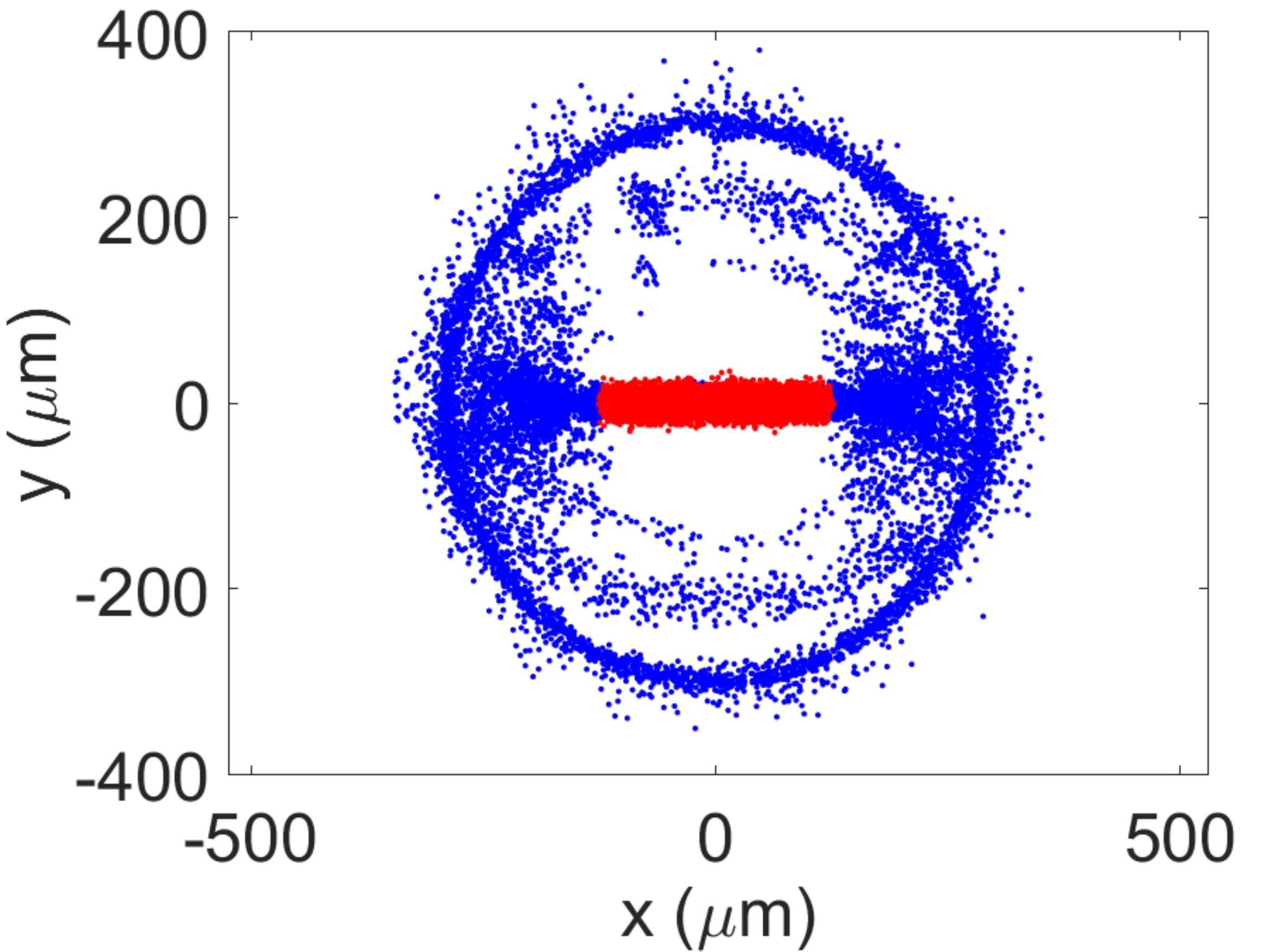} &
  	\includegraphics[draft=false,width=2.0in]{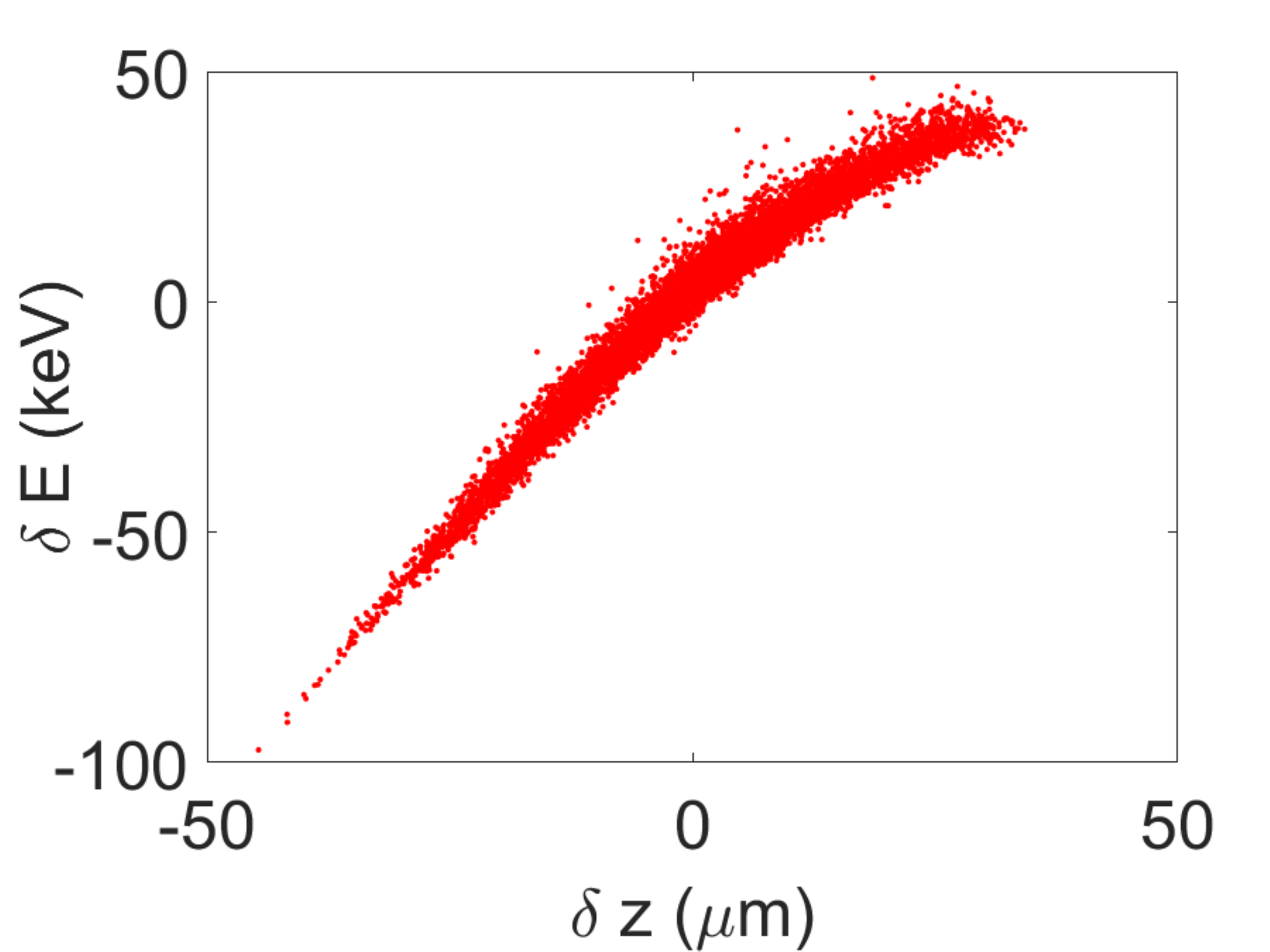} &
  	\includegraphics[draft=false,width=2.0in]{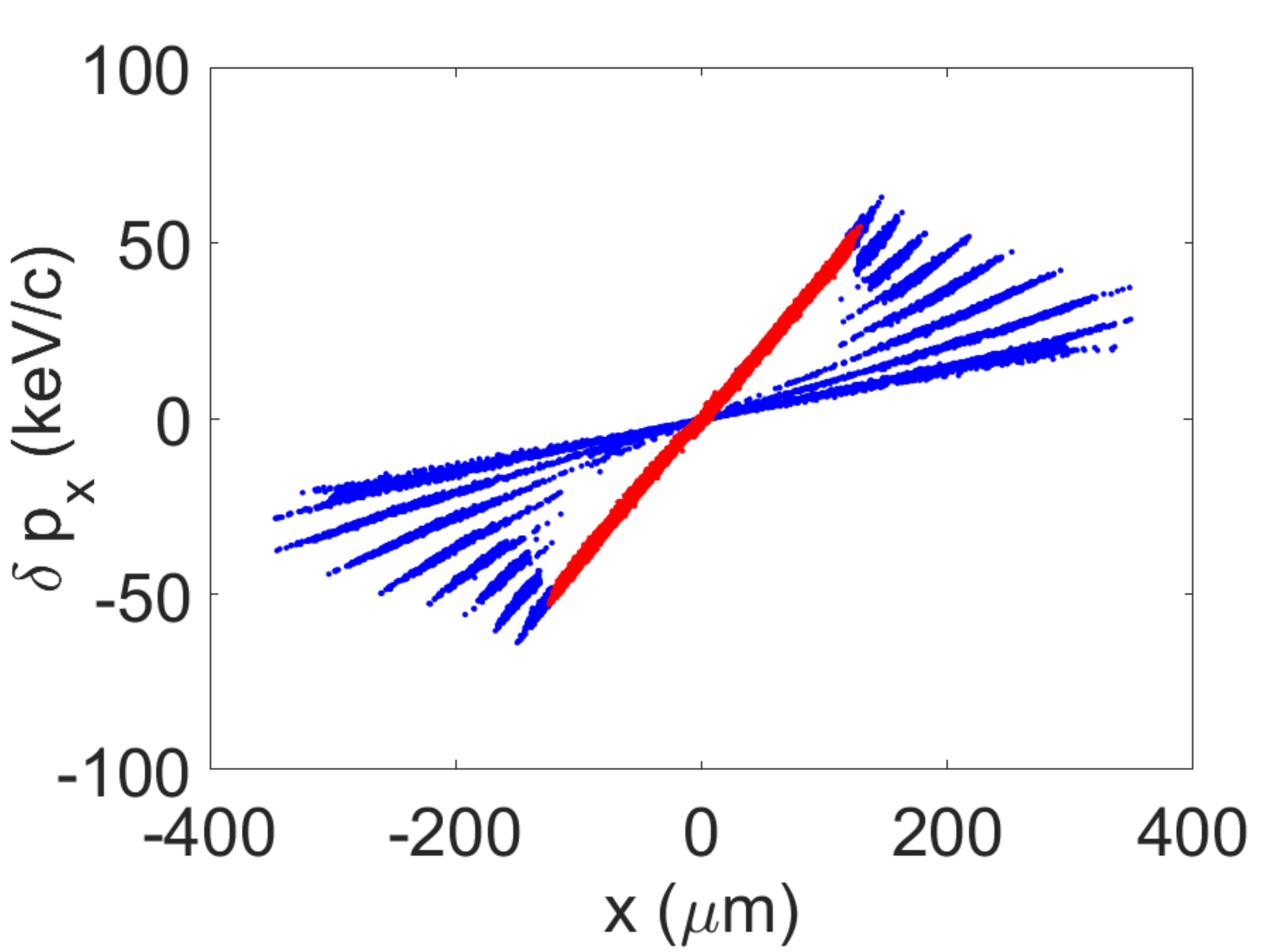} \\
  	(d) & (e) & (f)
  	\end{array}$
  	\caption{Acceleration of 1\,pC photoemitted bunch in the high energy THz gun: (a) bunch charge, (b) size and (c) energy in terms of travelled path and (d) shape, (e) transverse phase space, and (f) longitudinal phase-space 3\,mm from gun exit are depicted (red dots: output particles; blue dots: trapped particles). }
  	\label{highEGunFinalAcceleration}
  \end{figure}

  By again initializing a photo-emitted electron bunch, the bunch evolution in the proposed gun is simulated.
  We assume that a 40\,fs UV laser pulse generates 1\,pC charge over a 60\,{\textmu}m spot size, which is modeled using 20'000 macro-particles.
  The simulation results (Fig.\,\ref{highEGunFinalAcceleration}) demonstrate acceleration of 40\% of the bunch to 2.1\,MeV with 1.1\% energy spread, output  bunch length of 45\,fs, and normalized emittances equal to $(\varepsilon_{n,x},\varepsilon_{n,y},\varepsilon_{n,z})$=(0.13,0.13,0.09)\,mm$\cdot$mrad.
  The main reason for the particle loss is the deflection of the electron trajectories out of the considered vertical path and collision with the metallic layers, which could again be mitigated by focusing coils (see the supplementary material in \cite{fallahi2016short}).
  
\begin{figure}
  	\centering
  	$\begin{array}{cc}
  	\includegraphics[draft=false,width=3.0in]{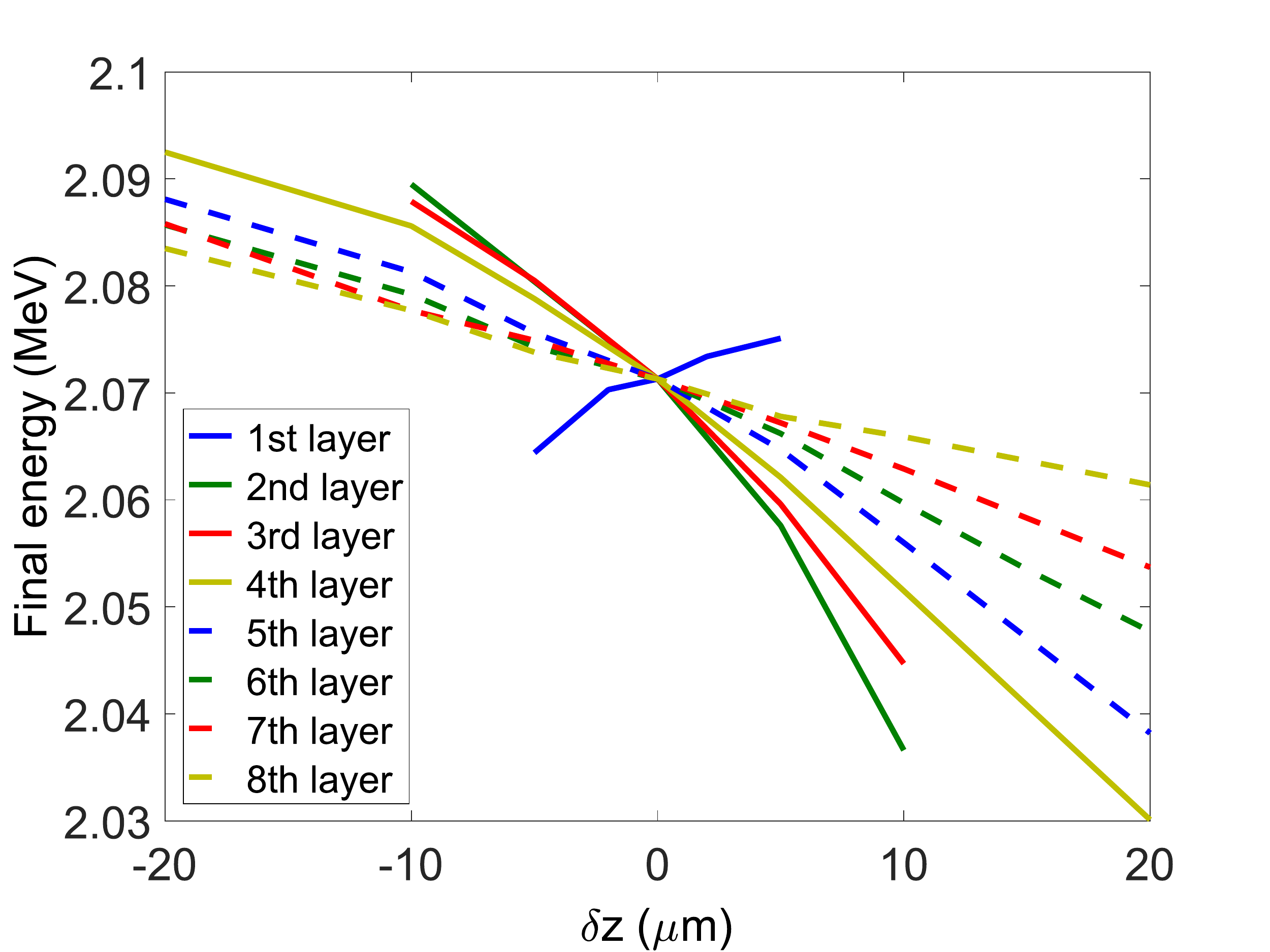} & \includegraphics[draft=false,width=3.0in]{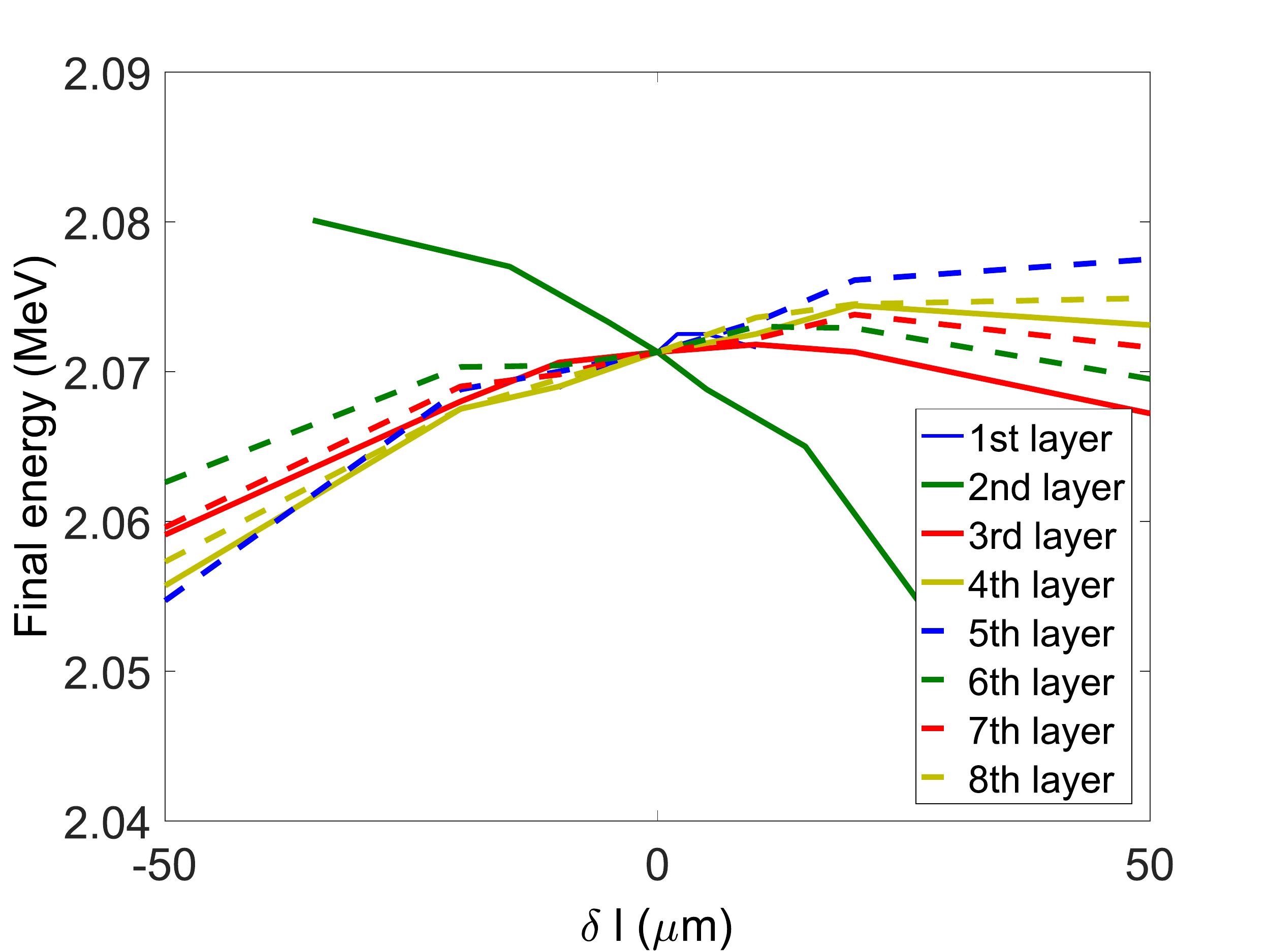} \\ (a) & (b)
  	\end{array}$
  	\caption{Sensitivity of the multilayer gun to variations in (a) layer thicknesses and (b) length of quartz inclusions.}
  	\label{sensitivity}
  \end{figure}

  Our detailed investigations of the introduced high energy ultrafast gun showed several advantages of such a scheme compared to conventional cascaded or travelling wave cavities.
  First, due to the inherent operation of the device with broadband excitations, the sensitivity of the gun outcome to the design parameters is minimal.
  This is illustrated in Fig.\,\ref{sensitivity} for the variations in dielectric lengths and layer thicknesses.
  It is observed that even 10\% change in the designed parameter is tolerated by the device.
  Second, the pulse heating due to the magnetic field is not only minimized by the single cycle operation, but also the magnetic field is canceled at the electron acceleration region.
  
  \section{Optimal Design of Single-Cycle Guns}
  
  After the thorough and detailed introduction of the ultrafast electron gun concept, a procedure for optimal electron gun design needs to be developed.
  This section presents an optimized design strategy for these electron guns.
  The discussion is started by a complete definition of the design problem.
  We start with the description of the optimal single-cycle ultrafast gun concept and proceed with the problem definition.
  Subsequently, the design process is presented as main part of this section.
  Techniques for fine tuning the design to enhance the output bunch characteristics are outlined.
  The whole process is explained in the framework of a gun design for 400\,keV electron beam.
  Throughout this design process, it is implicitly shown that the concept of single-cycle ultrafast electron guns can apply THz beams with energies in the level of 100-400 micor-joules to accelerate electrons, which is the state-of-the-art technology in THz radiation sources.
  Next, we present the outcome of the design process used for designing an ultrafast electron gun with higher electron beam energy than the first design, i.e. 800\,keV.
  This design shows the eligibility of this concept to perform as linac injectors in compact accelerator facilities.
  
  \subsection{Problem Definition}
  
  \subsubsection{Optimal Ultrafast Electron Gun Concept}
  
\begin{figure}
  	\centering
  	\includegraphics[draft=false,width=5.0in]{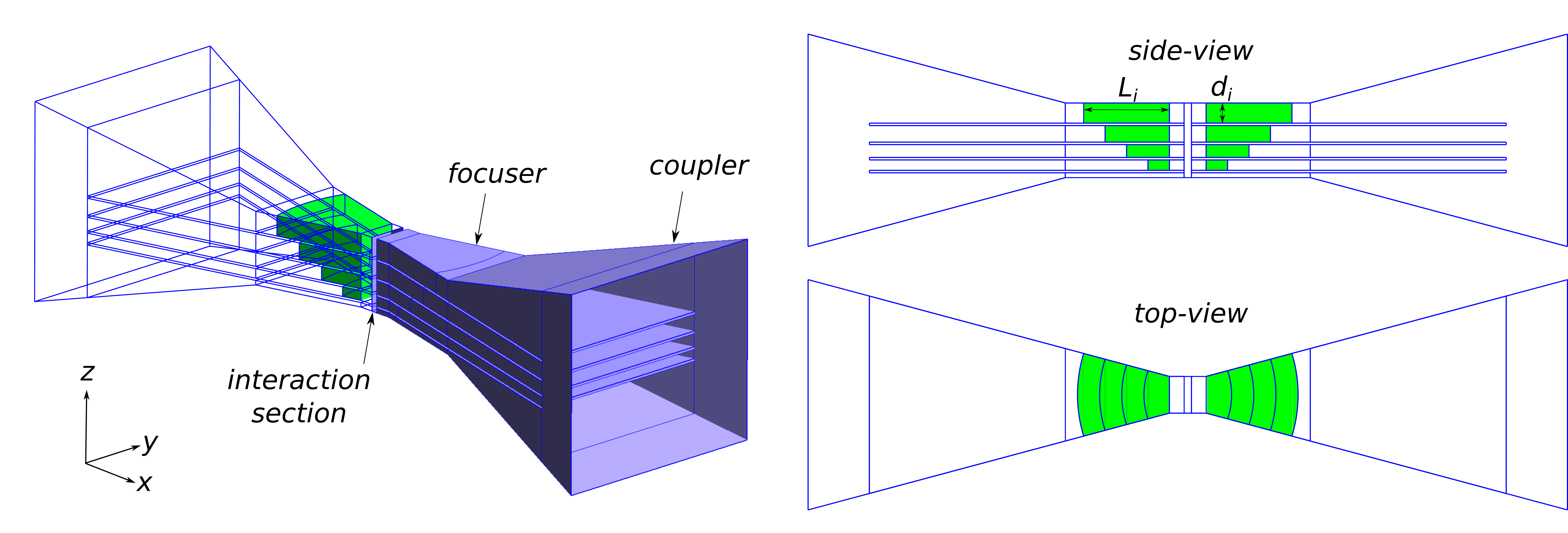}
  	\caption{Schematic illustration of a general configuration for single-cycle ultrafast electron guns.}
  	\label{optimalSingleCycleGunConcept}
  \end{figure}

  Fig.\,\ref{optimalSingleCycleGunConcept} schematically illustrates a single-cycle ultrafast electron gun, which consists of three principal sections, interaction region, focusing section, and coupler.
  It is assumed that two linearly polarized single-cycle Gaussian beams symmetrically impinge on the device from both sides.
  The coupler section transfers the energy of the Gaussian beam into the multilayer focusing section, where two metallic walls on both sides focus the beam into the interaction region.
  The interaction region in each layer can be considered as a rectangular waveguide, whose TE\textsubscript{01} mode is excited by the incoming fields from the focusing section.
  At the interaction region, the transverse and longitudinal magnetic fields of the two counter-propagating TE\textsubscript{01} modes cancel each other, whereas the vertical electric field will be constructively added.
  The superposition of these two beams results in a purely accelerating field along the $z$-axis in Fig.\,\ref{optimalSingleCycleGunConcept}.
  
\begin{figure}
  	\centering
  	\includegraphics[draft=false,width=5.5in]{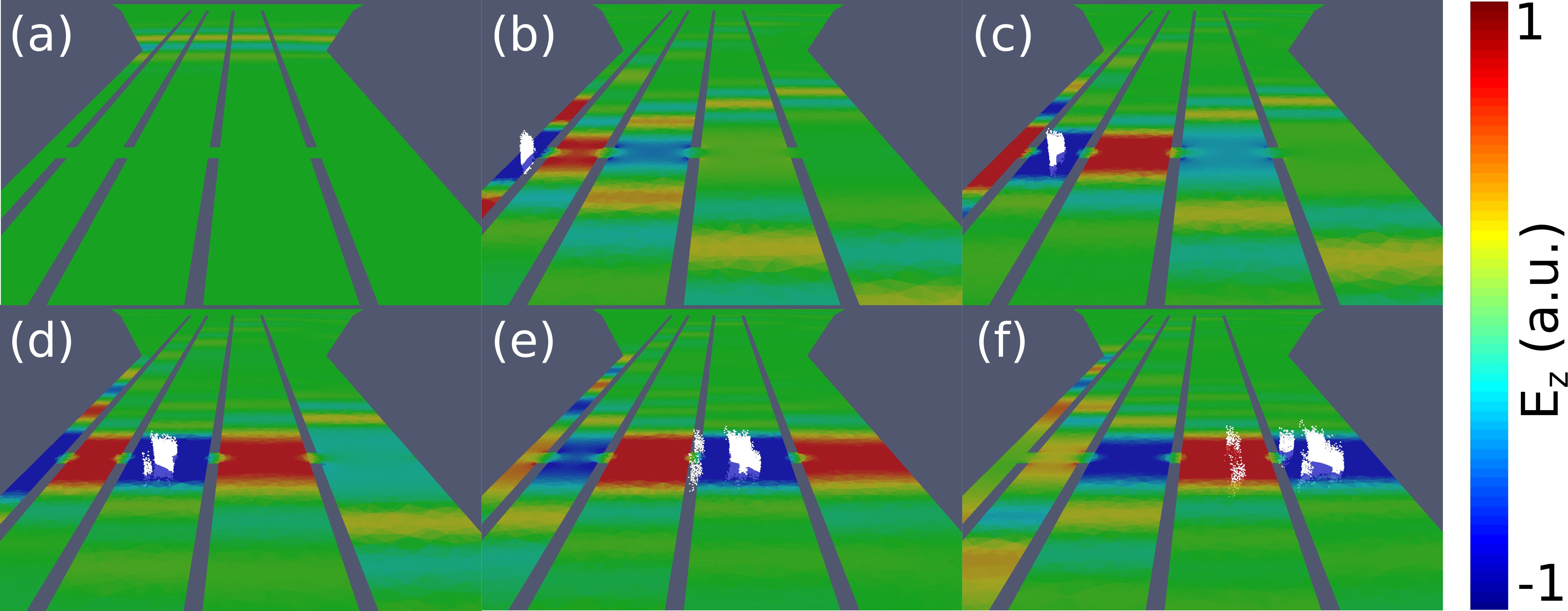}
  	\caption{Illustration of the phase-front matching of the single-cycle pulse with the accelerating electron bunch: (a)-(f) snapshots of the field profile and bunch distribution when the bunch center resides in the first five layer.}
  	\label{acelerationProcess}
  \end{figure}

  Starting in the coupler section, horizontal metallic plates, called here separators, divide the incoming Gaussian beam into several regions with thickness $d_i$.
  The energy in each region is then guided to each sub-waveguide of the gun.
  The travelling pulse entering each focusing section is subsequently delayed by dielectric inclusions, whose lengths, $L_i$, are designed to control the arrival of pulses into the interaction region.
  Proper design of the two sets of parameters $d_i$ and $L_i$ assures continuous interaction of travelling electrons with accelerating cycle of the pulse.
  In other words, the device substantiates phase-front matching of the incoming pulses with travelling electrons.
  The acceleration process in these ultrafast electron guns is visualized in Fig.\,\ref{acelerationProcess}.
  Snapshots of field profiles are superposed on particle profiles, which demonstrates continuous interaction of particles with the accelerating cycle of the field.
  
  Various significant improvements are considered compared with the geometries presented in the previous section.
  Previously, the dielectric contrast in each layer was realized by the two materials teflon and quartz.
  The requirements for mechanical stability of the thin separators necessitated filling the focusing sections with rigid materials.
  However, as will be observed in the fine tuning section, thick separators are advantageous for reducing the energy spread of the output bunch.
  Once thick metallic plates are used to divide the input energy among different sections, the need for filled focusing spaces is remedied.
  Therefore, one can rely on the dielectric contrast between vacuum and quartz to reach the goal with respect to phase-front matching.
  The second change is the open slot in the interaction region devised for incoupling of the photoinjector laser.
  The old configurations accounted for a back-illuminated photo-cathode structure.
  Nonetheless, our experimental investigations revealed some difficulties in extracting large amount of charge from thin metallic coatings in such type of cathodes \cite{huang2016terahertz}.
  The open narrow slot in the interaction region enables electron output as well as easy input coupling of the photoinjector laser from the front side, without dramatically disturbing the accelerating field profile.
  Furthermore, the coupler section in Fig.\,\ref{optimalSingleCycleGunConcept} takes flat separators into account.
  This differs with the structure shown in the supplementary material of \cite{fallahi2016short}, where minute inclinations are considered to gain a uniform accelerating gradient over the layers.
  Since a constant accelerating gradient is not a crucial requirement for the operation of these devices, the burden caused by these oblique separators in the fabrication process can simply be avoided through the assumption of flat separators.
  
  \subsubsection{Design Problem}
  
  Based on the above concept, ultrafast electron guns with unprecedented high accelerating fields can be implemented.
  For this purpose, a design process needs to be followed to achieve optimal operation.
  In other words, a design problem should be defined and systematically solved.
  Let us suppose that the gun is made out of a material that supports stable operation with maximum electric field $E_{\mathrm{max}}$ in the single-cycle operation regime.
  The desire to achieve high accelerating gradients and high quality bunches often inspires operation of accelerating devices close to damage threshold.
  The largest surface field in the proposed device exists over the photocathode surface, where the two incoming pulses interfere constructively in the proximity of a metallic surface.
  In the next layers, the above interference effect occurs in the vacuum region.
  Hence, the maximum accelerating gradient is equal to the maximum normal field strength, i.e. $E_{\mathrm{max}}$.
  On the other hand, as observed in Fig.\,\ref{acelerationProcess}, the superposition of two fields with opposite signs at the separators considerably reduces the field strengths around the edges in the gun geometry.
  Consequently, despite the field enhancement due to the edge effects, the field strengths at these regions do not exceed the photocathode surface field.
  The design problem consequently aims at a device which realizes peak accelerating fields equal to $E_{\mathrm{max}}$, using a minimum required energy in the two incoming Gaussian beams.
  
  \subsection{Design Process}
  
\begin{figure}
  	\centering
  	$\begin{array}{ccc}
  	\includegraphics[draft=false,height=1.5in]{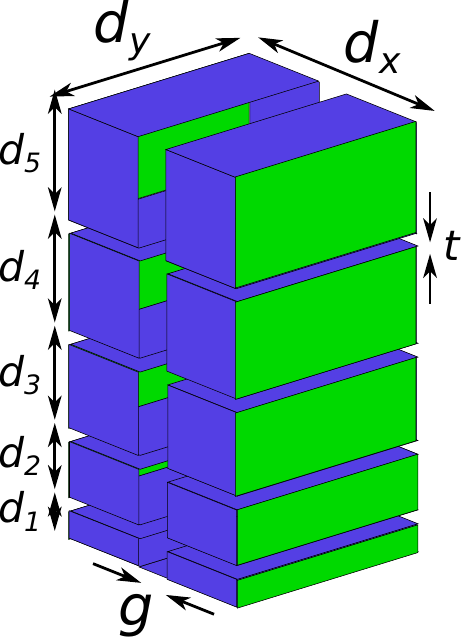} &
  	\includegraphics[draft=false,height=1.5in]{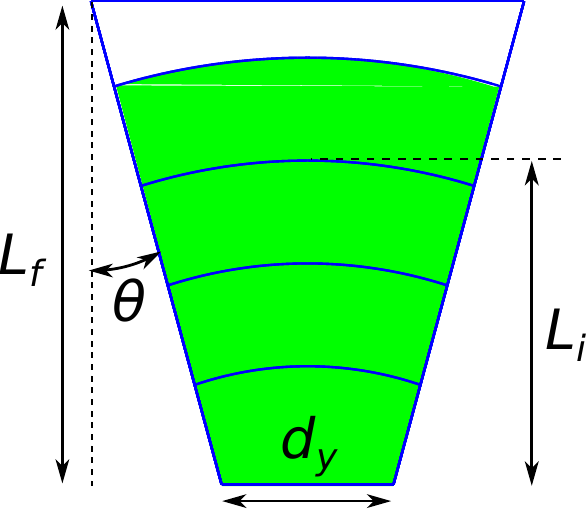} &
  	\includegraphics[draft=false,height=1.5in]{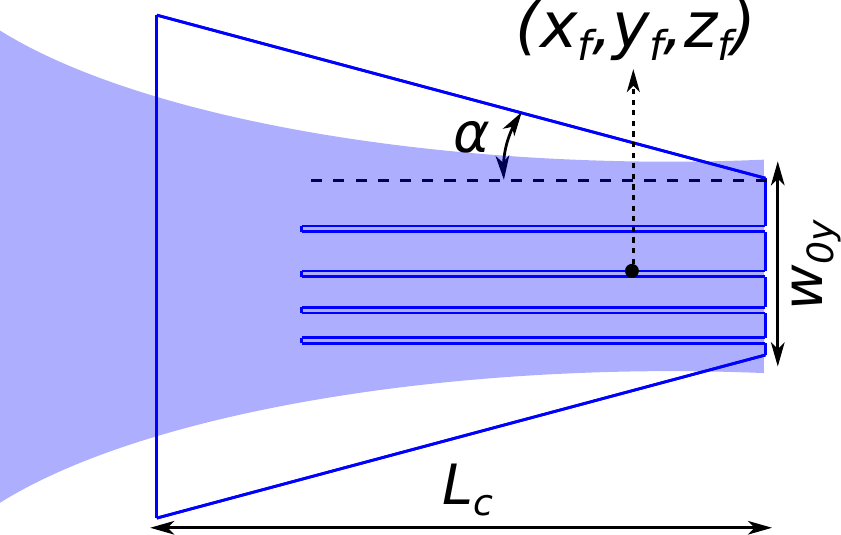} \\
  	(a) & (b) & (c)
  	\end{array}$
  	\caption{Parameterizations of the ultrafast electron gun: (a) Interaction region, (b) focuser section (top-view), and (c) coupler section (side-view) and the incoming Gaussian beam with critical dimensions.}
  	\label{gunParameterization}
  \end{figure}

  To explain how such an optimum design can be achieved, we take an exemplary problem into account.
  We aim to design a 400\,keV electron gun fed by single-cycle THz pulses centered at 300\,GHz with a copper photocathode.
  From the previous investigations and scaling laws, the value of $E_{\mathrm{max}}$ is assumed to be around 600\,MV/m \cite{huang2016terahertz,wu2017high,dal2016rf}.
  Once the accelerating gradient and operation frequency is fixed to 600\,MV/m and 300\,GHz, the interaction section, shown and parameterized in Fig.\,\ref{gunParameterization}a, is designed using an analytical formulation.
  
  To determine the transverse sizes of the interaction section, $d_x$ and $d_y$, the accelerating field profile of the two counter-propagating TE\textsubscript{01} modes needs to be considered, which reads as
  \begin{equation}
  E_z = A \cos \frac{\pi}{d_y} y \left( e^{-jk_x x}+e^{+jk_x x} \right) = 2 A \cos \frac{\pi}{d_y} y \cos k_x x,
  \label{acceleratingField}
  \end{equation}
  with $k_x=\sqrt{k_0^2-(\pi/d_y)^2}$ and $k_0$ being the vacuum wave number.
  For electron guns, a symmetric accelerating field over the bunch dimensions is usually favored, since it enables bunches with symmetric properties.
  As will be seen later, the considered slot for the photoinjector laser and the injection of fields along the $x$-axis break the symmetry of acceleration.
  Nonetheless, it is beneficial to keep the symmetry of the device as much as possible.
  Therefore, the conditions $k_x=\pi/d_y$ at center frequency and $d_x=d_y$ which yield $\partial/\partial x = \partial / \partial y$ in \eref{acceleratingField}, are assumed to maintain a symmetric field profile.
  For 300\,GHz operating frequency, it leads to $d_x=d_y=0.71$\,mm. The gap size, $g$, must be on one hand large enough to support particle transfer between consecutive layers and on the other sufficiently small to avoid destructive interference of fringing fields in the two layers.
  In addition, a large gap leads to field leakage outside of the interaction region and in turn weakens the accelerating field.
  As a rule of thumb, setting $g\approx \lambda_x/10$, with $\lambda_x=2\pi/k_x$, provides a proper compromise between the aforementioned effects.
  This leads to $g=0.12$\,mm for the example considered.
  
  To design the layer thicknesses, we initially consider an ideal scenario, in which the effect of fringing fields, transverse fields, inhomogeneous fields among different layers, and the broad frequency spectrum of the excitation are neglected.
  In this case, an electron synchronized with the incoming pulse, will be affected by the following field:
  \begin{equation}
  E_z = -A \eta_i \left| \sin \omega t \right| = -E_{\mathrm{max}} \eta_i \left| \sin \omega t \right|,
  \label{acceleratingFieldSynchronized}
  \end{equation}
  where $\omega$ is the center frequency of the pulse, and $\eta_i$ is a field scaling factor defined for each layer.
  Note that the accelerating field profile in the ultrafast gun is fundamentally different from the conventional cascaded cavity gun technology, where fields are also position dependent.
  The different field profiles lead to various advantages and shortcomings compared to conventional technologies, which will be the subject of future investigations.
  Due to Fresnel reflection from the quartz ($n=2.1$) wafers in each layer above the first one and additional fringing field effects from adjacent layers, the maximum accelerating gradient is smaller than $E_{\mathrm{max}}$ considered for the first layer.
  The former effect reduces the field in the interaction section to $0.87E_{\mathrm{max}}$.
  If we consider 10\% degradation for the later effect, the following equation for $\eta_i$ leads to a reasonable estimate for the accelerating gradient in different layers:
  \begin{equation}
  \eta_i = \left\{ \begin{array}{cl} 1 & i = 1 \\ 0.8 & i > 1 \end{array} \right. .
  \label{etaEstimate}
  \end{equation}
  
\begin{figure}
  	\centering
  	\includegraphics[draft=false,width=3.0in]{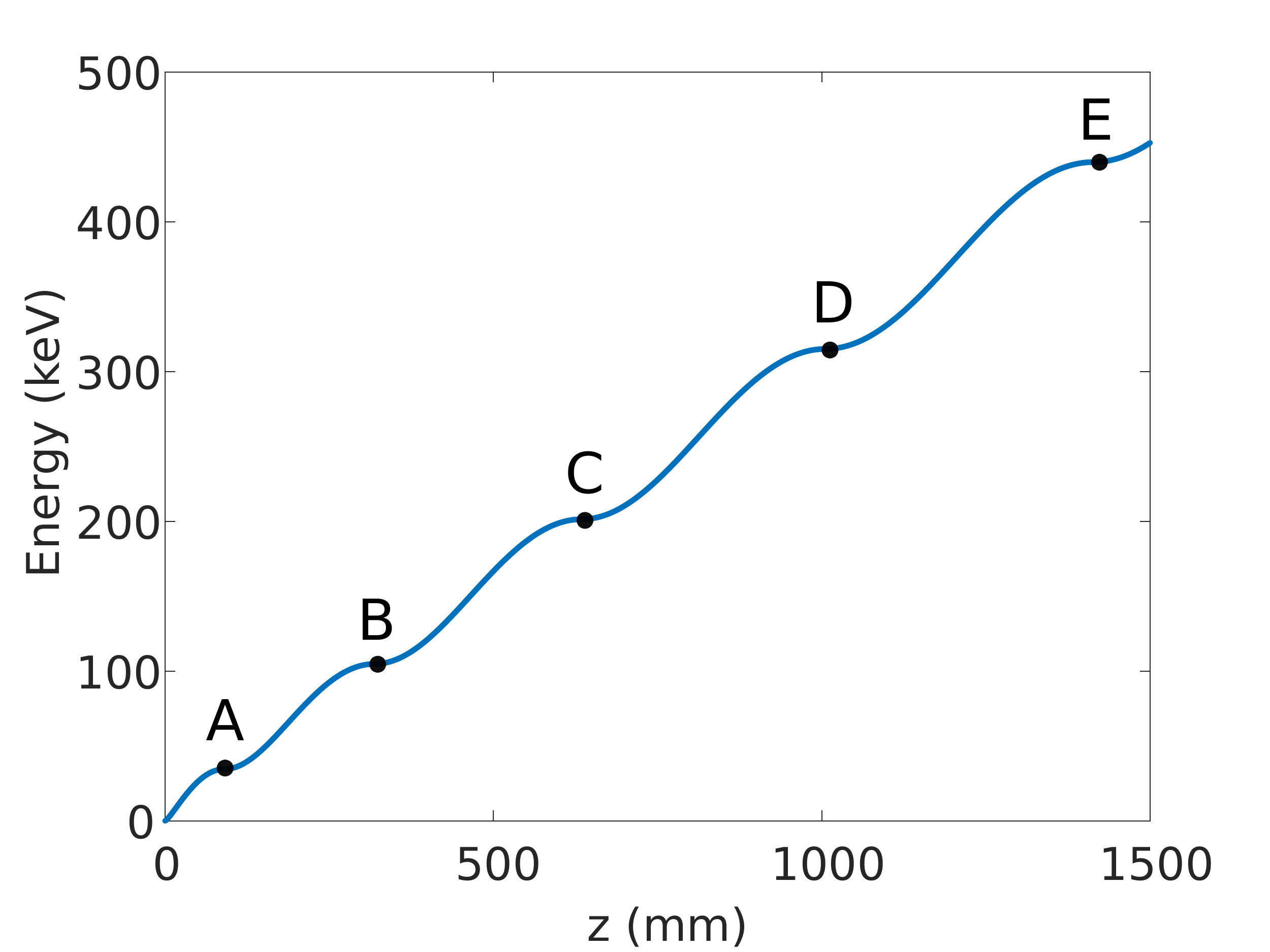}
  	\caption{Energy of an electron versus travelled distance accelerated by the field evaluated from \eref{etaEstimate}: Points A-E are highlighted as transition points between consecutive cycles.}
  	\label{electronEnergyInitialDesign}
  \end{figure}

  The energy of an electron in terms of distance during the first five half-cycles for the considered example is shown in Fig.\,\ref{electronEnergyInitialDesign}.
  The temporal energy change indicates that five layers are required to obtain 400\,keV electron beam.
  Moreover, the highlighted points A-E represent the transition positions from one cycle to the next one.
  Therefore, the corresponding $z$ coordinates should coincide with the center of the transition region between each layer.
  The positions of points A-E in Fig.\,\ref{electronEnergyInitialDesign} are obtained as
  \begin{equation}
  \{z_A,z_B,z_C,z_D,z_E\}=\{100,320,650,1020,1420\}\,\text{{\textmu}m}.
  \label{turningPointPositions}
  \end{equation}
  
  If we assume $t=40$\,{\textmu}m thick separators, the corresponding thickness of each layer reads as
  \begin{equation}
  \{d_1,d_2,d_3,d_4,d_5\}=\{80,180,290,330,360\}\,\text{{\textmu}m}.
  \label{layerThicknesses}
  \end{equation}
  
  Now that the interaction section is designed, it is straightforward to estimate the required amount of energy for the presumed acceleration.
  The power propagating in a TE\textsubscript{01} mode of a rectangular waveguide is obtained from:
  \begin{equation}
  P=\frac{E_0^2}{2\eta_0} \frac{d_y d_i}{2} \sqrt{ 1 - \left(\frac{\lambda}{2d_y}\right)^2 },
  \label{TE01Power}
  \end{equation}
  where $\eta_0$ is the intrinsic impedance of free space, $\lambda$ is the operation wavelength, and $E_0$ stands for the maximum electric field in the waveguide.
  For the designed dimensions, i.e. $d_y=\lambda/\sqrt{2}$ and $E_0=E_{\mathrm{max}}/2$, the power flow of each mode can be obtained from $P=E_{\mathrm{max}}^2 \lambda d_i/32 \eta_0$.
  If we neglect the dispersion in coupler and focuser and the resultant broadened pulse, the total energy due to each beam reaching the interaction region can be estimated as,
  \begin{equation}
  \mathcal{E} = P \tau \sqrt{ \frac{\pi}{4 \ln 2}} = \frac{E_{\mathrm{max}}^2 \lambda d_i}{32 \eta_0} \tau \sqrt{ \frac{\pi}{4 \ln 2}},
  \label{beamEnergy}
  \end{equation}
  with $\tau=1/f$ being the pulse duration of the single-cycle pulse.
  For the considered example in this section, the energy due to each beam in each layer is evaluated as
  \begin{equation}
  \{\mathcal{E}_1,\mathcal{E}_2,\mathcal{E}_3,\mathcal{E}_4,\mathcal{E}_5\}=\{9,19,31,35,38\}\,\text{{\textmu}J}.
  \label{beamEnergyInitialDesign}
  \end{equation}
  Therefore, total energy of $\mathcal{E}_t=132$\,{\textmu}J should be coupled into the interaction region from each side to realize the desired acceleration.
  
  The obtained value for $\mathcal{E}$ represents an estimate for the total amount of energy interacting with electrons.
  Several effects in the coupling and focusing process contribute to the energy loss before reaching the interaction region.
  The most dominant ones are reflection from thick separators in the coupler and intense pulse-broadening due to dispersion in the focuser and coupler.
  Note that in the above calculation the Fresnel losses are taken into account, where $E_0=E_{\mathrm{max}}/2$ is considered in all layers.
  According to our experience, 50\% more energy is usually needed to realize the accelerating gradients assumed in each layer.
  In other words, two 200\,{\textmu}J energy beams should excite the ultrafast electron gun to realize 400\,keV energy gain.
  This dramatic loss in energy motivates utilization of advanced coupling and focusing techniques instead of the simple horn couplers considered in this study.
  
  Designing the focusing section revolves around determination of the length of dielectric inclusions and the horn angle, $\theta$, in Fig.\,\ref{gunParameterization}b.
  The length of dielectrics are obtained from the difference in arrival time between two neighboring layers, which should be equal to $\tau/2$.
  If we again neglect the dispersion effects of the waveguide, the values of $L_i$ can be recursively obtained from the following equation:
  \begin{equation}
  L_i=L_{i-1}+\frac{c \tau}{2(n-1)},
  \label{dielectricLengths}
  \end{equation}
  with $n=2.1$ being the quartz refractive index.
  As observed in Fig.\,\ref{gunParameterization}b, the quartz inclusions are assumed to be curved on one side.
  This curvature assists in better coupling of the input energy, since the beam portions close to the walls need to travel a longer path to reach the end of the focuser.
  Obtaining the best radius of curvature is an optimization problem and according to our experience, a curved surface which is perpendicular to the side walls at both ends is close to optimum.
  
  An optimum value for $\theta$ strongly depends on the spatial profile, i.e. the beam size, of the incoming Gaussian beam.
  More accurately, this angle should match with the divergence angle of the beam.
  Generally, beam confinement in vacuum using optical elements introduces considerably smaller dispersion to the pulse format compared with waveguides.
  This effect becomes even more important when propagation of a single-cycle pulse is involved.
  Therefore, it is always advantageous to focus the single-cycle beam close to its diffraction limit, namely $w_0=\lambda$, before entering the coupler.
  In this case, the divergence angle of the Gaussian beam and consequently the horn angle $\theta$ is equal to 18$^\circ$.
  Eventually, based on the same hypothesis the coupler angle $\alpha$ in Fig.\,\ref{gunParameterization}c is similarly set to 18$^\circ$.
  As will be seen in 800\,keV gun design process, there exist cases where the total length of the interaction region is larger than one wavelength.
  In such cases, the incoming Gaussian beams are focused to $w_0>\lambda$ beam sizes for optimum coupling to the interaction region.
  The total length of the coupler ($L_c$) and focuser ($L_f$) does not play a significant role in the device operation.
  It is merely important to keep both lengths as small as possible in order to minimize the pulse broadening effect.
  However, both dimensions should be long enough to provide large enough aperture for capturing the total power of the Gaussian beam.
  
  The explained profile matching condition additionally suggests the optimum position for the beam focus.
  The transverse dimension of the coupler should match with the beam size at the focus point.
  Since a much stronger focusing is desired along $y$ direction compared to axes, we consider the beam size as well as the structure dimensions along $y$ axis to determine the focal point of the excitations.
  This assumption yields $x_f=\pm2.4\,$mm for the optimum focal point in the 400\,keV gun.
  In addition, $y_f=0$ and $z_f=0.7\,$mm maintains a symmetric excitation along y axis and distributes energy uniformly among different layers.
  
\begin{figure}
  	\centering
  	$\begin{array}{cc}
  	\includegraphics[draft=false,width=3.0in]{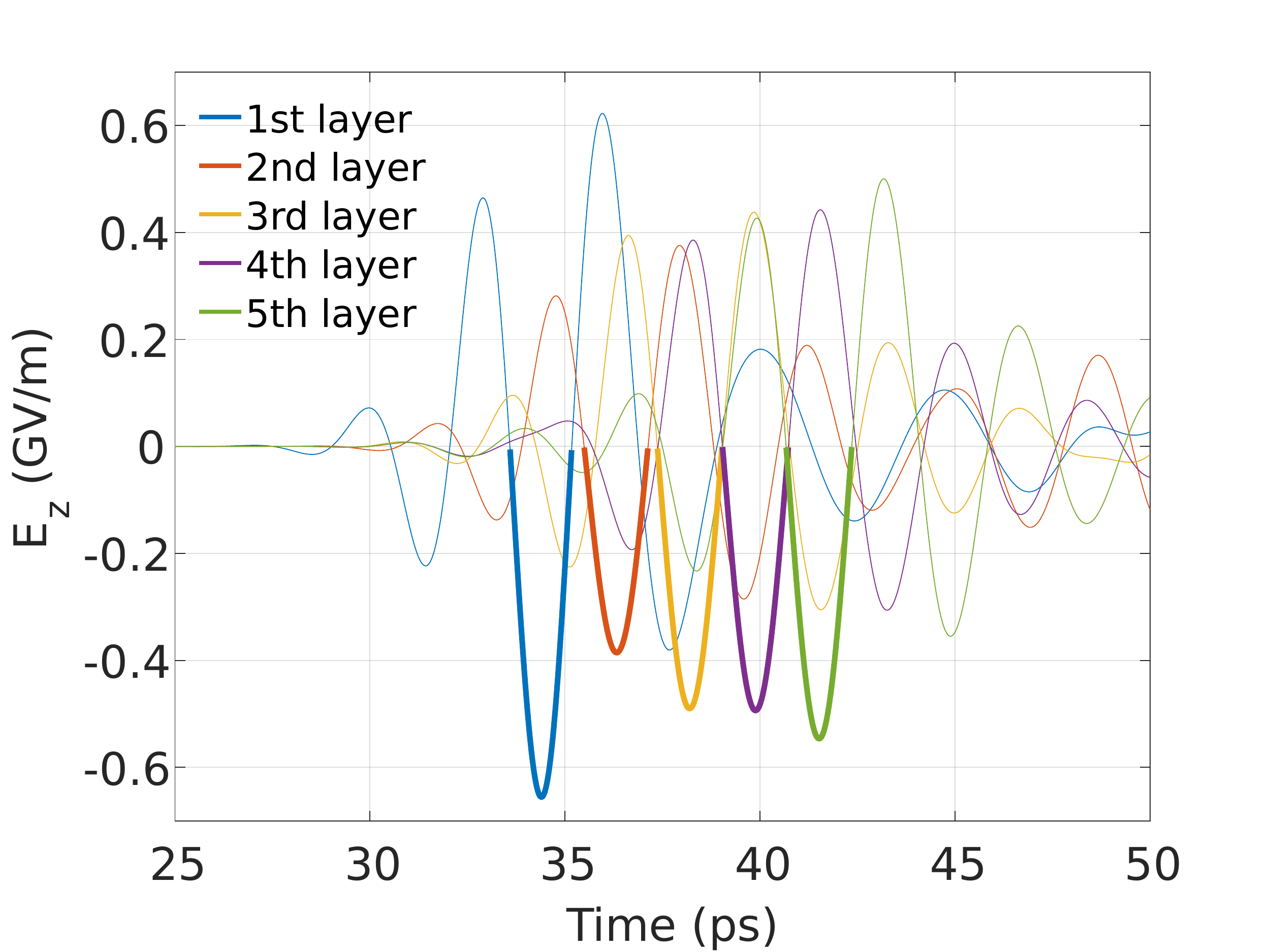} &
  	\includegraphics[draft=false,width=3.0in]{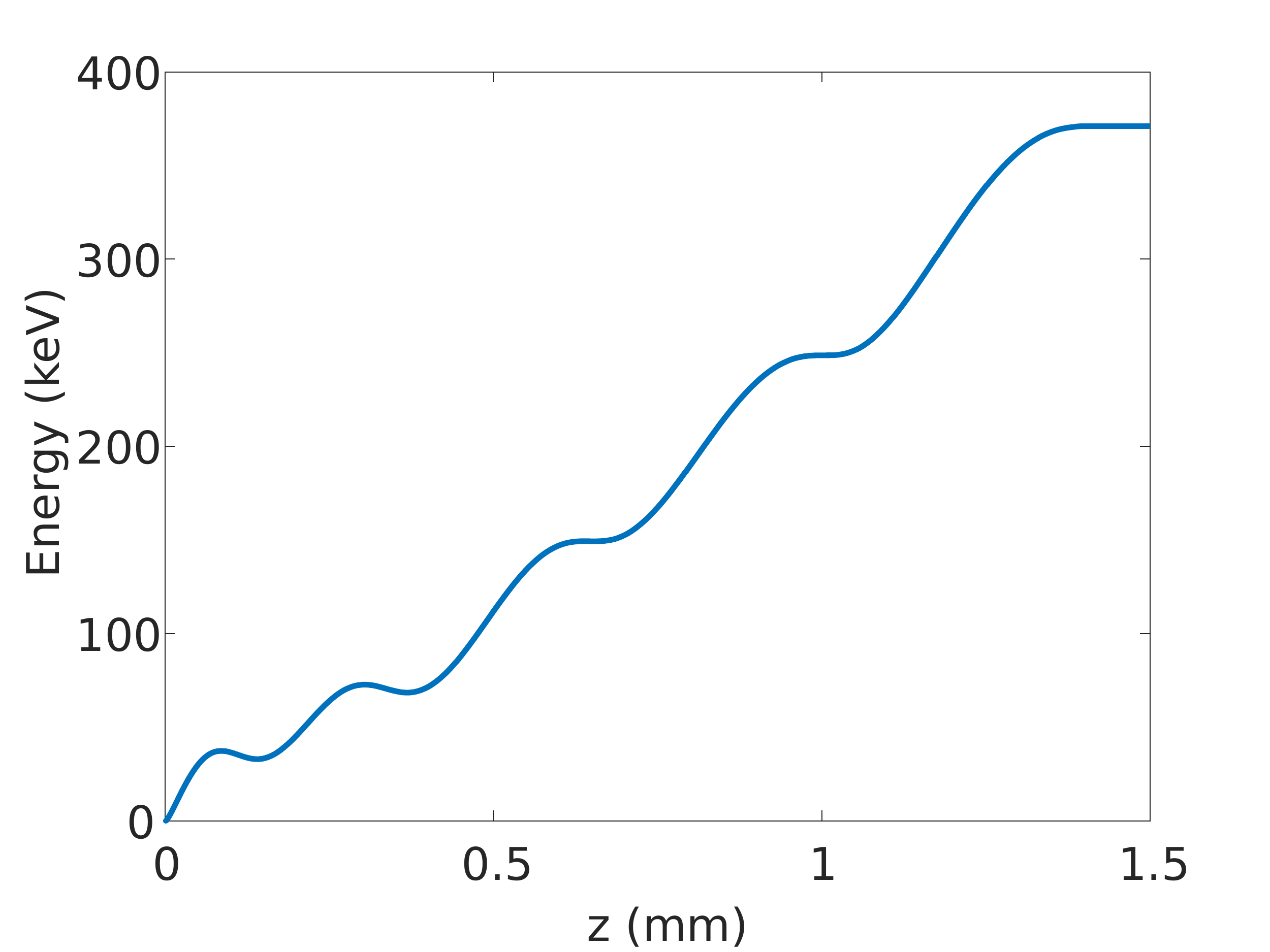} \\
  	(a) & (b)
  	\end{array}$
  	\caption{Simulation results of the initial design for the 400\,keV electron gun: (a) accelerating field in the middle of each layer versus time, and (b) energy of an electron injected at the 33.6\,ps time instant versus travelled distance.}
  	\label{400keVInitialDesign}
  \end{figure}

  When all device and excitation parameters are determined, we can simulate the described device and assess the design obtained through the initial optimization process.
  All the dimensions of the electron gun as well as the excitation beam are tabulated in Table\,\ref{400keVGunDesign}.
  \begin{table}
  	\caption{The design parameters for the 400\,keV gun and the optimized ones after fine tuning} \label{400keVGunDesign} \centering
  	{\footnotesize
  		\begin{tabular}{|c||c|c|}
  			\hline
  			parameter & designed value & fine-tuned value \\ \hline
  			$(d_x,d_y)$ & (0.71,0.71)\,mm & (0.71,0.71)\,mm \\ \hline
  			$(d_1,d_2,d_3,d_4,d_5)$ & (80, 180, 290, 330, 360)\,{\textmu}m & (80, 180, 290, 330, 360)\,{\textmu}m \\ \hline
  			$t$ & 40\,{\textmu}m & 40\,{\textmu}m \\ \hline
  			$g$ & 120\,{\textmu}m & 120\,{\textmu}m \\ \hline
  			$(L_1,L_2,L_3,L_4,L_5)$ & (0, 450, 900, 1350, 1800)\,{\textmu}m & (0, 400, 800, 1300, 1800)\,{\textmu}m \\ \hline
  			$(L_f,L_c)$ & (2.0,4.2)\,mm & (2.0,4.2)\,mm \\ \hline
  			$(\theta,\alpha)$ & $(18.0^{\circ},18.0^{\circ})$ & $(19.0^{\circ},14.0^{\circ})$ \\ \hline
  			$n$ & 2.1 (quartz) & 2.1 (quartz) \\ \hline
  			$(\tau,w_{0y},w_{0z})$ & (3.33\,ps, 1\,mm, 1\,mm) & (3.33\,ps, 1\,mm, 1\,mm) \\ \hline
  			$f$ & 300\,GHz & 300\,GHz \\ \hline
  			$\mathcal{E}$ & 200\,{\textmu}J & 200\,{\textmu}J \\ \hline
  			$(x_f,y_f,z_f)$ & ($\pm$2.4, 0.0, 0.7)\,mm & ($\pm$2.7, 0.0, 0.64)\,mm \\ \hline
  		\end{tabular}
  	}
  \end{table}
  The fields are evaluated using the in-house developed time domain Maxwell solver based on the discontinuous Galerkin time domain (DGTD) technique described in chapter 2.
  Next, the proper injection time of one electron is extracted from the field temporal variations in the first layer.
  The electron energy gain and trajectory within the simulated fields are then computed using the PIC algorithm (see chapter 2).
  In Fig.\,\ref{400keVInitialDesign}, the temporal signatures of the accelerating field at the center of each layer and the electron energy versus time are depicted as outcomes of the simulation.
  It is seen that the design process in spite of neglecting many influential effects results in a function close to the target operation.
  The sensitivity analysis presented in the previous section, demonstrates operational stability even after 10\% changes in the parameter values, which is a result of the single-cycle, i.e. broadband operation of the device.
  As a consequence, finding the values for an operating electron gun is a straightforward task.
  However, the required transient simulations introduce serious challenges during device optimization.
  
  \subsection{Fine Tuning}
  
  As previously emphasized, the described process for the initial design is neglecting several effects.
  To acquire an optimum operation, fine tuning of the dimensions is an indispensable task.
  Owing to the computationally demanding full-vector simulations needed to verify each set of parameters, this effort is very time consuming.
  However, scrutiny of the field and electron energy variations considerably reduces the expense for fine tuning and optimizing the gun performance.
  
  For example, it is observed that the time delay between the accelerating cycles in the first and second layers are longer than expected (Fig.\,\ref{400keVInitialDesign}a).
  Therefore, to correct this time delay the length of the dielectric inclusion in the second sub-waveguide, i.e. $L_2$, should be reduced.
  Once this time delay is corrected, the lengths $L_i$ in sub-waveguides of the upper layers should be similarly reduced to maintain the synchronization of electrons with the pulse front.
  In addition, care must be exercised when injecting one single electron and optimizing the acceleration scheme.
  In practice, a bunch over an extended time will be injected into the device.
  Hence, injecting electrons at the zero-crossing, as done in Fig.\,\ref{electronEnergyInitialDesign}, incurs losing almost half of the photo-emitted electrons.
  To avoid this effect, the center of the bunch must experience the field with a phase larger than zero.
  This results in losing a fraction of the input beam and smaller optimum thicknesss of the gun layers.
  
  Continuous acceleration between the layers may initially be envisaged and favored to maximize the energy transfer from the THz beam to the electrons.
  However, this leads to a concomitant increase in energy spread of the output bunch.
  To reduce the energy spread, it is advantageous to impose time delays between layers which are slightly larger than a half cycle.
  This assumption causes higher energy electrons in front of the bunch gain less energy than the ones in the back, which in turn reduces the ultimate energy spread.
  For this purpose, thick separators between acceleration layer enable more efficient velocity bunching compared with thinner ones.
  Similar subtleties exist in the coupling mechanism of the Gaussian beam.
  The presented qualitative measures for setting the angle of horn couplers and focus position of the beam are based on the beam waist ($w_0$), which encompasses around 86\% of the total incident power.
  More efficient coupling is obtained by accounting for a larger effective spot size, while determining the values of $\alpha$, $\theta$, and $(x_f,y_f,z_f)$.
  
\begin{figure}
  	\centering
  	$\begin{array}{cc}
  	\includegraphics[draft=false,width=3.0in]{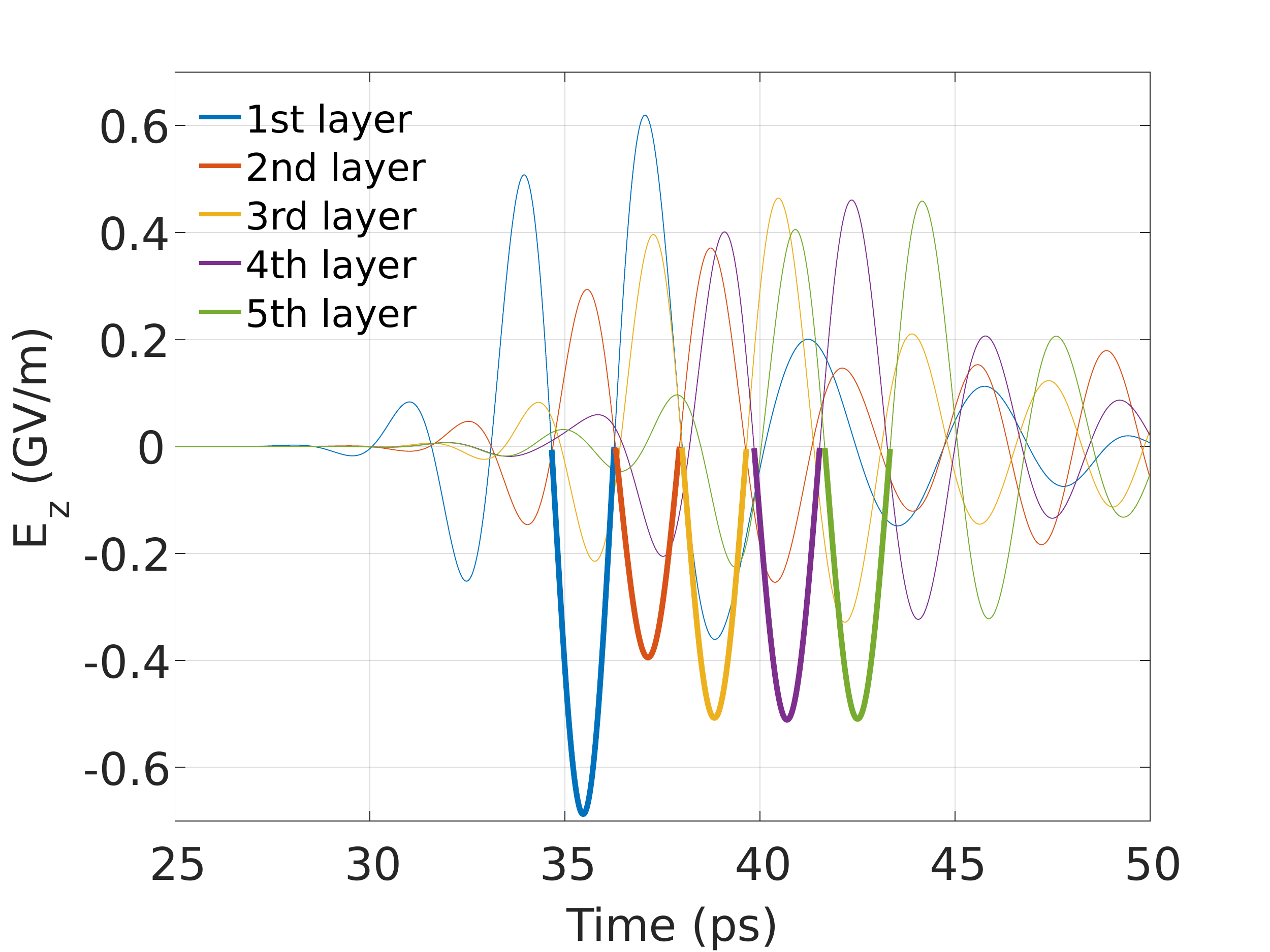} &
  	\includegraphics[draft=false,width=3.0in]{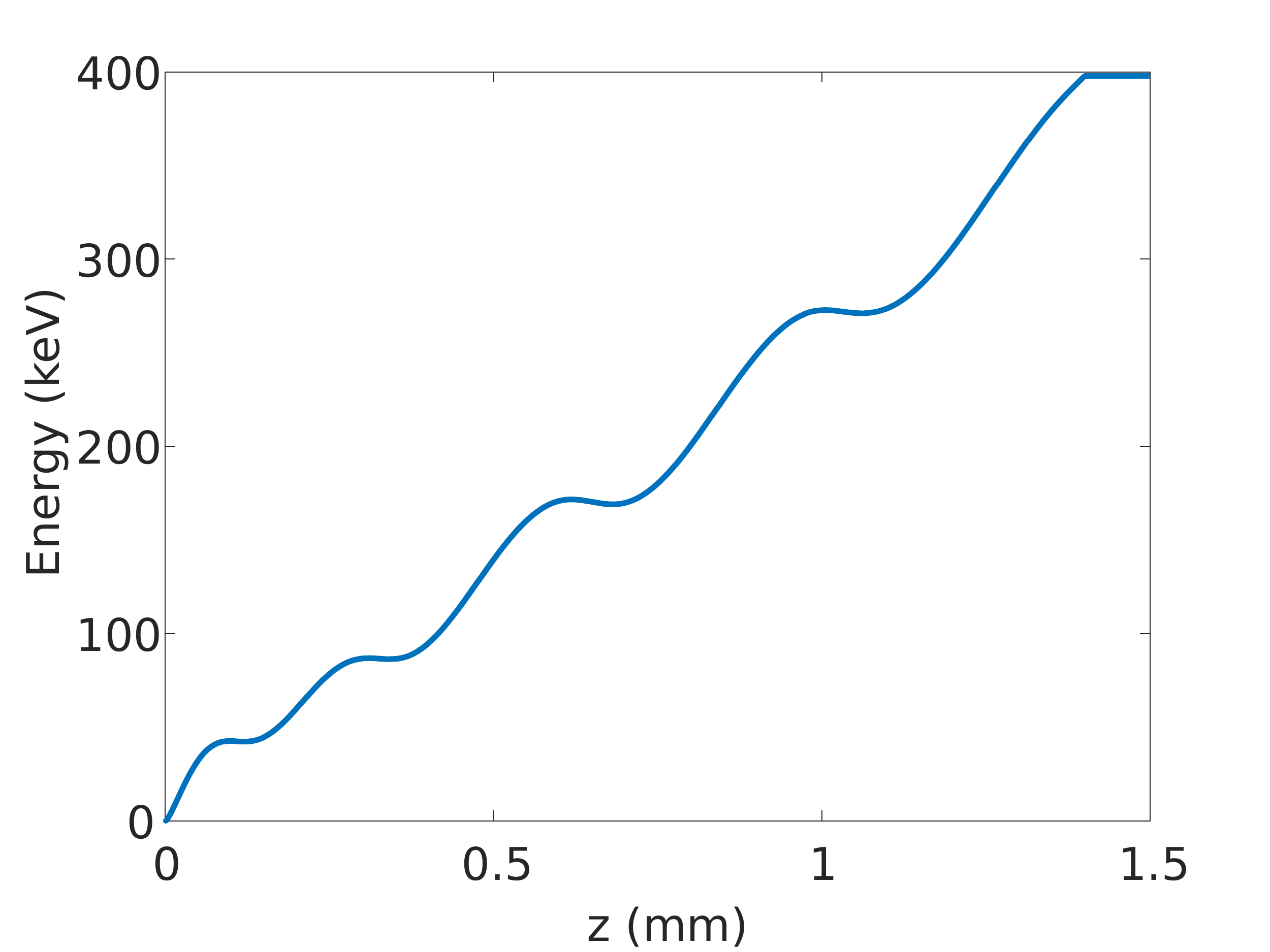} \\
  	(a) & (b)
  	\end{array}$
  	\caption{Simulation results of the optimized design for 400\,keV electron gun: (a) accelerating field in the middle of each layer versus time, and (b) energy of an electron injected at instant with 80 MV/m accelerating field.}
  	\label{400keVFinalDesign}
  \end{figure}

  After improving the design based on the above considerations and performing a series of iterative optimization on the various involved parameters, the values tabulated in Table\,\ref{400keVGunDesign} in conjunction with the updated field signature and electron energy gain depicted in Fig.\,\ref{400keVFinalDesign} are obtained.
  The new results evidence a pronounced enhancement in the device operation through the fine tuning procedure.
  We comment that the energy gain in Fig.\,\ref{400keVFinalDesign} is obtained for one electron injected at the time when the accelerating field is 80\,MV/m, which demonstrates acceleration up to 400\,keV electron energy.
  
  It is often helpful for the design process to define a figure of merit for the performance.
  A suitable measure is the ratio of energy gain for one electron to the total input energy.
  For instance, the fine-tuned device results in the ratio 1.0\,keV/{\textmu}J = 1.59$\times10^{-10}$, while the initial design leads to the ratio 0.92\,keV/{\textmu}J = 1.48$\times10^{-10}$.
  Note that this measure verifies the overall coupling, focusing and acceleration schemes with the exclusion of bunch evolution.
  Another measure to separately assess the coupling and focusing sections can be defined as the ratio between the energy of one cycle with maximum field in the interaction region to the total input energy.
  The nominator is obtained in a similar fashion as \eref{beamEnergy}, but with the ultimate maximum accelerating gradients ($E_i$), and the denominator is simply the input energy of the Gaussian beam ($E_t$).
  Therefore, this ratio can be written as
  \begin{equation}
  \mathcal{R} = \sum\limits_i \frac{E_i^2 \lambda d_i}{32 \eta_0 \mathcal{E}_t} \tau \sqrt{\frac{\pi}{4 \ln 2}}.
  \label{ratioDefinition}
  \end{equation}
  This ratio for the fine-tuned device and the initial design are obtained as 0.47 and 0.46.
  In other words, although the fine tuning enhanced the total acceleration scenario, it only slightly varied the coupling of the THz beam.
  This has occurred because of the strong interdependence between all the involved parameters in the overall performance.
  Examining the above introduced parameters for each design properly guides the fine tuning process of the electron gun dimensions.
  
  After optimization and finalizing the design for acceleration of a single particle, a more accurate assessment of the structure is achieved through investigation of the full bunch acceleration scenario.
  For this purpose, similar to the approach in the previous section, we assume a copper cathode excited by a UV laser pulse at 250\,nm wavelength with FWHM pulse duration equal to 47\,fs and FWHM spot size diameter 47\,{\textmu}m.
  The UV laser energy is supposed to be large enough such that 1\,pC of charge is released, which is modeled by 20'000 macro-particles.
  The 6D phase-space distribution of such a photocathode is obtained using the ASTRA bunch generator module.
  In addition, we use a point-to-point algorithm to account for Coulomb repulsion between electrons, i.e. space-charge effects.
  
\begin{figure}
  	\centering
  	$\begin{array}{cc}
  	\includegraphics[draft=false,width=3.0in]{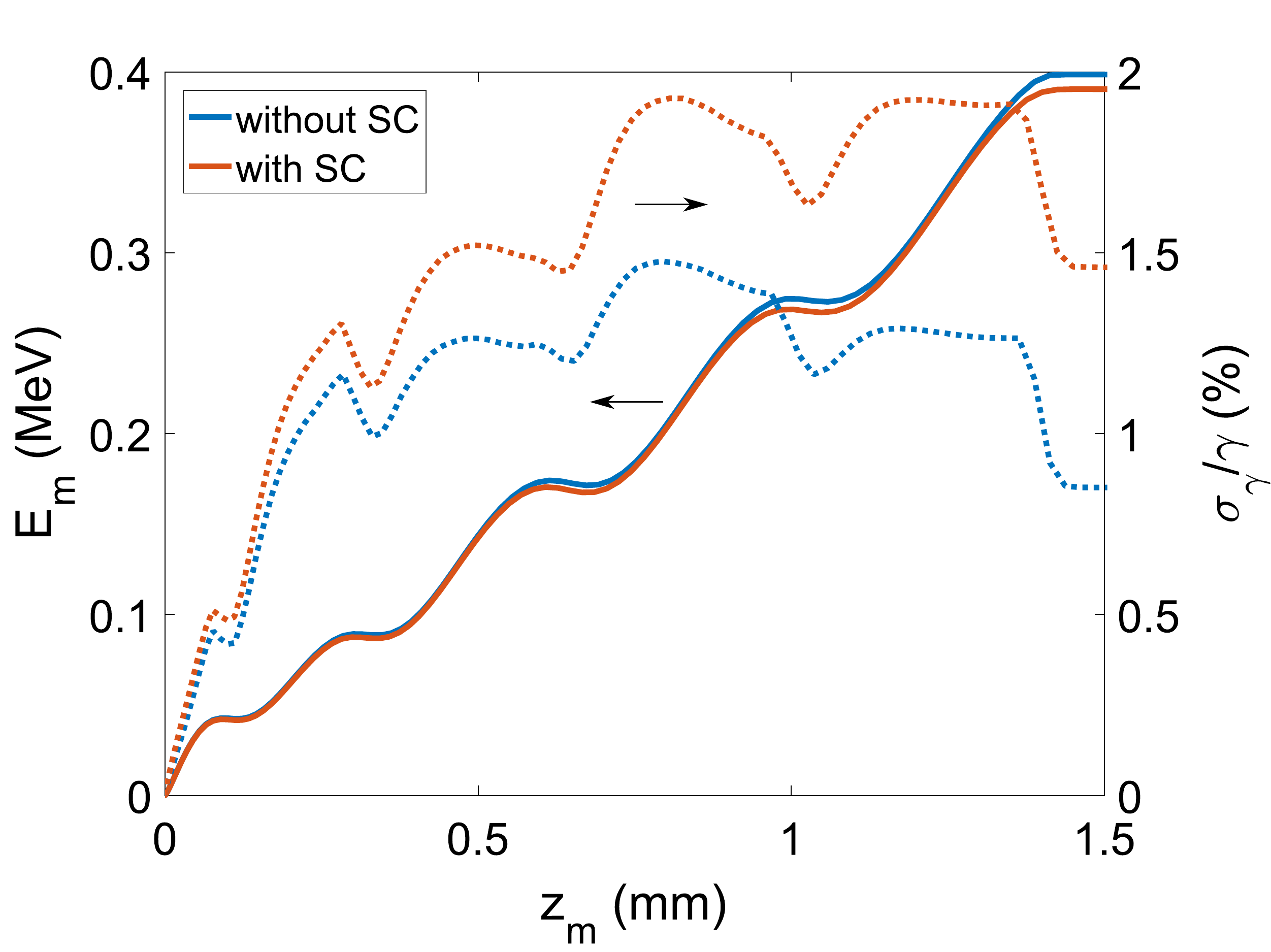} &
  	\includegraphics[draft=false,width=3.0in]{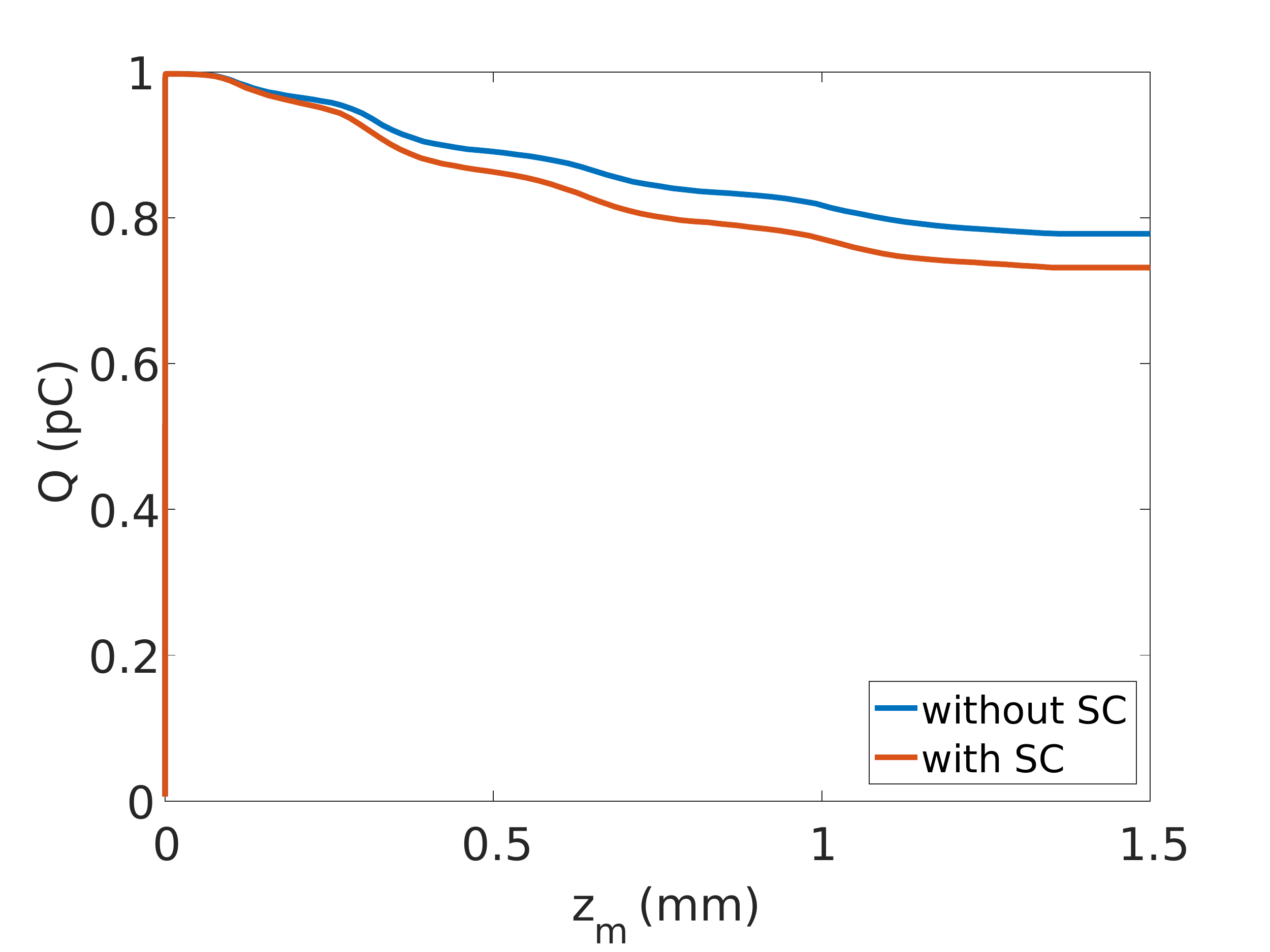} \\
  	(a) & (b) \\
  	\includegraphics[draft=false,width=3.0in]{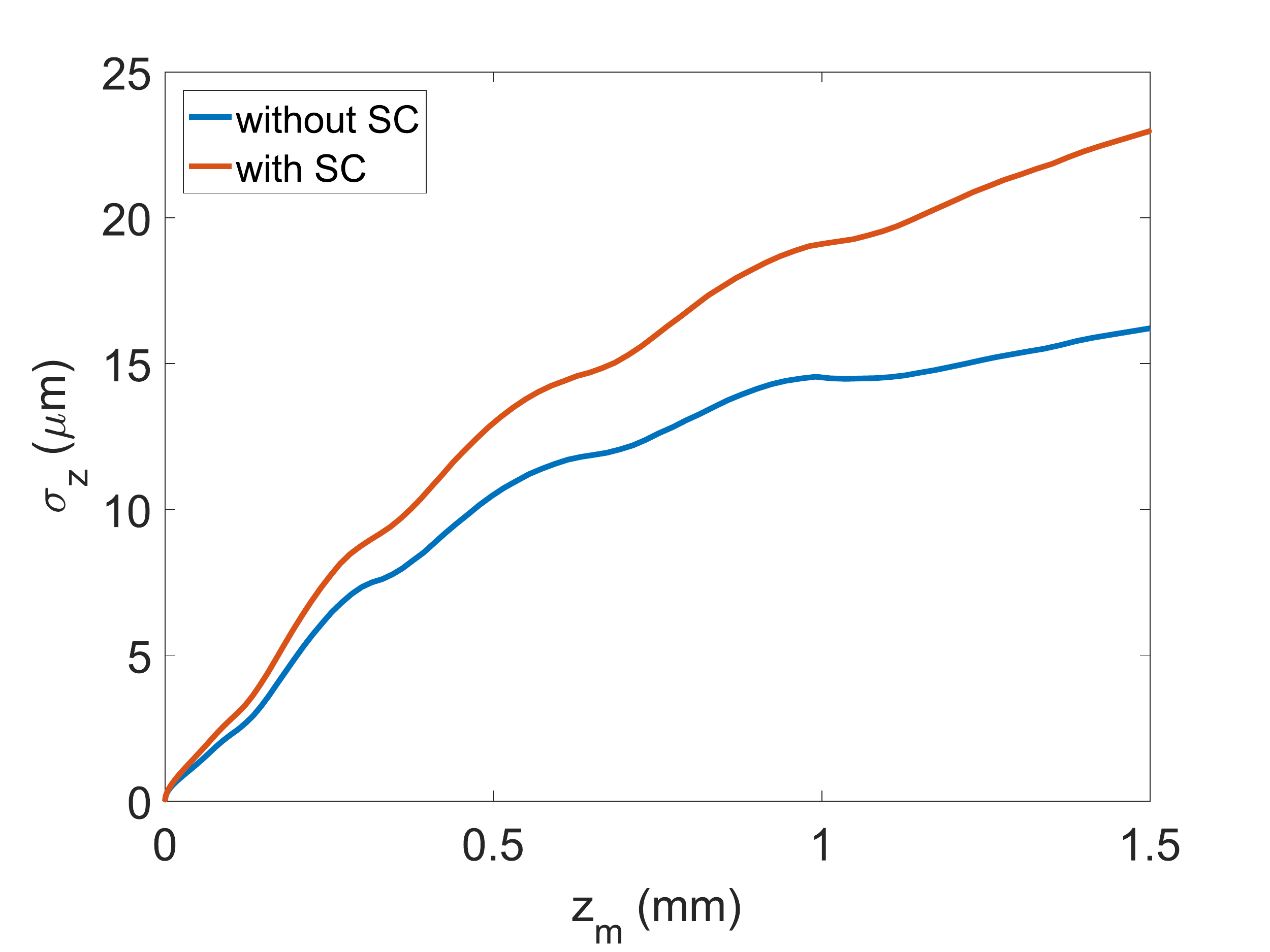} &
  	\includegraphics[draft=false,width=3.0in]{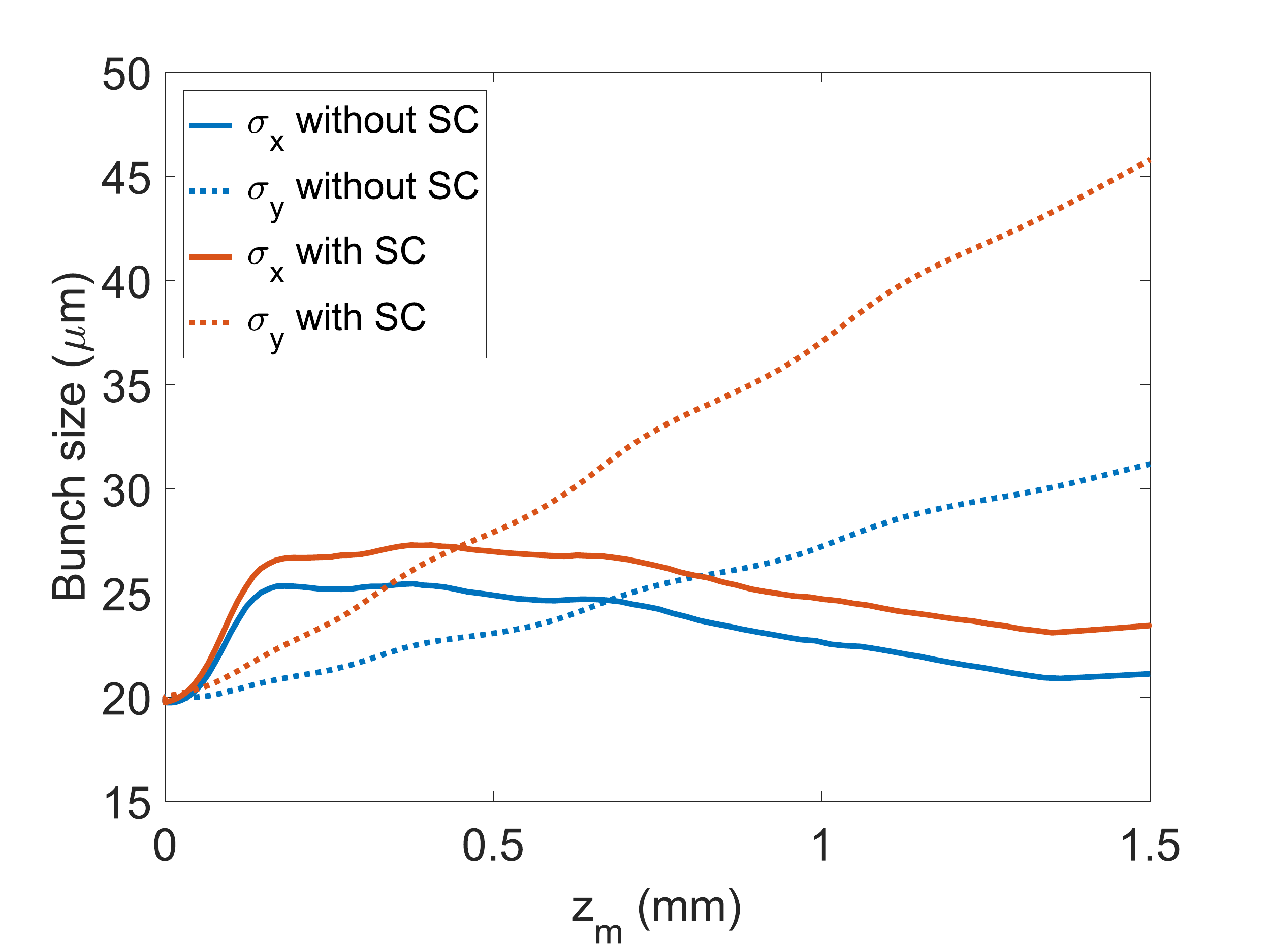} \\
  	(c) & (d)
  	\end{array}$
  	\caption{Acceleration of 1\,pC photoemitted bunch in the fine-tuned 400\,keV THz gun: (a) mean energy and energy spread of the bunch, (b) bunch charge, (c) bunch length and (d) bunch size at the gun exit are depicted in terms of the traversed distance.}
  	\label{400keVGunBunchAcceleration}
  \end{figure}

  Fig.\,\ref{400keVGunBunchAcceleration} shows the evolution of the bunch properties throughout the acceleration process.
  The bunch parameters are depicted for the two cases with and without consideration of space-charge effects.
  The bunch parameters obtained without space-charge are valid in the low-charge regime ($<100$\,fC), whereas acceleration of 1\,pC injected charge suffers from strong space-charge forces that increase position and energy spread.
  It is observed that the final mean energy of the bunch is about 390\,keV with an energy spread of about 1.5\%, which happens due to the large spot size of the injected bunch compared to the THz wavelength (1\,mm) and space-charge forces.
  The compensation of energy spread when the bunch traverses the separators can be seen in Fig.\,\ref{400keVGunBunchAcceleration}a.
  This effect is the main advantage of thick separators enabling velocity bunching in this region.
  Due to the collisions of the electrons with the metallic boundaries emanating from the transverse momentum of electrons, 73\% of the photo-emitted electrons are extracted from the gun.
  This effect shows the limitation on bunch size and correspondingly the amount of charge which can be accelerated with a desirable quality using the proposed THz gun.
  The normalized emittances of the beam at the output with and without space-charge are evaluated as $(\varepsilon_{xn},\varepsilon_{yn},\varepsilon_{zn}) = (0.20,0.25,0.21)\,$mm$\cdot$mrad and (0.17,0.13,0.11)\,mm$\cdot$mrad, respectively.
  
  \subsection{800\,keV Gun}
  
  The described design procedure, as used for designing an exemplary 400\,keV gun, can be generally followed to design any type of gun excited with short pulses in THz and microwave regimes.
  An electron gun with 400\,keV output energy may be a suitable option for applications like electron diffractive imaging.
  However, for injecting to a linear accelerator, where velocity matching with the phase of the accelerating field is essential, electron guns with higher output energies are required.
  In what follows, the design for an optimum 800\,keV electron injector is given following the same design procedure.
  The 800\,keV gun consists of eight layers in the interaction section with design parameters tabulated in Table\,\ref{800keVGunDesign}.
  \begin{table}
  	\caption{The parameter values for the 800\,keV gun} \label{800keVGunDesign} \centering
  	{\footnotesize
  		\begin{tabular}{|c||c|}
  			\hline
  			parameter & ultimate value \\ \hline
  			$(d_x,d_y)$ & (0.71,0.71)\,mm \\ \hline
  			$(d_1,d_2,d_3,d_4,d_5,d_6,d_7,d_8)$ & (80, 180, 280, 320, 360, 390, 420, 450)\,{\textmu}m \\ \hline
  			$t$ & 40\,{\textmu}m \\ \hline
  			$g$ & 120\,{\textmu}m \\ \hline
  			$(L_1,L_2,L_3,L_4,L_5,L_6,L_7,L_8)$ & (0, 400, 850, 1300, 1800, 2250, 2700, 3150)\,{\textmu}m \\ \hline
  			$(L_f,L_c)$ & (3.5,2.5)\,mm \\ \hline
  			$(\theta,\alpha)$ & $(20.0^{\circ},14.0^{\circ})$ \\ \hline
  			$n$ & 2.1 (quartz) \\ \hline
  			$(\tau,w_{0y},w_{0z})$ & (3.33\,ps, 1\,mm, 2\,mm) \\ \hline
  			$f$ & 300\,GHz \\ \hline
  			$\mathcal{E}$ & 400\,{\textmu}J \\ \hline
  			$(x_f,y_f,z_f)$ & ($\pm$2.4, 0.0, 1.34)\,mm \\ \hline
  		\end{tabular}
  	}
  \end{table}
  
\begin{figure}
  	\centering
  	$\begin{array}{cc}
  	\includegraphics[draft=false,width=3.0in]{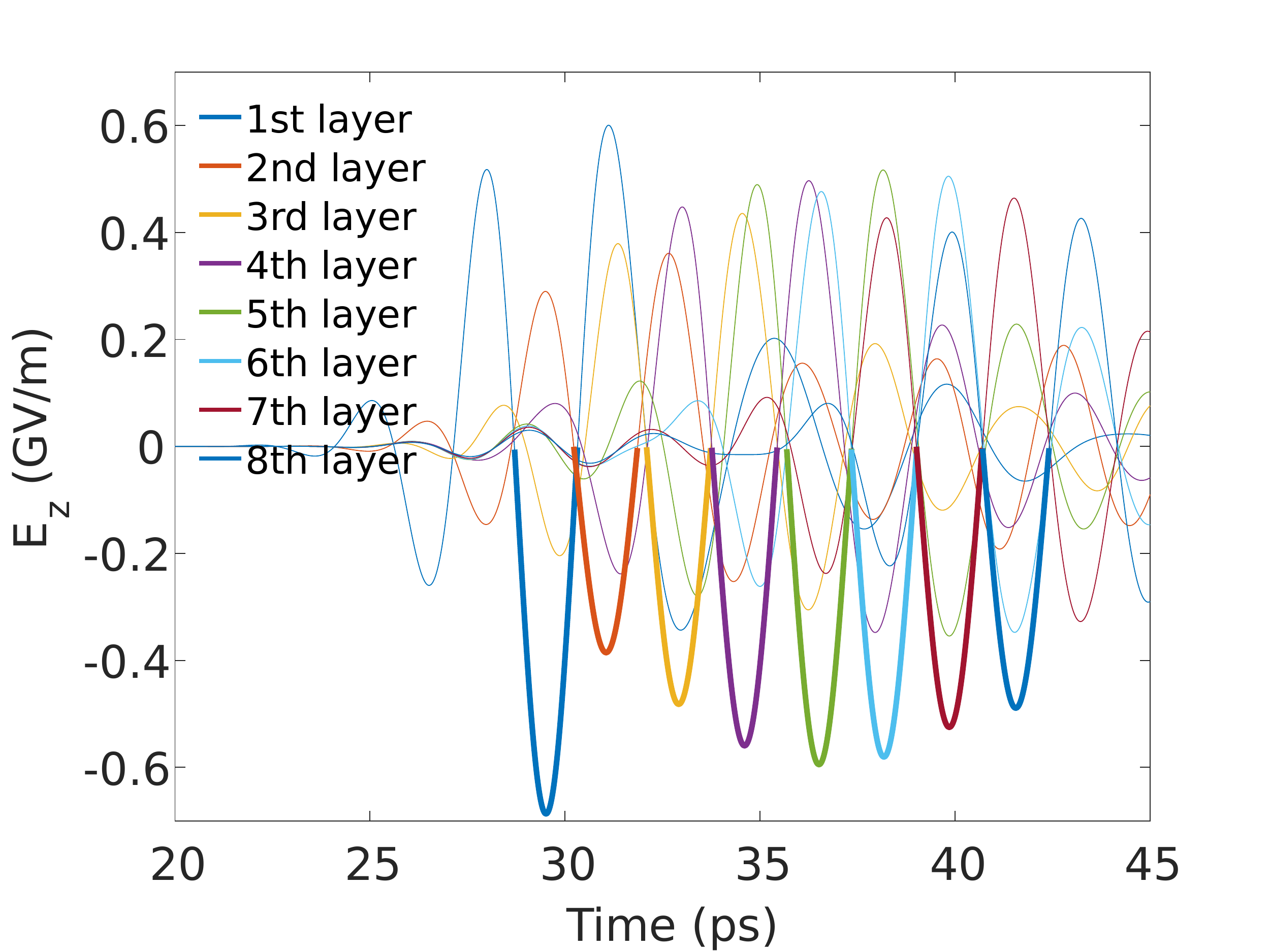} &
  	\includegraphics[draft=false,width=3.0in]{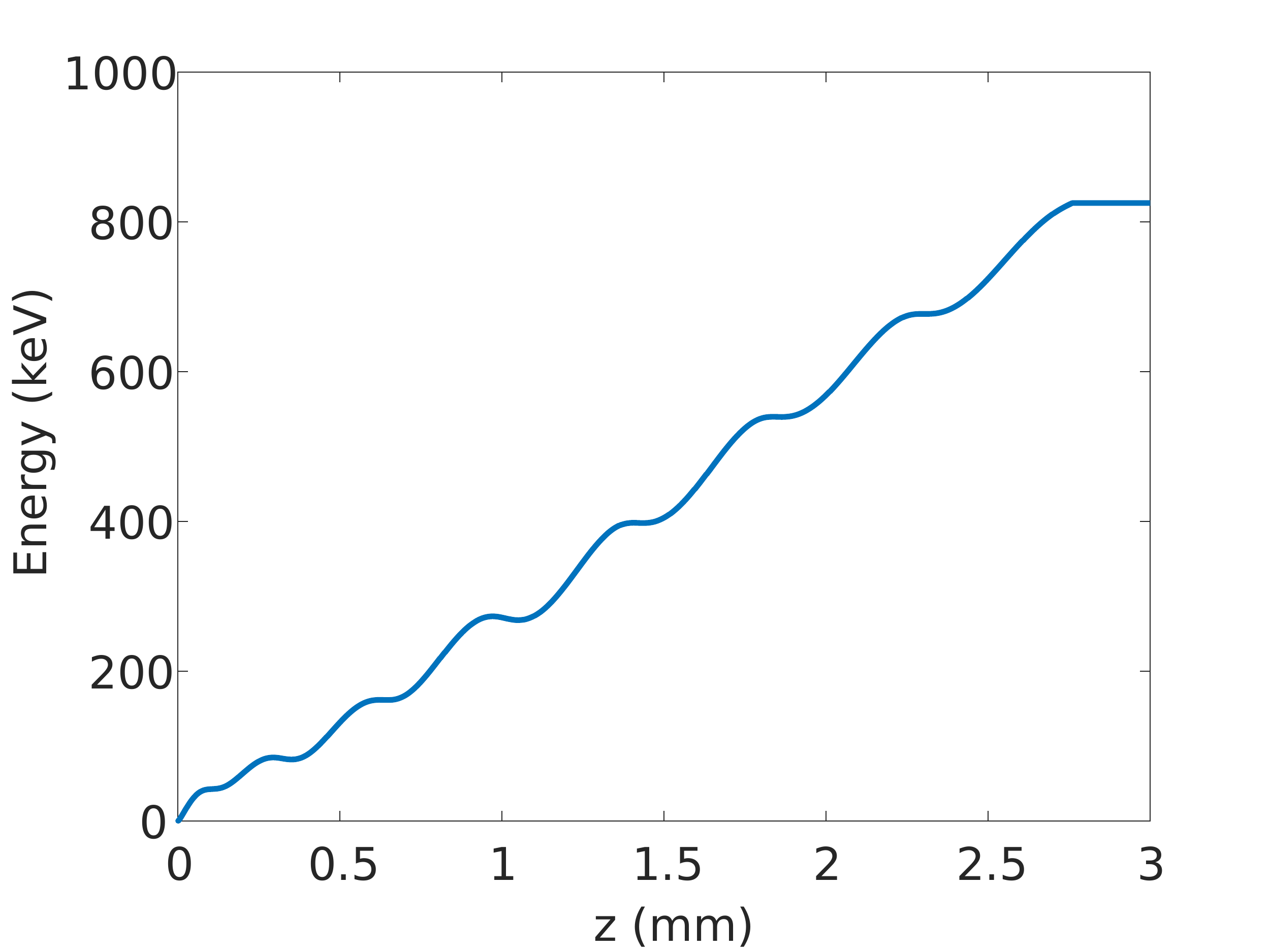} \\
  	(a) & (b)
  	\end{array}$
  	\caption{Simulation results of the optimized design for 800\,keV electron gun: (a) accelerating field in the middle of each layer versus time, and (b) energy of an electron injected at the instant with 200\,MV/m accelerating field.}
  	\label{800keVFinalDesign}
  \end{figure}

\begin{figure}
  	\centering
  	$\begin{array}{ccc}
  	\includegraphics[draft=false,width=2.0in]{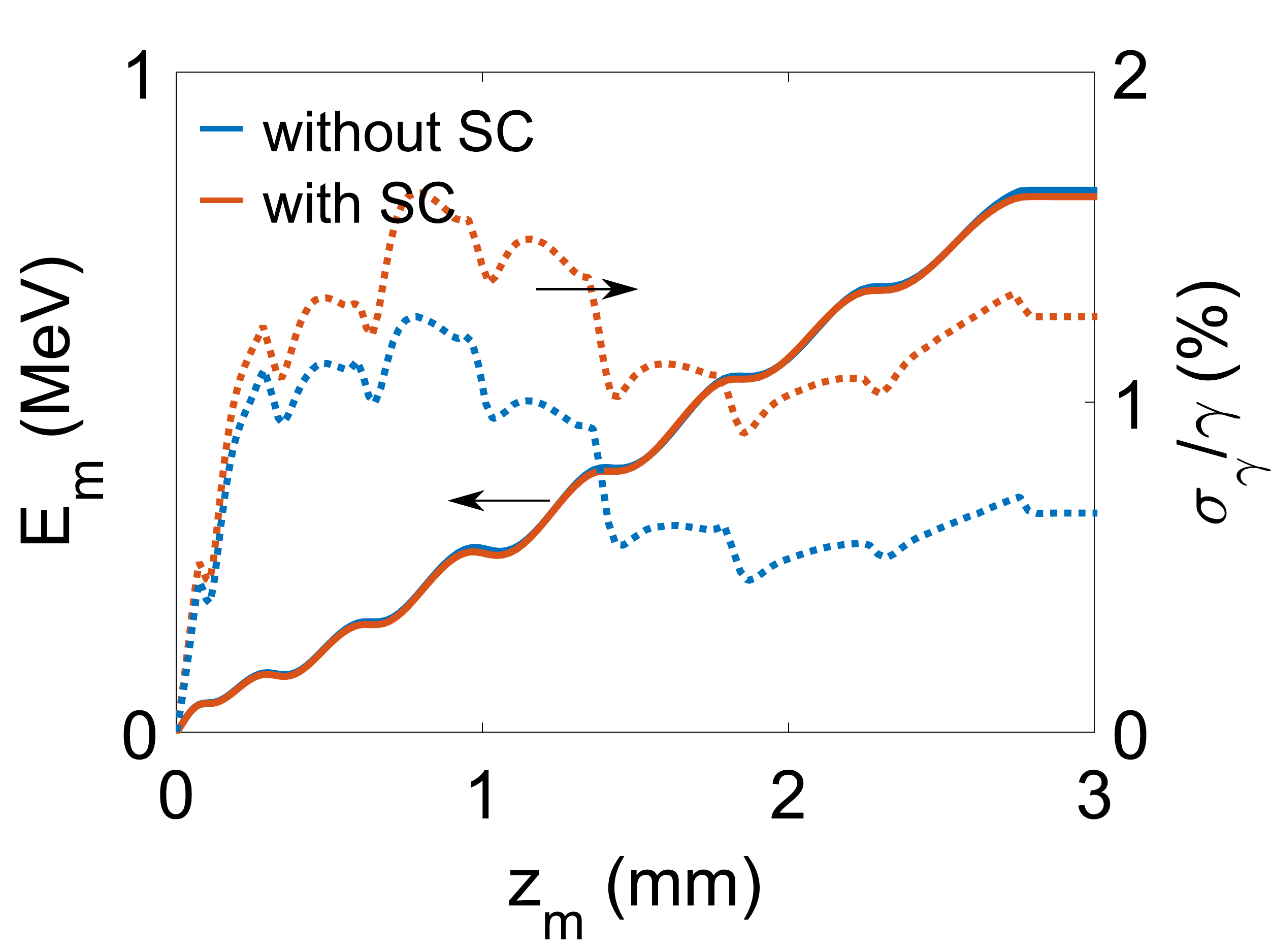} &
  	\includegraphics[draft=false,width=2.0in]{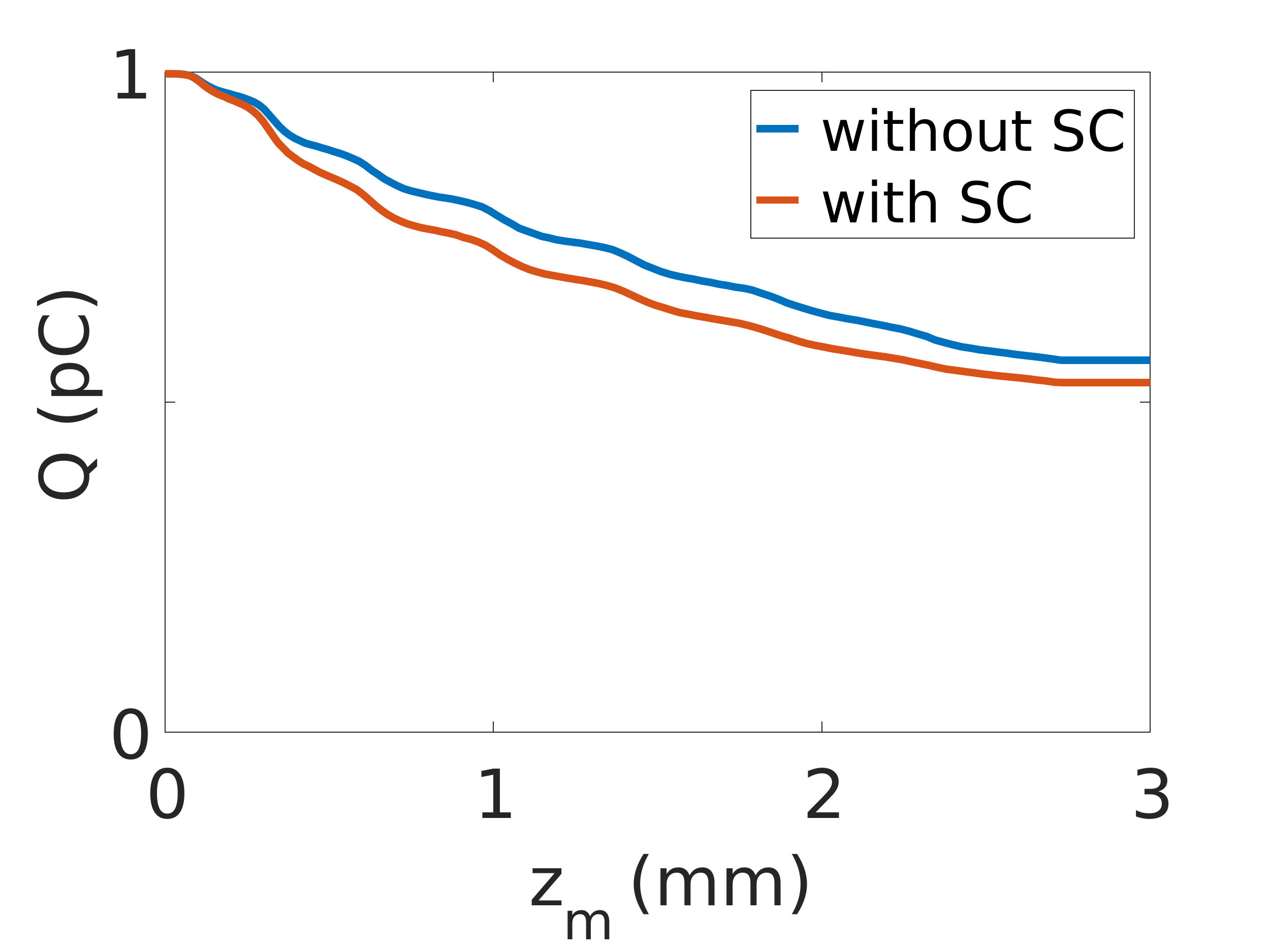} &
  	\includegraphics[draft=false,width=2.0in]{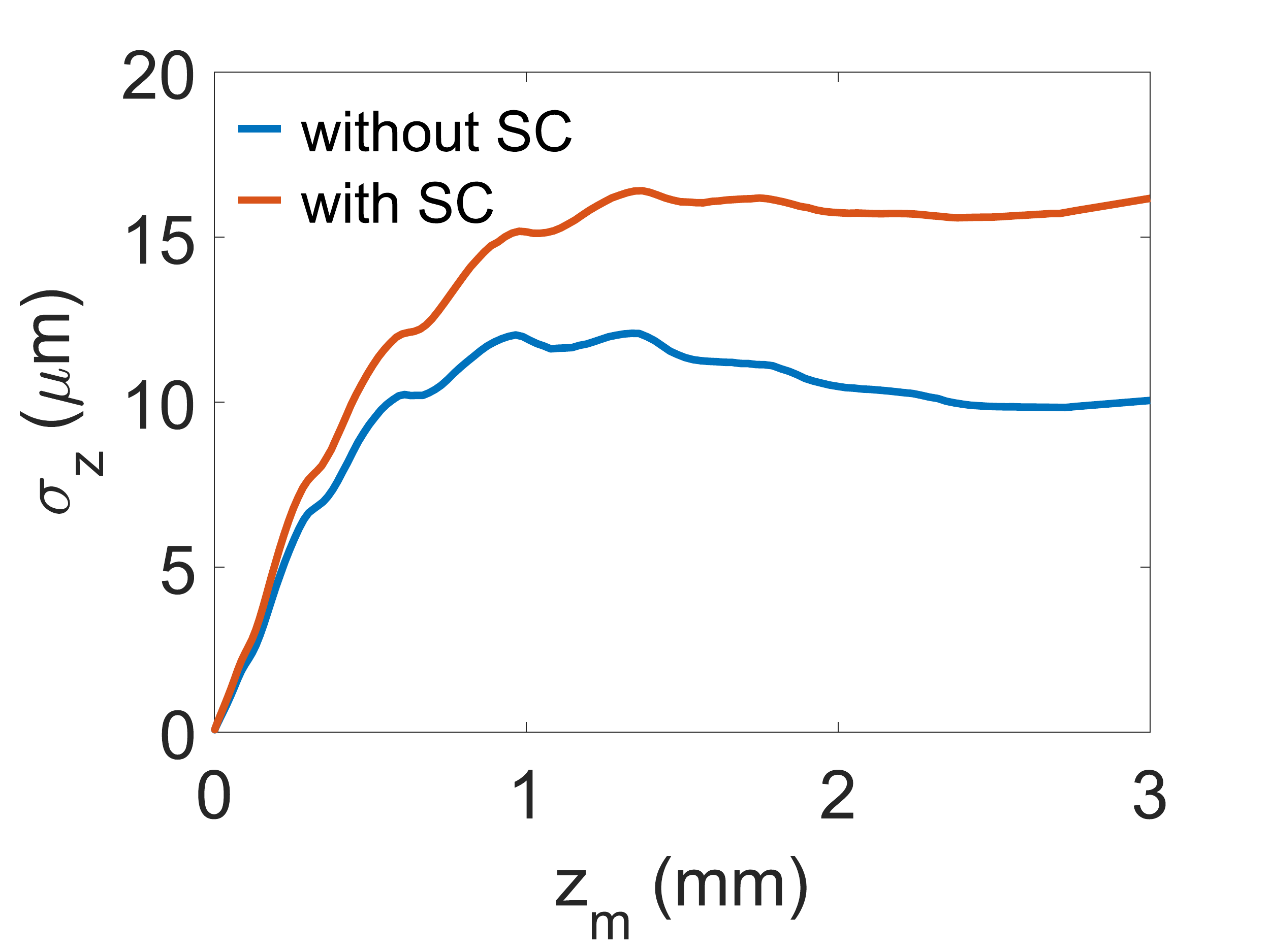} \\
  	(a) & (b) & (c) \\
  	\includegraphics[draft=false,width=2.0in]{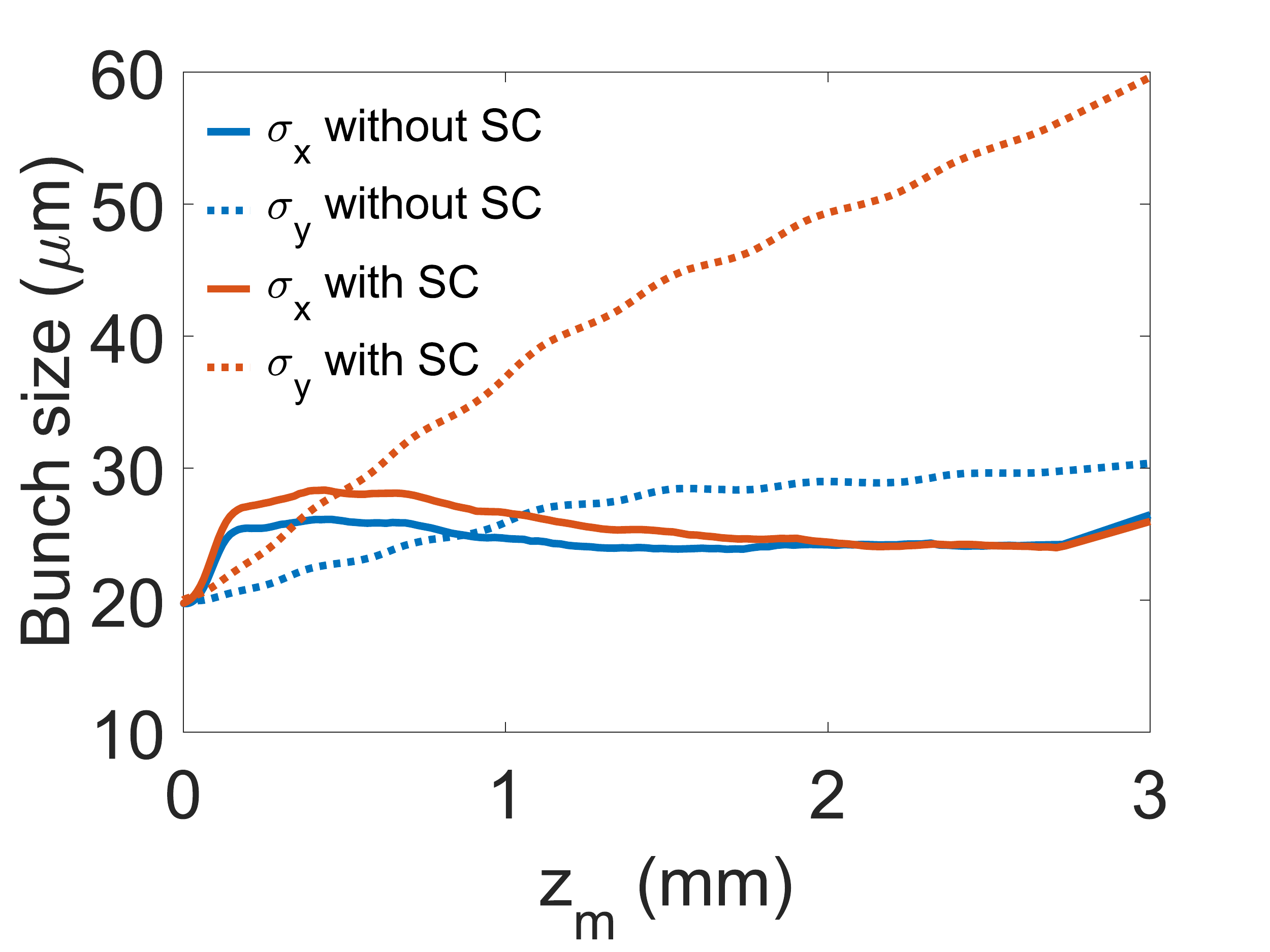} &
  	\includegraphics[draft=false,width=2.0in]{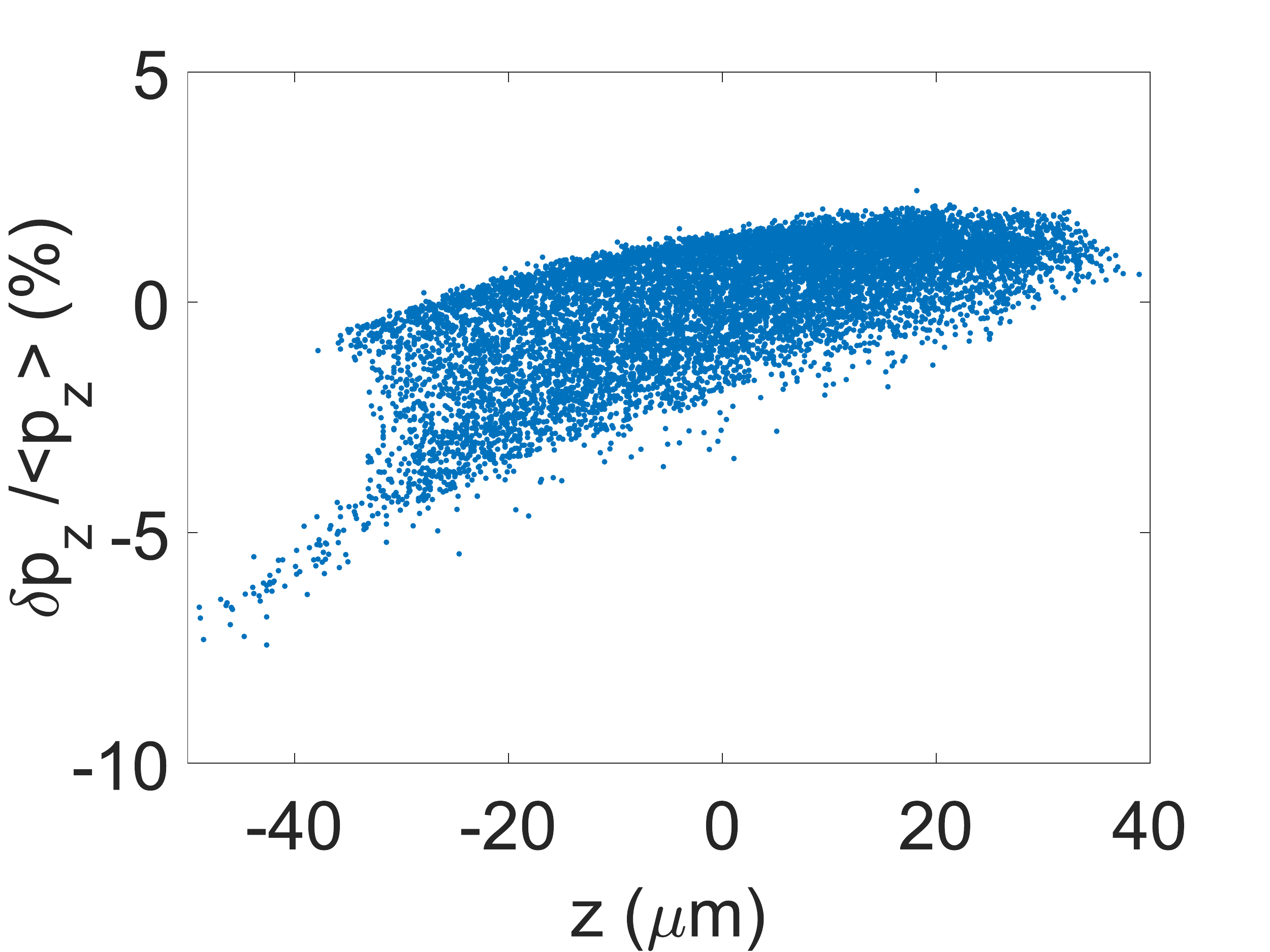} &
  	\includegraphics[draft=false,width=2.0in]{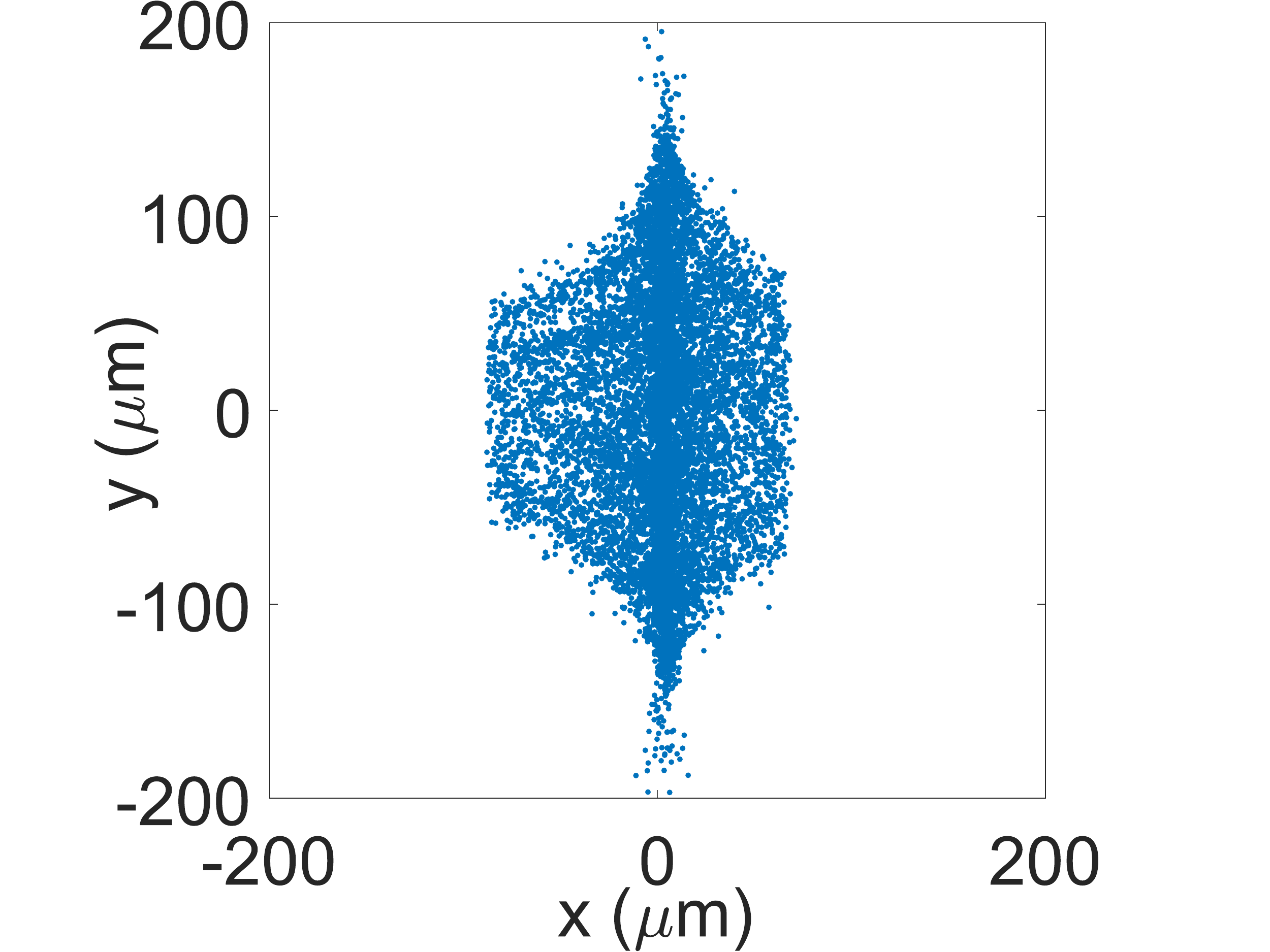} \\
  	(d) & (e) & (f)
  	\end{array}$
  	\caption{Acceleration of 1\,pC photoemitted bunch in the fine-tuned 800\,keV THz gun: (a) mean energy and energy spread of the bunch, (b) bunch charge, (c) bunch length and (d) bunch size, (e) transverse beam profile and (f) longitudinal phase-space at the gun exit are depicted in terms of the traversed distance.}
  	\label{800keVGunBunchAcceleration}
  \end{figure}

  The results of the single particle acceleration simulation are depicted in Fig.\,\ref{800keVFinalDesign}.
  As demonstrated by the simulations, by just adding three layers and doubling the beam input energy, the energy gain of the particles is doubled to 820\,keV.
  We comment that in this example the incident Gaussian beam is considered to be elliptical to better match with the acceleration length, which is relatively long in comparison with the transverse size of the interaction section.
  In practice, such an elliptical Gaussian beam is realized using standard optical elements.
  Otherwise, the coupler parameters need to be adjusted based on the input Gaussian beam dimensions.
  According to these results, the ratio between energy gain for one electron and the total input energy is equal to 1.03\,keV/{\textmu}J = $1.65\times10^{-10}$, which is slightly better than for the 400\,keV gun and additionally the $\mathcal{R}$ parameter defined in \eref{ratioDefinition} is calculated as 0.52.
  Eventually, the bunch acceleration is inspected by injecting the same photoemission electron bunch as in the 400\,keV gun study.
  The results of this analysis are illustrated in Fig.\,\ref{800keVGunBunchAcceleration}, which demonstrates realization of a 520\,fC electron beam with 811\,keV mean energy, 1.2\% energy spread, and bunch dimensions of about $(25\,\text{{\textmu}m} \times 58\,\text{{\textmu}m} \times 16\,\text{{\textmu}m})$.
  The output normalized emittances are evaluated as $(\varepsilon_{xn},\varepsilon_{yn},\varepsilon_{zn}) =(0.13,0.69,0.43)$\,mm$\cdot$mrad for the space-charge included simulation and (0.09,0.29,0.15)\,mm$\cdot$mrad without accounting for space-charge effects.
  In addition, the transverse beam profile and longitudinal phase-space of the electron bunch at the output of the gun including space-charge effects are also illustrated.
  
  \section{Experimental Test of the Parallel-plate Low-energy Gun}
  
  Here, we implement and test the first version of a low-energy THz gun.
  Leveraging the gun's simple geometries and flexible machining requirements, we integrate it into a practical, compact machine that is powered by a 1\,kHz, few-millijoule laser and operates without external synchronization.
  Our first results demonstrate high field (350\,MV/m) acceleration up to 0.8\,keV, as well as percent-level energy spread in sub-kiloelecton volt, several tens of fC bunches.
  These results, which are already suitable for time-resolved low-energy electron diffraction (LEED) experiments, confirm the performance of a THz-driven gun technology that is scalable to relativistic energies \cite{fallahi2016short}.
  
\begin{figure}
  	\centering
  	$\begin{array}{cc}
  	\begin{array}{c}
  	\includegraphics[draft=false,height=2.0in]{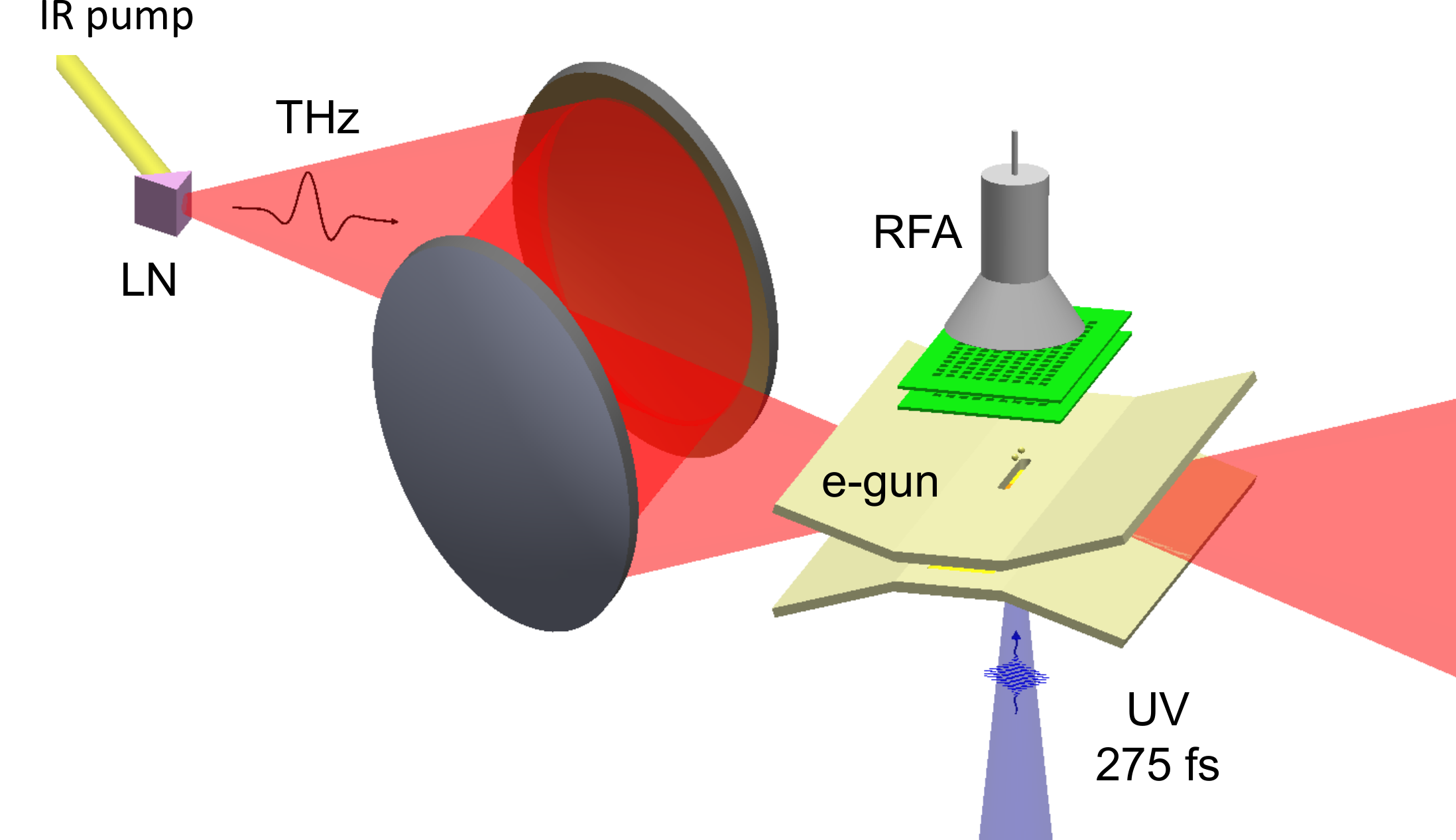} \\
  	(a)
  	\end{array} &
  	\begin{array}{c}
  	\includegraphics[draft=false,height=0.8in]{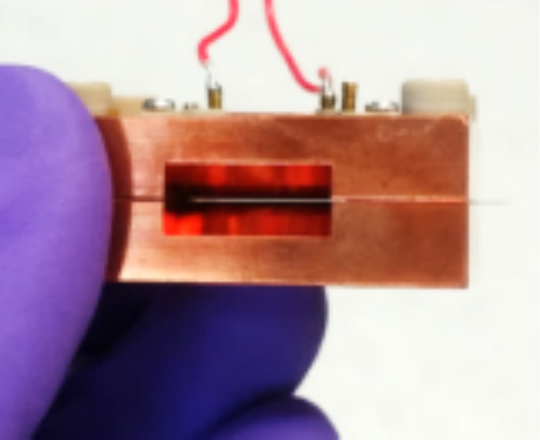} \\
  	(b) \\
  	\includegraphics[draft=false,height=1.0in]{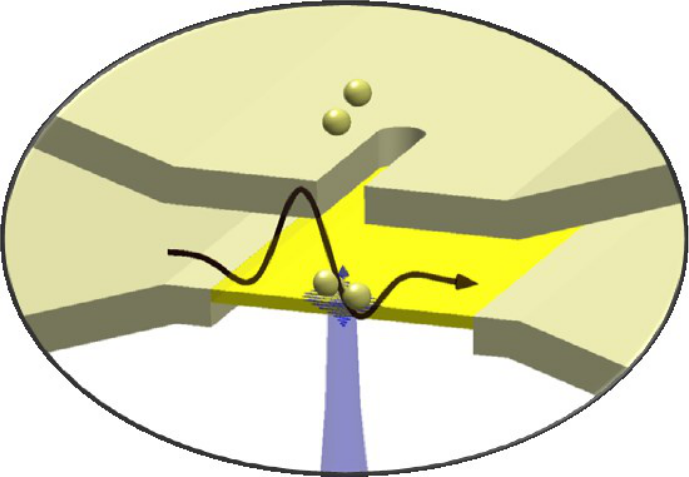} \\
  	(c)
  	\end{array}
  	\end{array}$
  	\caption{PPWG THz gun concept: (a) a single-cycle THz pulse, generated via optical rectification in lithium niobate (LN), is coupled into the THz gun, which takes the form of a parallel-plate waveguide for field confinement. A UV back-illuminated photocathode emits an electron bunch, which is accelerated by the THz field. The bunch exits through a slit on the top plate, and a retarding field analyzer (RFA) measures its energy spectrum. (b) Photograph of the THz gun, and (c) cross section of the gun, showing the UV photoemitted electrons accelerated by the THz field and exiting through the slit.}
  	\label{THzGunImplementationConcept}
  \end{figure}

\begin{figure}
  	\centering
  	$\begin{array}{c}
  	\begin{array}{ccc}
  	\includegraphics[draft=false,width=3.0in]{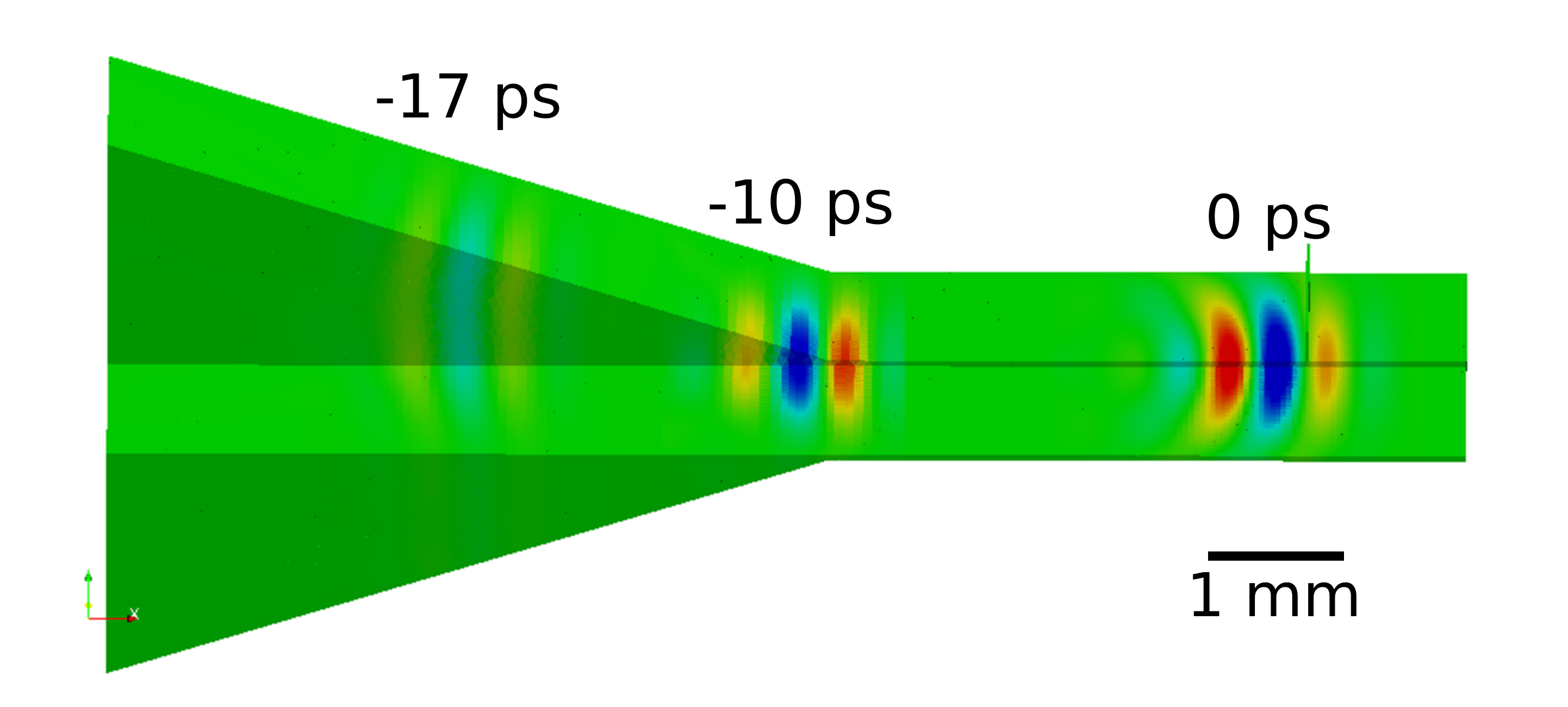} &
  	\includegraphics[draft=false,height=1.2in]{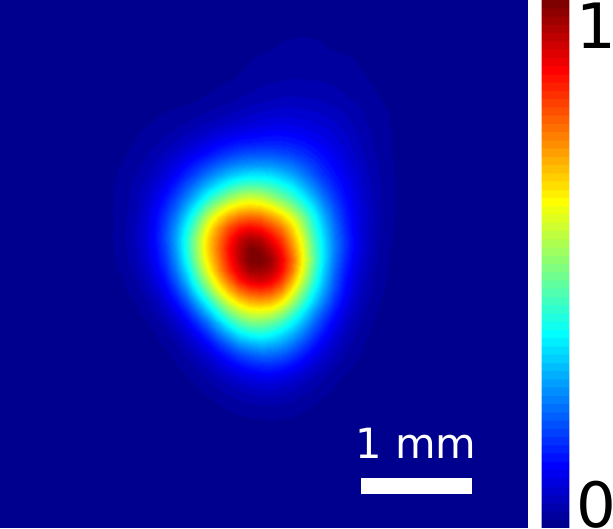} &
  	\includegraphics[draft=false,height=1.2in]{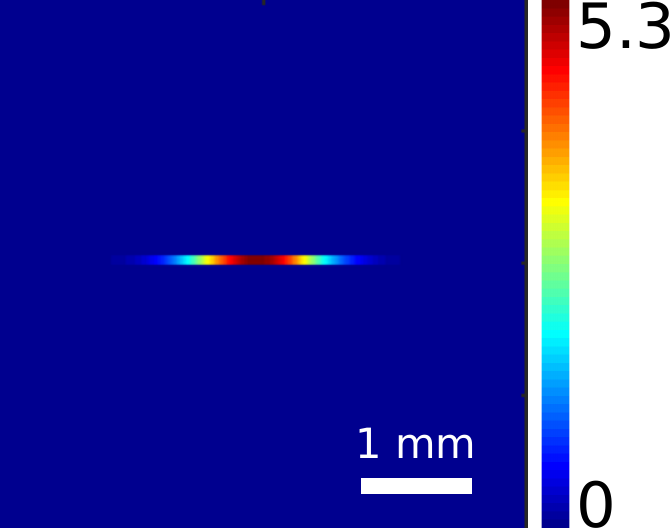} \\
  	(a) & (b) & (c)
  	\end{array} \\
  	\begin{array}{cc}
  	\includegraphics[draft=false,width=3.0in]{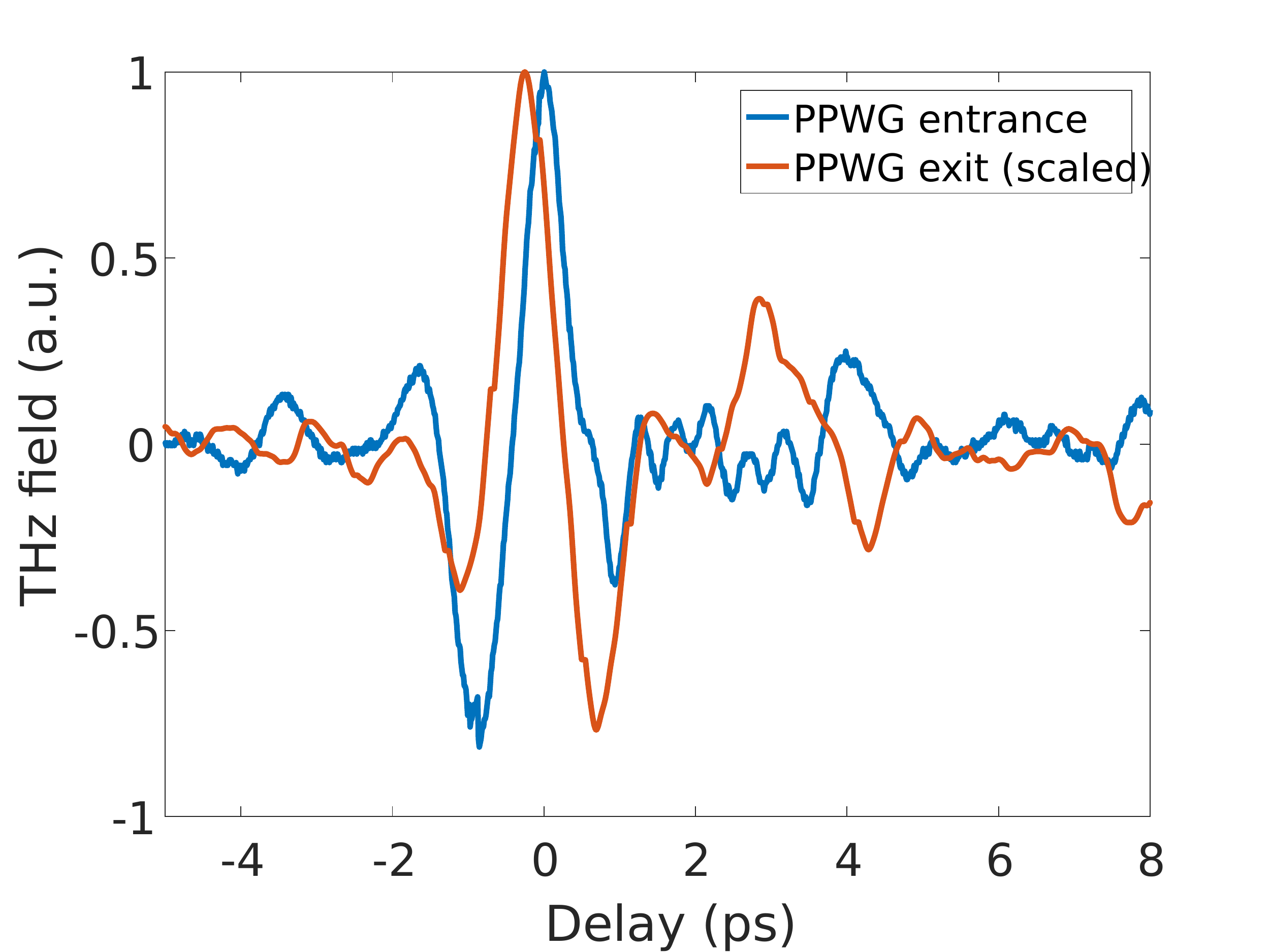} &
  	\includegraphics[draft=false,width=3.0in]{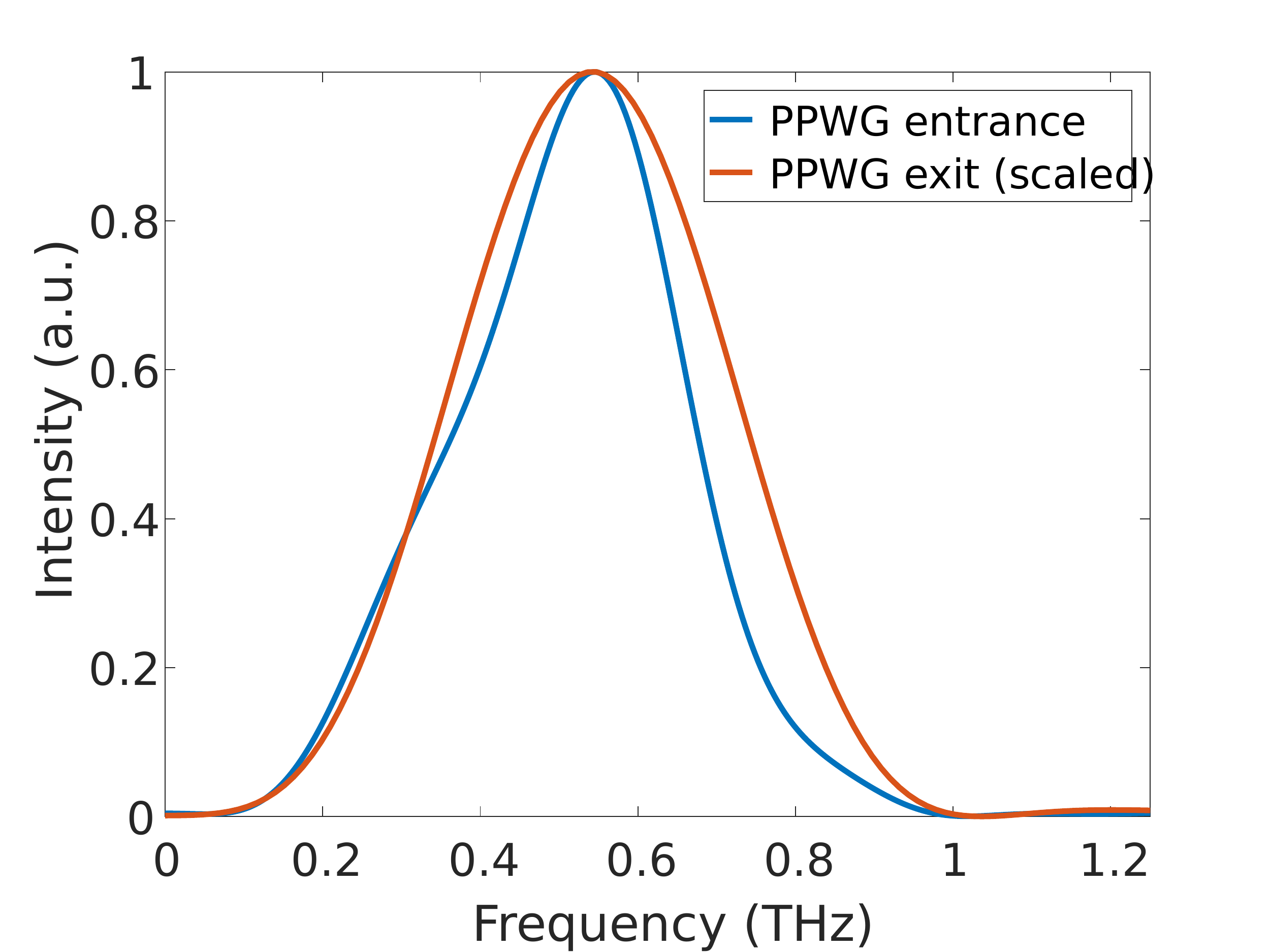} \\
  	(d) & (e)
  	\end{array}
  	\end{array}$
  	\caption{Design and characterization of the gun: (a) Several snapshots of the single-cycle THz wave coupling into the PPWG, based on EM simulations, (b) THz beam intensity at the free-space focus (measured), (c) calculated THz beam intensity in the gun. The colorbars of the two beam profiles show a 5.3 $\times$ intensity enhancement in the PPWG. (d) Temporal profiles measured via EO sampling of the THz electric field at PPWG-center and PPWG-thru (scaled) and (e) power spectrum of the THz beam after and before the PPWG.}
  	\label{THzGunImplementationAnalysis}
  \end{figure}

  The THz gun (Fig.\,\ref{THzGunImplementationConcept}a-c) takes the form of a copper parallel-plate waveguide (PPWG) with a subwavelength spacing of 75\,{\textmu}m.
  We exploit this structure's transverse electromagnetic mode for unchirped, uniform enhancement of the THz field \cite{iwaszczuk2012terahertz}.
  A free-space vertically polarized THz beam is coupled into the PPWG by a taper.
  EM simulations (Fig.\,\ref{THzGunImplementationAnalysis}a) were utilized to optimize the taper and calculate the coupling efficiency.
  Inside, a copper film photocathode serves as the bottom plate of the PPWG.
  There, a UV pulse back-illuminates the film, producing electrons inside the PPWG by photoemission.
  Concurrently, the THz field accelerates the electrons vertically across the PPWG.
  The electrons exit the gun through a slit on the top plate (anode) and are spectrally characterized by a retarding field analyzer (RFA) or counted by a Faraday cup.
  Both UV and THz pulses are generated from the same 1030\,nm pump laser, ensuring absolute timing synchronization.
  
  The THz pulse, generated in lithium niobate by the tilted pulse-front method, is focused into the gun with a maximum impinging energy of 35.7\,{\textmu}J.
  Electro-optic (EO) sampling at PPWG-center (location of the center of the gun with the gun removed) and PPWG-thru (focus of an image relay following propagation through the PPWG) reveals single-cycle durations of $\tau_\mathrm{THz}=1.2$\,ps (Fig.\,\ref{THzGunImplementationAnalysis}d and \ref{THzGunImplementationAnalysis}e), confirming that the PPWG induces minimal dispersion.
  Fig.\,\ref{THzGunImplementationAnalysis}b shows the THz beam profile at the free-space focus.
  Inside the waveguide, the horizontal ($x$) beam profile remains unaltered, while the vertical ($z$) profile is distributed uniformly across the 75\,{\textmu}m spacing.
  Based on this fact, Fig.\,\ref{THzGunImplementationAnalysis}c shows the calculated beam profile inside the gun.
  Taking into account the energy, waveform, beam profile, and coupling efficiency, the THz pulse has a calculated peak field of 153\,MV/m in free space and 350\,MV/m in the gun.
  The UV emitter pulse, generated by frequency quadrupling of the pump laser, has a wavelength of 258\,nm, an estimated duration of $\tau_\mathrm{UV}=275$\,fs (roughly 12.5\% of the THz period), and a focused beam waist of 60\,{\textmu}m ($x$) and 20\,{\textmu}m ($y$) on the photocathode.
  
\begin{figure}
  	\centering
  	$\begin{array}{cc}
  	\includegraphics[draft=false,width=3.0in]{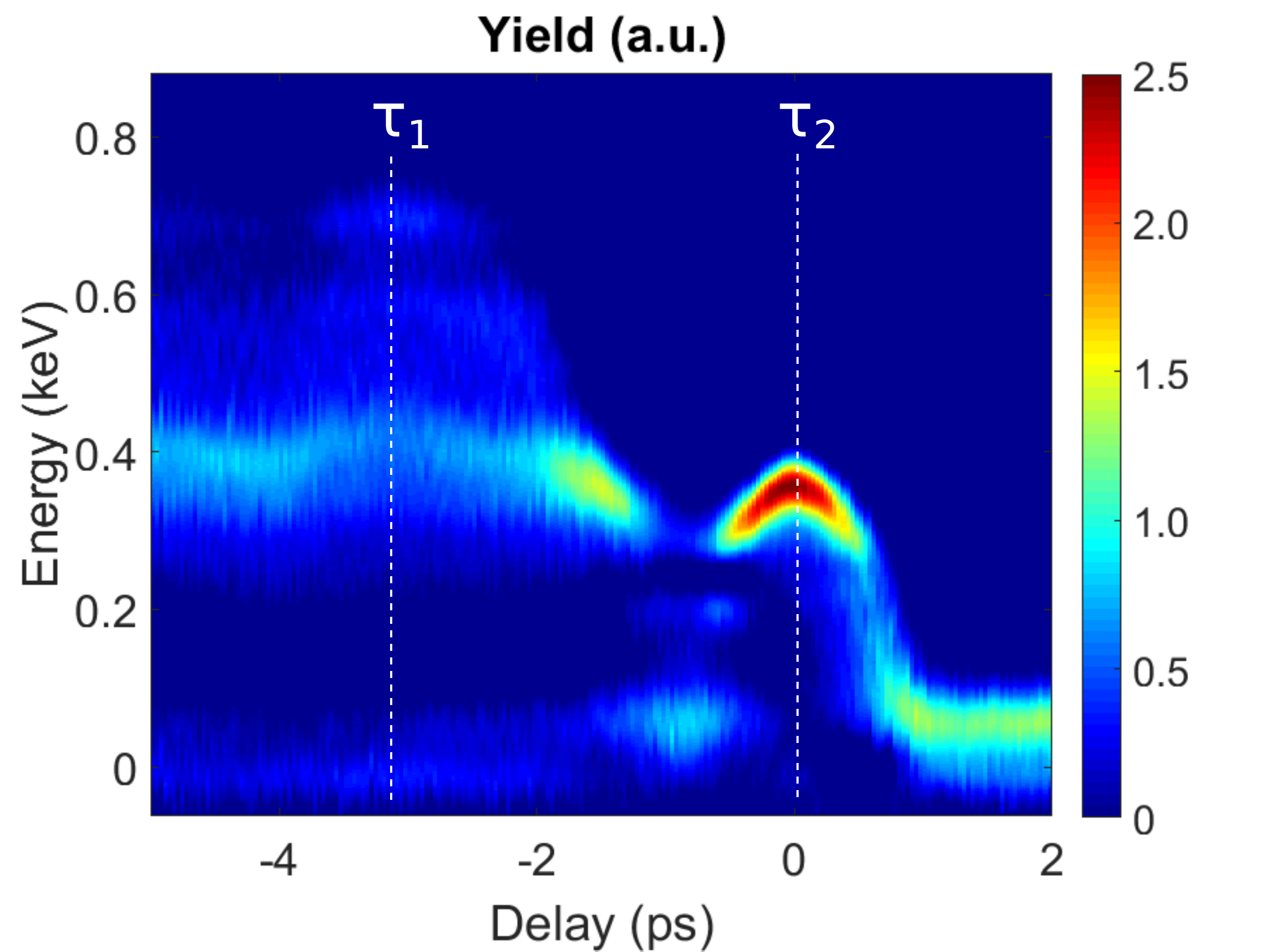} &
  	\includegraphics[draft=false,width=3.0in]{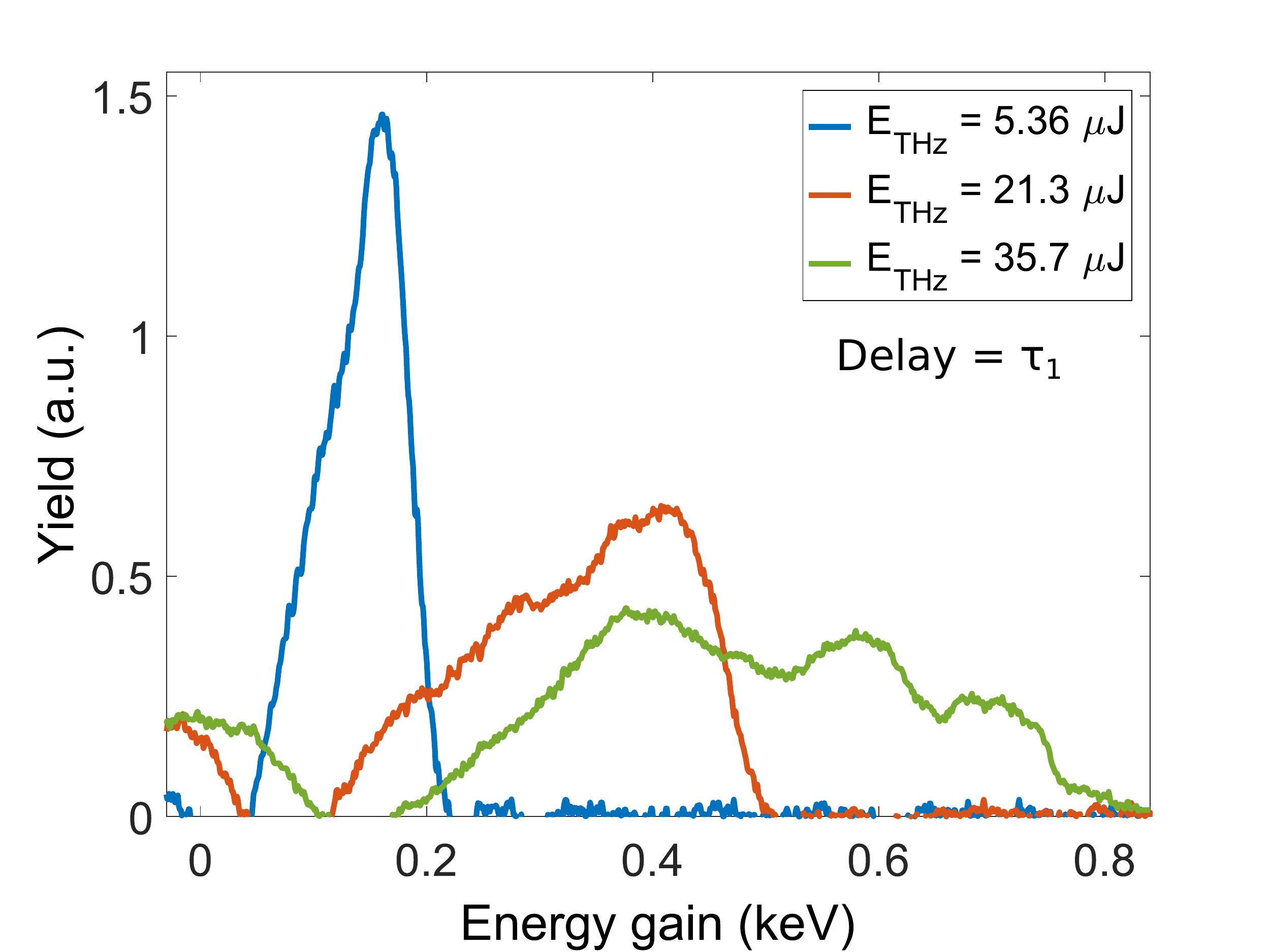} \\
  	(a) & (c) \\
  	\includegraphics[draft=false,width=3.0in]{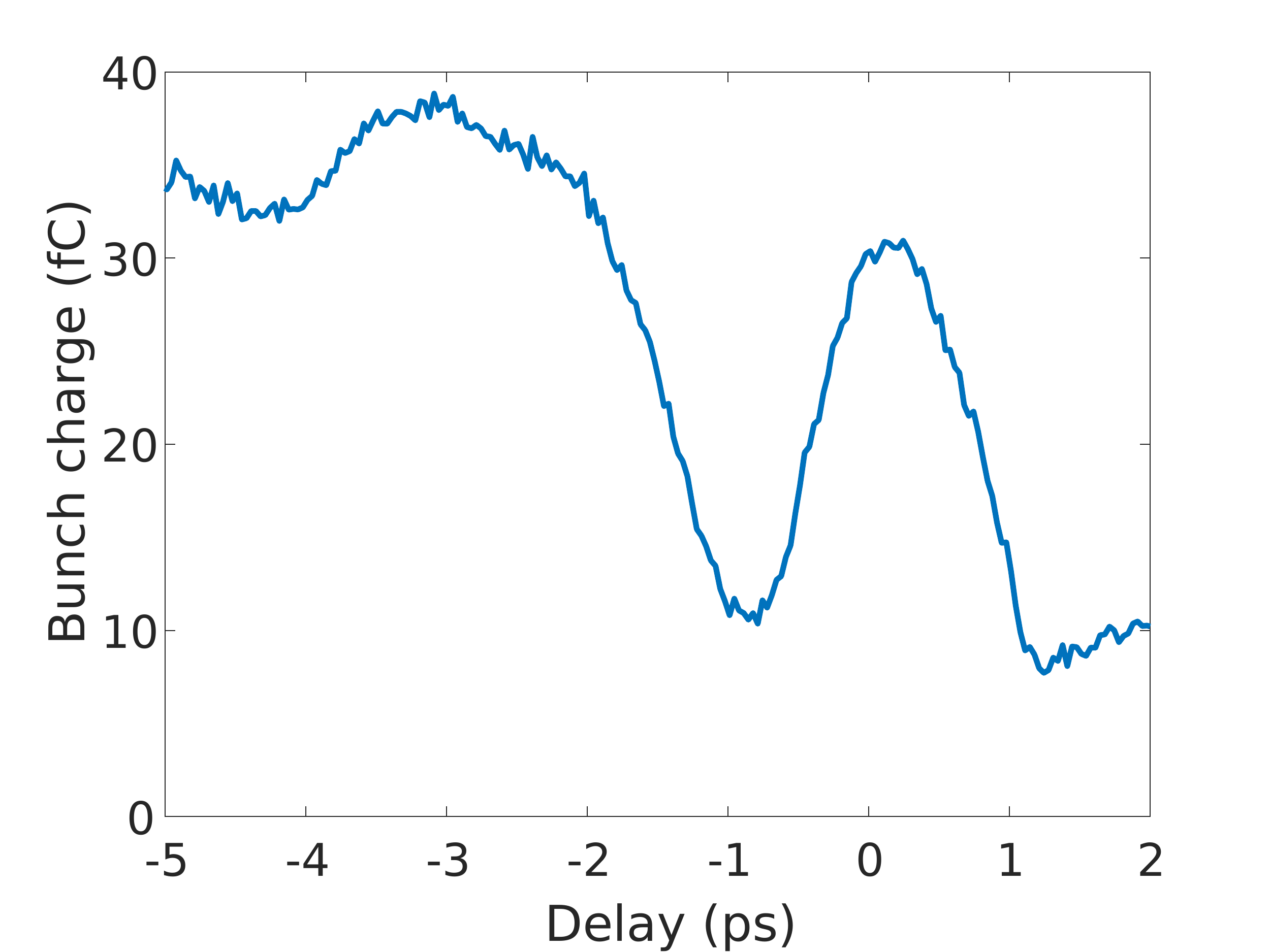} &
  	\includegraphics[draft=false,width=3.0in]{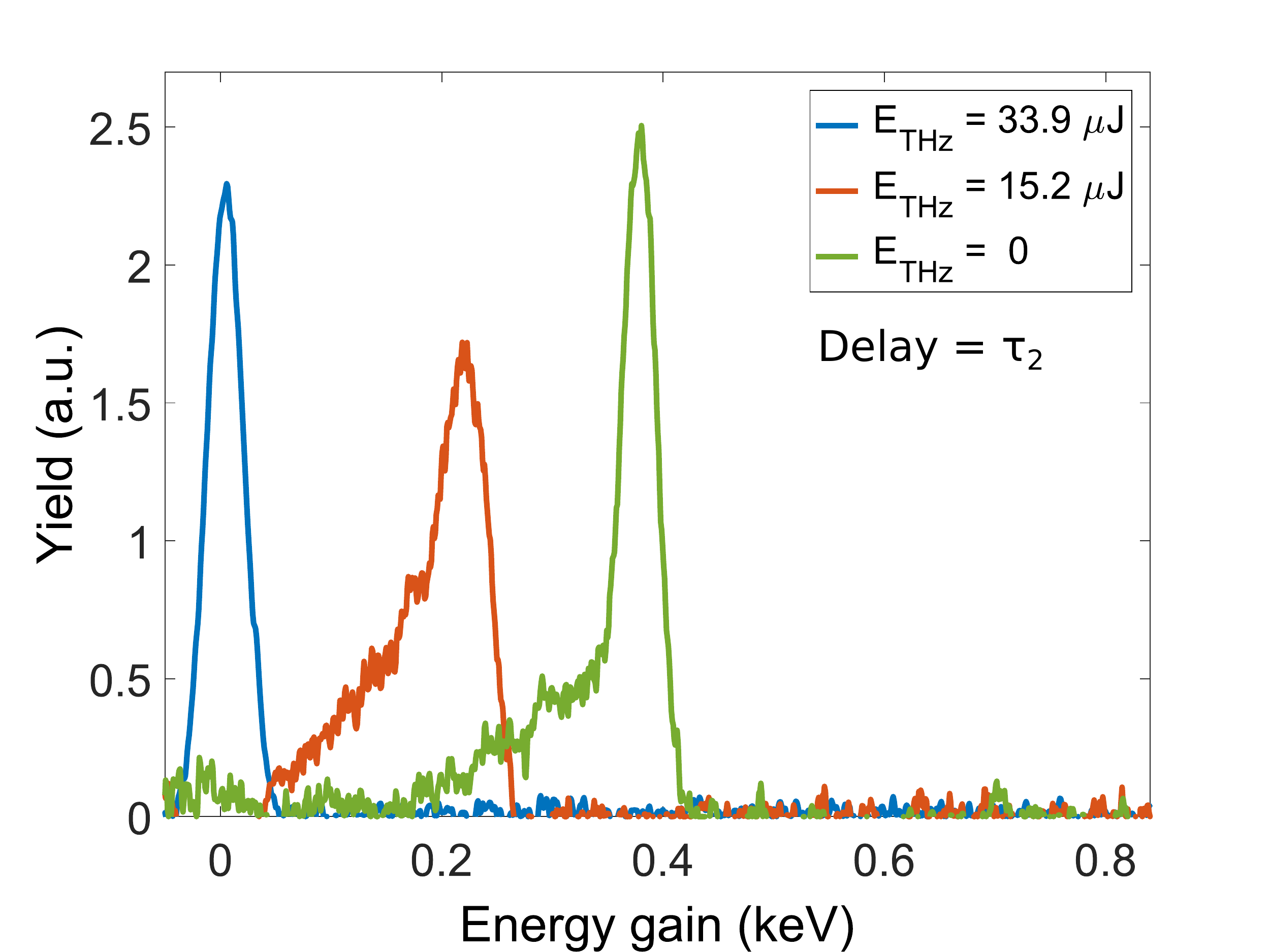} \\
  	(b) & (d)
  	\end{array}$
  	\caption{THz-driven electron energy gain and bunch charge modulation: (a) measured spectrogram showing the energy gain spectra as a function of delay between UV and THz pulses, at maximum THz energy; (b) measured bunch charge as a function of delay; (c) and (d) electron energy spectra for three different THz energies at delay positions (c) $\tau_1 = -2\,$ps and (d) $\tau_2 = 0.9\,$ps.}
  	\label{THzGunImplementationModulation}
  \end{figure}

  An electron's final momentum gain, $p_e$, depends on its emission time and can be expressed as $p_e=q\int_{t_\mathrm{emit}}^{t_\mathrm{exit}} E_\mathrm{THz}(t)dt$, where $t_\mathrm{emit}$ is the emission time and $t_\mathrm{exit}$ is the time the electron exits the PPWG.
  In cases where $t_\mathrm{exit} \gg \tau_\mathrm{THz}$ (the electron exits long after passage of the THz pulse), $p_e$ can be simplified to $p_e=q\int_{t_\mathrm{emit}}^\infty E_\mathrm{THz}(t)dt = qA(t_\mathrm{emit})$, where $A(t_\mathrm{emit})$ is the THz vector potential at the time of emission.
  To determine the optimum emission time for acceleration, we record the electron energy gain ($U_e$) spectra and bunch charge ($Q$) versus delay in Fig.\,\ref{THzGunImplementationModulation}a and \ref{THzGunImplementationModulation}b, respectively.
  The UV emitter can precede ($<-2$\,ps), overlap (-2\,ps to 2\,ps), or succeed ($>$2\,ps) the THz pulse.
  In the overlap region, $U_e$ maps out the phase and amplitude of $A_\mathrm{THz}(t_\mathrm{emit})$, similar to THz streaking in gases \cite{fruhling2009single}.
  One exception is that between -0.25\,ps and 0.4\,ps, emission occurs in the positive half-cycle of the THz field (opposing Lorentz force), causing a suppression of charge and energy gain.
  Two delays are selected to be the operating points of the gun.
  The first delay, $\tau_1=-2$\,ps, produced the highest peak acceleration, while the second delay, $\tau_2=0.9$\,ps, produced the most monoenergetic spectra.
  The total bunch charge was 40\,fC at $\tau_1$ and 32\,fC at $\tau_2$.
  
  When the photoemission precedes the THz pulse ($\tau<-2$\,ps), a large energy spread centered at $\sim 0.45$\,keV is observed.
  The origin of these broadened spectra, enduring for long decay times, is attributed to multiple complex mechanisms encompassing thermal \cite{herink2014field} or time-of-flight effects.
  When the emission succeeds the THz pulse ($\tau>2$\,ps), there is no net acceleration from that pulse.
  The constituency of electrons slightly elevated to 50\,eV is attributed to the aforementioned decay effects probed by a back-reflected THz pulse arriving at 18\,ps.
  
  In Fig.\,\ref{THzGunImplementationModulation}c and \ref{THzGunImplementationModulation}d, we take a closer look at the energy spectra from the two operating points, $\tau_1$ and $\tau_2$, for three different THz energies, $W_\mathrm{THz}$.
  Each spectrum exhibits a unimodal distribution with an average energy gain increasing with $W_\mathrm{THz}$.
  Except for the $W_\mathrm{THz}=35.7$\,{\textmu}J spectrum at $\tau_1$, the spectral shapes are asymmetric, with a pedestal toward lower energies and a maximum yield toward higher energies, followed by a sharp cutoff.
  The high yield near the cutoff indicates that most electrons are emitted at the optimal THz phase and concurrently experience the same acceleration.
  The pedestal can be attributed to electrons emitted away from the optimal phase, resulting in a lower energy gain.
  Such dependence of energy gain on emission phase is also evident in RF guns \cite{kim1989rf}.
  
\begin{figure}
  	\centering
  	$\begin{array}{cc}
  	\includegraphics[draft=false,width=2.0in]{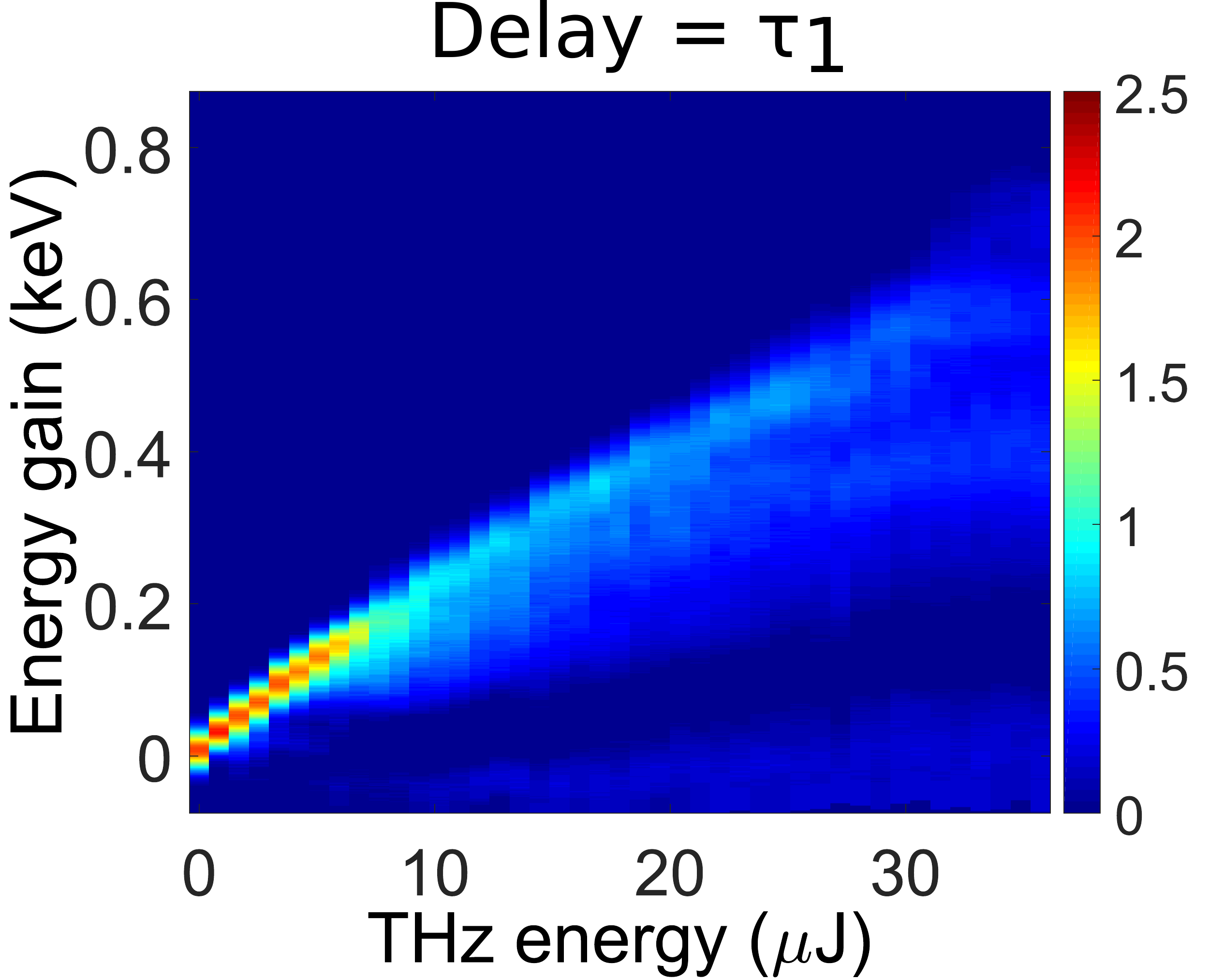} &
  	\includegraphics[draft=false,width=2.0in]{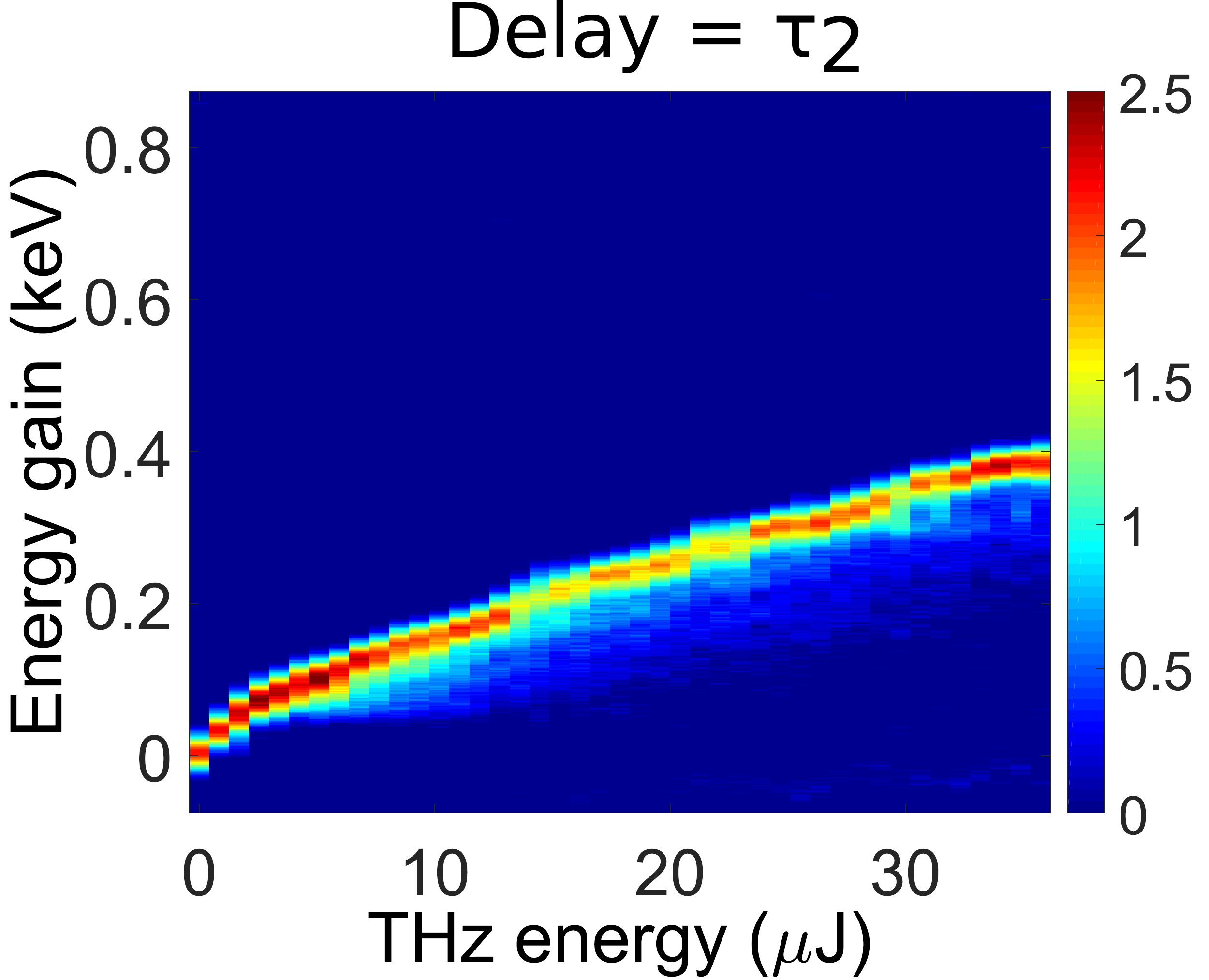} \\
  	(a) & (d) \\
  	\includegraphics[draft=false,width=2.0in]{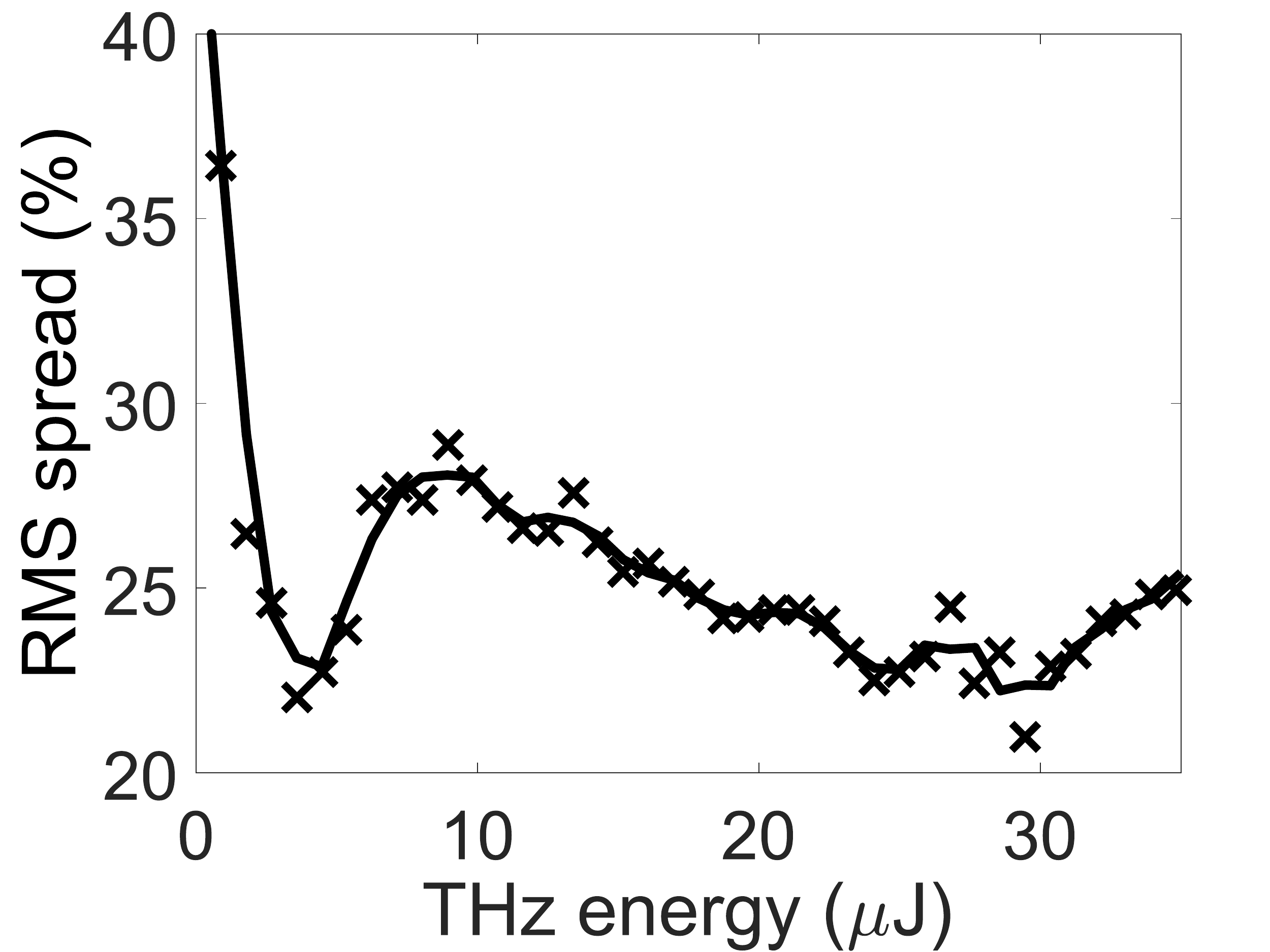} &
  	\includegraphics[draft=false,width=2.0in]{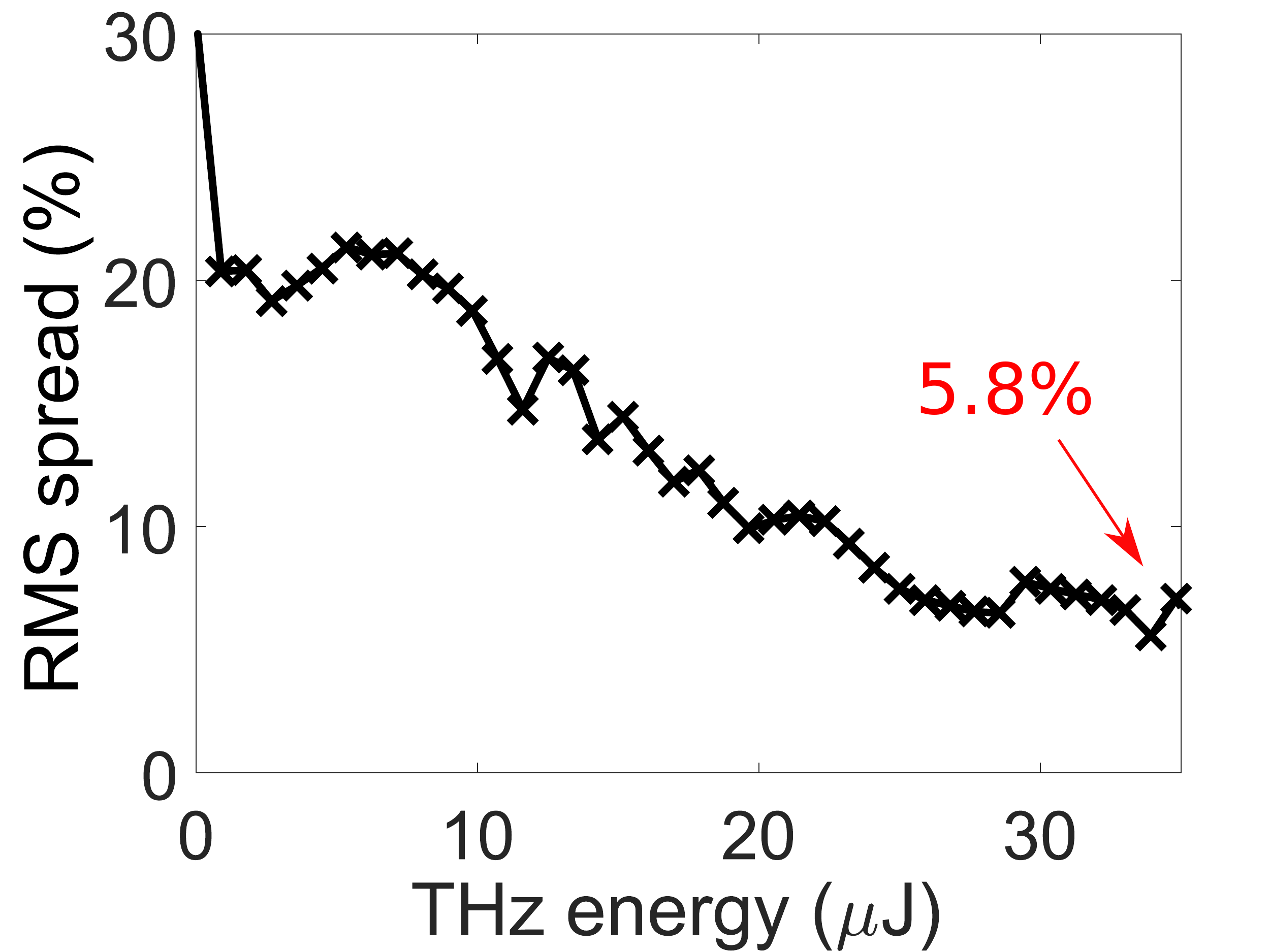} \\
  	(b) & (e) \\
  	\includegraphics[draft=false,width=2.0in]{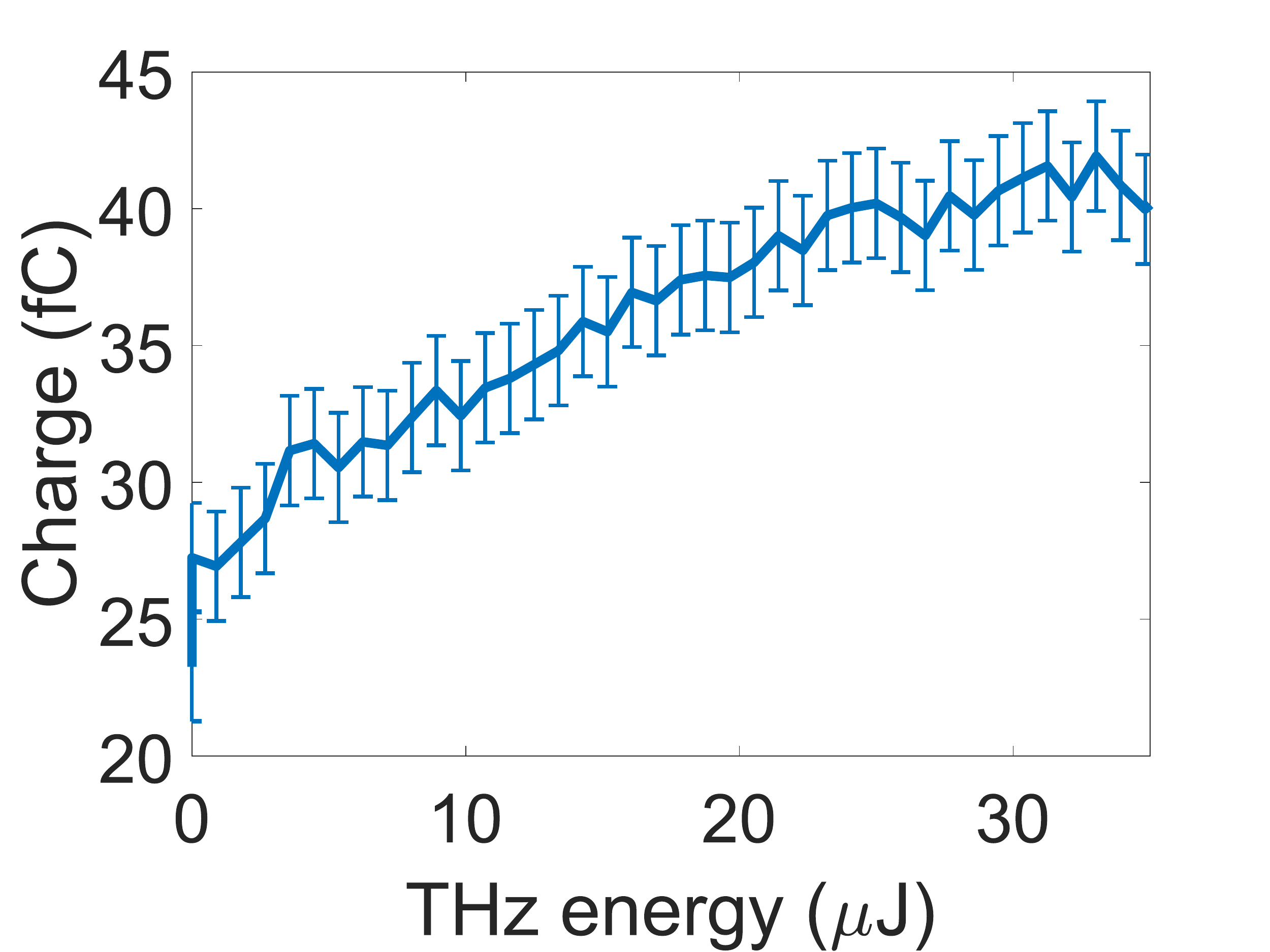} &
  	\includegraphics[draft=false,width=2.0in]{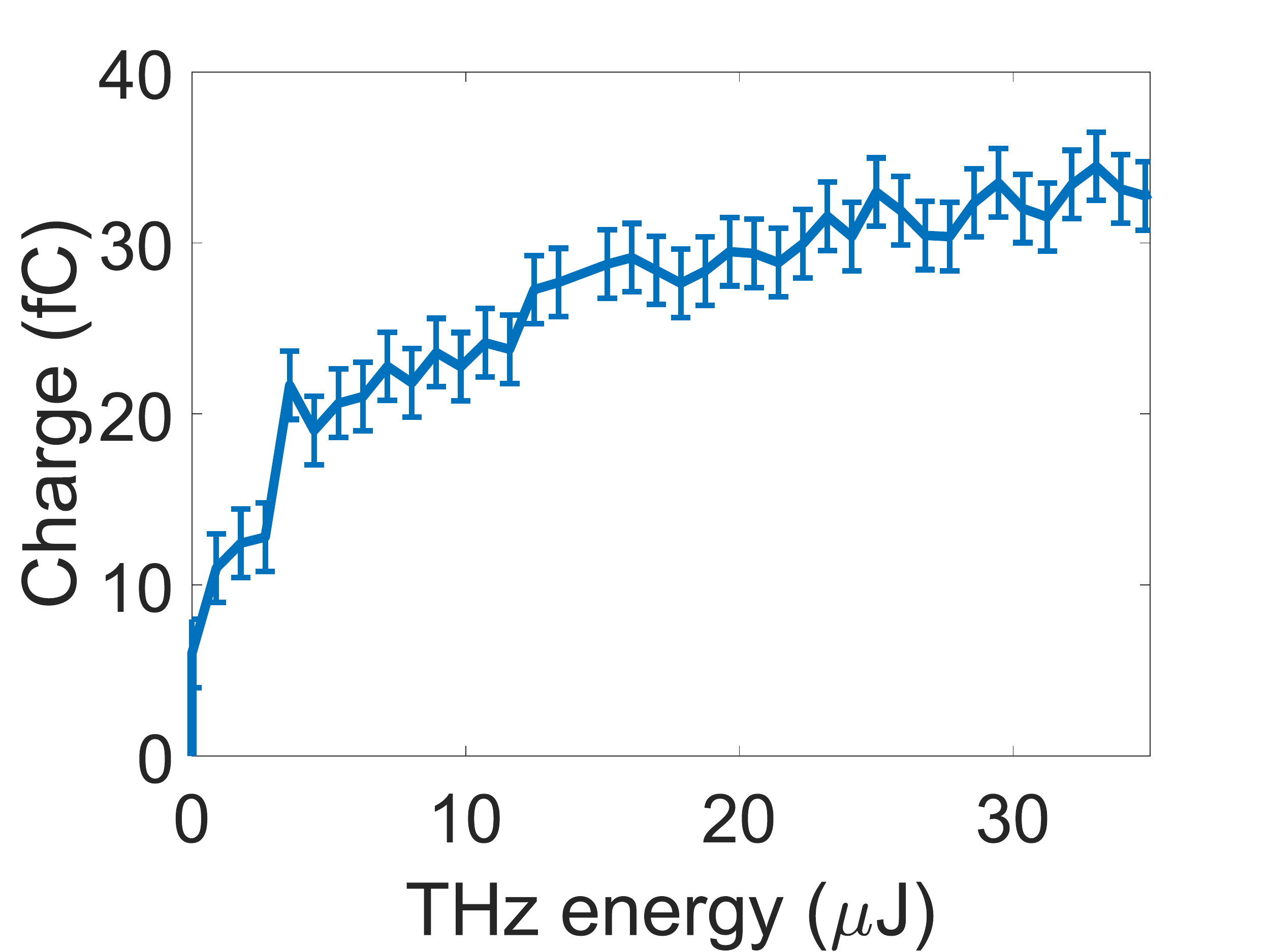} \\
  	(c) & (f)
  	\end{array}$
  	\caption{THz scaling at $\tau_1$ and $\tau_2$; delay positions defined earlier in Fig.\,\ref{THzGunImplementationModulation}a. (a) and (b) Energy gain plotted on a spectrogram to highlight its scaling as a function of accelerating THz energy or THz field. Error bar radius is equal to the absolute RMS energy spread. (c) and (d) Relative RMS energy spread, $\sigma_U$, of the accelerated bunch. (e) and (f) Total detected bunch charge exiting the gun, $Q$. Error bar radius is equal to the RMS instrument noise.}
  	\label{THzGunImplementationSpectrogram}
  \end{figure}

  We continue investigations at $\tau_1$ and $\tau_2$ by plotting $U_e$ versus $W_\mathrm{THz}$ on a spectrogram, as shown in Fig.\,\ref{THzGunImplementationSpectrogram}a and \ref{THzGunImplementationSpectrogram}b.
  At both delays, $U_e$ scales mostly linearly with $W_\mathrm{THz}$ or, equivalently, with $E^2_\mathrm{THz}$.
  This scaling law can be explained by $U_e=p_e^2/2m \propto E^2_\mathrm{THz}$, which is valid when $t_\mathrm{exit} \gg \tau_\mathrm{THz}$.
  Alternatively, if $t_\mathrm{exit} \ll \tau_\mathrm{THz}$, the energy gain would be dominated by $U_e=q \int_{z_\mathrm{emit}}^{z_\mathrm{exit}} E_\mathrm{THz}(z)dz$, leading to a $U_e \propto E_\mathrm{THz}$ scaling law, as is typical in RF guns \cite{kim1989rf,harris2011design} and would be the case in this study for a larger field, reduced PPWG spacing, or relativistic electrons.
  
  At $\tau_1$, increasing the THz energy results in an increase in absolute energy spread (Fig.\,\ref{THzGunImplementationSpectrogram}a).
  Consequently, the relative energy spread, $\sigma_U$, remains roughly constant at around 20\%-30\% (Fig.\,\ref{THzGunImplementationSpectrogram}c).
  The bunch charge increases monotonically with THz energy (Fig.\,\ref{THzGunImplementationSpectrogram}e).
  We obtain a peak energy gain of 0.8\,keV at $W_\mathrm{THz}=35.7$\,{\textmu}J (Fig.\,\ref{THzGunImplementationSpectrogram}a).
  
  At $\tau_2$, the absolute energy spread remains constant with THz energy (Fig.\,\ref{THzGunImplementationSpectrogram}b).
  Correspondingly, the relative energy spread monotonically decreases with THz energy, to a minimum of 5.8\% centered near 0.4\,keV (Fig.\,\ref{THzGunImplementationSpectrogram}d).
  The pedestal regions are neglected in the energy spread calculations, since over time those electrons separate from the main bunch.
  Half of this spread comes from THz shot-to-shot fluctuations (2\%), while another large contribution comes from the spread in electron emission time: $\Delta t_\mathrm{emit}=\tau_{UV}=275\,\text{fs}=T_\mathrm{THz}/8$.
  By stabilizing the laser and shortening $\tau_{UV}$ via an optical parametric amplifier \cite{ziegler1998tunable}, the energy spread can be further reduced.
  In Fig.\,\ref{THzGunImplementationSpectrogram}f, the bunch charge increases with THz energy below 7\,{\textmu}J, indicating that the emission is space charge limited \cite{rosenzweig1994initial}.
  Above 7\,{\textmu}J, the bunch charge plateaus, indicating that the THz field overcomes the space charge force and extracts all the emitted electrons.
  
\begin{figure}
  	\centering
  	$\begin{array}{cc}
  	\includegraphics[draft=false,width=3.0in]{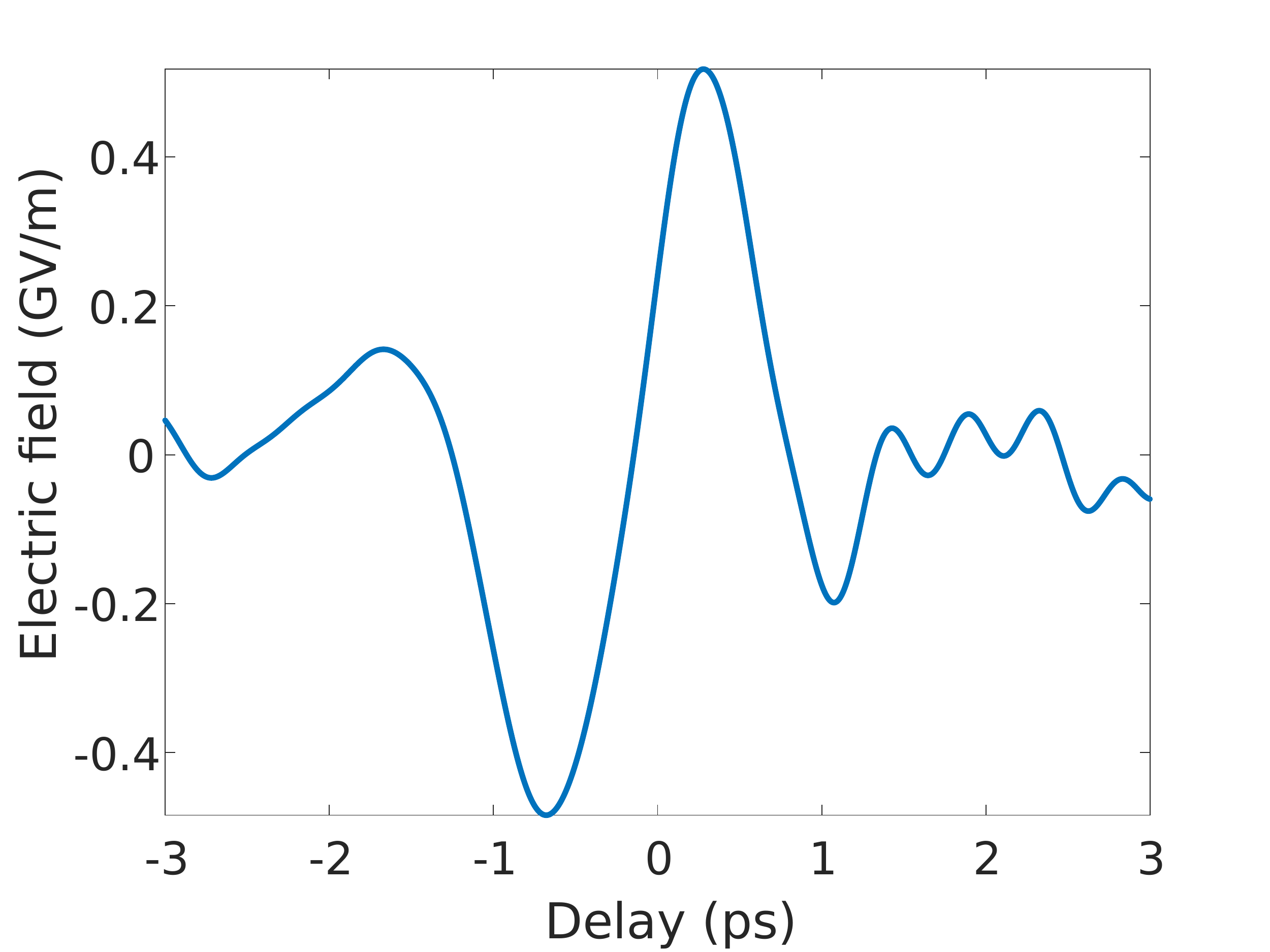} &
  	\includegraphics[draft=false,width=3.0in]{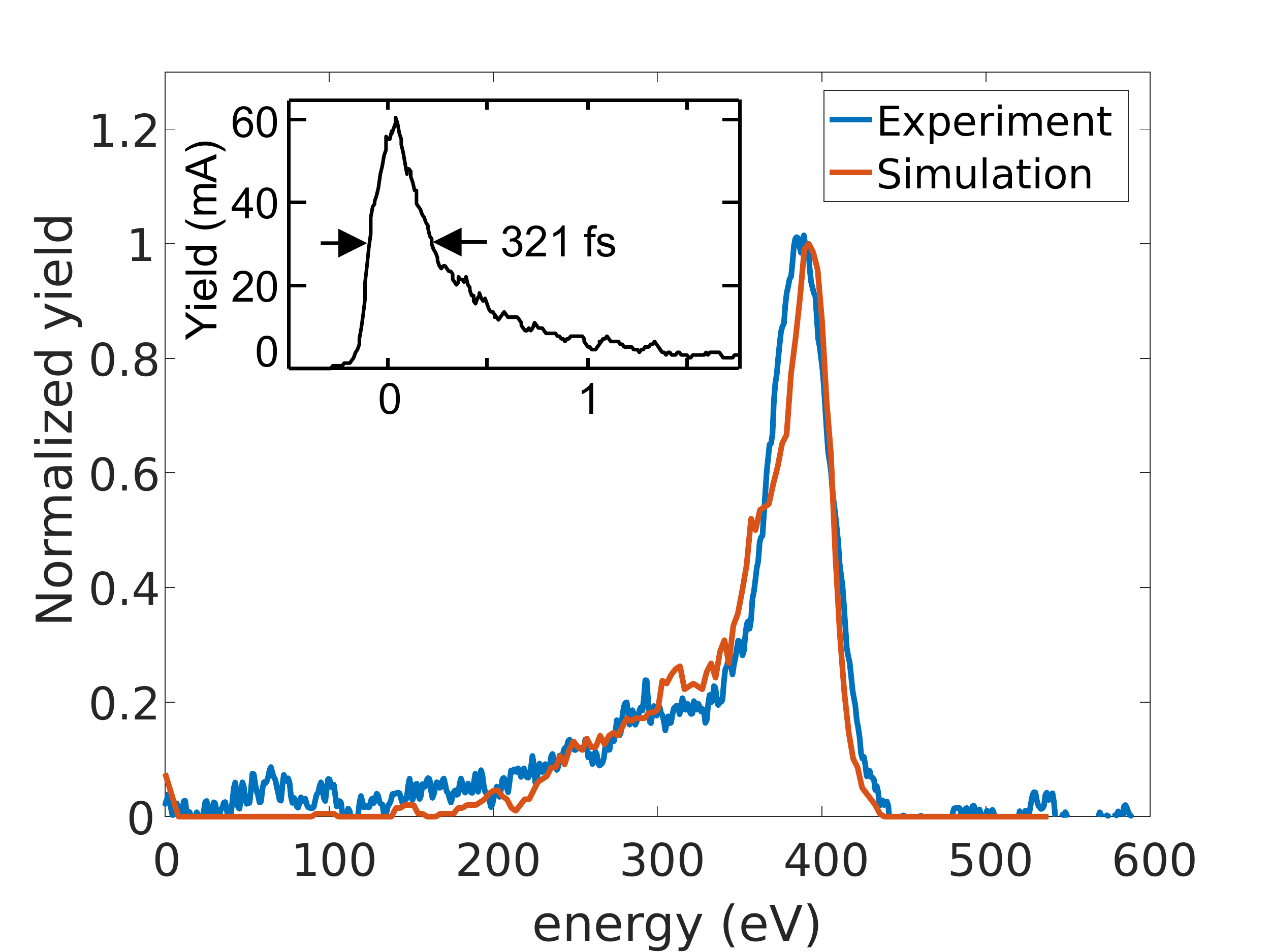} \\
  	(a) & (c) \\
  	\includegraphics[draft=false,width=3.0in]{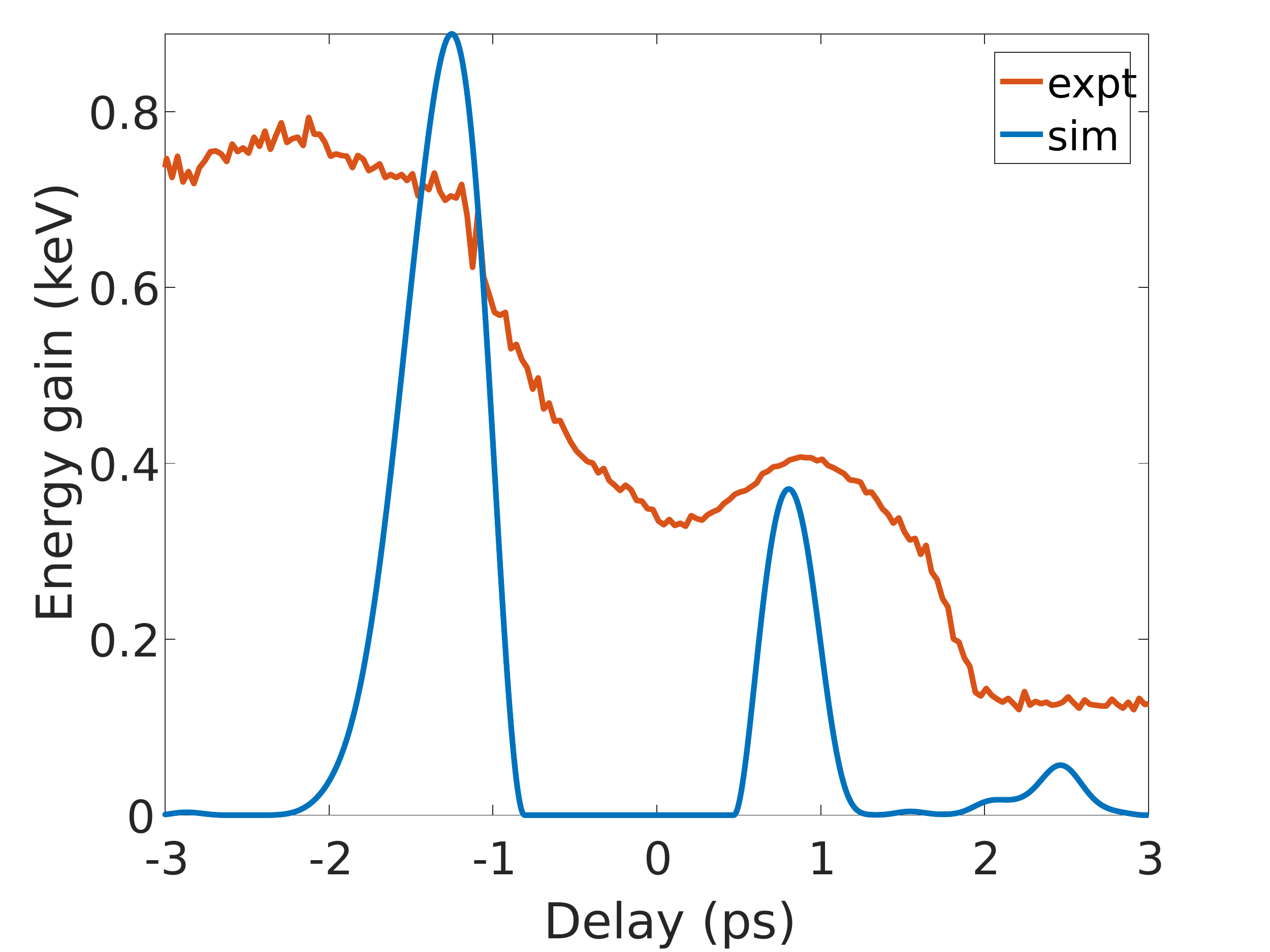} &
  	\includegraphics[draft=false,width=3.0in]{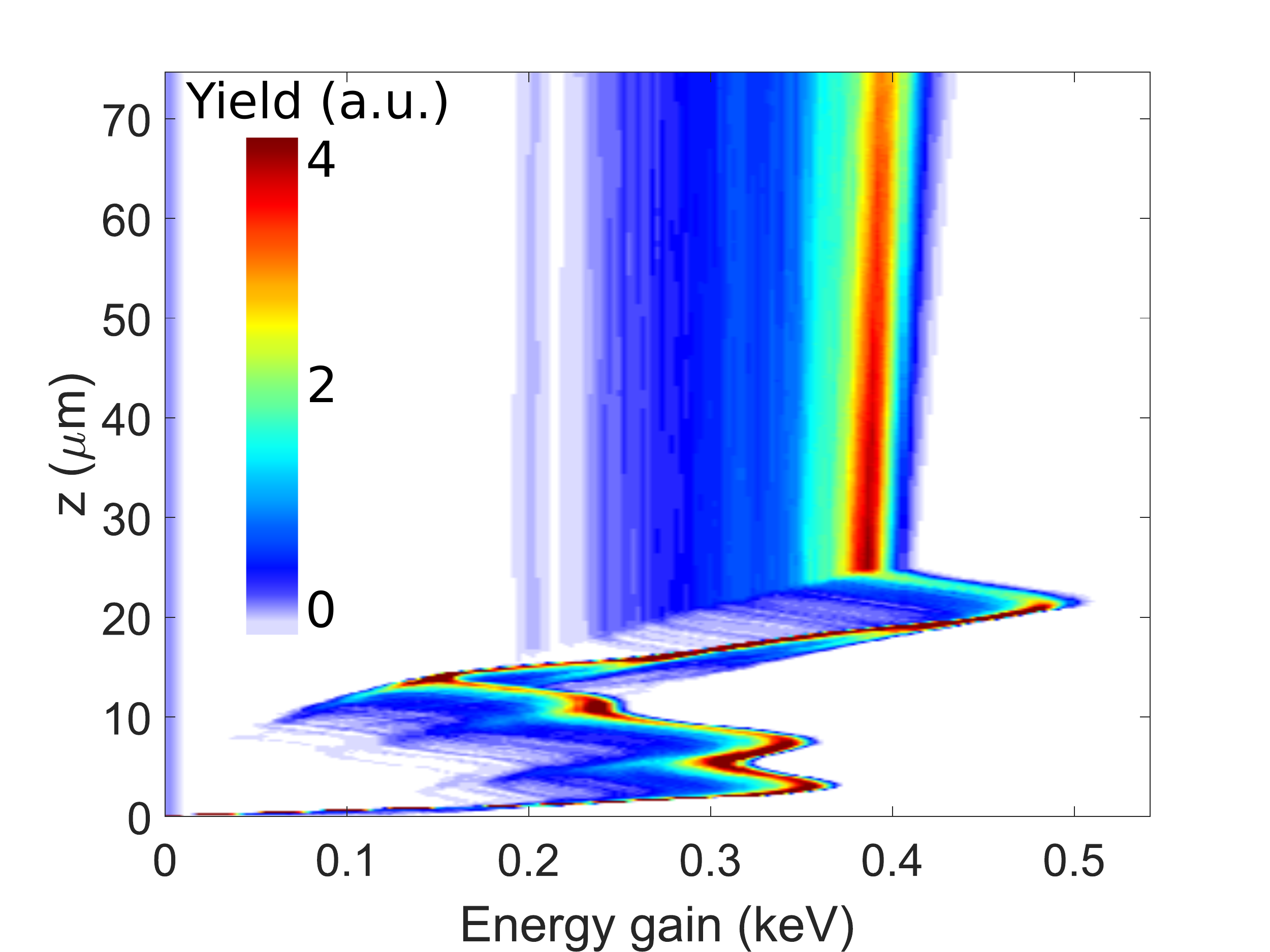} \\
  	(b) & (d)
  	\end{array}$
  	\caption{Numerical analysis of THz gun. (a) THz electric field measured by EO sampling with a fitted field strength; (b) the single-electron energy gain, calculated analytically, is overlaid with the peak energy gain obtained from the experiment in Fig.\,\ref{THzGunImplementationModulation}a; (c) simulated energy spectrum (black line) of the bunch at the gun exit for emission at $\tau_2$, showing excellent agreement with the experiment (blue line). Inset: Temporal profile of the electron bunch at the gun exit, showing a FWHM pulse duration of 321\,fs, elongated by space-charge. (d) Simulated evolution of the energy spectrum along $z$. The THz pulse is passed by the time the
  		electrons reach 25\,{\textmu}m.}
  	\label{THzGunImplementationNumerical}
  \end{figure}

  In Fig.\,\ref{THzGunImplementationNumerical}b, we show the calculated single-electron energy gain versus delay, utilizing the shape of the measured THz waveform with a fitted field strength (Fig.\,\ref{THzGunImplementationNumerical}a).
  We compare it with the measured peak energy gain from Fig.\,\ref{THzGunImplementationModulation}a.
  Since the experimentally measured peak energy gain represents the gain of the electron emitted at the optimal delay and spatial position, comparing it with our analytical expression for single-electron energy gain is justified.
  Several experimental features are represented in this simple analytical model: (1) a suppression region around 0\,ps, (2) relative energy gain levels, and (3) delay between the two peaks.
  This model provides an alternate method for quantifying the THz field strength inside the gun.
  Our fitted peak field was 480\,MV/m.
  
  To better understand the bunch dynamics under the influence of self-fields and the THz field, the multi-electron particle tracking simulation results in Fig.\,\ref{THzGunImplementationNumerical}d show the evolution of the energy spectrum of the 32\,fC bunch emitted at $\tau_2$ as it propagates along $z$.
  Immediately following emission, the bunch experiences a strong accelerating field, growing in energy to 350\,eV over the first 3\,{\textmu}m.
  During the time that the THz pulse interacts with the bunch (corresponding $z$ distance: 0 to 25\,{\textmu}m), the energy undergoes four acceleration/deceleration cycles, caused by the four oscillation cycles in the THz field following $\tau_2$.
  The THz pulse is passed by the time the bunch reaches 25\,{\textmu}m, verifying that $t_\mathrm{exit} \gg \tau_\mathrm{THz}$, and the bunch drifts to the exit while continuing to experience energy spreading due to space charge forces.
  At the gun exit ($z_\mathrm{exit}=75$\,{\textmu}m), the simulated energy spectrum has excellent overlap with the experimental spectrum (Fig.\,\ref{THzGunImplementationNumerical}c).
  The sharp cutoff, pedestal height, pedestal length, and central lobe width are all reproduced flawlessly by the model.
  The simulated temporal profile at the gun exit (Fig.\,\ref{THzGunImplementationNumerical}c inset) exhibits a pulse duration of 321\,fs, longer than the initial 275\,fs due to space-charge.
  All the numerical analyses incorporated space-charge, imitated the experimental conditions, and used the THz field profile shown in Fig.\,\ref{THzGunImplementationNumerical}a.
  
  \section{Segmented THz Electron Accelerator and Manipulator}
  
  The continuous progress in high power THz sources has shown the feasibility of millijoule level single-cycle THz beam generation \cite{fulop2012generation}.
  Nevertheless, the technology is still in its infancy and not sufficiently mature to provide a millijoule-level THz pulse with low cost and in a straightforward fashion like a conventional technology.
  This challenge was the main obstacle to experimentally test the proposed multilayer structure as ultrafast THz guns.
  Thus, instead of using the multilayer structure as an electron injector, we test the device in the framework of electron beam manipulation as reported in \cite{zhang2018segmented}.
  
  Previously, proof-of-principle demonstrations of THz beam manipulation include electron emission \cite{wimmer2014terahertz,li2016high} and acceleration \cite{Nanni2015,walsh2017demonstration,huang2015toward,fallahi2016short,hebling2011optical,huang2016terahertz,fakhari2017thz} as well as compression and streaking \cite{fabianska2014split,kealhofer2016all}.
  These experiments, although limited in charge, beam quality, energy gain and energy spread, have set the stage for development of practical, compact terahertz-based devices that can support sufficient charge and field gradients to realistically be used to boost performance of existing accelerators or as components of future compact accelerators and x-ray sources.
  Here, we demonstrate the first such device based on a layered, transversely pumped, waveguide structure.
  This segmented terahertz electron accelerator and manipulator (STEAM) device can dynamically switch between accelerating, streaking, focusing and compressing modes, can support multiple picocoulombs of charge and features intrinsic synchronization.
  Using only a few microjoules of single-cycle terahertz radiation, we demonstrate over 70\,MV/m peak acceleration fields, 2\,kT/m focusing gradients (which are an order of magnitude beyond current electromagnetic lenses and comparable to active plasma lenses), the highest reported terahertz streaking gradient of 140\,{\textmu}rad/fs (making it well-suited for characterization of ultrafast electron diffractometer bunches down to 10\,fs) as well as compression to $\sim 100$\,fs.
  All these demonstrations strongly benefit from very small temporal jitter achieved through laser-driven terahertz sources.
  By increasing terahertz pulse energies to state-of-the-art millijoule levels \cite{fulop2014efficient}, it is expected that accelerating gradients approaching 1\,GV/m can be achieved and sustained.
  Such gradients surpass those possible in radio-frequency accelerators by an order of magnitude and enable major improvements in electron bunch qualities such as emittance and bunch length.
  The picosecond duration of the terahertz pulses is an essential ingredient for reaching the GV/m regime, as experiments have shown that maximum accelerating gradients, which are limited by field-induced breakdown, scale with the sixth power of the field duration \cite{dal2016rf,dal2016experimental,dobert2005high,Kilpatrick1957}.
  Demonstration of the terahertz-driven STEAM device thus establishes a new compact, strong-field and extremely high-gradient accelerator technology.
  
  \subsection{Concept and Implementation}
  
\begin{figure}
  	\centering
  	\includegraphics[draft=false,width=6.0in]{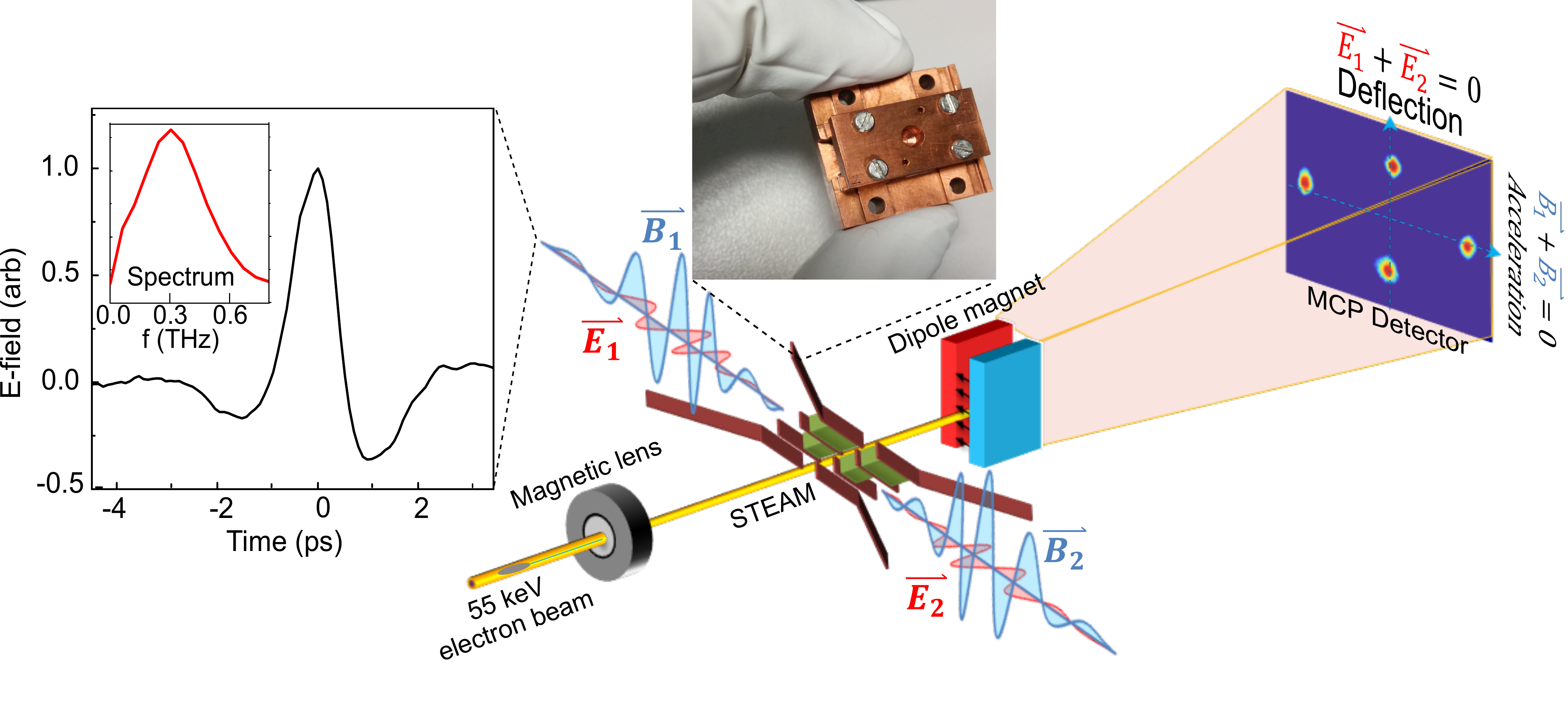}
  	\caption{STEAM experimental setup. A fraction of the infrared optical beam is converted to 257\,nm beam through fourth-harmonic generation. The 257\,nm laser pulse is directed onto a gold photocathode generating electron bunches, which are accelerated to 55\,keV by a DC electric field. This laser also drives two optical rectification stages, each generating single-cycle terahertz pulses with energy up to 30\,{\textmu}J. The two counter-propagating terahertz beams interact with the electron beam inside the segmented structure. Subsequently, the electron beam is detected by the camera. Top inset: photograph of the STEAM device. left inset: the time-domain waveform of the terahertz pulse measured by electro-optic sampling and its corresponding frequency-domain spectrum.}
  	\label{STEAMSetup}
  \end{figure}

  The experimental setup (Fig.\,\ref{STEAMSetup}) consisted of a 55\,keV photo-triggered DC gun, a terahertz-powered STEAM device for electron acceleration or manipulation and a diagnostic section that included a second STEAM device used as a streak camera, all of which were driven by the same infrared laser source.
  Ultraviolet pulses for photoemission were generated by two successive stages of second-harmonic generation, while single-cycle terahertz pulses were generated by difference frequency generation.
  Terahertz pulses from two independent setups were coupled into the STEAM device (Fig.\,\ref{STEAMSetup}) transversely to the electron motion by two horn structures that focused the counter-propagating terahertz fields beyond the diffraction limit into the interaction zone.
  The electrons experience both the electric and magnetic fields of the terahertz pulses according to the Lorentz force law $\vec{F} = q(\vec{E}+\vec{v} \times \vec{B})$, where $q=-e$ is the electron charge, $\vec{E}$ is the electric field, oriented parallel to the electron velocity $\vec{v}$, and $\vec{B}$ is the magnetic field, oriented vertically in the lab frame.
  The electric field is thus responsible for acceleration and deceleration, while the magnetic field induces transverse deflections.
  
  Efficient interaction of the electrons with the fields was accomplished by means of segmentation, which divided the interaction volume into multiple layers, each isolated from the others by thin metal sheets (Fig.\,\ref{STEAMSetup}).
  Dielectric slabs of varying length were inserted into each layer to delay the arrival time of the terahertz waveform to coincide with the arrival of the electrons, effectively phase-matching the interaction.
  Due to the transverse geometry, the degree of dephasing experienced in each layer was determined by the traversal time of the electrons, which was dependent on the electron speed and the layer thickness.
  A reduction in dephasing can thus be accomplished by reducing the layer thickness and increasing the number of layers, at the cost of increased complexity.
  The ability to tune the thickness and delay of each layer independently is a key design feature of the STEAM device that enables acceleration of sub-relativistic electrons for which the speed changes significantly during the interaction (for example, from 0.43c to 0.51c for our maximum acceleration case).
  
  The use of two counter-propagating drive pulses enabled two key modes of operation, which are specified with respect to the interaction point, that is, the centre of the interaction region of each layer: (1) an \emph{electric} mode, used for acceleration, compression and focusing, in which the pulses were timed to produce electric superposition and magnetic cancellation of the transverse fields at the interaction point; and (2) a \emph{magnetic} mode, used for deflection and streaking, where the magnetic fields superposed and the electric fields cancelled.
  
  The function of the device was thus selected by tuning the relative delay of the two terahertz pulses and the electrons, all of which were controlled by means of motorized stages acting on the respective infrared pump beams.
  In focusing and streaking modes, the electron beams were sent directly to a microchannel plate (MCP) detector.
  For acceleration measurements, an electromagnetic dipole was used to induce energy-dependent deflections in the vertical plane, so that both deflection and energy change could be measured simultaneously.
  To measure the compression, a second STEAM device in streaking mode was added downstream of the first to induce time-dependent deflections in the horizontal plane.
  The breakdown threshold of the device is determined by field emission from the metallic parts.
  Owing to the 4-5 orders of magnitude shorter field-exposure times of single-cycle terahertz pulses compared with radio-frequency excitations, previous studies suggest that a factor of 3-10 higher breakdown threshold for pulsed terahertz-driven devices can be expected \cite{huang2016terahertz,dal2016rf,dal2016experimental}.
  The remainder of this section gives a detailed description of the results obtained for acceleration, compression, focusing, deflection and streaking for this STEAM device.
  
  \subsection{Electric Mode}
  
\begin{figure}
  	\centering
  	$\begin{array}{cc}
  	\begin{array}{cc}
  	\includegraphics[draft=false,width=2.0in]{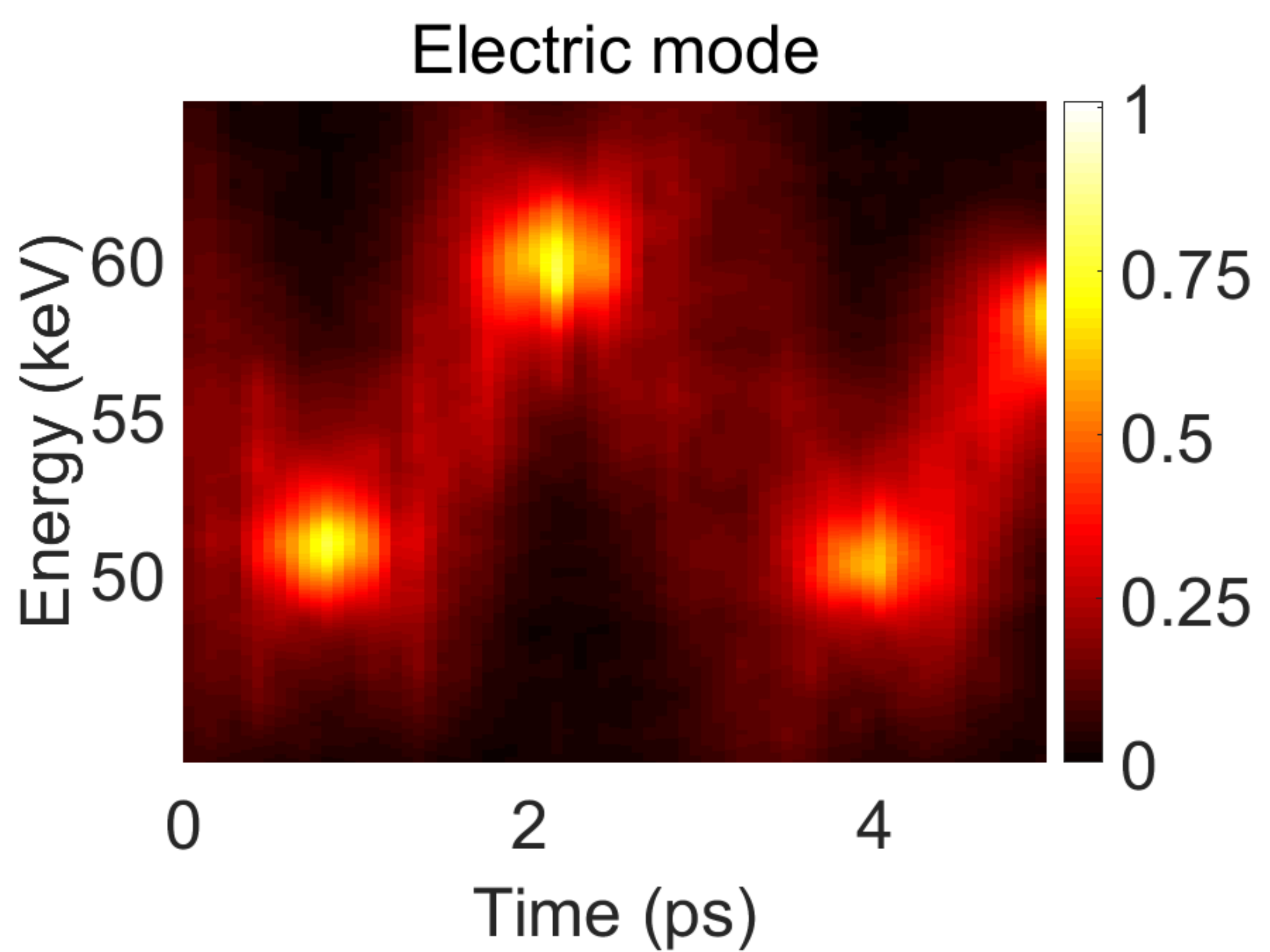} &
  	\includegraphics[draft=false,width=2.0in]{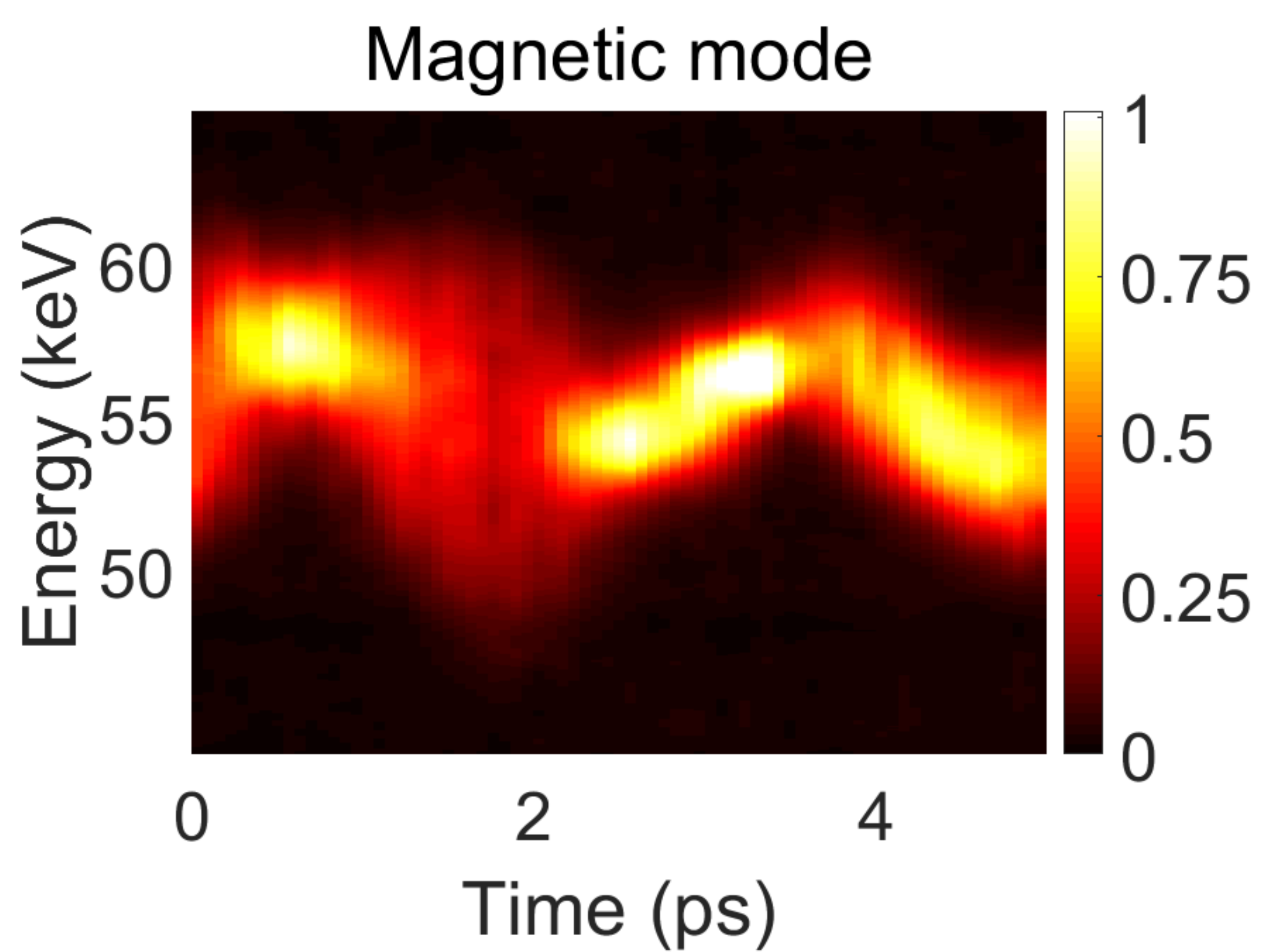} \\
  	(a) & (b) \\
  	\includegraphics[draft=false,width=2.0in]{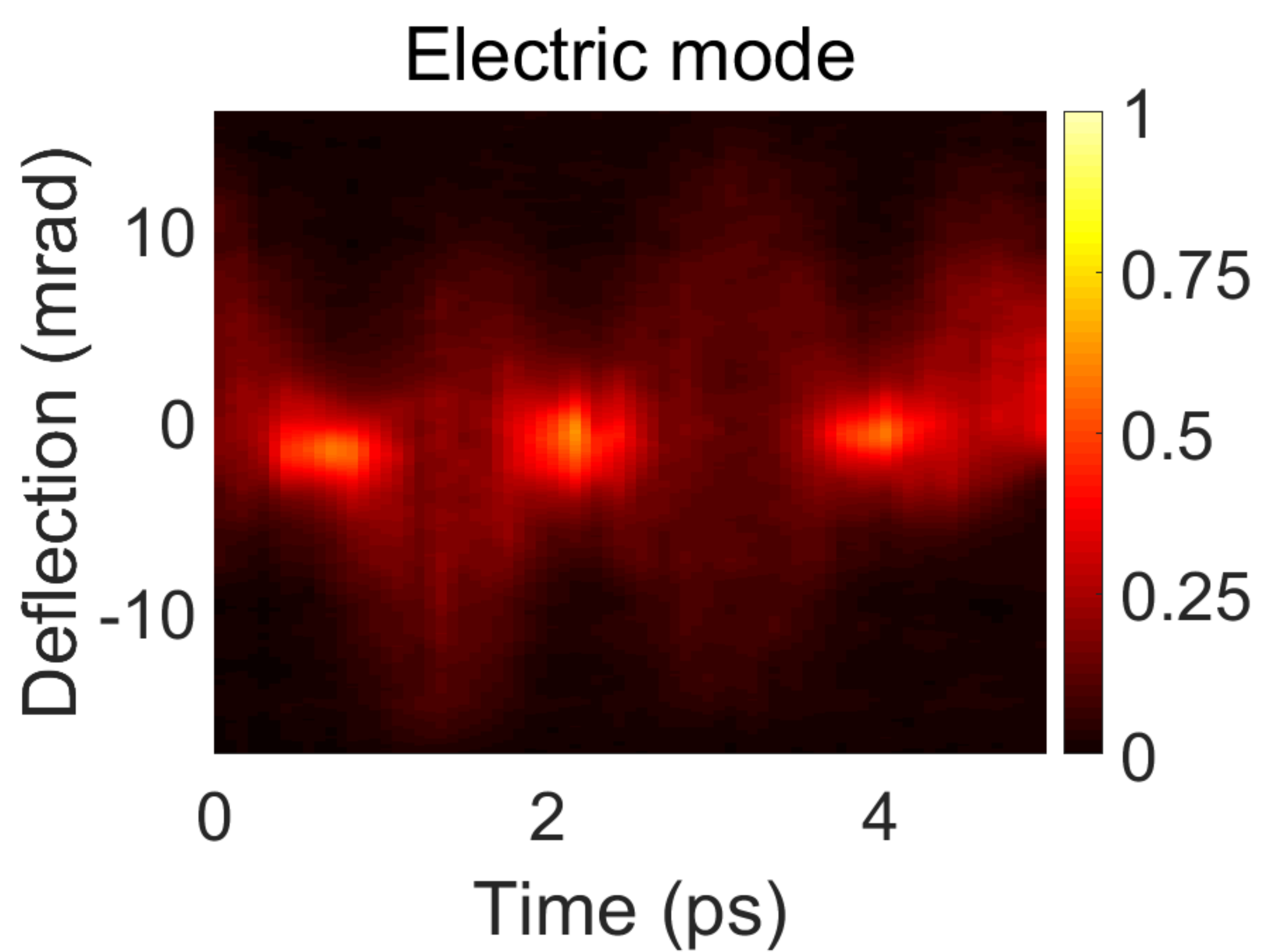} &
  	\includegraphics[draft=false,width=2.0in]{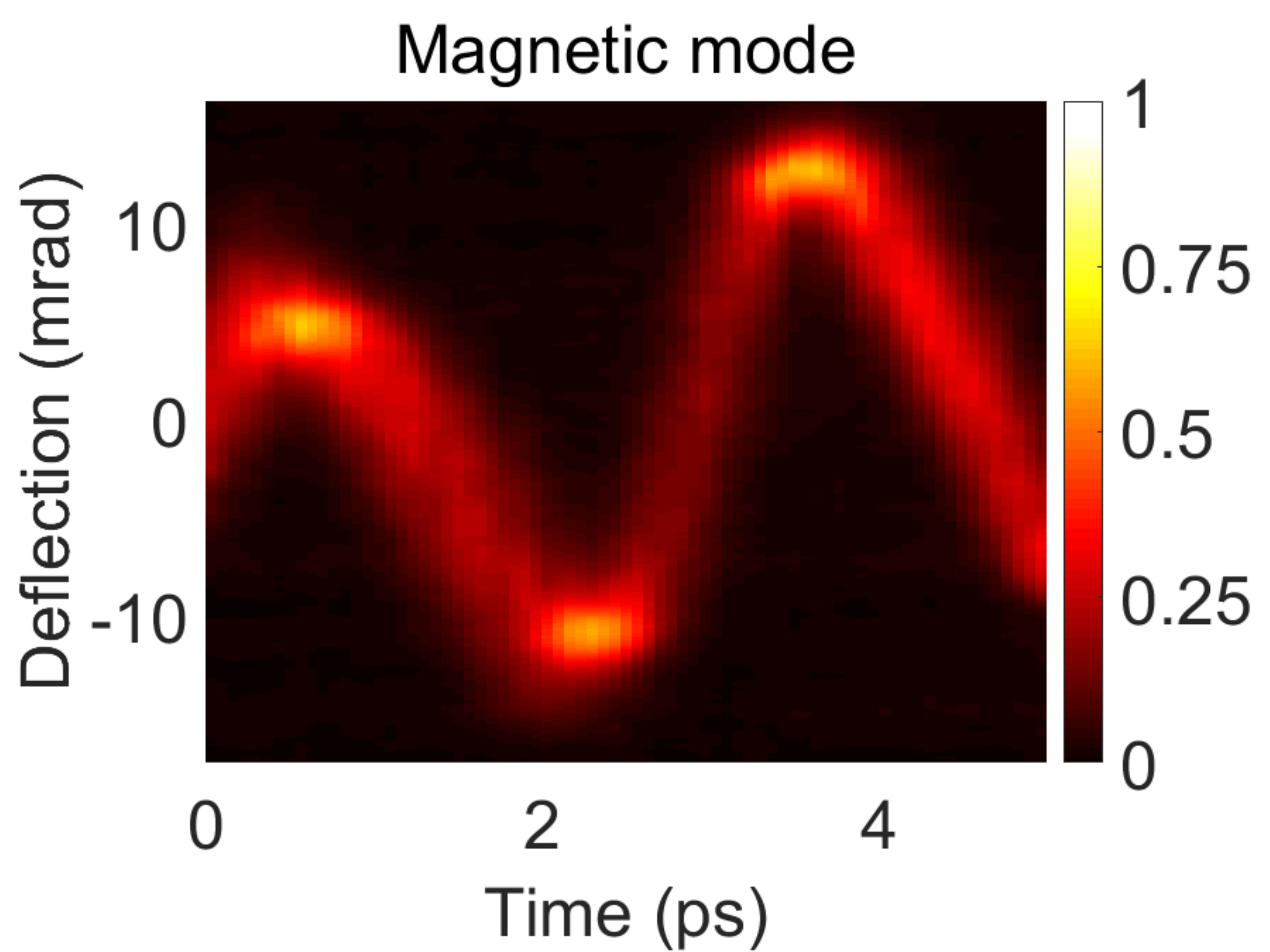} \\
  	(c) & (d)
  	\end{array} &
  	\begin{array}{c}
  	\includegraphics[draft=false,width=1.6in]{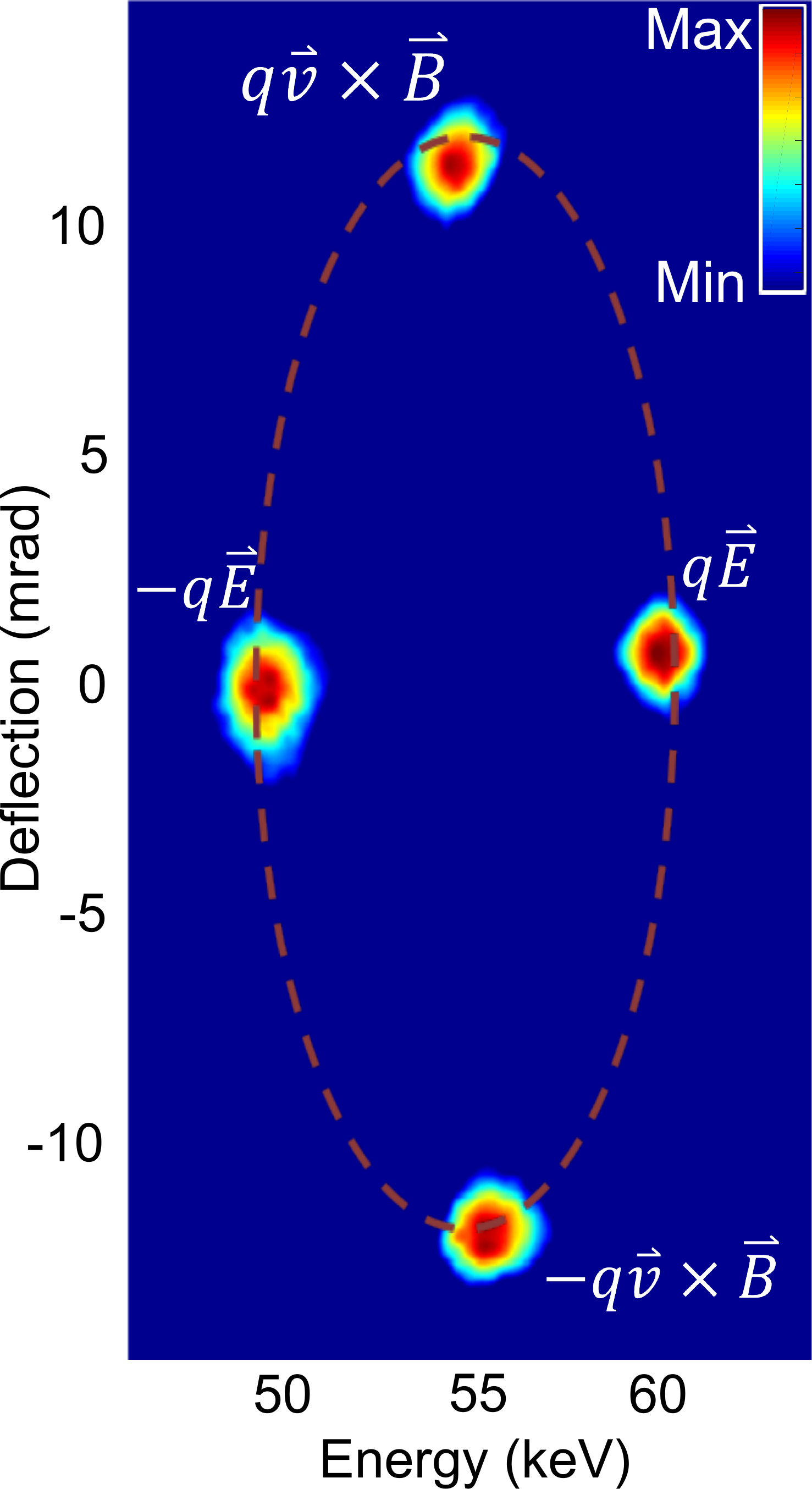} \\
  	(e)
  	\end{array}
  	\end{array}$
  	\caption{Operational modes of the STEAM device: (a) Measured energy modulation and (c) time-dependent deflection of electron pulse as a function of electron-terahertz delay for constructive interference of the electric fields entering the device and cancellation of the magnetic fields. (b) Measured beam energy and (c) time-dependent deflection deflection measured for constructive interference of the magnetic fields, that is, electric field cancellation scenario. (e) Measured shape of electron beam on MCP detector for maximum acceleration, deceleration, and right and left deflection points plotted in one image. Intensity was normalized and image contrast was tuned to show the relative positions more clearly. The red dashed line represents the predicted locus of beam positions corresponding to a sweep of the relative phases of the two THz waveforms. This demonstration was performed using a Yb:KYW laser with $\sim2\times0.5$\,{\textmu}J terahertz radiation coupled into the device and a bunch charge of $\sim1$\,fC.}
  	\label{STEAMelectricModeResults}
  \end{figure}

  In the electric mode, the relative timing of the terahertz pulses was adjusted so that the transverse electric fields constructively interfered at the interaction point.
  In this configuration, the magnetic fields (B fields) were 180$^\circ$ out of phase with each other and thus cancelled, minimizing unwanted deflections.
  The acceleration was sensitive to the terahertz phase at the interaction.
  Fig.\,\ref{STEAMelectricModeResults} shows energy and deflection diagrams that were obtained by recording the vertical and horizontal projections (respectively) of the electron-beam distribution on the MCP as a function of the electron-terahertz delay.
  Although the terahertz pulses injected into the device were nearly single cycle, several cycles of acceleration and deceleration were observed, due to dispersion induced by the horn couplers.
  
\begin{figure} \centering
  	$\begin{array}{cc}
  	\includegraphics[draft=false,width=3.0in]{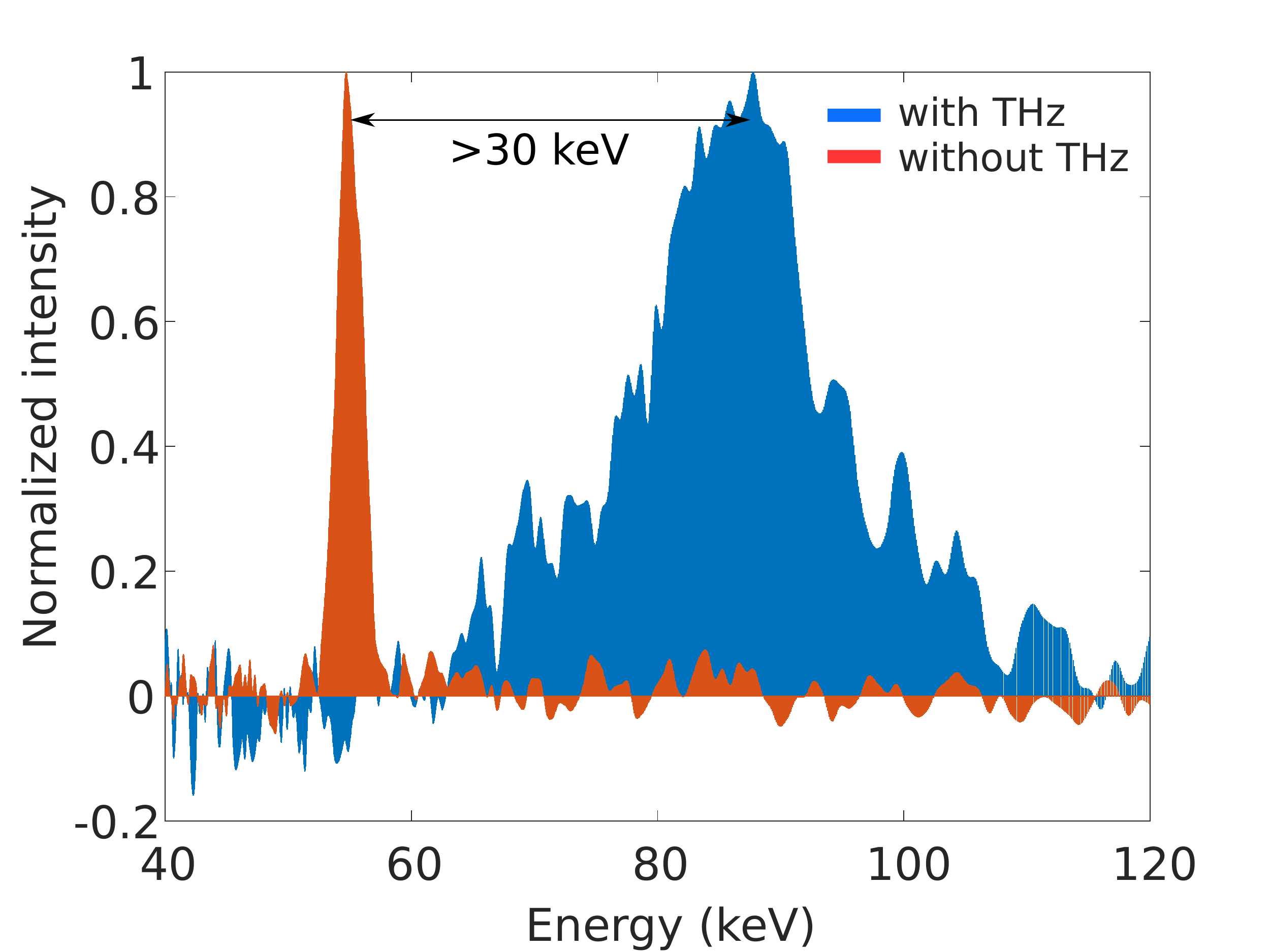} &
  	\includegraphics[draft=false,width=3.0in]{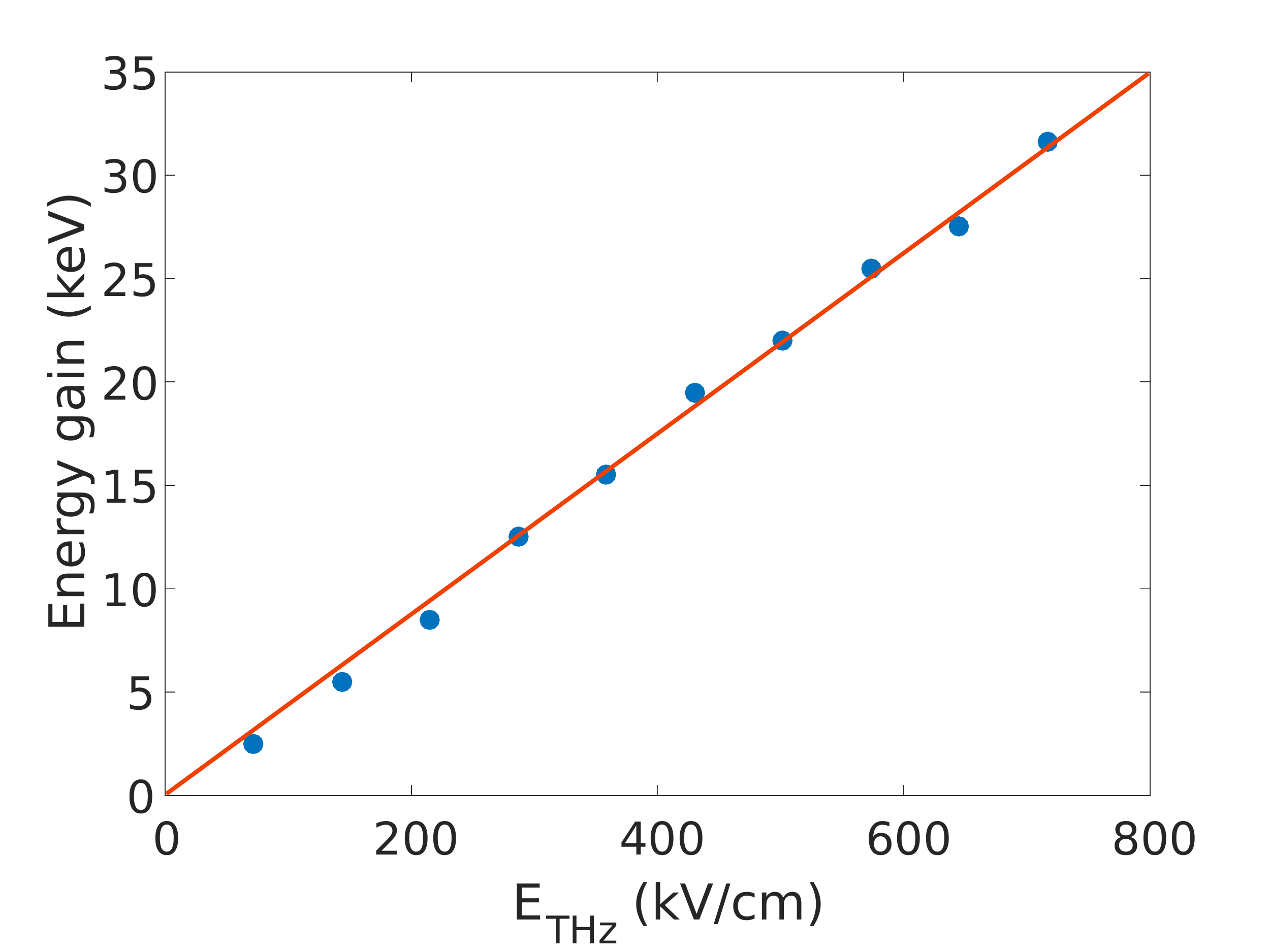} \\
  	(a) & (b) \\
  	\includegraphics[draft=false,width=3.0in]{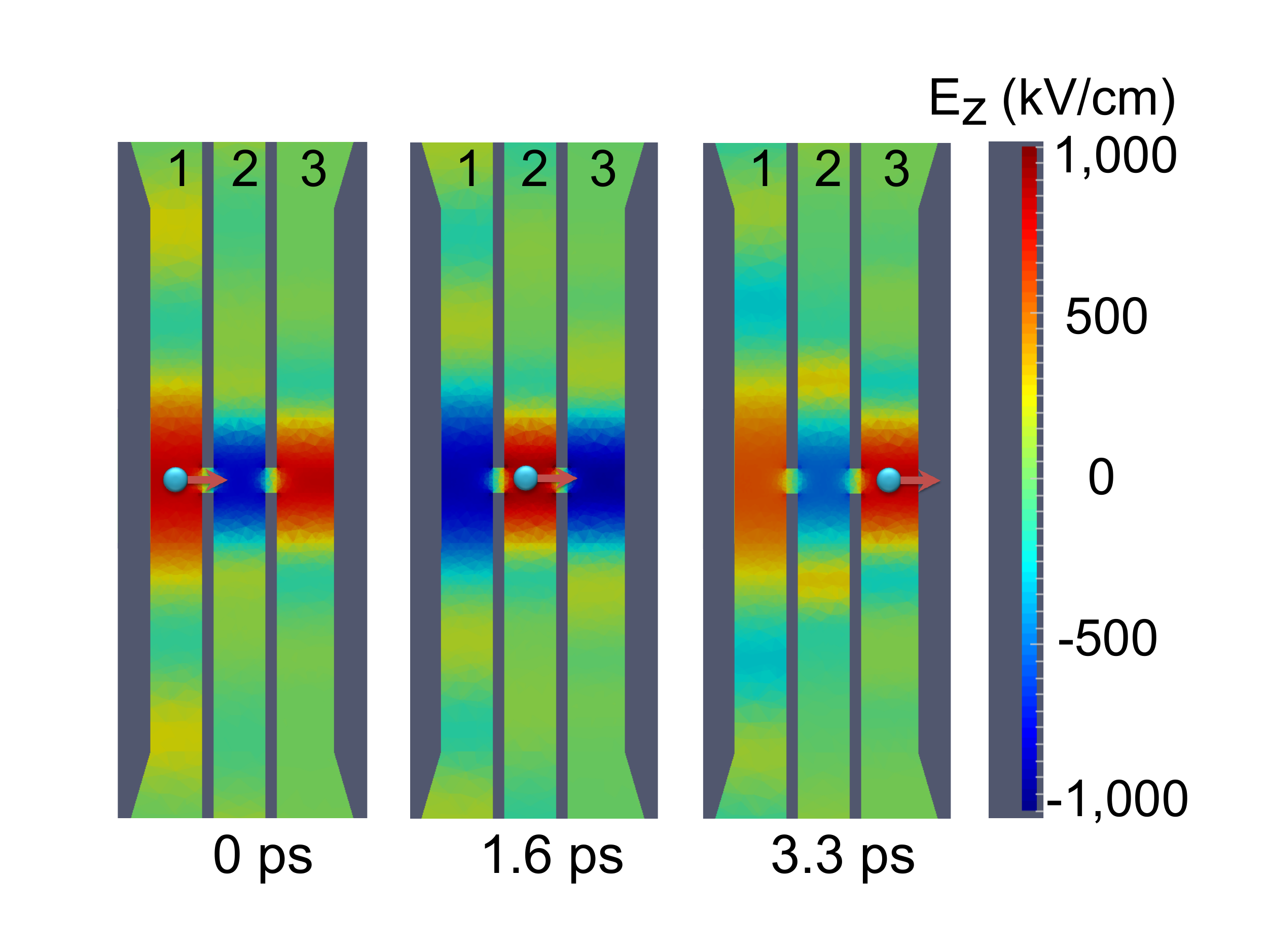} &
  	\includegraphics[draft=false,width=3.0in]{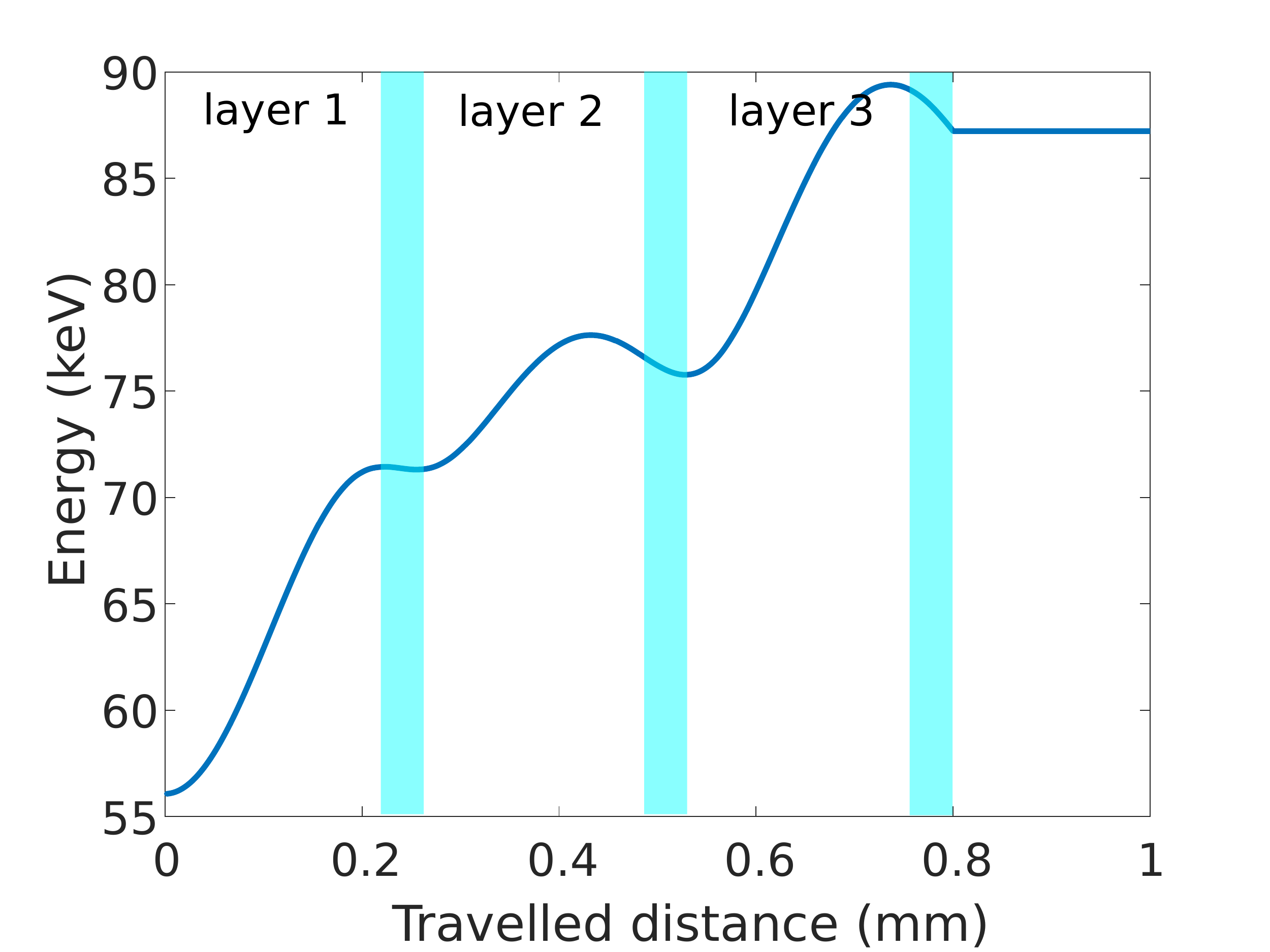} \\
  	(c) & (d)
  	\end{array}$
  	\caption{STEAM acceleration: (a) Measured electron energy spectra for initial input beam (blue curve) and accelerated beam (red curve) that shows an energy gain of more than 30\,keV. An increased energy spread is observed due to the long length of the initial electron bunch, as well as the slippage between the terahertz pulse and the electron bunch. (b) Relative energy versus input terahertz field strength with the red circle indicating the energy spectra plotted in (a). The linear relationship supports a direct, field-driven interaction. (c) Temporal evolution of the electric field inside each layer with the red arrow indicating the electron propagating. (d) Calculated acceleration along the electron propagation direction with $\sim2\times6$\,{\textmu}J terahertz radiation and beam diameter of 3\,mm. This illustration was performed using the Yb:YLF laser with $\sim2\times6$\,{\textmu}J terahertz radiation coupled into the device and a bunch charge of $\sim5$\,fC.}
  	\label{STEAMacceleration}
  \end{figure}

  Maximum acceleration and deceleration occurred at the electron injection points (Fig.\,\ref{STEAMelectricModeResults}a) where the deflection was minimized (Fig.\,\ref{STEAMelectricModeResults}c) and the beam spatial distribution was also preserved (Fig.\,\ref{STEAMelectricModeResults}e, left and right beams).
  The peak field is calculated to have reached $\sim 70$\,MV/m with the Yb-doped yttrium lithium fluoride (Yb:YLF) laser, based on comparisons of the measured terahertz energy transmitted through the device and the electron energy gain with simulation (described below).
  The energy gain scaled linearly with the applied field (Fig.\,\ref{STEAMacceleration}b) and reached a record of more than 30\,keV (five times larger than previous studies \cite{Nanni2015}) for a bunch charge of $\sim 5$\,fC.
  In contrast to previous results showing simultaneous acceleration and deceleration, the energy spectrum can be seen to move cleanly to a higher energy (Fig.\,\ref{STEAMacceleration}a), indicating that injected bunches were shorter than half the driver period.
  In fact, the bunches were measured (by the STEAM device in streaking mode) to have a duration of 670\,fs, and thus occupied about 20\% of the 3.33\,ps period accelerating field.
  The increase in energy spread is attributed in part to the variation of the electric field over the bunch temporal profile.
  Although bunches with charge up to 20\,fC were coupled into the device, space-charge effects and the long travel distances from the DC gun lead to longer bunch duration and larger energy spread.
  For demonstrating terahertz-driven acceleration, the charge was limited to 1-5\,fC during this experiment.
  Use of a terahertz-based re-buncher before the accelerator is thus anticipated for future experiments to reduce energy spread.
  
  The performance of the device was simulated using a finite-element based code \cite{fallahi2014field}.
  Fig.\,\ref{STEAMacceleration}c shows snapshots of the electrons traversing the device and staying in phase with the field.
  Fig.\,\ref{STEAMacceleration}d shows the electron energy as a function of distance.
  The energy gain can be seen to occur in three uneven steps corresponding to the three layers.
  The unevenness and the presence of deceleration at some points are evidence of dephasing due to the fact that the structure was designed for higher terahertz energies.
  Simulations predict that mega-electron-volt electron beams with up to 1\,pC of charge are achievable by increasing the number of layers and extending terahertz pulse energies to the millijoule level \cite{fallahi2016short}, which is within the reach of current terahertz-generation methods \cite{fulop2012generation}.
  
  At timings off from the optimum acceleration, the electrons experienced strong temporal gradients of the E-field resulting in large energy spreads (Fig.\,\ref{STEAMelectricModeResults}a).
  At the zero crossing of the field, the gradient is maximized and the electrons see symmetric acceleration and deceleration but no net energy gain.
  In this mode, the E-field imparts a temporally varying energy or \emph{chirp} resulting in a velocity gradient that causes either compression or stretching (depending on the sign of the gradient) of the electron bunch as it propagates \cite{van2007electron}.
  This technique, known as \emph{velocity bunching}, is an ideal application of terahertz technology, as the submillimetre-scale gradients allow bunch compression down to the femtosecond range.
  To test this concept, the applied terahertz energy was varied and a second STEAM device (\emph{streaker}) acting as a streak camera (described in the next section) was added to measure the bunch temporal profile at a point 200\,mm downstream of the first device (\emph{buncher}).
  
\begin{figure}
  	\centering
  	$\begin{array}{c}
  	\begin{array}{cc}
  	\includegraphics[draft=false,width=2.7in]{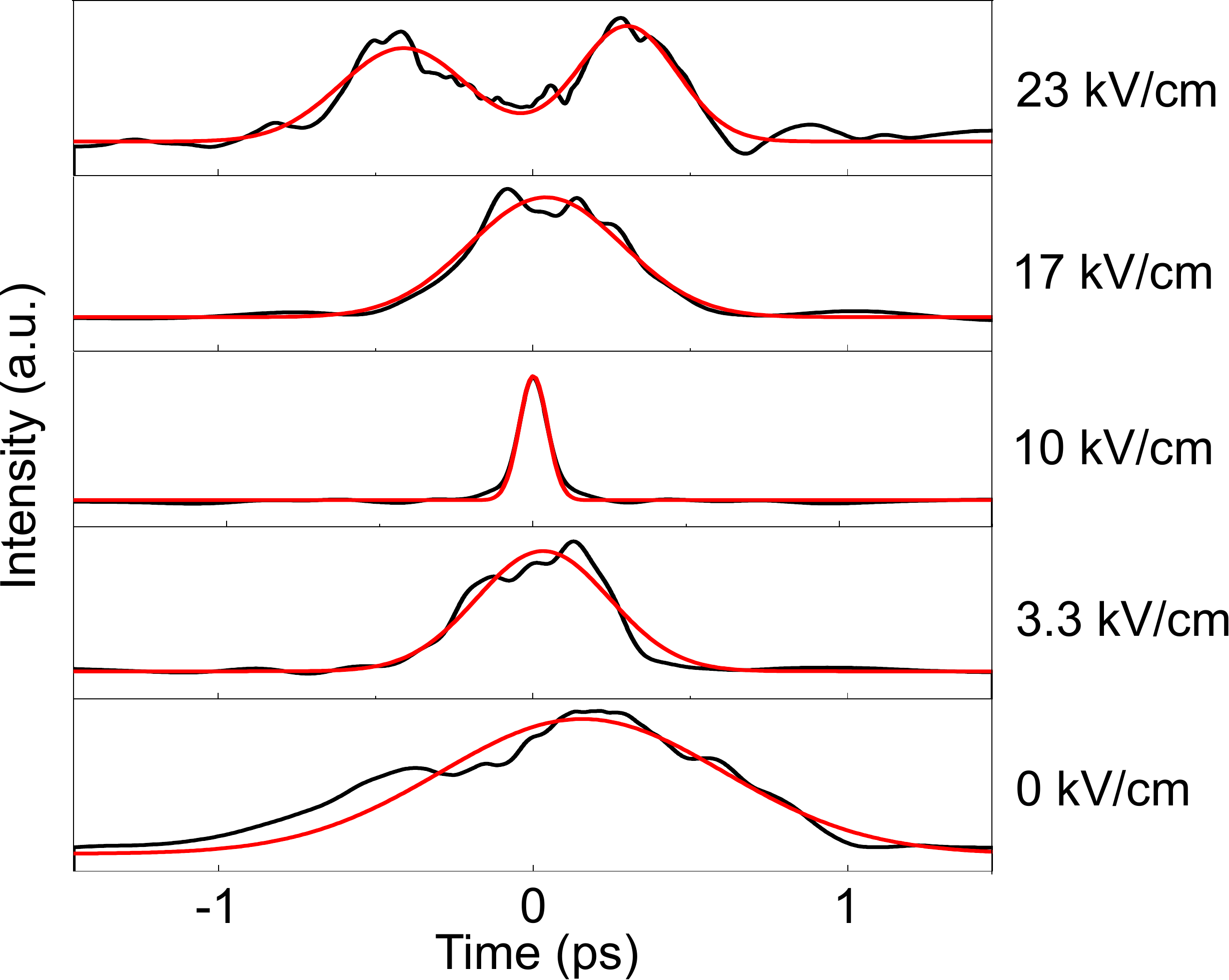} &
  	\includegraphics[draft=false,width=3.0in]{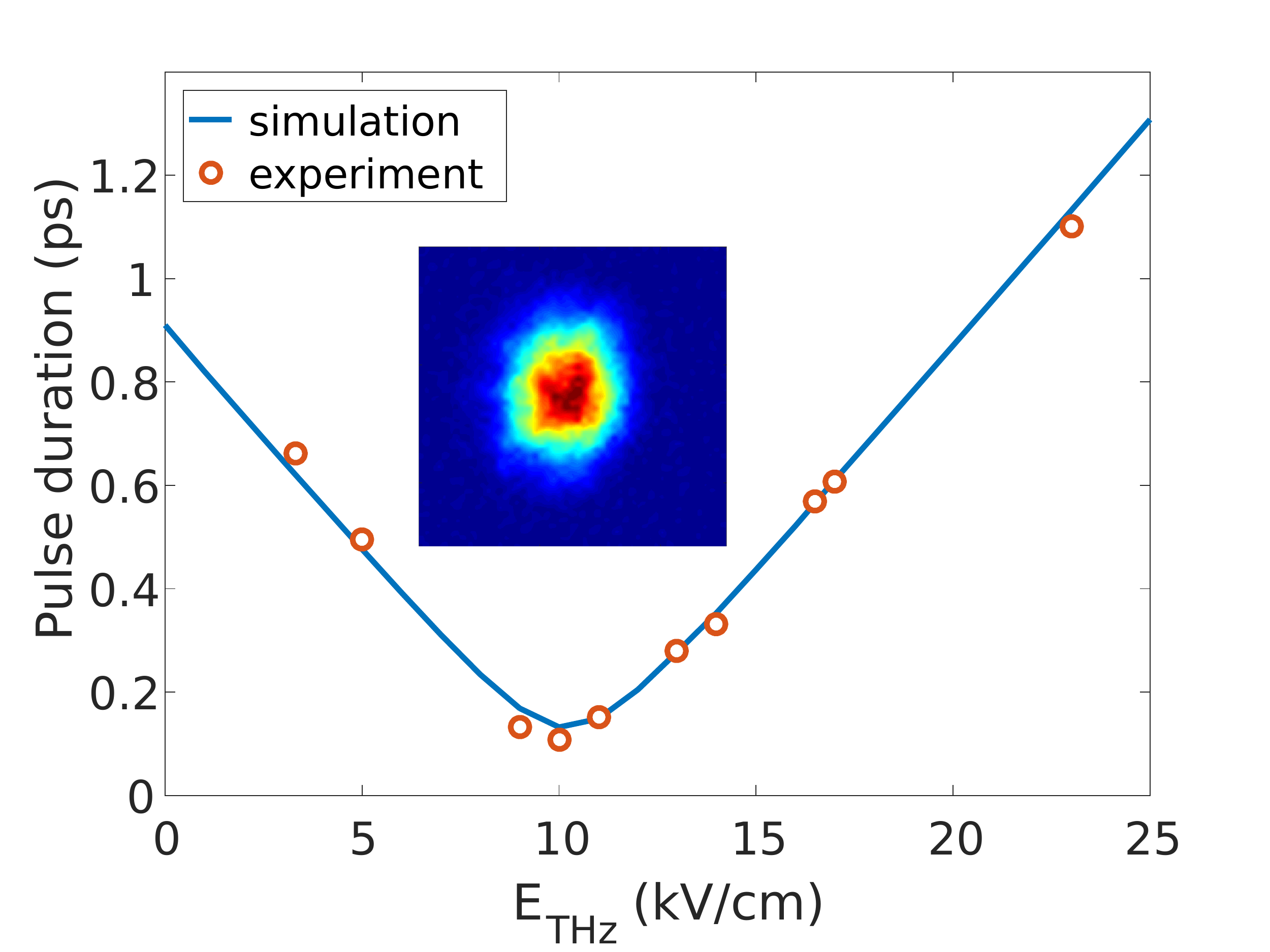} \\
  	(a) & (b)
  	\end{array} \\ \\
  	\includegraphics[draft=false,width=4.5in]{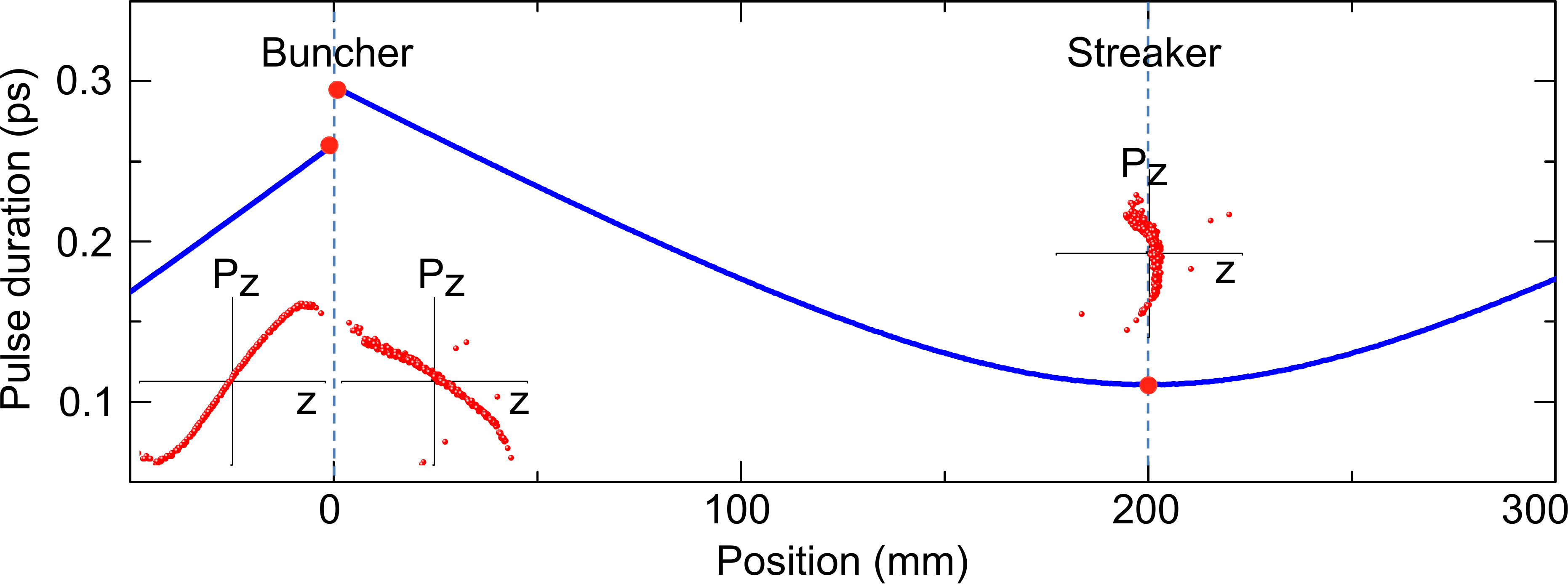} \\
  	(c)
  	\end{array}$
  	\caption{STEAM compression: (a) Measured temporal profiles of the electron pulses as the terahertz field in the buncher is increased (red arrow). The red lines represent Gaussian fits. (b) Measured electron bunch FWHM duration versus incident terahertz field strength (black squares) and corresponding simulation results (red line). Inset: the electron beam spatial profile on the detector at the optimal compressed condition. (c) Simulated bunch length versus position. Insets: from left to right, longitudinal phase-space distribution before the rebunching cavity, after the rebunching cavity and maximally compressed position (marked with red dots). This demonstration was performed with the Yb:KYW laser using one STEAM device as a rebunching cavity and one as an electron streak camera with a bunch charge of $\sim 1$\,fC.}
  	\label{STEAMcompression}
  \end{figure}

  Fig.\,\ref{STEAMcompression}a shows the electron bunch temporal profiles measured at the streaker for various field strengths applied to the buncher.
  The initial decrease in bunch duration with increasing field confirms that the electrons arrive at the buncher with a space-charge-induced energy chirp.
  A minimum duration of $\sim 100$\,fs FWHM was achieved, after which the duration increases again (Fig.\,\ref{STEAMcompression}b), implying that for high fields, the electrons temporally focus before the streaker and are overcompressed by the time they are measured.
  The minimum bunch duration can thus be reduced by using stronger fields and a shorter propagation distance.
  As observed on the MCP detector (Fig.\,\ref{STEAMcompression}b, inset), a good electron beam profile is maintained at the optimal compressed condition.
  Fig.\,\ref{STEAMcompression}c shows the evolution of the bunch duration with distance simulated for the minimum bunch duration case.
  The phase-space distributions in the insets show the reversal of the velocity correlation by the buncher and the eventual compression at the streaker location.
  
\begin{figure}
  	\centering
  	$\begin{array}{cc}
  	\includegraphics[draft=false,width=2.0in]{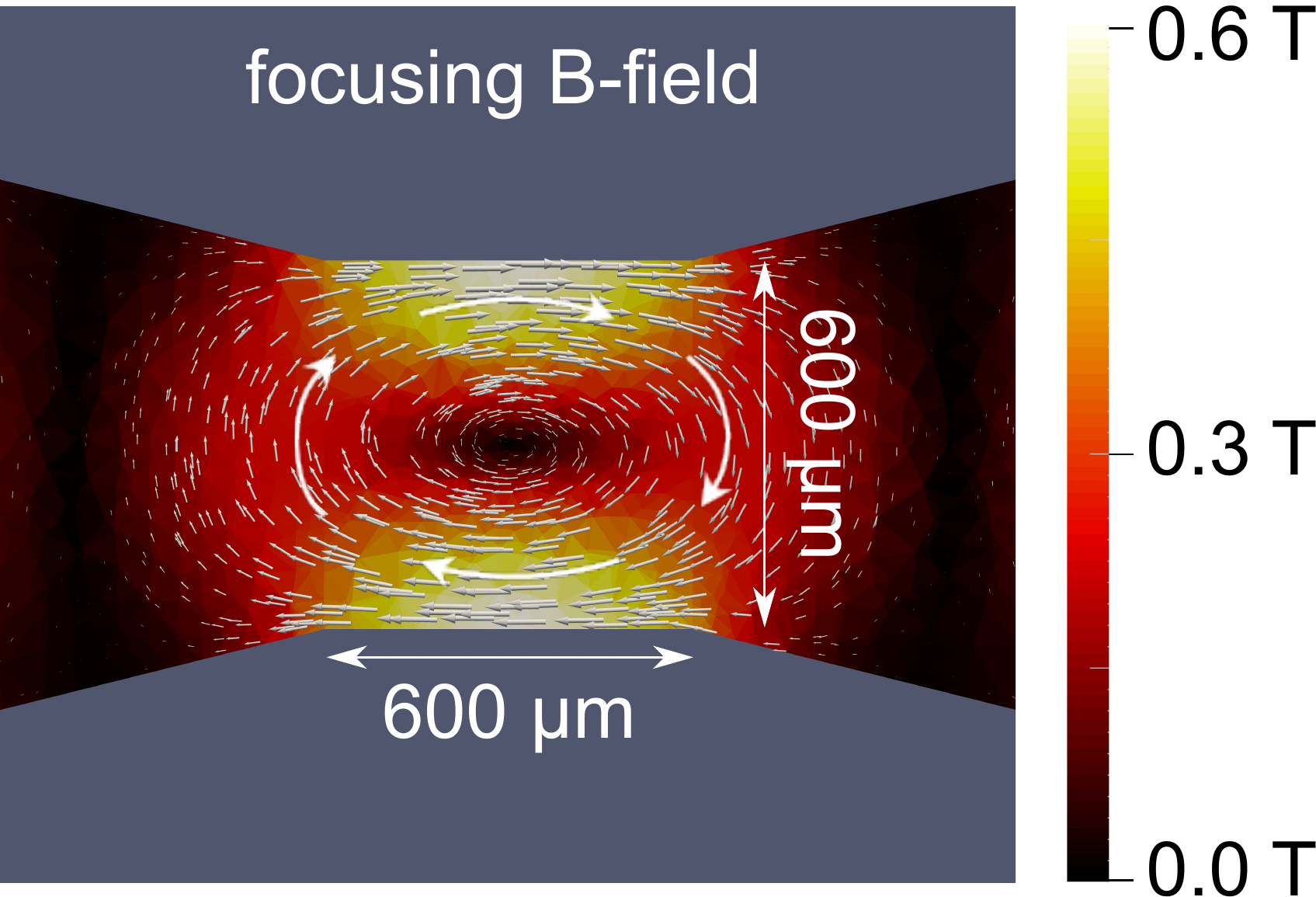} &
  	\includegraphics[draft=false,width=2.0in]{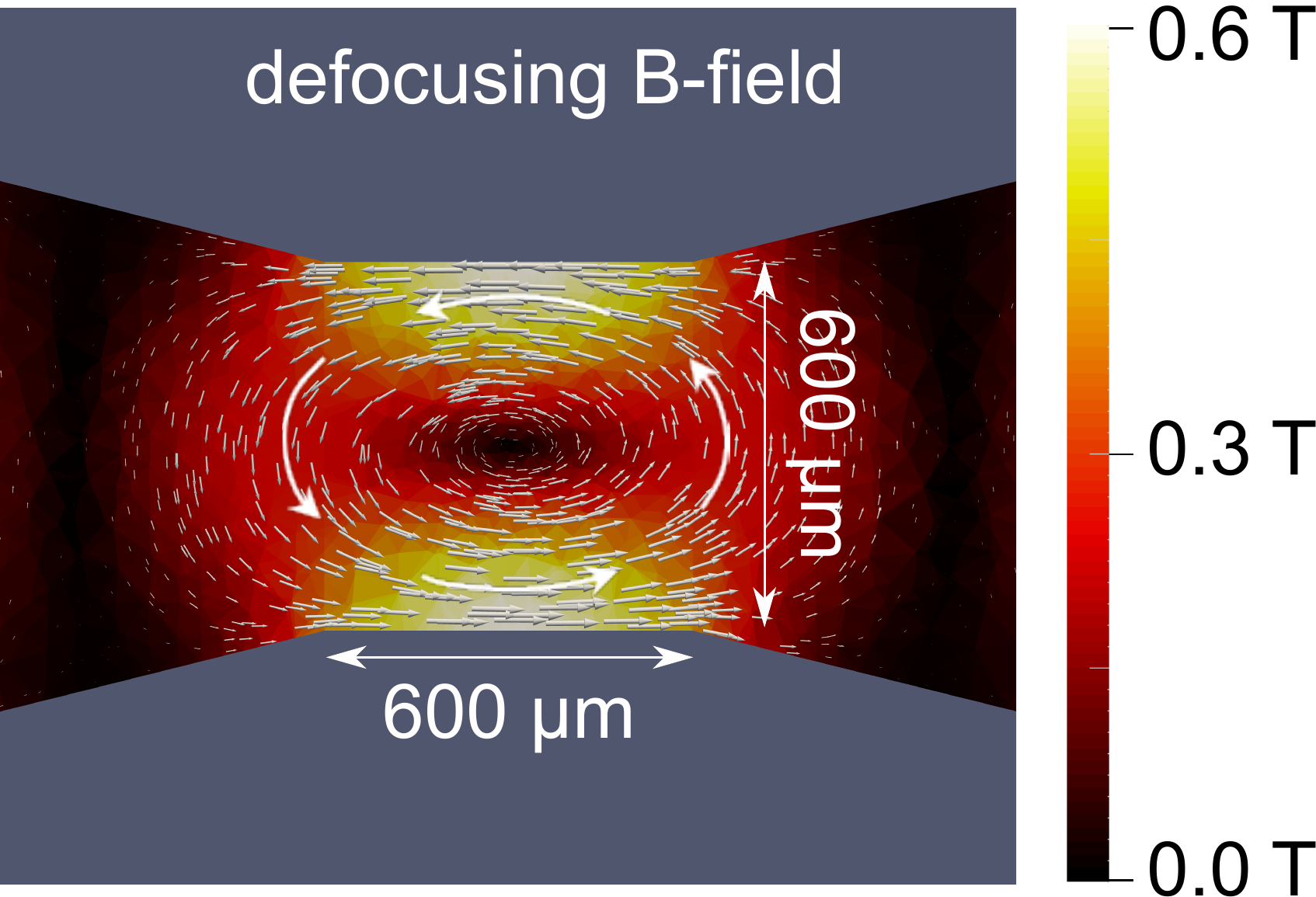} \\
  	(a) & (b)
  	\end{array}$
  	\caption{Computed spatial magnetic field distributions of the (c) focusing and (d) defocusing fields. Simulation was performed with $\sim2\times6$\,{\textmu}J THz radiation and a beam diameter of 3\,mm.}
  	\label{STEAMfocusingMechanism}
  \end{figure}

\begin{figure}
  	\centering
  	$\begin{array}{cc}
  	\includegraphics[draft=false,width=3.0in]{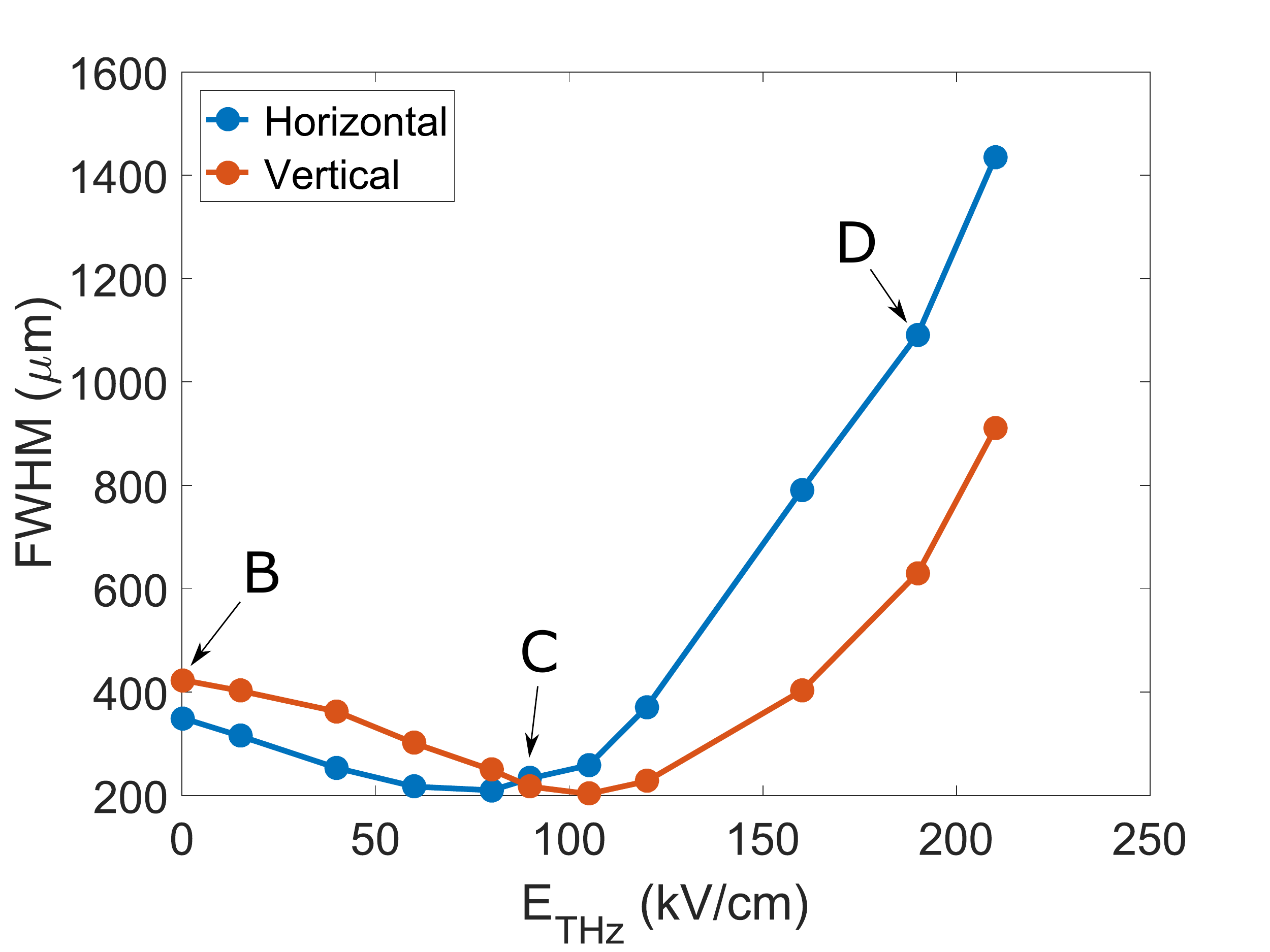} &
  	\includegraphics[draft=false,width=3.0in]{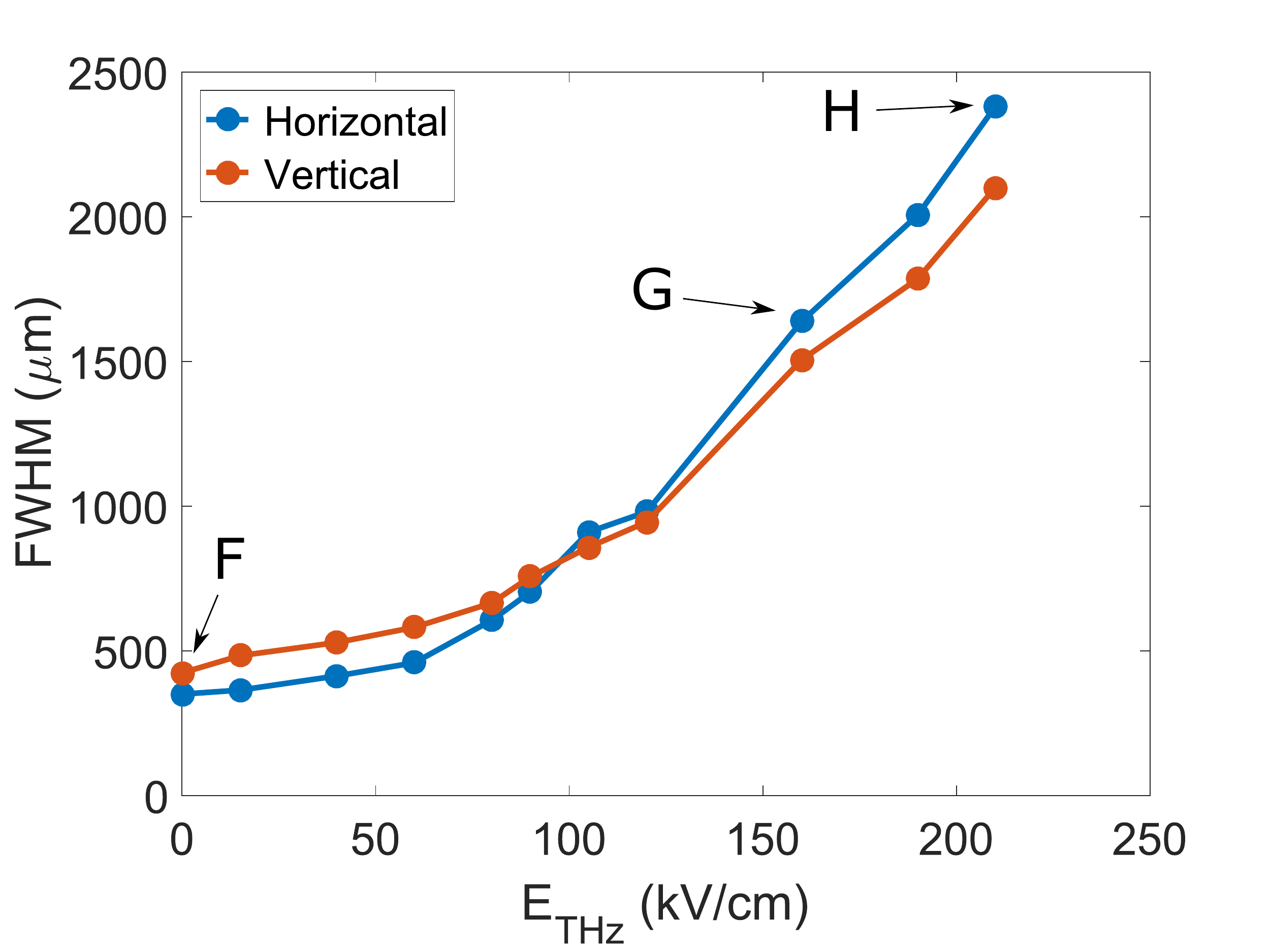} \\
  	(a) & (e) \\ \\
  	\begin{array}{ccc}
  	\includegraphics[draft=false,width=0.8in]{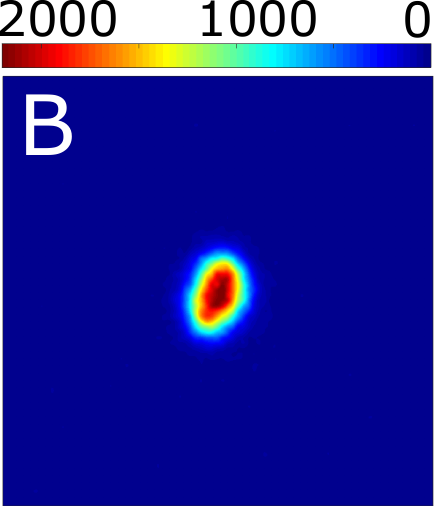} &
  	\includegraphics[draft=false,width=0.8in]{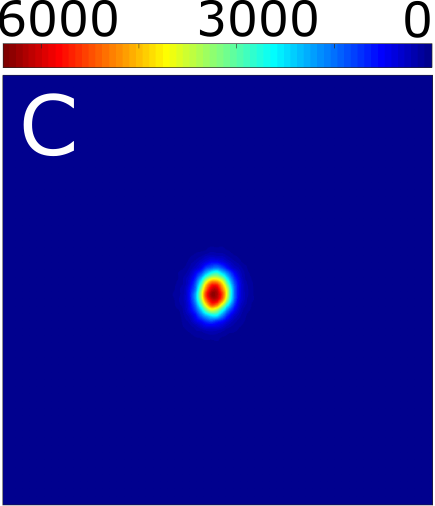} &
  	\includegraphics[draft=false,width=0.8in]{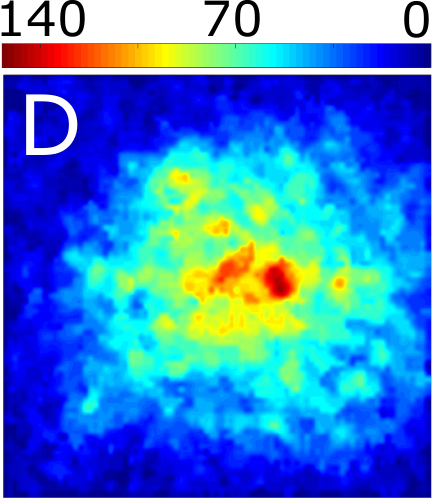} \\
  	(b) & (c) & (d)
  	\end{array} &
  	\begin{array}{ccc}
  	\includegraphics[draft=false,width=0.8in]{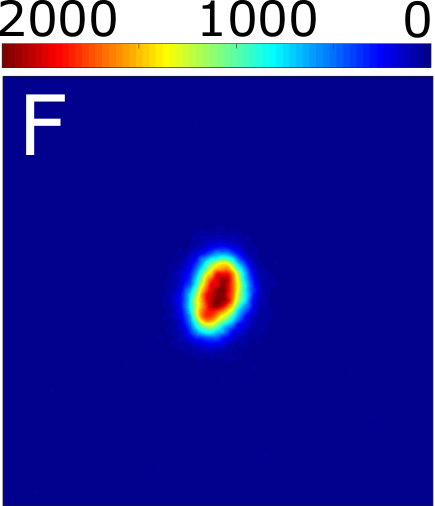} &
  	\includegraphics[draft=false,width=0.8in]{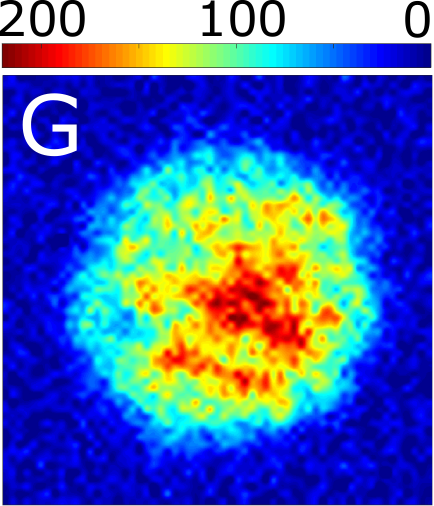} &
  	\includegraphics[draft=false,width=0.8in]{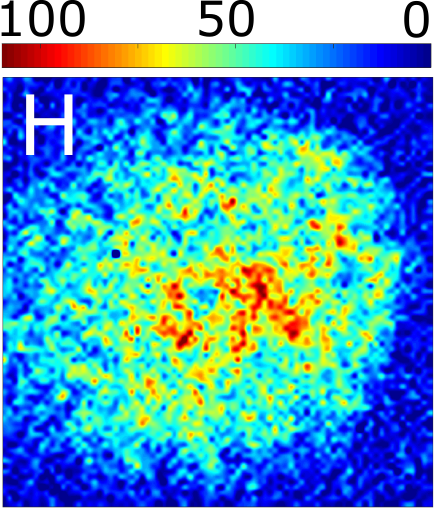} \\
  	(f) & (g) & (h)
  	\end{array}
  	\end{array}$
  	\caption{Terahertz lens for electron pulse focusing and defocusing: (a) and (e) Measured transverse electron beam size at the MCP as a function of terahertz field for electron pulse focusing (a) and defocusing (b). This demonstration was performed with the Yb:KYW laser and a bunch charge of $\sim 1$\,fC. Spatial profiles of (b-d) initial and focused as well as (f-h) initial and defocused electron beams on the MCP detector. The colour bars show the number of counts on the detector.}
  	\label{STEAMfocusing}
  \end{figure}

  By imposing the electrons to pass through the zero crossing in the electric mode (corresponding to the maximum magnetic field rotating around the interaction region), the STEAM device can also operate as a focusing or defocusing element, as can be seen by the horizontal spreading of the beam profile in Fig.\,\ref{STEAMelectricModeResults}c.
  Due to the strong terahertz field that leads to over-focusing at the fixed MCP position, both focusing and defocusing schemes are observed here as an increase of the beam size.
  This focusing effect is a consequence of the well-known Panofsky-Wenzel theorem \cite{panofsky1956some}, which uses Gauss' law to show that longitudinal compressing and decompressing fields must be accompanied by transverse defocusing and focusing fields, respectively.
  The magnetic field always cancels at the interaction point, while it still has a time-varying transverse distribution in the antinode region that contributes to the defocusing and focusing (illustrated in Fig.\,\ref{STEAMfocusingMechanism}a and b).
  The focusing was tested by monitoring the beam spatial profile at the MCP for varying terahertz pulse energies.
  Fig.\,\ref{STEAMfocusing}a-d shows the results for the focusing configuration, which corresponded to the longitudinal decompression condition.
  At best, the electron beam diameter was reduced by 2$\times$ compared with its input value.
  For higher field strengths, however, the device focal length became shorter than the 180\,mm distance to the MCP, causing the measured beam size to increase again.
  Similar to photon beams, a focusing optic with higher focusing power results in a smaller beam at focus, provided that the input beam size is constant.
  The defocusing configuration is obtained by shifting the electron timing to the longitudinal compression condition, which occurs at an adjacent zero crossing of opposite sign.
  In this case, the electron beam diameter increases monotonically with the terahertz field (Fig.\,\ref{STEAMfocusing}e-h), as expected.
  For both cases, the focusing performance is significantly beyond what is offered by conventional electrostatic \cite{cesar2016demonstration} and proposed dielectric  focusing structures and is comparable to those of plasma lenses \cite{van2015active}.
  Peak focusing gradients of $>2$\,kT/m were calculated based on $\sim 2 \times 6$\,{\textmu}J of coupled terahertz energy.
  A small (less than a factor of two) asymmetry is noticeable for the focusing strengths in the horizontal and vertical planes.
  This asymmetry is due to the asymmetry of the interaction region, which leads to stronger gradients in the vertical direction (Fig.\,\ref{STEAMfocusingMechanism}a and \ref{STEAMfocusingMechanism}b).
  
  \subsection{Magnetic Mode}
  
  In the magnetic mode, the relative timing of the terahertz fields is different from that of the electric mode by a half period, resulting in reinforcement of the magnetic and cancellation of the electric fields at the interaction region.
  In this configuration, electron acceleration is minimized (Fig.\,\ref{STEAMelectricModeResults}b), and the magnetic field dominates the interaction causing a transverse deflection of the electron beam that depends on the terahertz phase at the interaction (Fig.\,\ref{STEAMelectricModeResults}d).
  When electrons sweep the positive cycle of the magnetic field, the deflection is maximized and the beam profile is also best preserved.
  In this mode, electon beams can be precisely steered (Fig.\,\ref{STEAMelectricModeResults}e, top and bottom beams) by varying the terahertz pulse energy.
  Here, we achieved continuous control of the beam angle over a range of 70\,mrad, which was limited by the aperture of the device.
  Increasing the aperture enables greater range at the cost of a weaker deflection field, as the field confinement is affected.
  
\begin{figure}
  	\centering
  	$\begin{array}{cc}
  	\includegraphics[draft=false,height=2.0in]{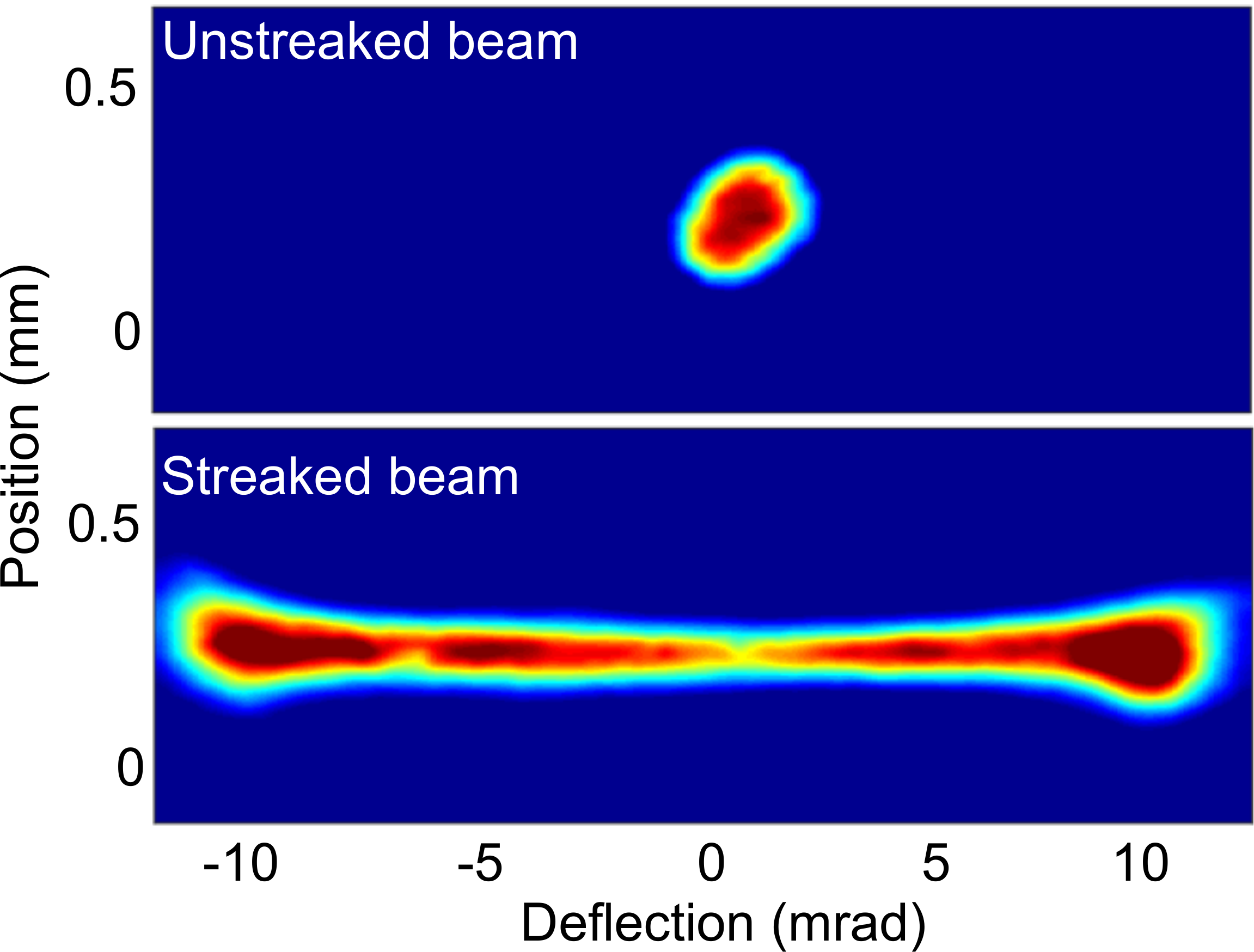} &
  	\includegraphics[draft=false,height=2.0in]{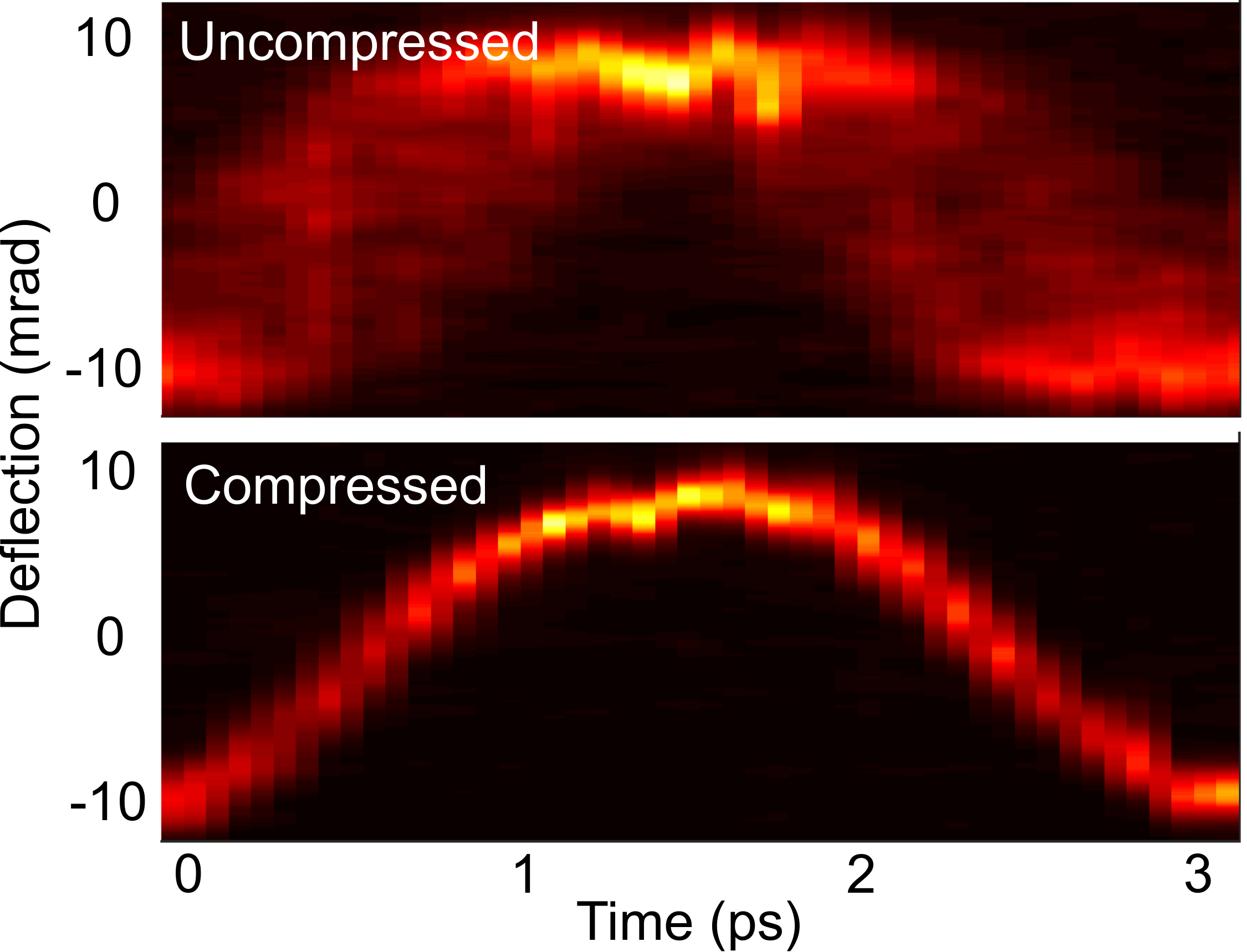} \\
  	(a) & (b)
  	\end{array}$
  	\caption{Terahertz streak camera: (a) Measured images of the electron beam on the MCP detector with and without the terahertz deflection field. (b) Time-dependent deflection diagrams measured by varying the delay between the arrival time of the electron bunch and the deflecting terahertz pulse for initially compressed and uncompressed electron bunches. This demonstration was performed with the Yb:KYW laser and a bunch charge of $\sim 1$\,fC, except for the maximum streaked beam (a), where a bunch charge of $\sim 10\,$fC was used to obtain a long pulse for demonstration.}
  	\label{STEAMstreaking}
  \end{figure}

  Electrons sweeping the zero-crossing cycle of the terahertz magnetic field, however, experience a strong deflection as a function of delay time enabling the measurement of the temporal bunch profiles of very short bunches by mapping (or \emph{streaking}) them onto the spatial dimension of a detector.
  To test this concept, a first STEAM device was used in compression mode (as described above) to provide electrons of varying bunch durations at a second, downstream STEAM device, which analyzed the temporal profiles by streaking.
  Fig.\,\ref{STEAMstreaking}a shows raw images of a temporally long electron beam with the terahertz streaking field switched on and off.
  Streaking \emph{deflectograms}, generated by plotting a lineout of the spatial charge distribution along the streaking dimension as a function of delay relative to the terahertz field, are shown in Fig.\,\ref{STEAMstreaking}b for compressed and uncompressed electron bunches.
  The degree of streaking, indicated by the vertical extent of the deflectogram, depends clearly on the bunch duration and on the phase of the terahertz field, as expected.
  For a terahertz energy of $\sim 2 \times 6$\,{\textmu}J coupled into the device, a maximum deflection rate of $>140$\,{\textmu}rad/fs was achieved, corresponding to a temporal resolution below 10\,fs.
  The resolution was limited here by the 350\,{\textmu}m size of the unstreaked beam.
  These results represent a new record in terahertz-based streaking gradient as well as the use of terahertz magnetic fields for deflection and streaking.
  
  \section{Conclusion}
  
  In this chapter, we have presented various structures that can serve as miniaturized electron guns excited by single-cycle THz pulses.
  Devices for both low energy ({\textmu}J level) and high energy (mJ level) THz pulses are proposed.
  The maximum normal electric field on the surface was allowed to be as high as 0.8\,GV/m, although the scaling equations predict much higher thresholds for extremely short pulse excitations.
  This leads to potential additional improvements in terms of accelerating gradient and justifies the use of ultrafast structures to achieve compact accelerators.
  The presented ultrafast electron guns are promising devices to realize short bunches for applications in electron diffractive imaging, microscopy and compact radiation sources.
  
  Motivated by the concept of the ultrafast electron guns, a systematic procedure to design and optimize an ultrafast electron gun driven by single-cycle THz pulses is developed.
  Based on the described process, a 400\,keV five-layer THz gun is designed which utilizes two 200\,{\textmu}J single-cycle pulses at 300\,GHz to realize a maximum accelerating gradient of about 650\,MV/m at the cathode surface.
  The output of the gun is a 0.75\,pC electron bunch with 50\,fs pulse duration and emittances in the range of 0.1\,mm$\cdot$mrad.
  Subsequently, an upgraded THz gun with three more layers is presented, which is fed by two 400\,{\textmu}J pulses and which boosts the energy gain to 810\,keV.
  The space-charge effect is the dominant factor causing particle loss, when pC-level bunches are injected into the gun.
  Due to this effect, the charge yield of the 800\,keV electron gun is about only 50\% or only 0.52\,pC bunches are generated from the gun.
  
  Experimental test of the proposed structures is an ongoing effort.
  These attempts began with the demonstration of high-field ($>$300\,MV/m), quasi-monoenergetic (few percent spread) THz acceleration of several tens of fC electron bunches to sub-keV energies in an ultracompact, robust device.
  No degradation in performance was observed over one billion shots.
  While the operating pressure was 40\,{\textmu}Torr, no change in performance was observable up to 10\,mTorr.
  This first result of a jitter-free, all-optical THz gun, powered by a few-millijoule laser, performs in accordance with underlying simulations and is encouraging for future developments.
  In its current state, it can be used for time-resolved LEED.
  
  Eventually, the operation of a multilayer device is tested through the demonstration of a novel segmented terahertz electron accelerator and manipulator setting new records in terahertz acceleration, streaking and focusing with a very compact device.
  The segmented structure makes it possible to phase match the electron-terahertz interaction for non-relativistic beams, making it ideal for use as a high-gradient photogun \cite{fallahi2016short}.
  The independent control over the counter-propagating terahertz pulse timing gives the STEAM device the ability to switch dynamically between acceleration, compression, focusing, deflection and streaking modes.
  As has been theoretically shown in the first section of this chapter, the use of terahertz pulses also brings other advantages, including negligible heat loads, high repetition rates and compactness while still supporting substantial charge in the picocoulomb regime.
  Furthermore, the three-layer STEAM device studied here indicates the path forward towards relativistic electron energies by staging more layers for higher operation efficiency.
  
  Using only $\sim 2 \times 6$\,{\textmu}J of terahertz energy, the STEAM device has demonstrated peak accelerating gradients of 70\,MV/m, compression of a bunch from over 1\,ps to 100\,fs, focusing strength of $\sim 2$\,kT/m and streaking gradients of $>140$\,{\textmu}rad/fs, leading to a temporal resolution below 10\,fs.
  By scaling to millijoule-level terahertz energies, which are already available in some terahertz wavelength ranges, the field strengths in the device can be increased by over an order of magnitude, far exceeding those of conventional radio-frequency devices.
  The exceptional performance and compactness of this terahertz-based device makes it very attractive for pursuing electron sources, such as ultrafast electron diffractometers, that operate in the few- and subfemtosecond range necessary for probing the fastest material dynamics \cite{gliserin2015sub,sciaini2011femtosecond}.
  In the pursuit of these sources, the demand is increasing for compact, high-gradient diagnostics and beam manipulation devices for novel and conventional accelerator platforms alike.
  In large-scale facilities, such as the European X-ray free-electron laser (XFEL), the Linac Coherent Light Source (LCLS) or the Swiss free-electron laser (SwissFEL), the STEAM devices can be used to add new, powerful and adaptable capabilities without major and therefore costly restructuring of the machine.
  More significant are the advantages in terms of cost and accessibility that come from using STEAM devices as the core components of an all-terahertz-powered compact, high-gradient accelerator with the ability to produce high-quality, controllable bunches of femtosecond or attosecond duration on a table top.
  The results presented here are a step in demonstrating the feasibility of that vision.
  
  \chapter{Terahertz Linac \label{chap:four}}
  
  \section{Introduction}
  
  In facilities operating based on relativistic particles, the largest and most energy consuming part is linear acceleration section.
  This section receives relativistic or sub-relativistic particles from the electron gun, transfers energy to the particles, and delivers high-energy or ultra-relativistic particles.
  Depending on the targeted application, final output energies of from few mega-electron-volt to tens of giga-electron-volts are needed, directly affecting the size and cost of the linear accelerator section.
  This acceleration regime differs fundamentally from the regime in electron guns.
  Particles in electron guns start from low energy (rest energy) to relativistic energies, leading to velocity variation from zero to relativistic values.
  In contrast, in linear accelerators particles are often relativistic and can be considered to travel with almost constant velocity.
  Due to fundamental distinctions between the two regimes, the operation of electron gun is sometimes referred to as \emph{injection}, whereas the term \emph{acceleration} adverts to mechanisms in linear accelerators.
  
  This chapter discusses the linear acceleration of particles using THz radiation.
  We demonstrate the capabilities of THz waveguides optimized for acceleration and/or compression of relativistic electron bunches by coherent THz pulses.
  The relativistic few femtosecond pico-Coulomb electron bunch achieved in the bunch compression scheme has applications in single-shot few-femtosecond electron diffraction \cite{sciaini2011femtosecond}.
  Dielectric-loaded cylindrical metallic waveguides is chosen for our studies for the ease of manufacturing and theoretical evaluation.
  The THz frequency range is chosen as the operation range because it appears to strike a compromise between the large wavelength and low accelerating gradient (due to breakdown limitations) of RF radiation and the small wavelength but high accelerating gradient of optical radiation.
  Note that a higher accelerating gradient is more favorable for bunch compression and acceleration, but space-charge effects make it difficult to confine a bunch of substantial charge well within a half-cycle if the wavelength is too small.
  The absence of plasma in a vacuum-core waveguide scheme precludes problems associated with the inherent instability of laser-plasma interactions.
  Although using a guiding structure leads to intensity limitations, it also increases acceleration efficiency due to a smaller driving energy required and a larger interaction distance.
  
  Observing the mechanism from a different viewpoint, to prevent emittance growth and increased energy spread, the electron bunch needs to occupy a small fraction of the optical cycle.
  Even for a long-infrared wavelength of 10\,mm, 1$^\circ$ of phase in the optical wavelength corresponds to only $\sim 28$\,nm.
  Another practical concern would be the timing precision between the optical cycle and the arrival of the electron bunch.
  For example, 1$^\circ$ phase jitter, commonly required for operational accelerators, requires $<$100\,as timing jitter between the optical pulse and the electron bunch, which is challenging to maintain over extended distances.
  Difficulties increase further when considering the available options for guiding the optical light to decrease the phase velocity to match the electron beam.
  A guided mode at a wavelength of 10\,{\textmu}m would require sub-micron precision for aligning the electron bunch and the optical waveguide.
  THz frequencies provide the best of both worlds.
  On one hand, the wavelength is long enough that we can fabricate waveguides with conventional machining techniques, provide accurate timing and accommodate a significant amount of charge per bunch.
  At 0.3\,THz, the wavelength is 1\,mm and 1$^\circ$ of phase precision corresponds to 10-fs timing jitter, which is readily achievable \cite{schibli2003attosecond}.
  
  Detailed theoretical analysis of electron bunch acceleration and compression in THz dielectric-leaded metallic waveguides is presented in the next section.
  Firstly, we furnish a technical discussion of the equations upon which our model rests.
  Next, we demonstrate the acceleration of a 1.6\,pC electron bunch from a kinetic energy of 1\,MeV to about 10\,MeV over an interaction distance of about 20\,mm, using a 20\,mJ pulse centered at 0.6\,THz in a dielectric-loaded metallic waveguide.
  The implications of using an arbitrarily distant injection point, as well as the prospects of dielectric breakdown and thermal damage for our optimized design are also analyzed.
  We then investigate the acceleration of 16\,pC and 160\,pC 1\,MeV electron bunches.
  Eventually, the dielectric-loaded metal waveguide design is optimized for simultaneous acceleration and bunch compression, achieving a 50 times (100\,fs 1.6\,pC electron bunch compressed to 2\,fs over an interaction distance of about 18\,mm) and 62 times (100\,fs to 1.61\,fs over an interaction distance of 42\,cm) compression for 1\,MeV and 10\,MeV electron bunches, respectively.
  The discussion proceeds with the first proof-of-principle experiment demonstrating linear acceleration in a THz waveguide.
  We observe a maximum energy gain of 7\,keV in 3\,mm using an optically generated 20\,{\textmu}J THz pulse centered at 0.45\,THz.
  Future THz accelerators designed for relativistic electron beams and using more intense THz sources will be able to reach GeV/m accelerating gradients.
  
  \section{Dielectric-Loaded Metallic Waveguides for THz Acceleration}
  
\begin{figure}
  	\centering
  	$\begin{array}{c}
  	\includegraphics[draft=false,height=2.0in]{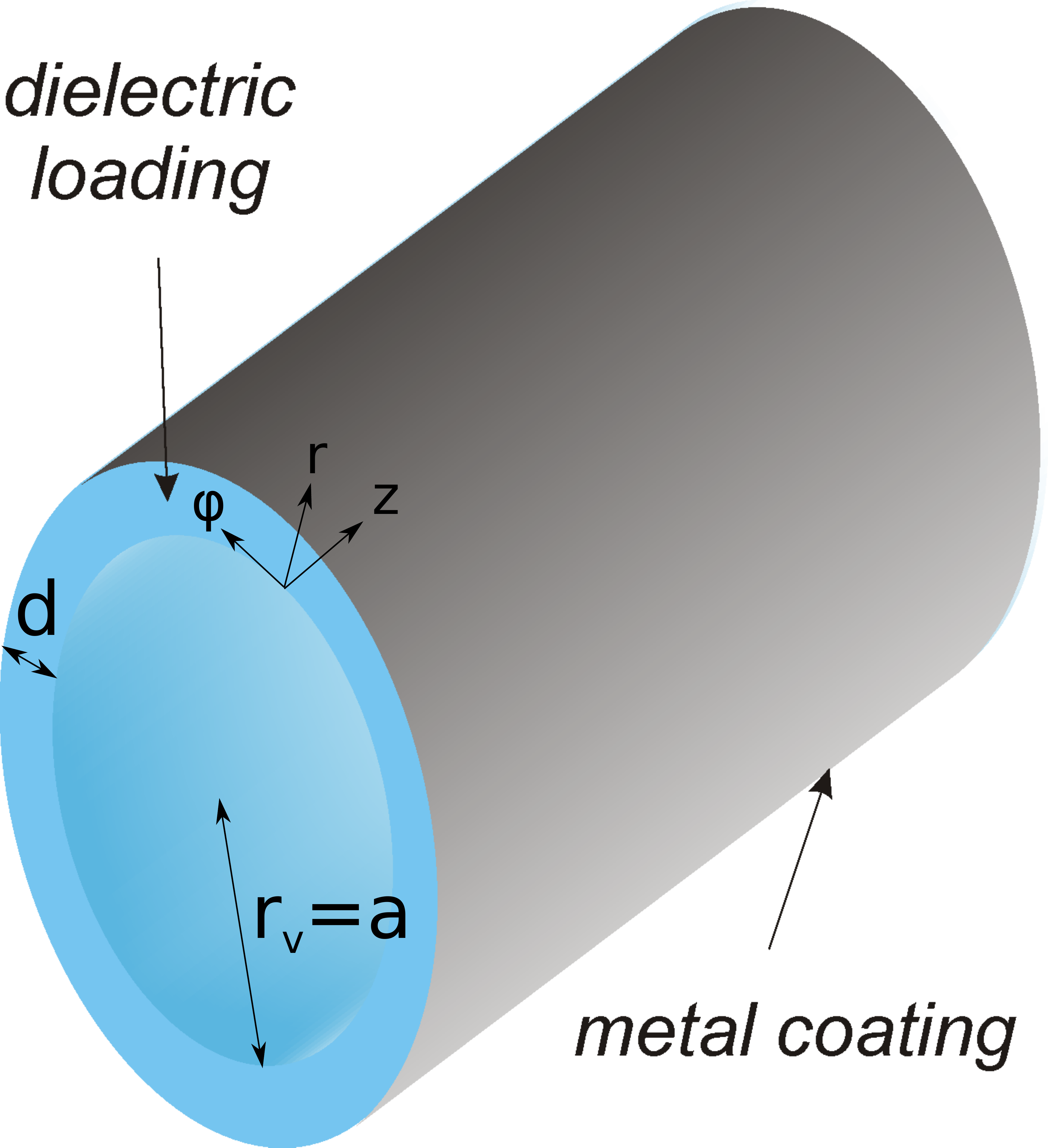} \\
  	(a) \\
  	\includegraphics[draft=false,width=6.0in]{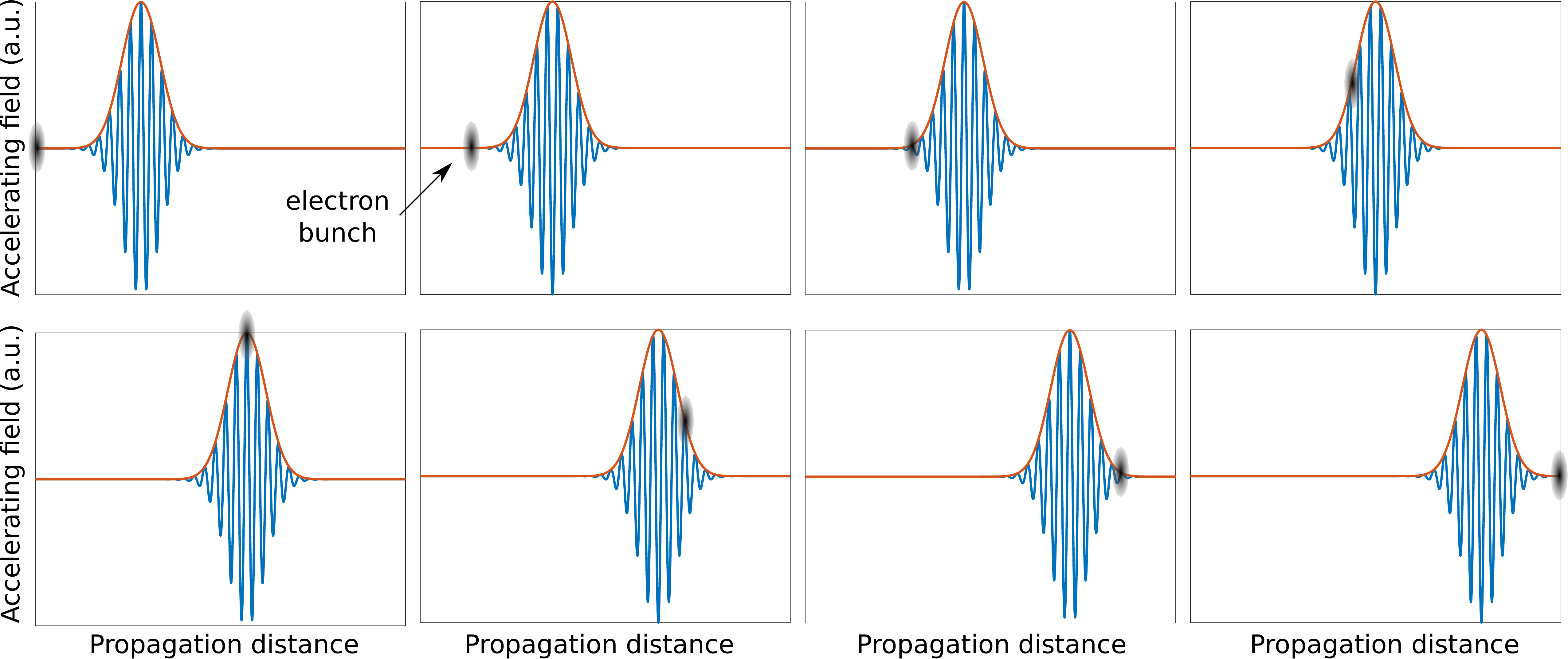} \\
  	(b) \\
  	\end{array}$
  	\caption{THz pulse acceleration in waveguides: schematic illustration of (a) the dielectric-loaded metallic waveguide, and (b) the acceleration scheme using the waveguide mode.}
  	\label{DLMWConcept}
  \end{figure}

  Fig.\,\ref{DLMWConcept} schematically illustrates the geometry of the dielectric-loaded metallic waveguide as well as the interaction between an electron bunch and guided field in this scheme.
  We present this visual example before any technical discussion to give some preliminary intuition of the electrodynamics that ensues when a 1\,MeV electron bunch (obtained, for instance, from an RF gun) is injected into a coherent THz pulse propagating in a dielectric-loaded cylindrical metal waveguide.
  
  \subsection{Relativistic electrodynamics in a waveguide}
  
  This section introduces the equations governing the behavior of an electron bunch in the vacuum-filled core of a waveguide, and discusses our approach in modeling this behavior.
  The equations are implemented in the software described in section 2.3, offering calculation of bunch dynamics within analytically driven field equations.
  The electron bunch is made up of $N$ interacting electrons that may be modeled classically as $N$ point charges propagating according to Newton's second law:
  \begin{equation}
  \frac{d \vec{p}_i(t)}{dt} = \vec{F}_i^d(t)+\sum\limits_{j=1,i \neq j}^N \left( \vec{F}_{i,j}^{pp}(t) + \vec{F}_i^{wf}(t)+\vec{F}_i^{rr}(t) \right), \qquad \text{with} \qquad i = 1,2,...,N
  \label{newtonLaw}
  \end{equation}
  where $\vec{p}_i(t)\equiv\gamma_i(t)m\vec{v}_i(t)$ is the momentum of electron $i$ at time $t$, with $m$, $\vec{v}_i$, and $\gamma_i\equiv 1/\sqrt{1-\beta_i^2}$ being its rest mass, velocity and Lorentz factor, respectively. $\beta_i \equiv | \vec{\beta}_i |$, $\vec{\beta}_i \equiv \vec{v}_i/c$ and $c$ is the speed of light in vacuum.
  
  According to \eref{newtonLaw}, each electron $i$ is subject to four kinds of forces: the force $\vec{F}_i^d$ exerted by the driving electromagnetic field, the sum of forces $\vec{F}_{i,j}^{pp}$ exerted directly by other electrons $j$, the force $\vec{F}_i^{wf}$ exerted by wakefields that result from electromagnetic fields of other electrons reflecting off the waveguide walls, and finally the radiation reaction force $\vec{F}_i^{rr}$ that the electron experiences as a result of its own radiation.
  In this study, we neglect $\vec{F}_i^{wf}$  because the relatively short propagation distances and bunch lengths make the effect of wakefields negligible.
  For acceleration studies involving long propagation distances, or multiple bunches of substantial charge, wakefields should be taken into consideration by implementing formulas derived in previous studies \cite{kim2010theory}.
  We also neglect the radiation reaction force since the employed scheme accelerates the electrons primarily via the $z$-directed component of the electric field, with minimal transverse wiggling.
  Consequently, radiation losses are negligible.
  Electrodynamic studies in which the radiation reaction force plays a significant role have commonly employed the Landau-Lifshitz formula \cite{landau1971classical} for the force.
  
  The force $\vec{F}_i^d$ exerted by the driving field on electron $i$ is given by the Lorentz force equation:
  \begin{equation}
  \vec{F}_i^d(t) = q \left[ \vec{E}_d(t,\vec{r}_i(t)) + \vec{v}_i(t) \times \vec{B}_d (t,\vec{r}_i(t)) \right],
  \label{LorentzForceEquation}
  \end{equation}
  where $q$ is the electron's charge and $\vec{r}_i$ its position.
  and $\vec{E}_d(t,\vec{r})$ and $\vec{B}_d(t,\vec{r})$ are respectively the electric field and magnetic flux density of the driving field.
  Similarly, we write the force $\vec{F}_{i,j}^{pp}$ that electron $j$ exerts on electron $i$ as
  \begin{equation}
  \vec{F}_{i,j}^{pp}(t) = q \left[ \vec{E}_j(t,\vec{r}_i(t)) + \vec{v}_i(t) \times \vec{B}_j (t,\vec{r}_i(t)) \right],
  \label{spaceChargeForce}
  \end{equation}
  where $\vec{E}_j(t,\vec{r})$ and $\vec{B}_j(t,\vec{r})$ are respectively the electric field and magnetic flux density due to electron $j$.
  These fields are derived by solving Maxwell's equations for a moving point charge in vacuum via the Li\'{e}nard-Wiechert potentials and the resulting electromagnetic fields are found as in \eref{lienardWichertPotentialICS}, which is rewritten and tailored to the formulation as follows \cite{JacksonClassical}
  \begin{align}
  \vec{E}_i(t,\vec{r}) & = \frac{q}{4\pi\epsilon_0}\frac{1}{\eta^3_{i,\tilde{t}_i}(\vec{r})R_{i,\tilde{t}_i}(\vec{r})} \left\{ \frac{\vec{u}_{i,\tilde{t}_i}(\vec{r})}{\gamma_i^2(\tilde{t}_i)R_{i,\tilde{t}_i}(\vec{r})} + \frac{1}{c}\left[\vec{n}_{i,\tilde{t}_i}(\vec{r}) \times \left( \vec{u}_{i,\tilde{t}_i}(\vec{r}) \times \frac{\vec{\dot{v}}_i(\tilde{t}_i)}{c} \right) \right] \right\} \\
  \vec{B}_i(t,\vec{r}) & = \frac{1}{c} \left(\vec{n}_{i,\tilde{t}_i}(\vec{r}) \times \vec{E}_i(t,\vec{r}) \right)
  \label{lienardWichertPotentialRepeated}
  \end{align}
  where $\epsilon_0$ is the permittivity of free space, $\vec{\dot{v}}_i$ the acceleration of particle $i$, $R_{i,t}(\vec{r}) \equiv | \vec{r} - \vec{r}_i(t) |$, $\vec{n}_{i,t}(\vec{r}) \equiv ( \vec{r} - \vec{r}_i(t) ) / R_{i,t}$, $\vec{u}_{i,t}(\vec{r}) \equiv \vec{n}_{i,t}(\vec{r}) - \vec{v}_i(t)/c $ and $\eta_{i,t}(\vec{r}) \equiv dt/d\tilde{t}_i = 1 - \vec{n}_{i,\tilde{t}_i}(\vec{r}) \cdot \vec{v}_i(\tilde{t}_i)/c $. $\tilde{t}_i = \tilde{t}_i(t,\vec{r})$ is the retarded time along particle $i$'s trajectory corresponding to time $t$ and observation point $\vec{r}$.
  Given $t$ and $\vec{r}$, the retarded time $\tilde{t}_i$ solves the implicit equation
  \begin{equation}
  \tilde{t}_i = t - \frac{R_{i,\tilde{t}_i}(\vec{r})}{c}.
  \label{retardedTime}
  \end{equation}
  
  If $\vec{F}_i^d$ is the only non-zero term on the right-hand side of \eref{newtonLaw}, the equation is simply an ordinary differential equation.
  With inter-particle interaction described by \eref{spaceChargeForce} and \eref{lienardWichertPotentialRepeated}, the right-hand side of \eref{newtonLaw} becomes a function of $\tilde{t}_i$ as well as $t$, and the equation is no longer an ordinary differential equation.
  Note that \eref{lienardWichertPotentialRepeated} considers both the velocity field (near-field) and the radiation field (far-field), which are given by the first and second term respectively.
  If the effect of the radiation field is insignificant and we assume that each particle always travels at its current velocity during each time step, \eref{lienardWichertPotentialRepeated} can be simplified to a function of only $t$, making \eref{newtonLaw} an ordinary differential equation and reducing the computation of inter-particle forces considerably.
  The formulas that should replace \eref{lienardWichertPotentialRepeated} are then the space-charge formulas obtained by Lorentz-boosting the Coulomb field of each electron from the electron's rest frame to the lab frame.
  These formulas are used in particle tracer programs like the General Particle Tracer (GPT) \cite{manual24pulsar}.
  
  We chose not to use externally-provided software packages in part to ascertain, by implementing \eref{lienardWichertPotentialRepeated}, the significance of non-uniform motion and electron radiation in inter-particle interaction.
  It turns out that for the regime investigated in this paper, the use of the exact formulas in \eref{lienardWichertPotentialRepeated} affects overall acceleration and bunch compression results negligibly, and for computational efficiency one may simply revert to the Lorentz-boosted Coulomb fields in modeling inter-particle interaction.
  
  We solve \eref{newtonLaw} using a fifth-order Runge-Kutta algorithm with adaptive step-size \cite{press1996numerical}.
  If the exact inter-particle fields in \eref{lienardWichertPotentialRepeated} are used, we adapt the Runge-Kutta algorithm to the problem by maintaining a history of $\vec{r}_i$ and $\vec{p}_i$, $i=1,...,N$, in a ring buffer.
  At each time $t$, cubic spline interpolation is applied to compute the retarded time needed in \eref{lienardWichertPotentialRepeated}.
  Gaussian-distributions of electrons in 6-dimensional phase space are generated by applying the Box-Muller transformation to the normalized output of the  \verb"rand()" in \verb"C", and computations of variance and covariance (required for emittance calculations) are performed using the corrected two-pass algorithm \cite{chan1983algorithms}.
  Multi-core processing capabilities are implemented using \verb"OpenMP".
  
  In this study, we are interested in simulating bunches on the order of pCs and tens of pCs, implying that we deal with $10^7-10^8$ electrons.
  To speed up the computational process, each particle $i = 1,...,N$ is treated as a macro-particle - with the charge and mass of a large number of electrons - instead of a single electron.
  We can verify that this approach is a good approximation if the solution converges as the number of macro-particles increases while the total number of electrons is kept constant.
  We have verified this for all results presented in this section.
  
  \subsection{Pulsed TM\textsubscript{01} mode in a dielectric-loaded metallic waveguide}
  
  We are interested in obtaining an analytical expression that models propagation of a coherent THz pulse in the waveguide.
  This involves integration over the continuous-wave (CW) solutions of the waveguide.
  The method we use to obtain these solutions is very similar to that detailed in \cite{yeh1978theory} for the optical Bragg fiber, so we only give an overview of the method here.
  For a general multilayer cylindrical waveguide, the continuous-wave solutions are obtained by solving the Helmholtz equation in cylindrical coordinates \cite{JacksonClassical}:
  \begin{equation}
  (\nabla^2 + k^2) \left\{ \begin{array}{c} E_z^{\mathrm{CW}} \\ H_z^{\mathrm{CW}} \end{array} \right\} = 0 \qquad \Rightarrow \qquad \frac{1}{r}\frac{\partial}{\partial r} \left( r \frac{\partial}{\partial r} \left\{ \begin{array}{c} \psi_e \\ \psi_h \end{array} \right\} \right) + \left( k^2 - \kappa^2 - \frac{l^2}{r^2} \right) \left\{ \begin{array}{c} \psi_e \\ \psi_h \end{array} \right\} = 0,
  \label{helmholtzEquationWaveguide}
  \end{equation}
  where $k \equiv \omega/c = 2\pi/\lambda$, $\omega$ being angular frequency, $c$ the speed of light in vacuum and $\lambda$ the vacuum wavelength.
  $E_z^\mathrm{CW} \equiv \psi_e (r) \exp (i(\omega t - \kappa z \pm l\phi))$ and $H_z^\mathrm{CW} \equiv \psi_h (r)\exp (i(\omega t - \kappa z \pm l\phi))$ are the complex CW $z$-directed electric and magnetic fields respectively, $\kappa$ is the propagation constant, $r$ the radial coordinate, $\phi$ the azimuthal coordinate, $z$ the direction of propagation along the waveguide, and $l$ a non-negative integer that determines the order of azimuthal variation.
  According to \eref{helmholtzEquationWaveguide}, a general solution for $\psi_e$ in layer $i$ of an $n$-layer cylindrical waveguide (the core counts as layer 1) is
  \begin{equation}
  \psi_e^i(r) = A_e^i J_l(h_i r) + B_e^i Y_l(h_i r), \qquad \text{with} \qquad r_{i-1} \leq r < r_i, \quad i=1,...,n,
  \label{generalSolution}
  \end{equation}
  where $r_0 \equiv 0$, $r_n \equiv \infty$ and $r_i$ for $0 < i < n$ is the radial position of the boundary between layers $i$ and $i+1$. $J_l$ and $Y_l$ are Bessel functions of the first and second kind respectively, $A_e^i$ and $B_e^i$ are constant complex coefficients within each layer and $h_i = \sqrt{\epsilon_{ri}(\lambda) \mu_{ri}(\lambda) k^2 - \kappa^2 }$, $\epsilon_{ri}$ and $\mu_{ri}$ being the dispersive relative permittivity and permeability respectively of the dielectric in layer $i$.
  The general solution for $\psi_h^i$ is identical in form to \eref{generalSolution} except that ``e" should be replaced by ``h" in all subscripts.
  In the core, it is usually expedient to express \eref{generalSolution} using the modified Bessel function of the first kind $I_l$, whereas in the final layer (which extends to infinity), it is usually expedient to express \eref{generalSolution} using either the modified Bessel function of the second kind $K_l$ for confined modes or the Hankel function of the second kind $H_l^{(2)}$ for leaky modes.
  These functions are all exactly represented by \eref{generalSolution} if we allow the coefficients and arguments of $J_l$ and $K_l$ to take on complex values.
  
  The transverse electromagnetic fields are obtained from the expressions for $E_z$ and $H_z$ via Ampere's law and Faraday's law.
  By matching boundary conditions among adjacent dielectric layers (continuity of $E_z$, $H_z$, $E_{\phi}$, $H_{\phi}$), we obtain a characteristic matrix which has a non-trivial null-space (zero determinant) if and only if a solution to \eref{helmholtzEquationWaveguide} exists.
  Given $l$, along with the dimensions and dielectric properties of the waveguide layers, we determine numerically the set of values $\{k, \kappa\}$ for which the characteristic matrix has a zero determinant.
  This set of values $\{k, \kappa\}$ constitute the dispersion curves of the waveguide for a mode of azimuthal order $l$, and the $4n$ coefficients $A_e^i$, $B_e^i$, $A_h^i$ and $B_h^i$, $i=1,...,n$, are the components of a $4n$-long vector in the corresponding null-space.
  Up to this point, our procedure is very similar to that detailed in \cite{yeh1978theory}, and we direct the reader there for more information.
  The real-valued $z$-directed electric field $E_z^i$ of a pulse in any layer $i$ is constructed by an inverse Fourier transform:
  \begin{align}
  E_z^i(l,t,r,z,\phi) & = \Re \left\{ \int_{-\infty}^{+\infty} F(\omega) E_z^{\mathrm{CW},i} (l,\omega,r,z,\phi) d\omega \right\}, \nonumber \\
  E_r^i(l,t,r,z,\phi) & = \Re \left\{ \int_{-\infty}^{+\infty} F(\omega) E_r^{\mathrm{CW},i} (l,\omega,r,z,\phi) d\omega \right\}, \qquad  \text{with} \qquad r_{i-1} \leq r < r_i, \quad i=1,...,n, \label{inverseFourierTransform} \\
  H_{\phi}^i(l,t,r,z,\phi) & = \Re \left\{ \int_{-\infty}^{+\infty} F(\omega) H_{\phi}^{\mathrm{CW},i} (l,\omega,r,z,\phi) d\omega \right\}, \nonumber
  \end{align}
  where $F(\omega)$ is the complex envelope in the frequency domain.
  In \eref{inverseFourierTransform}, the same inverse Fourier transform is also applied to field components $E_r$ and $H_{\phi}$ to obtain their real-valued pulsed versions.
  
  The structure we consider here is a vacuum core with a single layer of dielectric of relative permittivity $\epsilon_r = 5.5$ (a candidate for such a dielectric is diamond \cite{kubarev2009optical}) with an external copper coating (Fig.\,\ref{DLMWConcept}).
  The high thermal conductivity and breakdown properties of chemical-vapor-deposited diamond at THz frequencies are well-recognized, and has led to its use in waveguides for wakefield acceleration \cite{antipov2012experimental} and other applications involving intense terahertz radiation \cite{yoneda2001high}.
  For this reason, we use diamond for the dielectric throughout this study and assume a relative dielectric constant of $\epsilon_r = 5.5$.
  
  The spatial mode of interest is the TM\textsubscript{01} mode (i.e. $l = 0$ and radial variation is of the lowest order), for which only the $E_z$, $E_r$ and $H_{\phi}$ field components exist.
  The $E_z$ field peaks on axis whereas the transverse fields vanish, so an electron bunch concentrated at the waveguide axis will experience forces mainly along the direction of propagation.
  This facilitates longitudinal compression and acceleration of the bunch without significant transversal wiggling, which is undesirable since it tends to increase radiative losses.
  
  To excite the TM\textsubscript{01} mode of the cylindrical waveguide, it would be necessary to apply a radially-polarized (preferably TM\textsubscript{01}) beam to the waveguide.
  Studies on coupling linearly-polarized THz pulses into cylindrical metal waveguides show that the dominant modes excited are the TE\textsubscript{11}, TE\textsubscript{12} and TM\textsubscript{11} modes \cite{gallot2000terahertz}, so a linearly-polarized incoming beam is unlikely to serve our purpose.
  Although THz pulses generated by optical rectification are typically linearly-polarized, the direct generation of radially-polarized THz pulses has been demonstrated \cite{chang2007generation,winnerl2009terahertz}.
  Alternatively, a scheme to convert linearly-polarized THz pulses into radially-polarized pulses may be adopted \cite{grosjean2008linear}.
  
  Equation \eref{inverseFourierTransform} provides a rigorous way to compute the electromagnetic field at any point in space and time required for an electrodynamic simulation.
  However, performing a summation over a large number of frequency components at every time step for every macro-particle is computationally expensive.
  To obtain an analytical approximation for more efficient numerical simulation, notice that in the vacuum-filled core, the CW TM\textsubscript{01} mode is of the form:
  \begin{equation}
  E_z^{\mathrm{CW},1} = A_e^1 I_0(q_1 r) e^{i(\omega t - \kappa z)}, \qquad H_r^{\mathrm{CW},1} = A_e^1 \frac{i\kappa}{q_1} I_1(q_1 r) e^{i(\omega t - \kappa z)}, \qquad H_{\phi}^{\mathrm{CW},1} = \frac{k}{\eta_0 \kappa} E_r^{\mathrm{CW},1},
  \label{TM01Mode}
  \end{equation}
  where $q_i = \sqrt{\kappa^2 - \epsilon_{ri}(\lambda) \mu_{ri}(\lambda) k^2}$ is the radial wavevector and $\eta_0$ is the vacuum impedance.
  $I_0$ and $I_1$ are the modified Bessel functions of the first kind, of order 0 and 1 respectively.
  We need to make three more assumptions in the remainder of the formulation: Firstly, variations in propagation constant $\kappa$ across the spectrum are small enough that their effects on magnitude can be ignored.
  Secondly, variations in $\kappa$ are negligible above the second order.
  Thirdly, the imaginary part of $\kappa(\omega)$ is negligible beyond its 0'th order term, and the quadrature term produced by this imaginary part in $E_r$ does not contribute significantly to the field.
  Hence, Taylor-expanding $\kappa(\omega)$ about central angular frequency $\omega_0$ we have $\kappa(\omega_0 + \Delta \omega) \approx \kappa_0 - i\alpha + \kappa_1 \Delta \omega + \kappa_2 (\Delta \omega)^2/2$, where $\kappa_i$ denotes the real part of the $i$'th derivative of $\kappa(\omega)$ with respect to $\omega$ at $\omega=\omega_0$.
  $\alpha > 0$ to be physically valid and represents field attenuation per unit distance.
  
  To obtain the approximate analytical field solution, the rightmost expressions of \eref{TM01Mode} should be inserted into \eref{inverseFourierTransform}.
  Assuming a transform-limited Gaussian pulse at $z = 0$, we have $E_z(z = 0, t) \sim \exp (-(t/\tau)^2/2)\exp(i \omega_0 t)$, where $\omega_0 = k_0 c$ is the central frequency and $\tau$ is the half-width at 1/e intensity, related to the FWHM intensity $\tau_\mathrm{FWHM}$ as $\tau_\mathrm{FWHM} = 2 \sqrt{\ln 2} \tau \approx 1.665 \tau$.
  This is related to the spectral FWHM intensity width $\Delta \omega_\mathrm{FWHM}$ as $\Delta \omega_\mathrm{FWHM} = 4 \ln 2 / \tau_\mathrm{FWHM}$.
  Finally, we have
  \begin{align}
  E_z^1(t,z) & \approx \Re \left\{ I_0(q_1^0 r) e^{i(\omega_0 t - \kappa_0 z)} \int_{-\infty}^{+\infty} A_0 e^{-\frac{(\Delta \omega \tau)^2}{2}} e^{-i(\kappa_1 \Delta \omega + \kappa_2(\Delta \omega)^2/2)z} e^{i\Delta \omega t} d \Delta \omega \right\} \nonumber \\
  & = \frac{ |E_{z0}| I_0(q_1^0 r) }{ \left( 1 + (\kappa_2 z / \tau^2)^2 \right)^{1/4} } e^{-\frac{(t - \kappa_1(z-z_i))^2}{2 \tau^2 ( 1 + (\kappa_2 z / \tau^2)^2 )}} e^{-\alpha(z - z_i)} \cos(\psi_T), \qquad z \geq z_s,
  \label{TM01ModePulse}
  \end{align}
  where $A_0$ is an arbitrary complex constant and its role is replaced in the second line of \eref{TM01ModePulse} by $|E_{z0}|$, which represents the amplitude of the $z$-directed field at $t = 0$ and $z = z_i = z_s$, with $z_i$ being the initial position of the pulse peak.
  The precise relationship between $|E_{z0}|$ and the total pulse energy is complicated and must be obtained by integrating over the Poynting vector in both core and cladding regions.
  $q_1^0$ is $q_1$ evaluated at $\omega_0$.
  $z_s$ is the position of the start of the waveguide, where pulse attenuation begins, and before which \eref{TM01ModePulse} does not apply.
  Note that setting $z_s \neq 0$ implies that some special pulse, not transform-limited, is being coupled into the waveguide.
  We set $z_s = 0$ for all simulations here.
  The carrier phase $\psi_T$ is given by
  \begin{equation}
  \psi_T = \omega_0 t - \kappa_0 z + \frac{\left( t - \kappa_1 ( z - z_i ) \right)^2 \kappa_2 z / \tau^2}{2 \tau^2 ( 1 + (\kappa_2 z / \tau^2)^2 )} - \atan \left( \frac{\kappa_2 z}{\tau^2} \right) + \psi_0,
  \label{TM01ModePulseCarrierPhase}
  \end{equation}
  where $\psi_0$ is a real phase constant.
  The corresponding $E_z$, $E_r$ and $H_{\phi}$ fields are approximated as
  \begin{equation}
  E_r^1(t,r,z) \approx - \frac{\kappa_0}{q_1^0} \frac{I_1(q_1^0 r)}{I_0(q_1^0 r)} E_z^1(t,z) \tan(\psi_T), \qquad H_{\phi}^1(t,r,z) \approx \frac{k_0 E_r^1(t,r,z)}{\kappa_0 \eta_0}
  \label{TM01ModePulseFields}
  \end{equation}
  where $k_0=\omega_0/c$.
  Essentially, \eref{TM01ModePulse} and \eref{TM01ModePulseFields} furnish an approximate analytical description of a TM\textsubscript{01} pulse moving with an approximate phase velocity and group velocity of $v_{ph} = \omega_0/\kappa_0$ and $v_g = 1/\kappa_1$ respectively in the vacuum core of a cylindrical waveguide.
  If $z_s = 0$, the pulse at the start of the waveguide ($z = z_s = 0$) is a transform-limited pulse with a peak longitudinal electrical amplitude of $|E_{z0}|$.
  The primary reason for introducing $z_i$ in our formulas is to control when the pulse arrives at the start of the waveguide without having to compromise the intuitive convention of having $t = 0$ as the initial time (when the simulation begins and the initial electron bunch starts evolving according to \eref{newtonLaw}).
  
  \subsection{Acceleration of 1.6\,pC Electron Bunches}
  
  \subsubsection{Optimization procedure and acceleration results}
  
  Here, we optimize the dielectric-loaded metal waveguide for electron bunch acceleration and perform a rudimentary thermal damage and dielectric breakdown analysis to verify the realism of the scheme.
  We numerically demonstrate the acceleration of a 1.6\,pC electron bunch from a kinetic energy of 1\,MeV to one of 10\,MeV, using a 20\,mJ 10-cycle pulse centered at 0.6\,THz.
  Note that for a 10-cycle pulse, $\Delta \omega_\mathrm{FWHM}/\omega_0 = 4 \ln2 / (\omega_0 \tau_\mathrm{FWHM}) = 4 \ln2/ (2 \pi 10) = 4.41\%$.
  As will be seen in the results, some longitudinal compression is also inadvertently achieved in the process.
  
  Optimizing the dielectric-loaded metallic waveguide for bunch acceleration involves adjusting a large number of parameters, including operating frequency, choice of waveguide mode, waveguide dimensions, THz pulse energy and pulse duration, the type of dielectric, the type of external conductor and initial electron bunch properties.
  To make this optimization tractable, we fix all parameters in advance based on the available technology except for three degrees of freedom: (\emph{i}) the carrier-envelope phase $\psi_0$, (\emph{ii}) the initial position of the pulse $z_i$ (with initial position of electron fixed at $z = 0$, $z_i$ represents the injection time of the electrons into the pulse), and (\emph{iii}) the radius of the vacuum core $r_1$.
  In particular, we fix the phase velocity at $v_{ph}=c$ and the center frequency at $f_0 = 0.6\,$THz, which limits the dielectric thickness $d$ to specific values depending on $r_1$.
  However, because acceleration results can be very sensitive to small variations in the value of $v_{ph}$, we take the liberty of treating $v_{ph}$ as an optimization parameter (but ensuring that $v_{ph}\approx c$) after using $v_{ph}=c$ to determine properties of the TM\textsubscript{01} waveguide mode.
  Therefore, four degrees of freedom are ultimately considered.
  In practice, after the waveguide has been fabricated according to the optimal specifications, the operating frequency should be perturbed to vary the phase velocity until maximum electron acceleration is achieved.
  As long as the perturbation is small, the waveguide properties should be very close to those determined for $v_{ph}=c$ and $f_0=0.6\,$THz.
  The electron acceleration process is much more sensitive to small variations in $v_{ph}$ than to small variations in any other parameter caused by perturbing the operating frequency alone.
  
\begin{figure}
  	\centering
  	$\begin{array}{cc}
  	\includegraphics[draft=false,width=3.0in]{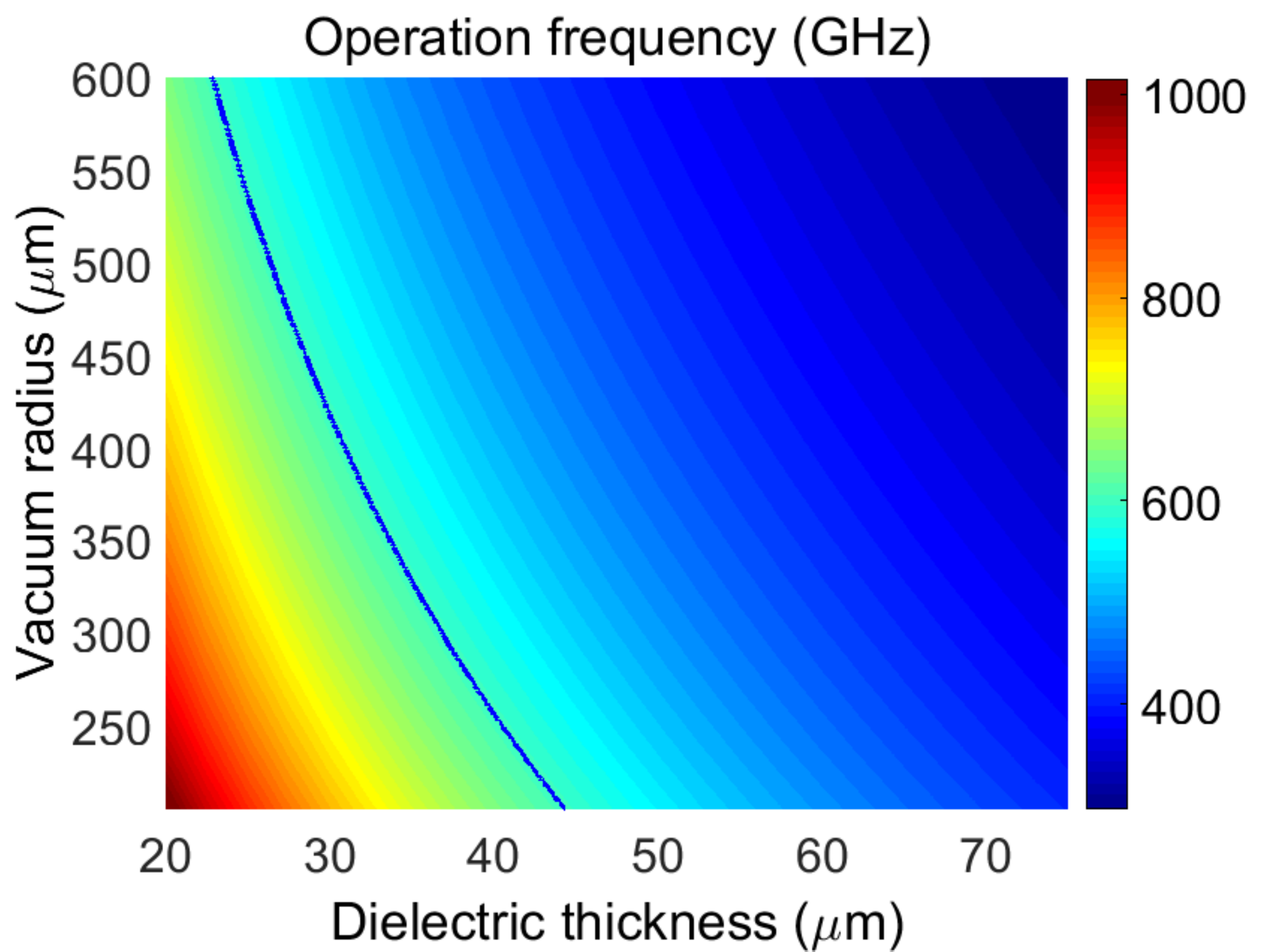} &
  	\includegraphics[draft=false,width=3.0in]{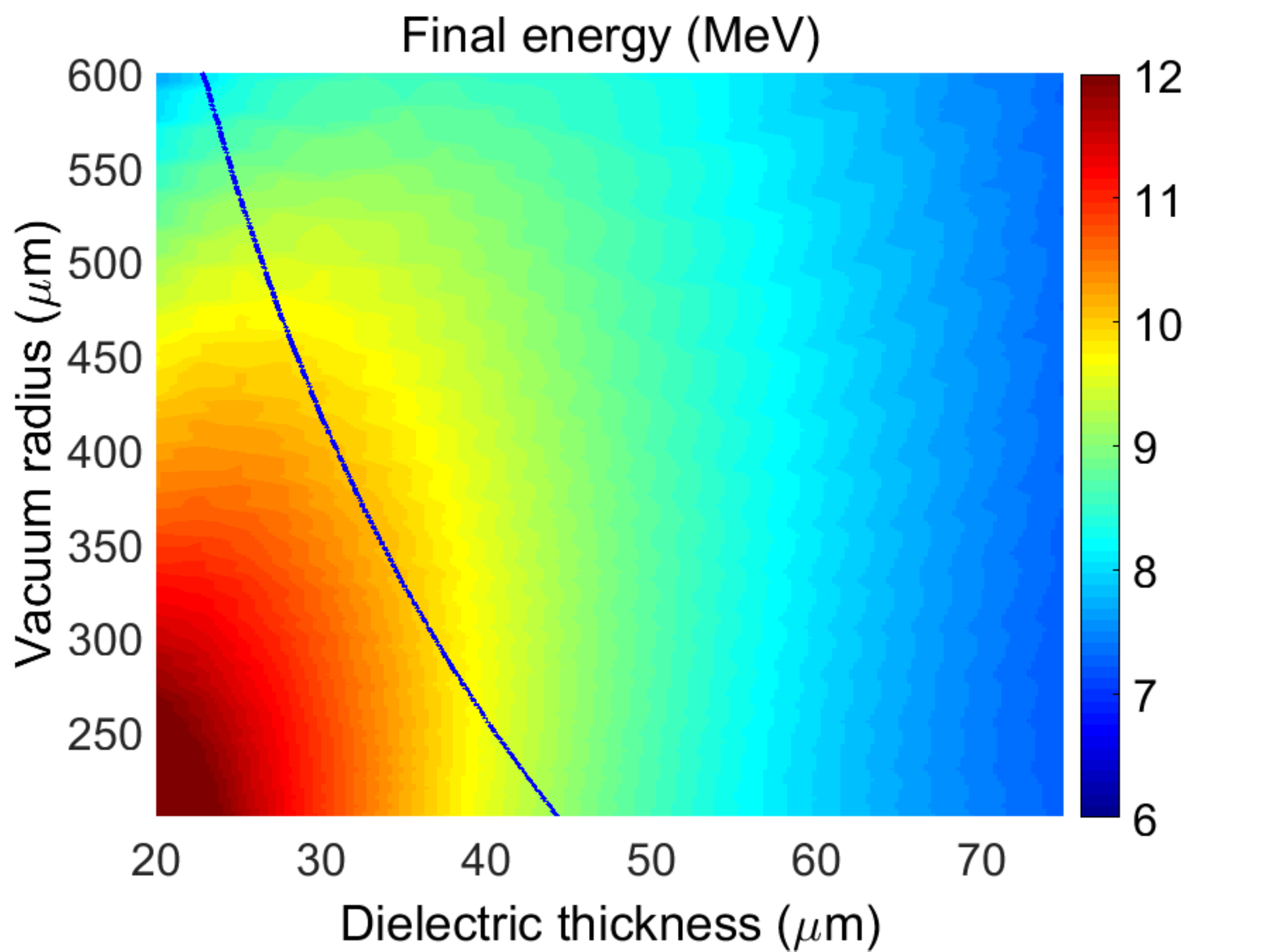} \\
  	(a) & (b) \\
  	\includegraphics[draft=false,width=3.0in]{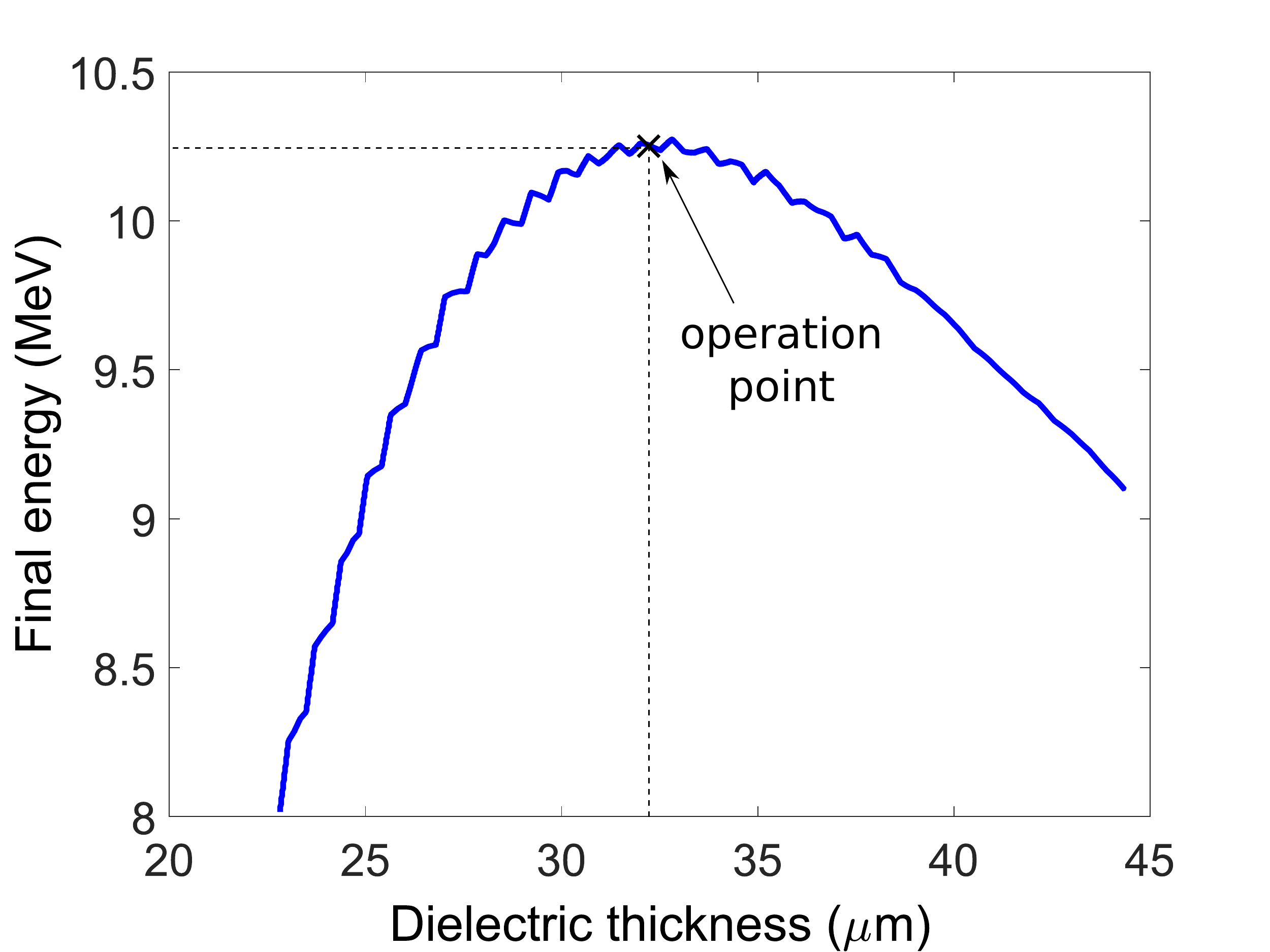} &
  	\includegraphics[draft=false,width=3.0in]{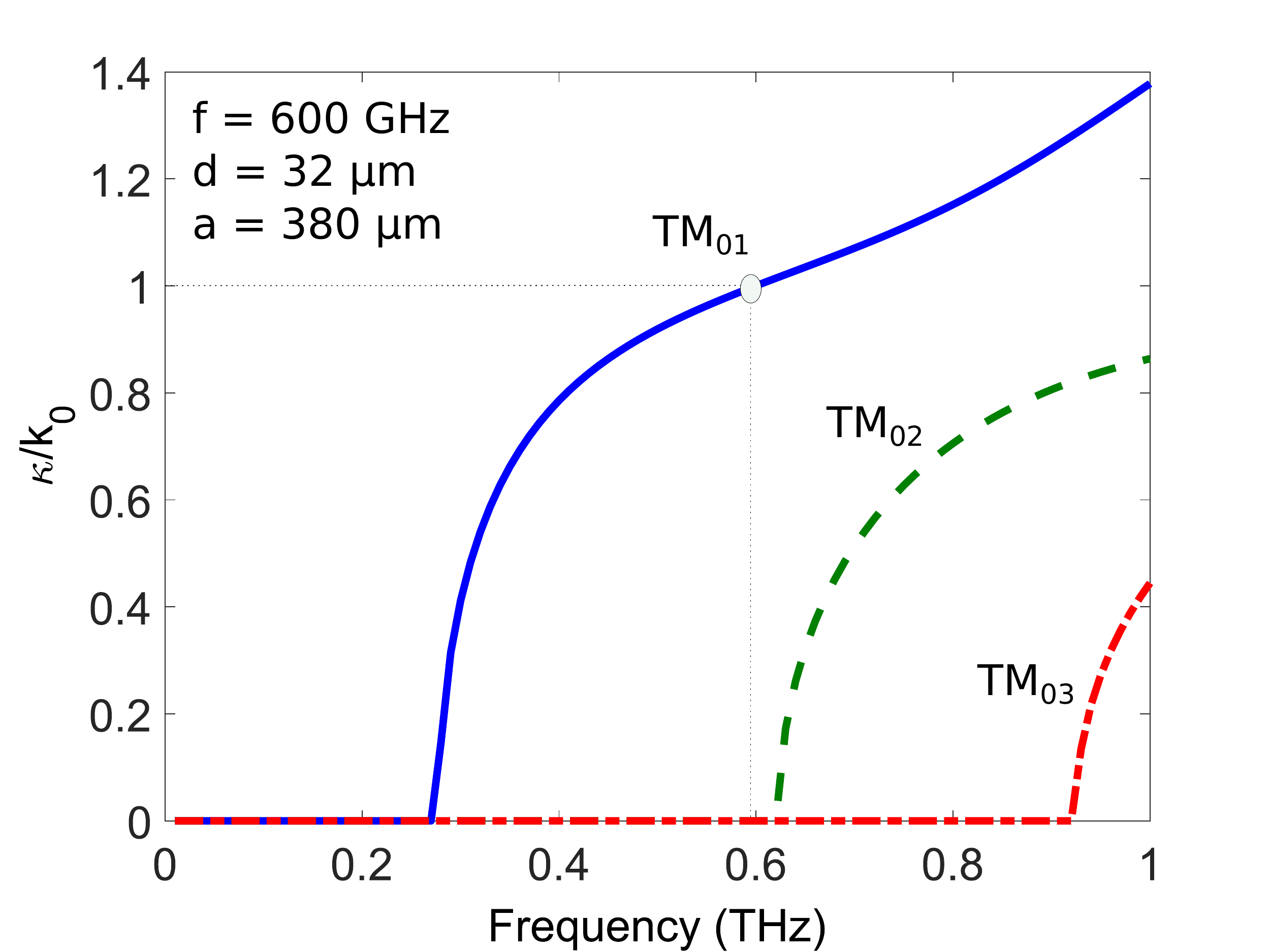} \\
  	(c) & (d) \\
  	\end{array}$
  	\caption{Determination of the optimal waveguide for electron acceleration: (a) Color map of operation frequency as a function of core radius $r_1$ and dielectric thickness d, (b) Color map of final kinetic energy of a single electron of initial kinetic energy 1\,MeV, optimized over $\psi_0$ and $z_i$, as a function of $r_1$ and $d$. The blue line in (a) and (b) correspond to an operation frequency of 0.6\,THz. The value of the color map in (b) along the 0.6\,THz operation line is plotted in (c), where the optimum value of $d$ is identified as $d = 32$\,{\textmu}m. (d) The dispersion curves corresponding to the final waveguide design.}
  	\label{waveguideDesign}
  \end{figure}

  Fig.\,\ref{waveguideDesign}a shows a color map of the operation frequency as a function of $r_1$ and $d$.
  As noted before, we define the operation frequency as the frequency of the TM\textsubscript{01} mode in the waveguide corresponding to $v_{ph}=c$.
  Fig.\,\ref{waveguideDesign}b shows a color map of the final electron kinetic energy for a single electron of initial kinetic energy 1\,MeV, optimized over $\psi_0$, $z_i$ and $v_{ph}$ (ensuring that $v_{ph}\approx c$), as a function of $r_1$ and $d$.
  We see that greater electron acceleration is generally achieved at higher operation frequencies.
  However, choosing a very small wavelength makes it challenging to accelerate a large number of electrons due to smaller waveguide dimensions.
  As pointed out previously, the emergence of promising techniques to generate radiation in the vicinity of $0.6\,$THz \cite{Fulop2010} encourages us to make that choice of frequency, which has been marked out by the black contour line in Fig.\,\ref{waveguideDesign}a.
  The same line is drawn in Fig.\,\ref{waveguideDesign}b, and the optimized final kinetic energy, read along that line, is reproduced in Fig.\,\ref{waveguideDesign}c, where an optimal choice of $d = 32$\,{\textmu}m, corresponding to a vacuum core radius of $r_1 = 380$\,{\textmu}m, is evident.
  In Fig.\,\ref{waveguideDesign}d, we plot the dispersion curves corresponding to the waveguide with $d = 32$\,{\textmu}m, $r_1 = 380$\,{\textmu}m, to show that at the operating frequency, the TM\textsubscript{01} dispersion curve of our waveguide design is sufficiently linear within the 4.41\% intensity FWHM spectral bandwidth.
  Hence, the electromagnetic fields are well approximated using \eref{TM01ModePulseFields} and \eref{TM01ModePulse}.
  
  The parameters of the final waveguide design are $d = 32$\,{\textmu}m, $r_1 = 380$\,{\textmu}m, $v_{ph} = 0.99c$, $v_g = 0.7c$, $\alpha= 5.21\,$1/m, $\tau_{\mathrm{FWHM}} = 16.7$\,ps, $\kappa_2 = 4.54\times10^{-22}$\,s$^2$/m.
  The 20\,mJ pulse yields a $|E_{z0}|$ of about 0.9\,GV/m.
  The initial parameters of the 1.6\,pC, 1\,MeV electron bunch with which we will demonstrate the acceleration are $\sigma_x = \sigma_y = \sigma_z = 30$\,{\textmu}m (a 100\,fs bunch), $\sigma_{\gamma \beta_x} = \sigma_{\gamma \beta_y} = \sigma_{\gamma \beta_z} = 0.006$, where $\sigma_{\gamma \beta_x}$, for instance, denotes the standard deviation of $\gamma \beta_x$.
  Producing a 1.6\,pC, 100\,fs electron bunch would be a challenge for typical RF guns, but strides are being made to realize a photocathode RF gun capable of delivering the bunch we have assumed as our input \cite{yang2012femtosecond}.
  Although a thorough examination of how performance is impacted by variations in the initial electron bunch properties is beyond the scope of this study, we expect the results to deteriorate with a larger initial energy spread.
  10'000 macro-particles, Gaussian-distributed in every dimension of phase space, were employed in the simulation.
  
\begin{figure}
  	\centering
  	$\begin{array}{ccc}
  	\includegraphics[draft=false,height=1.5in]{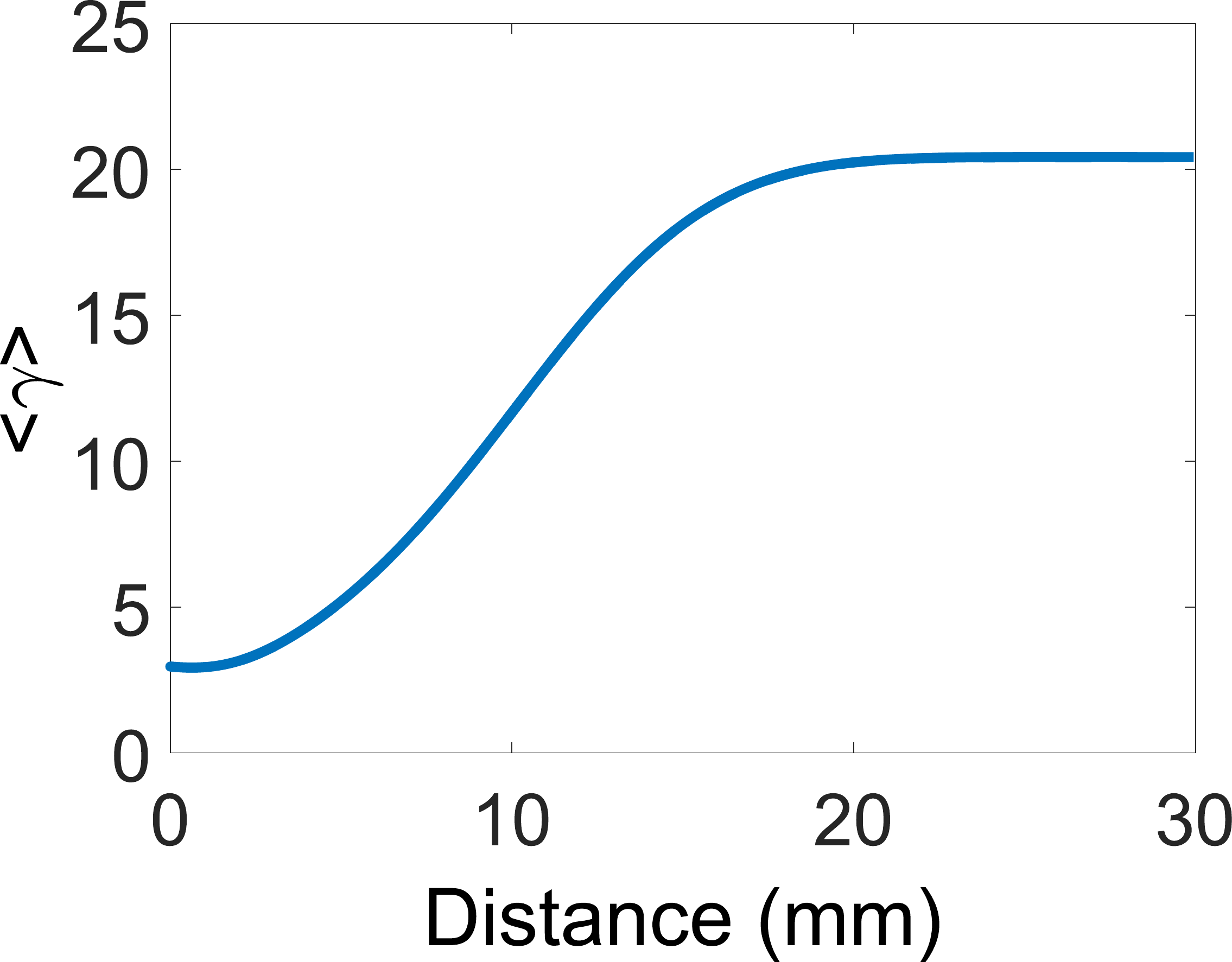} &
  	\includegraphics[draft=false,height=1.5in]{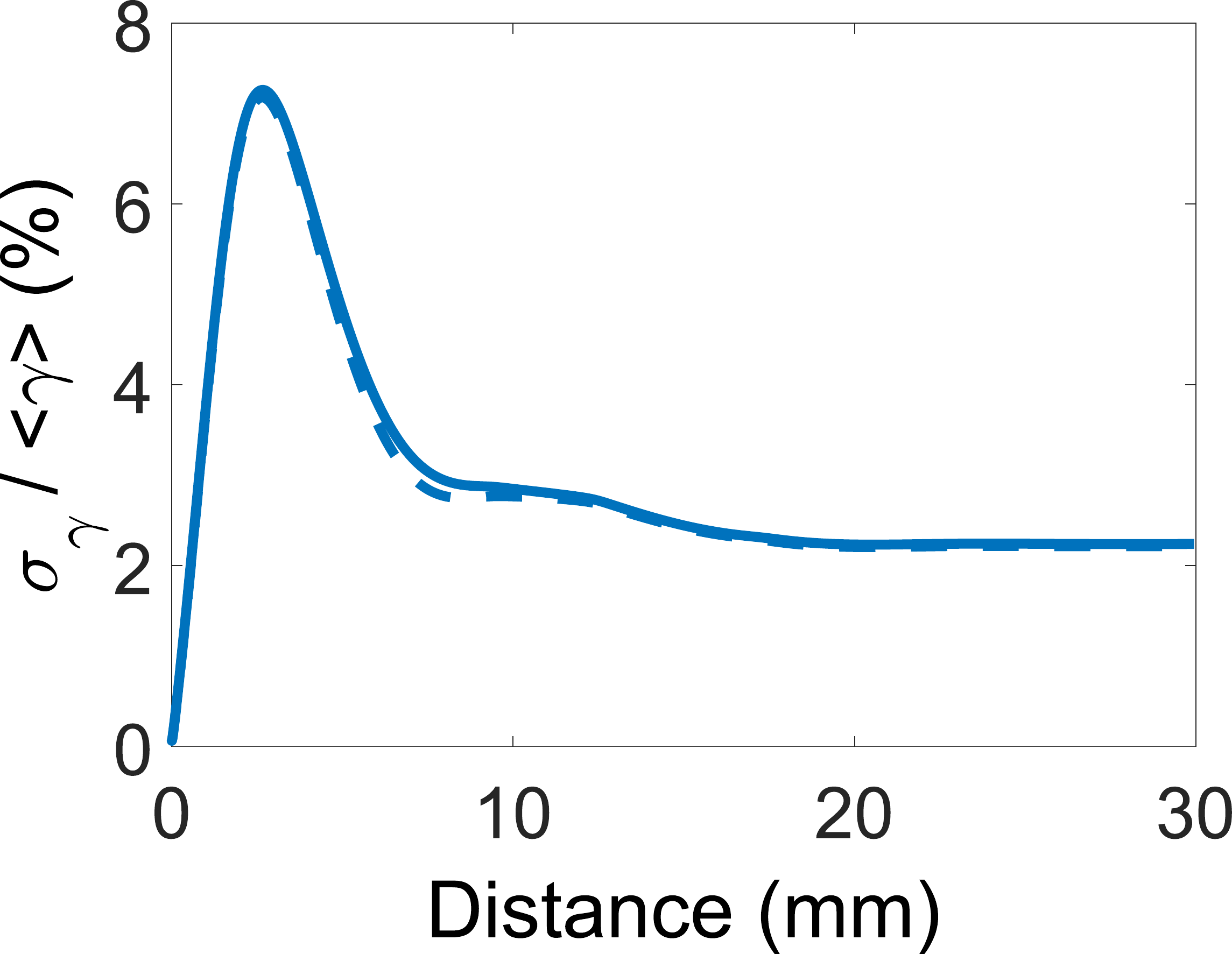} &
  	\includegraphics[draft=false,height=1.5in]{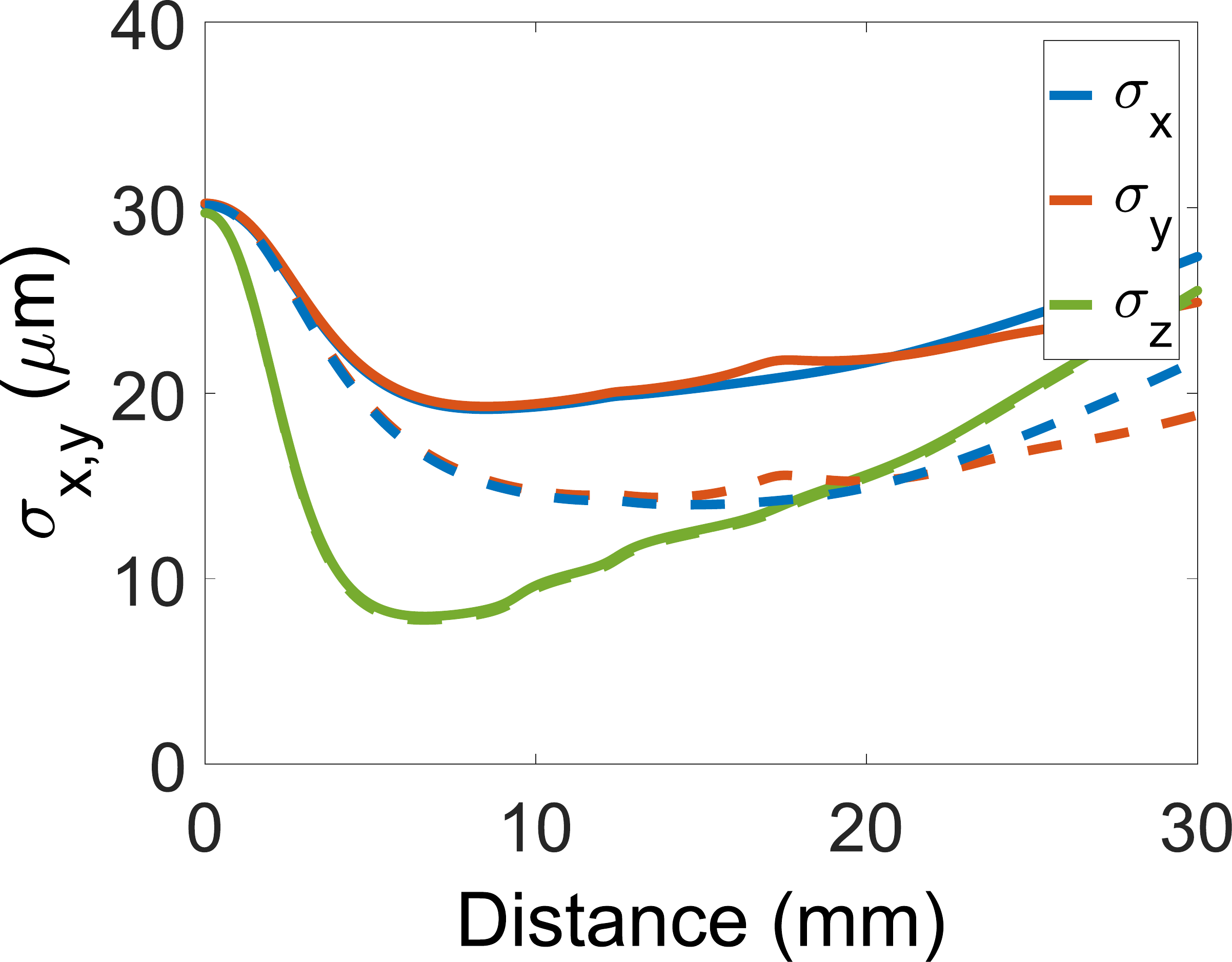} \\
  	(a) & (b) & (c) \\
  	\end{array}$
  	\caption{Evolution of bunch parameters with mean bunch position for acceleration of a 1.6\,pC electron bunch from 1\,MeV to 10\,MeV (kinetic energy) in about 20\,mm: (a) normalized mean energy, (b) relative energy spread, (c) transverse and longitudinal spread. The symbol $\sigma$ stands for the standard deviation of the variable in the subscript. Solid and dashed lines correspond to simulations with and without space charge, respectively. 10'000 macro-particles are used for the simulations. $\psi_0=1.34\pi$ and $k_0z_i = 10.96\pi$. A 20\,mJ, 10-cycle (16.7\,ps), 0.6\,THz-centered pulse is considered.}
  	\label{bunchEvolution1p6}
  \end{figure}

  Fig.\,\ref{bunchEvolution1p6} shows the evolution of bunch parameters as a function of mean particle position.
  We see from Fig.\,\ref{bunchEvolution1p6}a that the 1.6\,pC-bunch is accelerated from 1\,MeV to 10\,MeV of kinetic energy in about 20\,mm, without any of its other properties deteriorating prohibitively.
  The corresponding average accelerating gradient is about 450\,MeV/m.
  Note from Figs.\,\ref{bunchEvolution1p6}b-d that, depending on the extraction point, the final bunch can possess a smaller transverse and longitudinal spread compared to the initial distribution, but the final energy spread is degraded from the initial spread.
  
  \subsubsection{Injection point considerations}
  
\begin{figure}
  	\centering
  	$\begin{array}{cc}
  	\includegraphics[draft=false,width=3.0in]{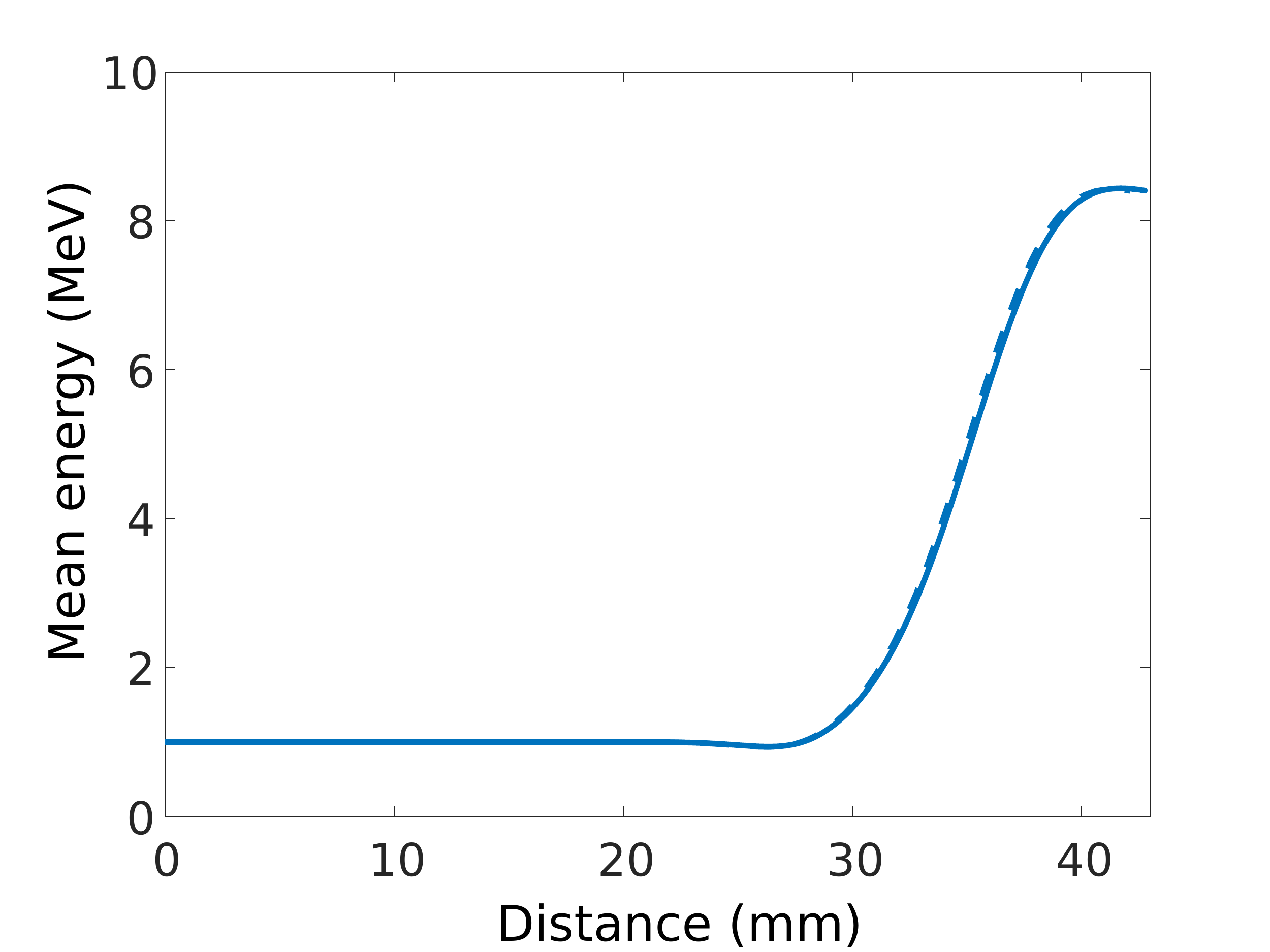} &
  	\includegraphics[draft=false,width=3.0in]{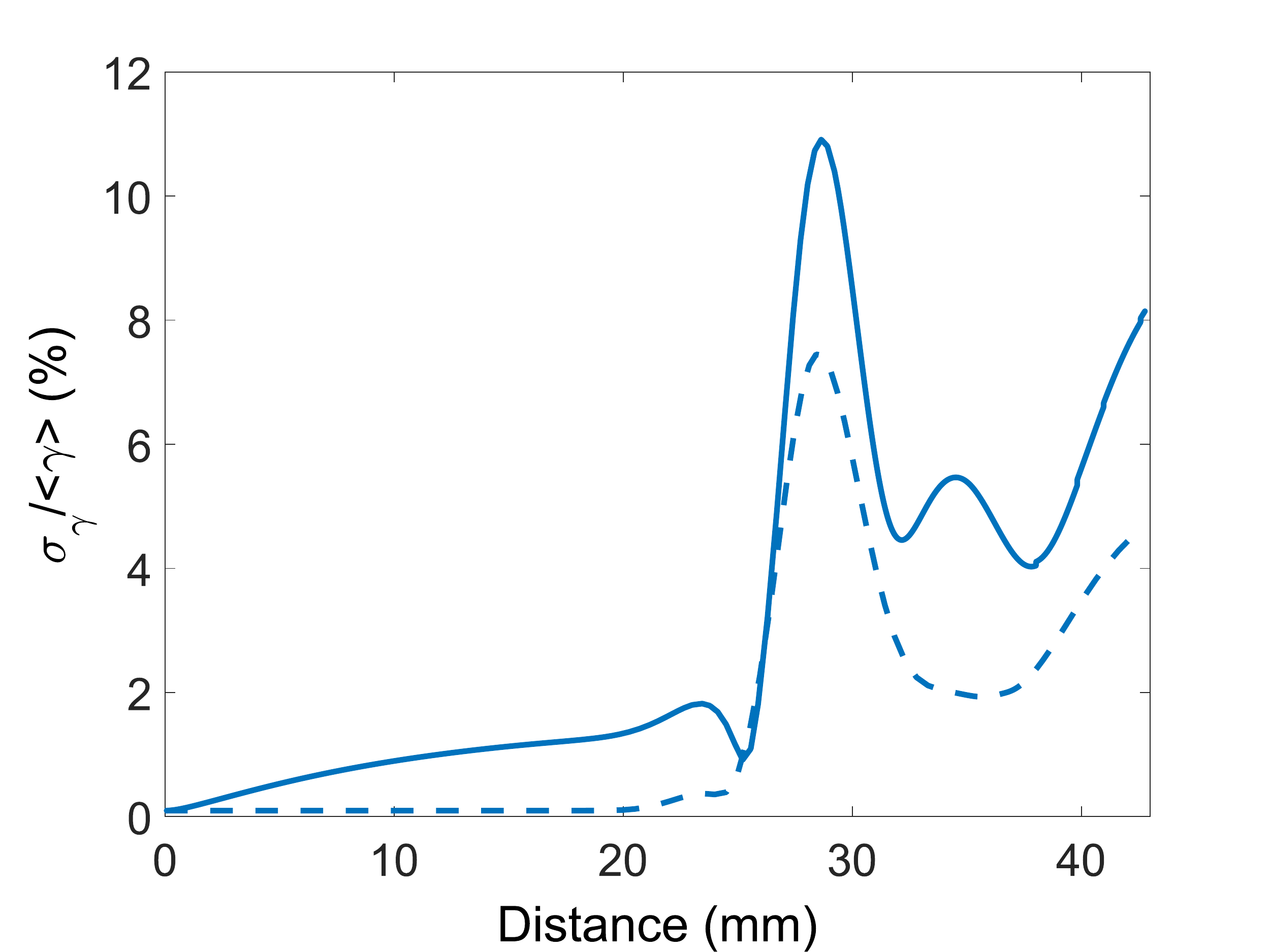} \\
  	(a) & (b)
  	\end{array}$
  	\caption{Evolution of bunch parameters with mean bunch position for acceleration of a 1.6\,pC electron injected at a distant point from the THz pulse peak: (a) normalized mean energy, (b) relative energy spread. Solid and dashed lines correspond to simulations with and without space charge, respectively. 10'000 macro-particles were used for the simulations.}
  	\label{injectionPoint}
  \end{figure}

  In our analysis, we have assumed the freedom to inject the electron bunch into any point of the electromagnetic field.
  According to our computations, the optimum injection point for the electron bunch is a point within the pulse (albeit in its tail).
  This may be challenging to realize if both the electron bunch and the electromagnetic pulse enter the waveguide from vacuum.
  The objective of this section is to consider injection of the electron bunch at a point with negligible electric field values and assess the amount by which our predictions would change.
  The optimum THz waveguide for this case is a waveguide with $r_1 = 338$\,{\textmu}m and $d = 33$\,{\textmu}m.
  In addition, $v_{ph} = 0.981c$, $\psi_0=1.49\pi$, $k_0z_i = 137.73$.
  We ensure that the electric field's amplitude at the injection point is negligible by making the amplitude $7.4\times10^{-15}|E_{z0}|$.
  The evolution of the electron bunch is shown in Fig.\,\ref{injectionPoint}, where we observe a final kinetic energy of 8.4\,MeV (instead of the 9\,MeV observed before).
  The smaller energy gain in this case is partly due to the dispersion and attenuation that the pulse suffers from before the injected bunch begins interacting with the pulse.
  A final energy close to what is predicted in the previous section should therefore be achievable if the electron bunch and THz pulse can interact before the pulse has travelled too far along the waveguide.
  
  \subsubsection{Thermal damage and dielectric breakdown considerations}
  
  Here, we assess the feasibility of the acceleration scheme in terms of its thermal damage and dielectric breakdown prospects.
  One concern is that the high energy injected into the waveguide and consequent energy dissipation would raise temperature of the copper coating beyond its melting point.
  Another concern is dielectric breakdown due to the high electric field values in the dielectric.
  
\begin{figure}
  	\centering
  	$\begin{array}{cc}
  	\includegraphics[draft=false,width=3.0in]{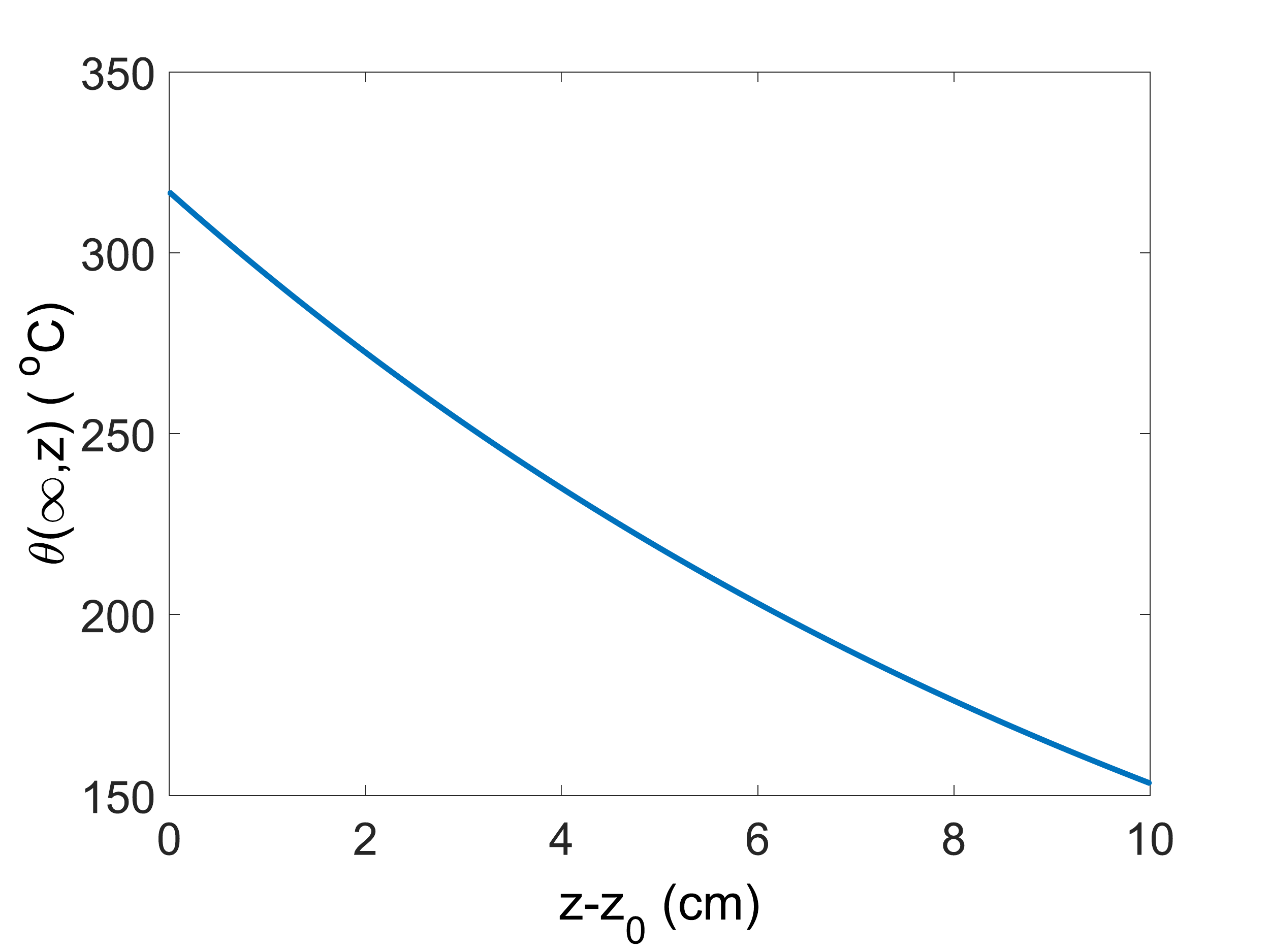} &
  	\includegraphics[draft=false,width=3.0in]{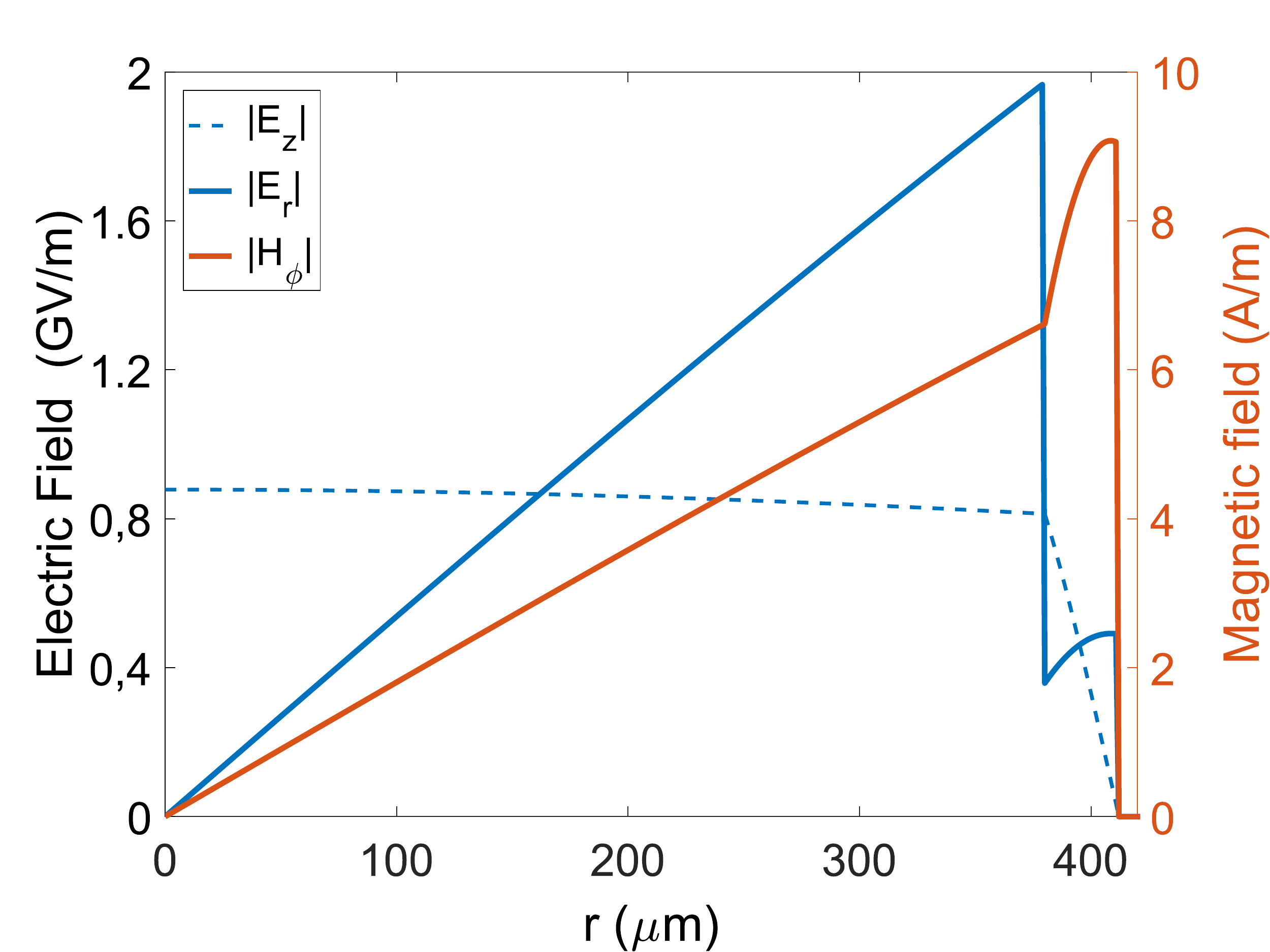} \\
  	(a) & (b)
  	\end{array}$
  	\caption{Plots used to assess thermal damage and dielectric breakdown prospects for the acceleration scheme. (a) Final temperature of copper cladding assuming initial temperature of 27\,$^\circ$C ($z = 0$ is the start of the waveguide), and (b) Field profile of the TM\textsubscript{01} mode in the transverse direction of the cylindrical waveguide under study. Discontinuities occur at material boundaries (once at the vacuum-diamond interface and again at the diamond-copper interface).}
  	\label{thermalStudy}
  \end{figure}

  The energy $dG$ transferred to a differential segment of copper at position $z$ ($z = 0$ being the start of the waveguide) is related to the associated temperature rise $\Delta \theta = \theta - \theta_0$ as
  \begin{equation}
  dG = dm_\mathrm{Cu}C\Delta \theta,
  \label{thermalStudy1}
  \end{equation}
  The differential mass $dm_\mathrm{Cu} = \rho_{Cu}(2\pi r_2 \delta_s)dz$, where $\rho_{Cu}$ is the density of copper and $\delta_s$ is the skin depth.
  $\theta_0$ is the original temperature of the copper and $C$ its specific heat capacity.
  Ignoring dispersion for simplicity (and because it is negligibly small here), we write the power propagating down the waveguide, averaged over the rapid carrier fluctuations, as
  \begin{equation}
  P(t,z_0) \approx P_0 e^{-\frac{(t-\kappa_1(z-z_i))^2}{\tau^2}} e^{-2\alpha z}, \qquad z \geq 0,
  \label{thermalStudy2}
  \end{equation}
  where $P_0$ is the average power that flows into the start of the waveguide when the pulse peak arrives there.
  Noting that $P = -dG/dt$ and that partial derivatives are relevant here because $z$ and $t$ are independent coordinates, \eref{thermalStudy1} can be written as
  \begin{equation}
  -\frac{\partial P(t,z)}{\partial z} \approx \rho_\mathrm{Cu} 2 \pi r_2 \delta_s C \frac{\partial \theta (t,z)}{\partial t}.
  \label{thermalStudy3}
  \end{equation}
  Solving \eref{thermalStudy1} for $\theta$ gives
  \begin{equation}
  \theta (t,z) \approx \theta_0 - \frac{P_0 e^{-2\alpha z}}{\rho_\mathrm{Cu} \pi r_2 \delta_s C} \int\limits_{-\infty}^t \left( \frac{\kappa_1 t}{\tau^2}-\alpha \right) e^{-\frac{t^2}{\tau^2}} dt.
  \label{thermalStudy4}
  \end{equation}
  $\theta(\infty,z)-\theta_0$ gives the net temperature rise after the pulse has passed entirely through point $z$:
  \begin{equation}
  \theta(\infty,z)-\theta_0 \approx \frac{P_0 \alpha \tau}{\rho_{cu} \sqrt{\pi} r_2 \delta_s C} e^{-2\alpha z}.
  \label{thermalStudy5}
  \end{equation}
  $\theta(\infty,z)$ is plotted in Fig.\,\ref{thermalStudy}a for $\theta_0=27\,^{\circ}$C.
  The relevant parameters for copper at 0.6\,THz are $\rho_\mathrm{Cu} = 8940\,$kg/m$^3$, $C = 385$\,J/kg/$^{\circ}$C and $\delta_s = 0.084$\,{\textmu}m.
  
  From the values in Fig.\,\ref{thermalStudy}a, the fact that the melting point of copper is 1084\,$^{\circ}$C and also that we have even neglected the conductivity of copper, we can conclude that the metal coating in the designed waveguide withstands the passage of the pulse without melting.
  
  Fig.\,\ref{thermalStudy}b shows a typical profile of the electromagnetic amplitude of a mode in the transverse direction of the waveguide.
  The breakdown electric field for diamond has been reported as 10-20\,MV/cm, depending on impurities.
  Reading off the plot we note that the maximum value of the electric field in the dielectric region is about 8\,MV/cm.
  This is close to the breakdown limit though still under it, showing that it would not be feasible to enhance the performance of our design by increasing the peak power of the accelerating pulse.
  Since we are relatively far from the melting point, an increase in available pulse energy should be used to increase pulse duration instead of peak power.
  
  \subsection{Acceleration of 16\,pC and 160\,pC Electron Bunches}
  
\begin{figure}
  	\centering
  	$\begin{array}{cc}
  	\includegraphics[draft=false,width=3.0in]{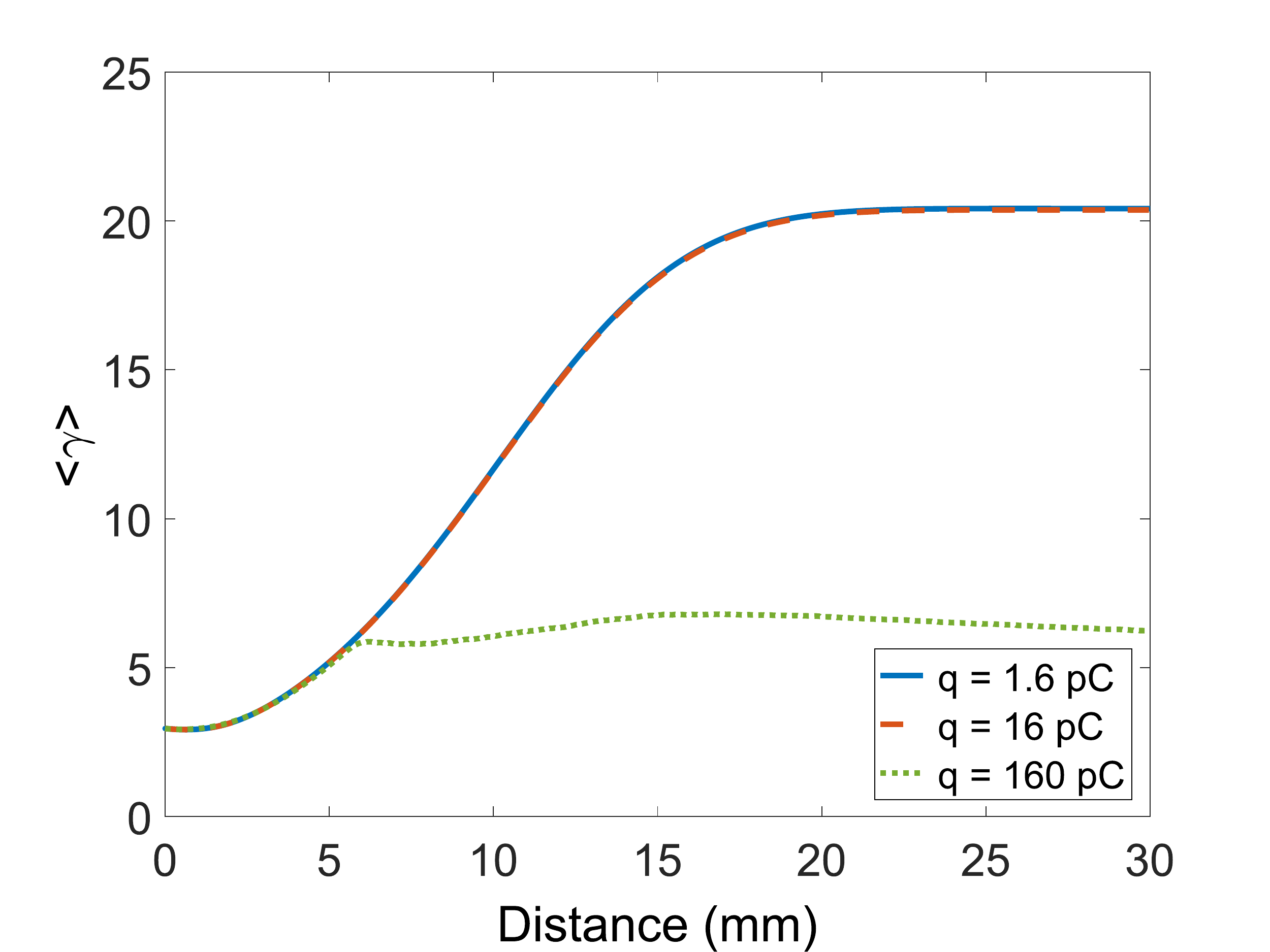} &
  	\includegraphics[draft=false,width=3.0in]{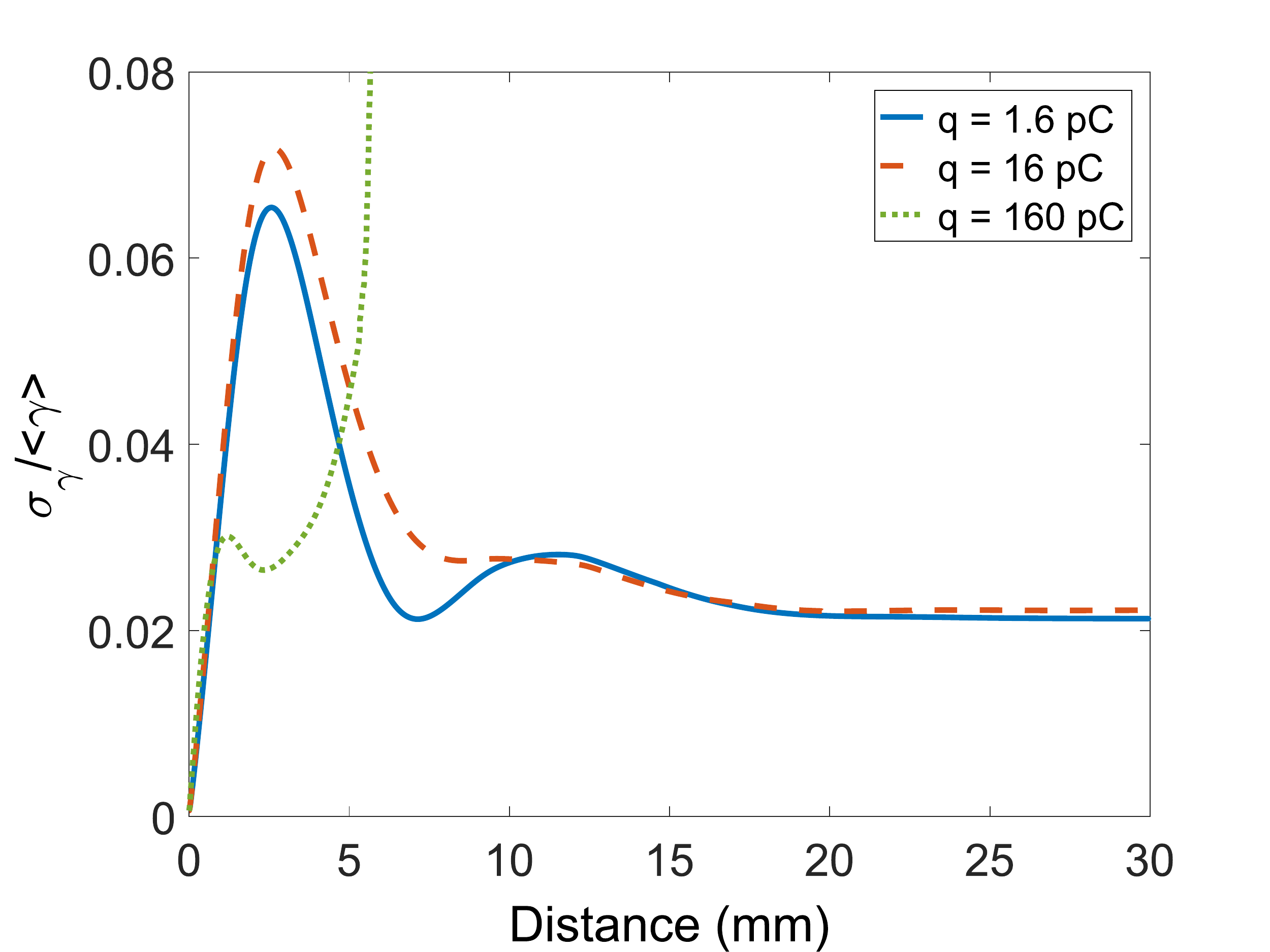} \\
  	(a) & (b) \\
  	\includegraphics[draft=false,width=3.0in]{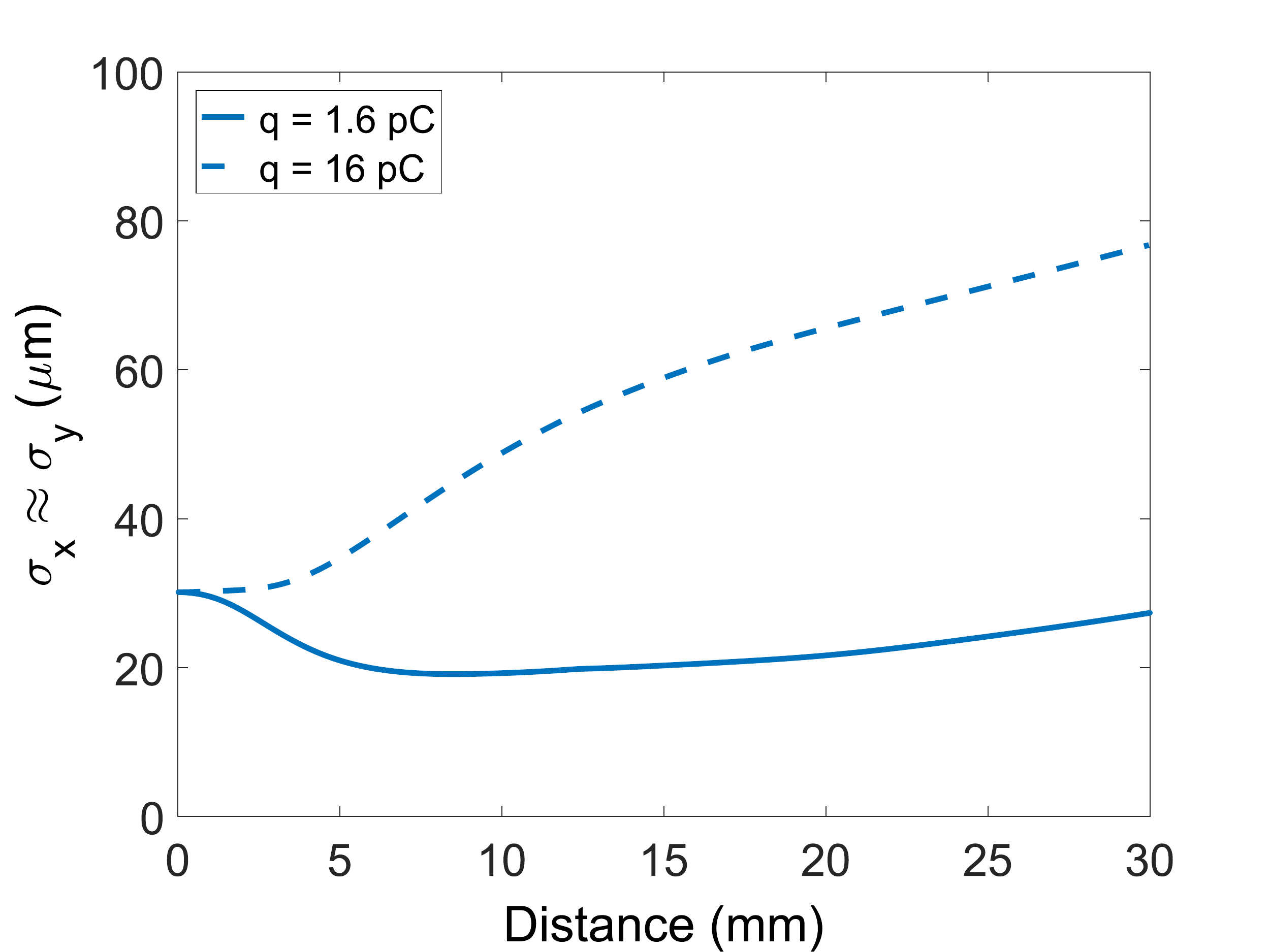} &
  	\includegraphics[draft=false,width=3.0in]{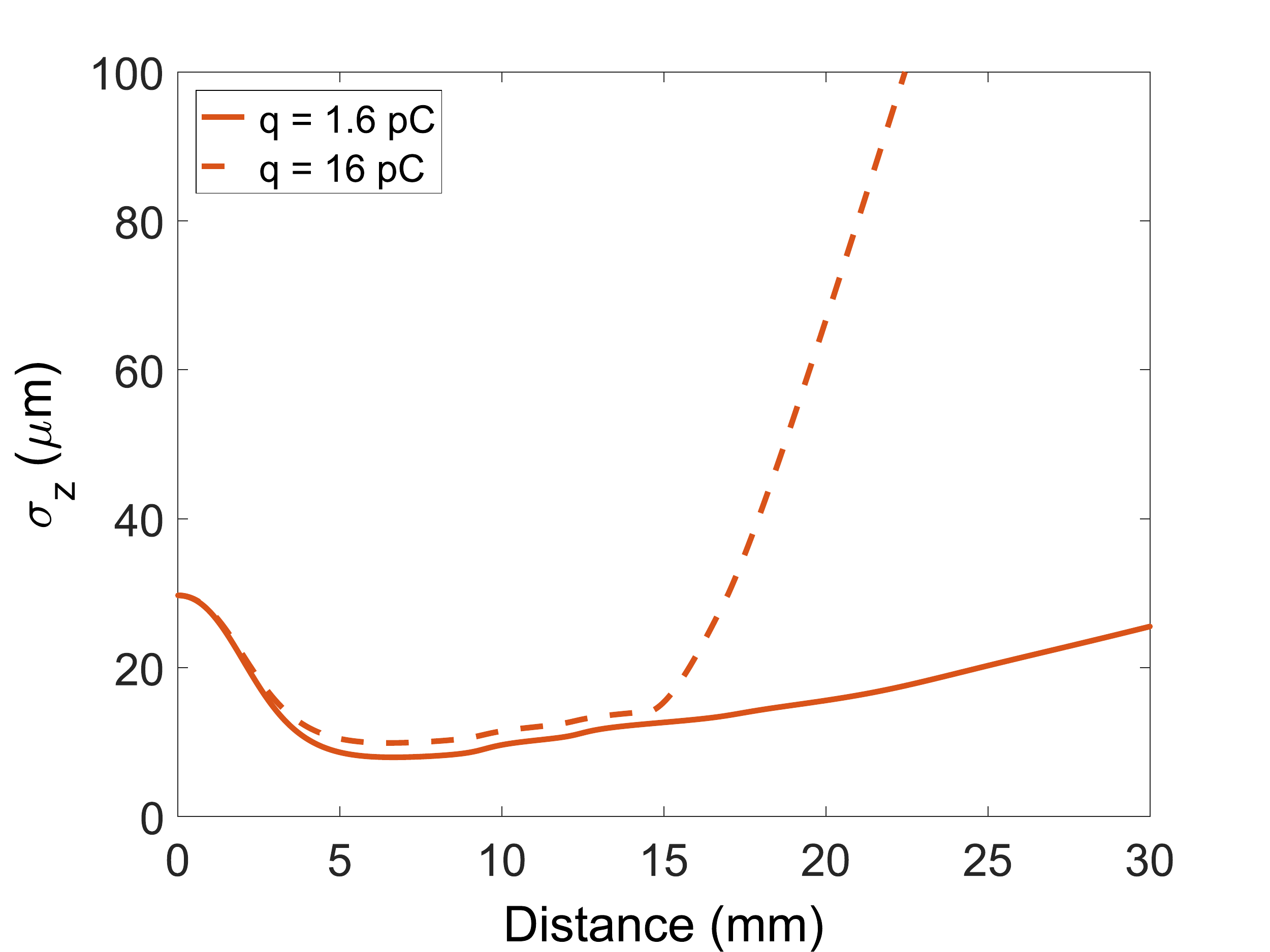} \\
  	(c) & (d) \\
  	\end{array}$
  	\caption{Evolution of bunch parameters with mean bunch position for optimized acceleration of 1.6\,pC, 16\,pC and 160\,pC electron bunches: (a) normalized mean energy, (b) relative energy spread, (c) transverse spread and (d) longitudinal spread. Acceleration of a 160\,pC bunch is not feasible and is shown only in (a) and (b). All results include space charge. 10'000 macro-particles are used for all simulations. $\psi_0=1.34\pi$ and $k_0z_i$ = 10.96$\pi$. A 20\,mJ, 10-cycle (16.7\,ps), 0.6\,THz-centered pulse is used in all cases.}
  	\label{bunchEvolution16}
  \end{figure}

  In this section, we explore the acceleration of electron bunches of greater charge.
  We see that it is feasible to use the dielectric-loaded metallic waveguide to accelerate electron bunches as large as 16\,pC, but that this is not possible when the charge increases to 160\,pC.
  All other bunch properties (including an initial kinetic energy of 1\,MeV) remain the same as the previous study.
  We use a 20\,mJ, 10-cycle, 0.6\,THz-centered pulse, and the same optimized waveguide and injection conditions as before.
  Fig.\,\ref{bunchEvolution16} shows the evolution of the electron bunch for 1.6\,pC, 16\,pC and 160\,pC-bunches.
  The effects of space charge are included in all computations.
  
  From Fig.\,\ref{bunchEvolution16}a and b, we observe that there is little difference in the mean kinetic energy and energy spread evolution of a 16\,pC-bunch and a 1.6\,pC-bunch.
  The energy spread of a 160\,pC-bunch, however, deteriorates prohibitively and the bunch is not significantly accelerated.
  Since this rules out the feasibility of accelerating a 160\,pC-bunch, we have omitted its plots from Fig.\,\ref{bunchEvolution16}c and \ref{bunchEvolution16}d.
  The inability of the waveguide to accelerate a 160\,pC-bunch is due to the overriding strength of the Coulomb repulsion, driving the electrons into the walls of the waveguide before significant acceleration takes place.
  Fig.\,\ref{bunchEvolution16}c explains how the 1.6\,pC and 16\,pC-bunches are able to have such similar energy and energy spread profiles during the acceleration: the greater Coulomb repulsion in the 16\,pC Coulomb is counter-balanced by larger transverse inter-particle spacing.
  Fig.\,\ref{bunchEvolution16}d shows that due to the larger amount of space charge, the 16\,pC expands rather rapidly compared to the 1.6\,pC-bunch after the pulse has slipped behind the bunch, so a 16\,pC-bunch accelerated via this scheme is likely to be useful for a shorter duration after being fully accelerated.
  
  \subsection{Concurrent Phase-limited Compression and Acceleration of 1.6\,pC Bunches}
  
  In this section, we optimize our waveguide design for simultaneous acceleration and bunch compression.
  We demonstrate phase-limited (longitudinal) bunch compression of 50 and 62 times for electron bunches of initial kinetic energy 1\,MeV and 10\,MeV, respectively.
  By \emph{phase-limited} we mean that the maximum compression results do not change substantially when space charge is removed from the simulations.
  
  As in previous sections, we use a 20\,mJ, 0.6\,THz-centered pulse.
  For each case (the 1\,MeV case and the 10\,MeV case), the waveguide and injection conditions are optimized in exactly the same manner as in acceleration scenario, except that in addition to $\psi_0$, $z_i$, $r_1$ and $v_{ph}$, we also optimize over pulse duration $\tau_\mathrm{FWHM}$ (keeping total energy constant at 20\,mJ), for a total of five optimization parameters.
  The initial conditions of the electron bunch, unless otherwise specified, are the same as before.
  
  To optimize for simultaneous acceleration and compression, the figure-of-merit found to be most useful is the ratio of energy to bunch-length of the electron bunch.
  Unlike the acceleration scheme, where we optimized using a single particle, here we optimized using 100 macro-particles and included the effects of space charge.
  The optimized results are then verified with simulations that use 10'000 macro-particles.
  
\begin{figure}[t]
  	\centering
  	$\begin{array}{ccc}
  	\includegraphics[draft=false,width=2.0in]{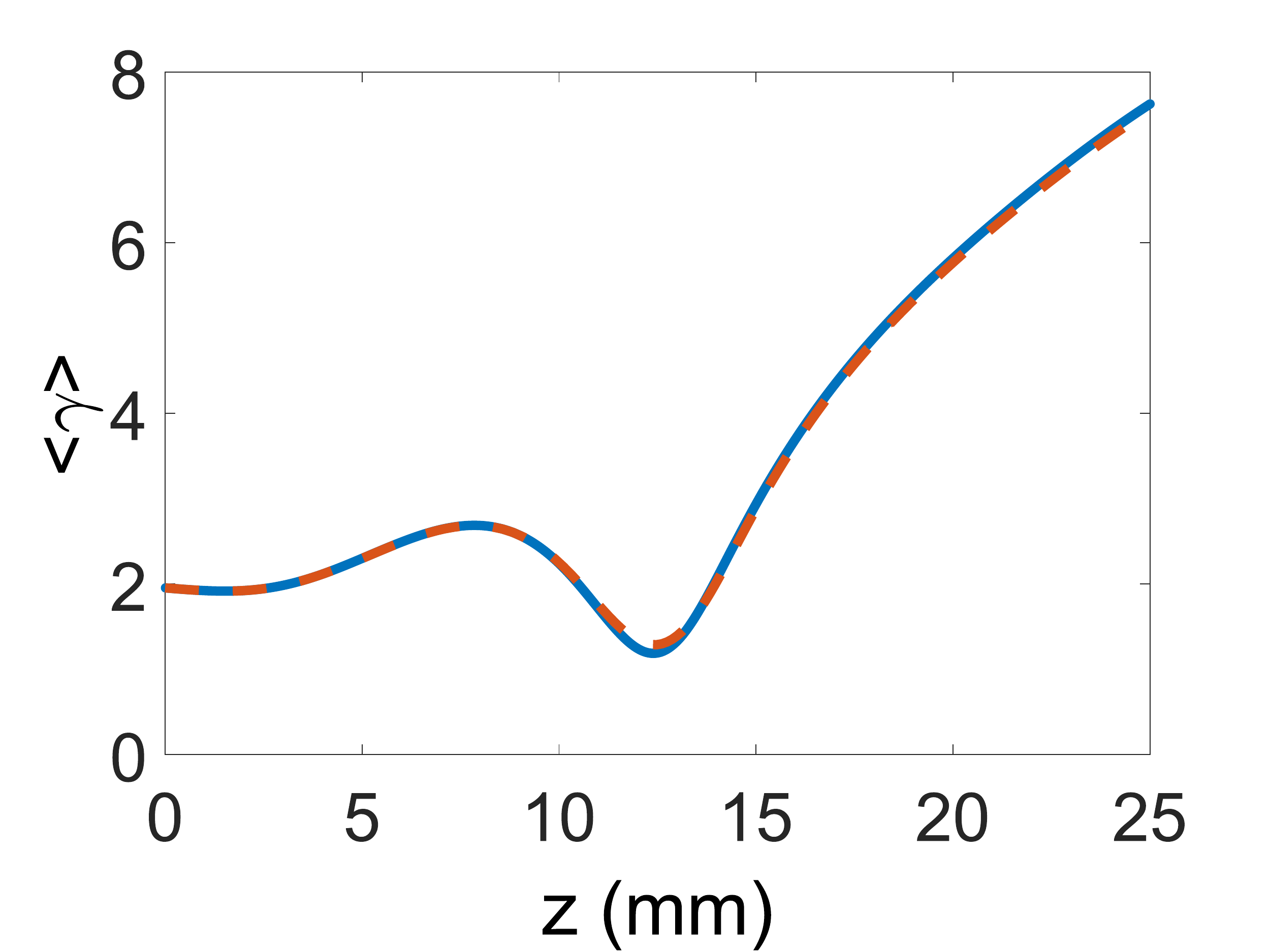} &
  	\includegraphics[draft=false,width=2.0in]{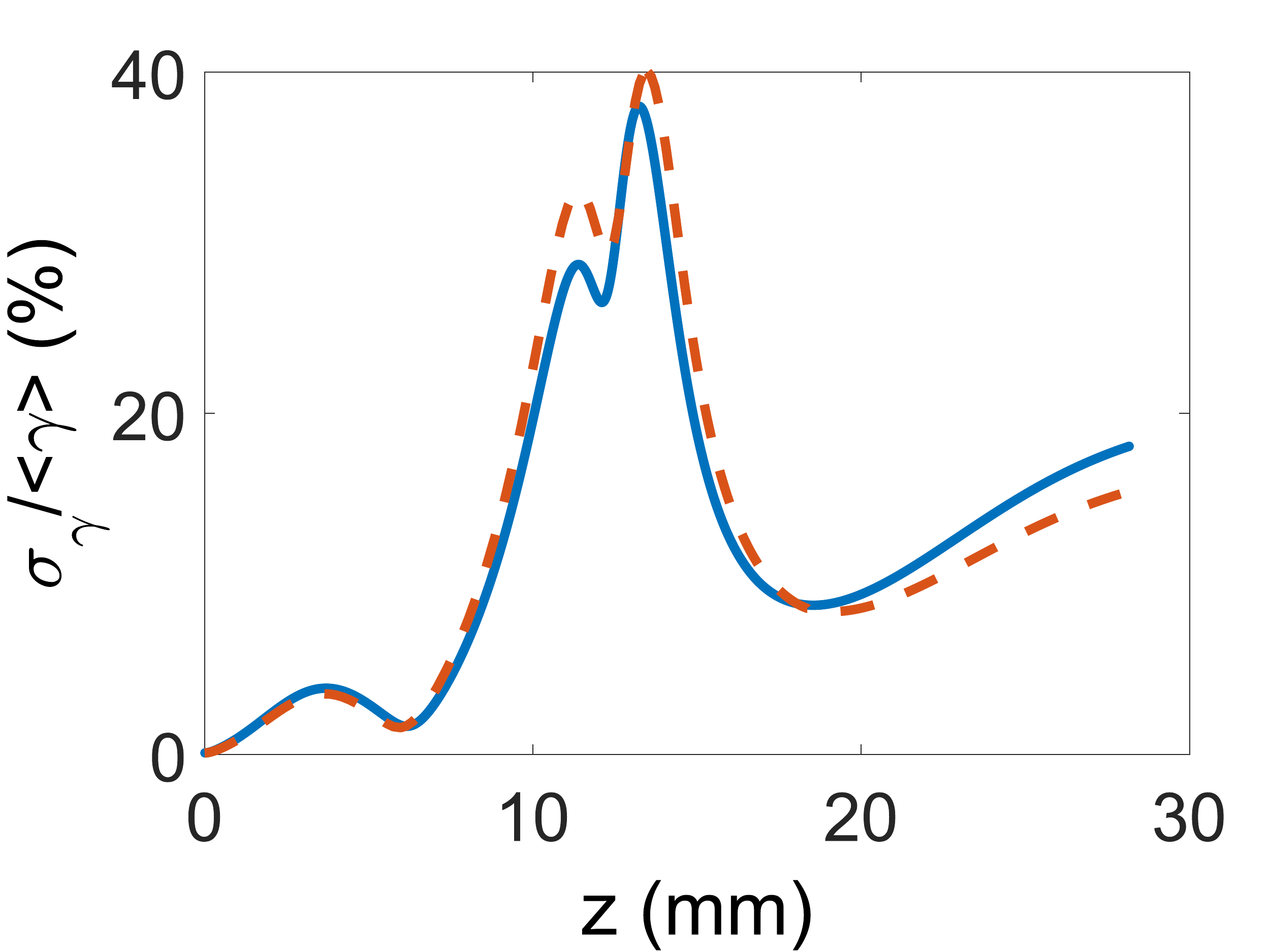} &
  	\includegraphics[draft=false,width=2.0in]{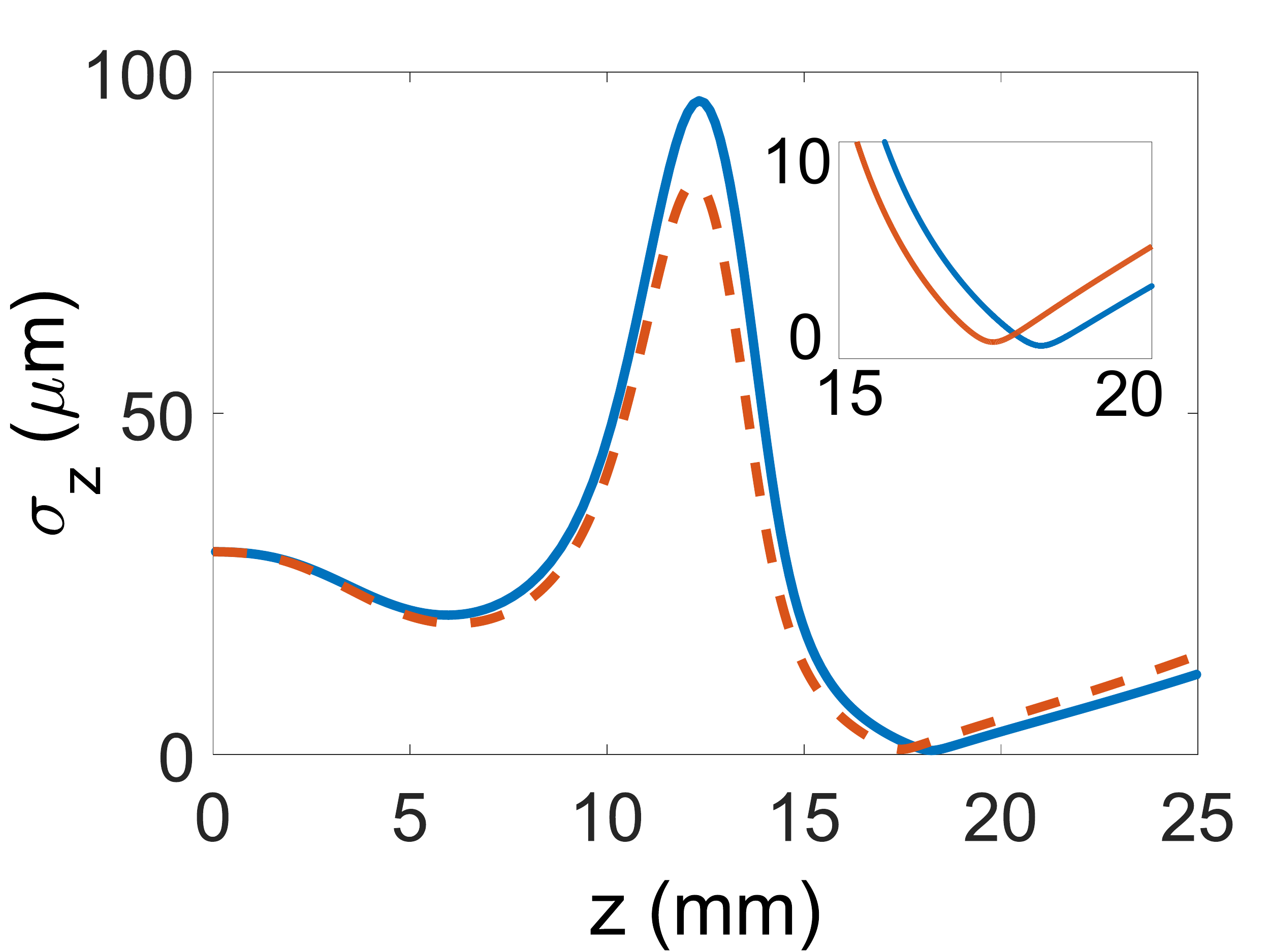} \\
  	(a) & (b) & (c) \\
  	\includegraphics[draft=false,width=2.0in]{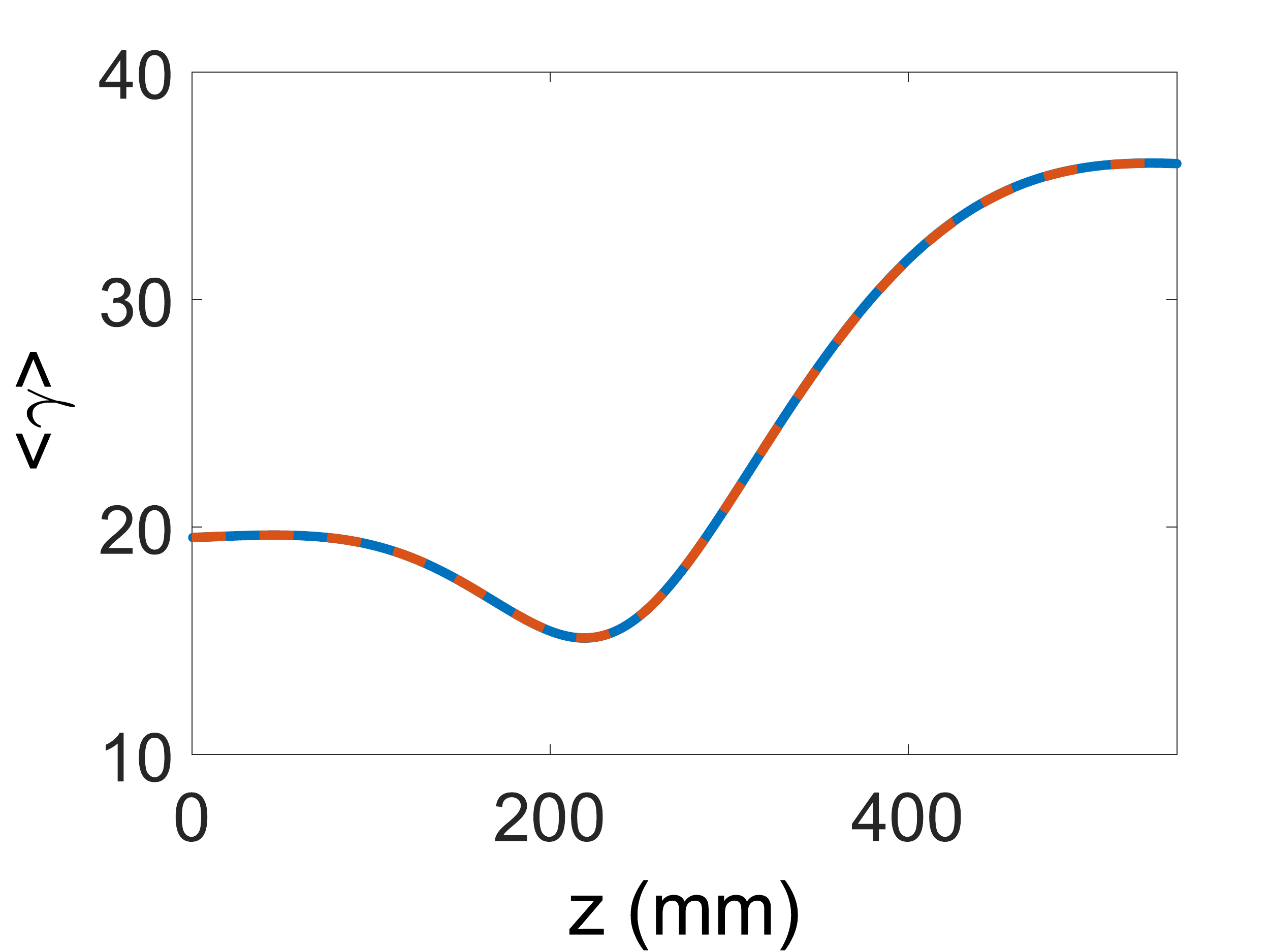} &
  	\includegraphics[draft=false,width=2.0in]{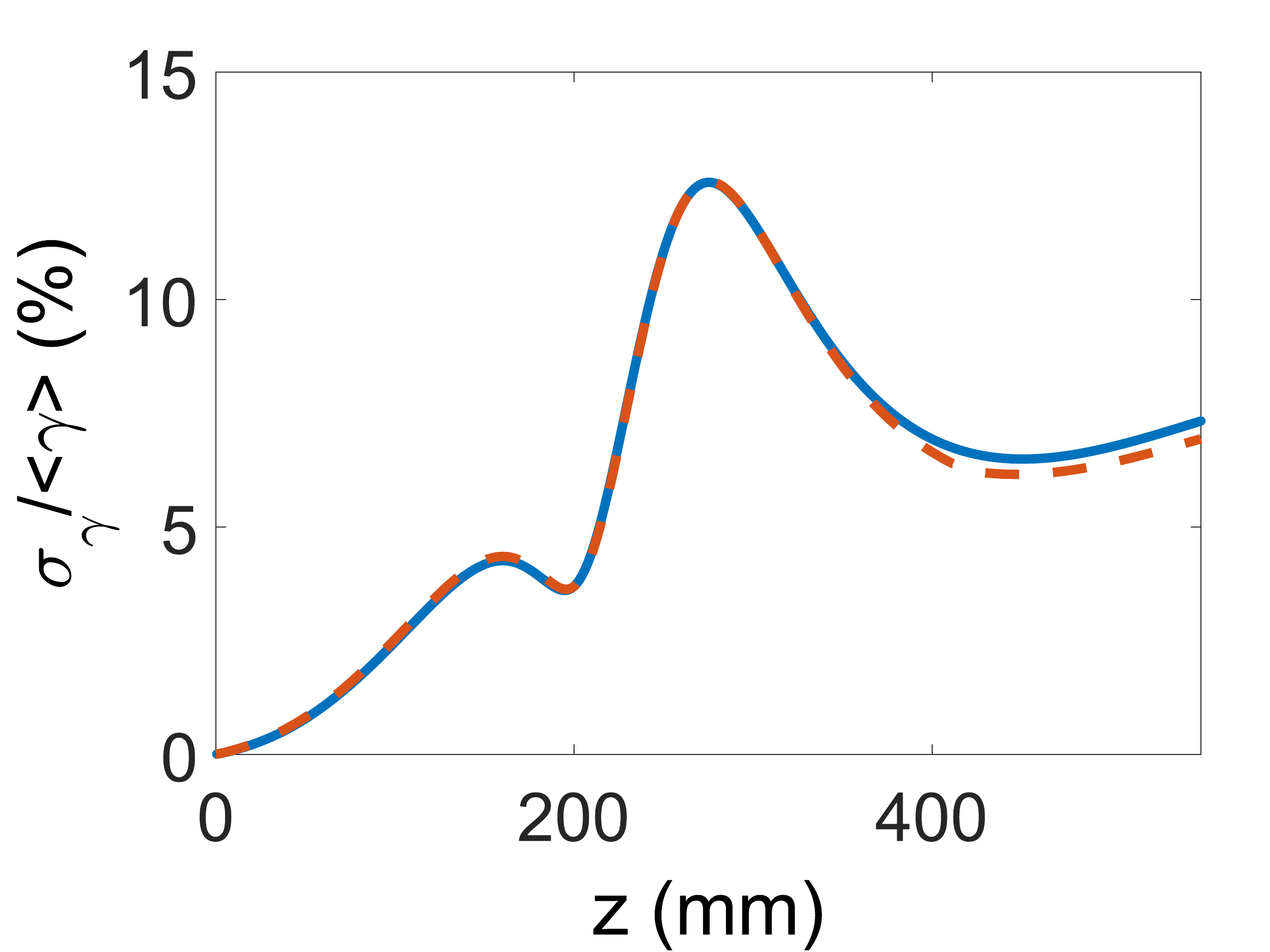} &
  	\includegraphics[draft=false,width=2.0in]{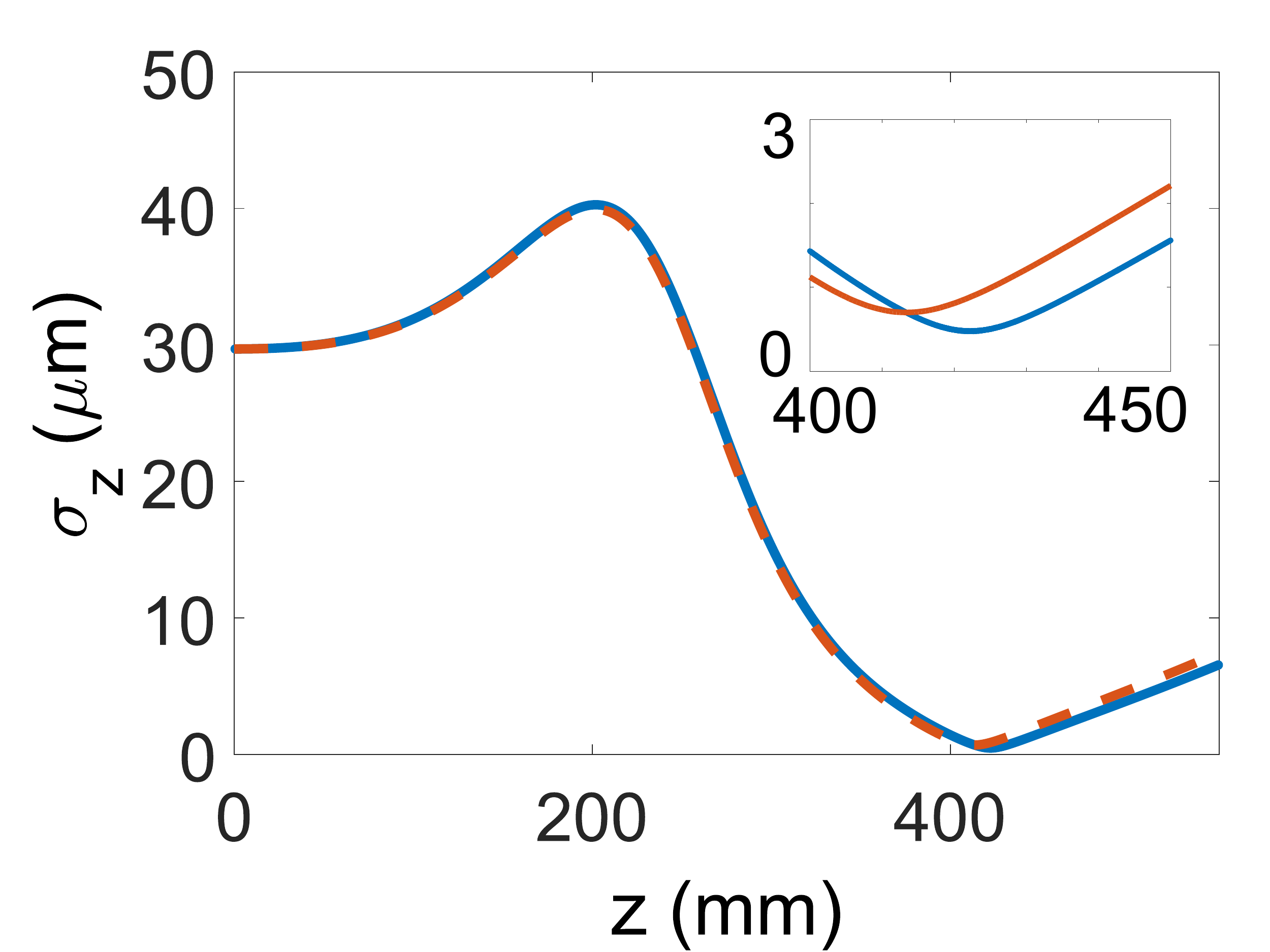} \\
  	(d) & (e) & (f)
  	\end{array}$
  	\caption{Concurrent compression and acceleration of a 1.6\,pC electron bunch under optimized conditions, with a compression factor of 50 and 62 achieved for initial kinetic energies of 1\,MeV and 10\,MeV, respectively. The evolution of (a) normalized mean energy, (b) relative energy spread and (c) longitudinal spread are shown for a 1\,MeV bunch subjected to a 20\,mJ, 7.86-cycle (13.1\,ps), 0.6\,THz-centered pulse ($\psi_0=0.73\pi$ and $k_0z_i = 13.3\pi$). Similarly, the evolution of (e) normalized mean energy, (f) relative energy spread and (g) longitudinal spread are shown for a 10\,MeV bunch subjected to a 20\,mJ, 102.3-cycle (170.5\,ps), 0.6\,THz-centered pulse ($\psi_0=1.02\pi$ and $k_0z_i = 206\pi$). Blue solid curves and red dashed curves indicate simulations with and without space charge, respectively. 10'000 macro-particles were used for all simulations.}
  	\label{bunchCompression}
  \end{figure}

  For the 1\,MeV case, our optimized parameters are $\psi_0=0.73\pi$, $k_0z_i = 13.3\pi$, $r_1 = 447$\,{\textmu}m, and $\tau_\mathrm{FWHM} = 13.1$\,ps (7.86 cycles).
  The evolution of the electron bunch parameters under these optimal conditions are presented in Fig.\,\ref{bunchCompression}a-c, where we observe a small net acceleration and a phase-limited compression of the electron bunch from 100\,fs (30\,{\textmu}m) to about 2\,fs over an interaction distance of about 18\,mm.
  Note that there is a limited time window during which the electron bunch remains maximally compressed.
  Conceptually, this is unavoidable due to the presence of space charge which causes the bunch to expand after the bunch has slipped from the THz pulse.
  
  For the 10\,MeV case, our optimized parameters are $\psi_0=1.02\pi$, $k_0z_i = 206\pi$, $r_1 = 597$\,{\textmu}m, $\tau_\mathrm{FWHM} = 170.5$\,ps (102.3-cycle).
  The evolution of the electron bunch parameters under these optimal conditions are presented in Fig.\,\ref{bunchCompression}d-f, where we observe a phase-limited compression of the electron bunch from 100\,fs to 1.61\,fs over an interaction distance of 42\,cm.
  Although the bunch is compressed by a slightly larger factor than in the 1\,MeV case, the much larger interaction distance suggests that the superior strategy to obtain a high energy, compressed bunch is to compress it before acceleration.
  
  \section{Experimental Test of the THz Linac}
  
  In this section, we report the experimental demonstration of electron acceleration using the axial component of an optically generated 10\,{\textmu}J THz pulse centred at 0.45\,THz in a waveguide.
  The THz pulse accelerates electrons in a circular waveguide consisting of a dielectric capillary with a metal outer boundary.
  The dielectric slows the group and phase velocity of the THz wave allowing it to accelerate low-energy electrons.
  We a observe a maximum energy gain of 7\,keV in 3\,mm.
  
  \subsection{THz Linac Design}
  
\begin{figure}
  	\centering
  	$\begin{array}{c}
  	\begin{array}{cc}
  	\includegraphics[draft=false,width=3.0in]{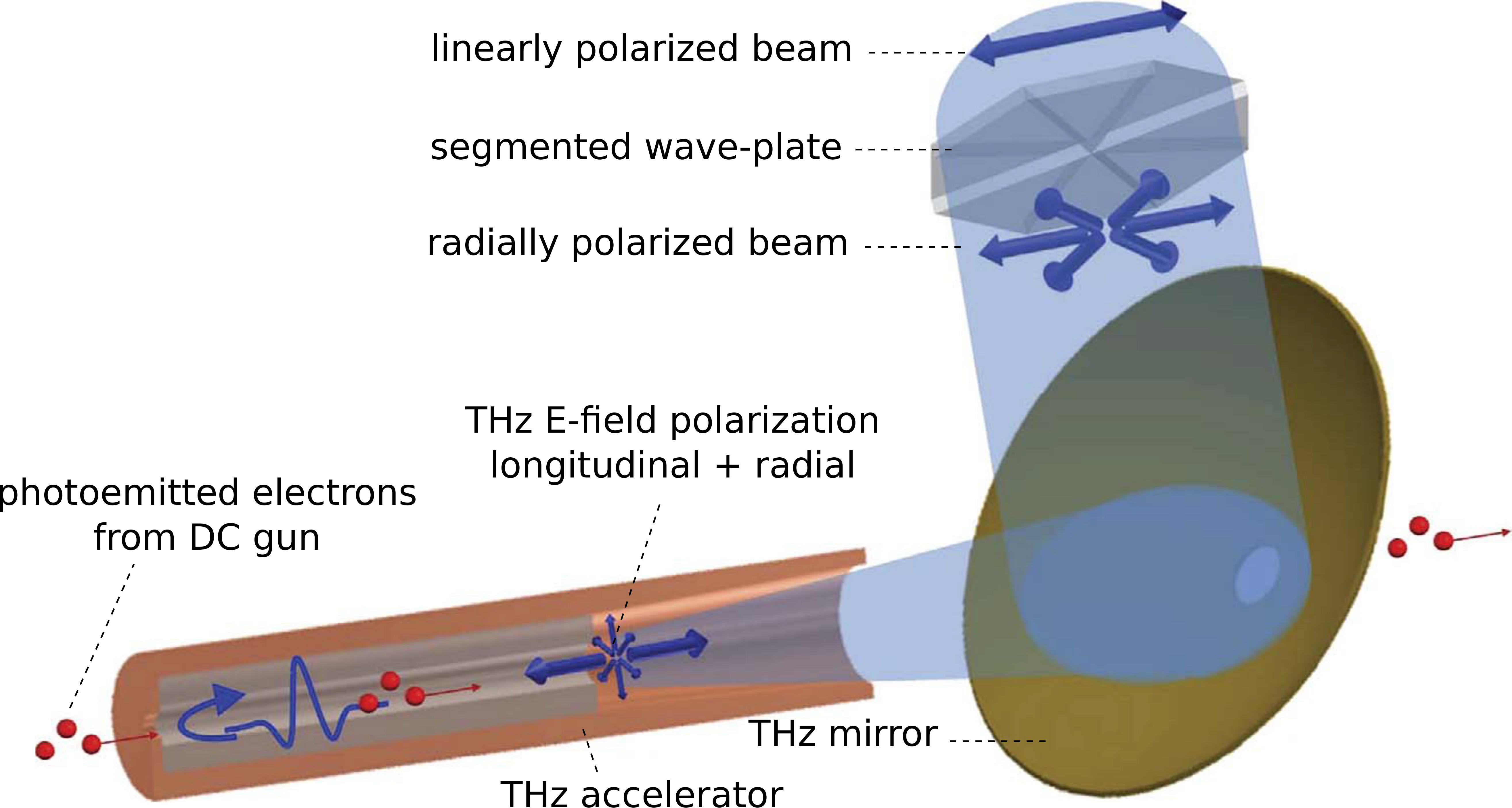} &
  	\includegraphics[draft=false,width=2.0in]{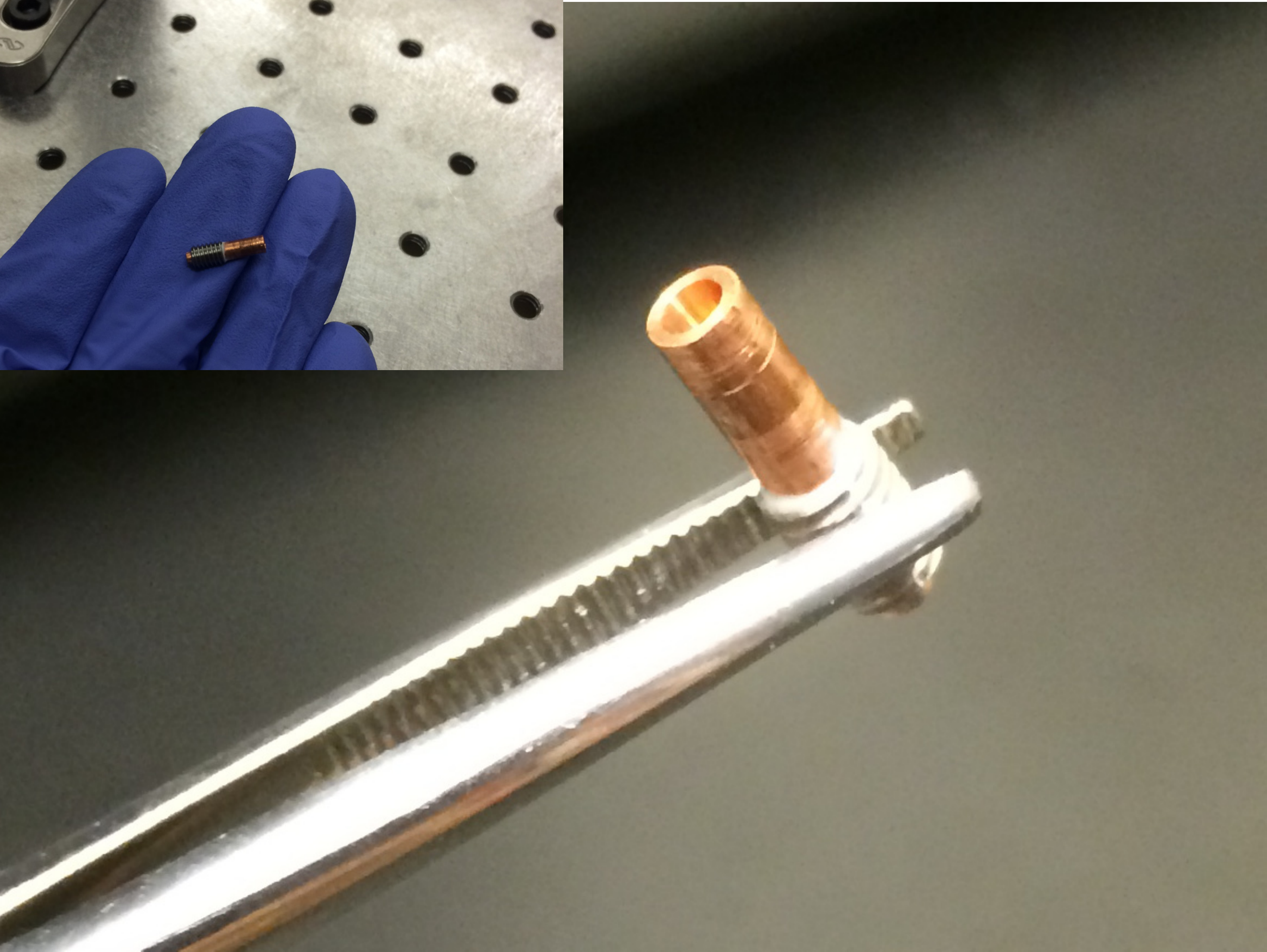} \\
  	(a) & (b)
  	\end{array} \\ \\
  	\begin{array}{ccc}
  	\includegraphics[draft=false,width=2.0in]{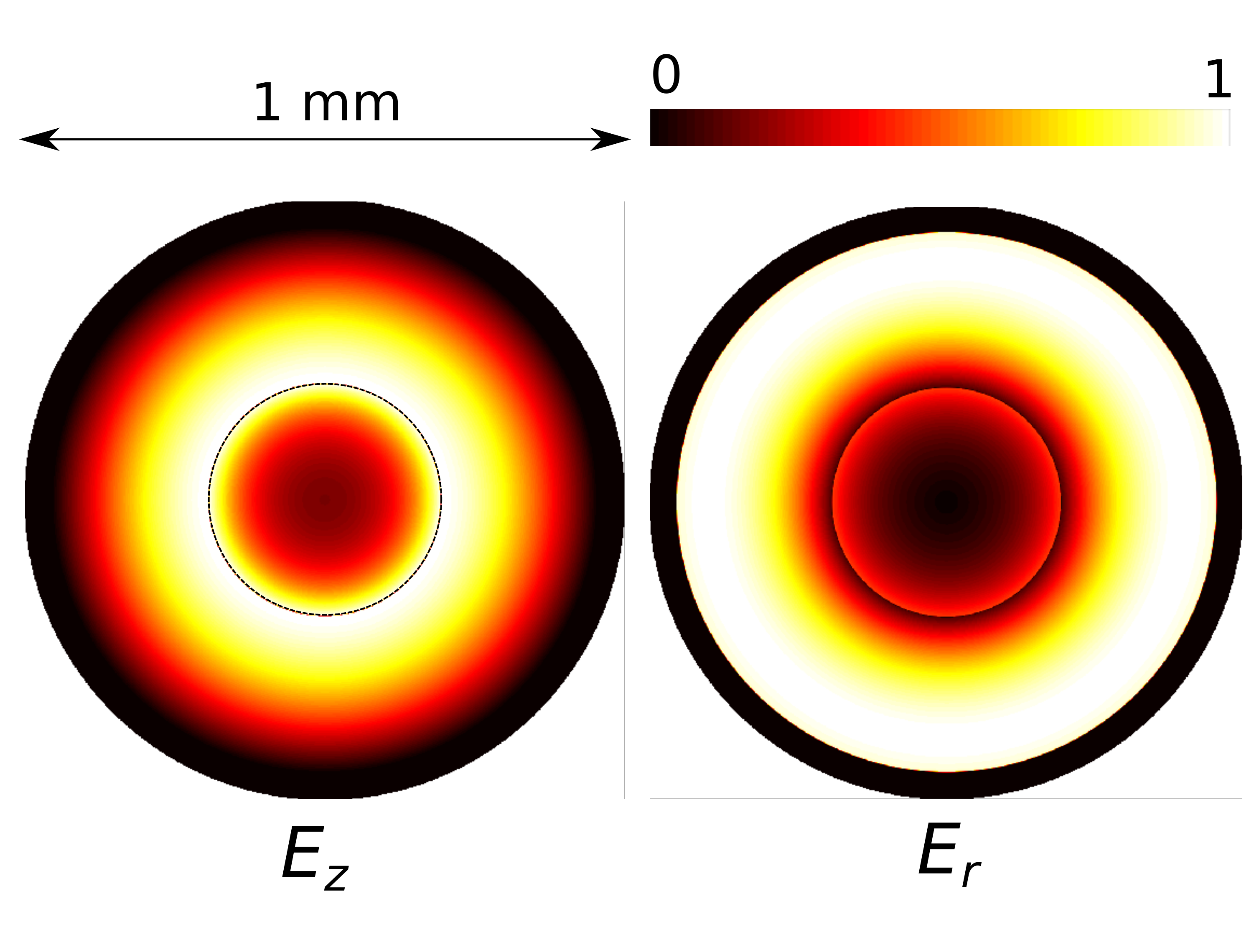} &
  	\includegraphics[draft=false,width=2.0in]{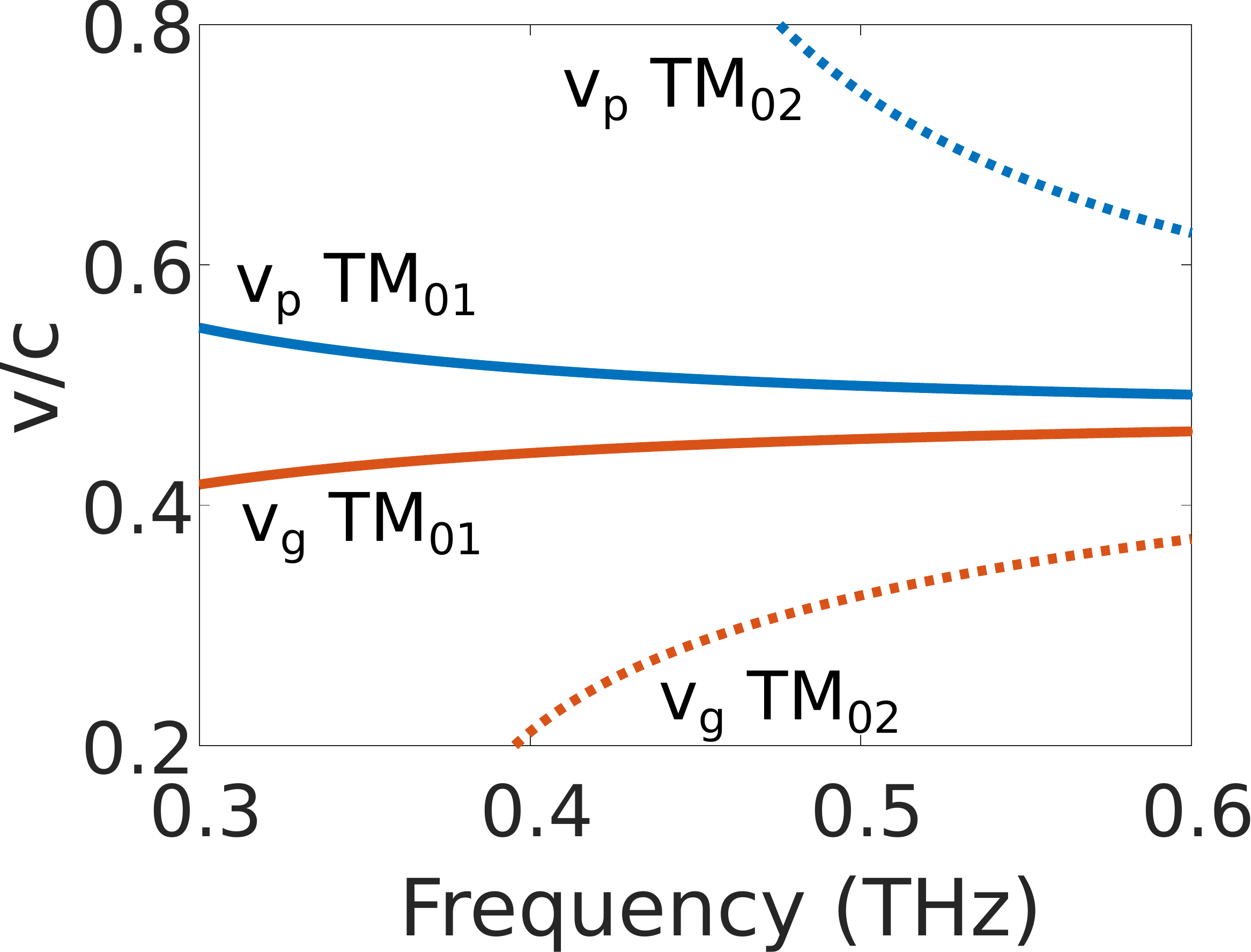} &
  	\includegraphics[draft=false,width=2.0in]{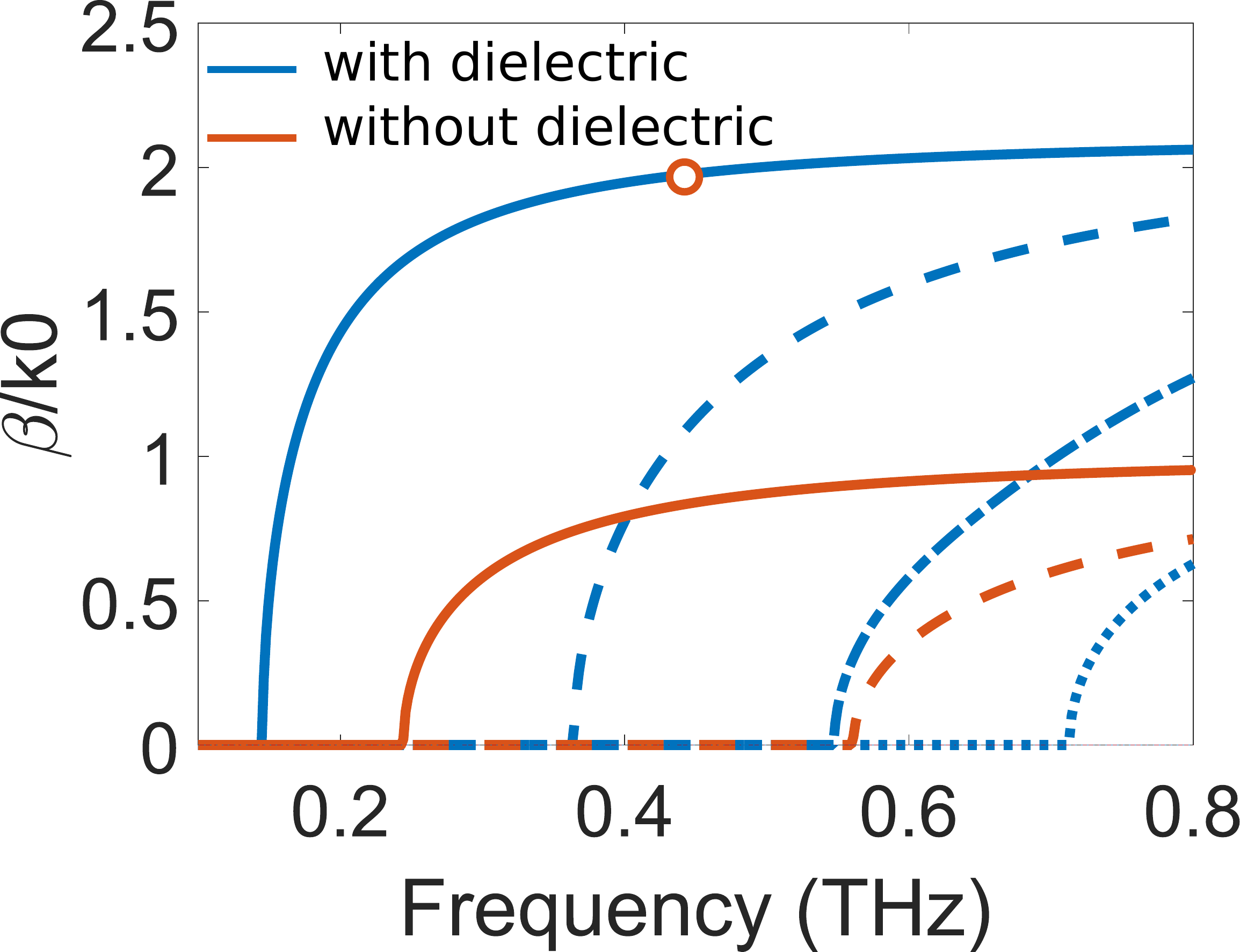} \\
  	(c) & (d) & (e)
  	\end{array}
  	\end{array}$
  	\caption{Terahertz-driven linac experiment: (a) Schematic of the THz linac. Top right: a linearly polarized THz pulse is converted into a radially polarized pulse by a segmented waveplate before being focused into the THz waveguide. The THz pulse is reflected at the end of the waveguide to co-propagate with the electron bunch, which enters the waveguide through a pinhole (lower left). The electron bunch is accelerated by the longitudinal electric field of the co-propagating THz pulse. The electron bunch exits the THz waveguide and passes through a hole in the focusing mirror (right) for the THz pulse. (b) Photograph of the compact millimetre scale THz linac. (c) Normalized magnitude of the longitudinal electric field and perpendicular magnetic field for the TM\textsubscript{01} mode at 450\,GHz in a circular copper waveguide with dielectric loading. The solid black line indicates the boundary between the vacuum core and the quartz capillary. (d) Phase and group velocities of the TM\textsubscript{01} and TM\textsubscript{02} modes in the designed waveguide. (e) The dispersion relation for the travelling modes with and without dielectric loading.}
  	\label{linacExperimentDesign}
  \end{figure}

  The THz pulse accelerates electrons in a circular waveguide consisting of a quartz capillary inserted into a hollow copper cylinder \cite{miyagi1984design} (Fig.\,\ref{linacExperimentDesign}a and \ref{linacExperimentDesign}b).
  The inner diameter of the copper waveguide is 940\,{\textmu}m with a dielectric wall thickness of $d=270$\,{\textmu}m.
  This results in a vacuum space with a radius of $r_v=a=200$\,{\textmu}m.
  The dielectric constant of the quartz capillary is nominally $\epsilon_r = 4.41$.
  The operational mode of the linac is a travelling TM\textsubscript{01} mode, (Fig.\,\ref{linacExperimentDesign}c).
  The dispersion relation for the operating mode is shown in Fig.\,\ref{linacExperimentDesign}d.
  At the centre frequency of the THz pulse (450\,GHz), the group velocity is $v_g/c=0.46$ and the phase velocity is $v_p/c = 0.505$.
  
  Due to the operational frequency's proximity to the cutoff of the waveguide, the accelerating mode is highly dispersive with the phase and group velocity shown in Fig.\,\ref{linacExperimentDesign}e as a function of frequency.
  The waveguide dimensions of the linac were chosen to optimize for this experimental setup with a low initial electron energy of 60\,keV, the THz pulse energy available and the transverse dimension of the electron beam.
  At the nominal 60\,keV, the electron velocity is $v/c=0.45$.
  This velocity is increased as the particles are accelerated, however, increasing the mismatch with the phase velocity.
  The overall dynamics result in an interaction length of 3\,mm.
  At this point, the electron bunch and THz pulse interaction is terminated by the presence of the taper, which rapidly reduces the intensity of the on-axis electric field as the waveguide diameter is increased.
  This slippage causes a peak on-axis electric field of 8.5\,MV/m to produce an accelerating gradient of 2.5\,MeV/m at the nominal initial energy of 60\,keV.
  
\begin{figure}
  	\centering
  	$\begin{array}{ccc}
  	\includegraphics[draft=false,width=2.0in]{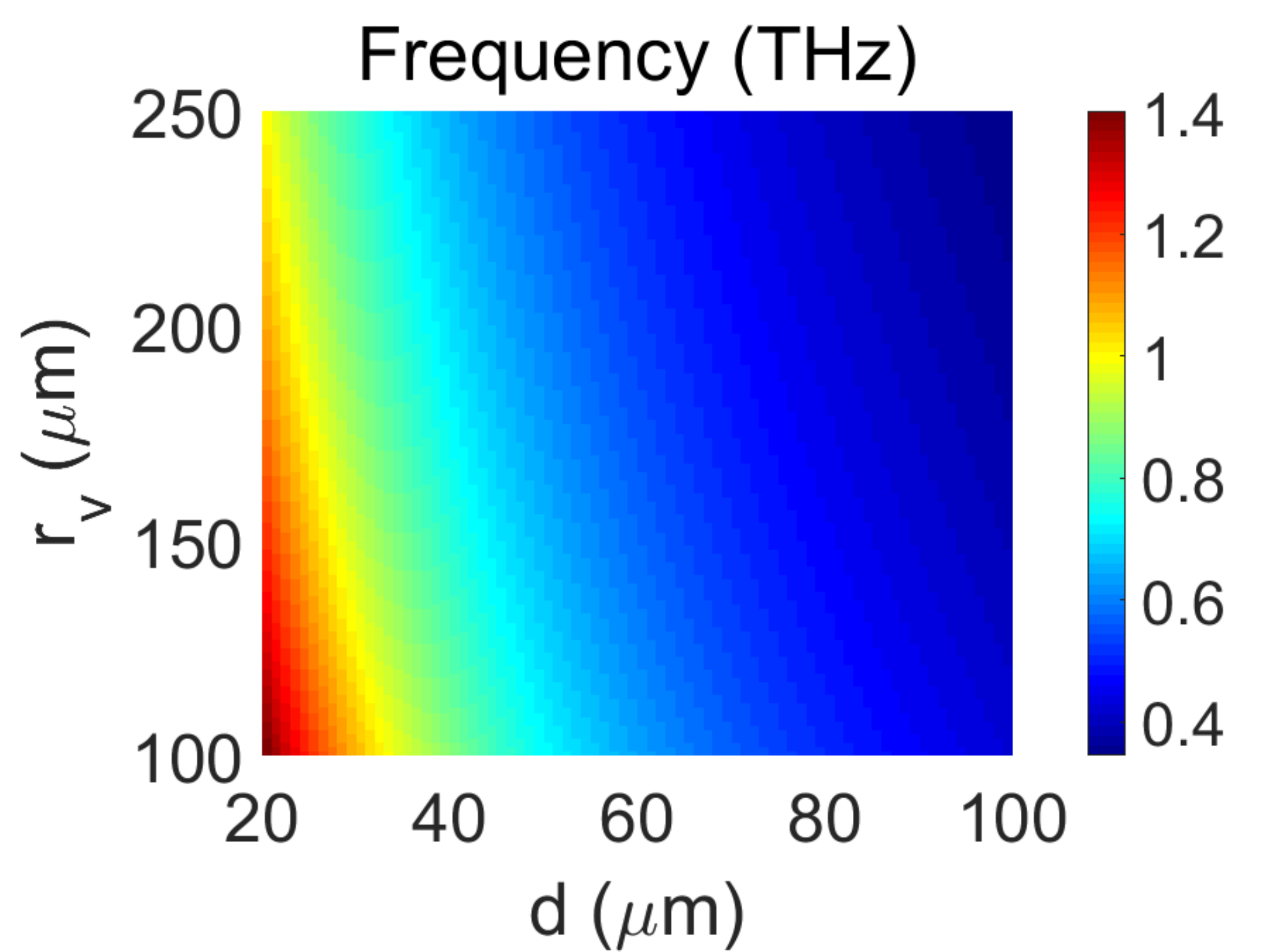} &
  	\includegraphics[draft=false,width=2.0in]{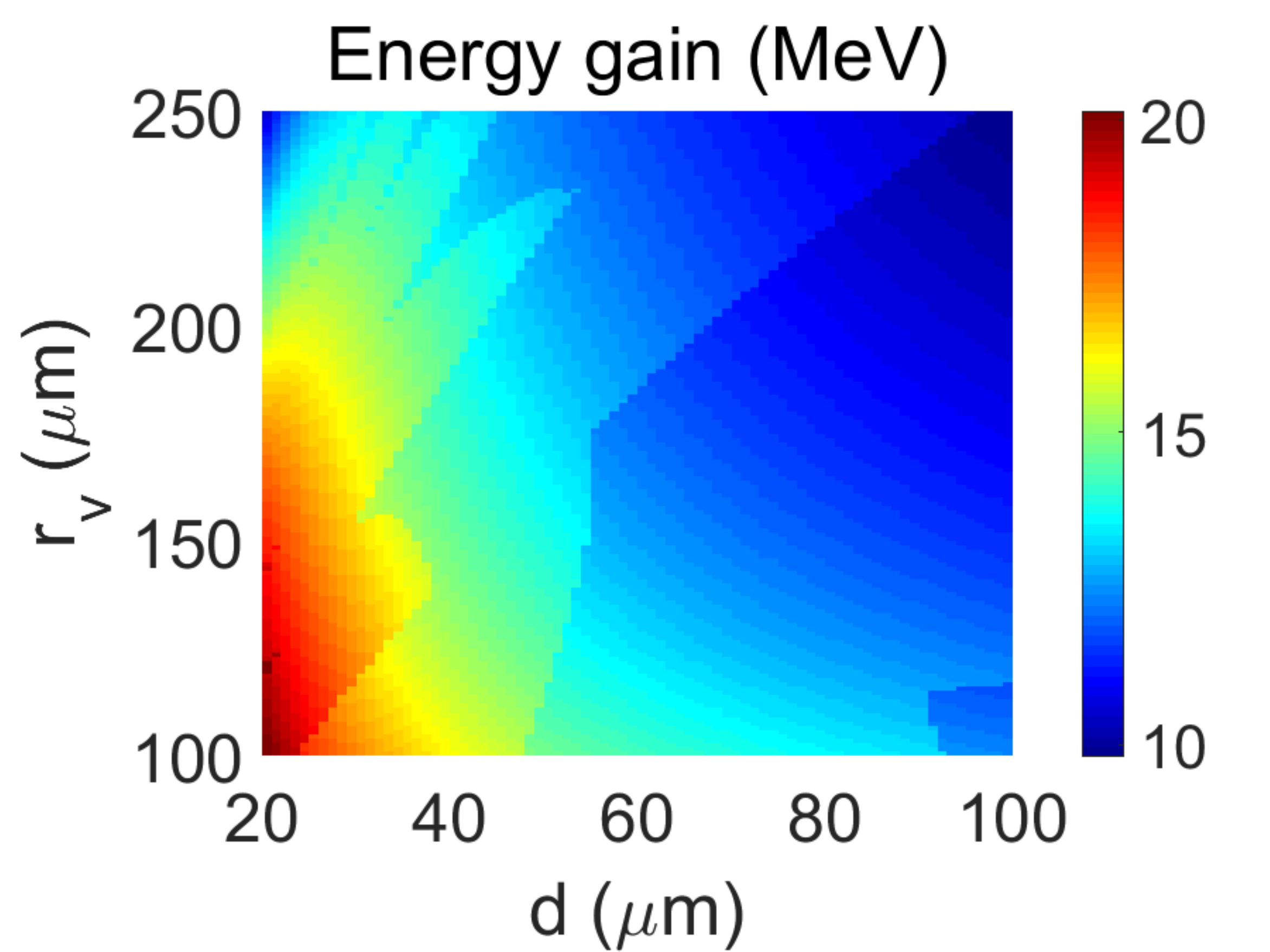} &
  	\includegraphics[draft=false,width=2.0in]{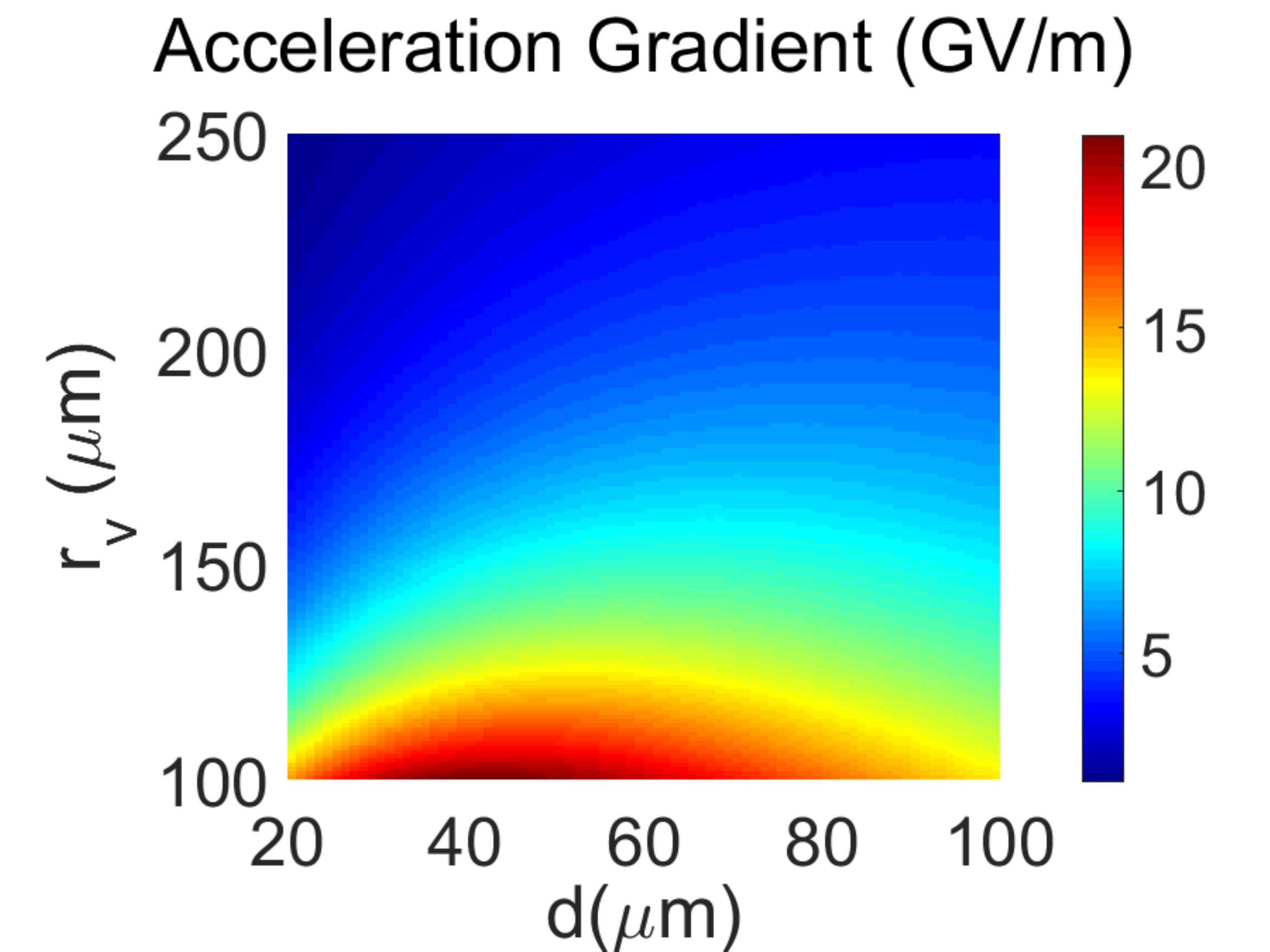} \\
  	(a) & (b) & (c) \\
  	\includegraphics[draft=false,width=2.0in]{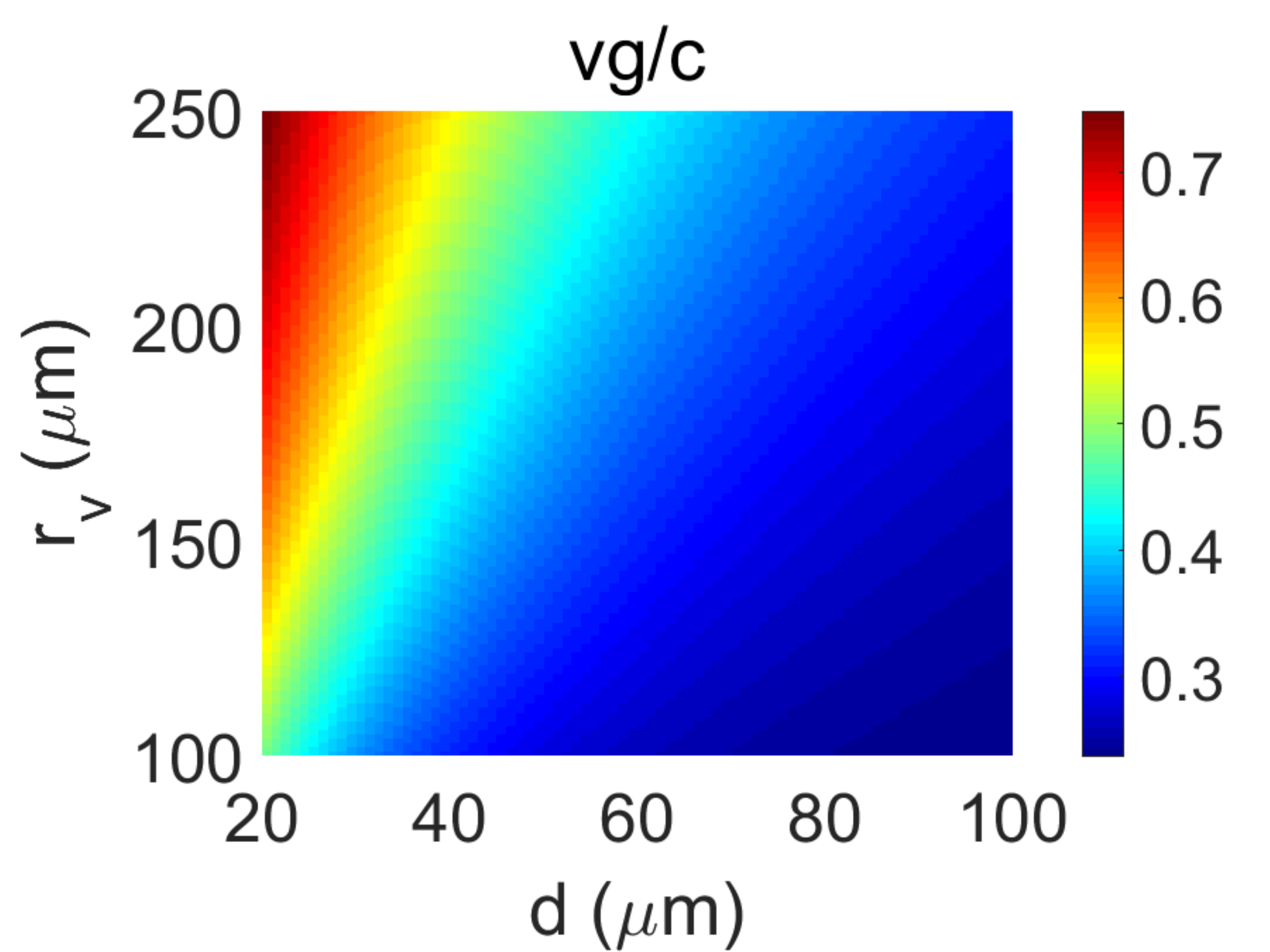} &
  	\includegraphics[draft=false,width=2.0in]{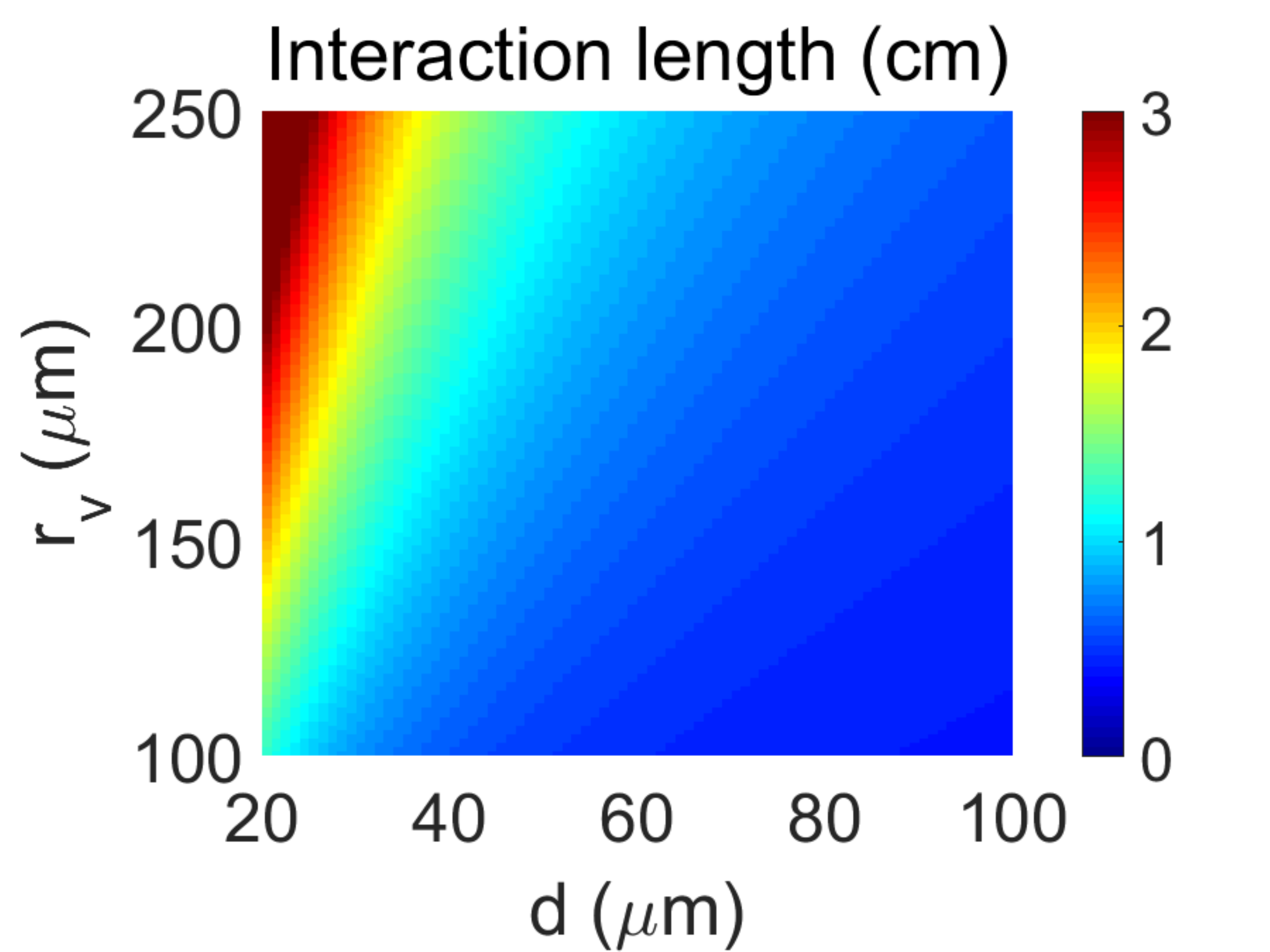} &
  	\includegraphics[draft=false,width=2.0in]{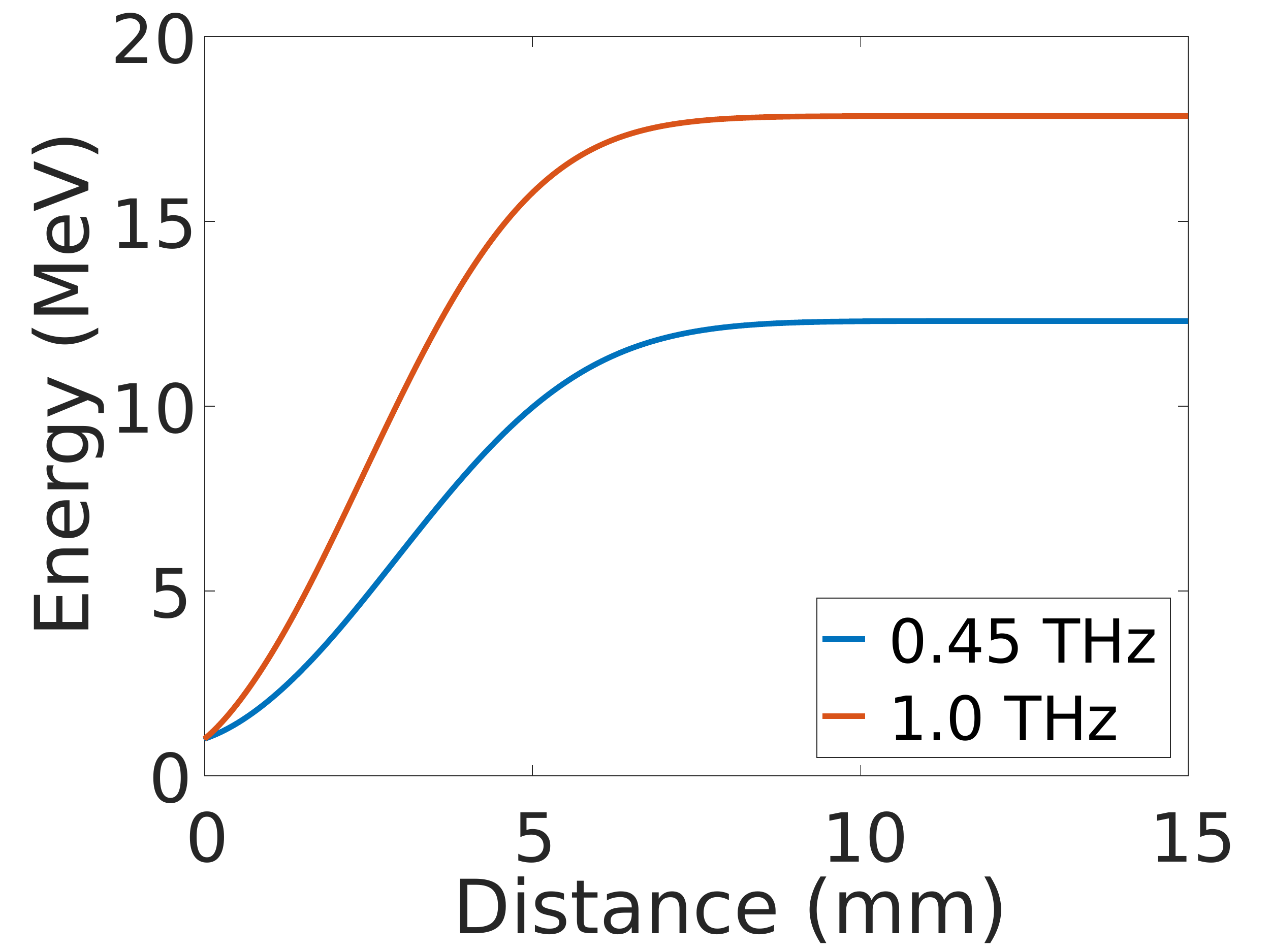} \\
  	(d) & (e) & (f)
  	\end{array}$
  	\caption{Relativistic THz linac design: Performance parameters as a function of vacuum radius and dielectric wall thickness for a relativistic THz linac operating in the TM\textsubscript{01} mode with a 10\,mJ single-cycle THz pulse and an initial electron energy of 1\,MeV. The phase velocity is $c$ for the nominal frequency of operation. The (a) frequency of operation, (b) energy gain, (c) accelerating gradient, (d) group velocity and (e) interaction length for the THz linac. (f) The electron energy as a function of distance for two cases, which operate with a frequency of (0.45,1)\,THz, a vacuum space with a radius of $r_v=(110,105)$\,{\textmu}m and a dielectric wall thickness of (90,32)\,{\textmu}m}
  	\label{linacExperimentOptimization}
  \end{figure}

\begin{figure}
  	\centering
  	$\begin{array}{ccc}
  	\includegraphics[draft=false,width=2.0in]{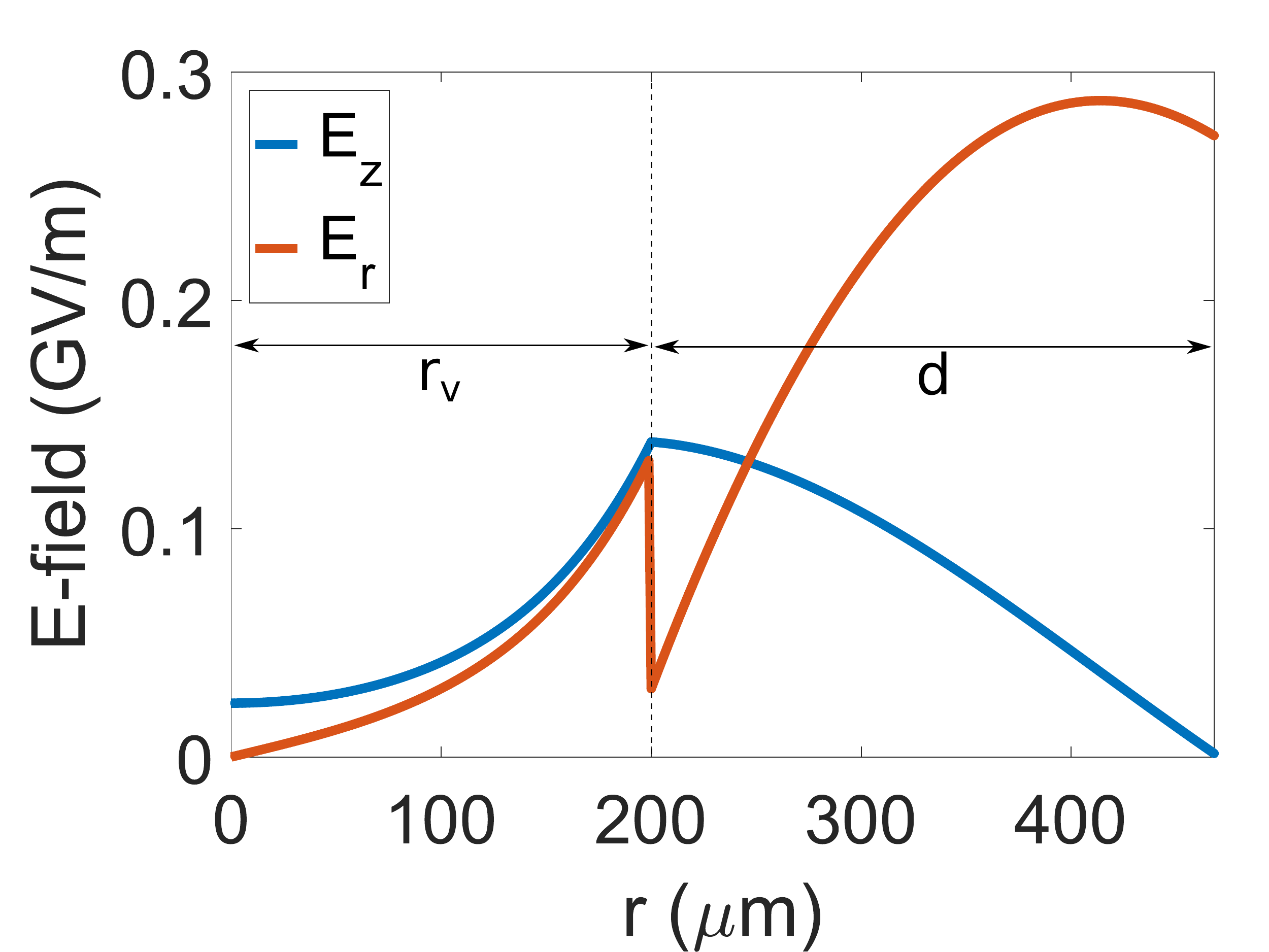} &
  	\includegraphics[draft=false,width=2.0in]{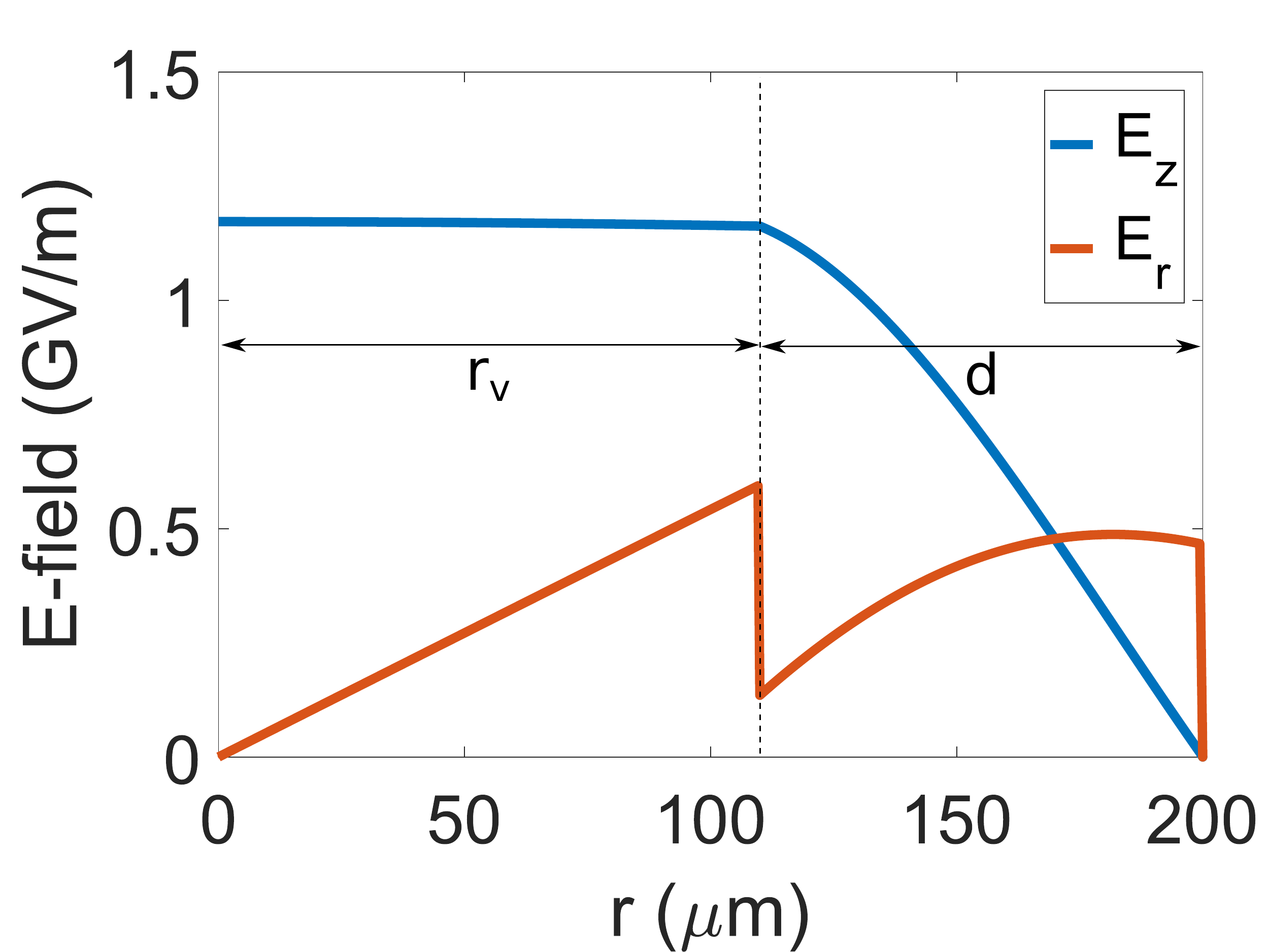} &
  	\includegraphics[draft=false,width=2.0in]{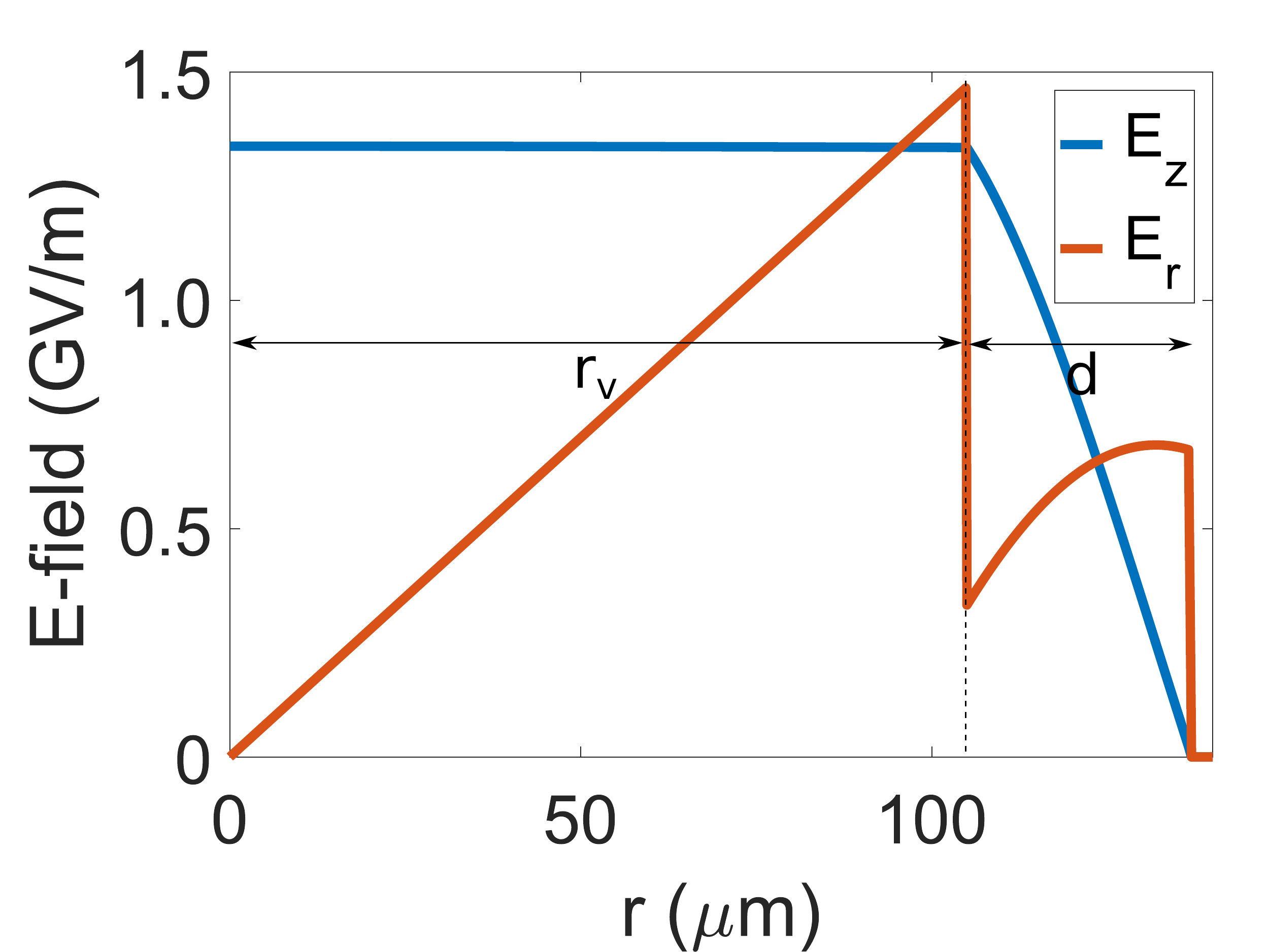} \\
  	(a) & (b) & (c)
  	\end{array}$
  	\caption{Electromagnetic field distribution: The field distribution in the THz waveguide for (a) the non-relativistic design described in Fig.\,\ref{linacExperimentDesign} with $r_v = 200$\,{\textmu}m and $d=270$\,{\textmu}m, (b) the relativistic design in Fig.\,\ref{linacExperimentOptimization}f for operation at 0.45\,THz with $r_v = 110$\,{\textmu}m and $d=90$\,{\textmu}m and (c) the relativistic design in Fig.\,\ref{linacExperimentOptimization}f for operation at 1\,THz with $r_v=105$\,{\textmu}m and $d = 32$\,{\textmu}m. All of the depicted curves consider 100\,MW total power for the propagating mode.}
  	\label{linacExperimentProfile}
  \end{figure}

  With increased THz energy and increased electron energy, one can consider a relativistic accelerator design in which the phase velocity of the TM\textsubscript{01} mode is equal to the speed of light.
  In addition, the design of the waveguide can be optimized to match the frequency of the available source.
  Fig.\,\ref{linacExperimentOptimization}a-e presents the frequency of operation, energy gain, accelerating gradient, group velocity and interaction length as a function of vacuum radius and dielectric thickness assuming a 10-mJ single cycle THz pulse and an initial electron energy of 1\,MeV.
  Fig.\,\ref{linacExperimentOptimization}f presents the electron energy as a function of distance for two cases, which operate with a frequency of (0.45,1)\,THz, a vacuum space with a radius of $r_v=(110,105)$\,{\textmu}m and a dielectric wall thickness of $d=(90,32)$\,{\textmu}m.
  The drastic increase in the accelerating gradients shown in Fig.\,\ref{linacExperimentOptimization} is not only due to the increased THz pulse energy.
  The performance of a travelling-wave THz accelerator structure greatly increases when the electron velocity approaches the speed of light.
  This effect originates from the less dispersion for the THz pulse, a longer interaction length, the decrease in the waveguide radius, the great reduction in the amount of dielectric material, and the improvement of the electric field profile.
  Fig.\,\ref{linacExperimentProfile} compares the field distribution for the non-relativistic waveguide design (Fig.\,\ref{linacExperimentDesign}), which was investigated experimentally and the relativistic designs highlighted in Fig.\,\ref{linacExperimentOptimization}f.
  Note the decrease in waveguide radius, the decrease in dielectric wall thickness and the improved relative amplitude of the longitudinal electric field in the accelerating region of the waveguide ($r<100$\,{\textmu}m) for the two relativistic ($v_p = c$) designs, which all contribute to improve the efficiency of the accelerator.
  The transverse fields in all cases are localized to the regions without the presence of the electron bunch alleviating concerns with the potential for ponderomotive effects.
  The use of multi-mJ THz pulses will increase the peak electric fields in these waveguides well above a GV/m; however, the long wavelength and short propagation distance prevent the onset of nonlinearities such as self phase modulation \cite{zalkovskij2013terahertz,harbold2002highly}.
  Further details and numerical studies for relativistic THz linacs using dielectric-loaded waveguides and few-cycle THz pulses were previously discussed \cite{Wong2013}.
  
  \subsection{Structure Testing}
  
\begin{figure}
  	\centering
  	$\begin{array}{ccc}
  	\includegraphics[draft=false,width=2.0in]{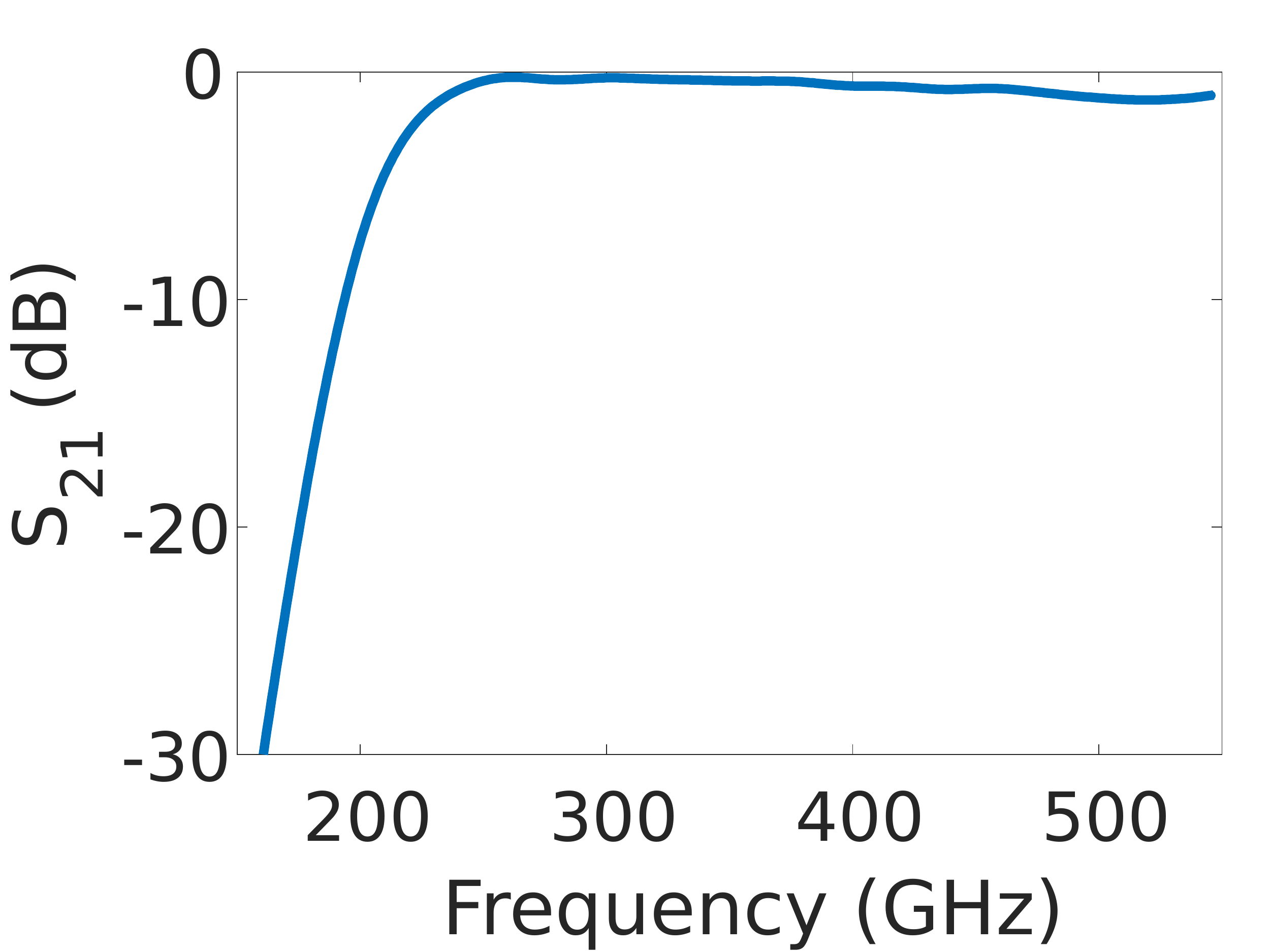} &
  	\includegraphics[draft=false,width=2.0in]{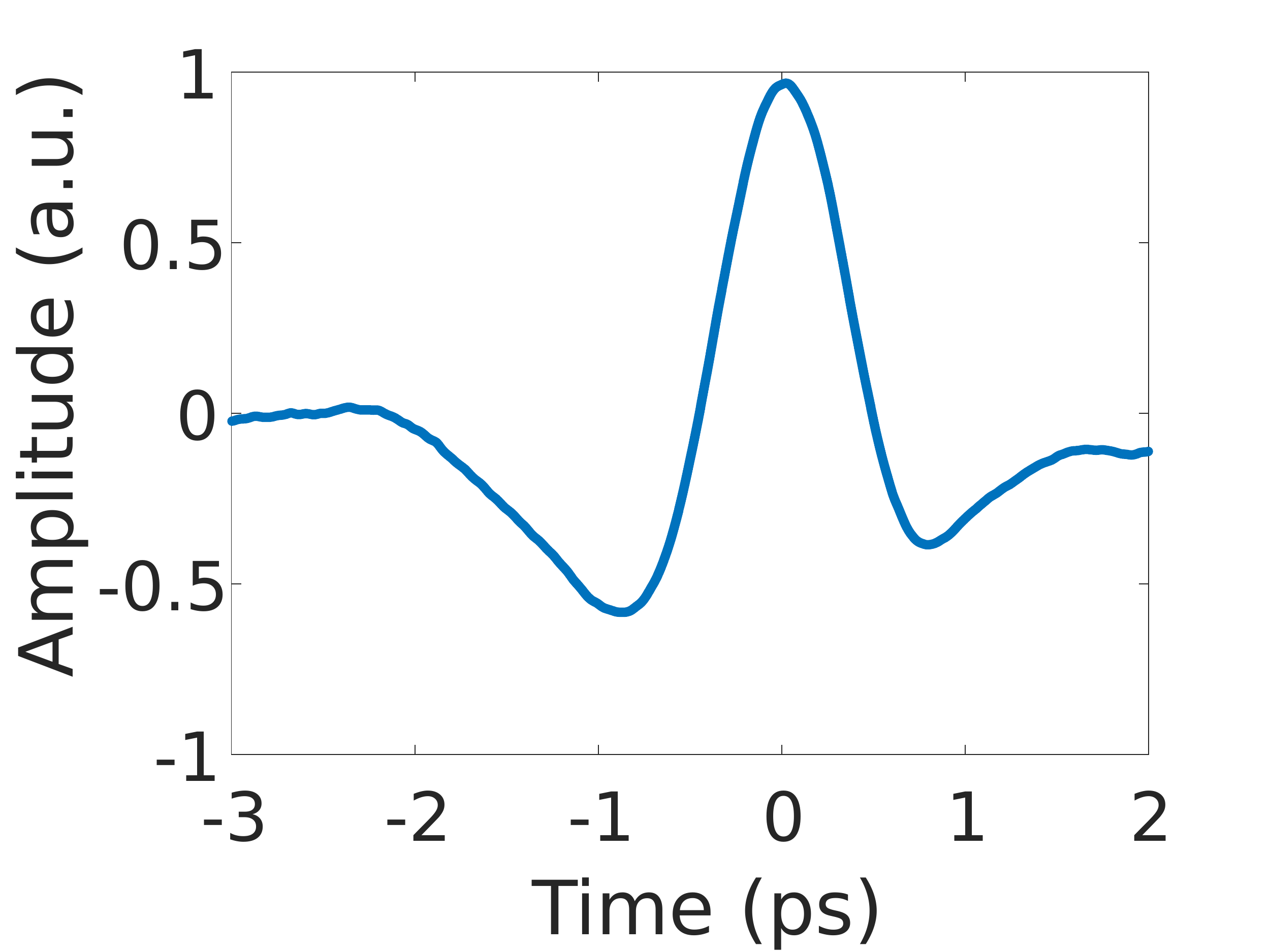} &
  	\includegraphics[draft=false,width=2.0in]{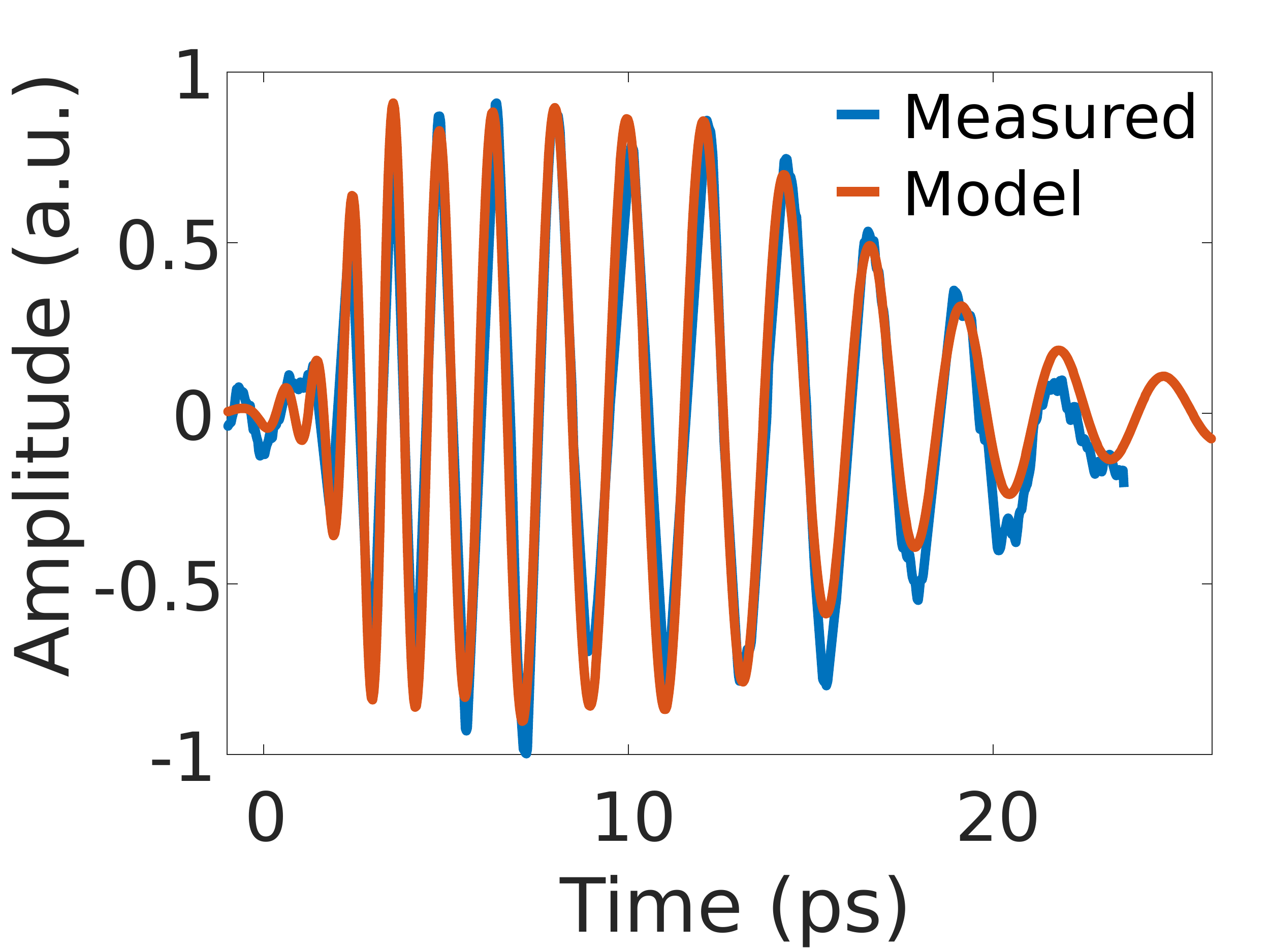} \\
  	(a) & (b) & (c)
  	\end{array}$
  	\caption{Coupling and dispersion of THz pulses in waveguides. (a) HFSS simulation of the coupling of the free-space radially polarized mode into the TM\textsubscript{01} mode through a dielectric-loaded taper. (b) The time-domain waveform of the THz pulse determined with electro-optic sampling before being coupled into the THz waveguide. (c) The measured versus the modelled time-domain waveform of the THz pulse at the exit of a 5-cm (including tapers) dielectric-loaded THz waveguide.}
  	\label{linacExperimentTest}
  \end{figure}

  A THz waveguide structure consisting of two tapers separated by a uniform waveguide section was built to test and optimize waveguide design by performing transmitted energy and polarization measurements.
  Measurements were performed with a Gentec-EO Pyroelectric Joulemeter Probe that is capable of measuring pulse energies exceeding 100\,nJ.
  Efficient excitation of the TM\textsubscript{01} mode, demonstrated in these measurements, is a critical requirement in developing a compact high-gradient THz linac.
  As a first step, the linearly polarized beam is converted to a radially polarized beam with a segmented waveplate.
  The vertical and horizontal polarization measured after the segmented waveplate were 53\% and 47\%, respectively.
  This radially polarized beam was coupled into a test waveguide structure that was 5\,cm in length, including two tapers to couple the THz pulse into and out of the waveguide.
  The long length was selected to demonstrate that ohmic losses are manageable even over significant interaction lengths.
  The simulated coupling for the THz pulse into the waveguide is shown in Fig.\,\ref{linacExperimentTest}a with a bandwidth of over 200\,GHz.
  The power coupled through the structure was measured with and without dielectric loading at 32\% and 54\%, respectively.
  The increased losses in the dielectric-loaded waveguide are due to increased ohmic losses.
  For the dielectric-loaded waveguide, 2\,mJ of energy was measured at the exit of the waveguide, which corresponds to a calculated peak on-axis electric field of 9.7\,MV/m.
  
  Electro-optic sampling was used to characterize the linear dispersion of the dielectric-loaded waveguide test structure, which has the same dimensions as the waveguide used in the acceleration experiment and the properties shown in Fig.\,\ref{linacExperimentDesign}d and e.
  Fig.\,\ref{linacExperimentTest}b shows the time-domain spectrum produced by the THz source.
  Fig.\,\ref{linacExperimentTest}c shows a comparison between the measured time-domain spectrum of THz pulse at the exit of the waveguide with the simulated propagation of the input THz pulse propagated with the dispersive parameters of the waveguide design provided in Fig.\,\ref{linacExperimentDesign}d and \ref{linacExperimentDesign}e.
  Excellent agreement indicates that the fabricated structure has the desired performance.
  
  \subsection{Operation of the THz Linac}
  
  A schematic view of the THz accelerator was shown in Fig.\,\ref{linacExperimentDesign}a with a photograph of the THz linac in Fig.\,\ref{linacExperimentDesign}b.
  Using 60-keV electrons, from a DC electron gun, an energy gain of 7\,keV is observed in a 3-mm interaction length.
  The single-cycle THz pulse (Fig.\,\ref{linacExperimentTest}b) is produced via optical rectification of a 1.2\,mJ, 1.03\,{\textmu}m laser pulse with a 1\,kHz repetition rate.
  The THz pulse, whose polarization is converted from linear to radial by a segmented waveplate, is coupled into a waveguide with 10\,MV/m peak on-axis electric field.
  A 25-fC input electron bunch is produced with a 60-keV DC photoemitting cathode excited by a 350-fs ultraviolet pulse.
  The accelerating gradient in the THz structures demonstrated in this work can be as high as GeV/m with a single-cycle THz pulse of 10\,mJ, which can be readily produced by a 250-mJ infrared pulse when using optimized THz generation \cite{Huang2013}.
  Laser systems producing such and even higher energy picosecond pulses with up to kHz repetition rates are on the horizon \cite{zapata2015cryogenic,calendron2018laser}.
  
  In this experiment, the THz waveguide supports a travelling TM\textsubscript{01} mode that is phase matched to the velocity of the electron bunch produced by the DC photoinjector.
  It is the axial component of the TM\textsubscript{01} mode that accelerates the electrons as they co-propagate down the waveguide.
  A travelling-wave mode is advantageous when considering the available single-cycle THz pulse because it does not require resonant excitation of the structure.
  A dielectric-loaded circular waveguide was selected due to the ease of fabrication in the THz band \cite{mitrofanov2011reducing}.
  The inner diameter of the copper waveguide is 940\,{\textmu}m with a dielectric wall thickness of 270\,{\textmu}m.
  This results in a vacuum space with a radius of 200\,{\textmu}m.
  The significant thickness of the dielectric is due to the low energy of the electrons entering the structure, and will decrease significantly at higher energy.
  One critical aspect for THz electron acceleration is proper interaction between the electron beam and the THz pulse.
  Coupling the radially polarized THz pulse into the single-mode dielectric waveguide was achieved with a centrally loaded dielectric horn.
  The design was optimized to maximize coupling with minimal fabrication complexity.
  Finite element electromagnetic simulations with HFSS \cite{ansoft20093} indicate excellent coupling of the THz pulse over a $\sim 200$\,GHz bandwidth, which is compatible with the bandwidth of the radially polarized mode converter.
  The accelerating waveguide is 10\,mm in length, including a single tapered horn for coupling the THz into the waveguide.
  Alignment between the THz waveguide and the DC gun is provided by a pin-hole aperture in a metal plate with a diameter of 100\,{\textmu}m that abuts the waveguide.
  The THz pulse is coupled into the waveguide downstream of the accelerator and it propagates along the full length of the waveguide before being reflected by the pin-hole aperture, which acts as a short at THz frequencies.
  After being reflected the THz pulse co-propagates with the electron bunch.
  The low initial energy of the electrons results in the rapid onset of a phase-velocity mismatch between the electron bunch and the THz pulse once the electrons have been accelerated by the THz pulse and this limits the interaction length to 3\,mm.
  
  \subsection{Observation of Acceleration}
  
\begin{figure}
  	\centering
  	$\begin{array}{cc}
  	\includegraphics[draft=false,width=3.0in]{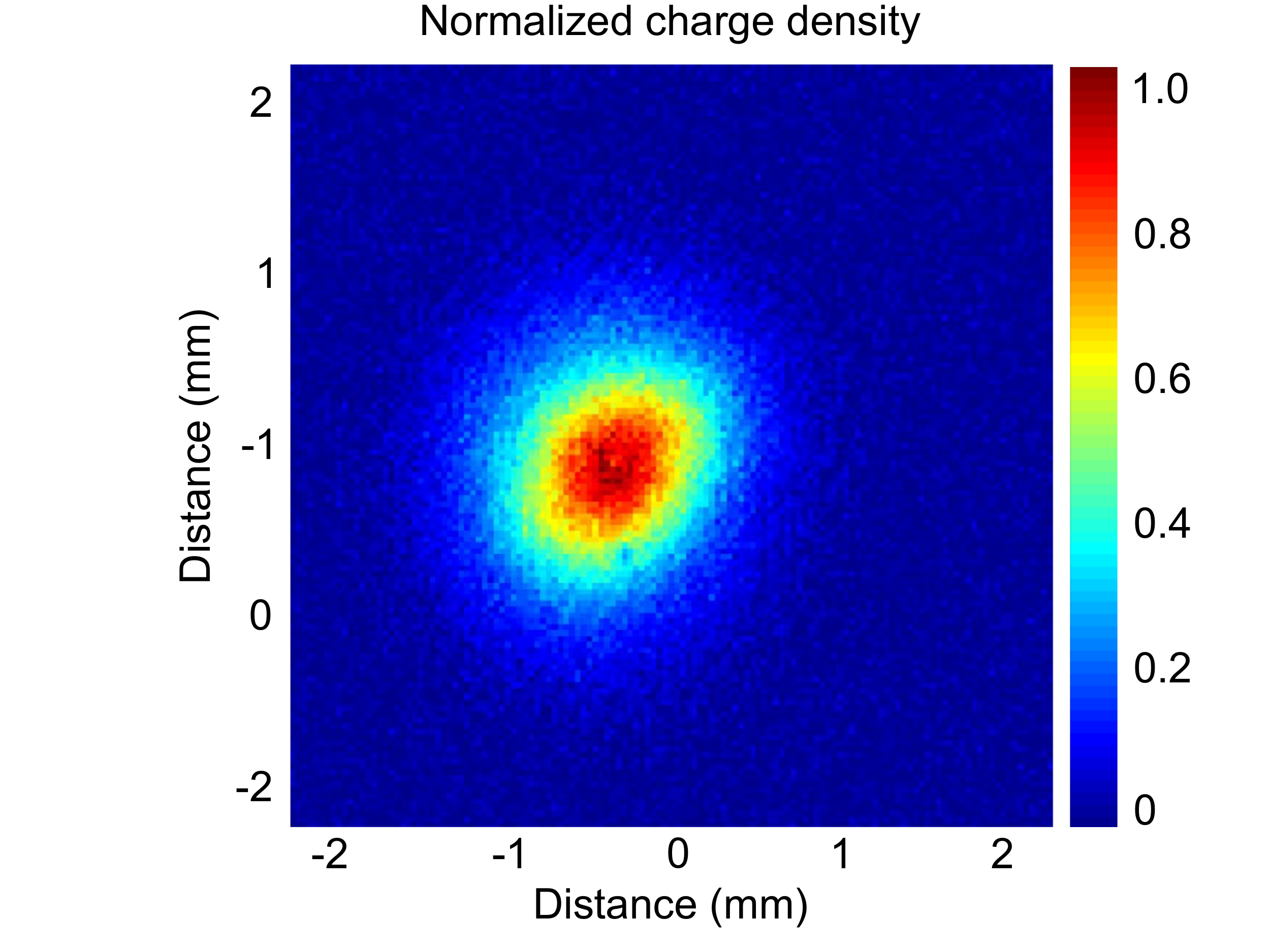} &
  	\includegraphics[draft=false,width=3.0in]{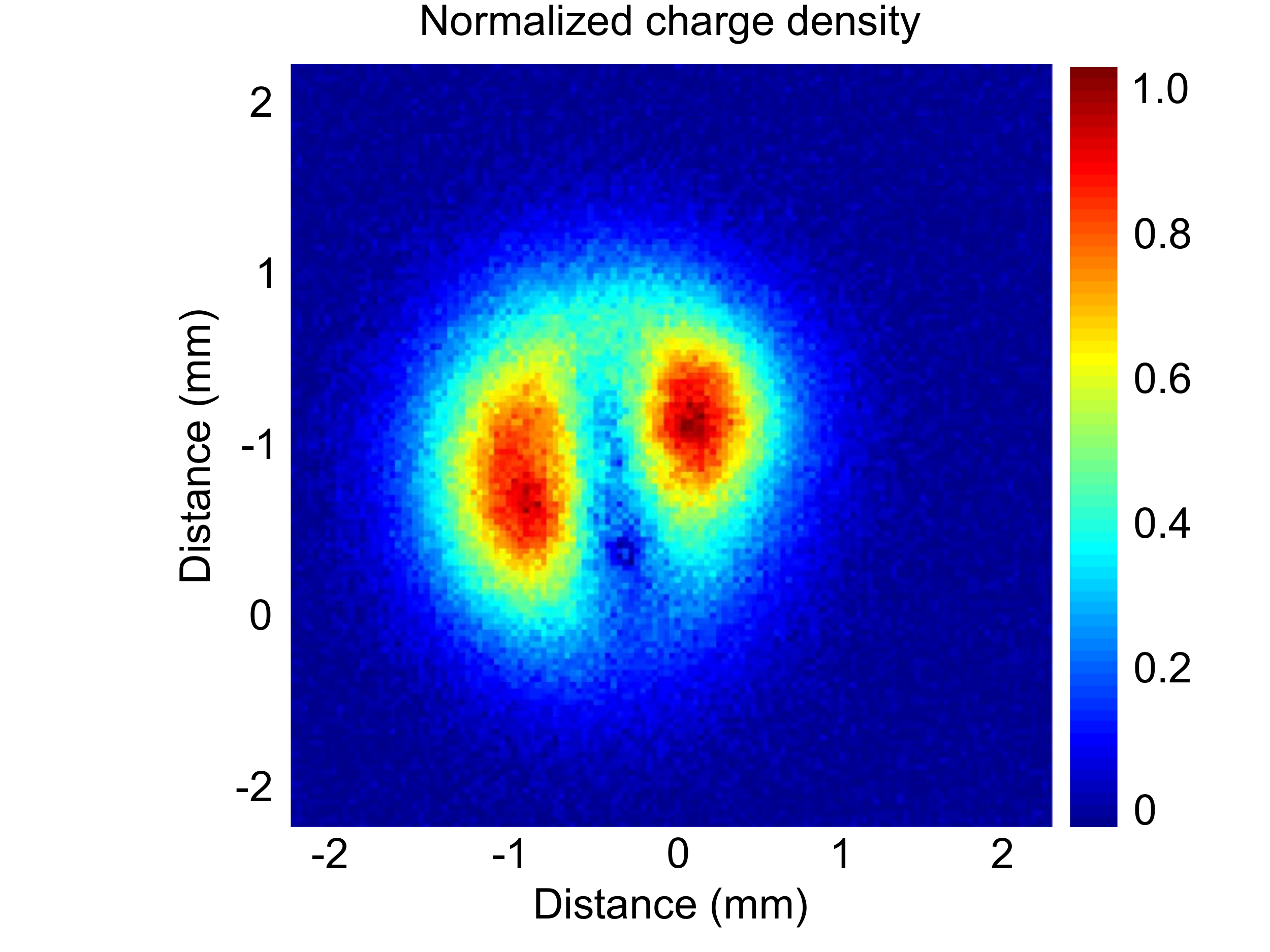} \\
  	(a) & (b) \\
  	\includegraphics[draft=false,width=3.0in]{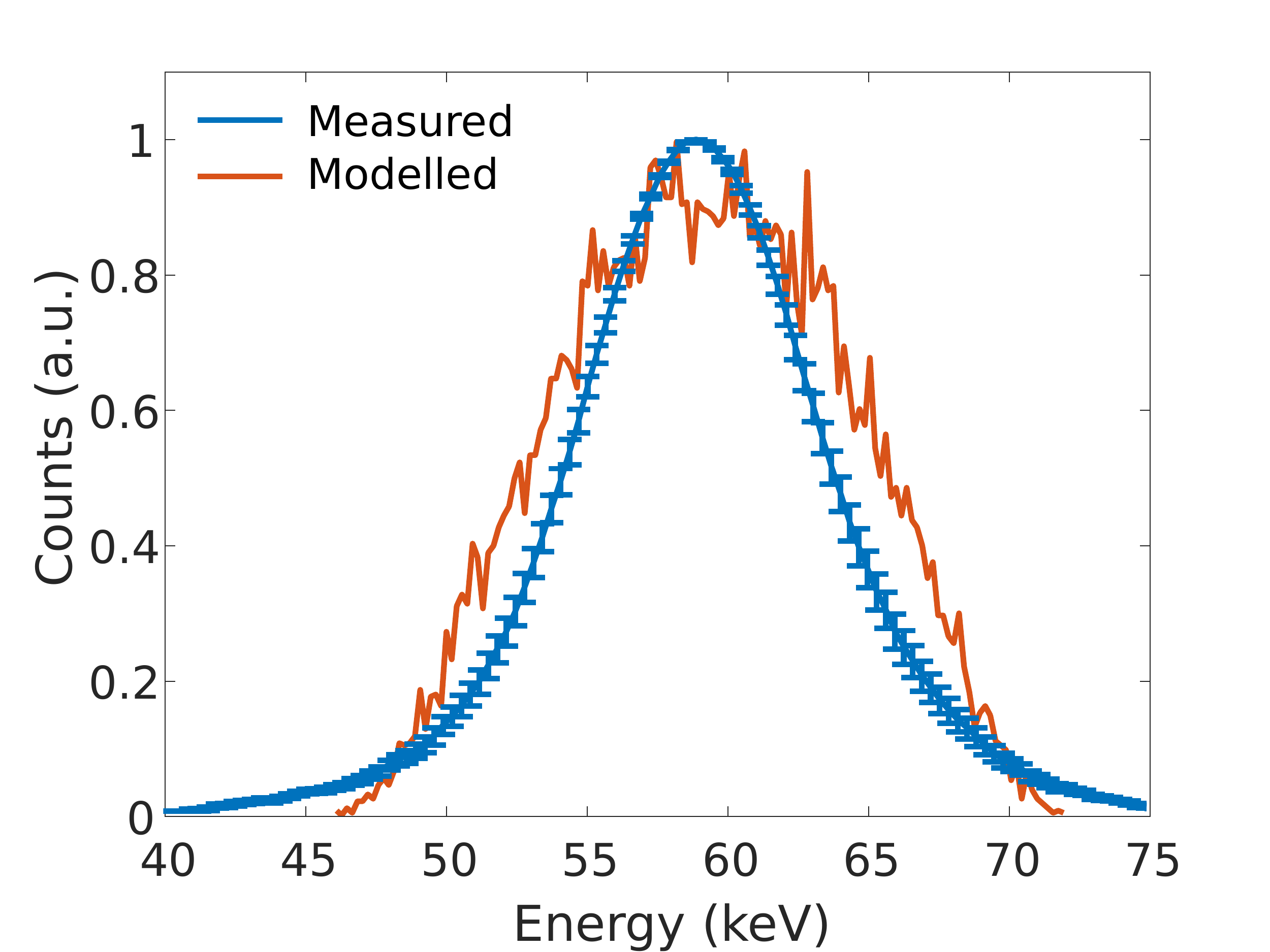} &
  	\includegraphics[draft=false,width=3.0in]{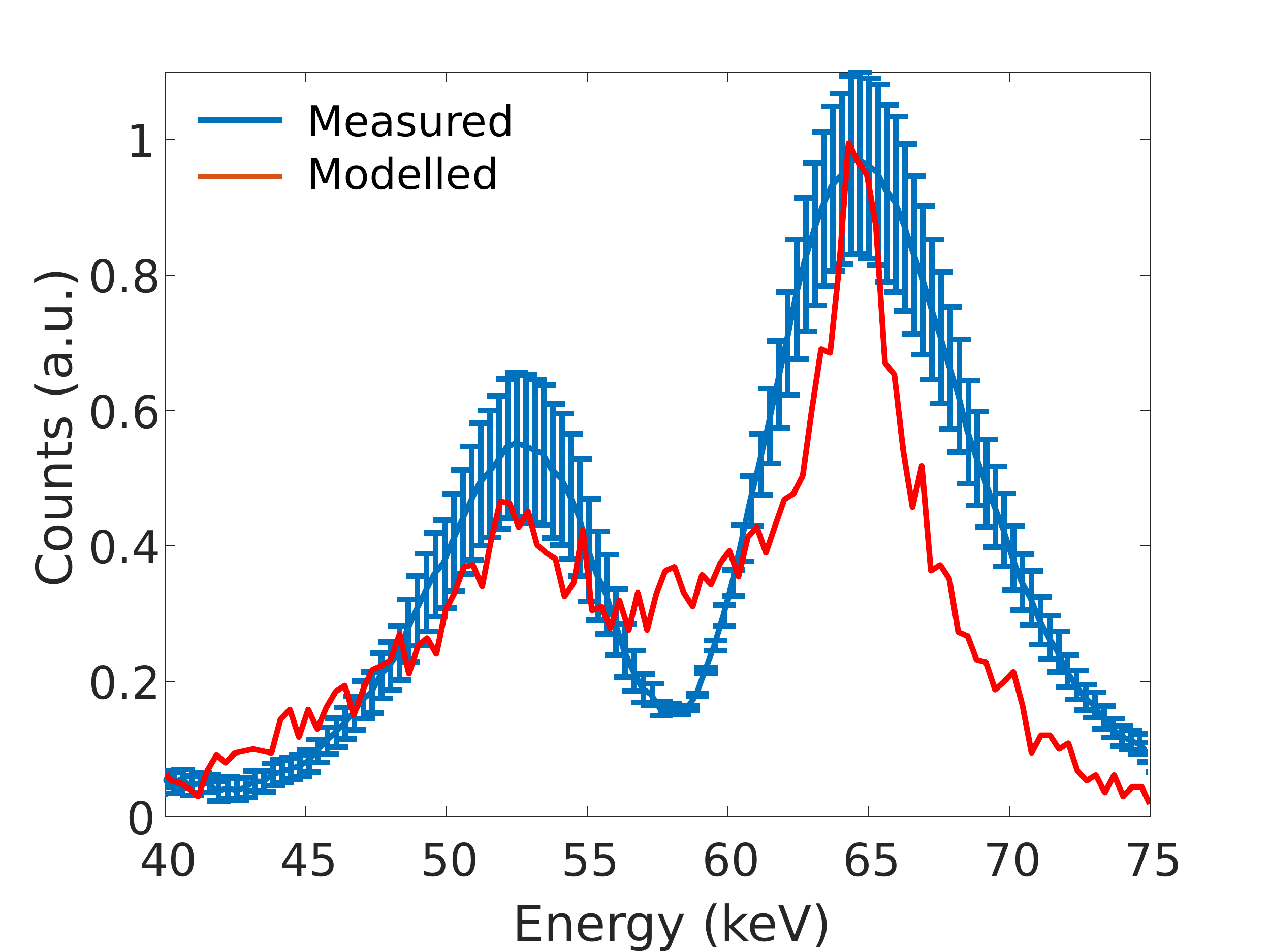} \\
  	(c) & (d) \\
  	\end{array}$
  	\caption{Transverse electron density of the electron bunch as recorded by a micro-channel plate (MCP) at 59\,kV for (a) THz off and (b) THz on. The bimodal distribution is due to the presence of accelerated and decelerated electrons, which are separated spatially by the magnetic dipole energy spectrometer. The images are recorded over a 3-s exposure at 1\,kHz repetition rate. (c) Comparison between simulated (red) and measured (black) energy spectrum of the electron bunch measured at the MCP due to the deflection of the beam by a magnetic dipole. At 59\,keV and with 25\,fC per bunch, the simulation predicts a $\sigma_\bot=513$\,{\textmu}m and $\Delta E = 1.25$\,keV. The observed $\Delta E/E$ appears larger due to the large transverse size of the beam. After the pinhole, the transverse emittance is 25\,nm$\cdot$rad and the longitudinal emittance is 5.5\,nm$\cdot$rad. (d) Comparison between simulated (red) and measured (black) electron bunch at MCP after acceleration with THz. Decelerated electrons are present due to the long length of the electron bunch, as well as the slippage between the THz pulse and the electron bunch. Error bars in (c) and (d) correspond to one s.d. in counts over the data set of 10 integrated exposures.}
  	\label{linacExperimentDemonstration}
  \end{figure}

  The electron beam energy is determined via energy-dependent magnetic steering with a dipole located after the accelerator.
  Fig.\,\ref{linacExperimentDemonstration}a and \ref{linacExperimentDemonstration}b shows images of the electron beam produced by the micro-channel plate detector.
  The measured energy spectrum from the electron bunch with and without THz is shown in Fig.\,\ref{linacExperimentDemonstration}c and \ref{linacExperimentDemonstration}d for an initial mean energy of 59\,keV.
  The curves are compared with PARMELA \cite{young2003particle} PIC simulation results used to model the DC gun and the THz linac.
  The full width of the electron bunch length after the pinhole is 200\,{\textmu}m, which is long with respect to the wavelength of the THz pulse in the waveguide, $\lambda_g = 315$\,{\textmu}m.
  The length of the electron bunch in combination with the phase-velocity mismatch between the electron bunch and the THz pulse results in the observation of both acceleration and deceleration of particles.
  With the available THz pulse energy, a peak energy gain of 7\,keV was observed by optimizing the electron beam voltage and timing of the THz pulse.
  The modelled curve in Fig.\,\ref{linacExperimentDemonstration}d concurred with experiments for an on-axis electric field of 8.5\,MV/m.
  Using this estimated field strength, at the exit of the linac, the modelled transverse and longitudinal emittance are 240 and 370\,nm$\cdot$rad, respectively.
  An increase in emittance from a transverse emittance of 25\,nm$\cdot$rad and a longitudinal emittance of 5.5\,nm$\cdot$rad after a pinhole located at the waveguide entrance is due to the long electron bunch length compared with the THz wavelength and can be easily remedied with a shorter ultraviolet pulse length.
  
  \subsection{Optimization of electron beam interaction with THz pulse}
  
\begin{figure}
  	\centering
  	$\begin{array}{cc}
  	\includegraphics[draft=false,width=3.0in]{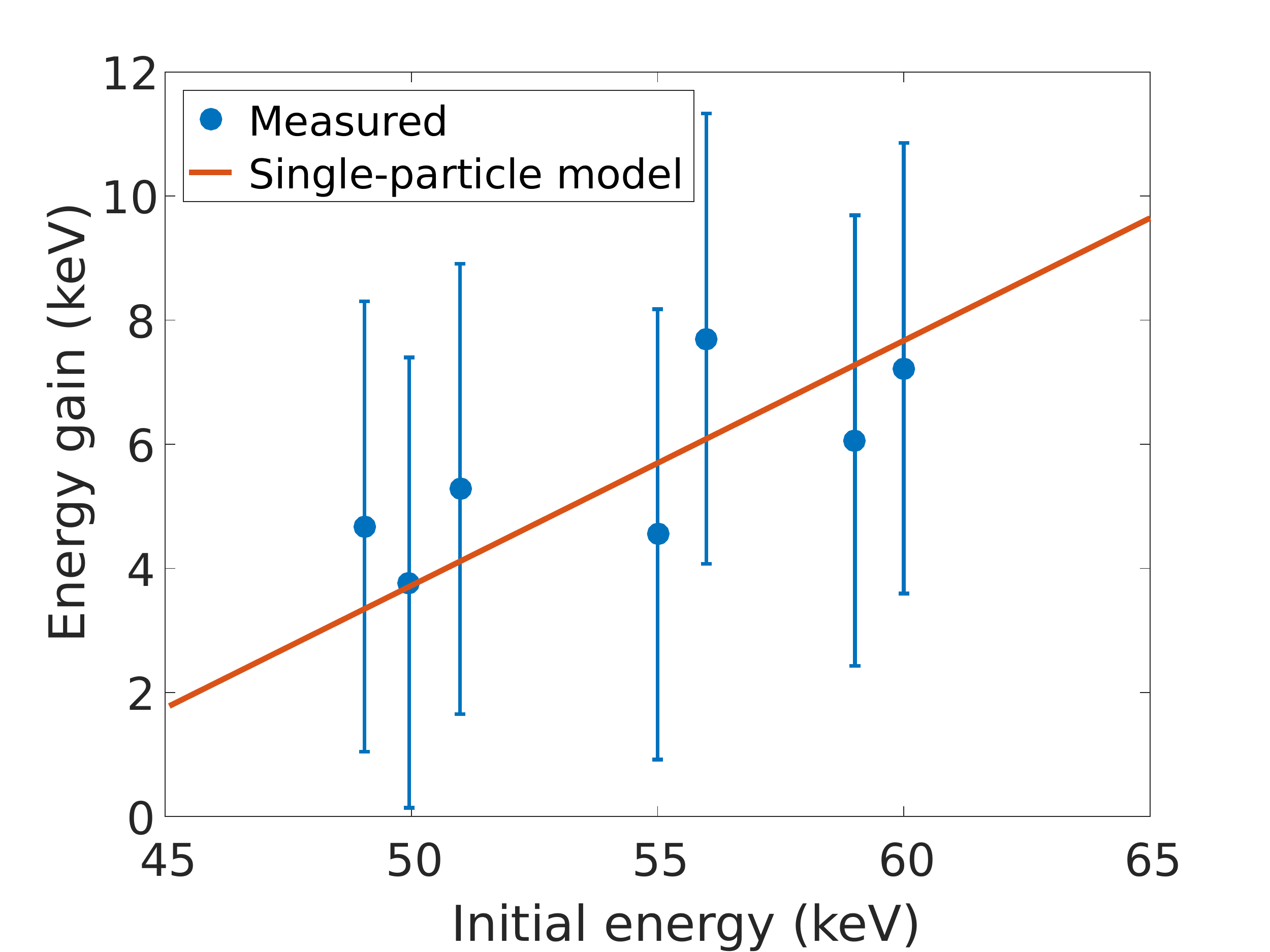} &
  	\includegraphics[draft=false,width=3.0in]{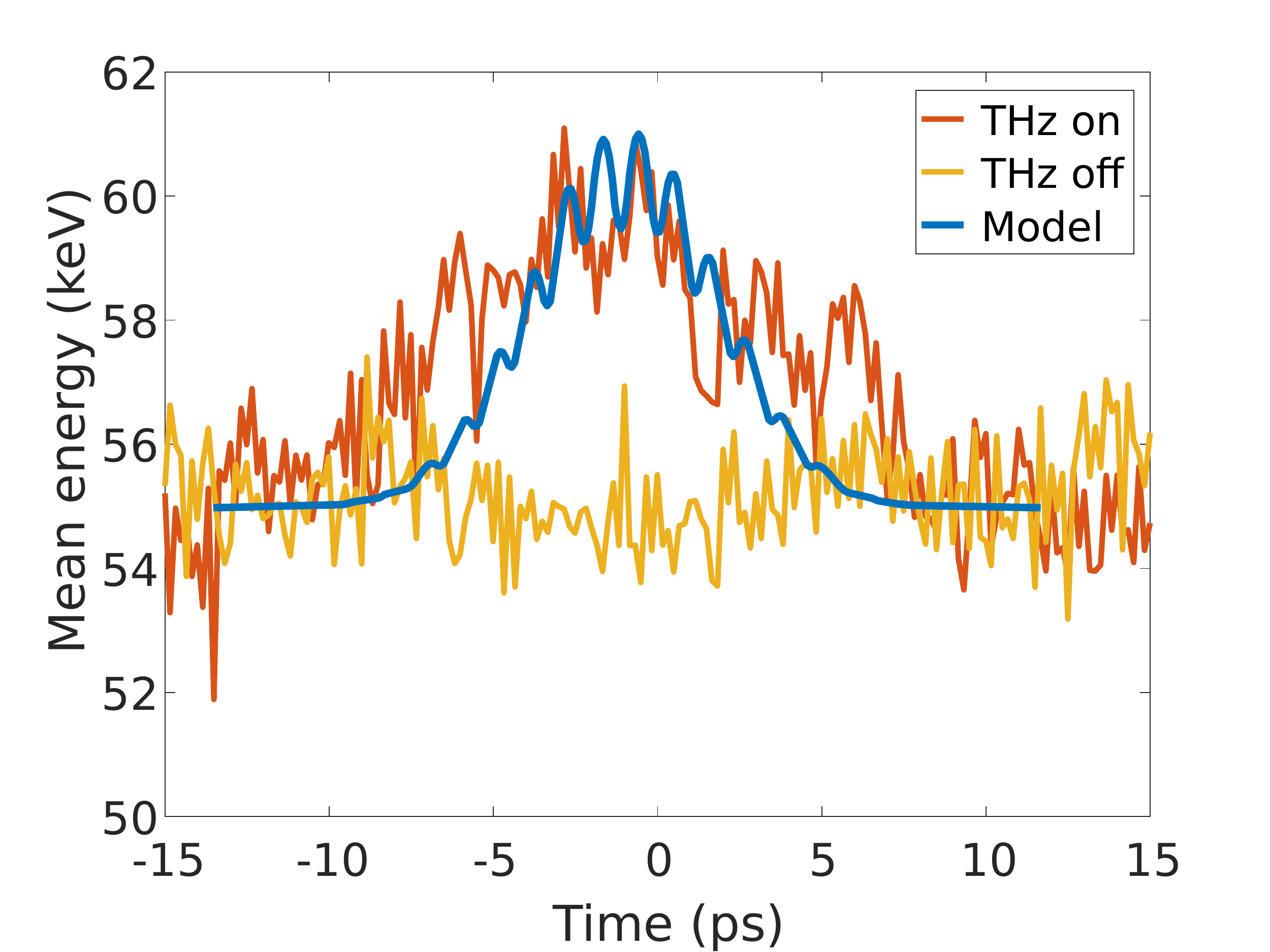} \\
  	(a) & (b)
  	\end{array}$
  	\caption{Acceleration gradient and terahertz phasing. (a) Scaling of energy gain for accelerated electrons as a function of the initial electron energy at the entrance of the THz linac. Black dots with one s.d. error bars are measured values and the red line is a single-particle model. (b) The temporal profile for the mean energy gain of accelerated electrons comparing the THz on and THz off signal against the simulated electron bunch. The initial electron energy was set at 55\,keV to ensure stable performance of the d.c. electron gun over the acquisition time of the data set.}
  	\label{linacExperimentPhasing}
  \end{figure}

  The energy gain achieved during the interaction of the electron bunch with the THz pulse is dependent on the initial energy of the electrons, because the set-up is operated in the non-relativistic limit where the velocity of the electrons varies rapidly.
  If the initial energy is decreased, the particle velocity decreases and the phase-velocity mismatch with the THz pulse increases reducing the interaction length and the acceleration of the particle.
  In Fig.\,\ref{linacExperimentPhasing}a, the achieved mean output energy of the electron bunch is shown versus the initial energy.
  Higher initial energy was experimentally found to be favourable for higher energy gain.
  This observation is in agreement with a single-particle model \cite{Wong2013} for a peak on-axis electric field of 8.5 MV/m and an effective accelerating gradient of 2.5\,MeV/m.
  In Fig.\,\ref{linacExperimentPhasing}b, the mean energy of accelerated electrons is shown as a function of the THz pulse delay for 55-keV initial energy.
  The large temporal range of observed acceleration results from waveguide dispersion, which broadens the single-cycle pulse temporally as it propagates.
  Accelerated electrons are observed over the full range of the phase-matched THz pulse due to the long length of the electron bunch as it enters the THz waveguide.
  Modelling indicates that at the entrance of the THz waveguide for this initial energy, the electron bunch full width is 1.5\,ps in length, which is already a significant fraction of the THz cycle (2.2\,ps).
  
  \section{Conclusion}
  
  The quest to realize an efficient, practical compact accelerator for electron bunches of substantial charge will likely involve a tradeoff between the large wavelengths but low accelerating gradient of RF accelerators, and the high accelerating gradient but small wavelengths available at optical frequencies.
  The trade-off between accelerating gradient and wavelength, together with the emergence of efficient methods to generate coherent pulses at THz frequencies, make electron acceleration at THz frequencies a promising candidate for the substantial acceleration and compression of pico-Coulomb electron bunches.
  In this chapter, we numerically demonstrated the acceleration of a 1.6\,pC electron bunch from a kinetic energy of 1\,MeV to one of 10\,MeV over an interaction distance of about 20\,mm, using a 20\,mJ pulse centered at 0.6\,THz in a dielectric-loaded metallic waveguide.
  We have also analyzed the implications of using an arbitrarily distant injection point, as well as the prospects of dielectric breakdown and thermal damage for our optimized design.
  
  Next, to show the feasibility of this scheme, optically generated THz pulses were used to accelerate electrons in a simple and practical THz accelerator.
  An energy gain of 7\,keV was achieved over a 3-mm interaction length with good modelled emittance.
  Performance of these structures improves with an increase in electron energy and gradient making them attractive for compact accelerator applications.
  With upgrades to pump laser energy and technological improvements to THz sources \cite{chen2011generation}, laboratory demonstration of GeV/m gradients in THz linacs is realistic.
  Multi-GeV/m gradients and $>10$\,MeV energy gain are achievable in dielectric-loaded circular waveguides with 10-mJ THz pulses and the injection of electrons at relativistic energies.
  The available THz pulse energy scales with infrared pump energy, with recently reported results of mJ THz pulse energies and $\sim 1\%$ infrared-to-THz conversion efficiencies \cite{Fulop2011}.
  Multiple stages of THz acceleration can be used to achieve higher energy gain with additional infrared pump lasers for subsequent stages.
  Timing jitter will improve on the jitter of conventional accelerators since the accelerating field and photoemitting pulse are produced by the same drive laser.
  Therefore, one expects the resulting electron bunch to have tighter synchronization than possible in today's RF-based accelerators, where the photoemitting laser pulse is synchronized to the RF drive by standard RF techniques (that is, phase locked loops operating at GHz speeds).
  The presented proof-of-principle THz linear accelerator demonstrates the potential for an all-optical acceleration scheme that can be readily integrated into small-scale laboratories providing users with electron beams that will enable new experiments in ultrafast electron diffraction and x-ray production.
  
  \chapter{Start-to-End Simulation of a THz-Driven Light Source \label{chap:five}}
  
\begin{figure}
  	\centering
  	\includegraphics[width=6.0in]{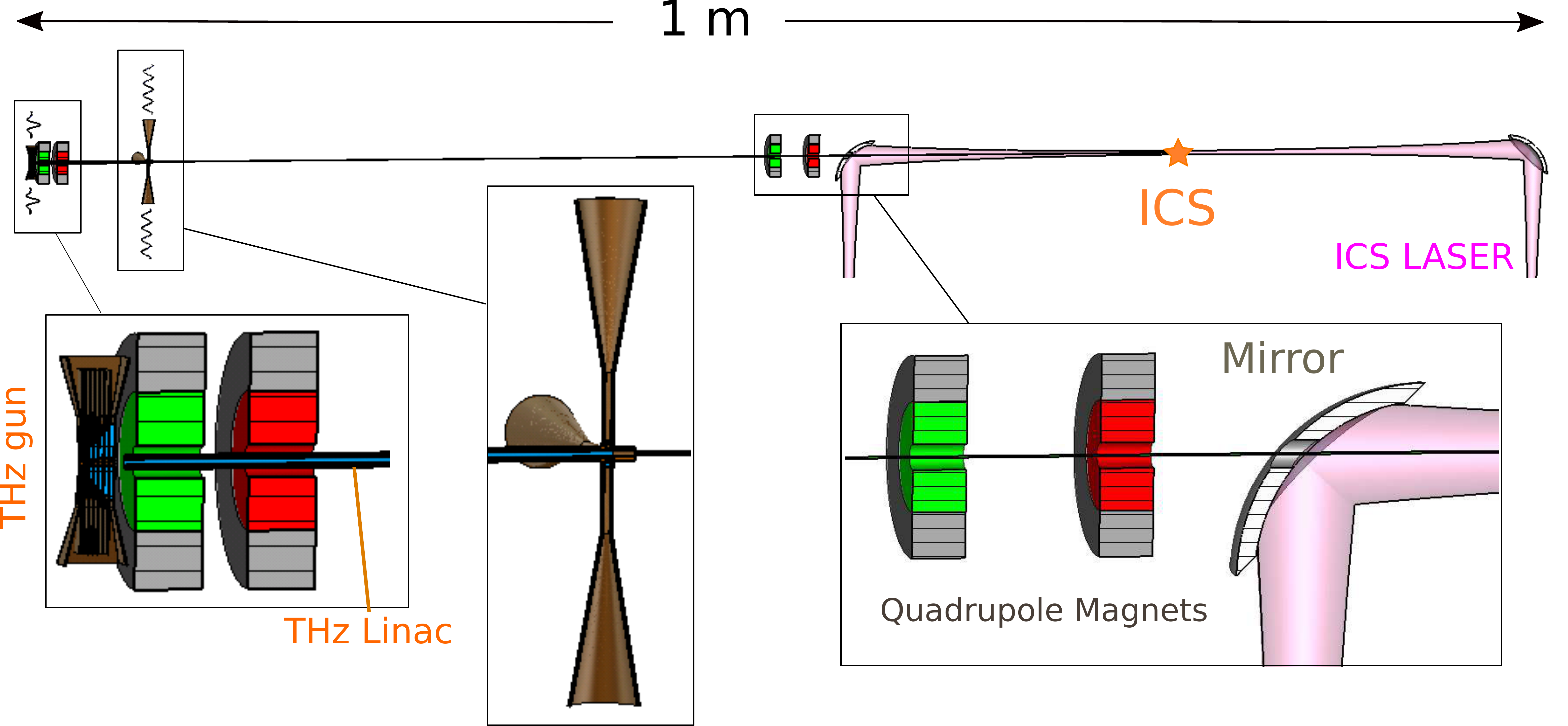}
  	\caption{Layout of the THz-driven compact x-ray source.}
  	\label{THzICSlayout}
  \end{figure}

  In chapter 2, a review on the inverse Compton scattering (ICS) sources, their importance and their applications were carried out.
  In this chapter, we present the start-to-end simulation of a THz-driven x-ray source, whose layout is schematically illustrated in Fig.\,\ref{THzICSlayout}.
  Electrons are generated via photo emission from a metallic cathode which is embedded in a THz gun driven by a THz source with just a single-cycle of an electromagnetic wave.
  Such single-cycle THz signals drive the ultrafast electron gun, which accelerates electrons up to 0.78\,MeV kinetic energy.
  The electrons then traverse through a dielectric-loaded metallic waveguide, which functions as a linear accelerator (linac) and boosts the electrons energy up to 19 MeV.
  A lattice of quadrupole magnets is used to keep the electron bunch collimated and transport it to the ICS interaction section.
  The main goal of this chapter is to present a design including full start-to-end simulations for a THz-driven light source, and to explore the requirements of the source.
  Certainly, technical implementation of the designed source needs extensive tolerance studies, which show the influence of deviations from designed values on the overall performance.
  However, such studies are not in the scope of this thesis and we refer the reader to cited documents in each section, where detailed sensitivity analyses are presented.
  
  This chapter is organized as follows:
  Next section describes the design of two laser-driven THz sources, which provide the required single-cycle and multi-cycle THz pulses.
  Simulation and design of the THz injector including the photo-emission process and the ultrafast single-cycle THz guns is the focus of the subsequent section.
  Next, we use analytic calculations for propagation of the electromagnetic wave in THz waveguides to evaluate the linear acceleration section and compute the electron bunch properties exiting the linac.
  To overcome difficulties in high power THz generation, a novel coupler is designed, which receives four THz multi-cycle inputs and simultaneously combines and couples the beams to a single TM\textsubscript{01} guided mode, used for accelerating electrons in the linac.
  The design and properties of such a coupler is thoroughly explained in this chapter.
  Eventually, the ICS interaction between the linac output and a counter-propagating laser pulse is analyzed to obtain the achievable photon flux of the designed x-ray source.
  
  \section{Terahertz generation}
  
  As described previously, the main challenge in the regime of THz accelerators is the so-called ``THz gap" problem, which refers to the lack of efficient radiation sources in this part of the spectrum.
  As a result, the acceleration of electrons for a THz-driven source shall be contingent on the development of sources of terahertz radiation with unprecedented performance.
  In order to achieve the requisite level of electron acceleration, it is envisaged that terahertz sources with pulse energies ranging between a milli-joule (mJ) and few tens of mJ's, in the frequency band around 0.3\,THz shall be necessary.
  In this section, we outline our strategy towards meeting the terahertz source requirements through initial designs.
  
  As depicted in Fig.\,\ref{THzICSlayout}, two broad categories of terahertz sources are necessary.
  Single-cycle or broadband terahertz pulses with 400\,{\textmu}J pulse energy accelerate electrons from rest up to 0.78\,MeV of electron energy via an electron gun.
  The second category comprises of multi-cycle or narrowband terahertz sources with pulse energies in excess of 10\,mJ which may be employed in linear accelerator structures to boost the energy of electron bunches emerging from the gun to the 20\,MeV level.
  
  A number of methods for the generation of terahertz radiation exist.
  These include solid-state devices, photoconductive switches \cite{yardimci2015high}, vacuum electronic devices \cite{gold1997review} free-electron lasers \cite{gallerano2004overview} and laser driven approaches employing nonlinear optics \cite{cronin2004cascaded,huang2015highly,fulop2014efficient,carbajo2015efficient,ravi2014limitations,ravi2015theory,ravi2016pulse}.
  Solid-state devices and photoconductive switches are limited in their scalability to large terahertz pulse power while free-electron lasers are relatively inaccessible to be employed for the purpose of the current machine.
  Consequently, laser driven approaches based on nonlinear optical frequency conversion are preferred for THz source implementation.
  These schemes provide the unique possibility of generating high energy, coherent terahertz sources with precisely controlled terahertz beam parameters such as energy, phase, and temporal pulse shape.
  Furthermore, the use of laser driven approaches helps to generate well synchronized driver pulses for gun, linac and ICS stage.
  In particular, terahertz sources driven by the same one micron (1\,{\textmu}m) laser technology \cite{zapata2015cryogenic} utilized for inverse Compton scattering and photoemission offers the possibility of a high level of synchronization.
  
  \subsection{Single-cycle THz generation}
  
  Laser source technology in the 1\,{\textmu}m wavelength regime is highly promising, since it offers the possibility of generating pulses at the Joule level with high repetition rates up to the kHz level \cite{baumgarten20161}.
  The choice of laser wavelength in turn places constraints on the choice of nonlinear material that may be used for terahertz generation.
  Lithium Niobate with a bandgap energy of 3.8\,eV is not only transparent to 1\,{\textmu}m light, but also circumvents the issue of multi-photon and tunneling ionization, which leads to nonlinear absorption as is the case for Gallium Arsenide and Zinc Telluride.
  
  The second order nonlinearity is the most important parameter for terahertz generation, making lithium niobate, with its very large second order nonlinear coefficient ($\chi^{(2)}_\mathrm{eff} = 336$\,pm/V), the most suited for generation of sub-terahertz transients.
  Furthermore, it possesses a moderate third order nonlinearity or equivalently, the nonlinear refractive index ($n_2=0.9 \times 10^{-19}$\,W/m$^2$).
  The relatively large terahertz absorption coefficient at room temperature maybe overcome by cryogenic cooling to 100\,K as has been demonstrated \cite{huang2015highly}.
  
\begin{figure}
  	\centering
  	$\begin{array}{c}
  	\begin{array}{cc}
  	\includegraphics[draft=false,width=3.5in]{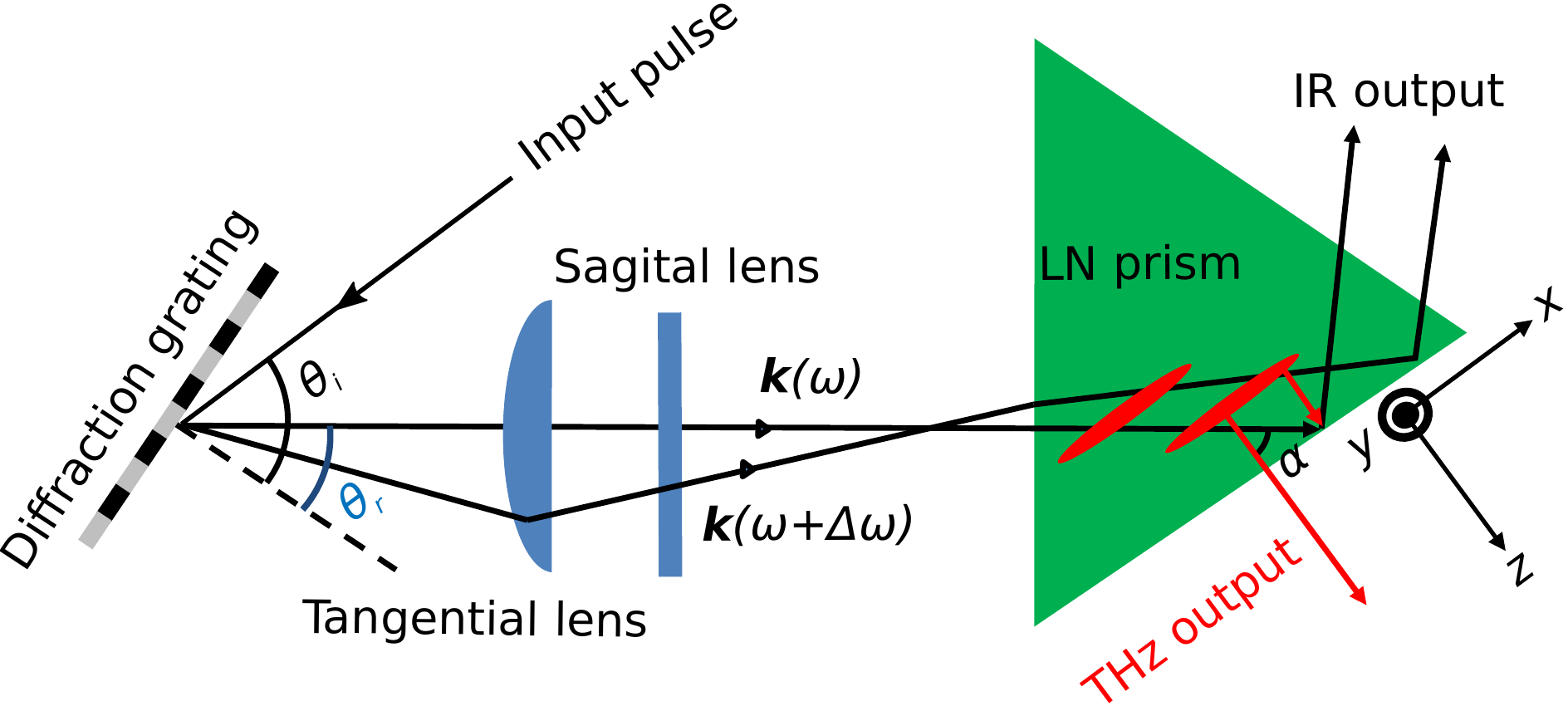} &
  	\includegraphics[draft=false,width=2.0in]{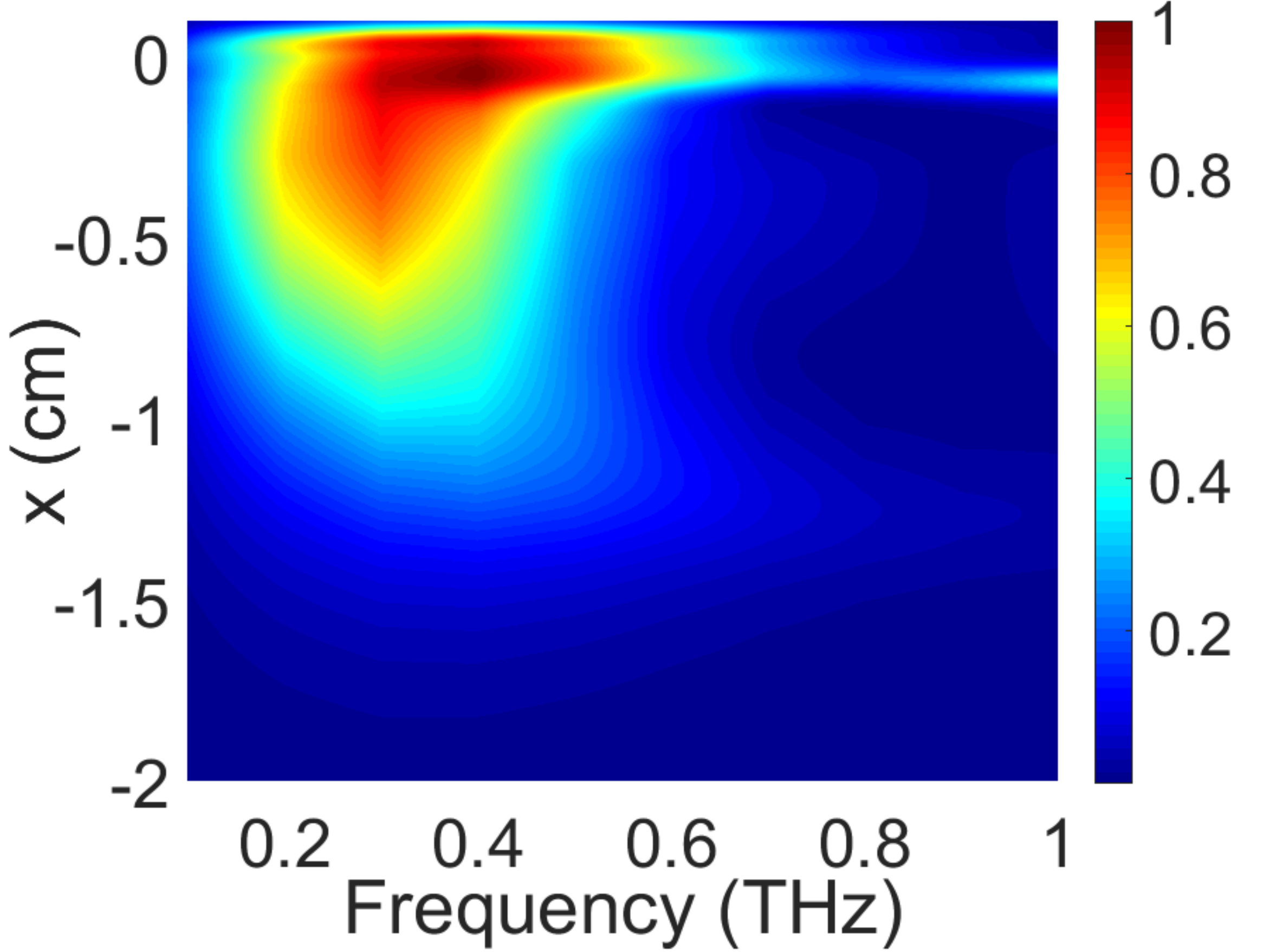} \\
  	(a) & (b)
  	\end{array} \\
  	\begin{array}{ccc}
  	\includegraphics[draft=false,width=2.0in]{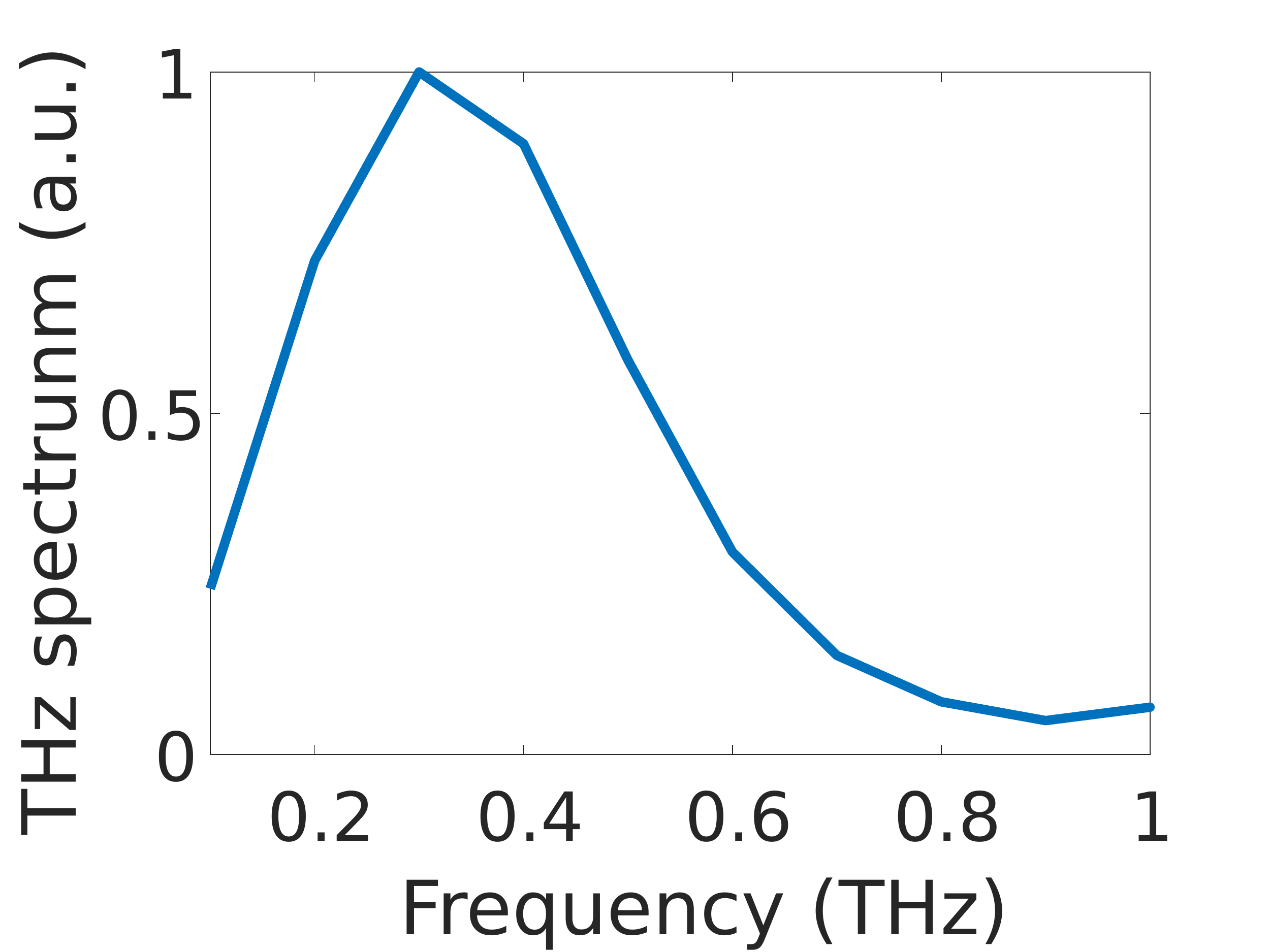} &
  	\includegraphics[draft=false,width=2.0in]{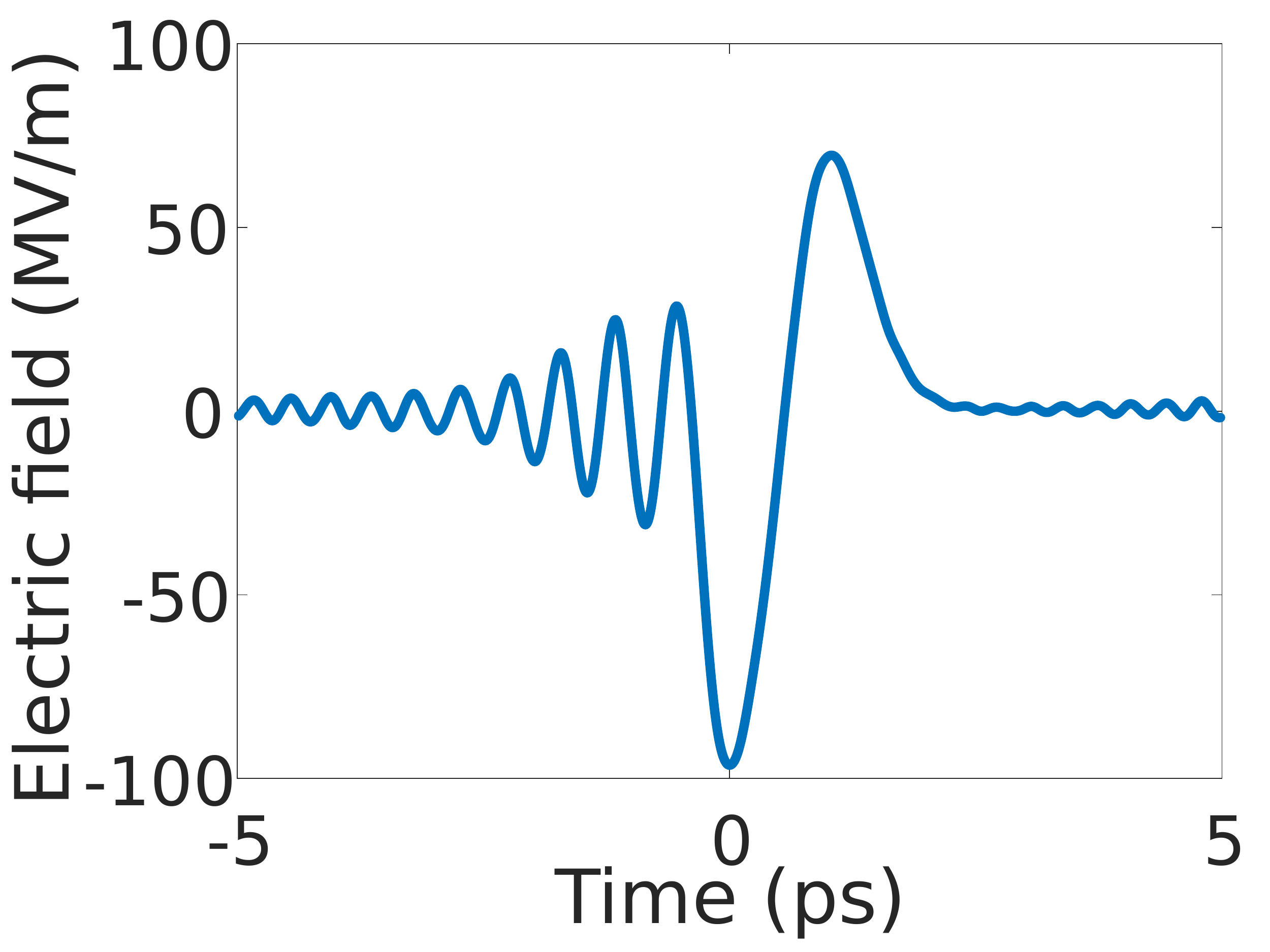} &
  	\includegraphics[draft=false,width=2.0in]{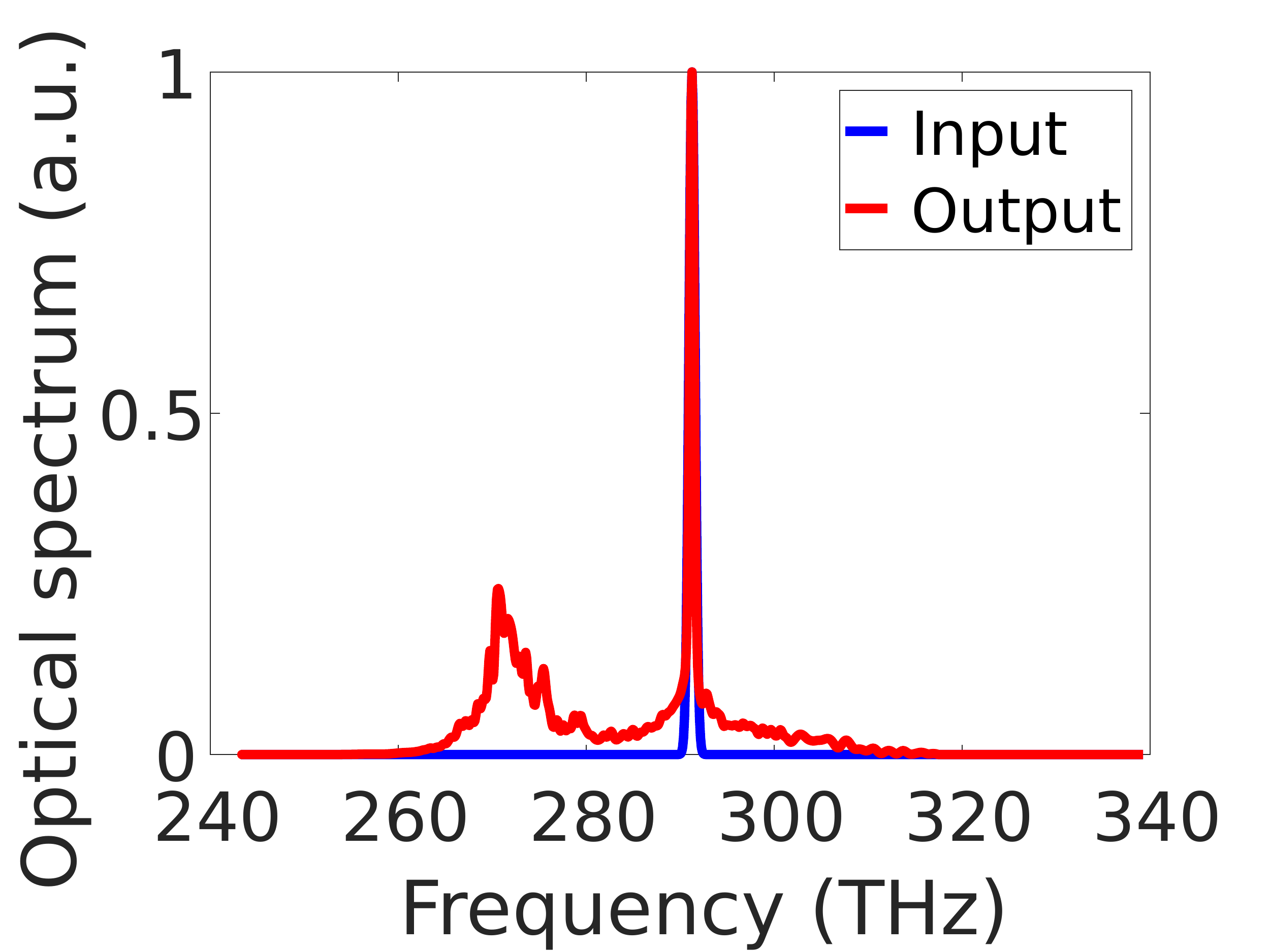} \\
  	(a) & (b) & (c)
  	\end{array}
  	\end{array}$
  	\caption{Single-cycle THz generation results: (a) Schematic illustration of the titlted-pulse-front scheme. (b) Spatially-dependent spectrum of the generated THz beam as a function of transverse position. (c) The average spectrum of the generated THz beam, and (d) average THz wave-form versus time. (e) Illustration of the cascading process as the red-shift and spectral broadening of the pump pulse.}
  	\label{ICSTHzTPF}
  \end{figure}

  Single-cycle terahertz generation occurs via an ensemble of difference frequency generation processes within the bandwidth of the optical pump pulse.
  This process is also known as optical rectification.
  As with any second order nonlinear process, the coherent superposition of transients generated at various points in space or phase-matching is necessary for efficient terahertz generation.
  In lithium niobate, there is a large disparity in the refractive index of the optical pump and generated terahertz radiation.
  Thus, collinear phase-matching is not feasible.
  In order to produce broadband phase-matching, Hebling \cite{hebling2002velocity} proposed the use of angularly dispersed beams to achieve non-collinear phase matching.
  The angularly dispersed beams maybe generated by reflecting an ultrafast optical pump pulse from a diffraction grating and imaging it onto a lithium niobate prism as shown in Fig.\,\ref{ICSTHzTPF}a.
  In time, the pulses appear tilted with respect to their propagation direction (as delineated by red ellipses in Fig.\,\ref{ICSTHzTPF}a) and are hence referred to as tilted-pulse-fronts (TPF).
  The phase-matching condition is given by $v_g = v_\mathrm{THz} \cos\alpha$, where $v_g$ and $v_\mathrm{THz}$ are the optical group velocity and terahertz phase velocity, respectively, and $\alpha$, known as the TPF angle is the angle between the pump and generated pulse.
  In the case of lithium niobate $\alpha \sim 62^\circ$ at 100\,K.
  
  To evaluate the performance of the required TPF setups, we employ our previously developed numerical model \cite{ravi2015theory} which solves for the coupled nonlinear interaction of the optical pump pulse and generated terahertz pulse in two spatial dimensions $(x,z)$ and time/frequency.
  The third spatial dimension ($y$) maybe ignored since diffraction effects of both terahertz and pump waves in this direction are negligible for large beams.
  The model accounts for various spatio-temporal distortions of the optical pulse such as angular dispersion and spatial chirp, which enables accurate quantitative predictions as verified by prior experiments \cite{wu2014terahertz,wu2016optical}.
  
  In order to generate terahertz radiation centered at approximately 0.3\,THz for the designed electron injector (or ultrafast terahertz gun), two optical pump pulses centered at 1.030\,{\textmu}m with a transform limited pulse duration of 800\,fs (FWHM) and $\sim 70$\,mJ pulse energy were assumed.
  Such pulses can be produced by Ytterbium:YLF lasers as demonstrated \cite{zapata2010power}.
  The input beam size is assumed to be of radius 1\,cm ($e^{-2}$ radius) with a peak fluence of 45\,mJ/cm$^2$.
  The beam undergoes demagnification by a cylindrical lens by a factor of 2 in the $x$-$z$ plane and by 1.25 in the $x$-$y$ plane by another cylindrical lens to yield a fluence of 100\,mJ/cm$^2$ on the input surface of the lithium niobate prism.
  The fluence was chosen to be within the laser-induced damage threshold although experimentally much higher fluences up to 180\,mJ/cm$^2$ have been used \cite{fulop2014efficient}.
  The output facet of the prism is coated with a terahertz anti-reflection coating which can enable up to 25\% improvement in transmission, for instance using Kapton tapes of appropriate thickness.
  
  The conversion efficiency predicted is 1.48\% at $T=100$\,K, after accounting for scaling to three spatial dimensions and effects of the anti-reflection coating.
  Thus, the generated terahertz pulse energy is expected to be in the range of 1\,mJ.
  
  The spatially dependent terahertz spectrum is plotted in Fig.\,\ref{ICSTHzTPF}b.
  As can be seen, the spectrum is highly spatially chirped due to the non-collinear nature of the process and the presence of spatio-temporal distortions in the optical pump pulse.
  This is one of the disadvantages of the tilted-pulse-front approach as the spatially chirped terahertz pulse results in poor propagation characteristics, thus degrading the transport of THz energy to the injector.
  As a result, we have assumed that only 50\% of the generated pulse energy is transported and coupled into the designed electron injector.
  
  The average spectrum, is presented in Fig.\,\ref{ICSTHzTPF}c, and is clearly centered at 0.3\,THz.
  The terahertz waveform is presented in Fig.\,\ref{ICSTHzTPF}d, and can be clearly seen to be close to a single-cycle waveform.
  The faster oscillations in the waveform were found to negligibly impact the acceleration process since they arise from higher frequency terahertz transients that appear to get ``cleaned-out'' in the process of being coupled to the designed electron injector.
  An important aspect of terahertz generation by second order nonlinear processes is known as cascading, which is the repeated energy down-conversion of the pump photons.
  This results in a pronounced red-shift and spectral broadening of the pump pulse as shown in Fig.\,\ref{ICSTHzTPF}e.
  
  \subsection{Multi-cycle THz generation}
  
  The proposed approach for multi-cycle terahertz generation is to beat two high energy, narrowband optical pulses separated by the desired terahertz frequency.
  Since the generated radiation is desired to be quasi-monochromatic, narrowband phase matching based on periodically poled lithium niobate crystals may be used for collinear phase-matching of optical and terahertz radiation.
  Multi-cycle terahertz generation in periodically poled crystals or the so-called quasi-phase matching technique has been the focus of extensive research efforts.
  In initial demonstrations, laser-to-terahertz conversion efficienies of only about $10^{-5}$ was achieved \cite{lee2000generation}.
  Later on, research of terahertz generation in PPLN with different pump pulse formats such as chirp and delay, pulse-train and cascaded optical parametric amplification have been conducted \cite{cirmi2017cascaded,ravi2016pulse,vodopyanov2006optical}.
  Recently, conversion efficiencies in excess of $10^{-3}$ and multi-cycle THz pulse energies of $>400$ micro joules have been demonstrated \cite{jolly2018millijoule,fulop2014efficient}.
  Here, we design a terahertz source that receives $\sim 2$\,J of laser pulse energy and generates $\sim 40$\,mJ terahertz pulses with center frequency at 300\,GHz and a duration of $\sim 554$\,ps.
  
\begin{figure}
  	\centering
  	$\begin{array}{cc}
  	\includegraphics[draft=false,width=3.0in]{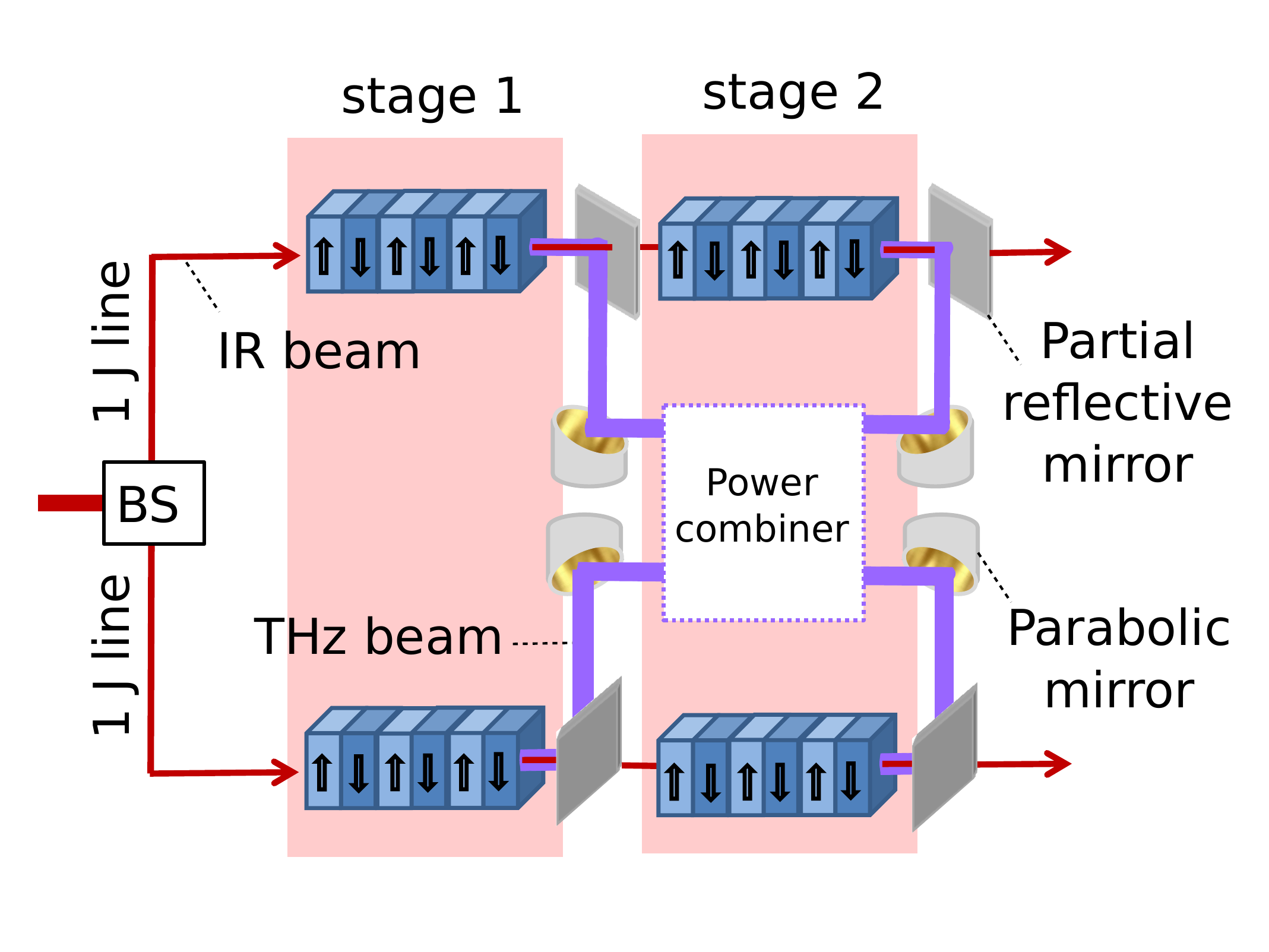} &
  	\includegraphics[draft=false,width=3.0in]{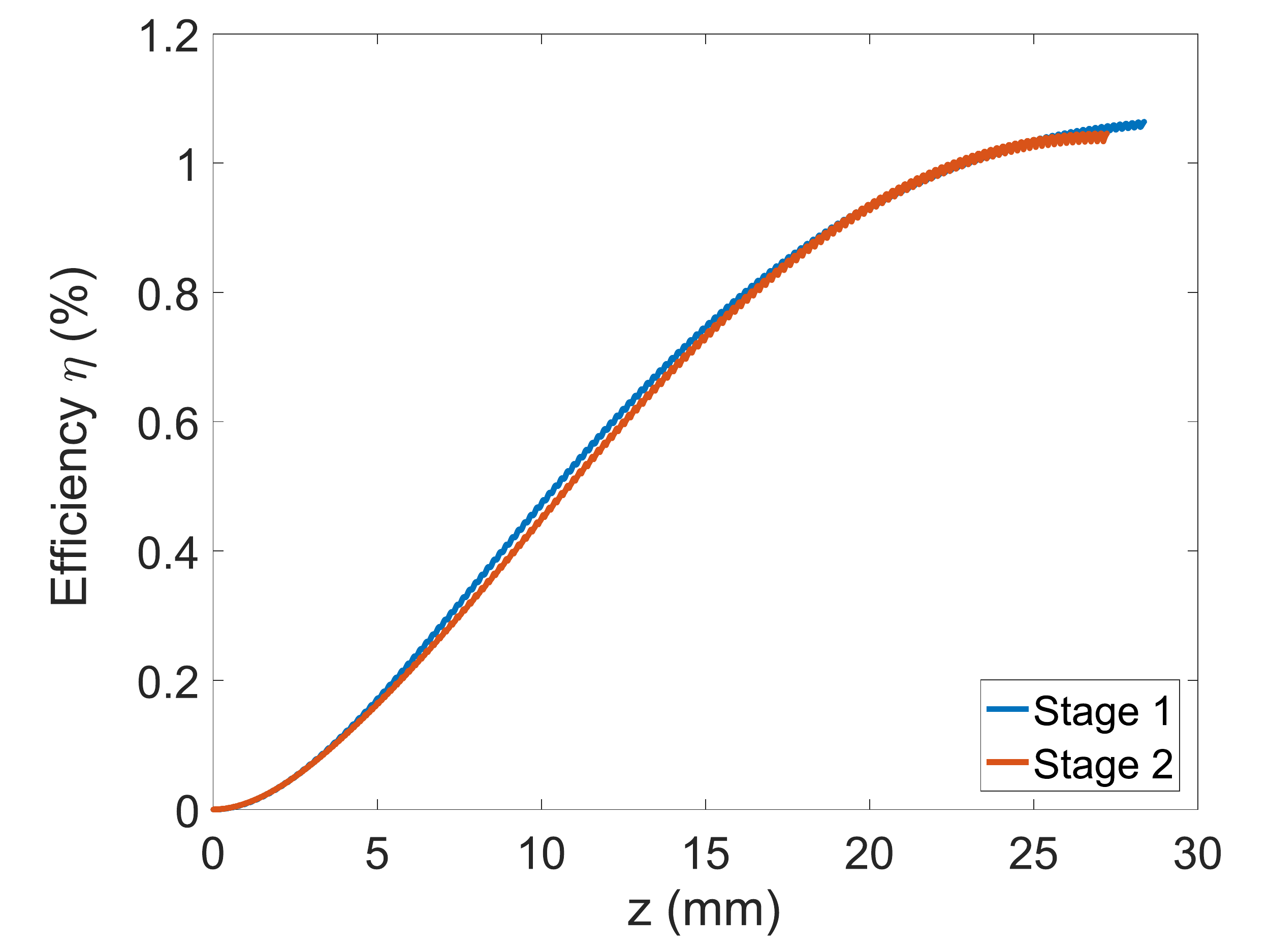} \\
  	(a) & (b) \\
  	\includegraphics[draft=false,width=3.0in]{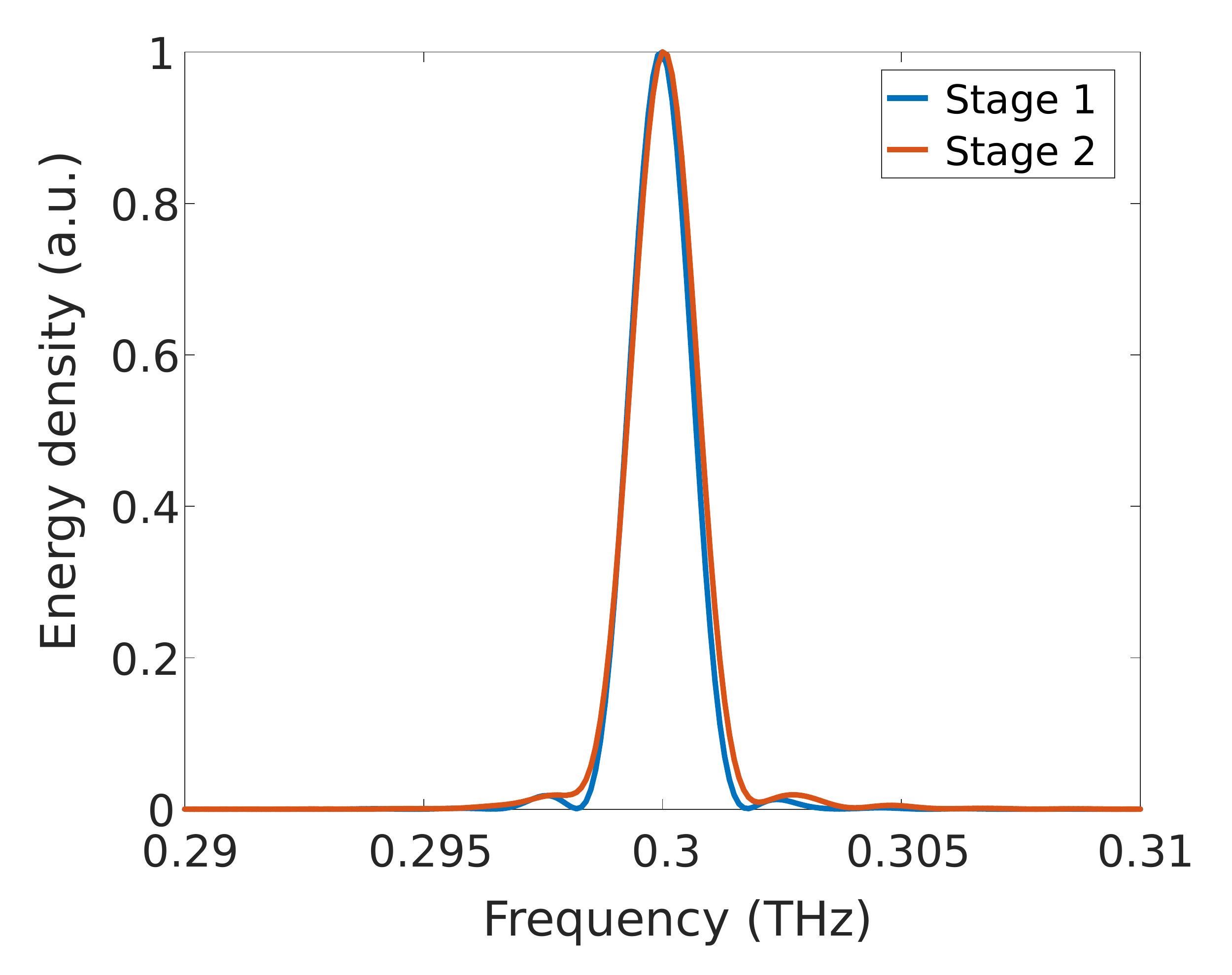} &
  	\includegraphics[draft=false,width=3.0in]{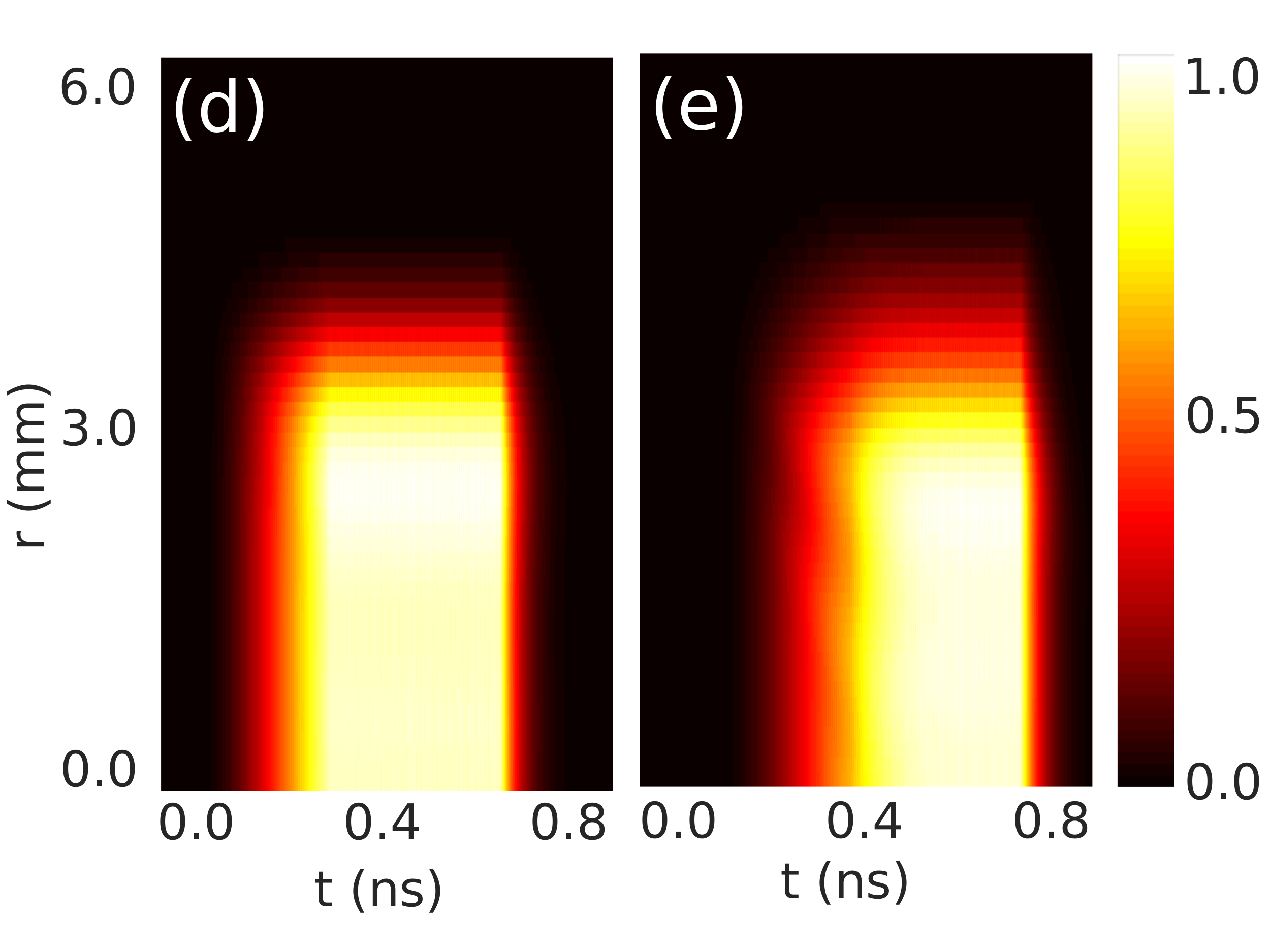}\\
  	(c) & (d) \text{and} (e)
  	\end{array}$
  	\caption{Multi-cycle THz generation: (a) Layout of the laser-driven THz source providing the required multi-cycle radiation. (b) Generation efficiency of each line versus the PPLN length, and (c) THz spectrum of the generated beam are depicted. (d) and (e) show the spatio-temporal beam profile of the THz beams produced in the first and second stages, respectively.}
  	\label{ICSTHzMulti}
  \end{figure}

  The terharetz source layout is illustrated in Fig.\,\ref{ICSTHzMulti}a.
  Using a beam splitter, a pair of optical pulses, with center frequencies separated by 0.3\,THz and total energy of $\sim 2$\,J is firstly split into two copies each having one joule of energy.
  Each of these copies of two pulses are directed to two separate PPLN crystals, where terahertz energy of about $\sim 10$\,mJ is generated at each stage.
  After passage through the crystal, the terahertz and optical pump pulses, can be separated by a specially designed output coupler \cite{wang2018high}.
  The terahertz beams are guided to the input of the beam combiner-coupler at the linac stage, whereas the isolated optical beams illuminate a second stage of PPLN crystals and generate another set of terahertz pulses.
  Subsequently, the two terahertz beams from the second stage are also transported to the input of the combiner-coupler at the linac entrance.
  The main reason for generating the required terahertz energy in four separate stages, is the finite damage threshold of PPLN crystals, and the limited aperture of even state-of-the-art PPLN crystals.
  In our calculations, we considered the previously obtained empirical data for the damage threshold of lithium-niobate to set the maximum beam energy and size at each crystal \cite{ravi2016pulse}.
  According to these results, the maximum beam fluence on the crystal can be obtained from
  \begin{equation}
  F_d^\mathrm{max} = 10\sqrt{\tau_\mathrm{FWHM}/(2\times10\,\text{ns})}.
  \label{maximumFluence}
  \end{equation}
  Furthermore, increasing the length of crystals to increase generation efficiency is not feasible due to terahertz absorption.
  Thus, a pair of two-stage systems increases the interaction length as well as pump energy budget.
  
  The proposed architecture is simulated by solving the coupled nonlinear wave equations for terahertz and optical radiation in cylindrical co-ordinates to account for the rotational symmetry of the beams produced by high energy laser systems \cite{wang2018high}.
  We have benchmarked and verified the developed framework for analysis and design of the multi-cycle THz sources through comparisons with experimental implementations \cite{carbajo2015efficient,ahr2017narrowband,jolly2018millijoule}.
  
  The important parameters of the THz source, and corresponding description are tabulated in table\,\ref{ICSTHzMultiParameters}.
  \begin{table}
  	\center
  	\caption{Parameters used in the simulation}
  	\label{ICSTHzMultiParameters}
  	\begin{tabular}{|l|l|}
  		\hline
  		\textbf{Parameters} & \textbf{Values} \\ \hline \hline
  		$n_2$ \cite{desalvo1996infrared}  & $1.25\times10^{-19}$\,W/m$^2$ \\ \hline
  		$\chi^{(2)}_\mathrm{eff}$ \cite{weis1985lithium,vodopyanov2006optical,hebling2008generation,ravi2016pulse}& 2 $\times$ 168\,pm/V \\ \hline
  		$\Omega_\mathrm{THz}$ & $2\pi\times0.3$\,THz \\ \hline
  		PPLN period $\Lambda$ & 374.1\,{\textmu}m \\ \hline
  		Temperature (T) & 80K \\ \hline
  		$\alpha$\cite{wu2015temperature}& $1.4/\text{cm}$ \\ \hline
  		$F_d$ & 1.62\, J/cm$^2$ \\ \hline \hline
  		\multicolumn{2}{|c|}{\textbf{Input Pump Pulse Format}} \\ \hline \hline
  		Super-Gaussian order $M$ & 5 \\ \hline
  		$\sigma$ & 3.5\,mm \\ \hline
  		$\tau$ & 554\,ps \\ \hline
  	\end{tabular}
  \end{table}
  The transverse distribution of the optical beam is modelled by a super-Gaussian beam ($E(r) \propto \exp (-r^{2M}/2\sigma^{2M}) $) of order $M=5$ and a rectangular temporal profile with pulse duration $\tau$ is assumed to be the optical driver of the THz source.
  In table\,\ref{ICSTHzMultiParameters}, $n_2$ stands for the nonlinear index of lithium niobate. $\alpha$ and $F_d$ represent the absorption coefficient and damage fluence of the material, respectively.
  
  The characteristics of the generated THz radiation are depicted in Fig.\,\ref{ICSTHzMulti}b-e.
  At the damage threshold limit, the input energy for the assumed super-Gaussian beam with $\sigma=3.5\,$mm is 1\,J, which dictates using 10\,mm$\times$10\,mm size PPLN crystals for each stage.
  Such PPLN crystals are available today \cite{ishizuki2012half}.
  The conversion efficiency of each stage is $\eta_1 = 1.06\%$ and, $\eta_2 = 1.03\%$, corresponding to $\sim 10$\,mJ THz at the output of each PPLN stage.
  The spatio-temporal profile of the terahertz beams generated by the two PPLN stages are shown in Fig.\,\ref{ICSTHzMulti}d and \ref{ICSTHzMulti}e.
  The illustrated plots correspond to the terahertz beam before the output couplers, which influence the beam profile.
  The final terahertz beam will be matched to the input of the input coupler of the power combiner at the linac stage.
  In this study, we assume a safety margin of 60\% for the loss of energy during the terahertz transport section.
  Specifically, the total generated terahertz energy is 40\,mJ whereas the required terahertz energy in the linac, as will be seen later, is 16\,mJ.
  
  \section{Terahertz Injector}
  
  The THz injector in the designed machine is assumed to be an ultrafast single-cycle electron gun \cite{fallahi2016short}.
  Another option for injecting relativistic electrons into the linac would be THz cavities \cite{fakhari2017thz}.
  Nevertheless, our investigations showed that the sensitivity of THz cavities to tolerances in dimensions is drastically higher than the ultrafast single-cycle electron guns.
  This difference emanates from the narrowband operational principle of cavities being based on the resonance effect, whereas the broadband nature of ultrafast guns reduces the sensitivity to geometrical errors.
  Furthermore, recent studies on break-down rates of various particle accelerators have shown that the pulse duration of the excitation plays the major role in determining the damage threshold of an acceleration device \cite{laurent2011experimental,dal2016experimental,wu2017high}.
  Therefore, using a device excited by a single-cycle pulse offers higher accelerating gradients than the multi-cycle based devices.
  This fact is crucially important in the injector, since high accelerating gradients bring electrons up to relativistic speeds in a short distance, preventing emittance growth due to strong space-charge forces in low energy regimes.
  
\begin{figure}
  	\centering
  	\includegraphics[width=6.0in]{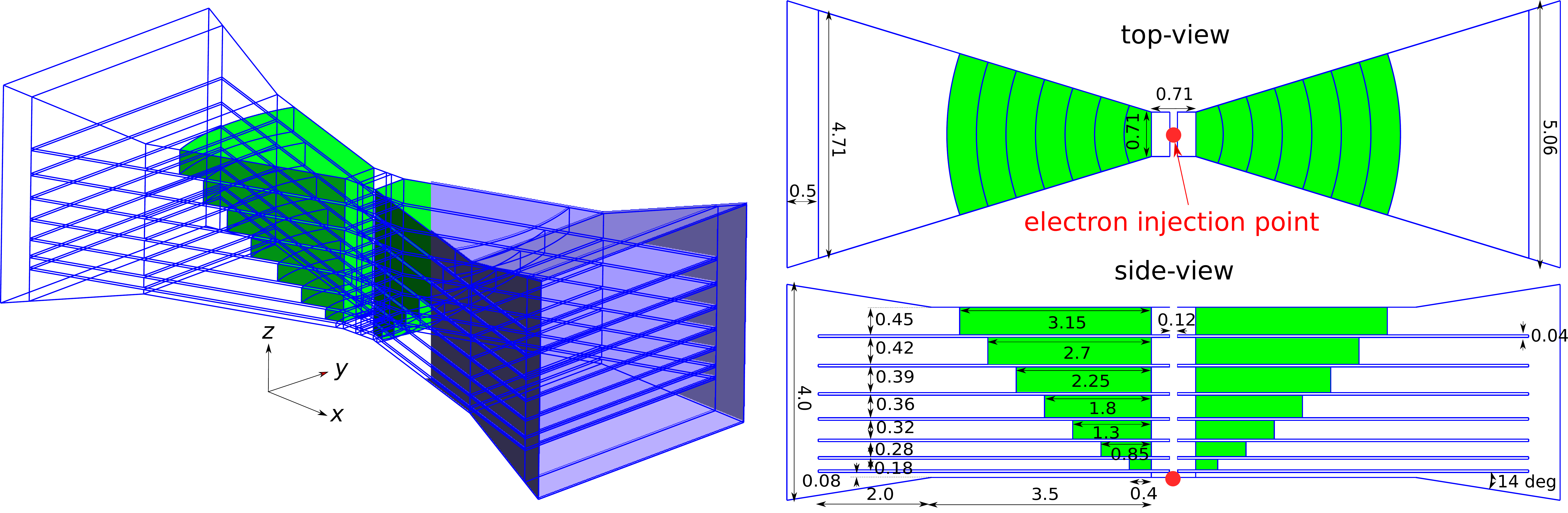}
  	\caption{Schematic illustration of the ultrafast single-cycle electron gun used in the ICS machine. All the dimensions are given in millimeters.}
  	\label{ICSTHzbutterflyGun}
  \end{figure}

  The concept of the ultrafast single-cycle guns have been introduced in chapter 4 with the operation principles experimentally tested and verified.
  The design we assume for our source is the same as the 800\,keV gun design presented in section 4.3.4.
  However, for the sake of clarity, we review again the parameters and properties of this gun here.
  Fig.\,\ref{ICSTHzbutterflyGun} schematically illustrates a single-cycle electron gun, which consists of three principal sections, namely interaction region, focusing section, and the coupler.
  Two linearly polarized single-cycle Gaussian beams, frequency-centered at 300\,GHz and generated using the sources designed in the previous section, symmetrically impinge on the device from both sides.
  The coupler section is designed to transfer the energy of the Gaussian beam into the multilayer focusing section, where the energy of the beam is guided into the interaction region.
  The interaction region in each layer behaves like a rectangular waveguide, whose TE\textsubscript{01} mode is excited by the incoming fields from the focusing section.
  In the TE\textsubscript{01} mode, the longitudinal field (component along the propagation axis) propagates with a $\pi/2$ phase delay with respect to the transverse components.
  Therefore, the $x$-components of the fields interfere destructively at the interaction region.
  Similarly, the opposing $y$-components of the beam, needed for opposite propagation directions, results in cancellation of the $y$-components of the fields.
  Consequently, the superposition of these two beams results in a purely accelerating field along the $z$-axis in Fig.\,\ref{ICSTHzbutterflyGun}.
  
  Horizontal metallic plates in the coupler section divide the incoming Gaussian beams into several portions with thickness $d_{i}$.
  The traveling pulse entering each focusing section is subsequently delayed by dielectric inclusions, whose lengths, $L_{i}$, are designed to control the arrival of pulses into the interaction region.
  Proper design of the two sets of parameters $d_{i}$ and $L_{i}$ assures continuous interaction of traveling electrons with the accelerating cycle of the pulse.
  In other words, the device realizes phase-front matching of the incoming pulses with traveling electrons.
  
  Two Gaussian beams focused at the points $(x_f,y_f,z_f) = (\pm 2.7,0.0,1.35)\,\text{mm}$, assuming that the electron injection point corresponds to the origin of the coordinate system, with 400\,{\textmu}J energy and elliptical spot size $(2\times1)$\,mm excite the electron gun.
  Each pulse has 3.33\,ps pulse duration with a central frequency of 300\,GHz.
  These values correspond to the generated beam from the single-cycle THz source described in section II.A.
  However, similar to the spatial beam shape, the temporal pulse format does not accurately describe the generated pulses at the output of the single-cycle THz source.
  The main reason for neglecting such small deviations, and considering an ideal single-cycle pulse are the changes introduced by the THz transport system, whose design is directly dependent on the implementation conditions.
  The THz transport and coupling systems can be tailored according to the implementation conditions.
  Here, we do not discuss the implementation details and merely consider a factor of two loss as a worst-case scenario due to mismatch between the THz source output and the excitations assumed for the gun.
  
  The ASTRA particle generator code is used to simulate the photo emission process.
  A 47-fs Full Width Half Maximum (FWHM) UV laser is assumed to illuminate the cathode with a spot size of 47\,{\textmu}m (FWHM).
  The amount of charge emitted by the cathode depends on the energy of the beam and quantum efficiency (QE) of the material used for cathode.
  The pulse energy required for emitting pico-Coulomb level of charge is at the micro-joule level, which is very small compared with the lasers used for other parts of the source.
  Therefore, we assume that the laser pulse is strong enough to produce 1\,pC of bunch charge, which is modeled by 20'000 macro-particles.
  The considered values for laser spot size and emitted charge are obtained by down-scaling the currently operating photocathodes \cite{dowell2010cathode}.
  The laser pulse duration needs to be small enough compared with the THz cycle, which is optimized to the aforementioned value of 47\,fs.
  
\begin{figure}
  	\centering
  	$\begin{array}{ccc}
  	\includegraphics[draft=false,width=2.0in]{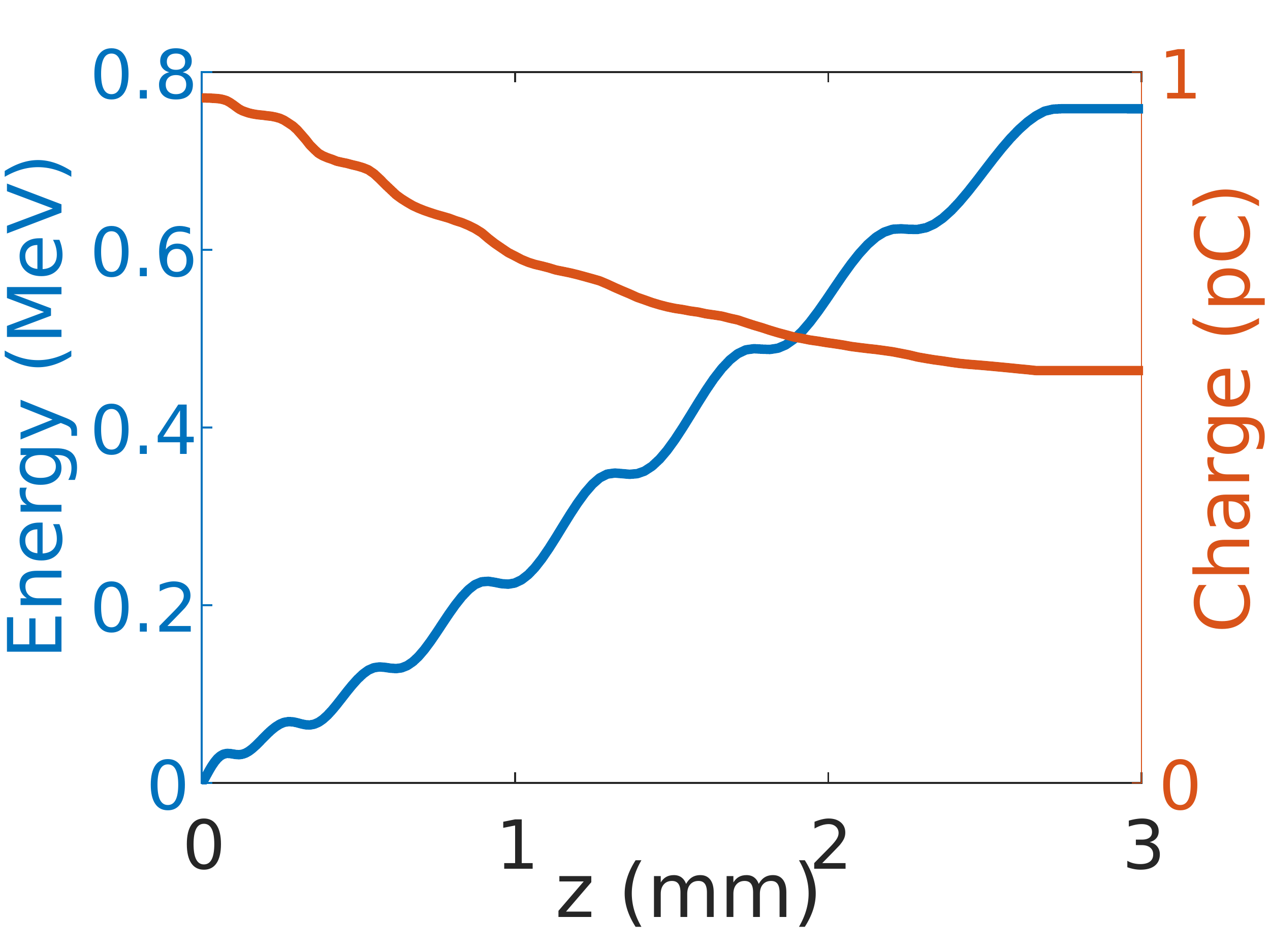} &
  	\includegraphics[draft=false,width=2.0in]{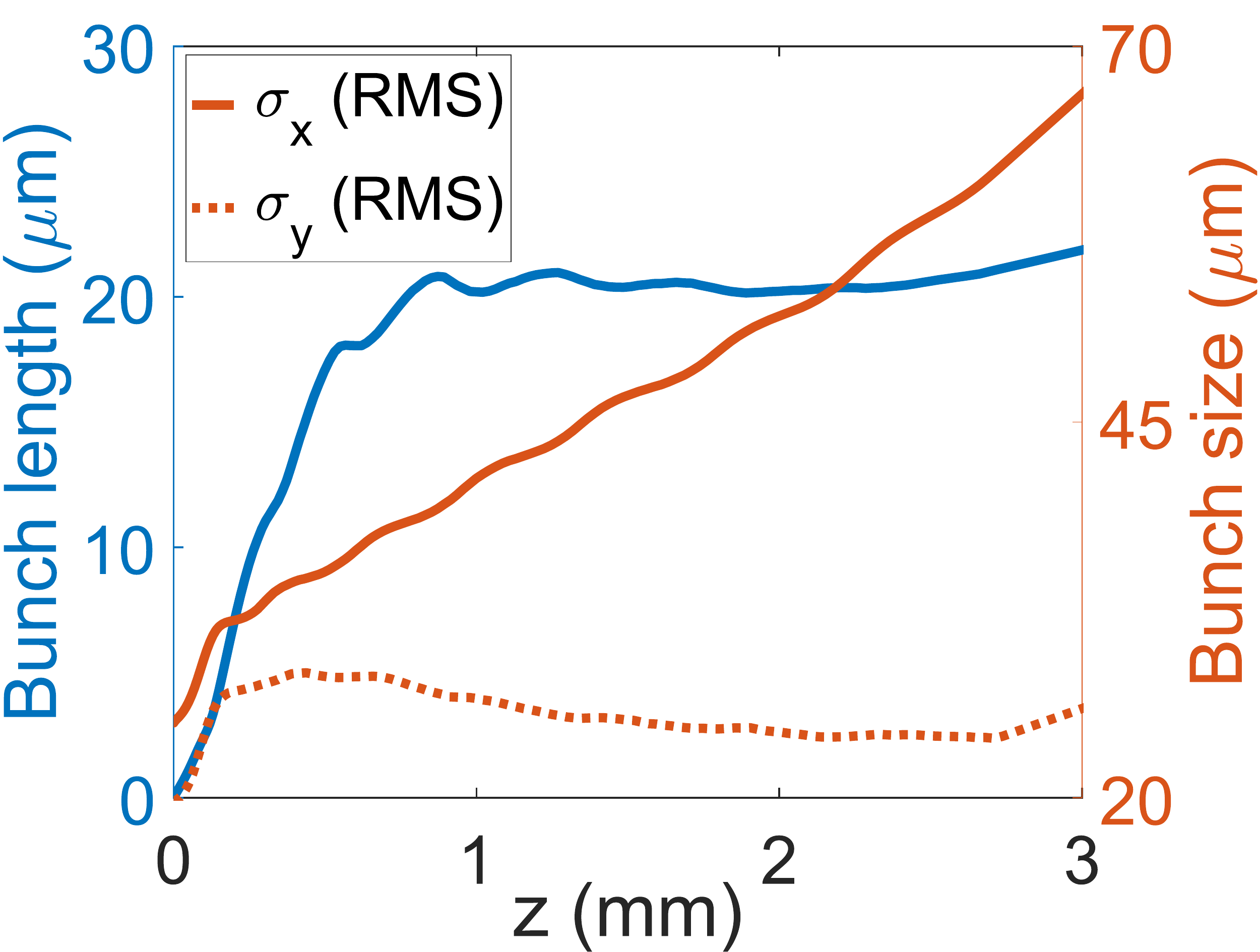} &
  	\includegraphics[draft=false,width=2.0in]{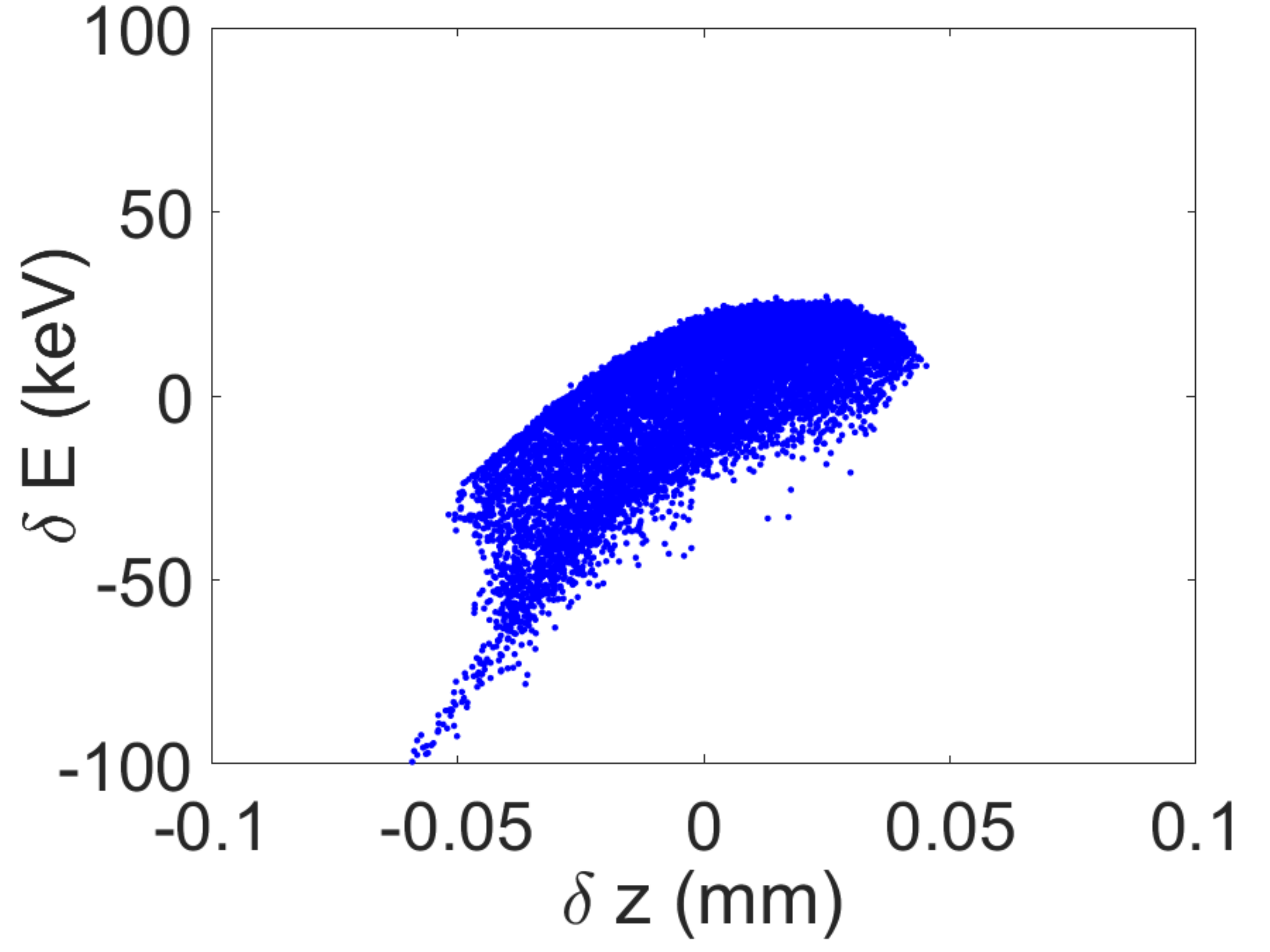} \\
  	(a) & (b) & (c) \\
  	\includegraphics[draft=false,width=2.0in]{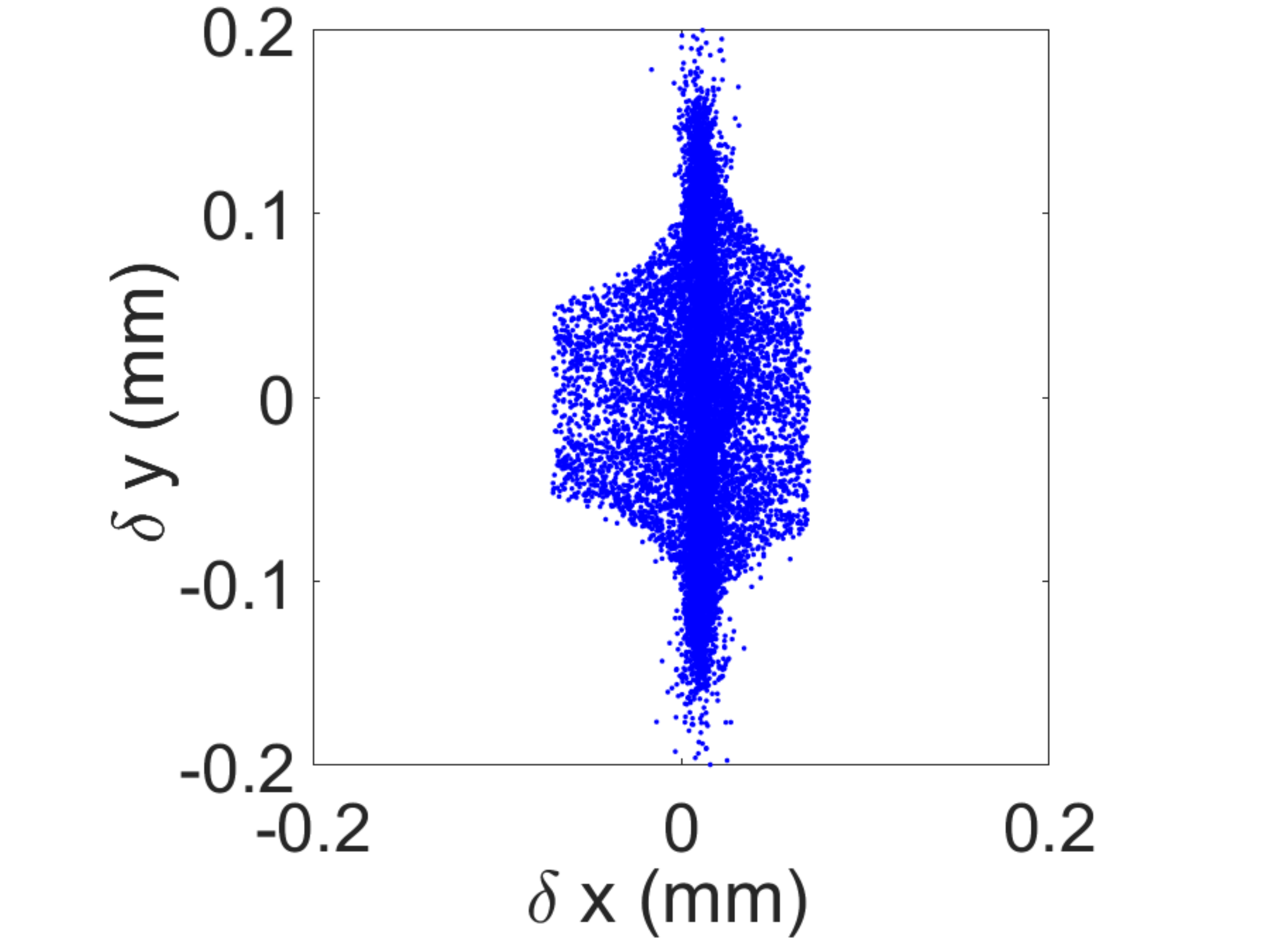} &
  	\includegraphics[draft=false,width=2.0in]{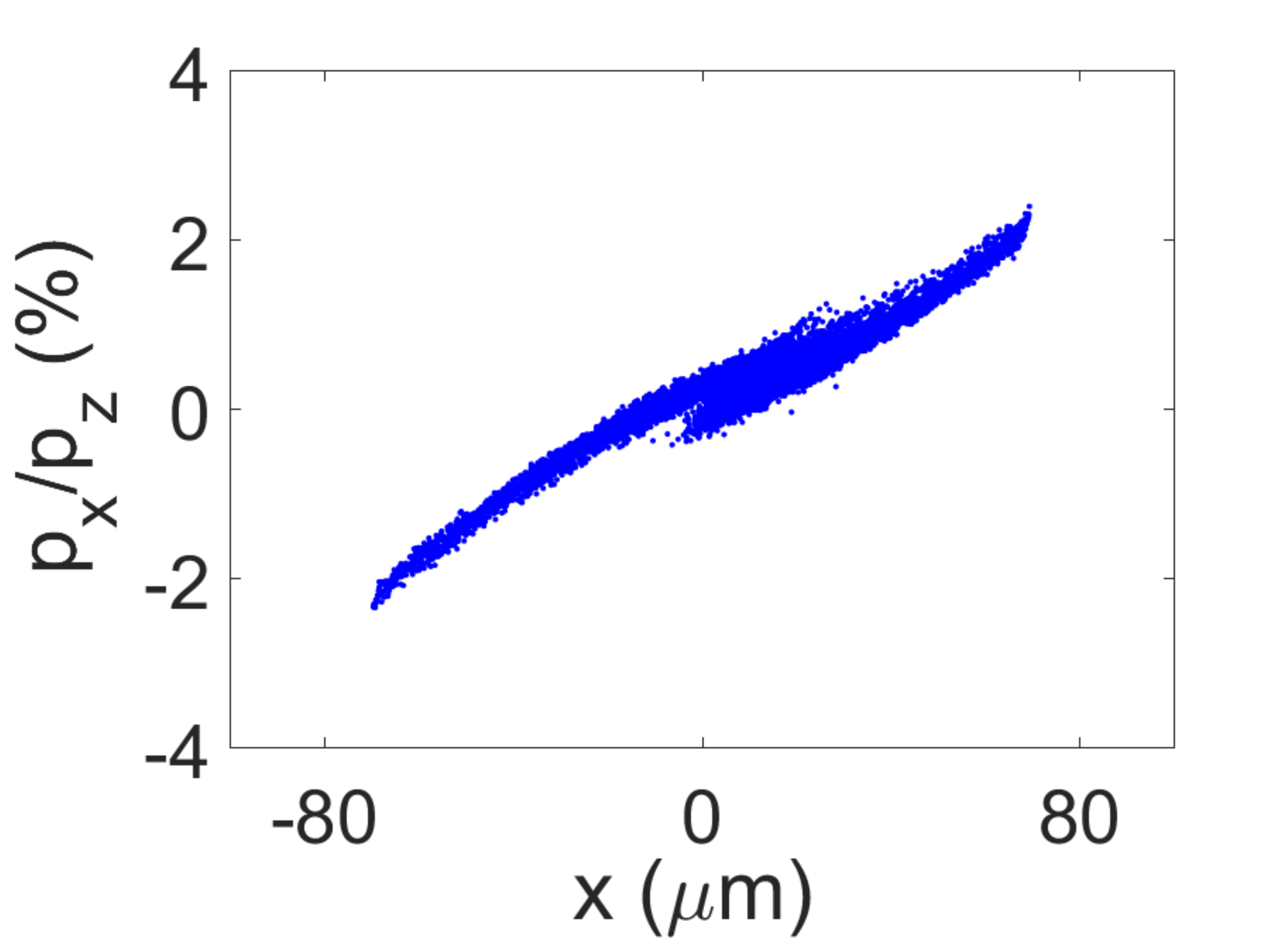} &
  	\includegraphics[draft=false,width=2.0in]{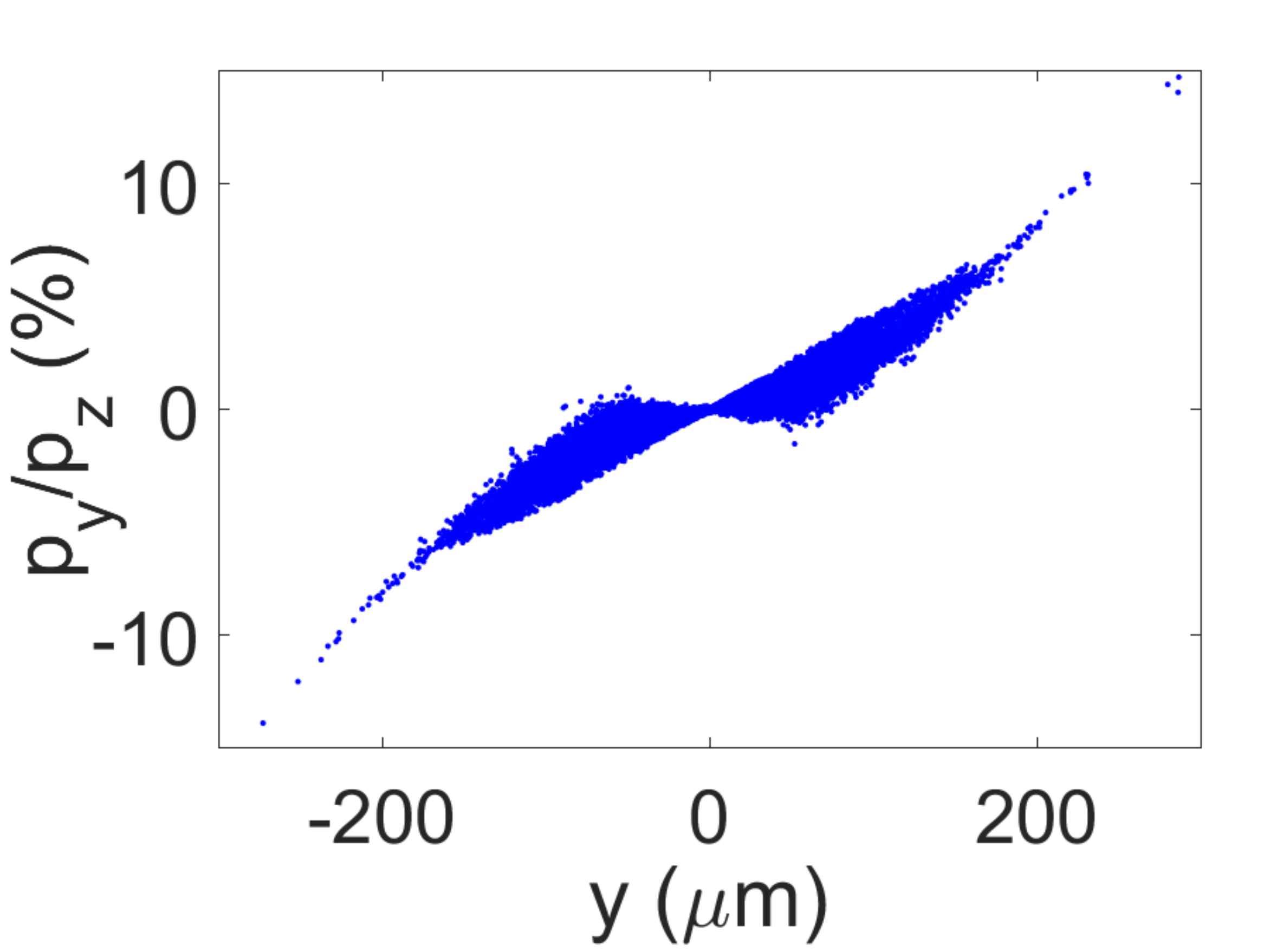} \\
  	(d) & (e) & (f)
  	\end{array}$
  	\caption{Simulation results for the THz injector: (a) bunch mean energy and output charge, (b) bunch length and bunch mean transverse size, (c) longitudinal phase space, (d) front view of the bunch, (e) transverse phase space ($x$-direction), and (f) transverse phase space ($y$-direction). }
  	\label{ICSTHzgunResults}
  \end{figure}

  Electron bunch acceleration is simulated with a DGTD/PIC code \cite{fallahi2014field} described in chapter 2.
  The performed simulations account for the space-charge effects through a point-to-point algorithm and neglects the effect of image charge on the cathode.
  The considered values for the total bunch charge and dimensions lead to $\sim$16\,MV/m screening field due to the image charge effect, which is much smaller than the accelerating field at the cathode surface around 800\,MV/m.
  The simulation results are shown in Fig.\,\ref{ICSTHzgunResults}a and \ref{ICSTHzgunResults}b.
  According to these results, it is possible to accelerate the electrons from rest up to 0.780\,MeV kinetic energy.
  However, we lose about 40$\%$ of the particles due to collisions with the metallic walls along the path.
  Fig.\,\ref{ICSTHzgunResults}c-f show the longitudinal phase space and the front view of the bunch exiting the gun as well as its transverse phase-space.
  As it can be seen from these figures, the cylindrically asymmetric configuration of the gun results in an asymmetric output beam with different transverse phase space distributions along $x$ and $y$ directions.
  The output normalized transverse emittances are evaluated as $(\epsilon_{x},\epsilon_{y})=(0.08,1.11)$\,mm$\cdot$mrad, while the longitudinal emittance is 0.66\,mm$\cdot$mrad.
  
  \section{Linear Acceleration}
  The linac geometry as shown in Fig.\,\ref{DLMWConcept}a is a dielectric-loaded metallic waveguide designed for receiving the 0.78\,MeV electron bunch from the gun and increasing the energy up to $\sim 20$\,MeV.
  The acceleration of particles with an electromagnetic wave is possible if and only if the particle velocity is synchronized with the field variations of the wave.
  Since the electron velocity is always less than the speed of light while a guided wave in an empty waveguide travels with phase velocities exceeding the speed of light, the phase velocity of the guided wave needs to be reduced to enable electron acceleration.
  Phase velocity reduction can be achieved either by corrugating the waveguide boundaries or increasing the electric permittivity of the material filling the waveguide.
  In THz range dimensions, it is easy and rather cheap to load the circular waveguides with quartz tubes, whose relative electric permittivity is $\epsilon_{r}=4.41$ as shown in Fig.\,\ref{DLMWConcept}a \cite{Wong2013}.
  It is possible to tune the phase velocity of the THz wave in the waveguide by adjusting the inner radius and thickness of the dielectric ($a$ and $d$ in Fig.\,\ref{DLMWConcept}a) and synchronize the field oscillations with a travelling particle.
  However, before accelerating the particles, the THz beam needs to be coupled into the linac.
  
  \subsection{Combiner-coupler design}
  
\begin{figure}
  	\centering
  	$\begin{array}{cc}
  	\includegraphics[draft=false,height=2.25in]{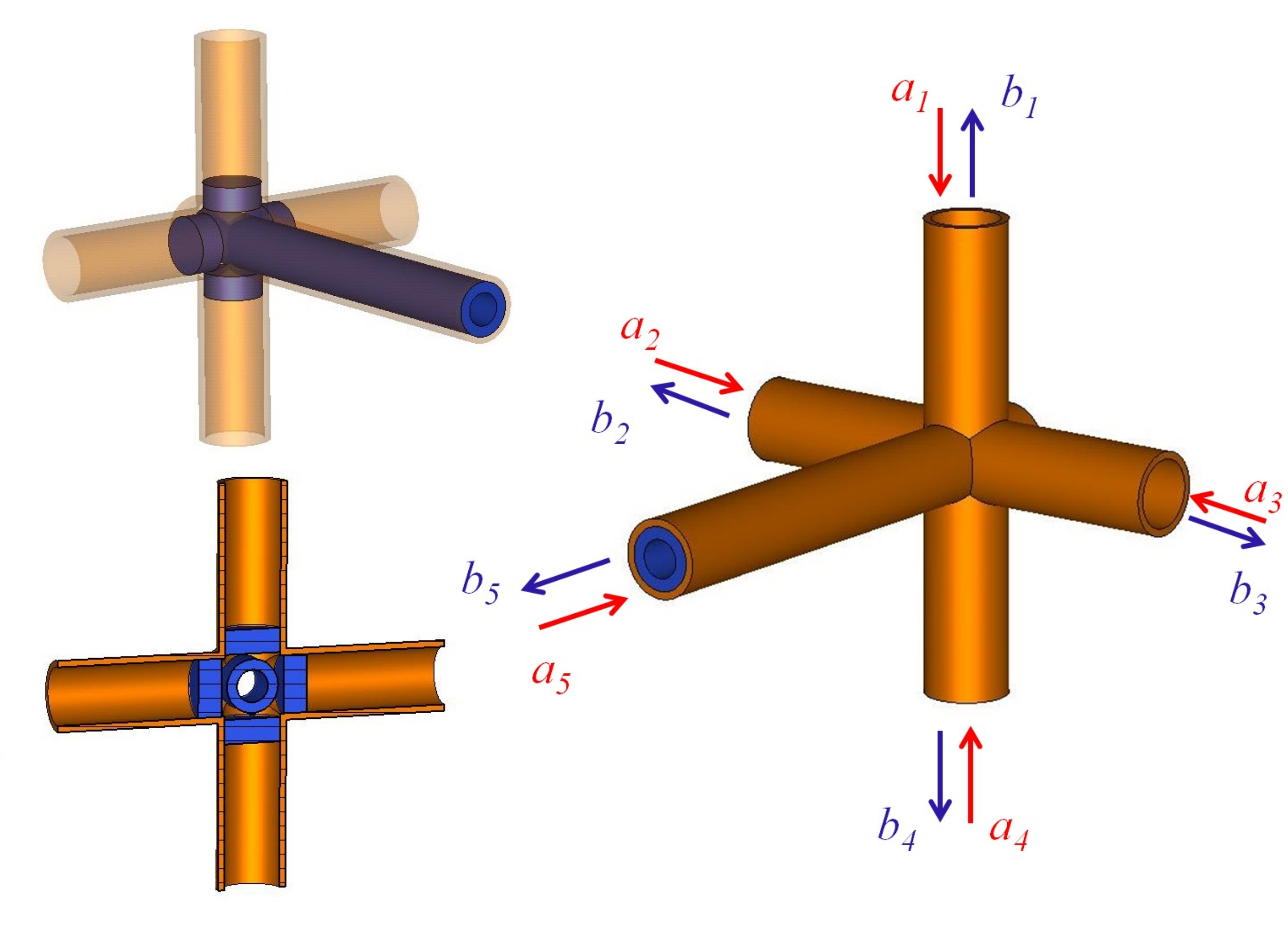} &
  	\includegraphics[draft=false,width=3.0in]{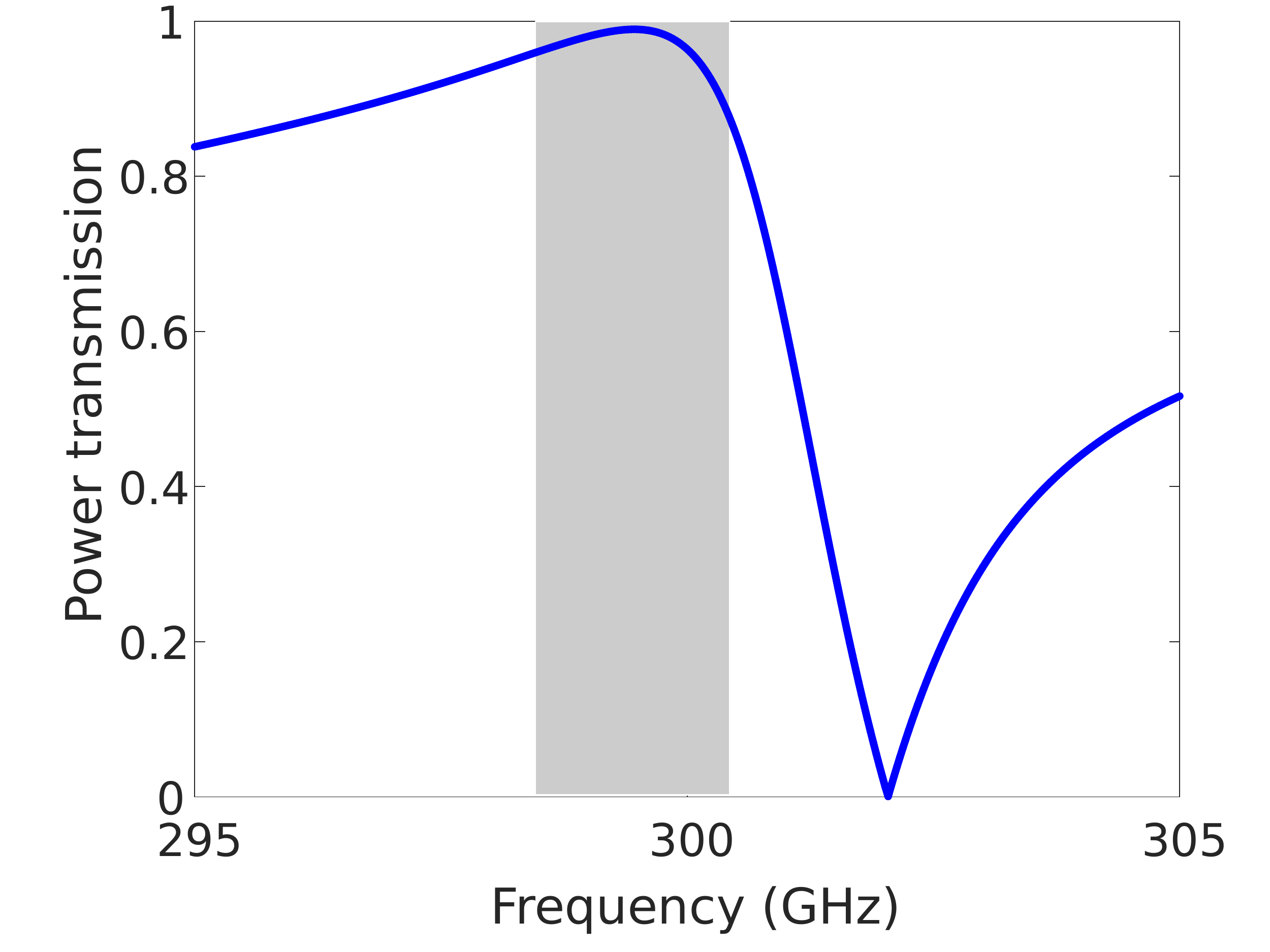} \\
  	(a) & (b)
  	\end{array}$
  	\caption{THz coupler-combiner: (a) Isometric and cross-sectional views of the combiner-coupler device attached to the linac as well as the parameter definitions for modeling the coupler as a five-port network. Ports 1 to 4 specify the inputs while the output is assumed to be port 5. (b) Coupling efficiency of the THz pulse to the waveguide. We have $99\%$ power transmission at 299.5 GHz.}
  	\label{ICSTHzCoupler}
  \end{figure}

  One of the novel features in the designed x-ray source is the four-port coupler designed for the linac that simultaneously combines the four beams generated by the THz source in section 6.1.2 and couples them into the linac.
  Since this coupler is not introduced in previous works, the overall performance of this element as well as its tolerance study is presented here.
  According to the discussed THz generation process, we would have four ``almost'' identical multi-cycle THz signals to be combined and coupled to the linac.
  Fig.\,\ref{ICSTHzCoupler}a shows the views of the designed combiner-coupler.
  
  The designed device has four ports for the four input signals.
  Each port has a small quartz inlet which is used to match the impedance in order to maximize the coupling efficiency.
  The inner radius of the ports is 360\,{\textmu}m while the quartz disks have a thickness of 300\,{\textmu}m.
  The coupling efficiency of the device is simulated by CST Microwave Studio-frequency domain solver and the results are displayed in Fig.\,\ref{ICSTHzCoupler}b.
  According to this plot, $99\%$ of the THz power at 299.5\,GHz will be coupled to the linac assuming four completely identical input signals.
  
  The above calculation of the coupling coefficient is performed assuming that four completely identical pulses excite the coupler from four directions.
  However, discrepancies will certainly exist in the beams of the THz source as observed in the simulation results of section 6.1.2.
  In order to evaluate the coupling efficiency when discrepancies between input signals exist, a scattering matrix analysis needs to be conducted.
  The designed coupler can be considered as a five-port network, in which the four input ports are named as port 1 to port 4 and the output port as port 5 (Fig.\,\ref{ICSTHzCoupler}a).
  According to the scattering matrix definition, the relation between the incident and reflected waves reads as:
  \begin {equation}
  \left[
  \begin{array}{c}
  	b_{1}\\
  	b_{2}\\
  	b_{3}\\
  	b_{4}\\
  	b_{5}
  \end{array}\right] = \left[
  \begin{array}{c c c c c}
  	S_{11} & S_{12} & S_{13} & S_{14} & S_{15}\\
  	S_{21} & S_{22} & S_{23} & S_{24} & S_{25}\\
  	S_{31} & S_{32} & S_{33} & S_{34} & S_{35}\\
  	S_{41} & S_{42} & S_{43} & S_{44} & S_{45}\\
  	S_{51} & S_{52} & S_{53} & S_{54} & S_{55}\\
  \end{array}\right] \left[
  \begin{array}{c}
  	a_{1}\\
  	a_{2}\\
  	a_{3}\\
  	a_{4}\\
  	a_{5}
  \end{array}\right] .
\end{equation}
Considering the symmetric structure of the coupler and assuming 100\% coupling for identical input signals, one can derive the following relationships between elements of the scattering matrix:
\begin {equation}
S_{51}=S_{52}=S_{53}=S_{54}=0.5, S_{55}=0.
\end{equation}
One can use the values of these elements in order to find the sensitivity of the coupler to discrepancies in the input signals.
For identical input:
\begin{equation}
a_{1}=a_{2}=a_{3}=a_{4}=a \Rightarrow b_{5}=2a,
\end{equation}
and
\begin{equation}
\mathrm{Transmission} =\frac{(2a)^{2}}{4\times a^{2}}=100\%
\end{equation}
While in the case of different input signals assuming constant input power:
\begin{equation}
\left\{
\begin{array}{l}
a_{1}=a+\delta_{1}\\
a_{2}=a+\delta_{2}\\
a_{3}=a+\delta_{3}\\
a_{4}=a+\delta_{4}\\
\end{array} \Rightarrow b_{5}=2a+0.5\sum_{i}\delta_{i}
\right.
\end{equation}
and
\begin{equation}
\mathrm{Transmission} =\frac{(2a+0.5\sum\delta_{i})^{2}}{4\times a^{2}}\leq 100\%
\end{equation}
To have a constant input power, the following equation should hold:
\begin{equation}
\sum_{i}(a+\delta_{i})^{2}=\mathrm{cons.}
\end{equation}

The above calculation assumes that all of the four input signals have the same frequency but different amplitudes.
If the input signals are different both in frequency and amplitude, numerical simulation of the coupling process is needed to find the exact value of the coupling efficiency.
We made such simulations using the THz signals discussed in section 6.1.2.
According to these simulations, a 10\% loss of the THz power as a result of input discrepancy is expected.
It is possible to decrease this loss by adjusting the coupler for slightly different THz signals generated in reality.
Due to this coupling loss and additional THz transport losses, we consider the a factor of two safety margin for operation of the source.
More accurately, the linac required less than 20\,mJ THz energy whereas the total generated THz radiation is about 40\,mJ.

\subsection{THz linac}

The linac design can be performed using the analytic solution of the guided modes in the dielectric-loaded metal waveguide \cite{Wong2013}.
The electric and magnetic field distributions of TM\textsubscript{01} mode for a dielectric-loaded metal waveguide can be found according to \eref{TM01Mode}-\eref{TM01ModePulseFields}
These equations return the field distribution inside the waveguide as well as the designed values for the inner and outer radii of the quartz layer to set the phase velocity equal to the speed of the electron bunch coming from the gun.
The electrons entering the linac have an average kinetic energy of 0.78\,MeV which corresponds to a normalized velocity of $\beta=0.92$.
However, we expect a final energy of about 20\,MeV, corresponding to the final velocity 99.97\% of the speed of light.
Therefore, it is not possible to synchronize the electrons with the wave throughout the whole linac.

\begin{figure}
\centering
$\begin{array}{cc}
\includegraphics[draft=false,width=2.5in]{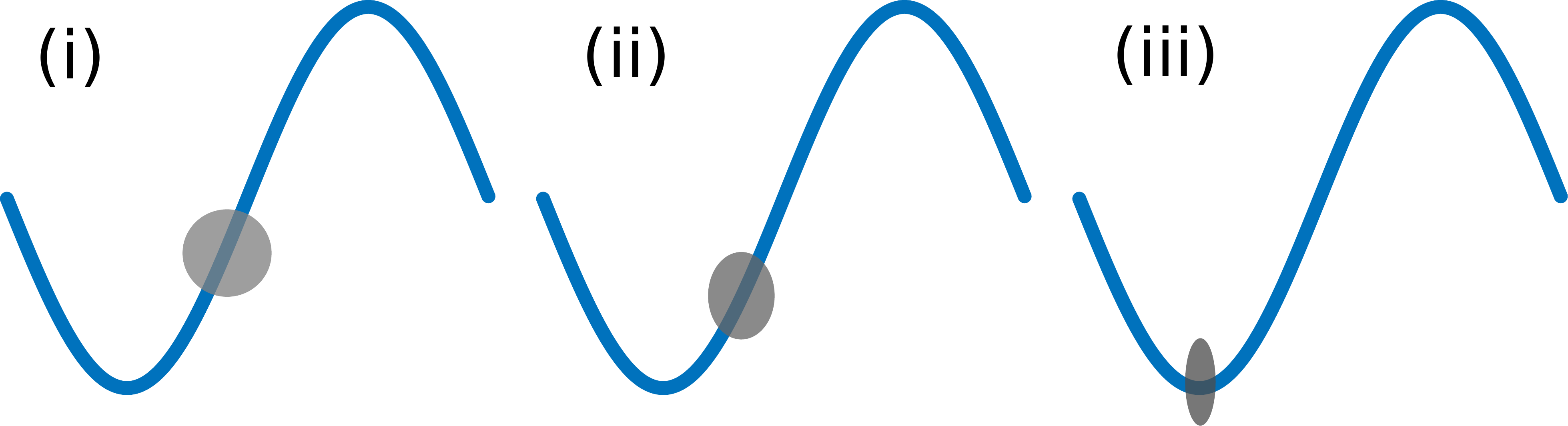} &
\includegraphics[draft=false,width=3.5in]{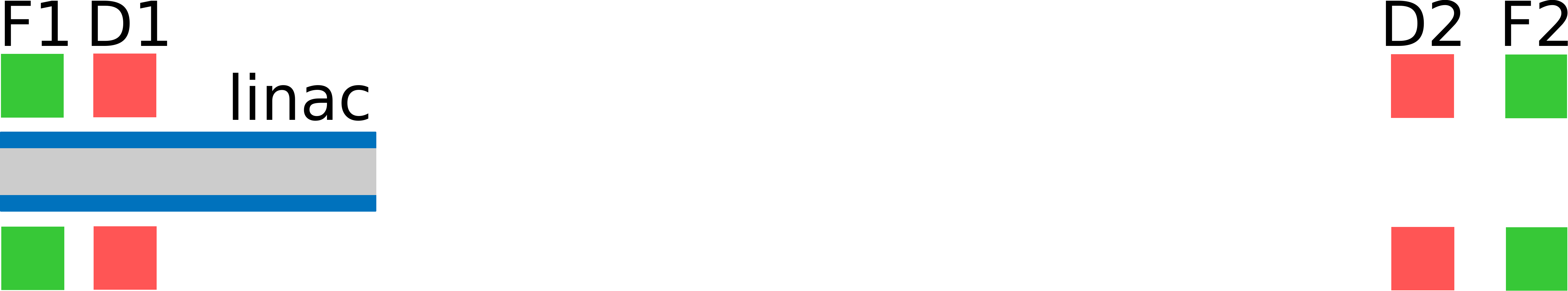} \\
(a) & (b)
\end{array}$
\caption{Linear acceleration setup: (a) Schematic view of the relative position of the electron bunch and the electric filed: (\emph{i}) at the beginning of linac electrons are injected in a phase between $\pi/2$ and $\pi$, (\emph{ii}) the bunch slips over the field, (\emph{iii}) when the electrons reach to the crest when their velocity is almost equal to the phase velocity, so their position relative to the wave no more varies, (b) Schematic view of the linac with the focusing (F) defocusing (D) magnetic lattice. Focusing quadrupole focuses the beam in x-plane while defocusing one focuses in y-plane.}
\label{ICSTHzLinearAcceleration}
\end{figure}

The solution to this problem is to set the phase velocity around the final expected velocity of the electrons and inject the electrons at a phase between $\pi/2$ and $\pi$.
This is indeed similar to the so-called off-crest electron injection typically done in conventional RF accelerators.
At the beginning of the linac, electrons travel slower than the field, therefore slippage over the field cycle takes place until electrons reach the crest where their velocity is almost equal to the phase velocity.
Due to their almost equal velocities, the electron bunch sits on the crest of the wave for the remaining path through the linac leading to a continuous increase in energy.
This mechanism is shown in Fig.\,\ref{ICSTHzLinearAcceleration}a.
Since a decreasing field gradient is influencing the bunch at the beginning of the waveguide (Fig.\,\ref{ICSTHzLinearAcceleration}a), the electrons at the tail of the bunch experiencing higher field values get accelerated stronger than the electrons at the head of the bunch.
Therefore, the bunch gets compressed as it passes through the linac.
The design parameters of the linac as well as the properties of the input THz signal is summarized in table\,\ref{linacParameters}.
\begin{table}
\caption{The designed parameters of the THz linac and quadrupole lattice} \label{linacParameters} \centering
\begin{tabular}  { | m{4cm} | m{9cm}| }
\hline
\textbf{Parameter} & \textbf{Designed value} \\
\hline
\hline
Quartz inner radius & 242.5 {\textmu}m \\
\hline
Quartz outer radius & 364.8 {\textmu}m \\
\hline
Frequency & 299.5 - 300.5 GHz\\
\hline
Phase velocity at 300\,GHz & $0.994c$\\
\hline
Group velocity at 300\,GHz & $0.281c$\\
\hline
THz pulse energy & 16.3 mJ\\
\hline
THz pulse duration & $\sim 554$\,ps\\
\hline
length & 7.5\,cm\\
\hline
on-axis electric field & 320 MV/m\\
\hline
\hline
\textbf{Magnet number} & \textbf{Field gradient (T/m) - Distance to gun exit (mm)} \\
\hline
\hline
$(F_{1}, F_{2})$ & (200, 190) - (6, 20)  \\
\hline
$(D_{1},D_{2})$ & (115, 120) - (520, 542) \\
\hline
\end{tabular}
\end{table}

\begin{figure}
\centering
$\begin{array}{cc}
\includegraphics[draft=false,width=3.0in]{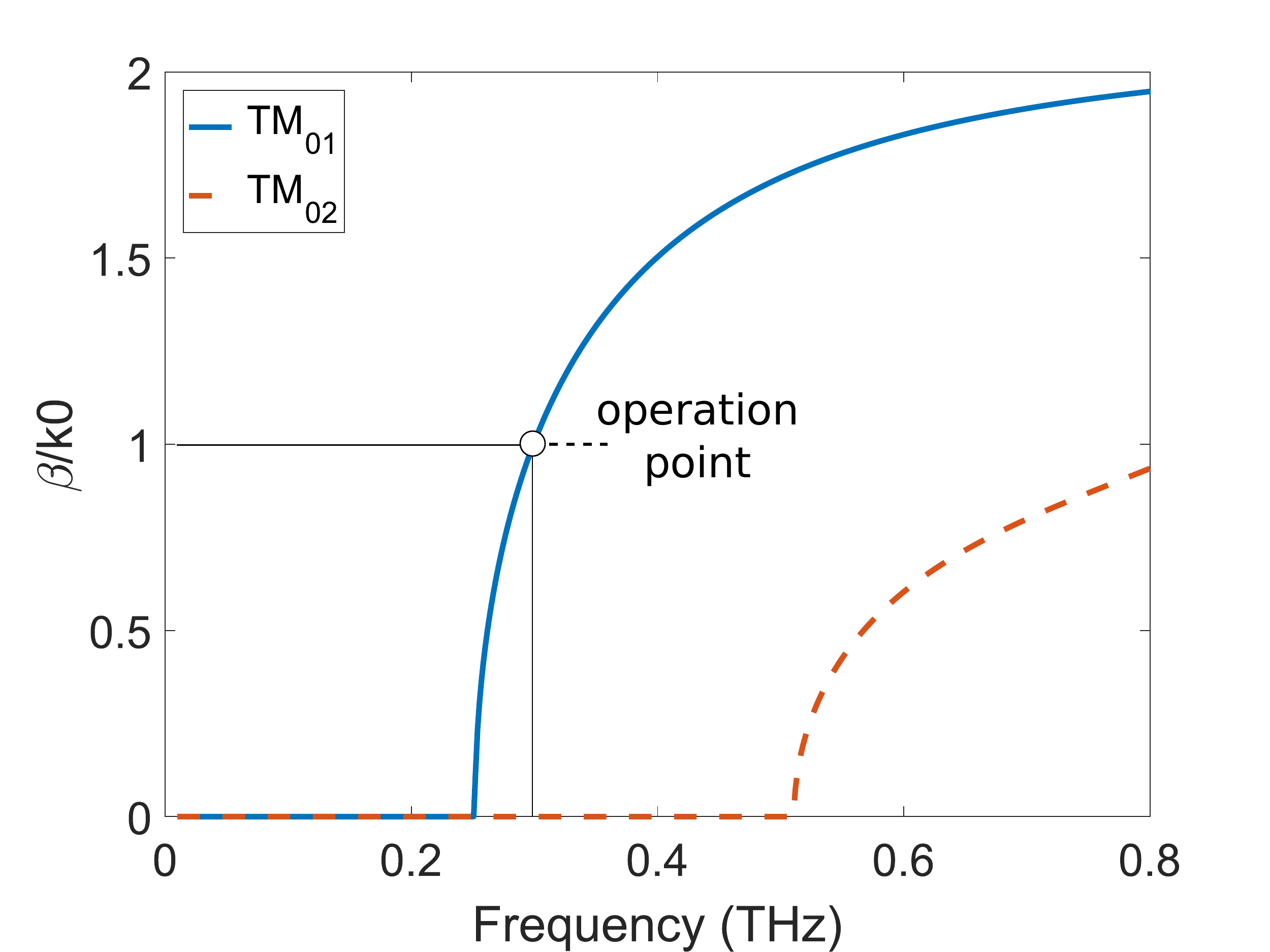} &
\includegraphics[draft=false,width=3.0in]{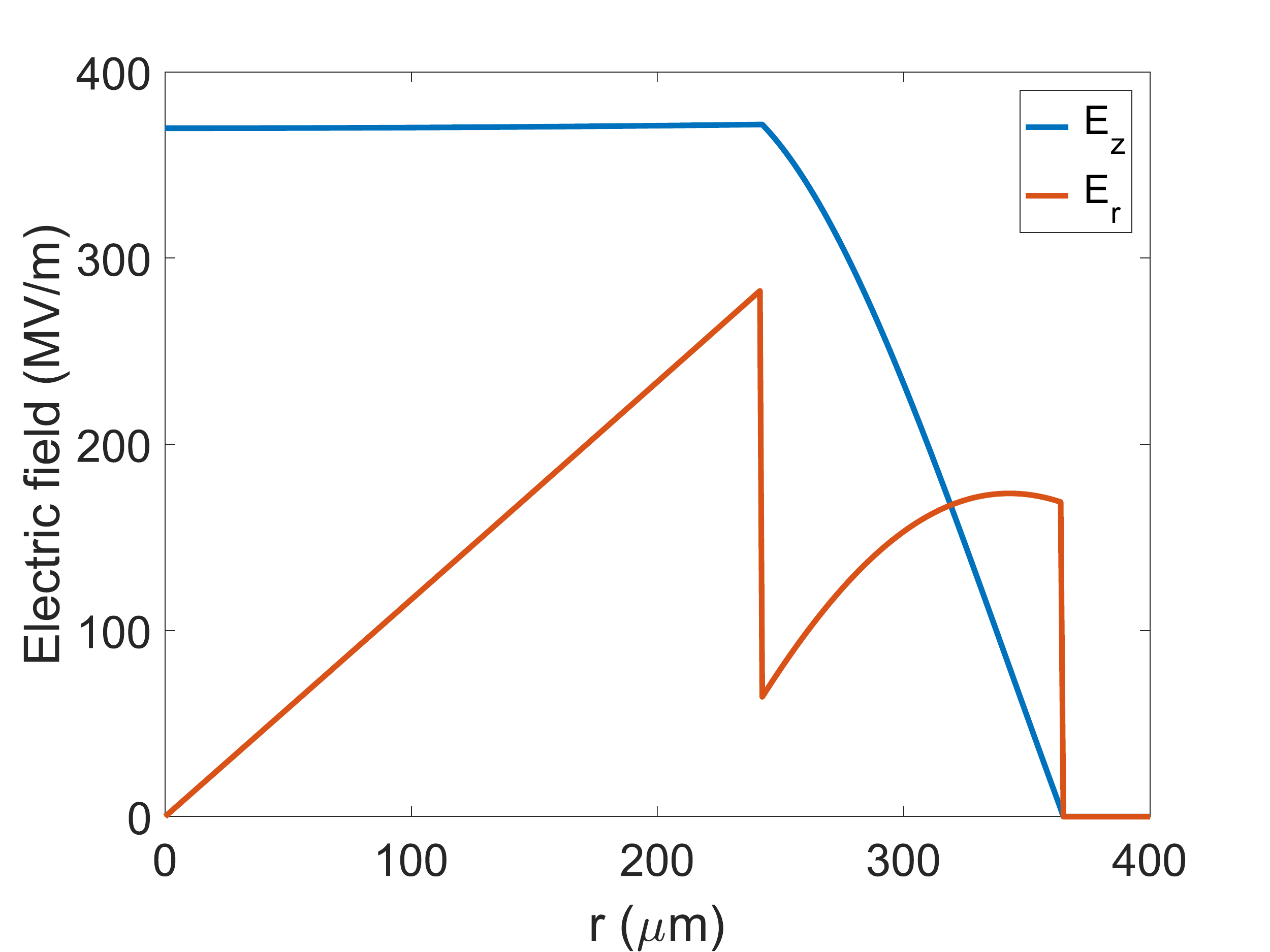} \\
(a) & (b)
\end{array}$
\caption{Linac performance: (a) the dispersion diagram of the dielectric-loaded metallic waveguide to be used for linear acceleration, and (b) the field distribution inside the waveguide.}
\label{ICSTHzLinacPerformance}
\end{figure}

In Fig.\,\ref{ICSTHzLinacPerformance}, we show the dispersion diagram of the travelling TM modes in the designed linac as well as the field distribution of the TM\textsubscript{01} mode considered for electron acceleration.
\begin{figure}
\centering
$\begin{array}{cc}
\includegraphics[draft=false,width=3.0in]{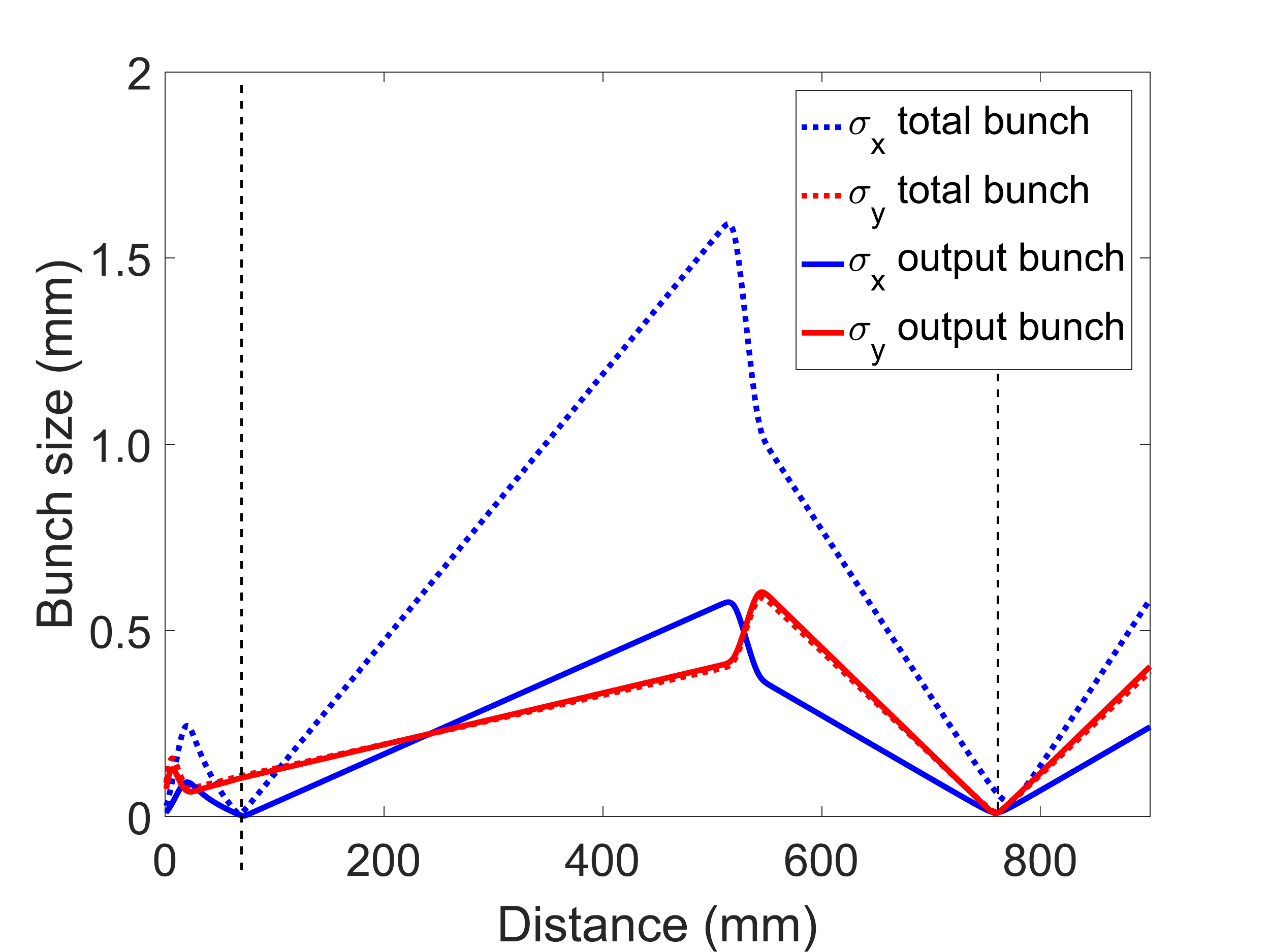} &
\includegraphics[draft=false,width=3.0in]{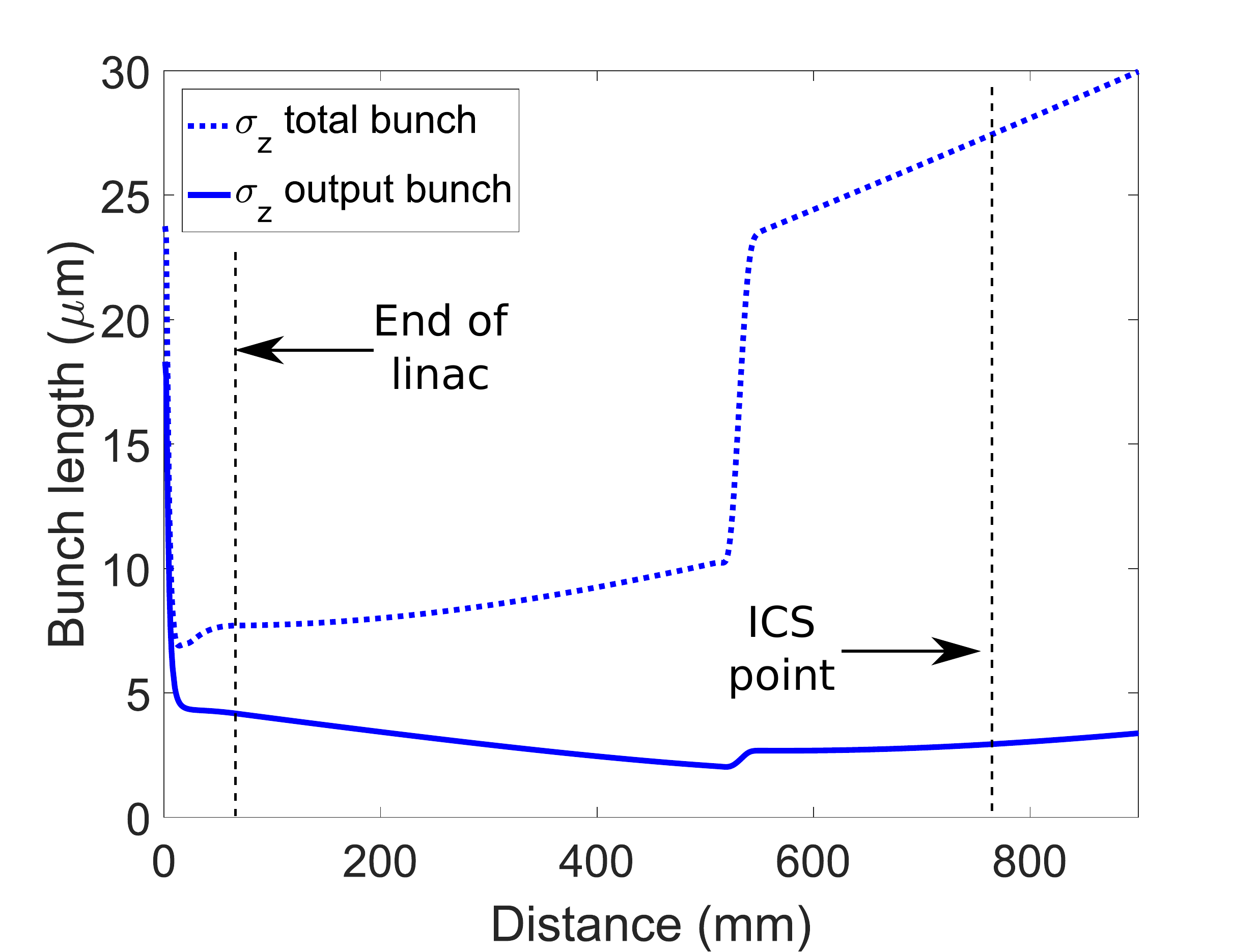} \\
(a) & (b) \\
\includegraphics[draft=false,width=3.0in]{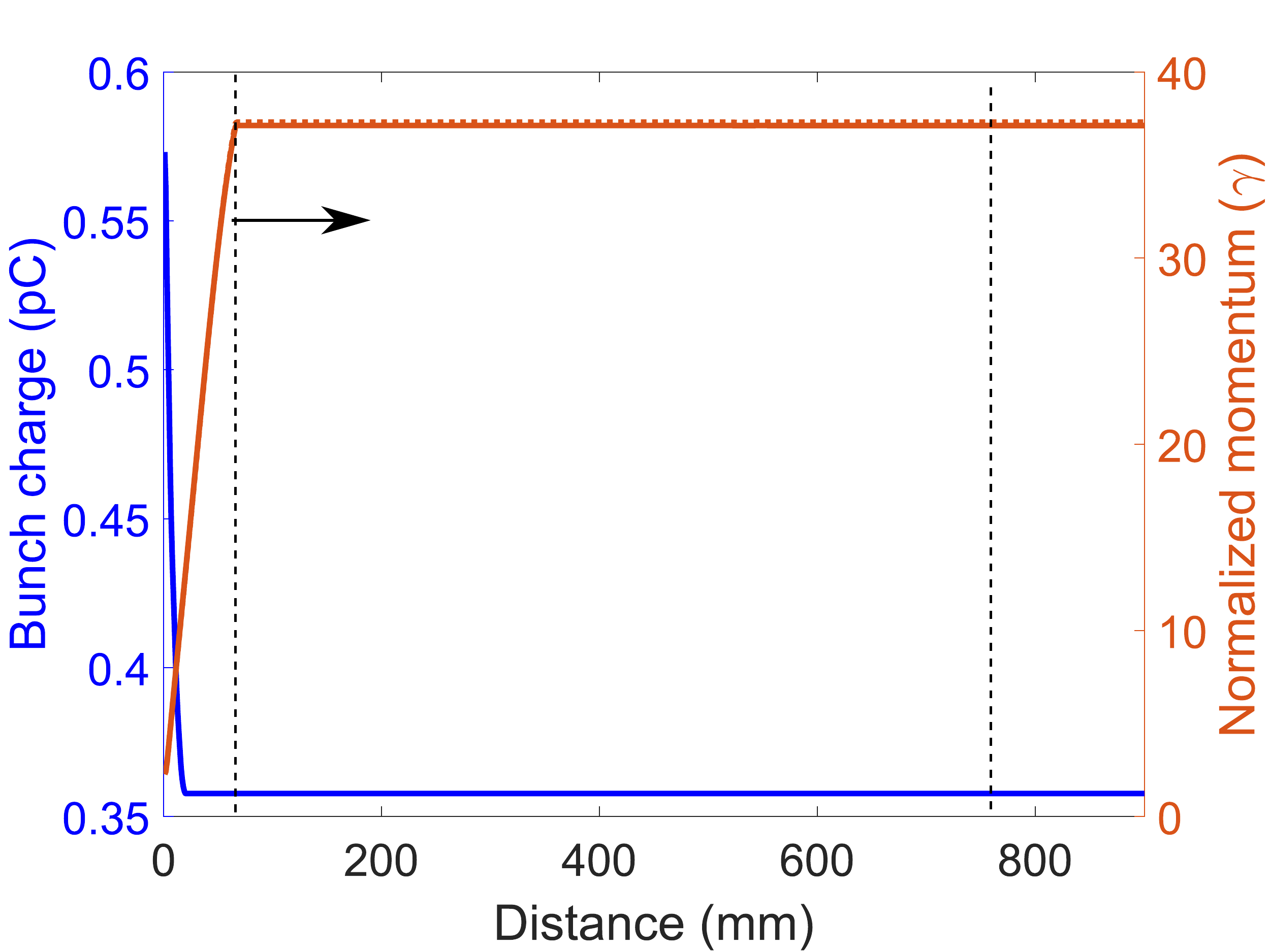} &
\includegraphics[draft=false,width=3.0in]{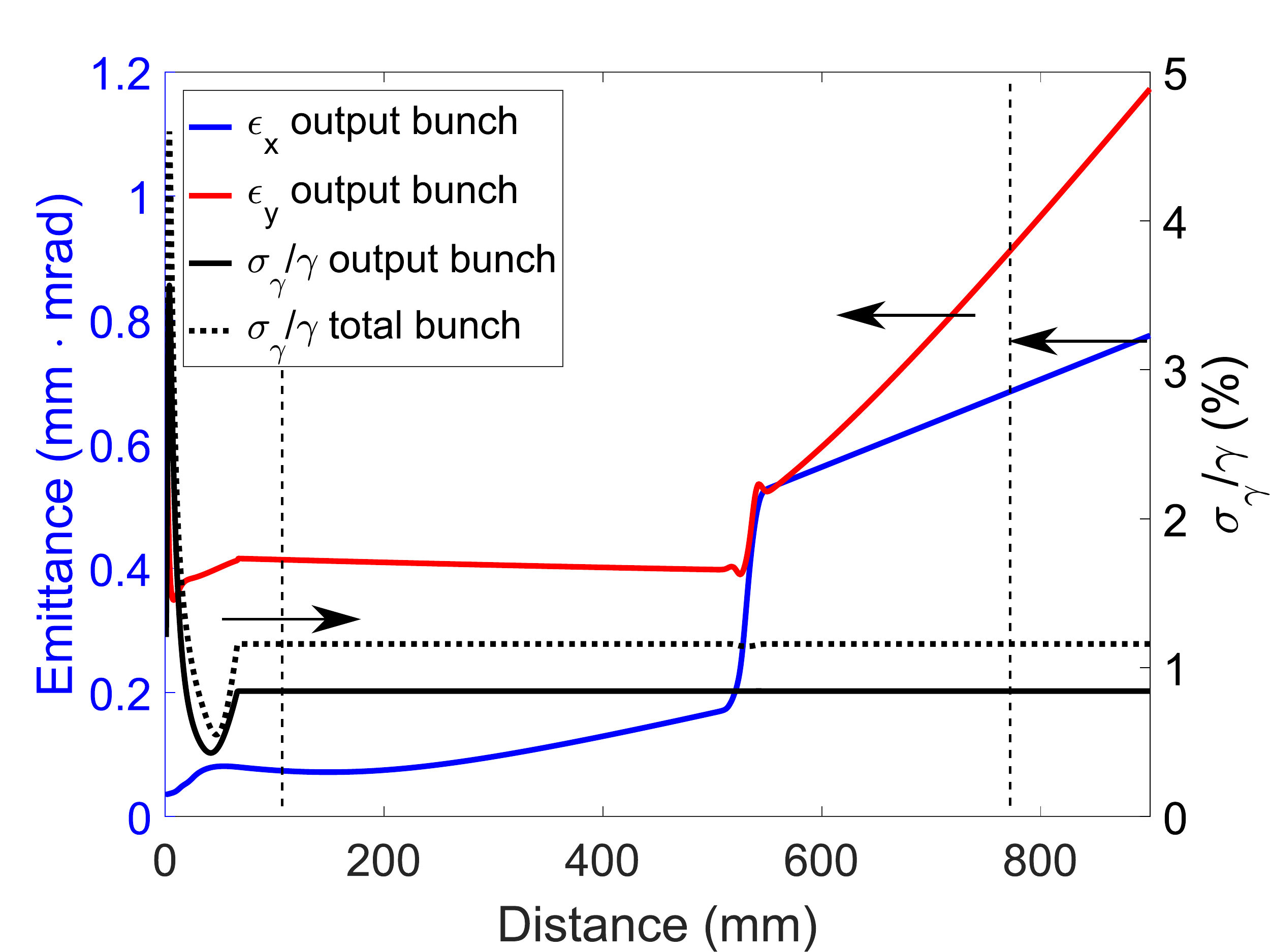} \\
(c) & (d)
\end{array}$
\caption{ASTRA simulation results for the linac: (a) the RMS transverse size along $x$ and $y$ directions, (b) the RMS bunch length, (c) the bunch charge and the average normalized momentum, i.e. average Lorentz factor, and (d) the transverse emittance and bunch energy spread. Dashed lines correspond to the case in which particles hitting the linac boundaries are not removed. The vertical dashed line at Distance=75\,mm defines the end of the linac. The vertical dashed line at Distance=760\,mm defines the ICS point.}
\label{ICSTHzlinacResults}
\end{figure}
From the dispersion diagram, it is seen that the dimensions are designed such that the phase velocity of the propagating mode is matched for relativistic electron acceleration, i.e. $v_p = 0.994c$.
Nevertheless, we observe a strong sensitivity of the linac output to the phase velocity, which dictates tunability of this parameter over a small range.
This is the reason for the small frequency interval for the linac operation in Table\,\ref{linacParameters}.
Such a small tunability for the frequency of the THz beam can be achieved at the laser-driven sources by adjusting the difference frequency.
The uniform accelerating field in the vacuum region of the waveguide (Fig.\,\ref{ICSTHzLinacPerformance}b) is the remarkable property of the TM\textsubscript{01} mode with a phase velocity close to the speed of light.

By examining the gun simulation results depicted in Fig.\,\ref{ICSTHzgunResults}, one simply deduces that the RMS transverse size of the beam equals to 70\,{\textmu}m at the gun exit and increases up to 300\,{\textmu}m after 7.5\,cm travel distance through the linac.
Considering that the inner radius of the dielectric loading is chosen as 242\,{\textmu}m, more than 85\% of the particles hit the waveguide wall and get lost if we send the beam directly to the linac without any focusing equipment.
In order to alleviate the difficulties caused by the strong transverse beam expansion, we devised a quadrupole focusing-defocusing lattice to be installed around the linac waveguide.
Furthermore, the electron bunch exiting the linac should also be focused to the ICS interaction point, which is again achieved by a quadrupole lattice after the linac.
Fig.\,\ref{ICSTHzLinearAcceleration}b shows the configuration of the linac with the quadrupole magnets focusing the beam along the waveguide.
The field gradients of the quadrupoles are shown in table\,\ref{linacParameters}.

ASTRA code is used to simulate the acceleration of the bunch in the linac.
The 12'000 macro-particles leaving the THz injector (corresponding to 0.6\,pC bunch charge) are initialized in the linac.
The traveling wave in the waveguide is modeled by the superposition of two standing waves with $\pi/2$ temporal and spatial phase differences, with space-charge effects being considered in all simulations.
In the linac simulations, as well as the gun simulations, the effect of wakefields are generally neglected.
This is motivated by the previous studies predicting negligible effects due to pico-Coulomb level bunch charge and less than 10\,{\textmu}m bunch lengths \cite{fakhari2017thz}.

The ASTRA simulation results are shown in Fig.\,\ref{ICSTHzlinacResults}.
Two sets of data are shown in these graphs.
The dashed lines ignore the particle collisions with the waveguide walls.
In other words, the total input charge is accelerated to the end, whereas the solid lines show the realistic results, in which the particles hitting the waveguide boundaries are removed from the simulation.
There are two vertical dashed lines in the curves that represent the end of the linac and the focusing point where the ICS occurs.
As observed in the graphs, the RMS transverse size of the beam is kept below 130\,{\textmu}m inside the linac.
The bunch exiting the linac enters the focusing lattice which reduces the transverse size down to 10.7\,{\textmu}m$\times$8.3\,{\textmu}m  at a distance 76\,cm away from the linac where the ICS occurs.
As explained before and can be seen in Fig.\,\ref{ICSTHzlinacResults}b, the bunch is compressed in the linac and furthermore after passing through the linac because of the negative energy chirp introduced to the bunch.
According to Fig.\,\ref{ICSTHzlinacResults}c about 50\% of the particles hit the quartz layer in the waveguide, which are considered to be absorbed and not leaving the linac.
The final mean energy of the electrons is 19\,MeV, which is the design parameter for the ICS process.

\section{Inverse Compton Scattering}

The last step to an x-ray source is the radiation generation through Inverse Compton Scattering (ICS).
The ICS process involves interaction of the accelerated electrons with a counter-propagating optical beam, whose field distribution is given by the Gaussian beam equations to high precision.
Modelling the ICS process is explained thoroughly in section 2.3.
Here, we use the developed software based on the described methodology to evaluate the final output of the machine.

As emphasized in chapter 2, if the linac bunch output is directly imported to the ICS calculations, the assumption of macro-particles results in coherent addition of radiations from electrons considered as one single macro-particle.
This leads to an overestimation of the ultimate photon flux.
To prevent this systematic error, an electron bunch with a real number of electrons is generated with the 6D phase-space of the bunch filled according to the cumulative bunch properties obtained in the last section.

\begin{figure}
\centering
$\begin{array}{ccc}
\includegraphics[draft=false,width=2.0in]{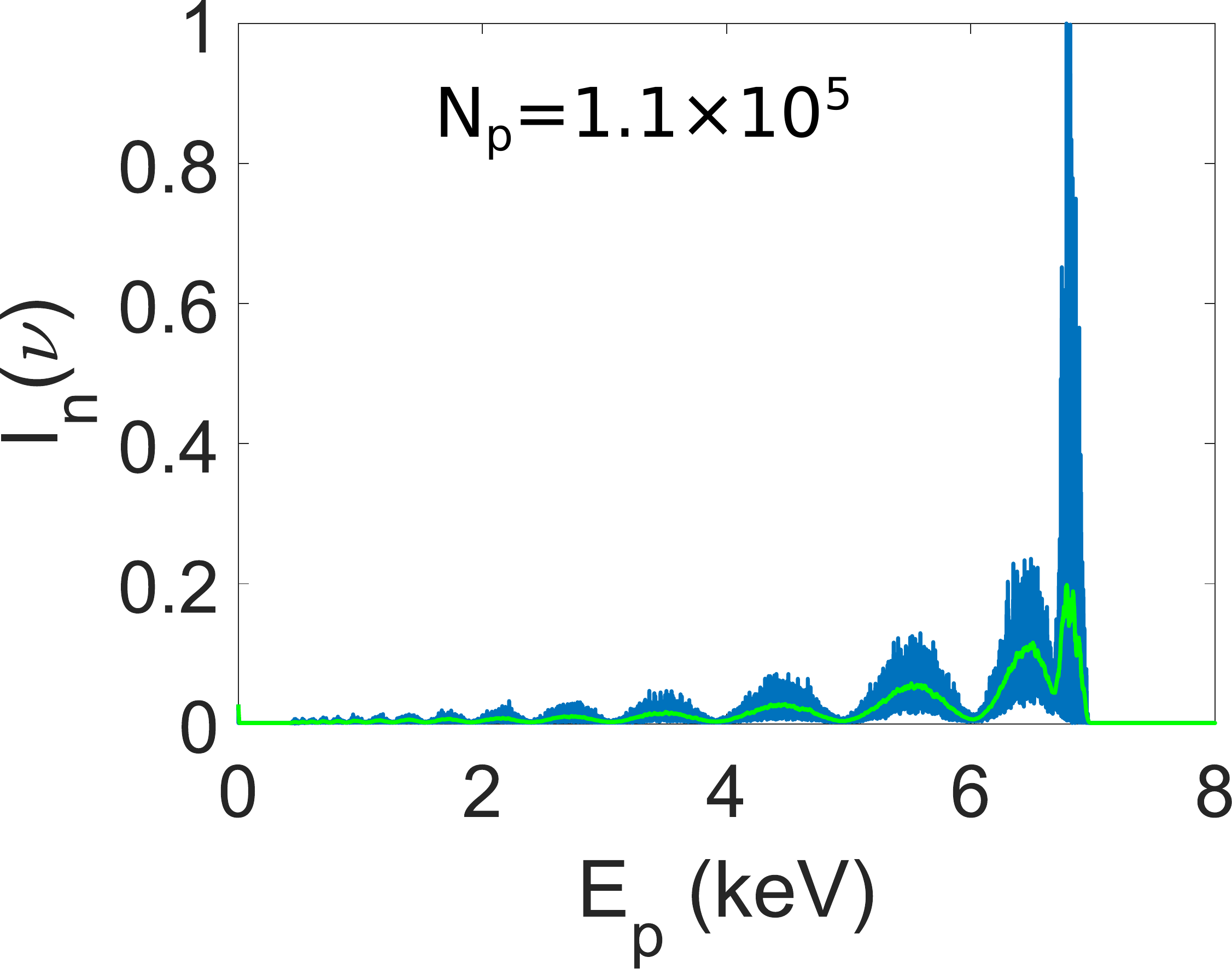} &
\includegraphics[draft=false,width=2.0in]{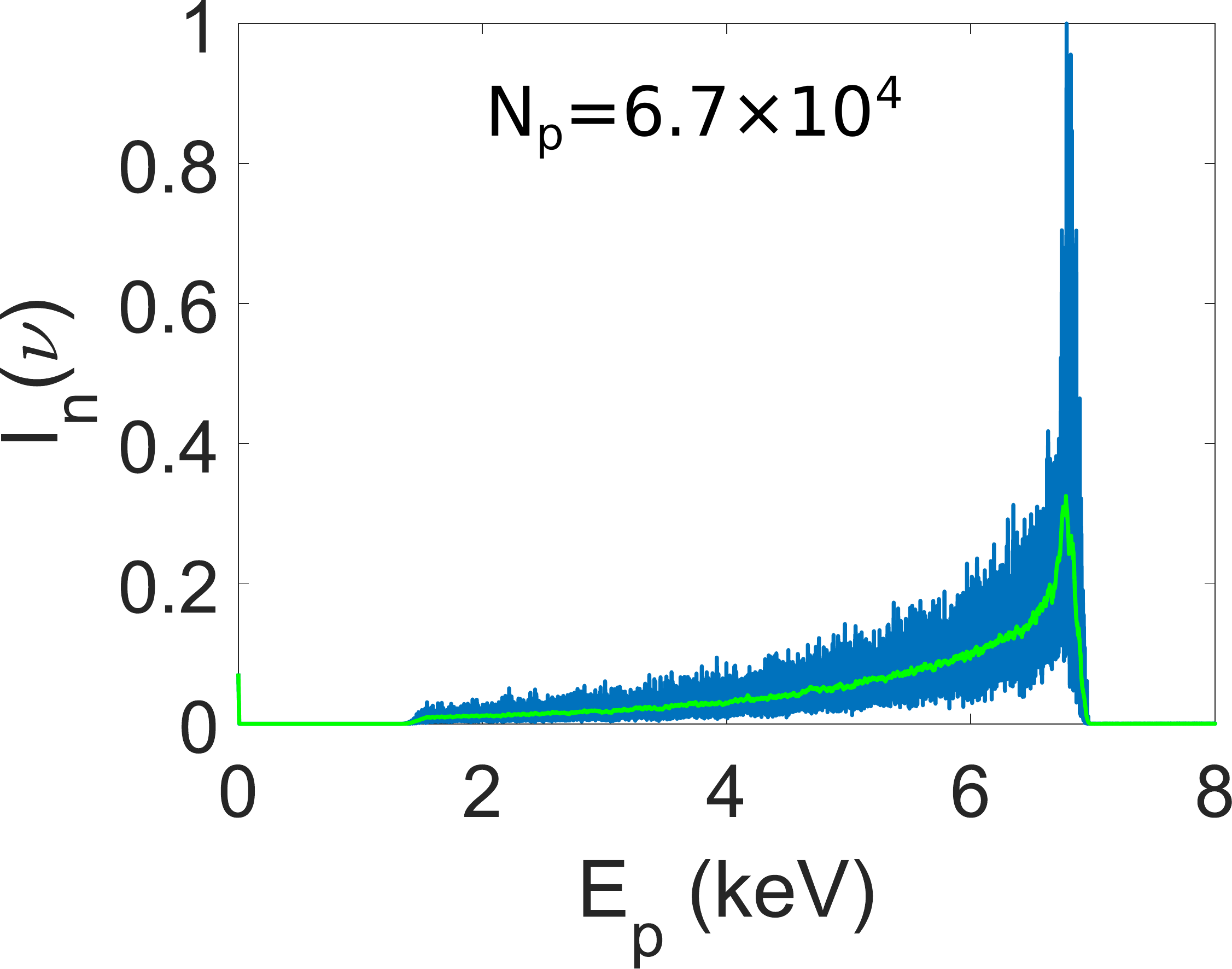} &
\includegraphics[draft=false,width=2.0in]{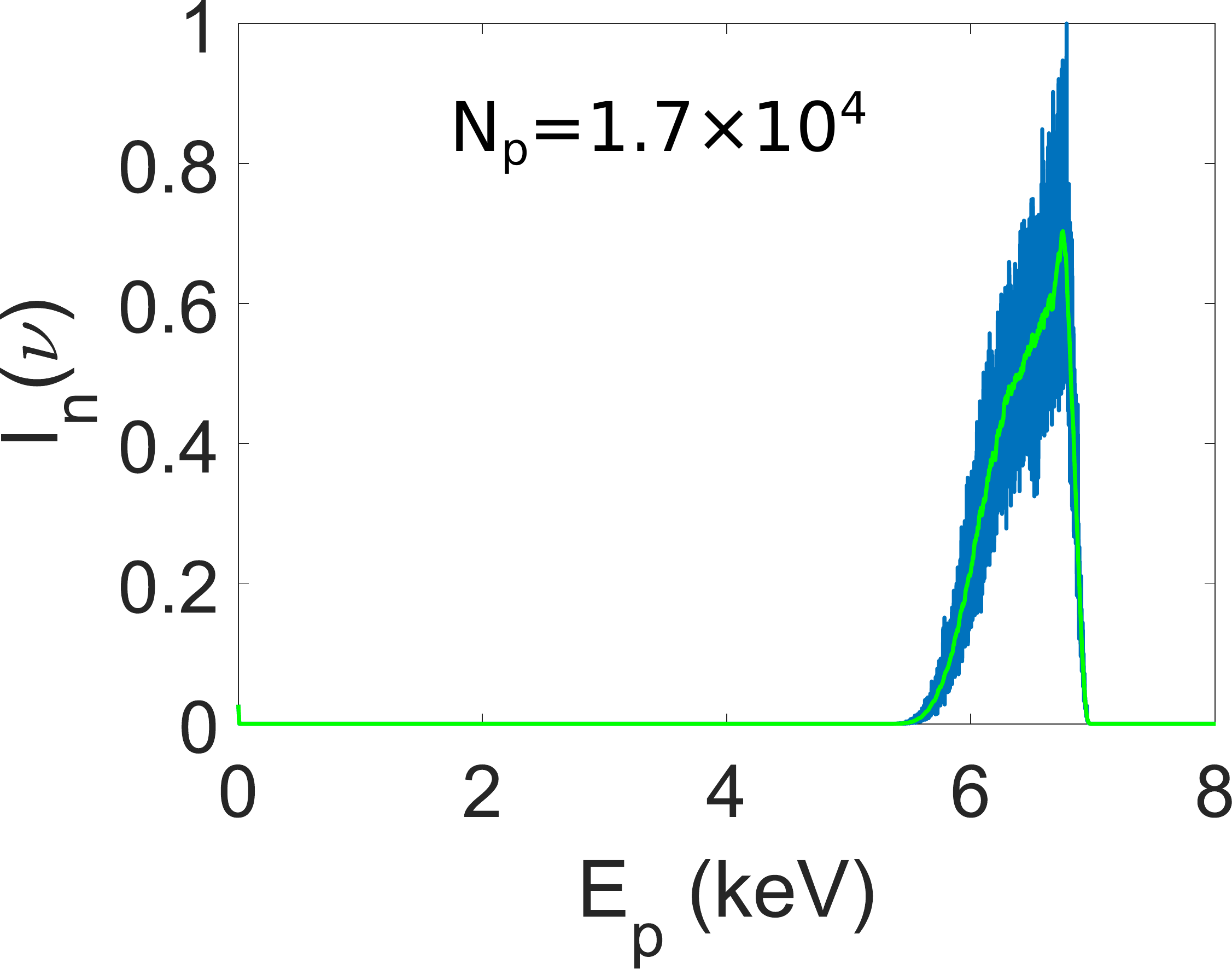} \\
(a) & (b) & (c) \\
\includegraphics[draft=false,width=2.0in]{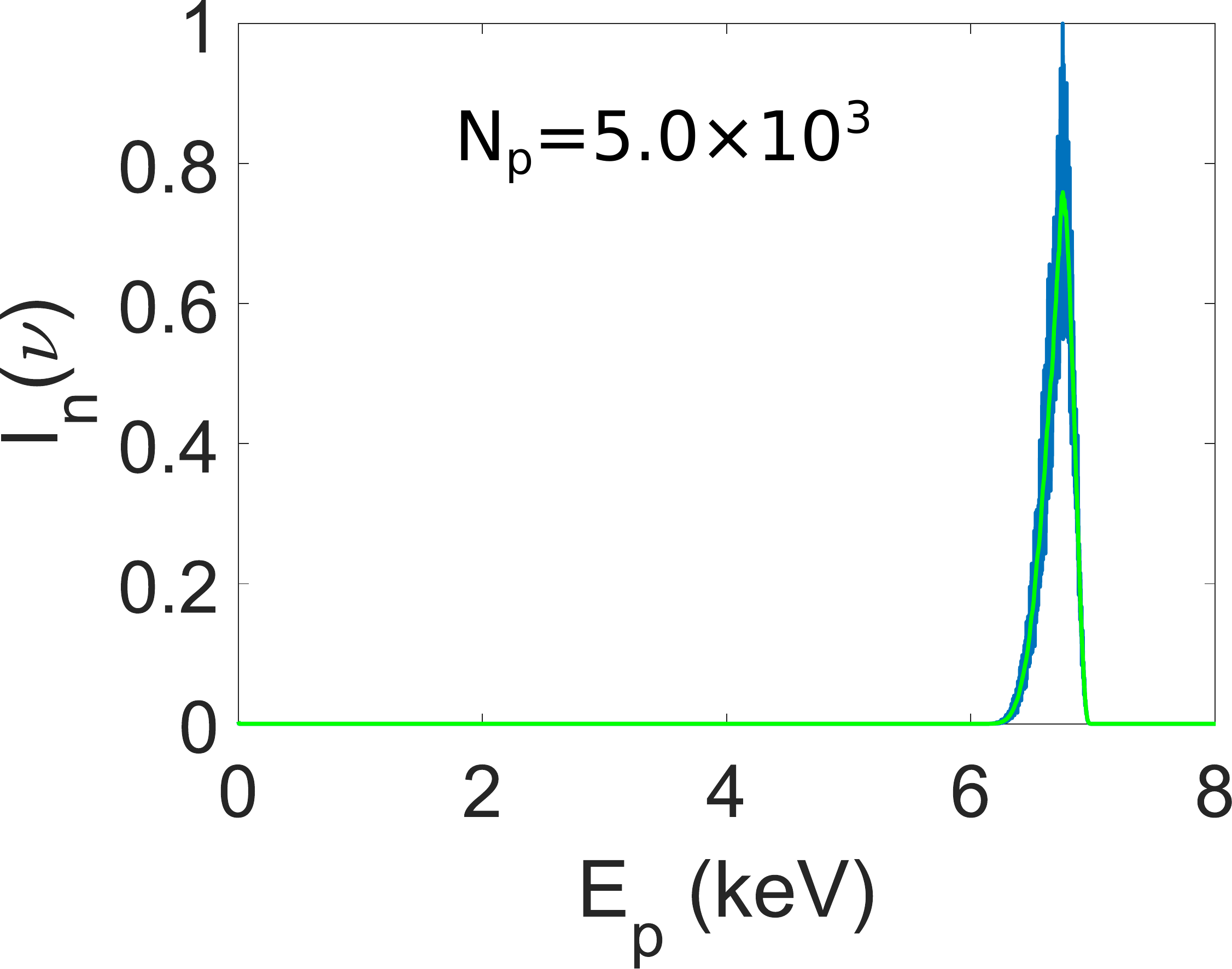} &
\includegraphics[draft=false,width=2.0in]{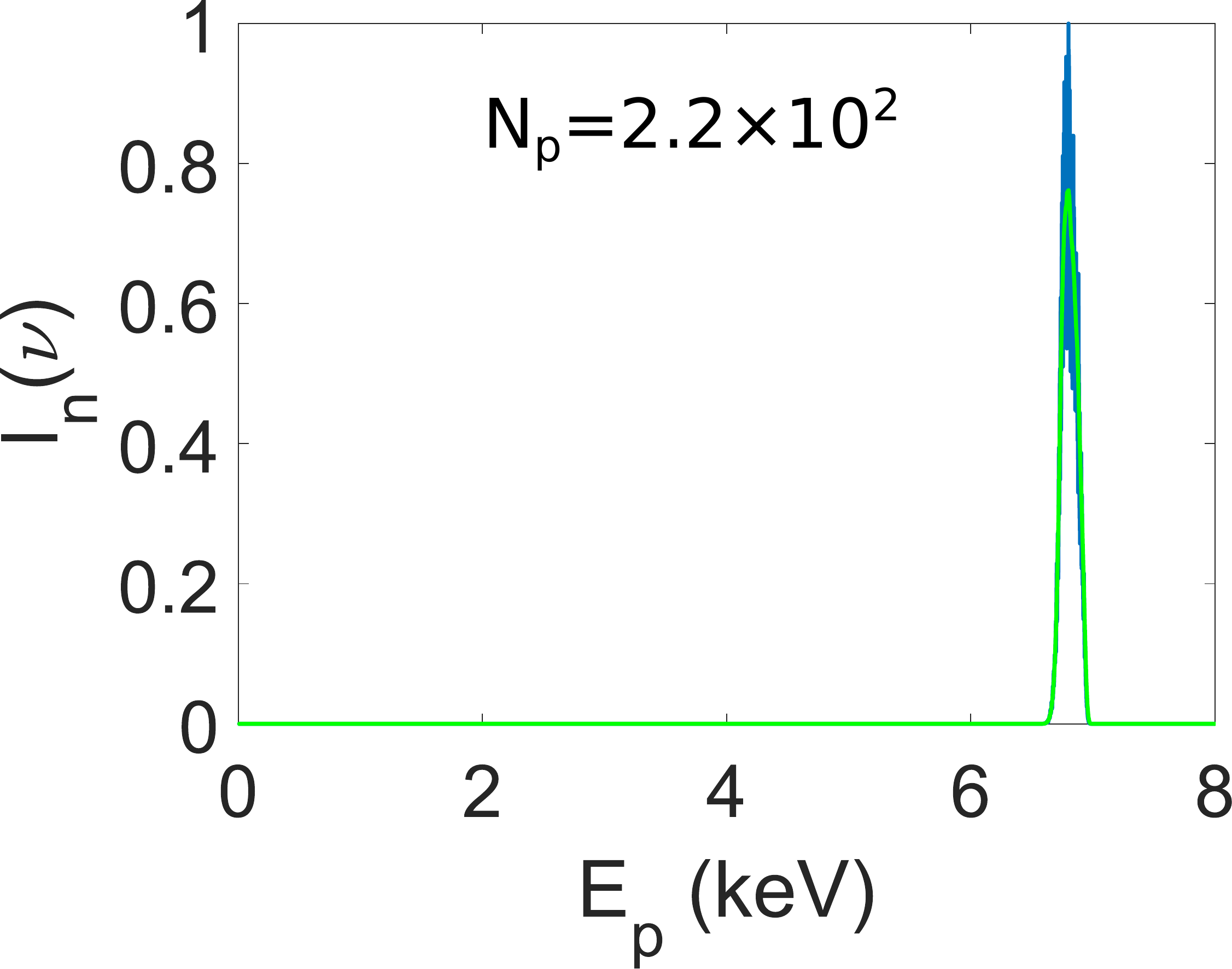} &
\includegraphics[draft=false,width=1.7in]{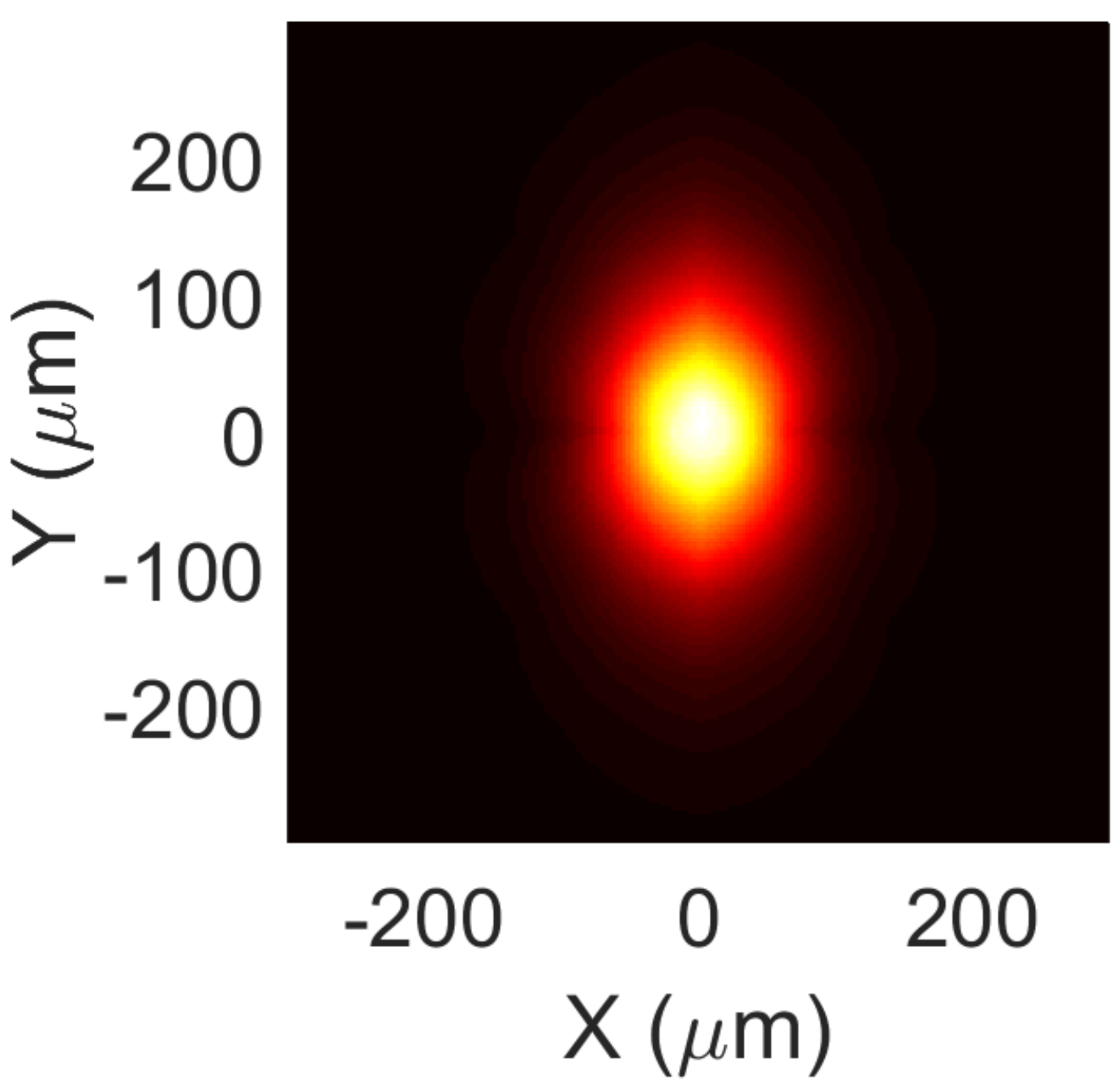} \\
(d) & (e) & (f)
\end{array}$
\caption{The normalized spectrum of the radiated x-ray beam captured within divergence angle (a) 100\,mrad, (b) 50\,mrad, (c) 10\,mrad, (d) 5\,mrad, and (e) 1\,mrad. The total number of photons ($N_p$) for photon energies between 2\,keV and 7\,keV are written in each figure. The green lines delineate the smoothed spectrum. (f) shows the beam profile on an assumed detector 6\,mm away from the ICS interaction point.}
\label{ICSTHzICSResult}
\end{figure}

Table\,\ref{ICSTHzICSparameters} presents the electron bunch properties as well as the assumed ICS laser beam parameters for the x-ray source.
\begin{table}
\centering
\begin{tabular}  { | m{6cm} | m{3cm}| }
\hline
Parameter & Designed value \\ \hline \hline
\multicolumn{2}{|c|}{\textbf{Electron bunch parameters}} \\ \hline
Total charge  & -0.358\,pC \\ \hline
Number of particles & 7'154 $\rightarrow$ 2'236'250 \\ \hline
RMS transverse beam size ($\sigma_x$,$\sigma_y$) & (10.7\,{\textmu}m,8.3\,{\textmu}m) \\ \hline
RMS longitudinal beam size ($\sigma_z$) & 2.9\,{\textmu}m \\ \hline
Mean bunch energy & 19.0\,MeV \\ \hline
RMS energy spread & 0.19\,MeV $\equiv$ 1\% \\ \hline
RMS emittance ($\varepsilon_x$,$\varepsilon_y$) & (0.67,0.88)mm$\cdot$mrad \\ \hline
\multicolumn{2}{|c|}{\textbf{ICS laser parameters}} \\ \hline
Wavelength & 1\,{\textmu}m \\ \hline
Temporal signature & secant hyperbolic \\ \hline
Pulse duration & 1.0\,ps \\ \hline
Spot size ($1/e^2$) & 20\,{\textmu}m \\ \hline
Pulse energy & 100\,mJ \\ \hline
\end{tabular}
\caption{Parameters of the ICS process.}
\label{ICSTHzICSparameters}
\end{table}
In Fig.\,\ref{ICSTHzICSResult}, the final output of the designed x-ray source is shown in form of the normalized energy spectrum of the emitted photons on an assumed detector residing 6\,mm away from the ICS interaction point.
The total number of photons ($N_p$) corresponding to each spectrum are also indicated in each figure.
The total radiated energy density over a plane with (300\,{\textmu}m$\times$300\,{\textmu}m) size is shown in Fig.\,\ref{ICSTHzICSResult}f and represents the radiated beam profile.
Fig.\,\ref{ICSTHzICSResult}a-e show the spectrum captured for different maximum divergence angles, namely 100\,mrad, 50\,mrad, 10\,mrad, 5\,mrad, and 1\,mrad, respectively.
Here, the divergence angle is defined as the half angle of the radiation cone.
As usual for an ICS spectrum, the bandwidths of the radiated x-rays increase with the captured solid angle.
As a result, the total amount of photons generated depends on the solid angle considered for the x-ray beam.
The number of photons generated between 2\,keV and 7\,keV energy range within a radiation cone with 50\,mrad angle is calculated as 6.7$\times 10^4$ photons.
This amount reduces if smaller acceptance angles are considered.
For example, within a 5\,mrad cone with spectrum shown in Fig.\,\ref{ICSTHzICSResult}d, 5.0$\times 10^3$ photons will be radiated.

\begin{figure}
\centering
\includegraphics[draft=false,width=3.0in]{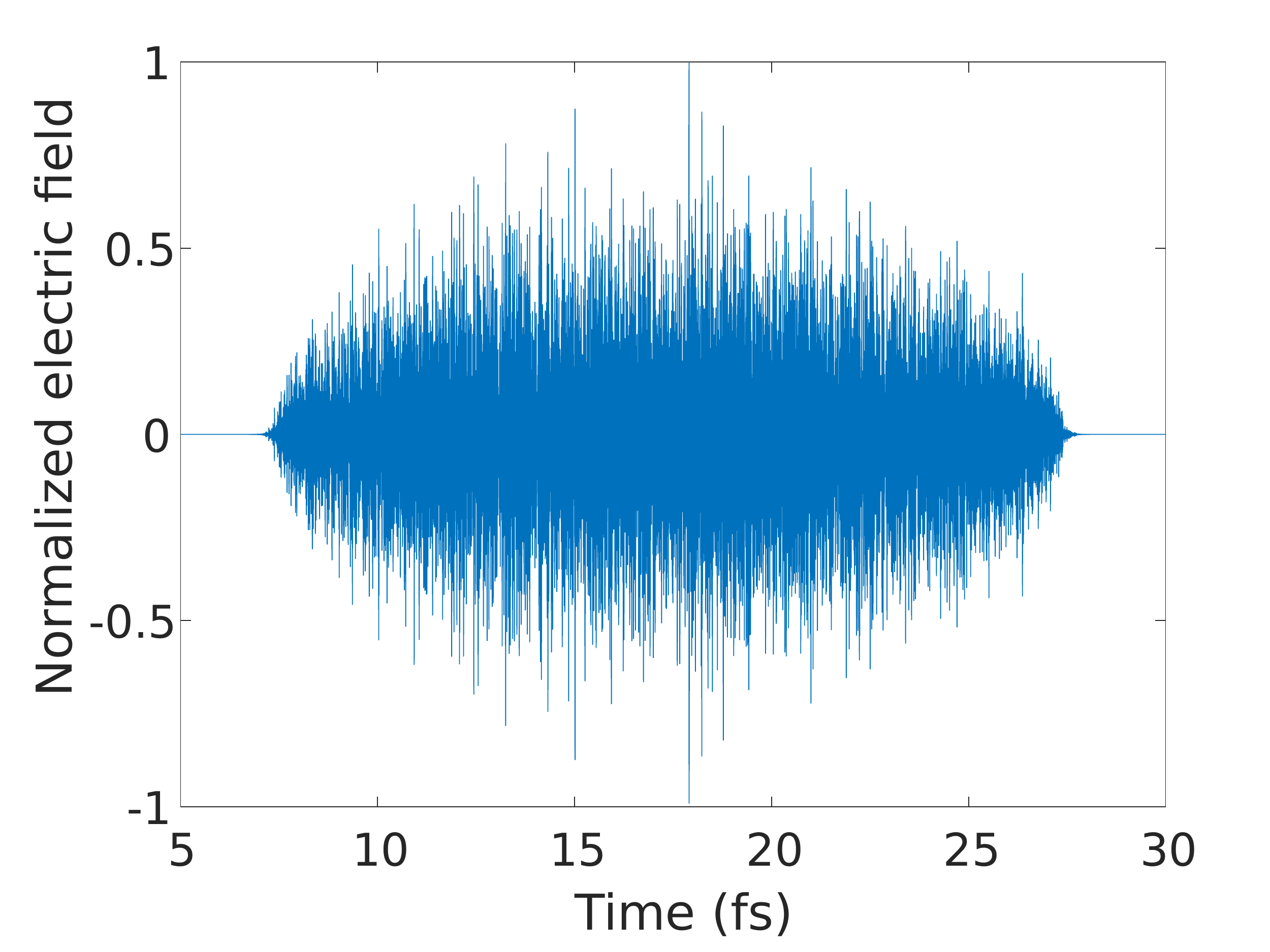}
\caption{The radiated pulse shape on the undulator axis.}
\label{ICSTHzICSTime}
\end{figure}

Eventually the temporal shape of the pulse is depicted in Fig.\,\ref{ICSTHzICSTime}, which shows and x-ray pulse with $\sim 10$\,fs duration.
Further increase in the ICS laser energy not only increases the photon flux of the radiated x-ray pulse, but also facilitates improving the radiated beam quality.
Moreover, the radiation bandwidth reduces when a stronger laser is utilized for ICS interaction.
Continuous reduction in the bandwidth ultimately enables stimulated radiation of the bunch and consequently enhances both temporal and spatial coherence of the radiation.

\section{Conclusion}

In conclusion, we have presented the start-to-end simulations of a compact THz-driven x-ray source, delivering $\sim 10^5$ photons per shot centered around $\sim 6.5$\,keV.
Full-wave simulation techniques for solving Maxwell's equations are utilized to model and design the section of the THz gun, accelerator and ICS source.
The THz generation is accomplished by 2\,J and 70\,mJ lasers radiating at 1\,{\textmu}m, which are readily available at 10\,Hz repetition rate.
The use of complete laser-driven schemes in the x-ray source enables synchronization of all involved elements to the sub-femtosecond precision level.
This advantage considerably assists in achieving high stability over high repetition rates, thereby increasing the photons-per-second flux to values comparable with existing large ICS source facilities.

Since the THz generation process is due to an instantaneous second order optical nonlinearity, there is certainly very little space for randomness in the generated THz fields derived from the optical fields.
Thus, if the shapes of the optical pulses are stable, the resulting THz waveform should not show fluctuations.
However, there will be energy fluctuations of the optical pulse on the order of 0.1-1\% without additional means for pulse energy stabilization.
Therefore, we should estimate the impact of such intensity noise on the electron beam timing.
This is most critical at the output of the gun, where the bunch needs to be injected phase synchronous with the multi-cycle THz field in the linac.
Since the motion of the electrons in the THz gun mostly follows a non-relativistic motion, We can invert the usual formula for the distance of a non-relativistic particle in an accelerating field as $s \propto E_\mathrm{THz} t^2$ to $t \propto s / \sqrt[4]{\mathcal{E}_\mathrm{THz}}$, with $\mathcal{E}_\mathrm{THz}$ being the THz energy.
Thus, the relative error in electron bunch arrival time is equal to a quarter of the relative optical pulse energy fluctuations.
The time of flight of the electron within the 3\,mm long THz electron gun is on the order of 10\,ps.
Therefore, for an energy stability of 0.1-1\% of the optical driver, which is already achievable without additional active stabilization, one should already achieve a stability in electron bunch arrival time at the THz linac of 2.5 - 25\,fs.
This is already well within 1\% of the THz period of 3.3\,ps.

The completely simulated x-ray source will be implemented as the first phase of the AXSIS (Frontiers in Attosecond x-ray Science: Imaging and Spectroscopy) project at Deutsches Elektronen-Synchrotron in Hamburg, Germany to guide the construction of more advanced coherent FEL-like THz-driven x-ray sources \cite{kartner2016axsis}.

\chapter{Outlook \label{conclusion}}

The research studies explained in this thesis are carried out in the framework of the AXSIS (Frontiers in Attosecond X-ray Science: Imaging and Spectroscopy) project.
In the AXSIS project, we look at the possibility of creating compact, fully coherent free-electron laser sources entirely laser driven for intrinsic synchronization of all components towards attosecond X-ray imaging and spectroscopy.
The start to end simulation presented in chapter 6 is the source design considered for the first implementation phase in this project, realizing an incoherent ICS source.
Using the developed framework in this thesis, particularly the developed full-wave software MITHRA, we aim to discover techniques to achieve coherent super-radiant radiation in a compact optical undulator based source.

To obtain a compact FEL machine, there are already several approaches followed based on laser plasma wake field acceleration of electrons, which provides very high gradient accelerating fields in the multi-GV/m range.
However, so far the energy spread of the generated electron bunches is not low enough for FEL operation.
The path we have considered toward such an operation regime is realization of an electron crystal at the ICS stage, leading to a compact, fully coherent, THz-driven X-ray source.
Use of THz-based acceleration technology allows for ultrashort electron bunches close to attosecond lengths, resulting in an attosecond X-ray source.
This achievement enables outrunning radiation damage effects to the crystal samples due to the necessary high X-ray irradiance required to acquire diffraction signals.
The highly synergistic project AXSIS starts from a completely clean slate rather than conforming to the specifications of a large free-electron laser (FEL) user facility, to optimize the entire instrumentation towards fundamental measurements of the mechanism of light absorption and excitation energy transfer.
A multidisciplinary team formed by laser-, accelerator,- X-ray scientists and as well as spectroscopists and biochemists optimizes X-ray pulse parameters, in tandem with sample delivery, crystal size, and advanced X-ray detectors.
Ultimately, the new capability, attosecond serial X-ray crystallography and spectroscopy, will be applied to one of the most important problems in structural biology, which is to elucidate the dynamics of light reactions, electron transfer and protein structure in photosynthesis.
This unique tool will fulfill the dream of observing chemical reactions and biological processes in real space and real time at the necessary time and length scales of atoms and molecules as is the goal of the large X-ray FELs but maybe at a cost level that is commensurate with University or Industrial Laboratory settings.
By developing this technique based on a compact laboratory-scale X-ray source, the availability of attosecond serial crystallography and spectroscopy is vastly extended to the general science community.

In addition to the compact X-ray source pursued in AXSIS, each developed element in the presented efforts has its specific applications, calling the need for further separate studies.
Currently, there is no full-wave numerical software that captures complete features of field and particle interaction.
Put differently, there is no high-order based algorithm that calculates bunch dynamics by including radiation and space-charge effects, while also calculating the effects of dielectric materials.
The DGTD/PIC code developed in this thesis provides a well-established framework to implement and test various algorithms and judge their efficacy.
A dedicated study on the already proposed techniques can lead to the discovery of a suitable algorithm for development of such a software.
The software MITHRA is the first full-wave numerical solver for free electron lasers and can thus be used for verification of various FEL schemes as well as testing the already developed techniques for fast evaluation of FEL radiation.
In order to have the software widely used by the FEL community, the implementation of various devices used in real FEL facilities is mandatory.
Future works on including the focusing systems, tapered undulators, and novel seeding mechanisms in the next versions of MITHRA are already planned.

Studies on laser-induced field emission accomplished in chapter three showed the feasibility of ultrafast electron sources with femtosecond-level emission times.
This possibility is not only a breakthrough for compact X-ray sources, but also can revolutionize the current technology in conventional RF injectors.
Nonetheless, the strong fields in RF injectors are a dramatically different operational environment compared to the DC fields in which the structures are currently tested.
The strong RF fields, further enhanced at the device corners, may trigger dark current in nanostructured photocathodes and detrimentally reduce the damage threshold of the cathode.
Incorporating structured cathodes in real injectors currently encounters as yet unresolved challenges, whose resolutions require novel ideas and detailed investigations.
Beyond development of electron source technology, these ultrafast emitters can be used for novel applications in ultrafast sensing and measurement down to optical sub-cycles, promising for attosecond and zeptosecond precision in measured dynamics.

The technology of ultrafast single-cycle THz guns potentially serves other domains where sub-relativistic electrons are beneficial.
Sub-relativistic electron beams have found widespread use by biologists, chemists and physicists investigating ultra-fast science through electron diffractive imaging and by physicians performing particle therapy using electrons.
The resolution of imaging using electron diffraction principles is directly determined by the emittance of the electron beam.
The low-emittance and bright electron beams produced within the compact setup of a THz gun will remarkably enhance the achievable resolution in this technology.
In chapter three operation of a STEAM device was presented.
The device state is already mature and well tested to be used in existing facilities.
In the future, further utilization of this device in accelerator facilities to measure and manipulate the electron beams with unprecedented high precision offered by the STEAM device is foreseen.

The studies here underpin a new research area of laser-driven THz acceleration, which was started by our group around six years ago and remains still in its infancy stage.
However, great progress with transformative impact has been made over the past years demonstrating the promises of this technique for numerous applications.
With new inventions for novel applications, new challenges emerge which call for further research and development.
This will be an ongoing effort for decades aiming at easy and low-cost access to high energy particles for universities and research centers around the world.

%
%
%
%
  \cleardoublepage                           
  \ihead{BIBLIOGRAPHY}

  \cleardoublepage 
%
  \ihead{LIST OF PUBLICATIONS}
  
  \chapter*{List of Publications}\label{publications}
  
  \newenvironment{outerlist}[1][\enskip\textbullet]
  {\begin{itemize}[#1]}{\end{itemize}
  	\vspace{-.6\baselineskip}}
  
  \newenvironment{innerlist}[1][\enskip\textbullet]%
  {\begin{compactitem}[#1]}{\end{compactitem}}
  
  \newcommand{\blankline}{\quad\pagebreak[2]}
  
  \blankline
  
  \textbf{Book Chapters}
  \begin{outerlist}
  	\item[] J. Smajic, M. Mishrikey, \textbf{A. Fallahi}, C. Hafner and R. Vahldieck, "Efficiency of Various Stochastic Optimization Algorithms in High Frequency Electromagnetic Applications," chapter in \textit{Nature Inspired Cooperative Strategies for Optimization}(\href{http://www.dmi.unict.it/nicso2007/}{NICSO 2007}), Springer
  	Berlin/Heidelberg, June 2008.
  \end{outerlist}
  
  \blankline
  
  \textbf{Journal Papers}
  \begin{outerlist}
  	\item[] M. Fakhari, \textbf{A. Fallahi}, L. Wang, K. Ravi, F. X. K\"artner, ``Start-to-End simulation of a compact THz-driven X-ray source'', submitted to \href{http://prb.aps.org/prab}{\emph{Physical Review Accelerators and Beams}}.
  	
  	\item[] G. Vashchenko, R. Assmann, U. Dorda, M. Fakhari, \textbf{A. Fallahi}, K. Galaydych, F. X. K\"artner, B. Marchetti, N. Matlis, T. Vinatier, W. Qiao, C. Zhou, ``Performance analysis of the prototype THz-driven electron gun for the AXSIS project'' \href{https://www.sciencedirect.com/journal/nuclear-instruments-and-methods-in-physics-research-section-a-accelerators-spectrometers-detectors-and-associated-equipment}{\emph{Nuclear Instruments and Methods in Physics Research Section A: Accelerators, Spectrometers, Detectors and Associated Equipment}}, April 2018.
  	
  	\item[] N. Matlis, F. Ahr, A.-L. Calendron, H. Cankaya, G. Cirmi, T. Eichner, \textbf{A. Fallahi}, M. Fakhari, A. Hartin, M. Hemmer, W. R. Huang, H. Ishizuki, S. W. Jolly, V. Leroux, A. R. Maier, J. Meier, W. Qiao, K. Ravi, D. N. Schimpf, T. Taira, X. Wu, L. Zapata, C. Zapata, D. Zhang, C. Zhou, F. X. K\"artner, ``Acceleration of electrons in THz driven structures for AXSIS'' \href{https://www.sciencedirect.com/journal/nuclear-instruments-and-methods-in-physics-research-section-a-accelerators-spectrometers-detectors-and-associated-equipment}{\emph{Nuclear Instruments and Methods in Physics Research Section A: Accelerators, Spectrometers, Detectors and Associated Equipment}}, February 2018.
  	
  	\item[] \textbf{A. Fallahi}, and F. X. K\"artner, ``Design Strategies for Single-Cycle Ultrafast Electroin Guns", accepted in special issue for ultrafast phenomenon in \href{http://iopscience.iop.org/0953-4075/47/23/234015}{\emph{Journal of Physics B}}.
  	
  	\item[] D. Zhang, \textbf{A. Fallahi}, M. Hemmer, X. Wu, M. Fakhari, Y. Hua, H. Cankaya, A.-L. Calendron, L. E. Zapata, N. H. Matlis, F. X. K\"artner, ``Segmented Terahertz Electron Accelerator and Manipulator (STEAM)", \href{http://www.sciencemag.org/}{\emph{Nature Photonics}} (2018) 1.
  	
  	\item[] A.-L. Calendron, J. Meier, M. Hemmer, L. E. Zapata, F. Reichert, H. Cankaya, D. N. Barre, Y. Hua, G. Chang, A. Kalaydzhyan, \textbf{A. Fallahi}, N. Matlis, F. X. K\"artner, ``Laser System Design for Table-top Free-electron Laser", \href{https://www.cambridge.org/core/journals/high-power-laser-science-and-engineering}{\emph{High Power Laser Science and Engineering}}, vol. 6, March 2018.
  	
  	\item[] H. Ye, S. Trippel, M. Di Fraia, \textbf{A. Fallahi}, O. M\"{u}cke, F. X. K\"artner, J. K\"{u}pper, ``Velocity-map imaging for emittance characterization of multiphoton-emitted electrons from a gold surface", \href{http://prb.aps.org/}{\emph{Physical Review Applied}}, vol. 9, no. 4, pp. 044018, April 2018.
  	
  	\item[] R. G. Hobbs, W. P. Putnam, \textbf{A. Fallahi}, Y. Yang, F. X. K\"artner, and K. K. Berggren, ``Mapping Photoemission and Hot-Electron Emission from Plasmonic Nanoantennas", \href{http://pubs.acs.org/journal/nalefd}{\emph{Nano Letters}}, vol. 17, no. 10, pp. 6069-6076, September 2014.
  	
  	\item[] K. Khaliji, \textbf{A. Fallahi}, L. Martin-Moreno, and T. Low ``Tunable plasmon-enhanced birefringence in ribbon array of anisotropic 2D materials", \href{http://prb.aps.org/}{\emph{Physical Review B}} vol. 95, no. 20, pp. 201401, May 2017.
  	
  	\item[] \textbf{A. Fallahi}, A. Yahaghi, and F. X. K\"artner ``MITHRA 1.0: A Full-wave Simulation Tool for Free Electron Lasers", \href{https://www.journals.elsevier.com/computer-physics-communications/}{\emph{Computer Physics Communications}}, March 2018.
  	
  	\item[] L. J. Wong, K.-H. Hong, S. Carbajo, \textbf{A. Fallahi}, M. Soljacic, J. D. Joannopoulos, F. X. K\"artner, and I. Kaminer ``Laser-Induced Linear Electron Acceleration in Free Space", \href{http://www.nature.com/srep/}{\emph{Nature Scientific reports}}, vol. 7, no. 11159, September 2017.
  	
  	\item[] A. Yahaghi, K. Ravi, \textbf{A. Fallahi}, and F. X. K\"artner ``Designing chirped aperiodically poled structures for high-energy single-cycle terahertz generation", \href{https://www.osapublishing.org/josab}{\emph{Journal of Optical Society of America B}}, vol. 34, no. 3, pp. 590-600, March 2017.
  	
  	\item[] M. Fakhari, \textbf{A. Fallahi}, and F. X. K\"artner ``THz Cavities and Injectors for Compact Electron Acceleration Using Laser-driven THz Generation Sources", \href{http://prb.aps.org/prab}{\emph{Physical Review Accelerators and Beams}} vol. 20, pp. 041302, April 2017.
  	
  	\item[] U. Dorda, R. Assmann, R. Brinkmann, K. Fl\"{o}ttmann, I. Hartl, M. H\"{u}ning, F. K\"{a}rtner, \textbf{A. Fallahi}, B. Marchetti, Y. Nie, J. Osterhoff, H. Schlarb, J. Zhu, A. R. Maier ``SINBAD - The accelerator R{\&}D facility under construction at DESY", \href{http://www.sciencedirect.com/science/journal/01689002}{\emph{Nuclear Instruments and Methods in Physics Research Section A: Accelerators, Spectrometers, Detectors and Associated Equipment}} vol. 829, pp. 233-236, September 2016.
  	
  	\item[] F. K\"{a}rtner, F. Ahr, A.-L. Calendron, H. Cankaya, S. Carbajo, G. Chang, G. Cirmi, K. D\"{o}rner, U. Dorda, \textbf{A. Fallahi}, A. Hartin, M. Hemmer, R. Hobbs, Y. Hua, W. R. Huang, R. Letrun, N. Matlis, V. Mazalova, O. M\"{u}cke, E. Nanni, W. Putnam, K. Ravi, F. Reichert, I. Sarrou, X. Wu, A. Yahaghi, H. Ye, L. Zapata, D. Zhang, C. Zhou, D. Miller, K. Berggren, H. Graafsma, A. Meents, R. Assmann, H. Chapman, P. Fromme ``AXSIS: Exploring the frontiers in attosecond X-ray science, imaging and spectroscopy", \href{http://www.sciencedirect.com/science/journal/01689002}{\emph{Nuclear Instruments and Methods in Physics Research Section A: Accelerators, Spectrometers, Detectors and Associated Equipment}}, vol. 829, pp. 24-29, September 2016.
  	
  	\item[] W. R. Huang, \textbf{A. Fallahi}, X. Wu, E. A. Nanni, H. Cankaya, A.-L. Calendron, D. Zhang, K. Ravi, K.-H. Hong, E. P. Ippen and F. X. K\"artner ``THz-driven sub-keV Electron Gun",  \href{https://www.osapublishing.org/optica}{\emph{Optica}}, vol. 3, no. 11, pp. 1209-1212, October 2016.
  	
  	\item[] \textbf{A. Fallahi}, M. Fakhari, A. Yahaghi, M. Arrieta, and F. X. K\"artner ``Short Electron Bunch Generation Using Single-Cycle Ultrafast Electron Guns", \href{http://prb.aps.org/prab}{\emph{Physical Review Accelerators and Beams}} vol. 19, no. 8, pp. 081302, August 2016.
  	
  	\item[] E. S. Torabi, \textbf{A. Fallahi}, A. Yahaghi ``Evolutionary Optimization of Graphene-Metal Metasurfaces for Tunable Broadband Terahertz Absorption", \href{http://www.ict.csiro.au/aps/}{\emph{IEEE Transactions on Antennas and Propagation}} vol. 65, no. 3, pp. 1464-1467, March 2017.
  	
  	\item[] \textbf{A. Fallahi}, and A. Enayati ``Modeling Pyramidal Absorbers Using the Fourier Modal Method", \href{http://ieeexplore.ieee.org/xpl/RecentIssue.jsp?punumber=15}{\emph{IEEE Transactions on Electromagnetic Compatibility}}, vol. 58, no. 3, pp. 820-827, June 2016.
  	
  	\item[] S. Carbajo, P. Alcorta, A. L. Calendron, H. Cankaya, R. Koustuban, F. Ahr, \textbf{A. Fallahi}, X. Wu, and F. X. K\"artner ``Effective Path Towards Relativistic Transients at mm-wavelengths", \href{http://arxiv.org/abs/1602.08136}{\emph{arXiv preprint arXiv:1602.08136}}.
  	
  	\item[] R. G. Hobbs, Y. Yang, \textbf{A. Fallahi}, P. D. Keathley, E. De Leo, F. X. K\"artner, W. S. Graves, and K. K. Berggren ``High-Yield, Ultrafast, Surface Plasmon-Enhanced, Au Nanorod Optical Field Electron Emitter Arrays", \href{http://pubs.acs.org/doi/abs/10.1021/nn504594g}{\emph{ACS nano}}, vol. 8, no. 11, pp. 11474-11482, November 2014.
  	
  	\item[] E. A. Nanni, W. R. Huang, K. Ravi, \textbf{A. Fallahi}, G. Moriena, R. J. Miller, and Franz X. K\"artner, ``Terahertz-driven Linear Electron Acceleration", \href{http://arxiv.org/abs/1411.4709}{\emph{Nature Communications}}, vol. 6, no. 8486, doi:10.1038/ncomms9486.
  	
  	\item[] W. R. Huang, E. A. Nanni, K. Ravi, K.-H. Hong, L. J. Wong, P. D. Keathley, \textbf{A. Fallahi}, L. Zapata, Franz X. K\"artner, ``Toward a Terahertz-driven Electron Gun", \href{http://www.nature.com/srep/}{\emph{Nature Scientific reports}}, vol. 5, pp. 14899, October 2015.
  	
  	\item[] \textbf{A. Fallahi}, T. Low, M. Tamagnone, J. Perruisseau-Carrier, ``Nonlocal Electromagnetic Response of Graphene Nanostructures", \href{http://prb.aps.org/}{\emph{Physical Review B}} vol. 91, no. 12, pp. 121405, March 2015.
  	
  	\item[] M. E. Swanwick, P. D. Keathley, \textbf{A. Fallahi}, P. R. Krogen, G. Laurent, J. Moses, Franz X. K\"artner, and L. F. Vel\'asquez-Garc\'ia, ``Nanostructured Ultrafast Silicon-Tip Optical Field-Emitter Arrays", \href{http://pubs.acs.org/journal/nalefd}{\emph{Nano Letters}} vol. 14, no. 9, pp. 5035-5043, July 2014.
  	
  	\item[] \textbf{A. Fallahi}, and F. X. K\"artner, ``Field-based DGTD/PIC Algorithm for General and Stable Simulation of Interaction Between Light and Electron Bunches", \href{http://iopscience.iop.org/0953-4075/47/23/234015}{\emph{Journal of Physics B}} vol. 47, no. 23, pp. 234015, November 2014.
  	
  	\item[] M. Tamagnone, \textbf{A. Fallahi}, J. R. Mosig, J. Perruisseau-Carrier, ``Fundamental limits and near-optimal design of graphene modulators and non-reciprocal devices", \href{http://www.nature.com/nphoton/index.html}{\emph{Nature Photonics}} vol. 8, pp. 556-563, April 2014.
  	
  	\item[] L. J. Wang, \textbf{A. Fallahi}, and F. X. K\"artner, ``Compact Electron Acceleration and Bunch Compression in THz Waveguides", \href{http://www.opticsinfobase.org/oe/home.cfm}{\emph{Optics Express}} vol. 21, no. 8, pp. 9792-9806, April 2013.
  	
  	\item[] \textbf{A. Fallahi}, and J. Perruisseau-Carrier, ``Manipulation of Giant Faraday Rotation in Graphene Metasurfaces", \href{http://apl.aip.org/}{\emph{Applied Physics Letters}} vol. 101, no. 23, pp. 231605, December 2012.
  	
  	\item[] \textbf{A. Fallahi}, and J. Perruisseau-Carrier, ``Design of Tunable Graphene Metasurfaces", \href{http://prb.aps.org/}{\emph{Physical Review B}} vol. 86, no. 19, pp. 195408, November 2012.
  	
  	\item[] \textbf{A. Fallahi}, and B. Oswald, ``The element Level Time Domain (ELTD) Method for the Analysis of Nano-Optical Systems: II. Dispersive Media", \href{http://www.elsevier.com/wps/find/journaldescription.cws_home/658453/description}{\emph{Photonics and Nanostructures Fundamental and Applications}} , vol. 10, no. 2, pp. 223-235, April 2012.
  	
  	\item[] \textbf{A. Fallahi}, B. Oswald, and P. Leidenberger, ``The element Level Time Domain (ELTD) Method for the Analysis of Nano-Optical Systems: I. Nondispersive Media", \href{http://www.elsevier.com/wps/find/journaldescription.cws_home/658453/description}{\emph{Photonics and Nanostructures Fundamental and Applications}}, vol. 10, no. 2, pp. 207-222, April 2012.
  	
  	\item[] B. Oswald, A. Fomins, \textbf{A. Fallahi}, P. Leidenberger, and P. Bastian, ``3-Dimensional Time-Domain Full-Wave Analysis of Optical Array Antennas", \href{http://www.aspbs.com/ctn/}{\emph{Journal of Computational and Theoretical Nanoscience}}, vol. 8, no. 8, pp. 1573-1589, August 2011.
  	
  	\item[] \textbf{A. Fallahi}, and B. Oswald, ``On the Computation of Electromagnetic Dyadic Green's function in Spherically Multilayered Media", \href{http://www.mtt.org/publications/Transactions/transactions.htm}{\emph{IEEE Transactions on Microwave Theory and Techniques}}, vol. 59, no. 6, pp. 1433-1440, June 2011.
  	
  	\item[] \textbf{A. Fallahi}, A. Yahaghi, H. Benedickter, H. Abiri, M. Shahabadi, and C. Hafner, ``Thin Wideband Radar Absorbers", \href{http://www.ict.csiro.au/aps/}{\emph{IEEE Transactions on Antennas and Propagation}}, vol. 58, no. 12, pp. 4051-4058, December 2010.
  	
  	\item[] \textbf{A. Fallahi}, and C. Hafner, ``Analysis of Semi-infinite Periodic Structures Using a Domain Reduction Technique", \href{http://www.opticsinfobase.org/josaa}{\emph{Journal of the Optical Society of America A}}, vol. 27, no. 1, pp. 40-49, January 2010.
  	
  	\item[] \textbf{A. Fallahi}, A. Yahaghi, H. Abiri, M. Shahabadi, and C. Hafner, ``Large Overlapping Subdomain Basis Functions For the Analysis of Frequency Selective Surfaces", \href{http://www.mtt.org/publications/Transactions/transactions.htm}{\emph{IEEE Transactions on Microwave Theory and Techniques}}, vol. 58, no. 8, pp. 2175-2187, August 2010.
  	
  	\item[] A. Yahaghi, C. Hafner, \textbf{A. Fallahi}, J. Smajic, and B. Cranganu-Cretu, ``Efficient Algorithms for the Optimization of Shielding Devices for Eddy-Currents", \href{http://iospress.metapress.com/content/w33112q7l7782m42/}{\emph{International Journal of Applied Electromagnetics and Mechanics}}, vol. 34, no. 3, pp. 141-154, November 2010.
  	
  	\item[] \textbf{A. Fallahi}, A. Yahaghi, H. Abiri, M. Shahabadi, C. Hafner, and R. Vahldieck, ``Analysis of Frequency Selective Surfaces on Periodic Substrates Using Entire Domain Basis Functions",
  	\href{http://www.ict.csiro.au/aps/}{\emph{IEEE Transactions on Antennas and Propagation}}, vol. 58, no. 3, pp. 876-886, December 2009.
  	
  	\item[] \textbf{A. Fallahi}, M. Mishrikey, C. Hafner, and R. Vahldieck, ``Analysis of Multilayer Frequency Selective Surfaces on Periodic and Anisotropic Substrates," \href{http://www.elsevier.com/locate/metmat}{\emph{Elsevier
  			Metamaterials}}, vol. 3, no. 2, pp. 63-74, October 2009.
  	
  	\item[] \textbf{A. Fallahi}, M. Mishrikey, C. Hafner, and R. Vahldieck, ``Efficient procedures for the optimization of frequency selective surfaces," \href{http://www.ict.csiro.au/aps/}{\emph{IEEE Transactions on Antennas and Propagation}}, vol. 56, no. 5, pp. 1340-1349, May 2008.
  	
  	\item[] \textbf{A. Fallahi}, M. Mishrikey, C. Hafner, R. Vahldieck, ``Radar Absorbers Based on Frequency Selective Surfaces on Perforated Substrates," \href{http://www.aspbs.com/ctn/}{\emph{Journal of Computational and Theoretical Nanoscience}} vol. 5, no. 4, pp. 704-710, April 2008.
  	
  	\item[] A. Enayati, K. Zamani, \textbf{A. Fallahi}, M. Shahabadi, ``Prediction and verification of negative refraction in a two-dimensional photonic crystal slab using a rigorous semi-analytical approach," \href{http://www.tandf.co.uk/journals/titles/09500340.asp}{\emph{Journal of Modern Optics}}, vol. 54, no. 12, pp. 1793-1802, August 2007.
  	
  	\item[] \textbf{A. Fallahi}, K. Zamani, A. Enayati, M. Shahabadi, ``Diffraction analysis of periodic structures using a transmission-line formulation: Principles and Applications," \href{http://www.aspbs.com/ctn/}{\emph{Journal of Computational and Theoretical Nanoscience}}, vol. 4, no. 3, pp. 649-666, May 2007.
  	
  \end{outerlist}
  
  \blankline
  
  \textbf{Conference Papers}
  
  \begin{outerlist}
  	\item[] D. Zhang, \textbf{A. Fallahi}, X. Wu, M. Fakhari, H. Cankaya, A.-L. Calendron, C. Zhou, W. Ronny Huang, D. Haynes, F. Lemery, F. Ahr, W. Qiao, N. H Matlis, F. X. K\"artner, ``THz-driven electron streak camera based on a multilayer structure'' In The European Conference on Lasers and Electro-Optics, M\"{u}nich, Germany, June 2017.
  	
  	\item[] M. Fakhari, \textbf{A. Fallahi}, F. X. K\"artner, ``A dielectric loaded higher order mode THz electron injector'' In 42nd International Conference on Infrared, Millimeter, and Terahertz waves (IRMMW-THz), Cacun, Mexico, August 2017.
  	
  	\item[] F. X. K\"artner, K. Ravi, S. W. Jolly, F. Ahr, D. Zhang, X. Wu, M. Fakhari, H. Cankaya, A.-L. Calendron, C. Zhou, F. Lemery, W. Qiao, R. W. Huang, S. Carbajo, D. N. Schimpf, A. R. Maier, M. Hemmer, L. Zapata, O. D. M\"{u}cke, G. Cirmi, \textbf{A. Fallahi}, N. H. Matlis, H. Ishizuki, T. Taira, ``Terahertz Accelerator Technology" in \emph{Nonliner Optics}, Waikoloa, United States, July 2017.
  	
  	\item[] M. Fakhari, R. Assmann, U. Dorda, \textbf{A. Fallahi}, K. Galaydych, F. X. K\"artner, B. Marchetti, N. Matlis, G. Vashchenko, T. Vinatier, A. Yahaghi, D. Zhang, and C. Zhou, ``Compact Electron Injectors Using Laser Driven THz Cavities" in \emph{7th International Particle Accelerator Conference} (IPAC'17), Copenhagen, Denmark, May 2017.
  	
  	\item[] C. Zhou, F. Ahr, R. Assmann, A.-L. Calendron, H. Cankaya, U. Dorda, M. Fakhari, \textbf{A. Fallahi}, K. Galaydych, F. X. K\"artner, B. Marchetti, N. Matlis, W. Qiao, G. Vashchenko, T. Vinatier, X. Wu, and D. Zhang, ``Electron Acceleration With a Ultrafast Gun Driven by Single-Cycle Terahertz Pulses" in \emph{7th International Particle Accelerator Conference} (IPAC'17), Copenhagen, Denmark, May 2017.
  	
  	\item[] U. Dorda, R. Assmann, \textbf{A. Fallahi}, K. Galaydych, F. X. K\"artner, W. Kuropka, B. Marchetti, D. Marx, N. Matlis, F. Mayet, G. Vashchenko, T. Vinatier, P. A. Walker, and J. Zhu, ``The dedicated accelerator R\&D facility SINBAD at DESY" in \emph{7th International Particle Accelerator Conference} (IPAC'17), Copenhagen, Denmark, May 2017.
  	
  	\item[] D. Zhang, F. Ahr, A.-L. Calendron, H. Cankaya, M. Fakhari, \textbf{A. Fallahi}, W. Huang, F. X. K\"artner, F. Lemery, N. H. Matlis, W. Qiao, X. Wu, and C. Zhou, ``THz Driven Electron Acceleration with a Multilayer Structure" in \emph{7th International Particle Accelerator Conference} (IPAC'17), Copenhagen, Denmark, May 2017.
  	
  	\item[] W. Huang, A.-L. Calendron, H. Cankaya, \textbf{A. Fallahi}, K.-H. Hong, F. X. K\"artner, E. Nanni, K. Ravi, X. Wu, and D. Zhang, ``Demonstration of an All-Optically Driven Sub-keV THz Gun" in \emph{7th International Particle Accelerator Conference} (IPAC'17), Copenhagen, Denmark, May 2017.
  	
  	\item[] G. Vashchenko, R. Assmann, U. Dorda, M. Fakhari, \textbf{A. Fallahi}, K. Galaydych, F. X. K\"artner, B. Marchetti, N. H. Matlis, W. Qiao, T. Vinatier, and C. Zhou, ``Characterization of the Electron Beam from the Thz Driven Gun for AXSIS" in \emph{7th International Particle Accelerator Conference} (IPAC'17), Copenhagen, Denmark, May 2017.
  	
  	\item[] K. Ravi, A. Yahaghi, \textbf{A. Fallahi}, and F. X. K\"artner, ``Aperiodically poled structures for high efficiency broadband terahertz generation," in \emph{CLEO Science and Innovations}, San Jose, USA, May 2017.
  	
  	\item[] H. Ye, S. Trippel, M. Di Fraia, \textbf{A. Fallahi}, O. D M\"ucke, J. K\"upper, and F. X. K\"artner, ``Velocity Map Imaging for Photocathode Characterization," in \emph{CLEO Science and Innovations}, San Jose, USA, May 2017.
  	
  	\item[] W. R. Huang, \textbf{A. Fallahi}, X. Wu, E. Nanni, H. Cankaya, A.-L. Calendron, D. Zhang, K. Ravi, K.-H. Hong, and F. X. K\"artner, ``Sub-keV Electron Gun Driven by Ultrafast THz Pulses" In 41st International Conference on Infrared, Millimeter, and Terahertz waves (IRMMW-THz), September 2016.
  	
  	\item[] M. Fakhari, \textbf{A. Fallahi}, and F. X. Kärtner, ``Laser driven THz accelerators" In 41st International Conference on Infrared, Millimeter, and Terahertz waves (IRMMW-THz), September 2016.
  	
  	\item[] K. Khaliji, \textbf{A. Fallahi}, and Tony Low, ``Tunable polarization rotation using black phosphorous monolayers" In 41st International Conference on Infrared, Millimeter, and Terahertz waves (IRMMW-THz), September 2016.
  	
  	\item[] \textbf{A. Fallahi}, A. Yahaghi, and F. X. Kärtner, ``Full-wave numerical modeling of an infrared free electron laser using FDTD/PIC" In 41st International Conference on Infrared, Millimeter, and Terahertz waves (IRMMW-THz), September 2016.
  	
  	\item[] A. Yahaghi, K. Ravi, \textbf{A. Fallahi}, and F. X. K\"artner, ``High power broadband terahertz pulse generation using chirped aperiodically poled structures" In 41st International Conference on Infrared, Millimeter, and Terahertz waves (IRMMW-THz), September 2016.
  	
  	\item[] W. R. Huang, \textbf{A. Fallahi}, X. Wu, E. Nanni, H. Cankaya, A.-L. Calendron, D. Zhang, K. Ravi, K.-H. Hong, and F. X. K\"artner, ``Sub-keV Electron Gun Driven by Ultrafast THz Pulses" In Applications of Lasers for Sensing and Free Space Communications, Boston, Massachusetts United States, November 2016.
  	
  	\item[] W. R. Huang, \textbf{A. Fallahi}, X. Wu, E. Nanni, H. Cankaya, A.-L. Calendron, D. Zhang, K. Ravi, K.-H. Hong, and F. X. K\"artner, ``Terahertz-driven, sub-keV electron gun" In \emph{CLEO} Science and Innovations, pp. SM4L-1, optical Society of America, 2016.
  	
  	\item[] L. Wong, K.-H. Hong, S. Carbajo, \textbf{A. Fallahi}, M. Soljacic, J. Joannopoulos, F. X. K\"artner, and I. Kaminer, ``Monoenergetic Relativistic Electron Pulses by Laser-Driven Linear Acceleration in Free Space," In \emph{CLEO} QELS-Fundamental Science, pp. FTh3M-3, Optical Society of America, 2016.
  	
  	\item[] A. Yahaghi, \textbf{A. Fallahi}, and F. X. K\"artner, ``Free Electron Laser Simulation Tool Based on FDTD/PIC in the Lorentz Boosted Frame." in \emph{7th International Particle Accelerator Conference} (IPAC'16), Busan, Korea, May 8-13, 2016.
  	
  	\item[] \textbf{A. Fallahi}, M. Fakhari, F. X. K\"artner, and A. Yahaghi, ``Ultrafast Electron Guns for the Efficient Acceleration using Single-Cycle THz Pulses." in \emph{7th International Particle Accelerator Conference} (IPAC'16), Busan, Korea, May 8-13, 2016.
  	
  	\item[] M. Tamagnone, \textbf{A. Fallahi}, J. Perruisseau-Carrier, ``Theoretical Limits of Graphene Terahertz Non-Reciprocal Devices," in \emph{9th European Conference on Antennas and Propagation}, Lisbon, Portugal, April, 12-17, 2015.
  	
  	\item[] A. Enayati, and \textbf{A. Fallahi}, ``Full-wave Modeling of Wedge Absorbers," in \emph{7th Antenna Test and Measurement Society Conference}, Bangalore, India, March 2014.
  	
  	\item[] M. Tamagnone, \textbf{A. Fallahi}, J. Perruisseau-Carrier, ``Theoretical upper bounds on the performance of graphene reconfigurable devices," in \emph{8th European Conference on Antennas and Propagation}, Hague, Netherland, April 2014.
  	
  	\item[] M. E. Swanwick, C. D. Dong, P. D. Keathley, \textbf{A. Fallahi}, F. X. K\"artner, L. F. Vel\'asquez-Garc\'ia, ``Pitch scaling of ultrafast, optically-triggered silicon field emitter arrays", in \emph{27th International Vacuum Nanoelectronics Conference}, Engelberg, Switzerland, July 2014.
  	
  	\item[] R. G. Hobbs, Y. Yang,, P. D. Keathley, E. De Leo, W. S. Graves, K. K. Berggren, \textbf{A. Fallahi}, and F. X. K\"artner, ``Ultrafast, surface plasmon-enhanced, au nanorod optical field electron emitter arrays", in \emph{27th International Vacuum Nanoelectronics Conference}, Engelberg, Switzerland, July 2014.
  	
  	\item[] H. Ye, J. S. Kienitz, S. Fang, S. Trippel, M. E Swanwick, P. D. Keathley, L. F. Vel\'asquez-Garc\'ia, G. Cirmi, G. M Rossi, \textbf{A. Fallahi}, O. D M\"ucke, J. K\"upper, F. X. K\"artner, ``Velocity Map Imaging of Electrons Strong-Field Photoemitted from Si-Nanotip Arrays," in \emph{19th International Conference on Ultrafast Phenomena}, Ukinawa, Japan, July 2014.
  	
  	\item[] W. S. Graves, K. K. Berggren, S. Carbajo, R. Hobbs, K.-H. Hong, W. R. Huang, F. X. K\"artner, P. D. Keathley, D. E. Moncton, E. A. Nanni, K. Ravi, M. Swanwick, L. F. Vel\'asquez-Garc\'ia, L. J. Wong, Y. Yang, L. Zapata, Y. Zhou, J. Bessuille, P. Brown, E. Ihloff, J. Derksen, \textbf{A. Fallahi}, F. Scheiba, X. Wu, D. Mihalcea, P. Piot, I. Viti, ``Compact XFEL Light Source," in \emph{35th International Free-Electron Laser Conference}, Manhattan, New York, USA, August 2013.
  	
  	\item[] P. D. Keathley, M. E. Swanwick, \textbf{A. Fallahi}, L. F. Vel\'asquez-Garc\'ia, and Franz X. K\"artner, ``Space Charge Effects in Strong-Field Emission From a Nanostructured Si Cathode," in \emph{CLEO 2014}, San Jose, California, June 2014.
  	
  	\item[] \textbf{A. Fallahi}, R. Hobbs, Y. Yang, P. D. Keathley, M. E. Swanwick, L. F. Vel\'asquez-Garc\'ia, F. X. K\"artner, ``Complete simulation of laser induced field emission from nanostructures using a DGTD, PIC and FEM code," in \emph{5th International Particle Accelerator Conference}, Dresden, Germany, June 2014.
  	
  	\item[] E. A. Nanni, W. S. Graves, K. H. Hong, W. R. Huang, K. Ravi, L. J. Wong, G. Moriena, \textbf{A. Fallahi}, D. Miller, and F. X. K\"artner, ``Linear electron acceleration in THz waveguides," in \emph{5th International Particle Accelerator Conference}, Dresden, Germany, June 2014.
  	
  	\item[] M. Tamagnone, \textbf{A. Fallahi}, J. Perruisseau-Carrier, ``Reconfigurable Graphene Metasurfaces: Determining and Approaching Theoretical Upper Bounds," in \emph{5th International Conference on Metamaterials, Photonic Crystals and Plasmonics}, Nanyang, Singapore, May 2014.
  	
  	\item[] M. Tamagnone, \textbf{A. Fallahi}, J. Perruisseau-Carrier, ``Theoretical upperbounds on the performance of graphene reconfigurable devices," in \emph{8th European Conference on Antennas and Propagation}, Hague, The Netherlands, April 2014.
  	
  	\item[] \textbf{A. Fallahi}, L. J. Wong, A. Sell, F. X. K\"artner, ``Optimization of Dielectric Loaded Metal Waveguides for Acceleration of Electron Bunches Using Short THz Pulses," in \emph{35th International Free-Electron Laser Conference}, Manhattan, New York, USA, August 2013.
  	
  	\item[] \textbf{A. Fallahi}, J. Perruisseau-Carrier, ``Faraday Rotation in Micro- and Nano-patterned Graphene Metasurfaces," in \emph{2013 IEEE International Symposium on Antennas and Propagation and USNC-URSI National Radio Science Meeting}, Orlando, Florida, USA, July 2013.
  	
  	\item[] \textbf{A. Fallahi}, J. Perruisseau-Carrier, ``Electromagnetic Properties of Graphene Metasurfaces and Applications," in \emph{7th European Conference on Antennas and Propagation (EUCAP 2013) - Convened Sessions - ANTENNAS AND RELATED TOPICS}, Gothenburg, Sweden, April 2013.
  	
  	\item[] J. Perruisseau-Carrier, \textbf{A. Fallahi}, E. Carrasco, ``Periodic and Quasi-Periodic Graphene-Based Reconfigurable Surfaces," in \emph{Proc. URSI International Symposium on Electromagnetic Theory - EMTS 2013}, Hiroshima, Japan, May 2013.
  	
  	\item[] \textbf{A. Fallahi}, and C. Hafner, ``Numerical Analysis of Semi-infinite Frequency Selective Surfaces," in \emph{Proc. Antennas and Propagation Society International Symposium, 2010. AP-S 2010. IEEE}, Toronto, Canada, July 2010.
  	
  	\item[] M. Mishrikey, \textbf{A. Fallahi}, Ch. Hafner, R. Vahldieck, L. Braginsky, V. Shklover, ``Scattering analysis of graded porous metamaterials using effective permittivity functions," \emph{2nd Intl. Congress on Advanced Electromagnetic Materials in Microwaves and Optics} Louvain-la-Neuve France, September 2008.
  	
  	\item[] \textbf{A. Fallahi}, C. Hafner, and R. Vahldieck, ``Calculation of the dispersion diagram for planar electromagnetic bandgap structures," in \emph{Proc. Antennas and Propagation Society International Symposium, 2008. AP-S 2008. IEEE}, California, USA, July 2008, pp. 1-4.
  	
  	\item[] M. Mishrikey, \textbf{A. Fallahi}, C. Hafner, and R. Vahldieck, ``Improved performance of thin film broadband antireflective coatings," in \emph{Proc. Optomechatronic Micro/Nano Devices and Components III} Lausanne, Switzerland, October 2007, vol. 6717, pp. 671702.
  	
  	\item[] \textbf{A. Fallahi}, M. Mishrikey, C. Hafner, R. Vahldieck, ``Analysis and optimization of frequency selective surfaces with inhomogeneous, periodic substrates," in \emph{Proc. Optomechatronic Micro/Nano Devices and Components III} Lausanne, Switzerland, October 2007, vol. 6717, pp. 67170N.
  	
  	\item[] \textbf{A. Fallahi}, C. Hafner, R. Vahldieck, ``MoM/RCWA analysis of frequency selective surfaces with inhomogeneous, periodic substrates," in \emph{Proc. 18th International Zurich Symposium on Electromagnetic Compatibility}, Munich, Germany, September 2007, pp. 309-312.
  	
  	\item[] \textbf{A. Fallahi}, J. Rashed-Mohassel, ``Dyadic Green Function for the Step Discontinuity in a Coaxial Cable," in \emph{Proc. International Conference on Mathematical Methods in Electromagnetic Theory}, Kharkiv, Ukraine, June 2006, pp. 402-404.
  	
  	\item[] \textbf{A. Fallahi}, K. Zamani, A. Enayati, M. Shahabadi, ``Application of Shank's transformation to the convergence improvement in the Fourier analysis of photonic crystals," in \emph{Proc. International Conference on Mathematical Methods in Electromagnetic Theory}, Kharkiv, Ukraine, June 2006, pp. 430-432.
  	
  	\item[] Z. Ghatan, \textbf{A. Fallahi}, B. Makki, M. Shahabadi, C. Lucas, ``A Novel 2D Genetic Algorithm for Band Gap Optimization of Two-Dimensional Photonic Crystals", in \emph{Proc. of IEEE Congress on Evolutionary Computation, CEC 2006}, Vancouver, Canada, July 2006, pp. 3231-3235.
  	
  	\item[] \textbf{A. Fallahi}, S. Fathololoomi, ``The Ways of Solving Optical Circuits", in Proc. \emph{4th Iranian Student Conference of Electrical Engineering ISCEE 2001}, Tehran, Iran, September 2001.
  	
  	\item[] \textbf{A. Fallahi}, A. Ghoreyshi, ``An introduction to Single Electronics", in Proc. \emph{3rd Iranian Student Conference of Electrical Engineering ISCEE 2000}, Tehran, Iran, November 2000.
  	
  \end{outerlist}
  
  \cleardoublepage                           
%
  \ofoot[]{}
  \ihead{CURRICULUM VITAE}
  
  \chapter*{Curriculum Vitae}
  \addcontentsline{toc}{chapter}{Curriculum Vitae}
  
  \renewcommand{\arraystretch}{1.3}
  \newlength{\CVindent}
  \newlength{\CVboxwidth}
  \newlength{\CVtextboxwidth}
  \setlength{\CVindent}{0.0cm} 
  \setlength{\CVboxwidth}{2.6cm}
  \setlength{\CVtextboxwidth}{\textwidth}
  \addtolength{\CVtextboxwidth}{-\CVindent}
  \addtolength{\CVtextboxwidth}{-\CVboxwidth}
  
  \vspace*{-0.5cm}
  \setlength{\tabcolsep}{0cm}

  \minisec{Personal data}
  \begin{flushleft}
  	\parbox[c][][c]{7cm}{\begin{tabular}{lll}
  			\hspace*{\CVindent} & \parbox{\CVboxwidth}{Name:} & Arya Fallahi\\
  			& Date of birth: & February 11, 1982\\
  			& E-mail: & \href{mailto:arya.fallahi@cfel.de}{arya.fallahi@cfel.de}
  	\end{tabular}}
  	
  	\vspace*{0.2cm} \minisec{Education}
  	\begin{tabular}{llp{\CVtextboxwidth}}
  		\hspace*{\CVindent} & \parbox{\CVboxwidth}{2012 -- 2018} &
  		\textbf{Research Scientist}
  		\newline Ultrafast Optics and X-ray Sources Division, DESY-Center for Free-Electron Laser Science (CFEL) \\
  		& 2010 -- 2012 & \textbf{Post-doctoral Scientist}
  		\newline Andvanced Modelling and Accelerator Simulations Group, Paul Scherrer Institut \\
  		& 2006 -- 2010 &
  		\textbf{PhD Student and Research Assistant}
  		\newline Labratoary for Electromagnetic Fields and Microwave
  		Electronics, ETH Zürich, Switzerland \\
  		& 2004 -- 2006 & \textbf{M.S. Studies in Fields and Waves}
  		\newline Department of Electrical and Computer Engineering,
  		University of Tehran, Iran \\
  		& 1999 -- 2004 & \textbf{B.S. Studies in Electrical Engineering}
  		\newline Department of Electrical Engineering,
  		Sharif University of Technology, Iran \\
  		& 1999 -- 2004 & \textbf{B.S. Studies in Applied Physics}
  		\newline Department of Physics, Sharif University of Technology,
  		Iran \\
  		& 1995 -- 1999 & \textbf{High School Studies}
  		\newline Prof. Hesabi High School, Tehran, Iran
  	\end{tabular}
  	
  \end{flushleft}
  

  \end{document}